%% file: pppiminus.tex
\documentclass[12pt]{cernart}
\usepackage{graphicx}
\usepackage{amsmath}
\usepackage{amssymb}
\usepackage{cite}
\usepackage{times}
\usepackage{multirow}
\usepackage{enumitem}
\usepackage{xcolor}
\usepackage{blindtext,rotating}
\usepackage{tocbasic}
\usepackage[toc,page]{appendix}
\usepackage[breaklinks=true]{hyperref}
\definecolor{darkgreen}{rgb}{0,0.5,0}
\hypersetup{
        colorlinks=true,
        linkcolor=blue,
        citecolor=darkgreen,
}

\makeatletter
\renewcommand\tableofcontents{%
  \null\hfill\textbf{\Large\contentsname}\hfill\null\par
  \vspace{5mm}
  \listoftoc*{toc}
}
\DeclareTOCStyleEntries[
]{tocline}{chapter,section,subsection,subsubsection,paragraph,subparagraph}
\DeclareTOCStyleEntry[entryformat=\textbf]{tocline}{section}
\DeclareTOCStyleEntry[entryformat=\textit]{tocline}{subsubsection}
\makeatother

\begin{document}
\setcounter{topnumber}{3}
\renewcommand{\topfraction}{0.999}
\renewcommand{\bottomfraction}{0.99}
\renewcommand{\textfraction}{0.0}
\setcounter{totalnumber}{6}
\renewcommand{\thefootnote}{\alph{footnote}}
\newcommand{\figw}{22.5cm}
\newcommand{\capw}{23cm}
\newcommand{\figwh}{11.25cm}
\newcommand{\capwh}{11cm}
\newcommand{\capsh}{5mm}
\newcommand{\rotAngle}{270}

\begin{titlepage}
\date{}

\title{\large{A comprehensive study of the inclusive production of negative pions in p+p collisions for interaction energies from 3~GeV to 13~TeV covering the non-perturbative sector of the Strong Interaction}}

\vspace{2mm}

	\author{H.~G.~Fischer \Iref{cern}$^,$\thanks{email: fischerhansgerhard@gmail.com}, M.~Makariev\Iref{inrne}$^,$\thanks{also at Faculty of Physics, Sofia University, Sofia, Bulgaria}, D.~Varga\Iref{wigner}, S.~Wenig\Iref{cern}$^,$\thanks{email: Siegfried.Wenig@cern.ch}}
\vspace*{2mm}

\noindent
\Institute{wigner}{Wigner Research Centre for Physics, Budapest, Hungary}
\Institute{cern}{CERN, Geneva, Switzerland}
\Institute{inrne}{Institute for Nuclear Research and Nuclear Energy, BAS, Sofia, Bulgaria}

\vspace{10mm}

\begin{abstract}
Over the past 60 years a rich sample of experimental results concerning the inclusive production of $\pi^-$ mesons has been obtained spanning a range from about 3~GeV to 13~TeV in interaction energy. This paper attempts a model-independent overview of these results with the aim at obtaining an internally consistent data description on a dense grid over the three inclusive variables transverse momentum, rapidity or Feynman $x_F$ and interaction energy. The study concentrates on the non-perturbative sector of the strong interaction by limiting the transverse momenta to $p_T <$~1.3~GeV/c. The three-dimensional interpolation which is mandatory and necessary for this aim is shown to provide a controlled systematic precision of better than 5\%. This accuracy allows for a critical inspection of each of the 40 experiments concerned in turn. It also allows precision tests of some of the physics concepts developed around inclusive processes like energy scaling, "thermal" production and the evolution of transverse momenta.
\end{abstract}
\vspace{5mm}

\clearpage
\end{titlepage}

\tableofcontents

\newpage

%
%
\section{Introduction}
\vspace{3mm}
\label{sec:intro}

Ever since the discovery, in rapid succession, of $\pi$ mesons, strange particles and hadron resonances in the 1950's and early 1960's, elementary particle production in hadronic interactions has been studied in an impressive series of experiments. These studies have closely followed the fast progress of available interaction energies due to the evolution of accelerator technology as well as particle detection and identification methods. The measurement of production cross sections has in fact been and still is a standard part of experimental work at any new accelerator facility coming into operation.

On a theoretical level, this evolution has been followed by an equally rapid development leading to the Standard Model of particle physics which still holds uncontested to date.

Within the framework of this model, hadronic collisions constitute an important part of the vast sector of the Strong Interaction described by Quantum Chromodynamics (QCD) and characterized by the strong coupling constant $\alpha_s(\mu)$ where $\mu$ is an energy scale parameter. The strong increase of $\alpha_s$ with decreasing $\mu$ leads to a breakdown of perturbation theory and a split of the description of the strong interaction into a perturbative and a non-perturbative or "soft" sector. The transition between these sectors is rather ill-defined. It depends on several parameters and the confidence in applying higher order perturbative calculations in $\alpha_s(\mu)$.

In view of the absence of a priory predictions in the soft sector a number of production "models" have been promoted which either depend on the application of parton interaction and fragmentation ideas -- in turn depending on data obtained from leptonic interactions -- or on rather general assumptions concerning the presence of statistical or thermal processes.

This paper will concentrate on the inclusive production of negative pions in the non-perturbative area by limiting the transverse momentum to less than 1.3~GeV/c which is well outside the next-to-next-to-leading order perturbative calculations. In addition, the approach will be exclusively based on experimental data in an effort to obtain an internally consistent description covering nearly the full production phase space with a dense coverage in the relevant kinematic variables and aiming at a level of about 5\% absolute precision. For this aim, all available experimental results will be scrutinized from interaction energy close to threshold up to LHC energies.

The paper is arranged as follows:

In a first part the about 4500 existing double differential cross sections from 36 experiments at interaction energies between 3 and 63~GeV are used to establish a three-dimensional interpolation scheme in rapidity, transverse momentum and interaction energy. This covers the complete production phase space (Sects.~\ref{sec:exp_sit} to \ref{sec:countdata}) with the exception of transverse momentum which is limited to $p_T<$~1.3~GeV/c in order to remain in the non-perturbative sector. 17 of the 36 experiments, mostly using bubble chambers, yield internally consistent results without additional corrections. Most of the remaining experiments, essentially using spectrometer detectors, may be brought into agreement with these reference data by single overall normalization factors. A detailed statistical analysis of the point-by-point deviations of the complete data sets from the global interpolation shows systematic offsets of less than 5\%.

This unprecedented precision allows for the elimination of complete data sets (Sect.~\ref{sec:na61}) or parts of data (Sects.~\ref{sec:allaby19} and \ref{sec:allaby24}) which fall far outside the global interpolation.

Data produced at  high-energy proton colliders from RHIC up to LHC energies are discussed in Sect.~\ref{sec:colliders}. Here the very limited phase space coverage allows for the extension of the energy scale and the comparison to the lower energy data only at central or very forward rapidities.

In a second part the high-precision global interpolation is used to establish final state pion distributions in various co-ordinates, first in longitudinal momentum (Sect.~\ref{sec:scaling}) and then in transverse momentum (Sect.~\ref{sec:transverse}). Integration over $p_T$ yields single differential distributions in different longitudinal momentum variables and finally the total $\pi^-$ yields (Sect.~\ref{sec:integrated}).

These distributions are used for a critical review of different attempts to bring the complex phenomenology into simple form using certain hypotheses. This concerns, in the longitudinal direction, the claims at energy scaling in forward direction as opposed to a central, non-scaling production mechanism. In transverse direction it is the hypothesis of a global, uniform and mass and energy-independent distribution in transverse mass as specified in the Statistical Bootstrap or "thermal" Model.

As none of these hypotheses, with the exception of Limiting Fragmentation in the extreme forward and backward regions, stands up to the experimental reality once a certain precision over the full phase space has been reached, an approach beyond the purely inclusive level is attempted by considering hadronic resonance decay as the source of the observed inclusive phenomena.

This is discussed in a third section of the paper. In a first step a well-measured baryonic resonance is used to establish the salient features of resonance decay as it feeds-down into final state hadrons (Sects.~\ref{sec:resonances} and \ref{sec:res_mass_spec}). In a second step several additional resonances are considered in their influence on measured quantities like hadronic "temperatures" and mean transverse momentum (Sects.~\ref{sec:res_invslope} and \ref{sec:meanpt}). In a third step an ensemble of 13 measured baryonic and mesonic resonances is invoked to show how all important features of the inclusive level emerge from their decay (Sect.~\ref{sec:beyond_inclusive}).

The paper closes with a detailed summary (Sect.~\ref{sec:conclusion}) and an outlook concerning basic experimental conditions to further the understanding of the non-perturbative sector of QCD (Sect.~\ref{sec:future_studies}).
%
%
\section{Inclusive physics in the non-perturbative sector}
\vspace{3mm}
\label{sec:inc}

%
%
\subsection{Definition of inclusive cross sections}
\vspace{3mm}
\label{sec:inc_def}

The Lorentz invariant production cross section is defined in the most general fashion as

\begin{equation}
     d\sigma = \frac{|M|^2}{F}dV
\end{equation}
where $|M|^2$ is the invariant matrix element which is incalculable in the non-perturbative sector, $F$ is M{\o}ller's invariant flux factor and $dV$ the invariant phase space element.

For a general n-body final state with unpolarized beam and target (summing over helicities) this can be written

\begin{equation}
	d\sigma = H(\vec{p}_1 .... \vec{p}_n, s) \prod_{n} \frac{d^3p_i}{E_i}
\end{equation}
where $H$ is a function of 3$n$-5 variables, i.e. 3$n$ momentum components minus 4 constraints from energy-momentum conservation minus one free angle of rotation around the beam axis. The total centre of mass system (cms) energy squared,

\begin{equation}
     s = 2m_p^2 + 2m_p\sqrt{m_p^2 + p_{\textrm{beam}}^2}
\end{equation}
with $m_p$ proton mass and $p_{\textrm{beam}}$ beam momentum in fixed target mode, presents an important additional parameter.

Accordingly for a restricted $m$-body inclusive cross section looking at $m$ particles of type $c$ only in the final state, one can define

\begin{equation}
    d\sigma_c = G(\vec{p}_1 ... \vec{p}_m, s) \prod_{m} \frac{d^3p_i}{E_i}
\end{equation}
$G$ is now a function of 3$m$-1 variables (no energy-momentum conservation for sub-group $c$). This reduces, for one-body inclusive reactions of the type

\begin{equation}
    a + b \rightarrow c + \textrm{anything}
\end{equation}
to

\begin{equation}
   \label{eq:incl}
   d\sigma_c = f(p,s) \frac{d^3p}{E}
\end{equation}
with 2 variables and the energy parameter $s$. The function $f(p,s)$ is also called "structure function".

This dramatic reduction to the simplest hyper-surface of the complex multidimensional phase space poses of course the question whether any relevant physics results can be drawn from the experimental study of single particle inclusive cross sections. Indeed this field seems to have been abandoned at least in the non-perturbative sector by the theory community. On the other hand there is active interest in the fields of neutrino and astroparticle physics where experimental results are important and mandatory for the enumeration of background contributions to the research of otherwise unconnected phenomena. Nevertheless, this paper will show that if an internally consistent data sample with a wide phase space coverage and tight systematic uncertainties can be provided, a number of important constraints concerning soft hadronic interactions may be obtained.

%
%
\subsection{The problem of inclusiveness}
\vspace{3mm}
\label{sec:inc_ness}

When regarding the available experimental data it becomes apparent that a problem is posed by the presence of weak decays leading to negative pion production. In a first set of experiments, hereafter called "reference" experiments, which have access to the complete or at least partial detection of decay vertices, the decay pions from $\Lambda$/$\Sigma$ and K$^0$ decays, the so-called "feeddown" pions, are eliminated from the inclusive sample. In a second set of measurements, mostly falling into the realm of "spectrometer" experiments, this subtraction is not performed. As it will be shown below this will lead, in certain regions of phase space, to up to 40\% differences in the differential cross sections and up to about 12--15\% differences in the integrated yields. The procedure of feeddown subtraction which touches exclusively the sector of strange particle production, is in itself completely arbitrary as on-vertex, strong decays of strange resonances like $\Sigma^{\ast}$ and K$^{\ast}$ are kept by definition in the inclusive sample. In addition, for certain applications in long-baseline or atmospheric neutrino physics, even the contribution of K$^0_L$ decays should be included.

%
%
\subsection{Variables}
\vspace{3mm}
\label{sec:inc_var}

Given the simple structure of the phase space element $d^3p$ contained in (\ref{eq:incl}), characterized by two parameters only (the azimuthal angle being integrated over) it is surprising to see the large variety of variables used in describing different experimental data. No agreement on a single set of coordinates has ever been achieved, not to speak about a common choice of binning in order to facilitate the comparison of different results.

The most natural choice of longitudinal $p_L$ and transverse $p_T$ momentum,

\begin{equation}
	\label{eq:pLpT}
	\begin{split}
		d^3p &= 2\pi dp_L dp_T \\
       		 &= \pi dp_L dp_T^2
     \end{split}
\end{equation}
as it was indicated early on by the evidence of "longitudinal phase space" with a strong, almost exponential cutoff in $p_T$ and a wide spread in longitudinal momentum characterized by a power-law like behaviour depending on the particle type, has been mostly used in early work.

The choice of total momentum and polar angle,

\begin{equation}
	\begin{split}
		d^3p &= 2\pi p^2 dpd(\cos{\Theta}) \\
             &= 2\pi p^2 \sin{\Theta}dp d\Theta
     \end{split}
\end{equation}
has been common to spectrometer experiments performed at fixed laboratory angle. Both the above definitions depend of course on the choice of the overall laboratory and cm systems as well as eventually target and projectile frames.

This problem is avoided by the choice of rapidity $y$ and transverse momentum,

\begin{equation}
	d^3p = \pi E dy dp_T^2
\end{equation}
with rapidity

\begin{equation}
	y = \frac{1}{2} \ln\frac{E + p_L}{E - p_L}
\end{equation}
and consequently

\begin{equation}
	 \frac{d^3p}{E} = \pi dy dp_T^2
\end{equation}

Constant rapidity corresponds, for light particles even in the soft sector, approximately to a constant polar angle and the invariant cross sections in different Lorentz frames are connected by a shift in rapidity.

%
%
\subsection{Dependence on interaction energy}
\vspace{3mm}
\label{sec:dep_ener}

The above definitions of different phase space coordinates are not related to the interaction energy $\sqrt{s}$. In fact the available range of longitudinal momentum increases roughly with $\sqrt{s}$ whereas the $y$ range grows logarithmically with $s$. The aim at comparing cross sections at different $\sqrt{s}$ has therefore lead to definitions of phase space variables renormalizing, if only approximately, to the available energy scale. From a physics point of view this has been driven by the concept of "scaling" which would postulate the independence of invariant cross sections on interaction energy over parts or all of the available phase space. Following a conjecture by Feynman \cite{feynman} one re-defines the  longitudinal momentum $p_L$ by

\begin{equation}
	\label{eq:xf_def}
  	x_F = \frac{2p_L}{\sqrt{s}}
\end{equation}
in the cm system and

\begin{equation}
  	d^3p = \pi \frac{\sqrt{s}}{2} dx_{F} dp_T^2
\end{equation}

This definition does not take into account energy-momentum conservation in the final state and is sometimes replaced by

\begin{equation}
	\label{eq:xfprime}
	x_F' = \frac{p_L}{p_L^{\textrm{max}}}
\end{equation}
where ${p_L^{\textrm{max}}}$ depends on the interaction energy and ensures basic constraints like charge and baryon number conservation which become important at interaction energies below about 10~GeV, see Sect.~\ref{sec:scaling_xf} below.

\begin{figure}[h]
 \begin{center}
   \includegraphics[width=13.5cm] {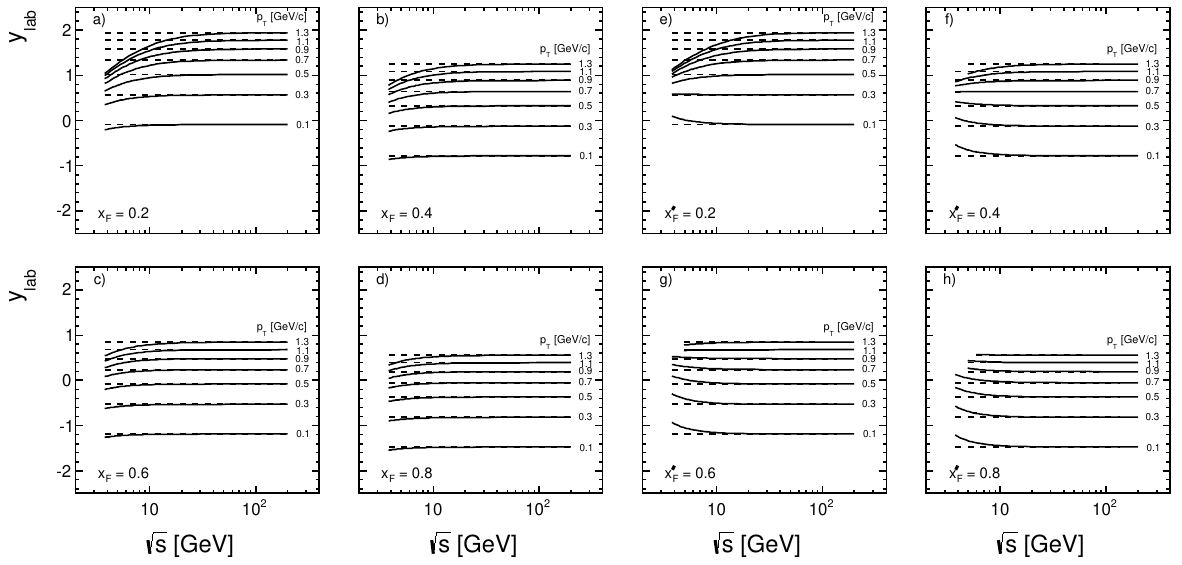} 
	\caption{Correlation between the $x_F$/$x_F'$ and $y_{\textrm{lab}}$ as a function of $\sqrt{s}$ for different values of $x_F$/$x_F'$ and $p_T$}
  \label{fig:xycor}
 \end{center}
\end{figure}

Also in rapidity space a renormalization has been proposed in order to allow for a convenient way to compare the forward/backward part of the rapidity distributions by taking out the growth of their width with $\sqrt{s}$. Here one defines as $s$-dependent beam rapidity in the cm system

\begin{equation}
	\begin{split}
		y_{\textrm{beam}} &= \frac{1}{2} \ln\frac{E_{\textrm{beam}} + p_{\textrm{beam}}}{E_{\textrm{beam}} - p_{\textrm{beam}}}  \\
                          &= \frac{1}{2} \ln\frac{1 + \beta_{\textrm{cm}}}{1 - \beta_{\textrm{cm}}}                                 \\
                          &\approx \ln\frac{\sqrt{s}}{m_p} \qquad\qquad \textrm{for}\quad \sqrt{s} \geq \textrm{20~GeV}
     \end{split}
 \label{eq:ybeam}
\end{equation}
using the proton mass $m_p$. The rapidity scale is here replaced by the shifted quantity

\begin{equation}
   y_{\textrm{lab}} = y_{\textrm{beam}} - y
   \label{eq:ylab}
\end{equation}
which suitably overlaps the forward part of the rapidity distributions for different interaction energies, leaving however a logarithmic upwards shift of the $y_{\textrm{lab}}$ values at central rapidity with $s$. In fact there is equivalence between the forward part of the $x_F$ scale and the $y_{\textrm{lab}}$ scale for large rapidities as shown in Fig.~\ref{fig:xycor}.

The area of equivalence at $x_F {\scriptstyle\gtrsim}$~0.2 is usually called "fragmentation region" in contrast to a "central production region" around $x_F$ = $y$ = 0. This juxtaposition of two different particle production mechanisms has been suggested by the approximate $s$-independence of cross sections in forward direction as opposed to the increase of yields in the central area as first observed at the CERN ISR \cite{cernisr1,cernisr2,cernisr3}. As will be shown below this assumption is arbitrary: in fact particle production may be split into two independent contributions from target and projectile ("factorization") which are governed by resonance formation and decay, resulting in a well defined overlap region which for pions has a width of about 0.05 units of $x_F$ \cite{pc_discuss}. The increase of yields in the central region has its origin, at ISR energy and above, in the contributions from strangeness (more generally, "heavy flavour") production. It depends in a rather complex way on $p_T$ and $y$/$x_F$ as well as on the particle type.

\begin{figure}[h]
 \begin{center}
   \includegraphics[width=8cm] {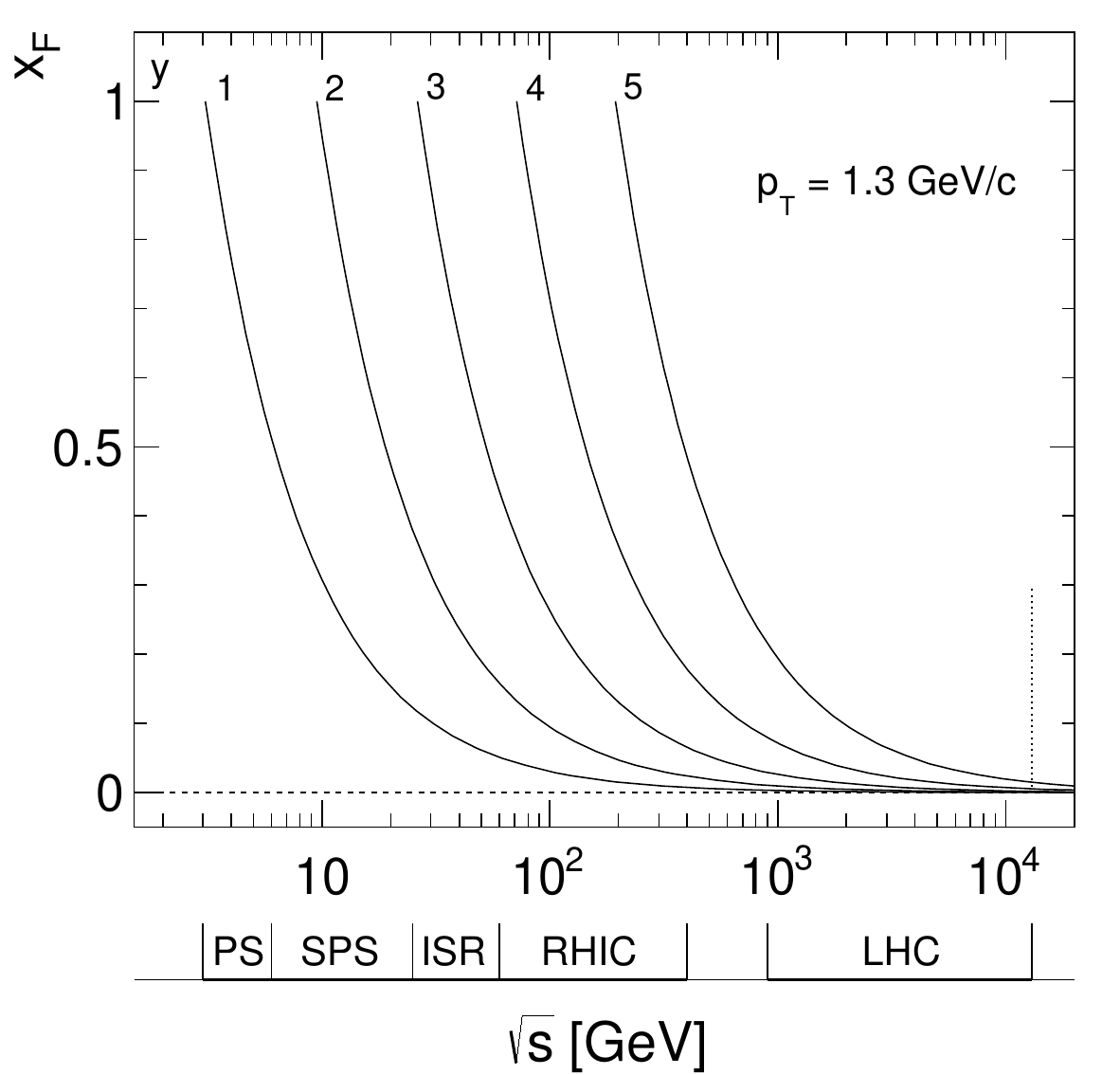} 
	\caption{Accessible $x_F$ range as a function of  $\sqrt{s}$  for different $y$ values, at $p_T$~=~1.3~GeV/c}
  \label{fig:xdycor}
 \end{center}
\end{figure}

Nevertheless, central production has been and still is regarded as being of special interest, in particular also in heavy ion interactions ("hot" central as opposed to "cool" forward regions). This is especially true for the experimental situation at the high energy colliders where by construction the "fragmentation" regions are inaccessible to experiment. This is demonstrated in Fig.~\ref{fig:xdycor} where the $x_F$ range is plotted as a function of the interaction energy  $\sqrt{s}$  for different rapidities, for the upper limit of $p_T$ at 1.3~GeV/c used in this paper.

In fact the ISR has been and will be in the foreseeable future the only proton collider allowing the experimental study of the full phase space in $x_F$ for soft interactions, whereas at the LHC the $x_F$ range reduces, for the eventually accessible rapidity range of about 5 units, to almost a delta function around $x_F$~=~0 with a coverage of less than 1\% of the total phase space which does not allow the separation of target and projectile contributions even for the asymmetric p+A interactions.

%
%
\subsection{Energy scaling}
\vspace{3mm}
\label{sec:energy_scaling}

In soft hadronic production, the concept of "scaling" has been proposed in the late 1960's following the experimental finding that the invariant cross sections, (\ref{eq:incl}), which should a priori depend on the particle momentum and the interaction energy separately, seemed to depend only on the renormalized "scaling" variable $x_F$, defined in (\ref{eq:xf_def}).

This result was relying on a rather small range of beam momenta between about 12 and 30~GeV/c together with the fact that over this range the total inelastic cross section $\sigma_{\textrm{inel}}$ only varies by a few percent. Nevertheless a connection with scaling in deep inelastic e+p scattering was immediately established leading to a number of predictions concerning the parton content of the final state hadrons ("counting rules") and the direct comparison with the partonic structure functions ("recombination models").

A scrutiny of all available data over a range of $\sqrt{s}$ from $\sim$3~GeV up to LHC energies on a level of precision of better than 10\%, as it is attempted here, reveals, however, a very intricate pattern of dependences on all three variables $p_L$, $p_T$ and $s$ which puts into doubt the very idea of energy scaling, not to mention assumed cross connections into the leptonic sector.

A major problem is here posed by the fact that the total inelastic cross section increases, over the $s$ range indicated above, by almost a factor of three. Which quantity should be used in comparison: the invariant cross section $f$ or the renormalized quantity $f/\sigma_{\textrm{inel}}$? The latter definition would assume that particle production happens by the same mechanism over the full increasing surface of the colliding nucleons. Actual estimations assume, however, that there should be a constant "central core" and an increasing rim area \cite{increase_rim1,increase_rim2}. What is the role of increasing heavy flavour production and where should it manifest itself? Would certain regions of phase space show different $s$-dependences?

In this paper the renormalized cross sections $f/\sigma_{\textrm{inel}}$ will be used for most $s$-dependent quantities. As the upper $s$-limit of full phase space coverage is given by the highest ISR energy at $\sqrt{s}$~=~63~GeV, the according increase of $\sigma_{\textrm{inel}}$ of 29\% might allow for a test of the scaling behaviour in different regions of phase space within the rather tight systematic error limits of this study.

%
%
\section{The experimental situation}
\vspace{3mm}
\label{sec:exp_sit}

\begin{figure}[h]
 \begin{center}
   \includegraphics[width=8cm] {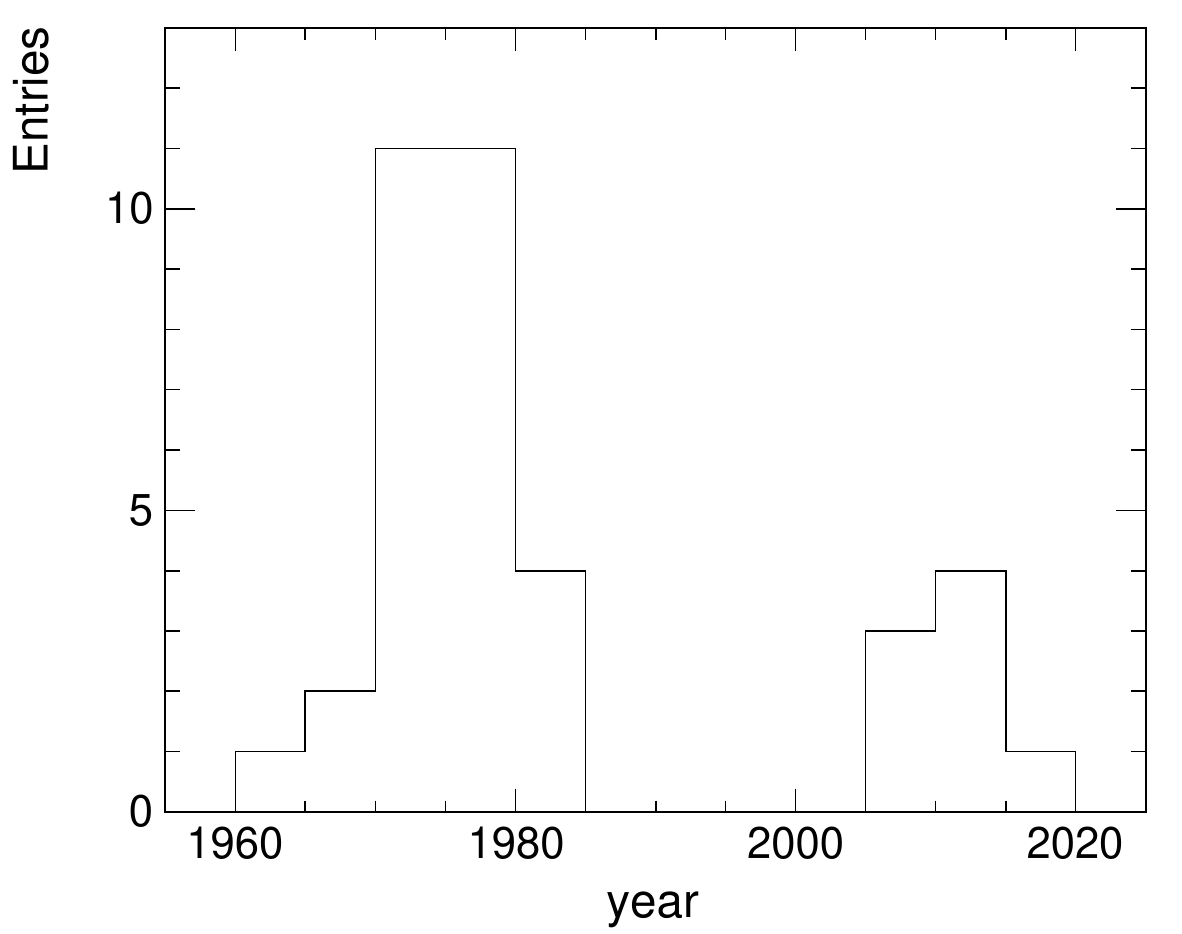} 
	\caption{Time distribution of published results on $\pi^-$ production in the range of $p_T <$~1.3~GeV/c}
  \label{fig:year}
 \end{center}
\end{figure}

A search for published results on double-differential $\pi^-$ cross sections in the region of non-perturbative QCD discussed here yields 36 experiments using a large variety of detector systems at virtually all accelerators coming into service after the late 1950's, with a range of interaction energies of 3~GeV~$< \sqrt{s} <$~13~TeV. A time distribution of the published data results in an interesting two-peak structure shown in Fig.~\ref{fig:year}.

A first peak around the mid-1970's is representative of a genuine interest in the general features of hadron production in the early days of particle physics, irrespective of a reliable theoretical background. The advent of QCD as part of the Standard Model in the 1970's and early 1980's quickly lead to the realization that the non-perturbative sector was not amenable to quantitative predictions which in turn reduced the experimental activity in this sector. This caused a gap of about two decades where practically no new measurements were undertaken.

A second, very recent peak around the first decade of this century has several contributions. The accessibility of high (RHIC) and very high (LHC) interaction energies revived interest in the more general features of particle production where the low-$p_T$ region may be seen as a by-product of the more general aim at "discovery potential". At the same time the increased importance of neutrino and astro-particle physics necessitates reference data of sufficient precision for the evaluation of hadronic background components, also and especially at high energies. And, surprisingly enough, some new efforts (NA49, NA61) at the CERN SPS have emerged with the aim at high precision measurements in the inclusive sector even at lower interaction energy. None of these efforts are however aimed at a more precise understanding of the soft sector of the strong interaction itself which after all represents the overwhelming contribution to the total cross section.

This may be easily verified by looking at the list of references to the published data. Here, three main interests in inclusive data may be identified:

\begin{enumerate}[label=(\arabic*)]
  \item \label{item1} Reference data for Heavy Ion collisions and the connected claim of the discovery of the Quark-Gluon Plasma (QGP) as a new state of matter.
  \item \label{item2} Reference data for studies in astro-particle and neutrino physics.
  \item \label{item3} Reference data for the development of so-called "microscopic" models of particle production which are multi-parameter descriptions of a non-predictable reality.
\end{enumerate}

This paper is motivated by a different approach. It is felt that it would be about time to try and overview the wealth of available data from the multitude of experiments mentioned above with the aim at establishing a reliable data base, covering the available phase space with an absolute precision of about 5\%. It could be hoped that such a precision would allow for a new assessment of the underlying production process as far as its principle features are concerned. For this aim, and in view of the fact that every single experiment has its proper systematic uncertainties, it is evident that each data set has to be examined separately with respect to the overall ensemble. The systematic uncertainties, as will be shown below, being on the level of +-30\% and more, this seems to be an impossible task. Fortunately the situation is helped by the fact that a sub-set of data with decisively smaller error margins may be identified. This subset, hereafter named "reference experiments", is formed by the early bubble chamber experiments which span the region from 3.8 to 27~GeV in $\sqrt{s}$. Here the systematic errors, especially concerning the overall normalization, are on a percent level and similar if not identical for the different data. To this set may be added the data from the NA49 experiment \cite{pp_pion} which has been shown, for all identified types of charged particles, to comply with the bubble chamber data. These detectors benefit from a wide phase space coverage allowing for a simultaneous data collection over the full range of variables, thus further minimizing the systematic uncertainties. Due to the fact that the CERN ISR is the only -- and probably last -- collider giving access to full phase space, and also due to the fact that its extremely stable operation in unbunched (DC) mode allows for precision determination of absolute normalization, the four existing ISR experiments have been added to this list, see Tab.~\ref{tab:pp_pim_data}.

\begin{table}[h]
\renewcommand{\tabcolsep}{0.3pc}
\renewcommand{\arraystretch}{1.1}
\begin{center}
\begin{tabular}{lccccccc}
\hline
   Experiment    & beam mom. &  $\sqrt{s}$   & detector & det. size & accelerator &  N $_{events}$ & N$_{\pi^-}$  \\
                 &  [GeV/c]  &    [GeV]      &         &  &  &  &     \\ \hline
 Gellert \cite{gellert}                          &    6.6   &   3.78  & HBC      & 72 in & LBL       & 130 k &  34 k    \\
 Blobel \cite{blobel}                            &    12    &   4.93  & HBC      & 2 m   & CERN PS   & 175 k &  75 k   \\
 Smith \cite{smith,landolt}                      &    12.9  &   5.10  & HBC      & 80 in & BNL       & 16.3 k &  7.4 k   \\
 Smith \cite{smith,landolt}                      &    18    &   5.97  & HBC      & 80 in & BNL       & 22.7 k &  13 k   \\
 B{\o}ggild \cite{boggild,landolt}               &    19    &   6.12  & HBC      & 2 m   & CERN PS   & 9.7 k  &  3 k   \\
 Smith \cite{smith,landolt}                      &    21    &   6.42  & HBC      & 80 in & BNL       & 22.4 k &  14 k   \\
 Smith \cite{smith,landolt}                      &    24    &   6.84  & HBC      & 80 in & BNL       & 17.6 k &  12 k   \\
 Blobel \cite{blobel}                            &    24    &   6.84  & HBC      & 2 m   & CERN PS   & 100 k  &  65 k   \\
 Smith \cite{smith,landolt}                      &    28.4  &   7.42  & HBC      & 80 in & BNL       & 16.2 k &  12 k   \\
 Sims \cite{sims}                                &    28.5  &   7.43  & HBC      & 80 in & BNL       & 83 k &  12 k    \\
 Zabrodin \cite{zabrodin}                        &    32    &   7.86  & HBC  & Mirabelle & Serpukhov & 80 k &  6.6 k   \\
 Ammosov \cite{ammosov}                          &    69    &  11.5   & HBC  & Mirabelle & Serpukhov & 7.85 k &  9.6 k \\
 Bromberg \cite{bromberg}                        &   102    &  13.9   & HBC      & 30 in & FNAL      & 3 k &  2.7 k    \\
 NA49 \cite{pp_pion}                             &   158    &  17.3   & TPCs     & 13 m  & CERN SPS  & 4.8 M &  2.5 M   \\
 ISR \cite{albrow,alper,capi,guettler}           &   281    &  23     & spectrometer &   & CERN ISR  &  &     \\
 Bromberg \cite{bromberg}                        &   400    &  27.4   & HBC      & 30 in & FNAL      & 2.2 k &  3.1 k   \\
 ISR \cite{albrow,alper,capi,guettler}           &   511    &  31     & spectrometer &   & CERN ISR  &  &  \\
 ISR \cite{albrow,albrow45,alper,capi,guettler}  &  1078    &  45     & spectrometer &   & CERN ISR  &  &       \\
 ISR \cite{albrow,albrow53,alper,capi,guettler}  &  1496    &  53     & spectrometer &   & CERN ISR  &  &       \\
 ISR \cite{albrow,alper,capi,guettler}           &  2114    &  63     & spectrometer &   & CERN ISR  &  &  \\

\hline
\end{tabular}
\end{center}
\caption{Reference experiments}
\label{tab:pp_pim_data}
\end{table}

\begin{figure}[h]
 \begin{center}
   \includegraphics[width=9cm] {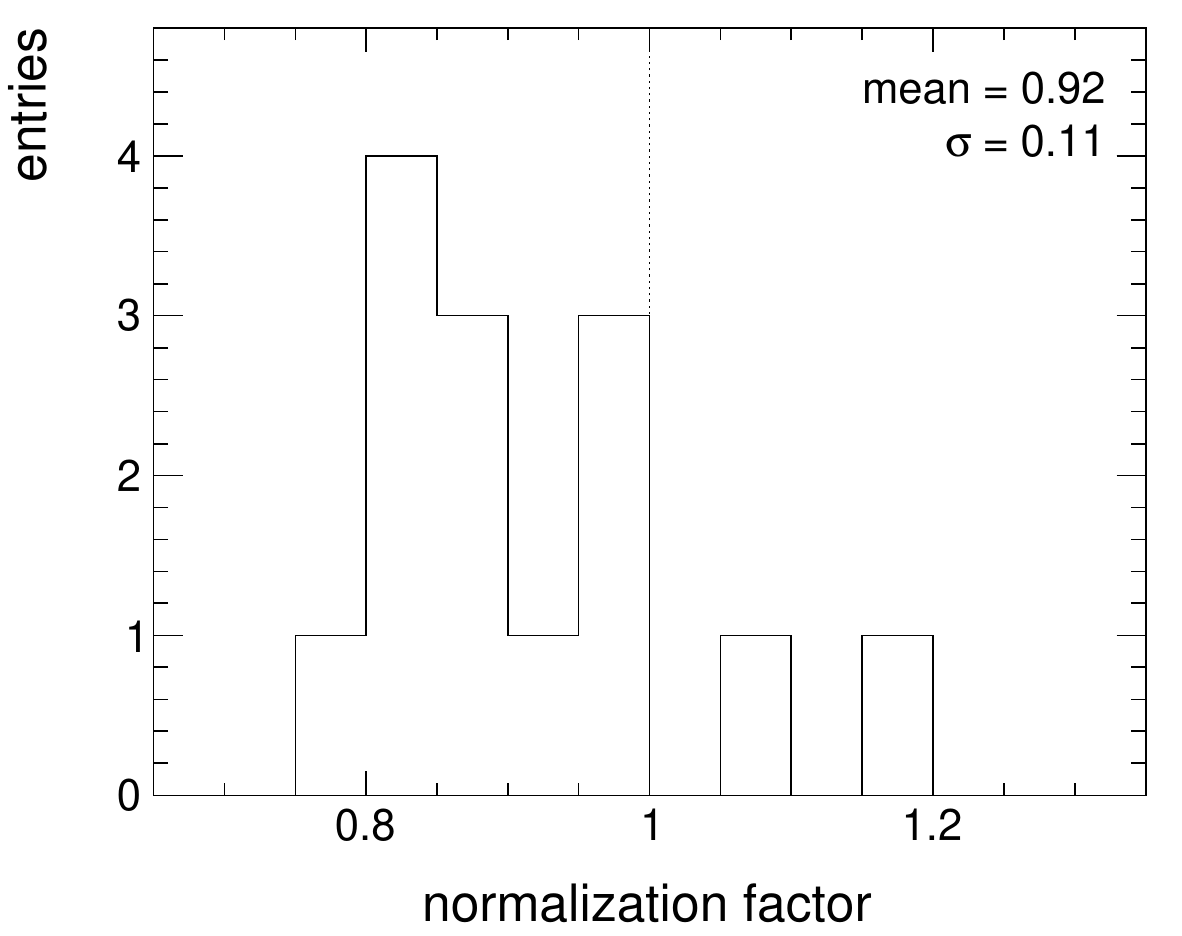} 
	\caption{Normalization factors introduced for the spectrometer experiments}
  \label{fig:factor}
 \end{center}
\end{figure}

In contrast, the group of counter experiments, hereafter named "spectrometer experiments" feature a limited phase space coverage, typically with solid angles in the milli- to microsteradian range, Tab.~\ref{tab:pp_pim_spect}. Consequently, there arise sizeable systematic uncertainties, especially concerning the overall normalization. For each experiment there has to be introduced, to first order, normalization factor in order to establish compatibility with the reference data, as shown in Fig.~\ref{fig:factor}.

In addition, in certain cases, additional deviations in certain phase space regions have to be taken into account. A blind use of these data would jeopardize the overall precision of any attempted general description. Moreover, as will be shown below, these experiments did not subtract out the pions from weak decays, thus introducing another source of systematic uncertainties of up to 40\%.

\begin{table}[h]
\renewcommand{\tabcolsep}{0.3pc}
\renewcommand{\arraystretch}{1.1}
\begin{center}
\begin{tabular}{lcccccc}
\hline
   Experiment         & beam mom.  &  $\sqrt{s}$   &   & accelerator &  N $_{points}$ & N$_{\pi^-}$  \\
                      &  [GeV/c]   &    [GeV]      &                &         &       &     \\ \hline
 Melissinos \cite{melissinos}      &     3.67  &   2.98   &   & BNL Cosmotron& 105 &  29 k    \\
 Akerlof \cite{akerlof,landolt}    &    12.5   &   5.02   &   &   ANL ZGS    &  83 &  33 k   \\
 Dekkers \cite{dekkers}            &    18.8   &   6.12   &   &   CERN PS    &  15 &   3 k   \\
 Allaby \cite{allaby1,landolt}     &    19.2   &   6.15   &   &   CERN PS    &  87 &  96 k   \\
 Allaby \cite{allaby2,landolt}     &    24     &   6.84   &   &   CERN PS    &  96 & 107 k   \\
 Beier \cite{beier}                &    24     &   6.84   &   &   BNL AGS    &  21 &  51 k  \\
 Anderson \cite{anderson,landolt}  &    29.7   &   7.58   &   &   BNL AGS    &  50 &  20 k   \\
 Abramov \cite{abramov}            &    70     &  11.5    &   &   Serpukhov  &   5 & 142 k   \\
 Brenner \cite{brenner}            &   100     &  13.9    &   &   FNAL       &  25 &   7 k   \\
 Brenner \cite{brenner}            &   175     &  18.2    &   &   FNAL       &  23 &   6 k   \\
 Johnson \cite{johnson}            &   100     &  13.8    &   &   FNAL       &  32 &  19 k    \\
 Johnson \cite{johnson}            &   200     &  19.4    &   &   FNAL       &  30 &  18 k   \\
 Johnson \cite{johnson}            &   400     &  27.4    &   &   FNAL       &  31 &  19 k \\
 Antreasyan \cite{antreasyan}      &   200     &  19.4    &   &   FNAL       &   1 &  100    \\
 Antreasyan \cite{antreasyan}      &   300     &  23.8    &   &   FNAL       &   1 &  100    \\
 Antreasyan \cite{antreasyan}      &   400     &  27.4    &   &   FNAL       &   1 &  100    \\

\hline
\end{tabular}
\end{center}
\caption{Spectrometer experiments}
\label{tab:pp_pim_spect}
\end{table}

In the lists of experiments given in Tabs.~\ref{tab:pp_pim_data} and \ref{tab:pp_pim_spect} above, one set of results, which is at the same time the most recent one that has been published, is missing: the NA61 experiment, Table~\ref{tab:na61}. This collaboration uses basically the same TPC detector as NA49. It aims at providing data over a range of beam momenta from 20 to 158~GeV/c thus covering a good fraction of the energy region, from the CERN PS/AGS to the CERN SPS, given in Tabs.~\ref{tab:pp_pim_data} and \ref{tab:pp_pim_spect} above. A detailed comparison with the preceding experiments reveals, however, very sizeable deviations from these references which precludes the inclusion of these new results into the global data interpolation. In addition, their first publication \cite{abgrall} does not use the particle identification capabilities of the NA49 detector but gives cross sections for negative hadrons (h$^-$) which have to be corrected for K$^-$ and $\overline{\textrm{p}}$ contributions using external model assumptions. The second paper \cite{aduszkiewicz} uses particle identification but suffers from a sizeably reduced phase space coverage.

\begin{table}[h]
\renewcommand{\tabcolsep}{0.5pc}
\renewcommand{\arraystretch}{1.1}
\begin{center}
\begin{tabular}{lccccccc}
\hline
   Experiment                      & beam mom. & $\sqrt{s}$ & detector & N $_{points}$ & N$_{h^-}$/N$_{\pi^-}$  \\
                                   &   [GeV/c] &    [GeV]   &          &               &              \\ \hline
 Abgrail \cite{abgrall}            &    20.0   &     6.27   &    TPC   &     201       &      23 k    \\
 Abgrail \cite{abgrall}            &    31.0   &     7.74   &    TPC   &     216       &     122 k    \\
 Abgrail \cite{abgrall}            &    40.0   &     8.77   &    TPC   &     225       &     232 k    \\
 Abgrail \cite{abgrall}            &    80.0   &    12.32   &    TPC   &     243       &     325 k    \\
 Abgrail \cite{abgrall}            &   158.0   &    17.27   &    TPC   &     253       &     510 k    \\
 Aduszkiewicz \cite{aduszkiewicz}  &    20.0   &     6.27   &    TPC   &      54       &      19 k    \\
 Aduszkiewicz \cite{aduszkiewicz}  &    31.0   &     7.74   &    TPC   &      76       &      75 k    \\
 Aduszkiewicz \cite{aduszkiewicz}  &    40.0   &     8.77   &    TPC   &      90       &     184 k    \\
 Aduszkiewicz \cite{aduszkiewicz}  &    80.0   &    12.32   &    TPC   &      86       &     296 k    \\
 Aduszkiewicz \cite{aduszkiewicz}  &   158.0   &    17.27   &    TPC   &     113       &     359 k    \\

\hline
\end{tabular}
\end{center}
\caption{The NA61 experiment}
\label{tab:na61}
\end{table}

As far as results from higher energy p+p colliders, basically RHIC at $\sqrt{s}$~=~200~GeV and LHC at $\sqrt{s}$ from 900~GeV to 13~TeV, are concerned, there is a drastic reduction of phase space coverage. As shown in Sect.~\ref{sec:dep_ener} (Fig.~\ref{fig:xdycor}) above, the accessible rapidity range for particle detection and identification only allows for the study of very central production and does not reach into the fragmentation region. This is apparent from Tab.~\ref{tab:rhic} where the 5 experiments providing data on $\pi^-$ production in the high energy region are listed.

\begin{table}[h]
\renewcommand{\tabcolsep}{0.5pc}
\renewcommand{\arraystretch}{1.1}
\begin{center}
\begin{tabular}{lcccccccc}
\hline
 Experiment & \multicolumn{6}{c}{$\sqrt{s}$ [TeV]} & &rapidity\\
  & 0.063  &    0.20   &   0.90  &    2.76  &    5.02   &   7.00 &     13.00 &  \\
  STAR   & & \cite{adams,abelev}         &&&&&&  0  \\
  PHENIX & \cite{adare} & \cite{adare}   &&&&&&  0  \\
  BRAHMS & & \cite{arsene,yang}          &&&&&& 0, 1.0, 1.2, 2.95, 3.3, 3.5  \\
  ALICE  & & & \cite{aamodt} & \cite{abelev2} & \cite{acharya} & \cite{adam} & \cite{acharya2} & 0  \\
  CMS    & & & \cite{chatrchyan} & \cite{chatrchyan} &  & \cite{chatrchyan} & \cite{sirunyan} & 0 \\
\hline
\end{tabular}
\end{center}
\caption{Collider experiments}
\label{tab:rhic}
\end{table}

All experimental results listed in the Tabs.~\ref{tab:pp_pim_data}-\ref{tab:rhic} will be discussed one by one in Sect.~\ref{sec:fd_corr} below with respect to a detailed three dimensional interpolation scheme introduced in Sect.~\ref{sec:data}.

There are however $\pi^0$ data from the LHCf experiment covering the $x_F$ range from 0.2 to 0.9 for $p_T$ values between 0.025 and 0.6~GeV/c. These unique data will be included in the comparison after extracting $\pi^0$ cross sections from $\pi^+$ and $\pi^-$ data at SPS and ISR energy.

%
%
\section{Data treatment}
\vspace{3mm}
\label{sec:data}

%
%
\subsection{Definition of a reference grid}
\vspace{3mm}
\label{sec:data_grid}

The aim of this paper is the establishment of a consistent data base exclusively from the measurements provided by the experiments introduced above. This data base should cover the three independent variables involved with the structure function $f(p,s)$ (\ref{eq:incl}) with a grid which is sufficiently fine-grained in order to allow for a precise interpolation into any choice of variables connected with the momentum $p$ and the cms energy squared $s$. Such a grid structure does not exist for the measured data: a wide choice of momentum variables have in fact been used. In addition, no common, well-defined binning scheme has been agreed on. Therefore in a first step, a convention concerning the chosen momentum components including a binning scheme has to be defined. In a second step, the available data have to be interpolated such that they fit into this grid system.

The following conventions will be used in this paper.

The momentum components are:

\begin{itemize}
  \item Transverse momentum $p_T$ in 26 steps of 0.05~GeV/c from $p_T$~=~0.05 to $p_T$=~1.3~GeV/c.
  \item Reduced rapidity $y_{\textrm{lab}} = y_{\textrm{beam}} - y$ (\ref{eq:ylab}) over the range from -1.2 to +3.8 in 26 steps of 0.2 units. As these components are not orthogonal, as an additional choice Feynman $x_F$ (\ref{eq:xf_def}) will be used with 21 steps from 0 to 0.7 with a variable step width.
  \item As each experiment is performed at a given cms energy $\sqrt{s}$ which is defined by the available accelerator rather than an agreed coverage at given $s$ values, the cms energy is represented as $\log_{10} (s)$ in 27 steps of 0.1 from 1 to 3.6. This defines  a range from $\sqrt{s}$~=~3.16 to 63~GeV corresponding to the highest ISR energy and covering the reference data, Tab.~\ref{tab:pp_pim_data}. As the higher energy data from RHIC and LHC are confined to the central rapidity plateau this range will be extended to 7.7 at $y$~=~0.
\end{itemize}

This grid offers, for the reference data, about 18~k points out of which 13~k or about 70\% are covered by the experimental data.

%
%
\subsection{Data interpolation and selection}
\vspace{3mm}
\label{sec:data_interp}

A general overview over the experiments introduced above and the totality of the more than 4000 data points concerned, in addition with an unprecedented precision, has as yet never been attempted. Several steps are mandatory in order to achieve a common and consistent data base:

\begin{enumerate}[label=(\alph*)]
	\item A first, two-dimensional interpolation of the double-differential cross sections for each data set at fixed $s$ into the binning grid in $p_T$/$y_{\textrm{lab}}$ or $p_T$/$x_F$ defined above.
	\item Extension into a three-dimensional interpolation in momentum and $\log{s}$ in order to connect the data at different interaction energies.
	\item Scrutiny of each data set in turn in order to check for individual deviations precluding the establishment of a consistent ensemble. This concerns the detection of overall inconsistencies for instance in absolute normalization, see Fig.~\ref{fig:factor}, or eventually the partial or complete exclusion of results.
\end{enumerate}

In the following argumentation the chosen procedures will be described in some detail. It has been, however, clear to the authors from the outset that human intervention and judgement on several levels was needed in order to achieve the desired result, in peculiar concerning the overall precision.

%
%
\subsubsection{Interpolation by algebraic fits}
\vspace{3mm}
\label{sec:data_interp_fits}

The task of describing data distributions in any coordinate is not facilitated by the fact that these are not predictable in the framework of non-perturbative QCD. Simple algebraic formulations for $p_T$ and $x_F$ distributions were nevertheless widely used in the past, like exponential $p_T$ and $m_T$, or power-law $x_F$ fits. A look at the complexity of the corresponding data distributions, once a certain precision is reached, should be sufficient to refrain from such solutions. Two examples may be mentioned here in this context.

A high statistics bubble chamber experiment \cite{sims} has published the original data at only three $p_T$ values, Fig.~\ref{fig:int2data28_28} at $p_{\textrm{beam}}$~=~28.5~GeV/c. The bulk of the data were fitted with double-Gaussian rapidity distributions, Fig.~\ref{fig:int2data28_32}. Whereas the original data are very well described by the global interpolation, the fitted data show substantial deviations with mean residuals at 1.5 and an rms deviation of 1.7.

Another typical example is given in Fig.~\ref{fig:pp_lowpt} \cite{pp_pion} where complex structures as a function of $p_T$ are visible in a certain $x_F$ range. These structures which are different for $\pi^+$ and $\pi^-$ mesons, are due to the decay of baryonic resonances (Fig.~\ref{fig:resonance}) but difficult to describe by generalized algebraic formulations.

%
%
\subsubsection{Errors}
\vspace{3mm}
\label{sec:data_interp_errors}

The data are not only subject to statistical errors, but most importantly to systematic deviations (Fig.~\ref{fig:factor}) which exceed in many cases the statistical uncertainties. Concerning the reference data, even the bubble chamber experiments with the biggest statistical relevance \cite{blobel} reach barely up to $p_T$ values of 1~GeV/c, although with systematic errors on the few percent level. Here the spectrometer experiments may give relatively high statistics data albeit only in restricted phase space areas and, furthermore, afflicted with important systematic uncertainties. The task to find an optimum compromise between these two types of experiments cannot be left to an automatized, "computer aided" approach. Suffice it to say that only the existence of "reference" data with small systematic errors combined with the fact that most of the systematic deviations in the "spectrometer" data may be resolved by only one normalization constant per experiment allows a consistent build-up of a global data set.

%
%
\subsubsection{Data treatment: tasks and solutions}
\vspace{3mm}
\label{sec:data_interp_treatment}

In a first step, each data set has to be scrutinized for internal consistency and the data points have to be interpolated to comply with the reference grid, Sect.~\ref{sec:data_grid}. Already at this stage, internal inconsistencies become visible where "visible" means inspection by eye. Some typical examples are given by Figs.~\ref{fig:alpery}, \ref{fig:alpers}, \ref{fig:allaby19}, \ref{fig:allaby24}, \ref{fig:beier}. As transverse momentum is the only common variable which may be extracted from all data sets, this first step is conducted in $p_T$.

In a second step, the interpolated $p_T$ distributions at a variety of longitudinal variables, have to extended into $y_{\textrm{lab}}$ distributions on the reference grid, Sect.~\ref{sec:data_grid}, again by interpolation.

In a third step, this two-dimensional interpolation has to be extended into a three-dimensional one by studying the dependence on $\log{s}$, again on the reference grid.

%
%
\subsubsection{An all-out optical approach}
\vspace{3mm}
\label{sec:data_interp_approach}

The authors do not see any way to realize the tasks mentioned above by "computer aided"  methods. Those methods are based on mathematical rather than physical constraints and indeed no application of comparable complexity has ever been tried concerning inclusive particle physics.

Instead an all-out visual approach to the problems has been opted for. Such an approach is tedious and time consuming but offers safety in fulfilling all constraints combined with controllable performance in terms of both statistical and systematical uncertainties.

This approach, often quoted under "eye scanning" has nowadays an odour of "imprecision" and "unscientific behaviour".There is no reason for this prejudice. There is no harm in using millimeter or semi-logarithmic paper together with flexible rulers (in fact the "spline" methods have been developed with such rulers in mind) as long as some basic constraints are fulfilled:

\begin{itemize}
	\item in this paper one aims at an overall precision on the less than 5\% level, in fact the achieved interpolations are shown to be reliable with rms deviations of 2--3\%.
	\item therefore the coordinate scales used must allow for the safe setting and reading of results on the percent level.
	\item the rulers must be used in a minimum educated way so as to find the way through the statistical error bars by using a maximum of available data points, at the same time looking for systematic data irregularities relevant with respect to the statistical errors.
	\item the boundary conditions as well as continuity and smoothness imposed by physics are readily fulfilled.
	\item the interpolation needs to be recursive in the sense that several successive stages in all three dimensions have to be passed before a final and optimized result may be claimed.
\end{itemize}

In the end the success of the global interpolation has to be controlled for each data set by the distribution of normalized residuals where all data points contribute, and by the mean deviations of all experiments from the interpolation. The residual distributions are given for each data plot, Figs.~\ref{fig:int2data6.6_12}--\ref{fig:int2dataisr53_63} and Figs.~\ref{fig:melissinos}--\ref{fig:johnson3} and the mean deviations are presented in Figs.~\ref{fig:summary_ref} and \ref{fig:precision} for the reference and spectrometer data respectively.

%
%
\subsection{Reference data}
\vspace{3mm}
\label{sec:data_ref}

The reference data, Tab.~\ref{tab:pp_pim_data}, have three components:

\begin{enumerate}[label=(\alph*)]
  \item 14 data sets obtained with Hydrogen Bubble Chambers (HBC).
  \item NA49 data using a large set of Time Projection Chambers (TPC).
  \item ISR data from basically 4 different spectrometer setups.
\end{enumerate}
%
%
\subsubsection{Bubble Chamber data}
\vspace{3mm}
\label{sec:bc_data}

\begin{figure}[h]
 \begin{center}
   \includegraphics[width=7cm] {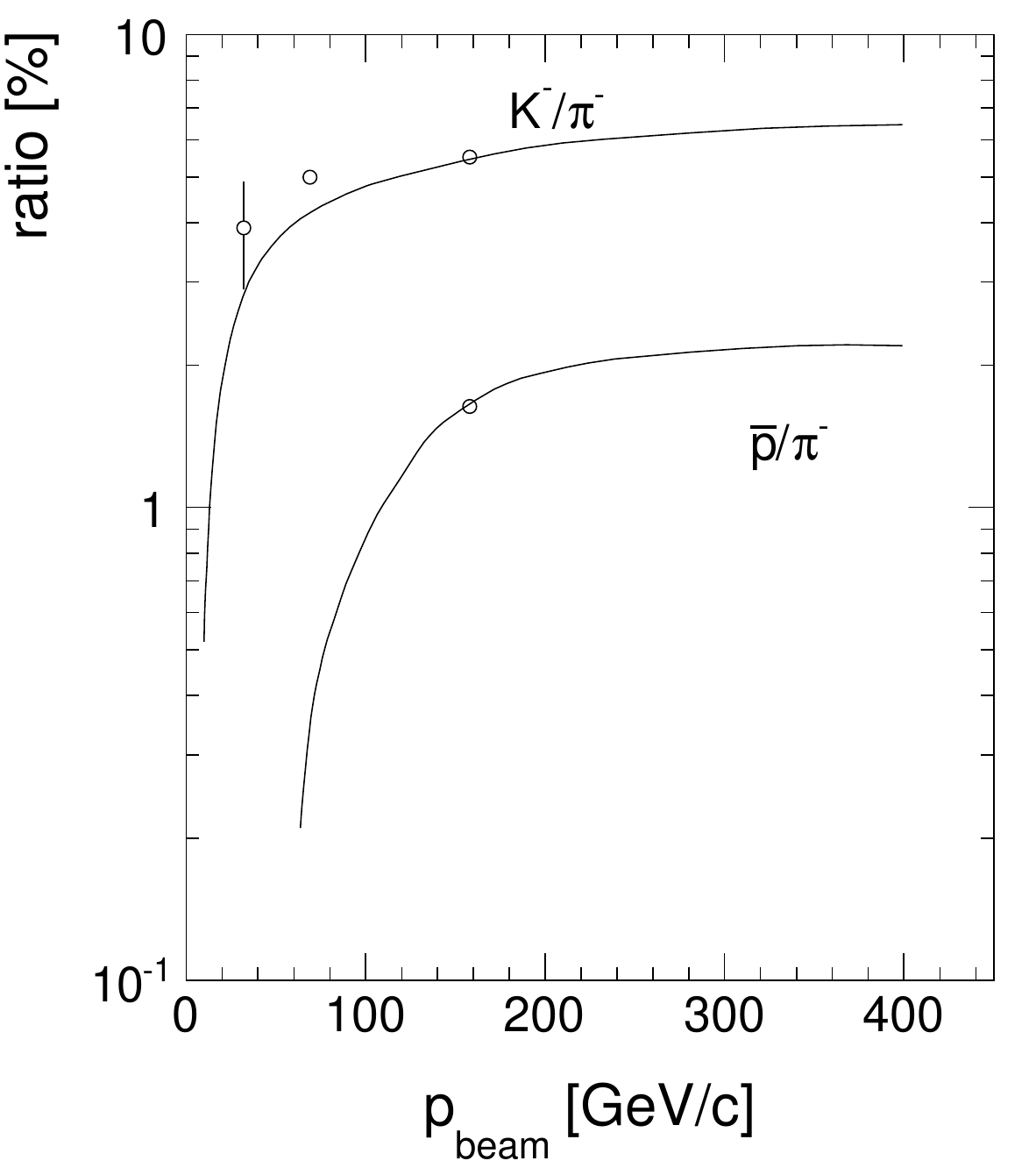} 
	\caption{K$^-$/$\pi^-$ and $\overline{\textrm{p}}$/$\pi^-$ ratios as a function of $\sqrt{s}$. The full lines correspond to the new determination of the energy dependences of K$^-$ \cite{pp_kaon} and $\overline{\textrm{p}}$ \cite{pp_proton} cross sections by the NA49 collaboration.The measurement of the K$^-$/$\pi^-$ ratio at 32~GeV/c and the k$^-$/$\pi^-$ and $\overline{\textrm{p}}$/$\pi^-$ ratios at 158~GeV/c beam momentum as well as the correction deduced from the measured K$^0_S$ yields at 69~GeV/c are given as data points}
  \label{fig:k2pi_ap2pi}
 \end{center}
\end{figure}

Bubble Chamber data feature by conception very small systematic errors. As all interactions are directly visible inside a fiducial volume, they are self-normalizing which is a decisive advantage over all other detection methods. In addition only small corrections, typically on a few percent level, are necessary, for instance for non-identified Dalitz decays. Only cuts on the dip angle (in direction of the optical axis of the camera system) and on non-resolved secondary vertices are applied. On the other hand, the identification of secondary particles is difficult if not completely absent. This is especially valid for $\pi^-$ as usually all negative particles are called pions. This necessitates corrections for K$^-$ and $\overline{\textrm{p}}$ yields which are strongly variable with $\sqrt{s}$ as shown in Fig.~\ref{fig:k2pi_ap2pi}.

\begin{figure}[h]
 \begin{center}
   \includegraphics[width=7cm] {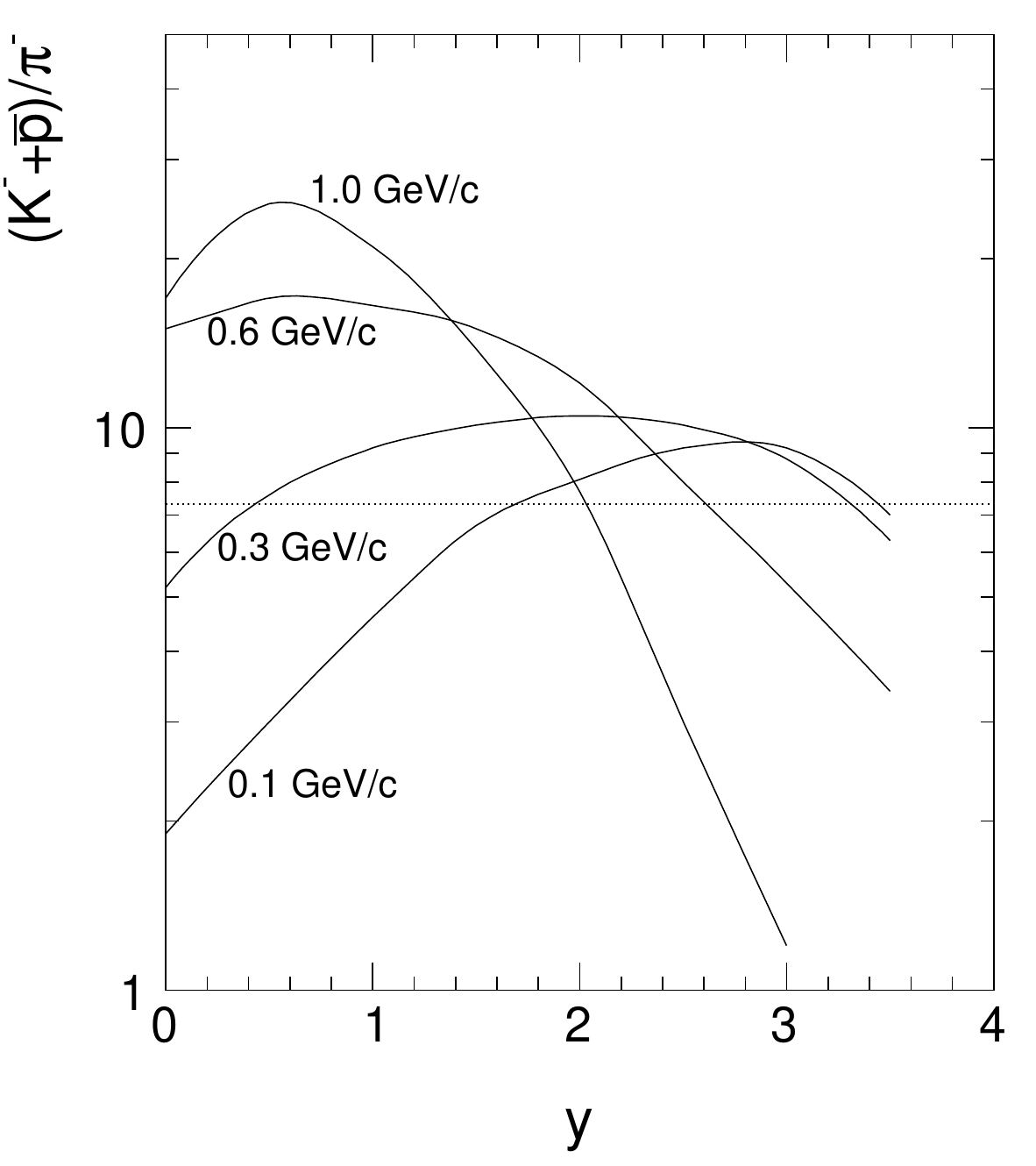} 
	\caption{(K$^-$+$\overline{\textrm{p}}$)/$\pi^-$ ratio as a function of rapidity for different $p_T$ for the NA49 data at $\sqrt{s}$~=~17.2~GeV, compared to the mean ratio at 7.3\%}
  \label{fig:kap2pi}
 \end{center}
\end{figure}

It is apparent from Fig.~\ref{fig:k2pi_ap2pi} that in comparison to an overall systematic uncertainty of about 2\% the K$^-$ and $\overline{\textrm{p}}$ contributions are of importance above $\sqrt{s}$~5~GeV and 15~GeV, respectively. In view of a correction of double-differential cross sections the dependence of the particle ratios on the kinematic variables has however to be taken into account. This is exemplified in Fig.~\ref{fig:kap2pi} where the (K$^-$+$\overline{\textrm{p}}$)/$\pi^-$ ratio is given as a function of rapidity for different transverse momenta as presented for the NA49 data at $\sqrt{s}$~=~17.2~GeV.

It should be noted that here the use of the pion mass for the heavier particles in the transformation from the lab to the cms frame has to be taken into account. Whereas these dependencies may be neglected in the lower energy range, the cross sections at the highest beam momenta at 102 and 400~GeV/c have to be properly corrected. In the absence of double differential measurements at these momenta, the data contained in Fig.~\ref{fig:kap2pi} have been scaled down or up, respectively, at 102 and 400~GeV/c according to the mean ratios.

%
%
\subsubsection{NA49 data}
\vspace{3mm}
\label{sec:na49_data}

The NA49 collaboration has obtained a sizeable data set of 4.8 million events from p+p interactions at 158~GeV/c beam momentum, using a set of two superconducting magnets and of 4 large Time Projection Chambers \cite{nim}. All charged particles have been identified via energy loss ($dE/dx$) measurements in the TPC system \cite{pp_pion,pp_proton,pp_kaon,pc_paper}. The large data volume from the TPC readout necessitated the introduction of an event trigger with an efficiency of 86\% of the total inelastic cross section. The corresponding normalization has been obtained with a topology-dependent correction for this trigger bias which amounts to less than 8\% for $\pi^-$. The data have been corrected for feed-down from K$^0_S$, $\Lambda^0$ and $\Sigma^-$ decay, especially necessitated by the fact that only a fraction of these decays are reconstructed on vertex as opposed to the bubble chamber data. Fiducial cuts in rapidity ($y >$~0) and azimuth (180$\pm$90~degrees) reduce the yield to 2.5 million measured $\pi^-$ which makes this sample the by far biggest one obtained to date. The systematic uncertainty is estimated to 2\% \cite{pp_pion}. This value has been verified by the comparison of the total charged particle yield with precision data from bubble chamber experiments \cite{pp_kaon}.

%
%
\subsubsection{ISR data}
\vspace{3mm}
\label{sec:isr_data}

As stated above, the availability of double differential data from four independent ISR experiments spanning most of the available phase space is unique for high energy proton colliders. The trigger efficiency of typically 95 to 98\% of the total inelastic cross section ensures high precision in absolute normalization as opposed to both lower energy spectrometer experiments (see Fig.~\ref{fig:factor}) and higher energy colliders. The ISR data have an overlap with the bubble chamber data at their lowest cms energy of 23~GeV and extend to 63~GeV with complete particle identification.

\begin{figure}[h]
 \begin{center}
   \includegraphics[width=7cm] {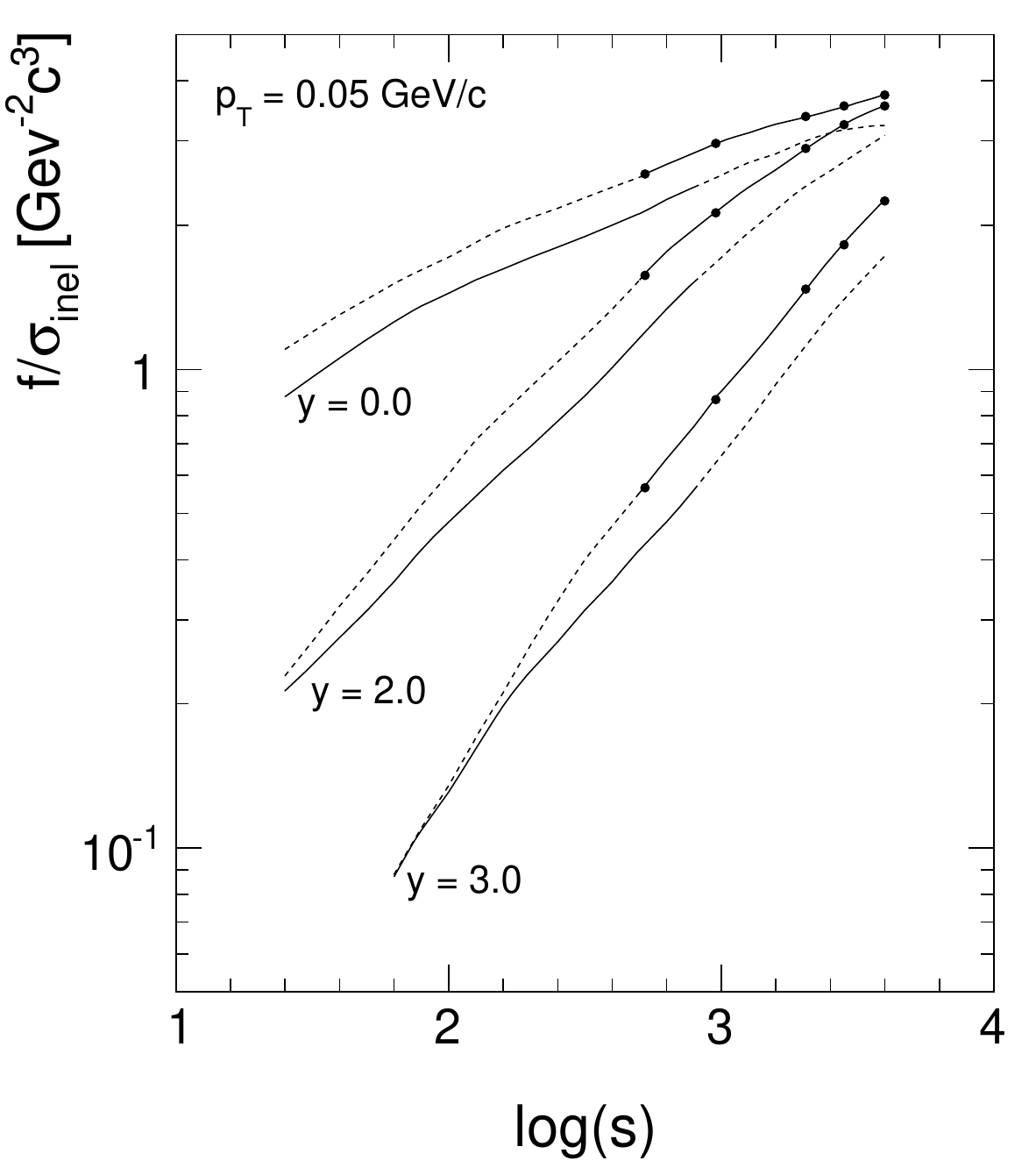} 
	\caption{$\pi^-$ cross sections $f/\sigma_{\textrm{inel}}$ for the reference sample of HBC and NA49 results compared to the ISR data (points) as a function of cms energy, at $p_T$~=~0.05~GeV/c. The full lines represent the original, feed-down subtracted reference data and the ISR data without subtraction; the broken lines show the reference data with feed-down added and the ISR data with feed-down subtraction, respectively}
  \label{fig:feedeff}
 \end{center}
\end{figure}

A problem is given by the fact that the ISR data have not been corrected for feed-down from V$_0$ decays. As the resulting contribution to the $\pi^-$ yield is mostly concentrated at low transverse momentum, the respective correction amounts to sizeable values of up to 40\% and has therefore to be carefully quantified. This is exemplified in Fig.~\ref{fig:feedeff} where the $\pi^-$ cross sections from lower energies are compared to the ISR results at different rapidities as a function of cms energy.

This figure demonstrates several facts:
\begin{enumerate}[label=(\alph*)]
  \item Mutually consistent and continuous energy dependences may be obtained by properly taking into account the feed-down contributions.
  \item These contributions are not confined to central rapidity but extend far into the forward direction.
  \item The relative feed-down yield is not concentrated in the central region but increases -- depending on energy -- with increasing rapidity.
\end{enumerate}

The feed-down problem is of a more general importance as practically all of the spectrometer experiments, Tab.~\ref{tab:pp_pim_spect}, are not feed-down corrected. In fact no high resolution vertex detectors were available in the 1970's and 1980's and the first active elements were placed at sizeable distances from the target in the fixed-target experiments. In this paper, a general enumeration of this correction has therefore been worked out over the full energy scale. This is presented in the following Section.

%
%
\section{Feed-down correction for weak decays}
\vspace{3mm}
\label{sec:fd_corr}

\begin{figure}[h]
 \begin{center}
   \includegraphics[width=14cm] {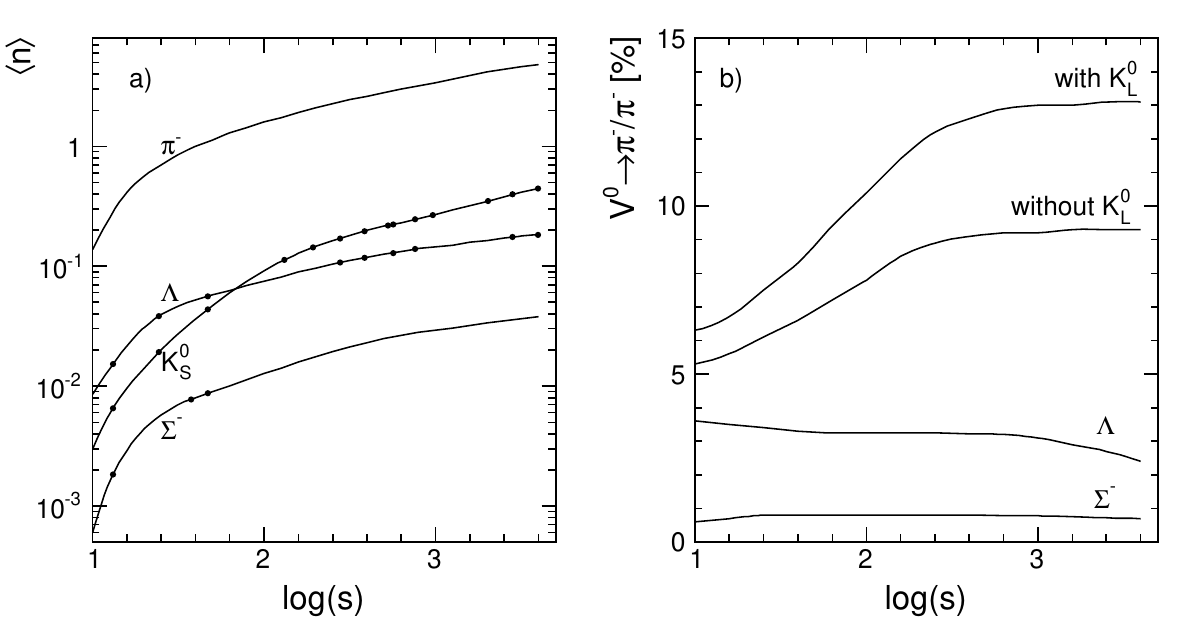} 
	\caption{a) Total yields $\langle n \rangle$ of K$^0_S$, $\Lambda$ and $\Sigma^-$ as a function of $\log{s}$, in comparison to $\pi^-$ without feed-down correction; b) percentage contribution of $V^0$ decays to the $\pi^-$ yield with and without K$^0_L$ and from $\Lambda$ and $\Sigma^-$}
  \label{fig:feedtot}
 \end{center}
\end{figure}

Contributions to the $\pi^-$ yield come from the weak decays of K$^0$ mesons and $\Lambda$ and $\Sigma^-$ baryons. A search for data on these strange particles reveals 18 experiments which cover the complete energy range considered here, from $\sqrt{s}$~=~3.6 to 63~GeV, see \cite{bierman,eisner,oh,jaeger,alpgard,blobel1,fesefeldt,ammosov1,chapman,brick,jaeger1,sheng,asai,kass,kichimi,busser,erhan,rensch}. The results have generally modest statistics and double differential cross sections are scarce with the exception of the high statistics bubble chamber experiments \cite{eisner,blobel1} at 6, 12 and 24~GeV/c beam momentum. Single differential $x_F$ and $p_T$ distributions are generally available in addition to the total yields $\langle n \rangle$ which are shown in Fig.~\ref{fig:feedtot}a.

The energy dependence of the total yields reflects several features of strange baryon and meson production in p+p interactions: approaching the strangeness threshold at low energy, associate $\Lambda$ + K$^+$ production will prevail by pion exchange while kaon pair production will be suppressed, whereas with increasing energy meson pair production will increase faster than the $\Lambda$+$\overline{\Lambda}$ yield. The scarce data on $\Sigma^-$, \cite{alpgard,blobel1,fesefeldt,kass}, permit an estimation of this contribution which is less than 1\% over the full energy range (Fig.~\ref{fig:feedtot}b), also using the approximate ratios for ($\Sigma^+$+$\Sigma^0$+$\Sigma^-$)/$\Lambda \sim$~1.0, $\Sigma^+$/$\Sigma^- \sim$~3.5 and $\Sigma^0$/$\Lambda \sim$~0.4.

Fig.~\ref{fig:feedtot}b presents the percentage contributions of the different $V^0$ decays to the total $\pi^-$ yield. The fact that this total contribution corresponds, above the SPS energy range, to 9\% and 13\% without and with K$^0_L$ decay respectively, gives a first indication of the primordial importance of particle decays to the final state hadron yields. This contribution will be shown below to correspond to a complex structure in the double-differential cross sections.

%
%
\subsection{Single differential cross sections}
\vspace{3mm}
\label{sec:fd_sdcs}

\begin{figure}[h]
 \begin{center}
   \includegraphics[width=14cm] {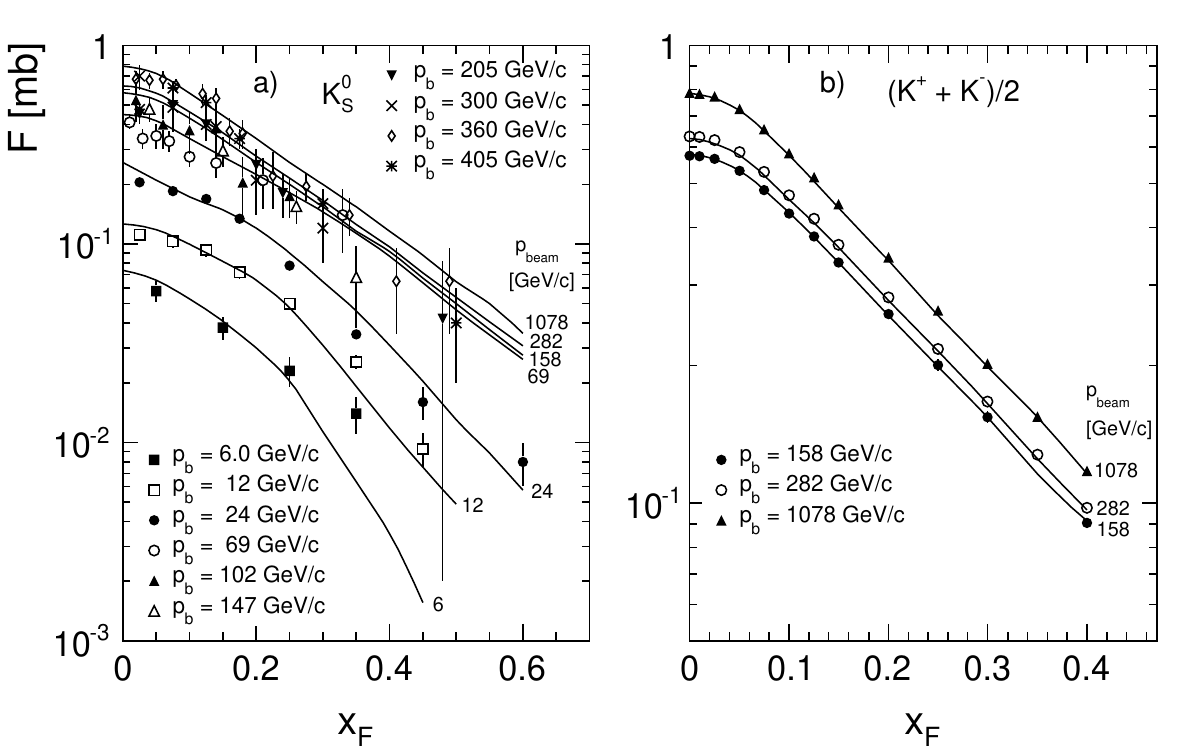} 
	\caption{$F(x_F)$ for K$^0_S$ as a function of $x_F$ for different beam momenta}
  \label{fig:kzero}
 \end{center}
\end{figure}

\begin{figure}[h]
 \begin{center}
   \includegraphics[width=13cm] {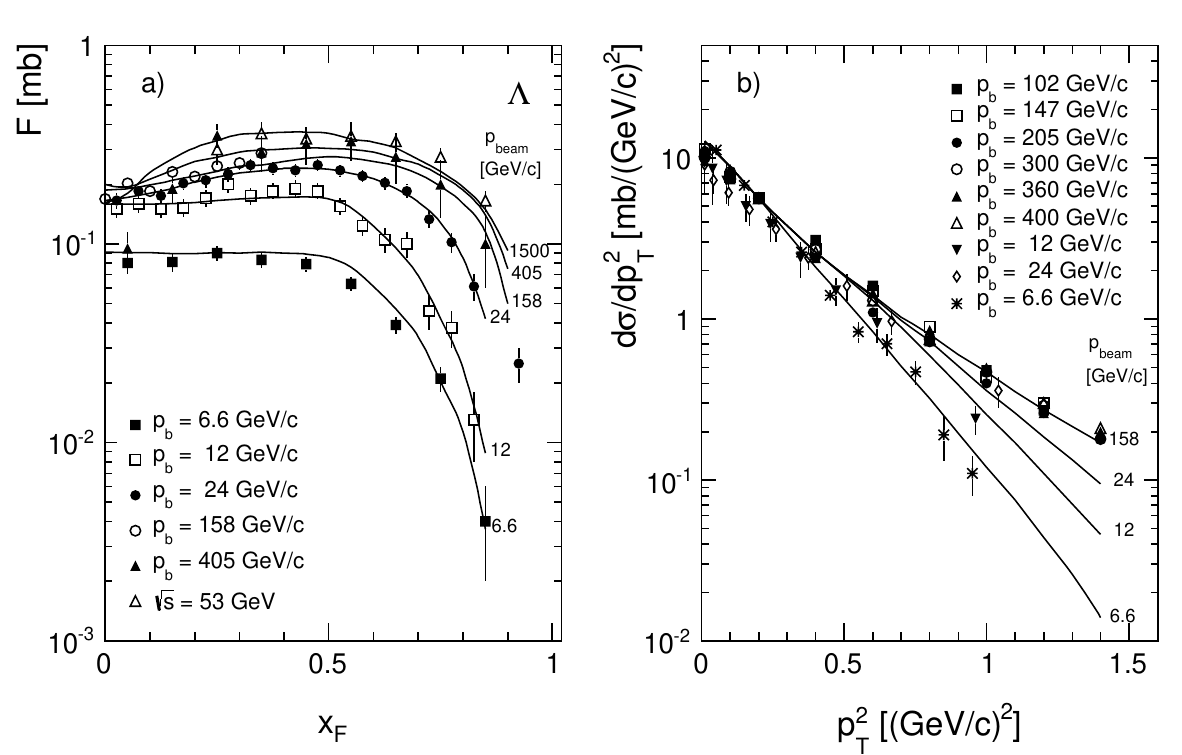} 
	\caption{a) $F(x_F)$ for $\Lambda$ as a function of $x_F$ for different beam momenta; b) $d\sigma/dp_T^2$ distributions as a function of $p_T^2$ from 6.6 to 400~GeV/c beam momentum renormalized at $p_T^2$~=~0.2~(GeV/c)$^2$}
  \label{fig:lambda}
 \end{center}
\end{figure}

Single invariant differential cross sections for K$^0_S$ and $\Lambda$ are presented in Figs.~\ref{fig:kzero} and \ref{fig:lambda} as a function of $x_F$ where the invariant cross section is:

\begin{equation}
   F(x_F) = \int{f(x_F,p_T) \cdot dp_T^2}
\end{equation}

The full lines shown in Figs.~\ref{fig:kzero} and \ref{fig:lambda} correspond to the interpolated yields used in the Monte Carlo calculation of the $\pi^-$ contributions at the indicated beam momenta.

For K$^0_S$, Fig.~\ref{fig:kzero}a, the data up to 24~GeV/c beam momentum define the interpolation with sufficient precision \cite{eisner}, \cite{blobel1}. In the higher energy range, $p_{\textrm{beam}}$~=~158~--~2100~GeV/c the averaged charged kaon yields \cite{pp_kaon} provide a precise reference as shown in Fig.~\ref{fig:kzero}b.

For $\Lambda$, Fig.~\ref{fig:lambda}, sufficient precision is given in the lower $p_{\textrm{beam}}$ range up to 24~GeV/c.

The measurements in the higher energy range are again characterized by sizeable statistical uncertainties in the range of typically 10 to 20\%. Nevertheless  the results may be interpolated with sufficient precision as shown by the full lines in Fig.~\ref{fig:lambda}a). As far as the $p_T^2$ distributions are concerned the data above 100~GeV/c beam momentum may be described within their sizeable error margins by a common $p_T^2$ dependence. At lower beam momenta precise data are available. They show a progressive steepening of the $p_T^2$ distributions, see Fig.~\ref{fig:lambda}b).

%
%
\subsection{Double differential cross sections}
\vspace{3mm}
\label{sec:fd_ddcs}

\begin{figure}[h]
 \begin{center}
   \includegraphics[width=7cm] {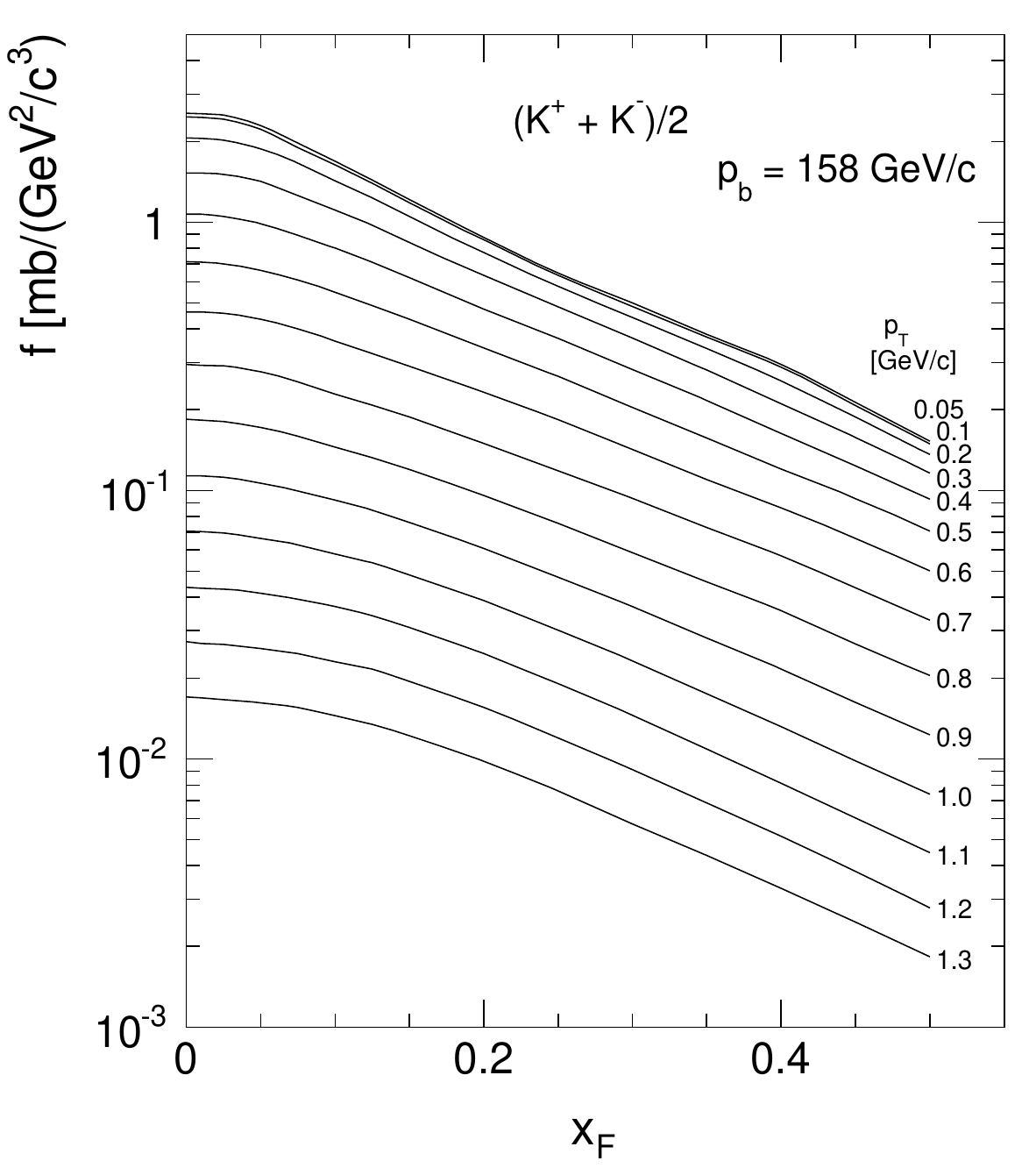} 
	\caption{$f(x_F,p_T)$ for K$^0_S$ as a function of $x_F$ at $p_{\textrm{beam}}$~=~158~GeV/c obtained from the average charged kaon yields \cite{pp_kaon}}
  \label{fig:kzero_dd}
 \end{center}
\end{figure}

For K$^0_S$ double differential cross sections are available at low energy from direct measurements \cite{eisner}, \cite{blobel1} and at higher energy from averaged charged kaon data. An example is shown in Fig.~\ref{fig:kzero_dd} at 158~GeV/c beam momentum \cite{pp_kaon}.

\begin{figure}[h]
 \begin{center}
   \includegraphics[width=8cm] {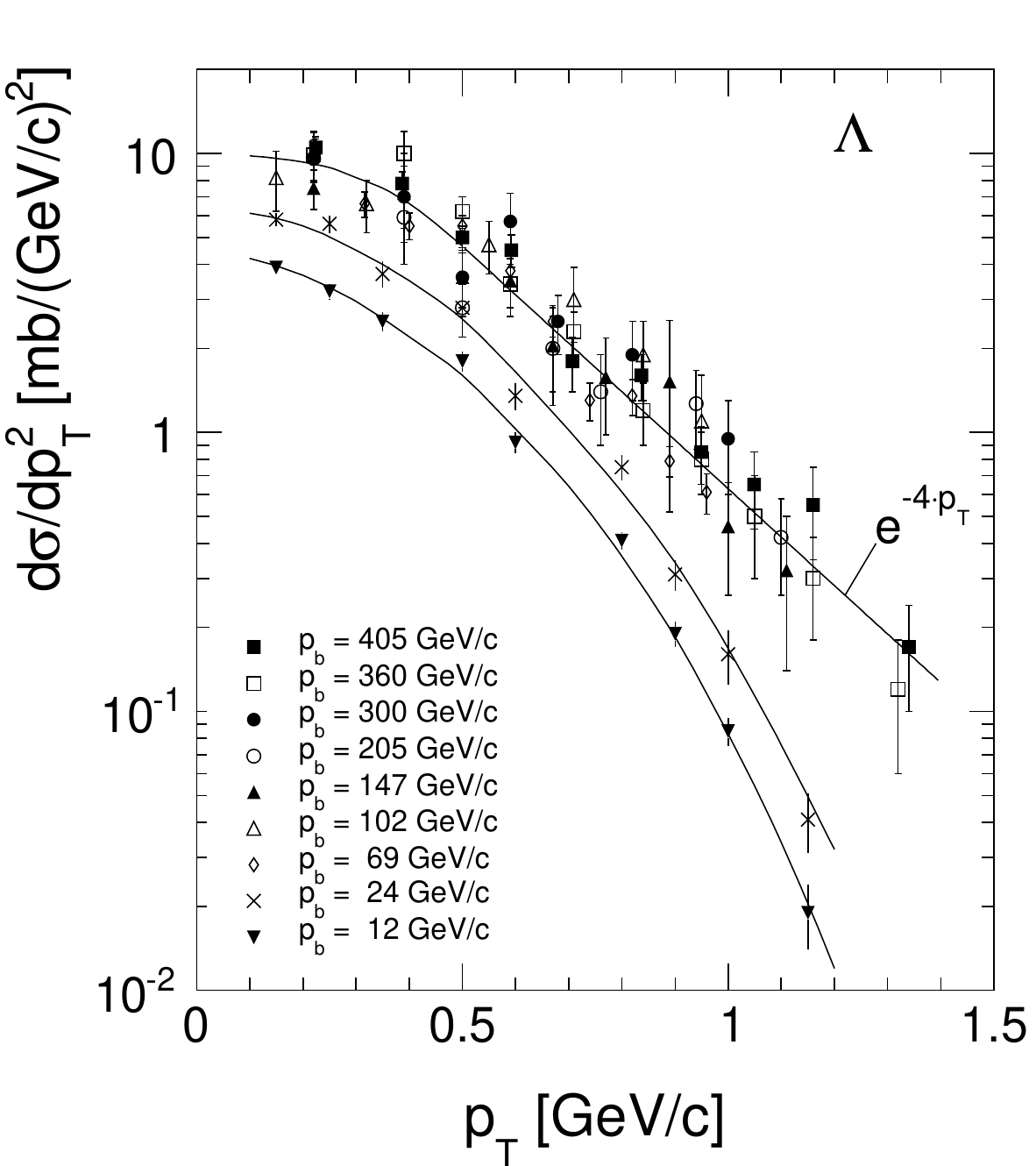} 
	\caption{$d\sigma/dp_T^2$ as a function of $p_T$ for $\Lambda$ at different beam momenta}
  \label{fig:lam_pt}
 \end{center}
\end{figure}

For $\Lambda$ the situation is experimentally less well defined. At low energy \cite{eisner}, \cite{blobel1} direct measurements are available. At higher energies only $x_F$ integrated transverse momentum distributions are given, see Fig.~\ref{fig:lam_pt}.

These data may be approximated by exponential functions of the form $Ae^{-Bp_T}$, with B in the range between 3.5 and 4.5. They may be related to the double differential cross section of protons by the fact that the ratio $(dn/dp_T)^\Lambda/(dn/dp_T)^p$ turns out to be independent of $x_F$ in the range 0~$< x_F <$~0.35 \cite{na61_lam}. This ratio is steadily increasing as a function of $p_T$ as it is typical for cascading decays, Fig.~\ref{fig:lam_dd} (Sect.~\ref{sec:res_casc_bw} below) where the final state proton is diluted in momentum with respect to the decaying resonance. This dilution depends on the $Q$-value of the decay as exemplified by the broken line in Fig.~\ref{fig:lam_dd} for the decay $\Lambda \rightarrow p+\pi^-$.

\begin{figure}[h]
 \begin{center}
   \includegraphics[width=8cm] {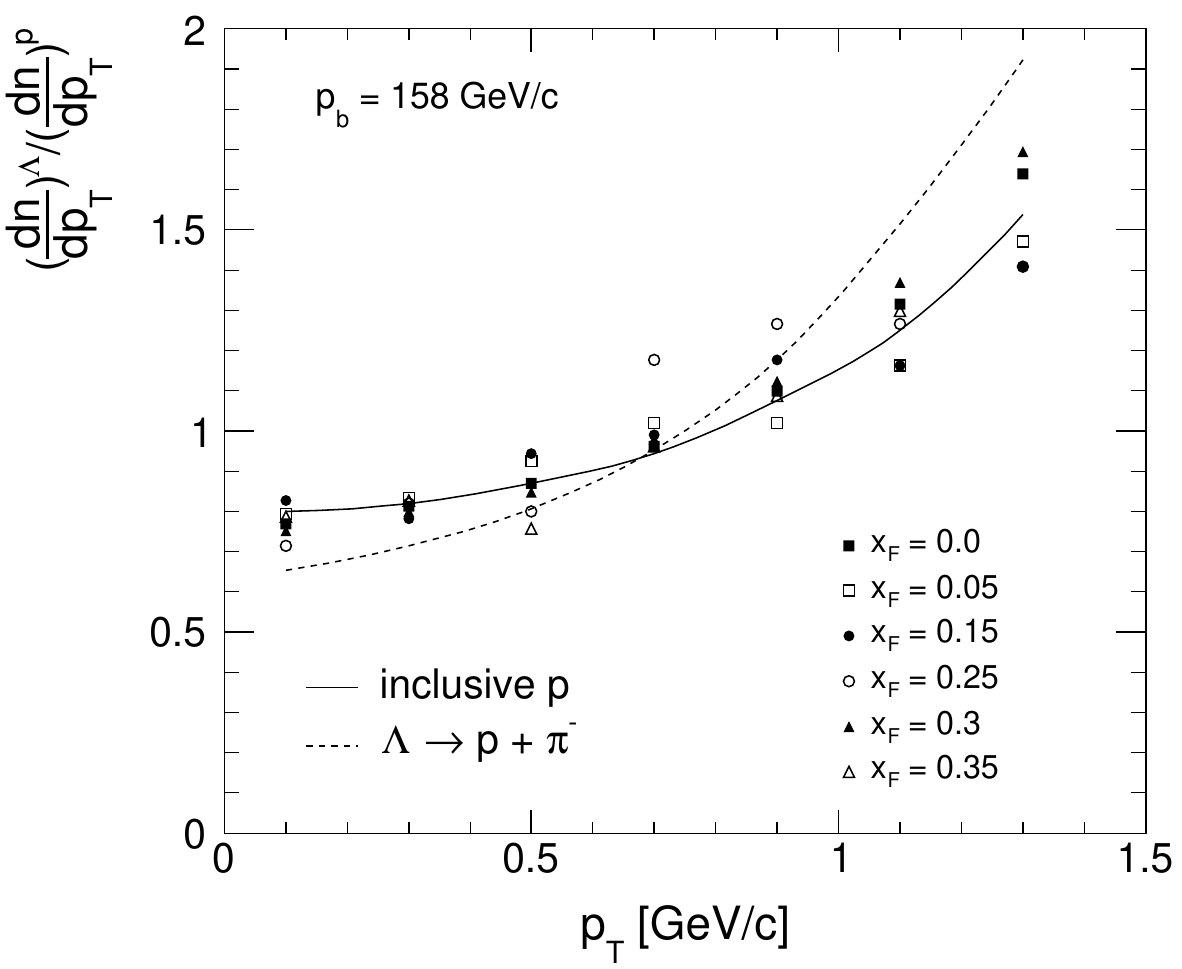} 
	\caption{Ratio $(dn/dp_T)^\Lambda/(dn/dp_T)^p$ as a function of $p_T$ at $p_{\textrm{beam}}$~=158~GeV/c. The dots at different $p_T$ values represent the measurements for 6 values of $x_F$ between $x_F$~=~0 and $x_F$~=~0.35. For comparison the ratio for $\Lambda \rightarrow p+\pi^-$ is given as the broken line}
  \label{fig:lam_dd}
 \end{center}
\end{figure}

\begin{figure}[h]
 \begin{center}
   \includegraphics[width=14.cm] {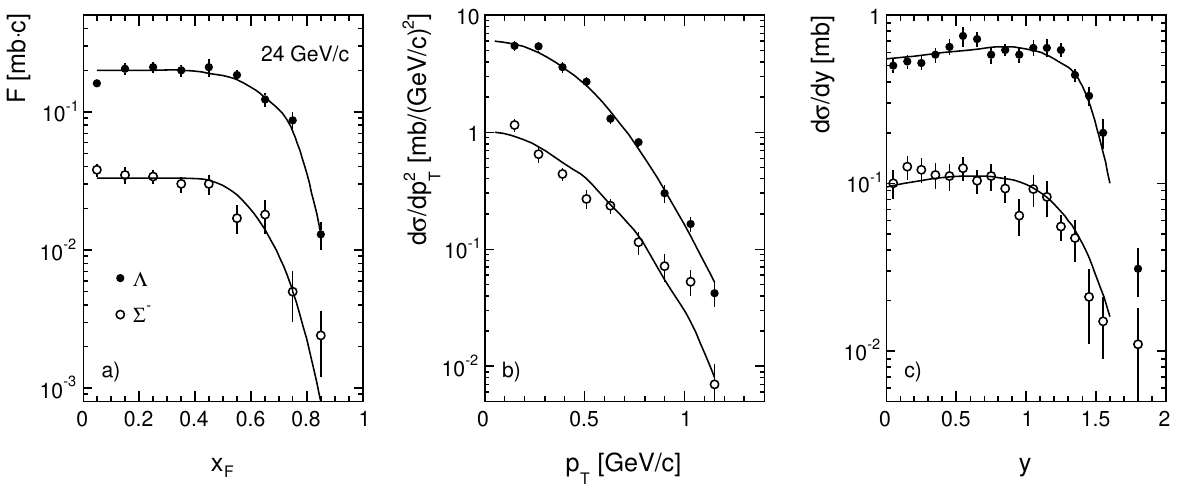} 
	\caption{a) $F(x_F)$, b) $d\sigma/dp_T^2(p_T)$ and c) $d\sigma/dy(y)$ for $\Sigma^-$ as compared to $\Lambda$}
  \label{fig:siglam}
 \end{center}
\end{figure}

For $\Sigma^-$ only \cite{blobel1} gives single differential cross sections at $p_{\textrm{beam}}$~=~12 and 24~GeV/c. These data are shown in Fig.~\ref{fig:siglam} as functions of $x_F$ and $p_T$ for $p_{\textrm{beam}}$~=~24~GeV/c.

The $\Sigma^-$ distributions are rather conformal with the ones for $\Lambda$ with the exception of the large $x_F$ region. This suggests the same treatment of the double differential cross sections in $p_T$ in reference to the proton data as shown above for the $\Lambda$.

%
%
\subsubsection{\boldmath K$_S^0$ decay}
\vspace{3mm}
\label{sec:feed_ddk0s}

\begin{figure}[h]
 \begin{center}
   \includegraphics[width=8cm] {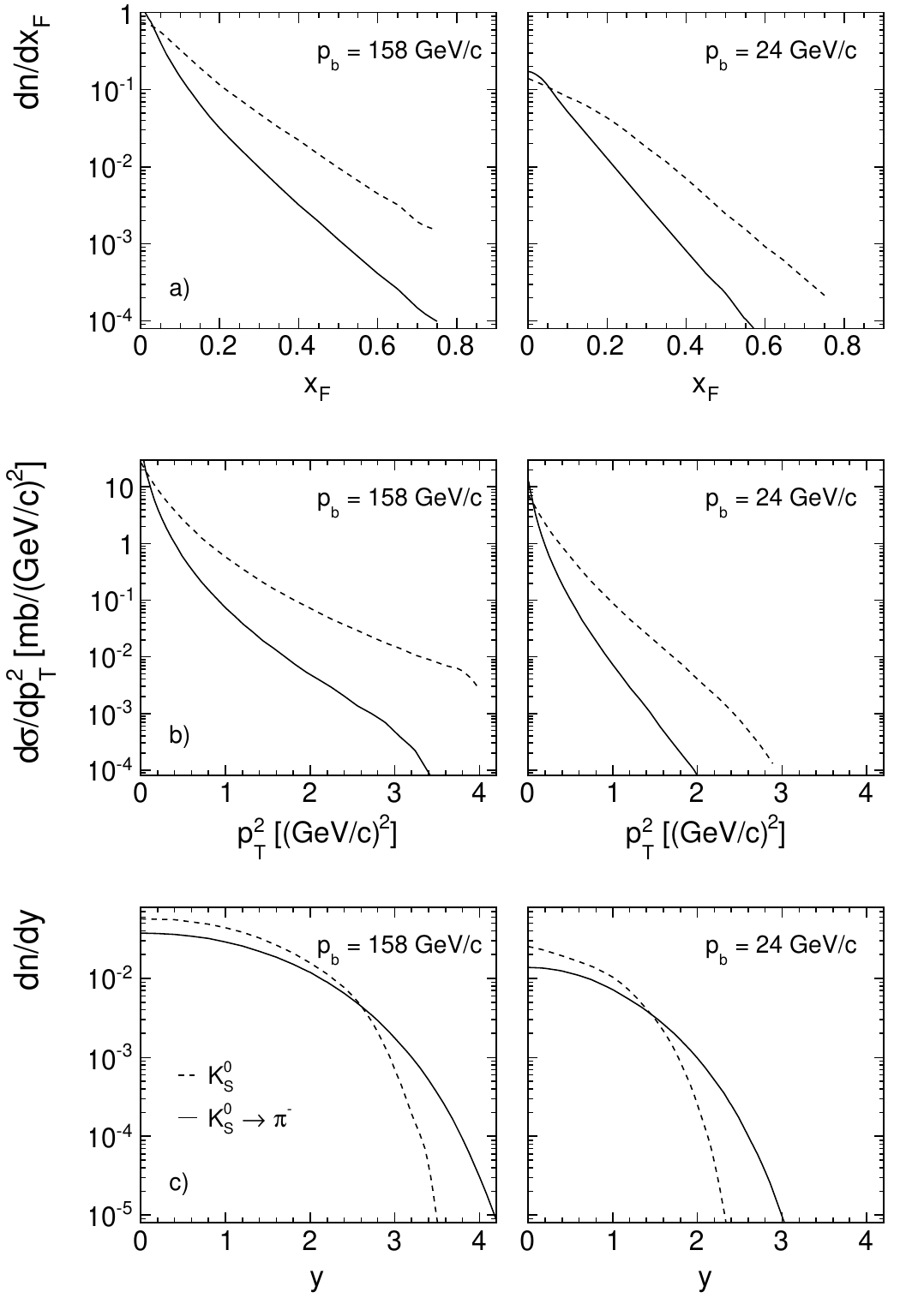} 
	\caption{a) $dn/dx_F$ as a function of $x_F$, b) $d\sigma/dp_T^2$ as a function of $p_T^2$ and c) $dn/dy$ as a function of rapidity for $p_{\textrm{beam}}$~=~158 and 24~GeV/c. The full lines correspond to the $\pi^-$ yields, the broken lines to the parent particle K$_S^0$}
  \label{fig:kazeros}
 \end{center}
\end{figure}

K$_S^0$ decays into $\pi^+$ + $\pi^-$ with a branching fraction of 69.2\%. This decay results in a rather involved relation between K$_S^0$ and the decay pions as functions of $x_F$, $p_T^2$ and $y$ as shown in Fig.~\ref{fig:kazeros} for the integrated quantities $dn/dx_F$, $d\sigma/dp_T^2$ and $dn/dy$.

\begin{figure}[h]
 \begin{center}
   \includegraphics[width=7.cm] {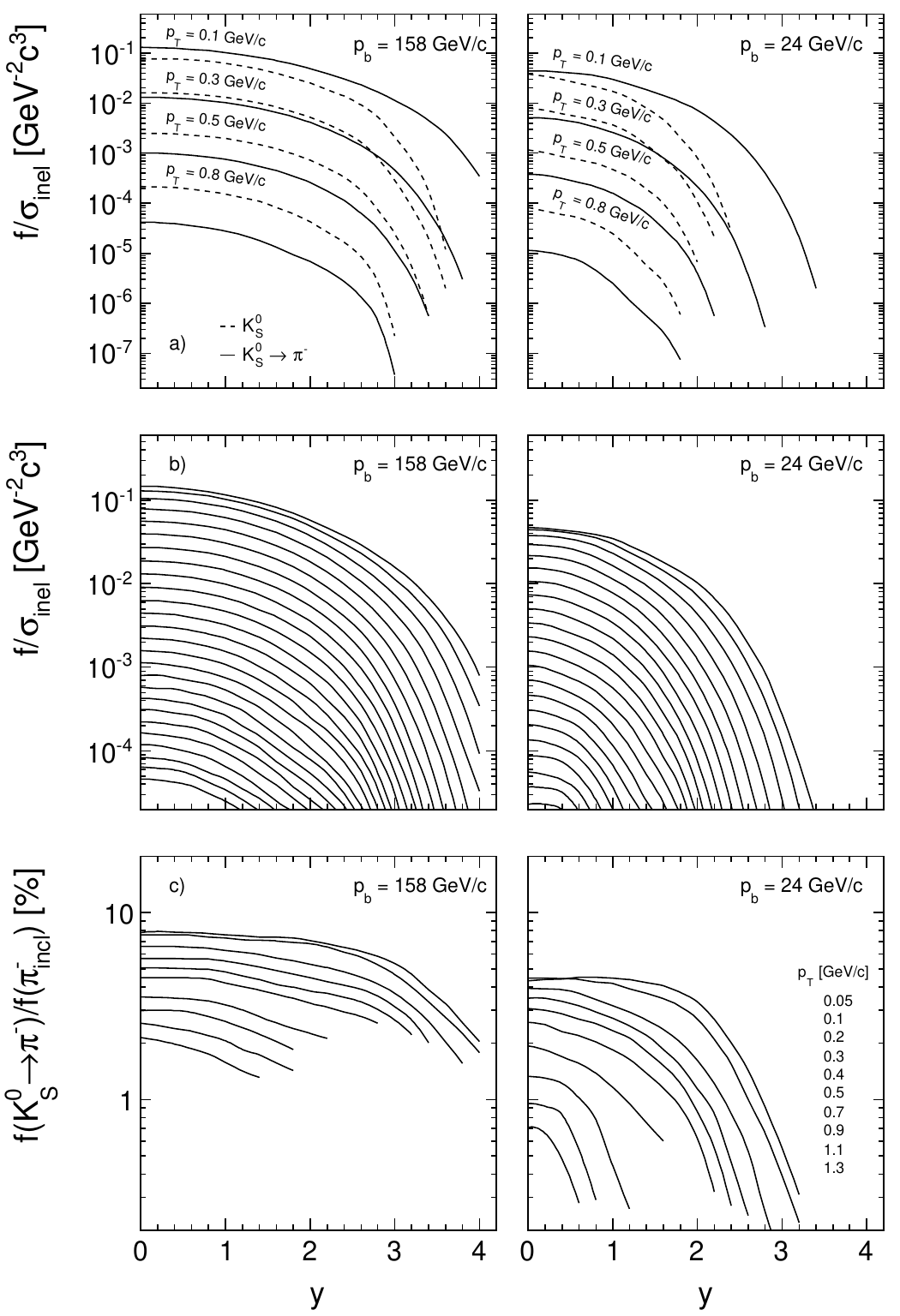} 
	\caption{a) invariant $\pi^-$ cross section $f/\sigma_{\textrm{inel}}$ as a function of $y$ for four values of $p_T$ at 158 and 24~GeV/c beam momentum. Full lines for $\pi^-$ and broken lines for the parent K$_S^0$. Subsequent $p_T$ values are multiplied by 1/3 for clarity; b) invariant $\pi^-$ cross section $f/\sigma_{\textrm{inel}}$ from K$_S^0$ decay as a function of rapidity for $p_T$ between 0.05 and 1.3~GeV/c; c) percentage ratio of $\pi^-$ from K$_S^0$ decay to the inclusive, feed-down subtracted $\pi^-$ yield}
  \label{fig:kazeros2pi}
 \end{center}
\end{figure}

The decay particle yields exceed the K$_S^0$ cross section at low $x_F$ and low $p_T^2$, whereas the situation inverts to large rapidity for $dn/dy$. On the double differential level this corresponds to a complex interplay between the K$_S^0$ and the decay pion in $p_T$ and rapidity as presented in Fig.~\ref{fig:kazeros2pi}.

The data in Figs.~\ref{fig:kazeros} and \ref{fig:kazeros2pi} are given for beam momenta of 158~GeV/c and 24~GeV/c.

Fig.~\ref{fig:kazeros2pi}c) demonstrates that the decay pion contributions come up to 8\% and 4\% respectively of the inclusive pion yield at low $p_T$. It is not only concentrated  at central rapidity but decreases only slowly as a function of $y$. Integrated over $p_T$ this corresponds to 5.2 and 2.8\%, respectively.

%
%
\subsubsection{\boldmath K$_L^0$ decay}
\vspace{3mm}
\label{sec:feed_ddk0l}

K$_L^0$ has three decay channels into negative pions:

\begin{equation}
	\begin{split}
      K_L^0 \rightarrow \pi^- + e^+ + \nu_e        \qquad (K^0_{e3})   \qquad & \textrm{with 20.3\%}  \\
      K_L^0 \rightarrow \pi^- + \mu^+ + \nu_{\mu}  \qquad (K^0_{\mu3}) \qquad & \textrm{with 13.5\%} \\
      K_L^0 \rightarrow \pi^- + \pi^+ + \pi^0      \qquad \qquad \qquad \ \, & \textrm{with 12.5\% }
    \end{split}
    \label{eq:k0ldecays}
\end{equation}

\begin{figure}[h]
 \begin{center}
   \includegraphics[width=7.5cm] {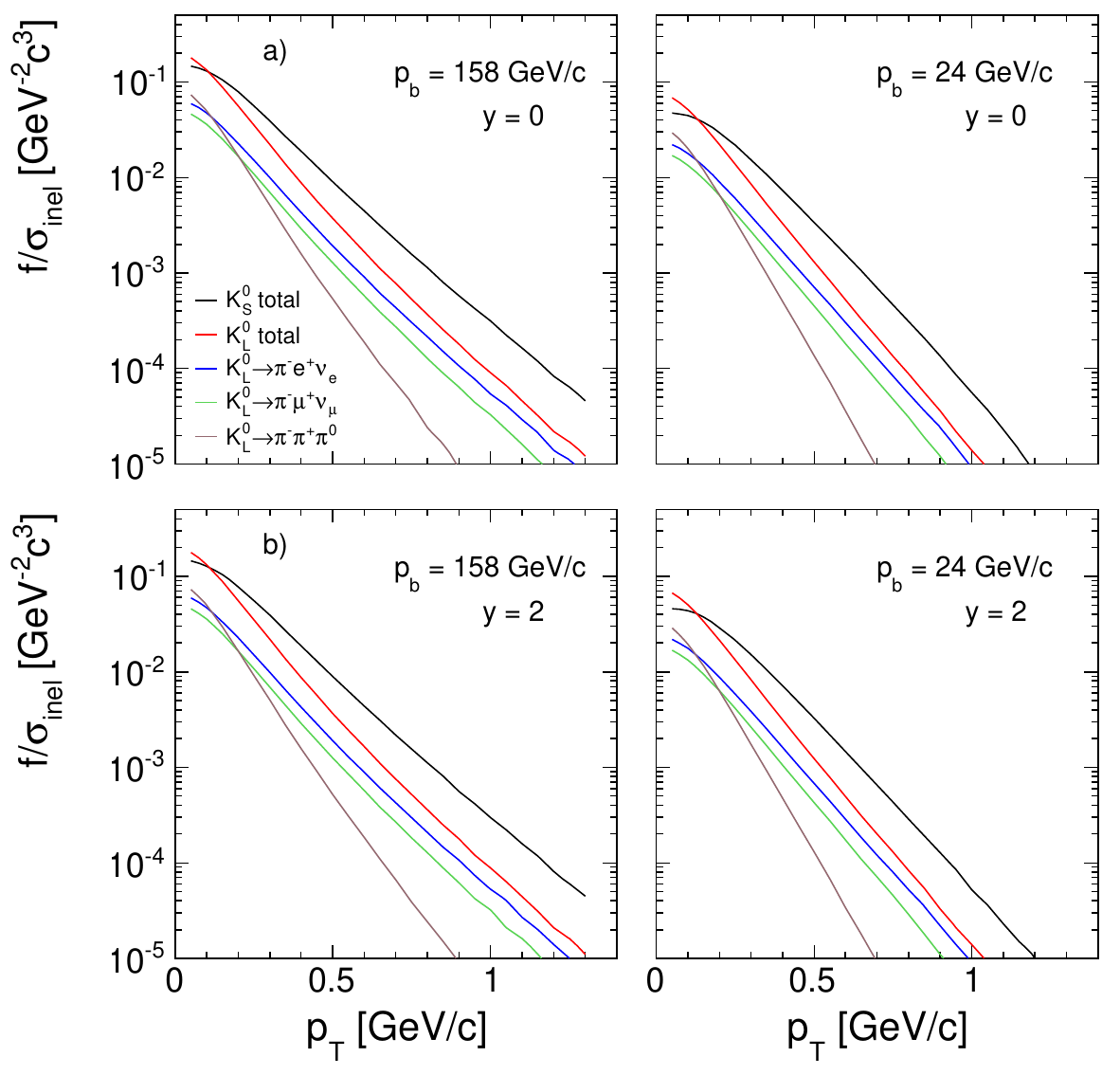} 
	\caption{Invariant $\pi^-$ cross sections as a function of $p_T$ for the different K$_L^0$ decays (\ref{eq:k0ldecays}) and the corresponding total yield compared to the K$_S^0$ decay a) $y$~=~0 and b) $y$~=~2 at $p_{\textrm{beam}}$~=~158 and 24~GeV/c }
  \label{fig:kazeroldecay}
 \end{center}
\end{figure}

This yields a total branching fraction of 46.3\% as compared to 69.2\% for the K$_S^0$ decay. Due to the different $Q$ values involved with these 3-particle decays the double differential cross sections have different $p_T$ distributions as shown in Fig.~\ref{fig:kazeroldecay} for two rapidity values.

\begin{figure}[h]
 \begin{center}
   \includegraphics[width=7cm] {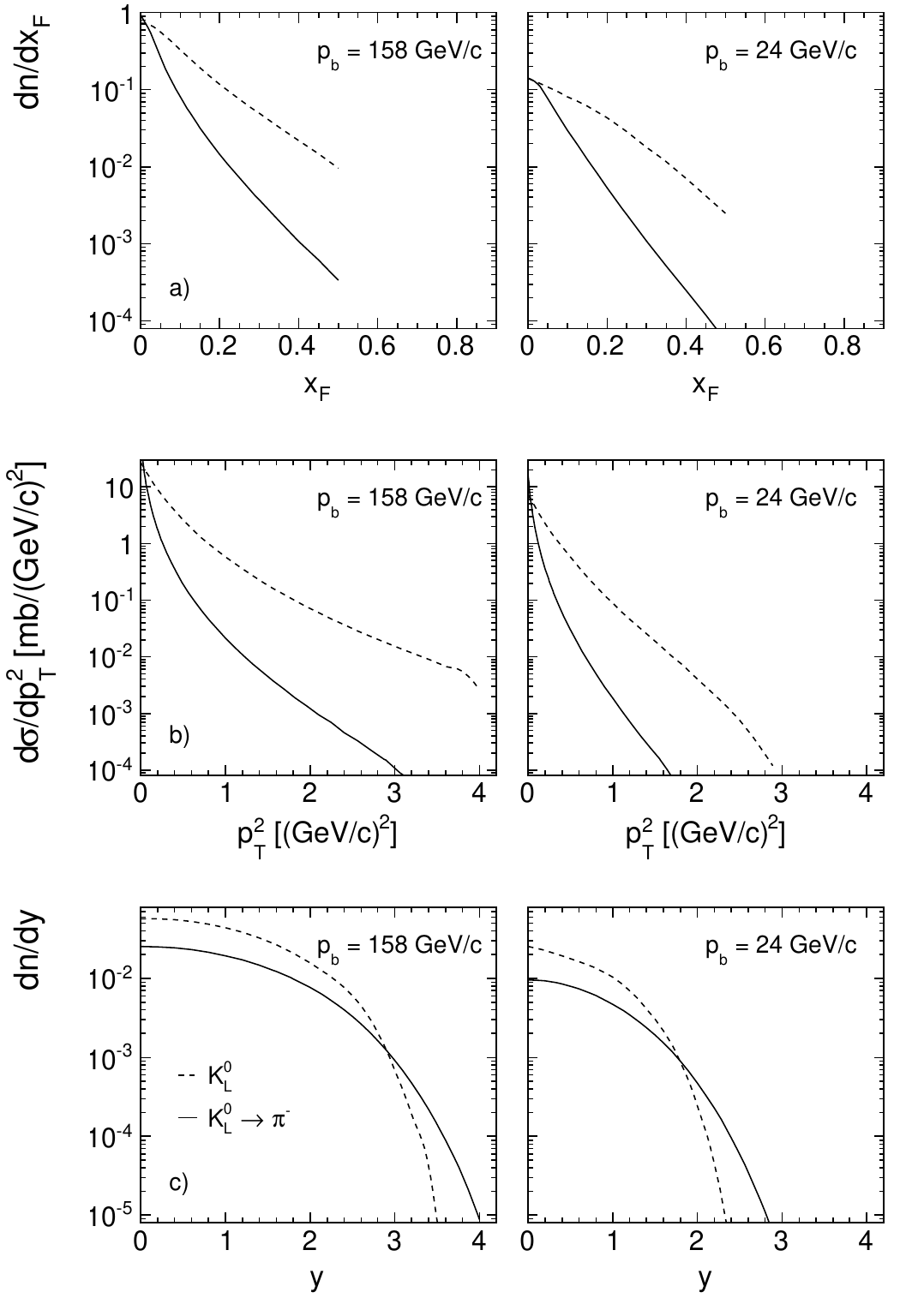} 
	\caption{a) $dn/dx_F$ as a function of $x_F$, b) $d\sigma/dp_T^2$ as a function of $p_T^2$ and c) $dn/dy$ as a function of rapidity for $p_{\textrm{beam}}$~=~158 and 24~GeV/c. The full lines correspond to the $\pi^-$ yields for the sum of the three decay channels, the broken lines to the parent particle K$_S^0$}
  \label{fig:kazerol}
 \end{center}
\end{figure}

The integrated quantities $dn/dx_F$, $d\sigma/dp_T^2$ and $dn/dy$ are presented in Fig.~\ref{fig:kazerol} for the sum of the three decay channels.

\begin{figure}[h]
 \begin{center}
   \includegraphics[width=7.5cm] {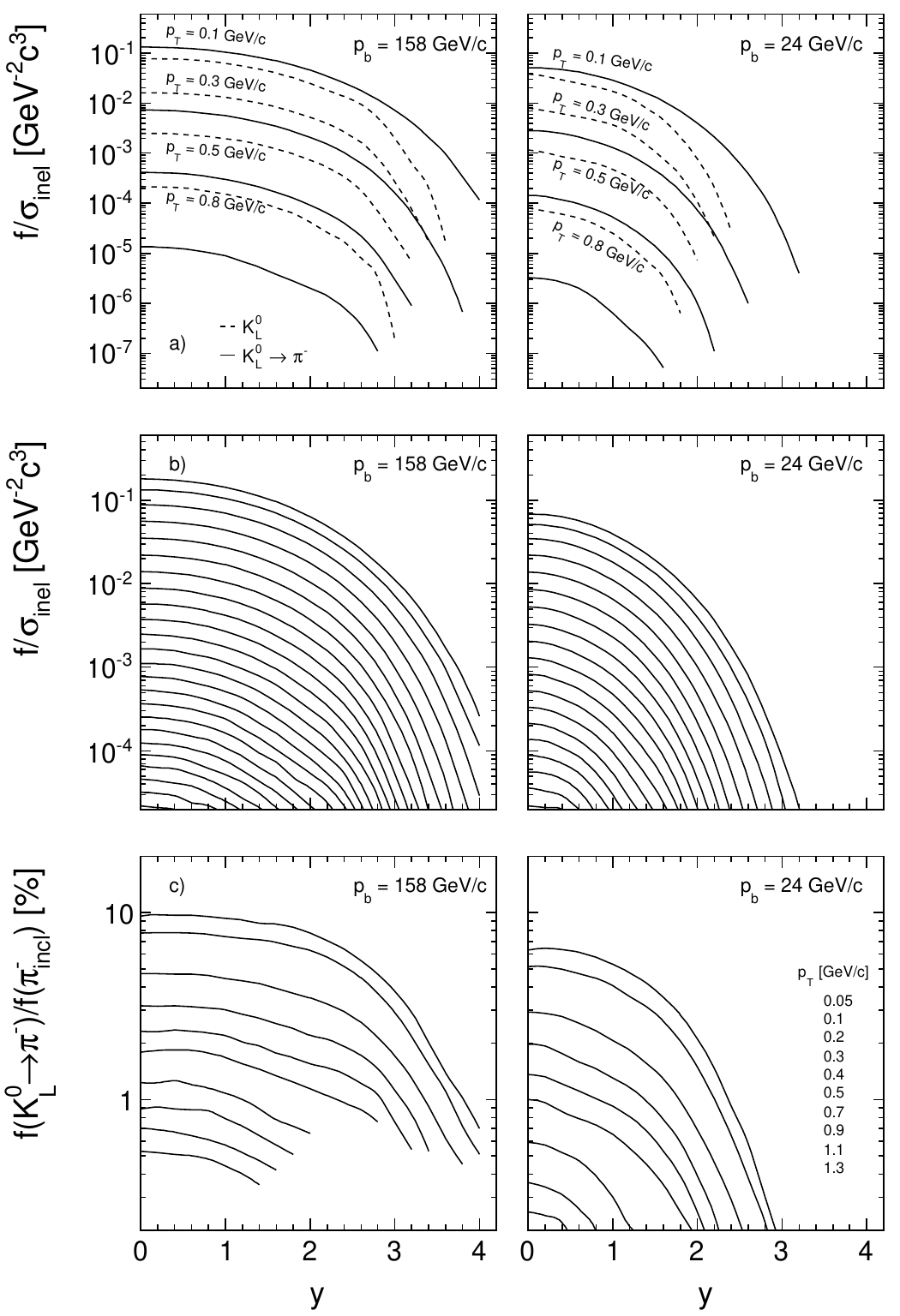} 
	\caption{a) invariant $\pi^-$ cross section $f/\sigma_{\textrm{inel}}$ as a function of $y$ for four values of $p_T$ at 158 and 24~GeV/c beam momentum. Full lines for decay $\pi^-$ and broken lines for the parent K$_L^0$. Subsequent $p_T$ values are multiplied by 1/3 for clarity. b) invariant $\pi^-$ cross section $f/\sigma_{\textrm{inel}}$ from K$_L^0$ decay as a function of rapidity for $p_T$ between 0.05 and 1.3~GeV/c given for the sum of the three decay channels c) percentage ratio of the total $\pi^-$ from K$_L^0$ decay to the inclusive, feed-down subtracted $\pi^-$ yield }
  \label{fig:kazerol2pi}
 \end{center}
\end{figure}

The similarity to the K$_S^0$ decays, Fig.~\ref{fig:kazeros}, is apparent. The interplay between the K$_L^0$ yield and the $\pi^-$ resulting from the sum of its decay channels is shown in Fig.~\ref{fig:kazerol2pi}.

The ratio of the decay pion contribution to the inclusive $\pi^-$ yield reaches a maximum of 10\% and 6\% at low $p_T$ respectively at 158 and 24~GeV/c beam momentum. This is higher than that from K$_S^0$ decay but it decreases more rapidly with increasing $p_T$ such that the  $p_T$ integrated yield amounts to 3.4 and 1.8\%, respectively.

%
%
\subsubsection{\boldmath $\Lambda$ decay}
\vspace{3mm}
\label{sec:feed_ddlam}

\begin{figure*}[h]
	\begin{minipage}[h]{0.45\linewidth}
 		\begin{center}
   			\includegraphics[width=7cm] {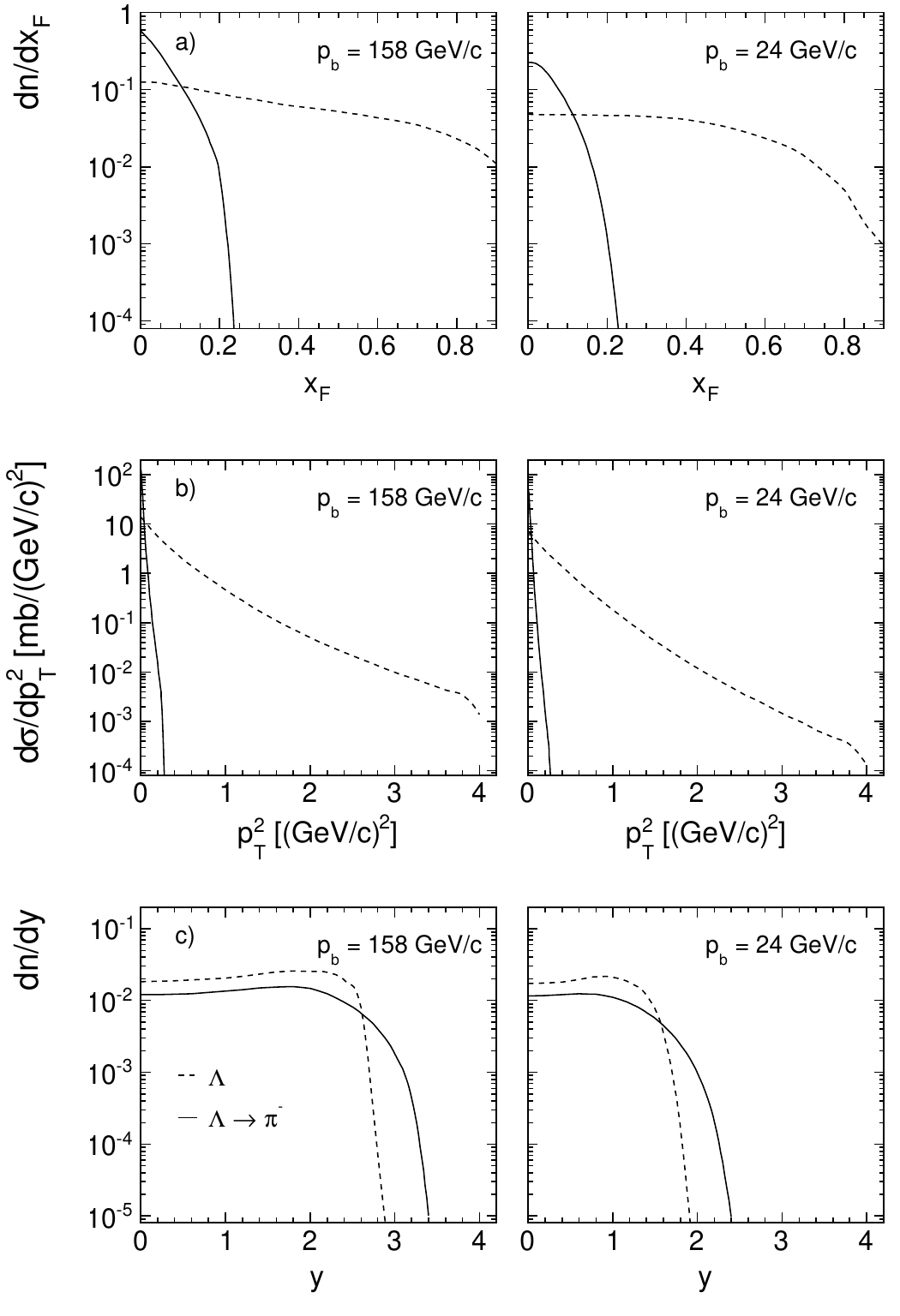} 
			\caption{a) $dn/dx_F$ as a function of $x_F$, b) $d\sigma/dp_T^2$ as a function of $p_T^2$ and c) $dn/dy$ as a function of rapidity for $p_{\textrm{beam}}$~=~158 and 24~GeV/c. The full lines correspond to the $\pi^-$ yields, the broken lines to the parent $\Lambda$}
  			\label{fig:lambdaS}
 		\end{center}
	\end{minipage}
	\begin{minipage}[h]{0.1\linewidth}
	\hspace*{0mm}
	\end{minipage}
	\begin{minipage}[h]{0.45\linewidth}
 		\begin{center}
   			\includegraphics[width=7cm] {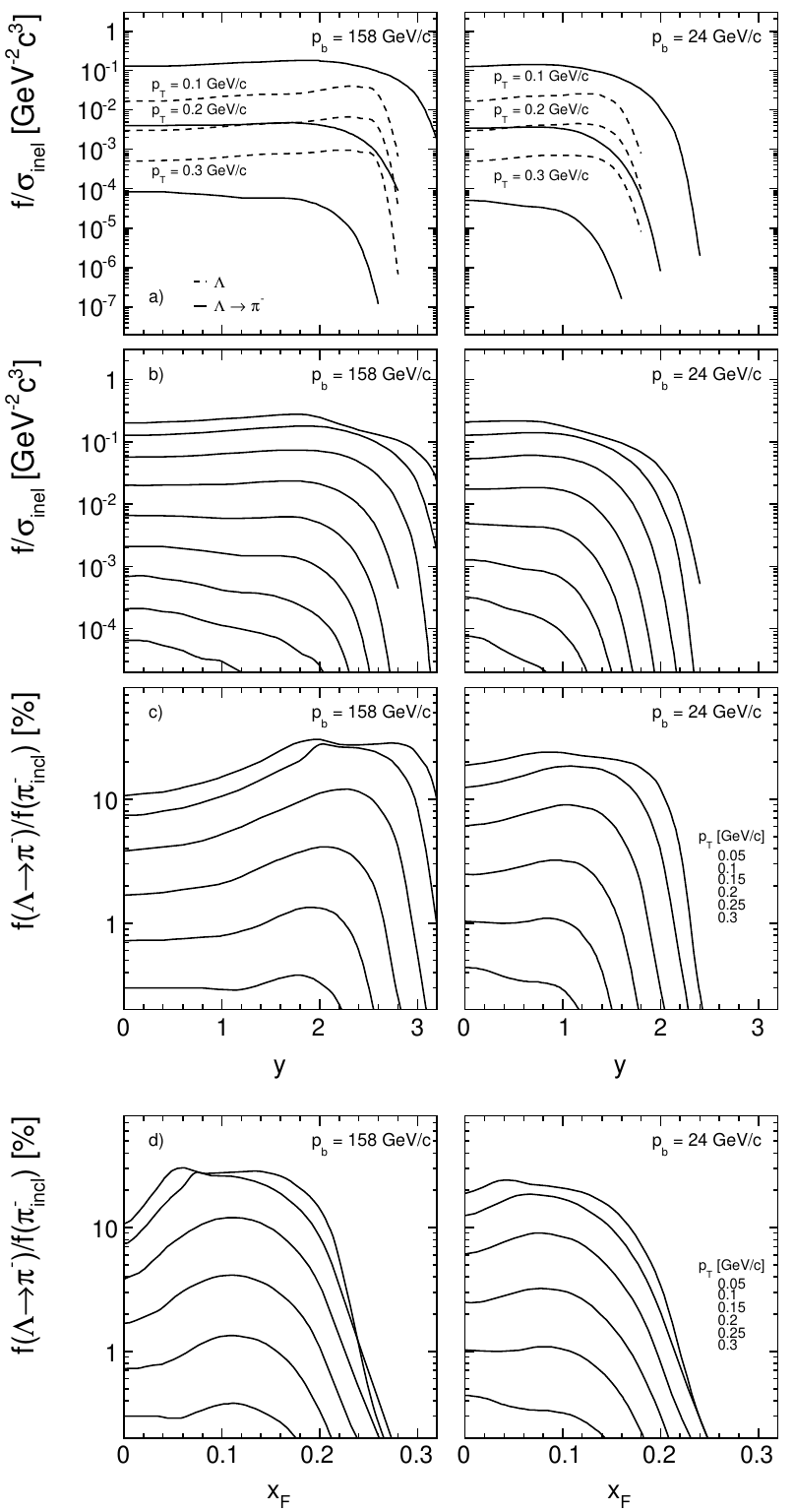} 
			\caption{a) invariant $\pi^-$ cross section $f/\sigma_{\textrm{inel}}$ from $\Lambda$ decay as a function of $y$ for four values of $p_T$ at 158 and 24~GeV/c beam momentum. Full lines for decay $\pi^-$ and broken lines for the parent $\Lambda$. Subsequent $p_T$ values are multiplied by 1/5 for clarity; b) invariant $\pi^-$ cross section $f/\sigma_{\textrm{inel}}$ from $\Lambda$ decay as a function of rapidity for $p_T$ between 0.05 and 1.3~GeV/c; c) percentage ratio of the total $\pi^-$ from $\Lambda$ decay to the inclusive, feed-down subtracted $\pi^-$ yield; d) same as c) but as a function of $x_F$}
  			\label{fig:lam2pi}
 		\end{center}
	\end{minipage}
\end{figure*}

$\Lambda$ decays with a branching fraction of 63.9\% into p + $\pi^-$ with a $Q$ value of only 0.038~GeV/c. This together with the large mass difference of the decay particles proton and $\pi^-$ leads to a rather sharp limitation of the $\pi^-$ in $x_F$ and $p_T^2$, Fig.~\ref{fig:lambdaS}, as compared to the K$^0_S$ decay, Fig.~\ref{fig:kazeros}.
On the double-differential level the relation between parent Lambda and decay pion is presented in Fig.~\ref{fig:lam2pi} as a function of rapidity and $x_F$.

The sizeable contribution of up to 30\% at low $p_T$ is followed by a rapid decrease with increasing $p_T$ such that it vanishes at about $p_T$~=0.3~GeV/c. The maxima in $y$ and $x_F$ are off $y$~=~$x_F$~=~0 due to the wide longitudinal momentum distribution of the $\Lambda$. In fact the $x_F$ value of the $\pi^-$ is approximately related to the $\Lambda$ by the mass ratio: ratio $x_F(\pi^-) \sim m_{\pi^-}/m_{\Lambda} * x_F(\Lambda)$.

%
%
\subsubsection{\boldmath $\Sigma^-$ decay}
\vspace{3mm}
\label{sec:feed_ddsigma}

\begin{figure*}[h]
	\begin{minipage}[h]{0.45\linewidth}
 		\begin{center}
   			\includegraphics[width=7cm] {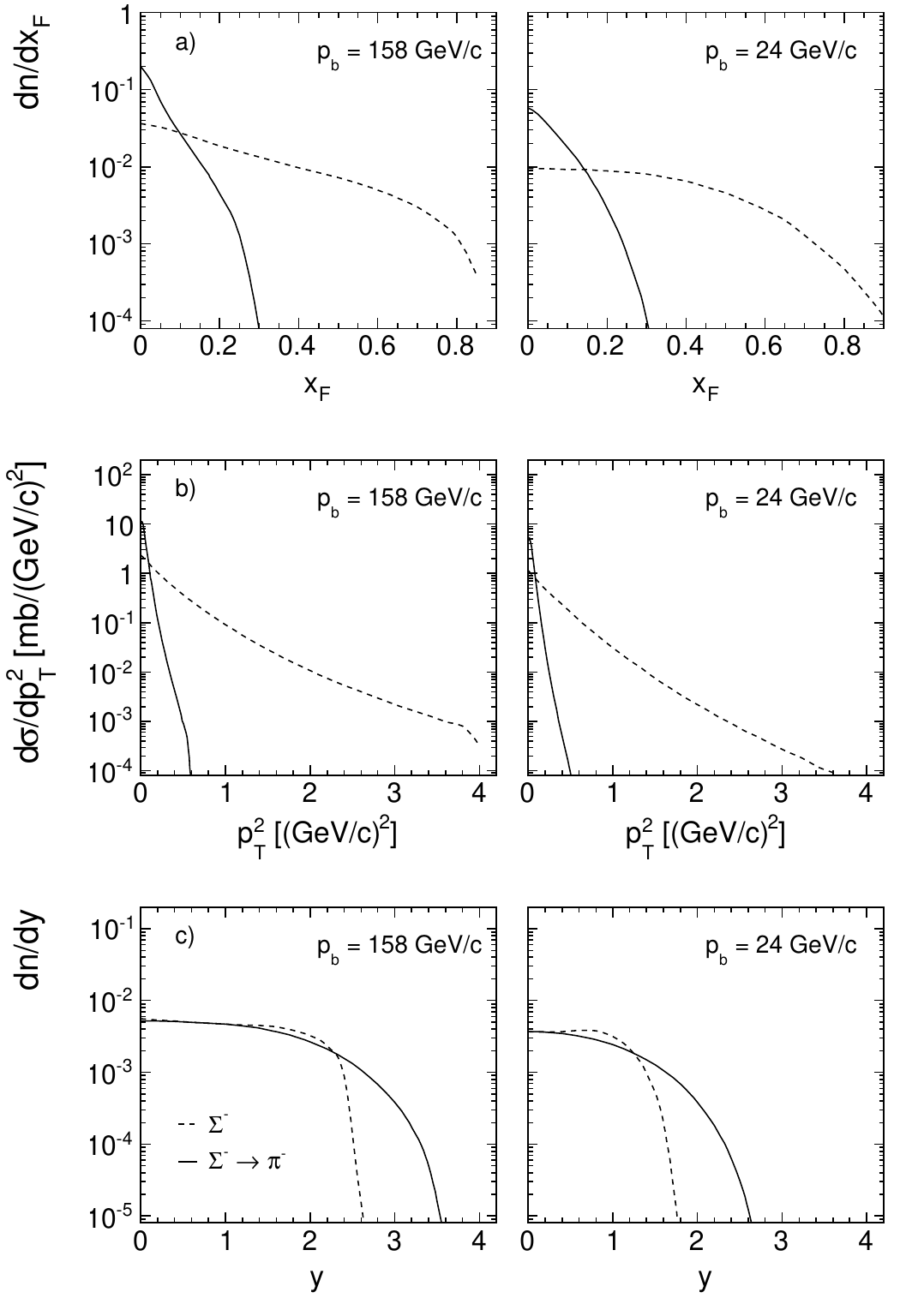} 
			\caption{a) $dn/dx_F$ as a function of $x_F$, b) $d\sigma/dp_T^2$ as a function of $p_T^2$ and c) $dn/dy$ as a function of rapidity for $p_{\textrm{beam}}$~=~158 and 24~GeV/c. The full lines correspond to the $\pi^-$ yields, the broken lines to the parent $\Sigma^-$}
  			\label{fig:sigmaS}
 		\end{center}
	\end{minipage}
	\begin{minipage}[h]{0.1\linewidth}
	\hspace*{0mm}
	\end{minipage}
	\begin{minipage}[h]{0.45\linewidth}
 		\begin{center}
   			\includegraphics[width=6.8cm] {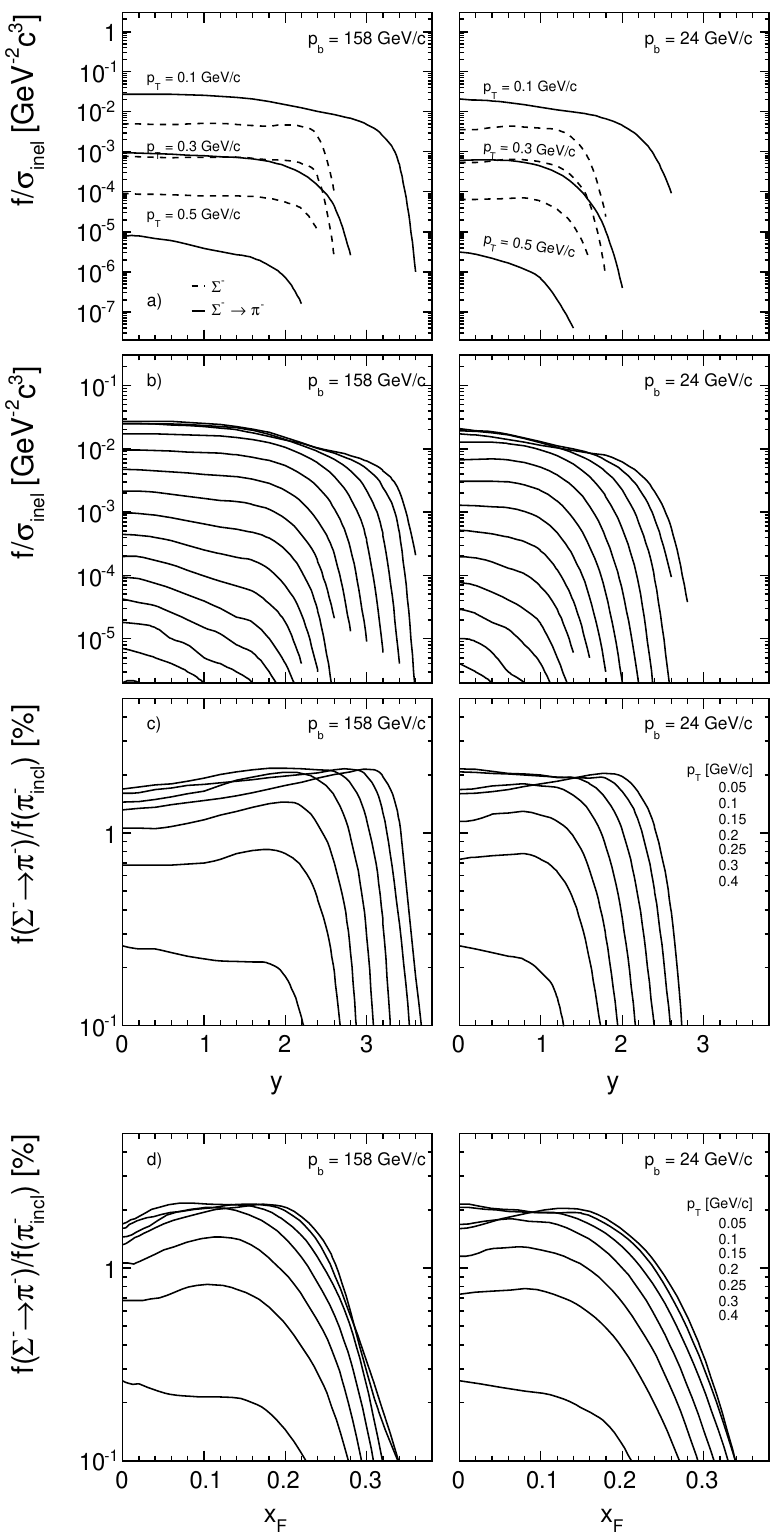} 
			\caption{a) Invariant $\pi^-$ cross section $f/\sigma_{\textrm{inel}}$ from $\Sigma^-$ decay as a function of $y$ for four values of $p_T$ at 158 and 24~GeV/c beam momentum. Full lines for decay $\pi^-$ and broken lines for the parent $\Lambda$. Subsequent $p_T$ values are multiplied by 1/5 for clarity; b) Invariant $\pi^-$ cross section $f/\sigma_{\textrm{inel}}$ from $\Sigma^-$ decay as a function of rapidity for $p_T$ between 0.05 and 1.3~GeV/c; c) Percentage ratio of the total $\pi^-$ from $\Sigma^-$ decay to the inclusive, feed-down subtracted $\pi^-$ yield; d) Same as c) but as a function of $x_F$}
  			\label{fig:sig2pi}
 		\end{center}
	\end{minipage}
\end{figure*}

$\Sigma^-$ decays with a branching fraction of 99.8\% into n + $\pi^-$ with a $Q$ value of 0.118~GeV/c. Compared to the $\Lambda$ decay this extends the range of the decay pions in $x_F$ and $p_T$ as shown in Fig.~\ref{fig:sigmaS}.

On the double-differential level the relation between parent $\Sigma$ and decay pion is presented in Fig.~\ref{fig:sig2pi} as a function of rapidity and $x_F$.

The contribution at low $p_T$ is much smaller than for the $\Lambda$ which is another consequence of the bigger $Q$ value. It vanishes at about $p_T >$~0.4~GeV/c. The maxima in $x_F$ are again in the region of 0.15.

%
%
\subsubsection{Total feed-down}
\vspace{3mm}
\label{sec:feed_tot}

\begin{figure*}[h]
	\begin{minipage}[h]{0.45\linewidth}
 		\begin{center}
   			\includegraphics[width=7.cm] {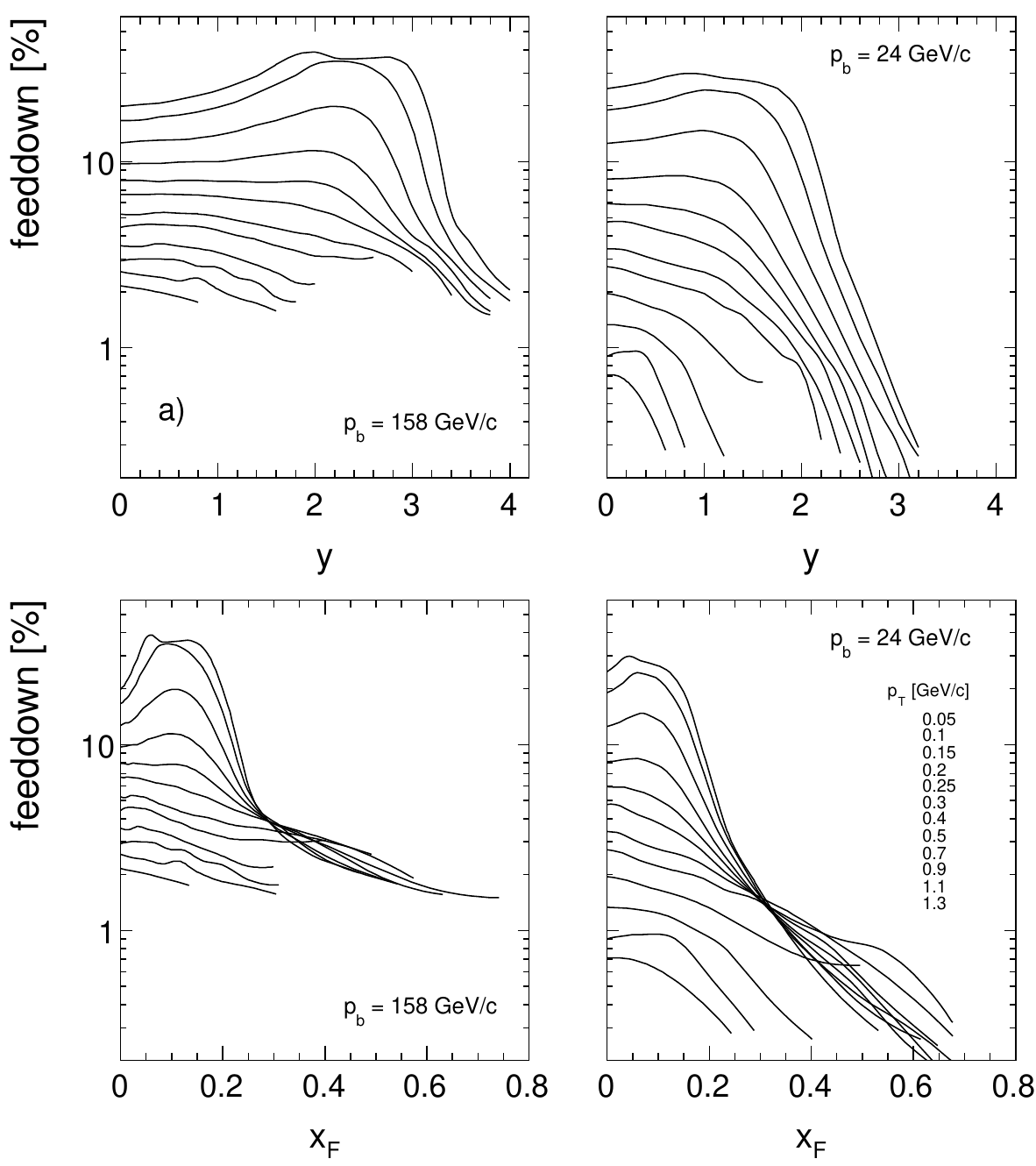} 
 		\end{center}
	\end{minipage}
	\begin{minipage}[h]{0.1\linewidth}
	\hspace*{0mm}
	\end{minipage}
	\begin{minipage}[h]{0.45\linewidth}
 		\begin{center}
   			\includegraphics[width=7cm] {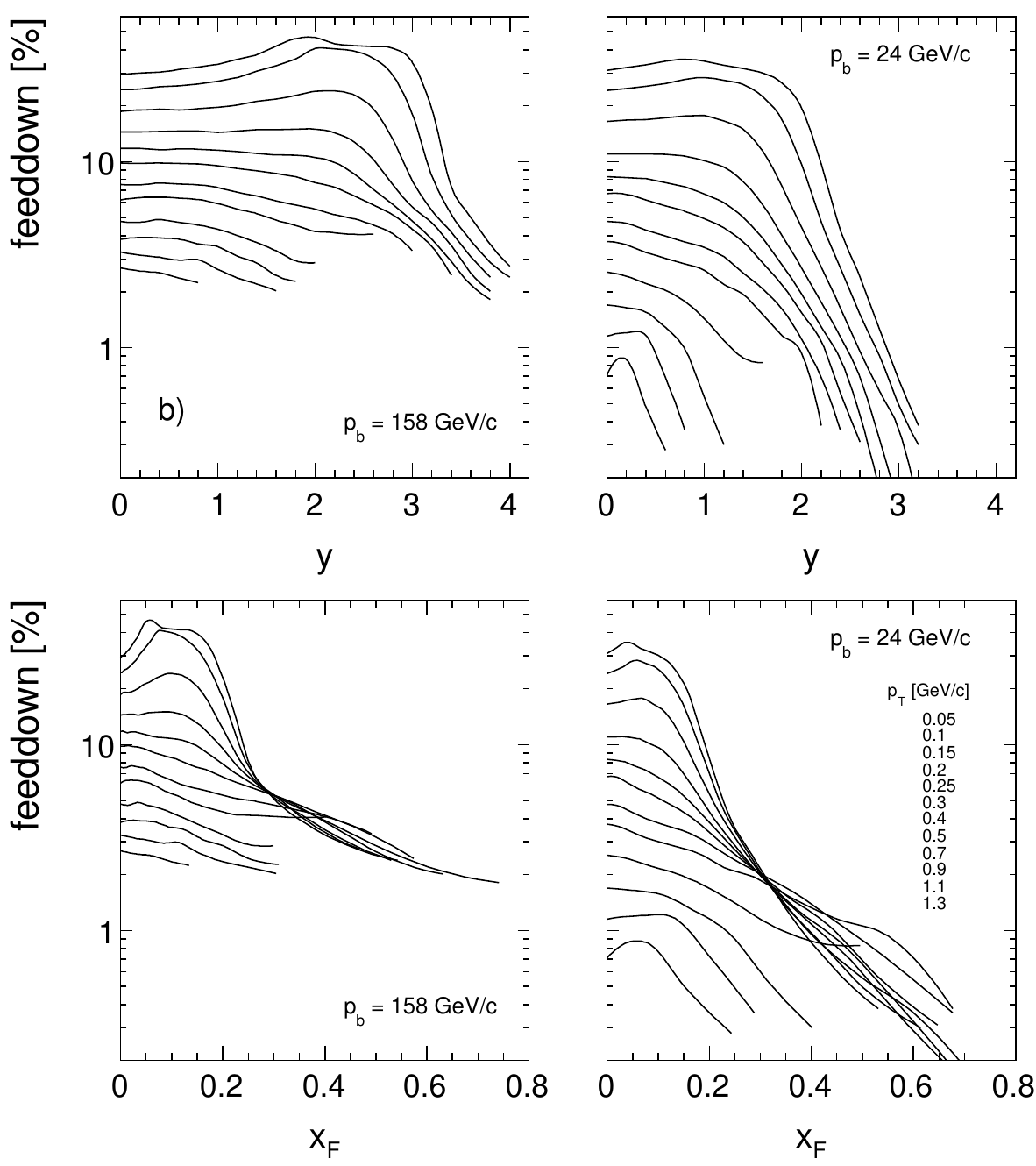} 
 		\end{center}
	\end{minipage}
	\caption{Total feed-down contribution from a) K$_S^0$, $\Lambda$ and $\Sigma^-$ b) ( K$_S^0$ + K$_L^0$ ), $\Lambda$ and $\Sigma^-$ 		as a function of $y$ and $x_F$ for beam momenta of 158 and 24~GeV/c}
  	\label{fig:totalfeed}
\end{figure*}

As a conclusion of this section on feed-down the sum of the components from K$_S^0$, (K$_S^0$+K$_L^0$), $\Lambda$ and $\Sigma^-$ decays are given in Fig.~\ref{fig:totalfeed} at 158 and 24~GeV/c beam momentum.

It is apparent that this correction spans a large region both in rapidity and in $x_F$ and, being concentrated at low $p_T$ nevertheless covers the complete  $p_T$ range at a non-negligible level. It reaches more than 30\% at 24~GeV/c and more than 40\% at 158~GeV/c beam momentum at $p_T$ below 0.2~GeV/c.

%
%
\section{A comment on inverse $m_T$ slopes ("Temperature")}
\vspace{3mm}
\label{sec:mtslopes}

In so called "thermal" models it is claimed that final state hadrons are characterized by a general, mass-independent transverse momentum spectrum if the invariant cross section is plotted against

\begin{equation}
     m_T = \sqrt{ p_T^2 + m^2 }
     \label{eq:mt}
\end{equation}
rather than $p_T$, in the scale $m_T - m$

The inverse slope of the $m_T$ distributions is assumed to be independent of $m$ and $m_T$ and is brought in connection with a thermal radiator of temperature $T$.

In \cite{hagedorn} it is admitted that this universality does in general not hold for resonance decays. In this sense it is interesting to regard the above study of weak decays of kaons and hyperons into negative pions in terms of inverse $m_T$ slopes.

\begin{figure*}[h]
	\begin{minipage}[h]{0.47\linewidth}
 		\begin{center}
   			\includegraphics[width=7.cm] {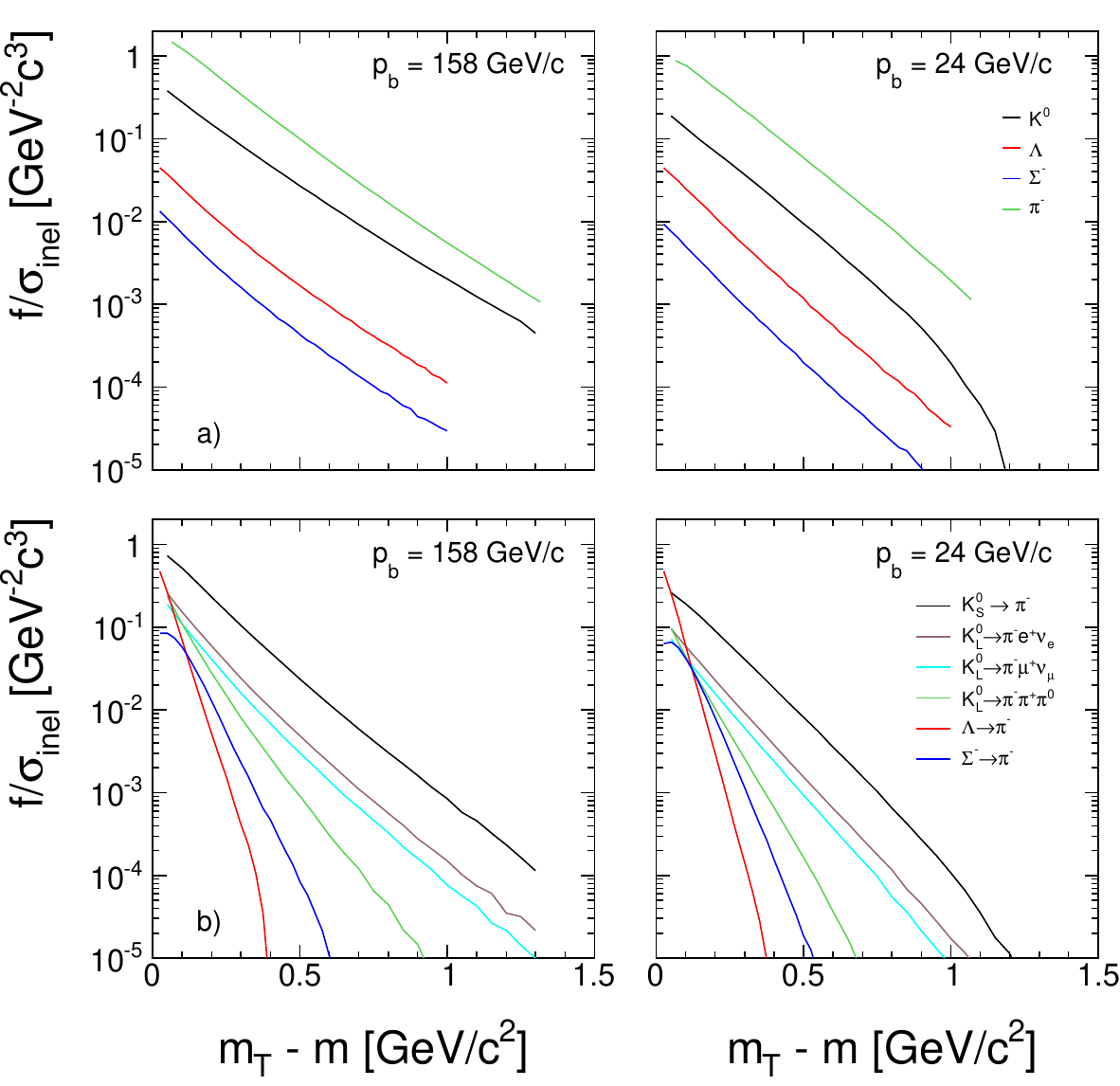} 
			\caption{a) $m_T-m$ distributions for inclusive $\pi^-$, K$^0$, $\Lambda$ and $\Sigma^-$ for the beam momenta of 158 and 24~GeV/c; b) corresponding distributions for the decay pions including the 3 body decays of K$_L^0$}
  			\label{fig:mtdist}
 		\end{center}
	\end{minipage}
	\begin{minipage}[h]{0.05\linewidth}
	\hspace*{0mm}
	\end{minipage}
	\begin{minipage}[h]{0.47\linewidth}
 		\begin{center}
   			\includegraphics[width=7.cm] {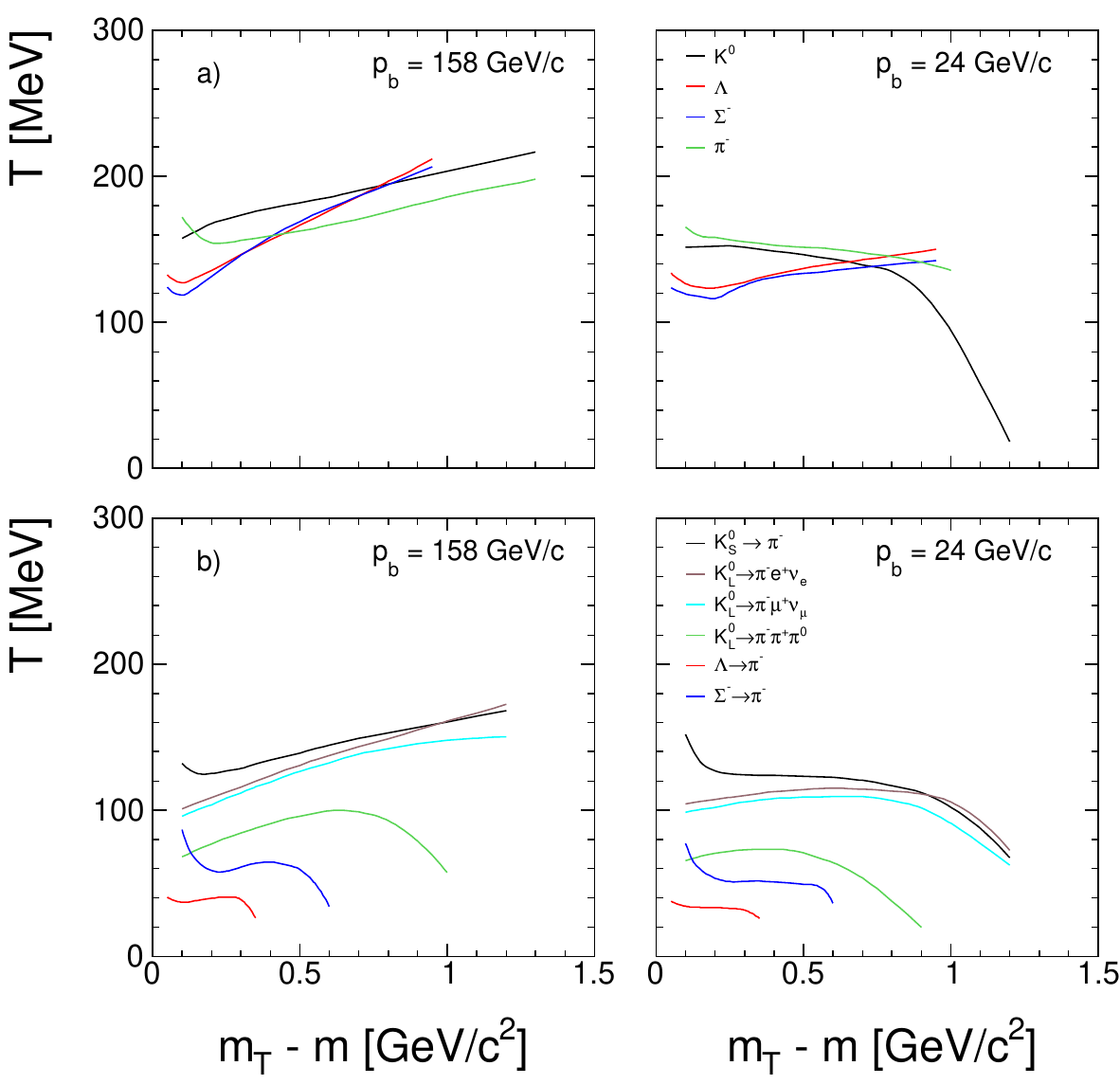} 
			\caption{a) Local inverse $m_T$ slopes for inclusive $\pi^-$, K$^0$, $\Lambda$ and $\Sigma^-$ as a function of $m_T$-m at 158 and 24~GeV/c beam momentum; b) inverse slopes for the decay $\pi^-$ from 2 and 3 body decays}
  			\label{fig:mtslopes}
 		\end{center}
	\end{minipage}
\end{figure*}

Some examples of $m_T-m$ distributions for K$_S^0$, $\Lambda$ and $\Sigma^-$ and their decay pions are shown in Fig.~\ref{fig:mtdist}.

Analyzing these distributions for their inverse slopes ("Temperature") one obtains the local inverse slopes presented in Fig.~\ref{fig:mtslopes}.

Several features are noteworthy here:

As far as the the $m_T-m$ distributions (Fig.~\ref{fig:mtdist}) are concerned, already for the parent particles including the inclusive $\pi^-$ production which is generally regarded as a prime example of "thermal" behaviour, the shapes are in general not exponential. There are marked differences between 158 and 24~GeV/c beam momentum and between the particle species.These differences increase for the decay pions where the $m_T$ distributions are in general much steeper. This is borne out quantitatively by the inverse slopes as a function of $m_T-m$ (Fig.~\ref{fig:mtslopes}). For the parent particles there is a wide range of "Temperatures" both regarding the $m_T$ and particle type dependences, superimposed to a strong variation with beam momentum. The deviations from "thermal" behaviour are much more pronounced for the decay pions where it becomes apparent that it is the $Q$-value of the respective 2 or 3 body decays that dominates the resulting inverse slopes. It should be stressed that these results are valid for central rapidity. A more detailed discussion of the $y$ and $x_F$ dependences will be presented in Sect.~\ref{sec:res_invslope} below, keeping in mind that transverse momentum arguments are in general not valid in rapidity as $y$ and $p_T$ are not orthogonal for $|y| >$~0. Several facts emerge from this study:

\begin{enumerate}
	\item The contribution from only three weekly decaying particles to the total $\pi^-$ yield amounts, at $\sqrt{s} >$~10~GeV, to 13\% and 10\% with and without K$_L^0$ decay, respectively (Fig~\ref{fig:feedtot})
	\item As is well known \cite{grassler,jancso} most if not all final state hadrons emerge from resonance decays, a fact established up to the highest ISR energy
	\item The role of resonance decays in particle production has therefore to be scrutinized in detail before taking reference to oversimplified "models"
\end{enumerate}

%
%
\section{Global interpolation as a function of \boldmath $x_F$, $y_{\textrm{lab}}$, $p_T$ and $\log(s)$ with and without feed-down correction}
\vspace{3mm}
\label{sec:interpolation}

The data interpolation scheme introduced in Sect.~\ref{sec:data} above has been established in the variables $p_T$, $y_{\textrm{lab}}$ and $\log(s)$ and $p_T$, $x_F$ and $\log(s)$ using the reference data (Sect.~\ref{sec:data_ref}). As the original data from bubble chambers and NA49 are feed-down subtracted, and the ISR data are given without feed-down subtraction, a total of four sets of interpolated results had to be produced in order to accomplish a complete and consistent picture with and without feed-down correction. The corresponding four large sets of cross sections with about 10$^\textrm{4}$ bins each are available on the web-site spshadrons of the NA49 pp/pA group \cite{webpage}. A sub-sample is shown as Table~\ref{tab:interp} for six values of $\log(s)$ as functions of $y_{\textrm{lab}}$ and $p_T$.

\begin{table}
	\renewcommand{\tabcolsep}{0.17pc}
	\renewcommand{\arraystretch}{0.93}
	\tiny
	\input{table.tex}
	\caption{Interpolated cross sections $f/\sigma_{\textrm{inel}}$ as functions of $p_T$ and $y_{\textrm{lab}}$ for six values of $\log(s)$, with feed-down subtraction. The uppermost $y_{\textrm{lab}}$ values at each energy correspond to $y$~=~0. The cut-off at lower $y_{\textrm{lab}}$ is given by the availability of data at high cms rapidity where the limit in $f/\sigma_{\textrm{inel}}$ varies, with increasing $\log(s)$, from 0.01 to 0.0001}
	\label{tab:interp}
\end{table}

Some corresponding plots are presented in Fig.~\ref{fig:interp} in the same variables.

\begin{figure}[h]
 \begin{center}
   \includegraphics[width=16cm] {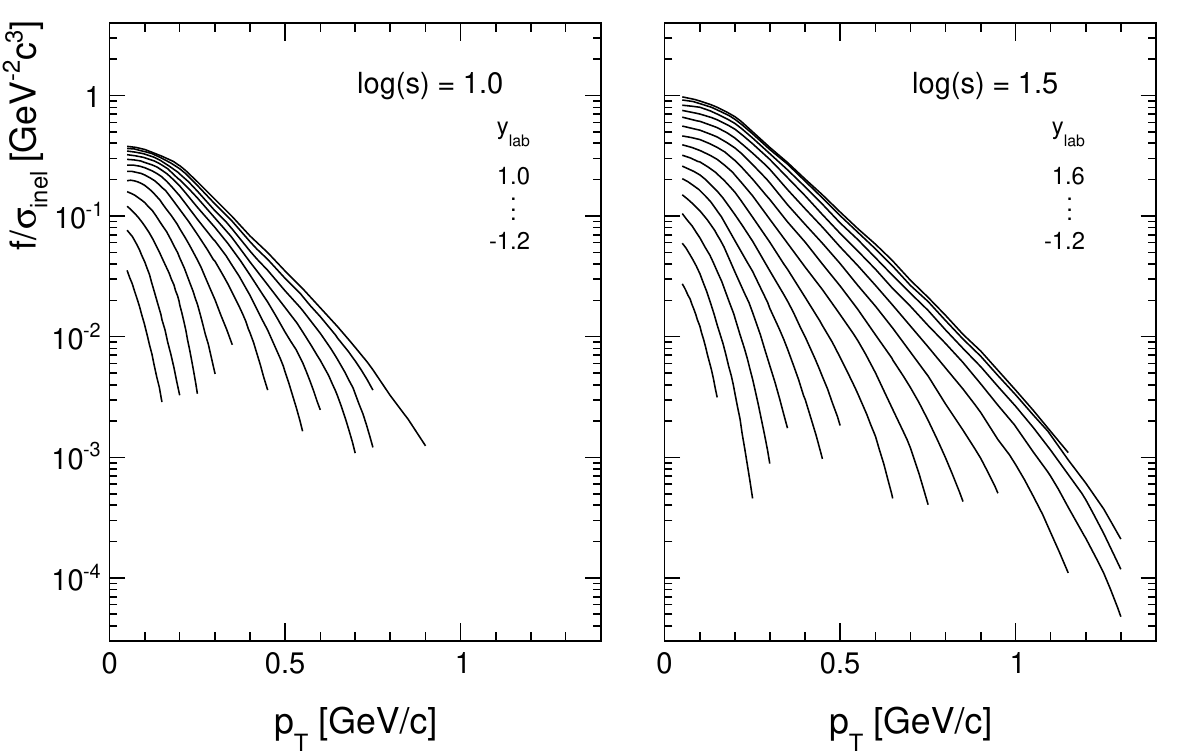} 
 \end{center}
\end{figure}
\begin{figure}[h]
 \begin{center}
   \includegraphics[width=16cm] {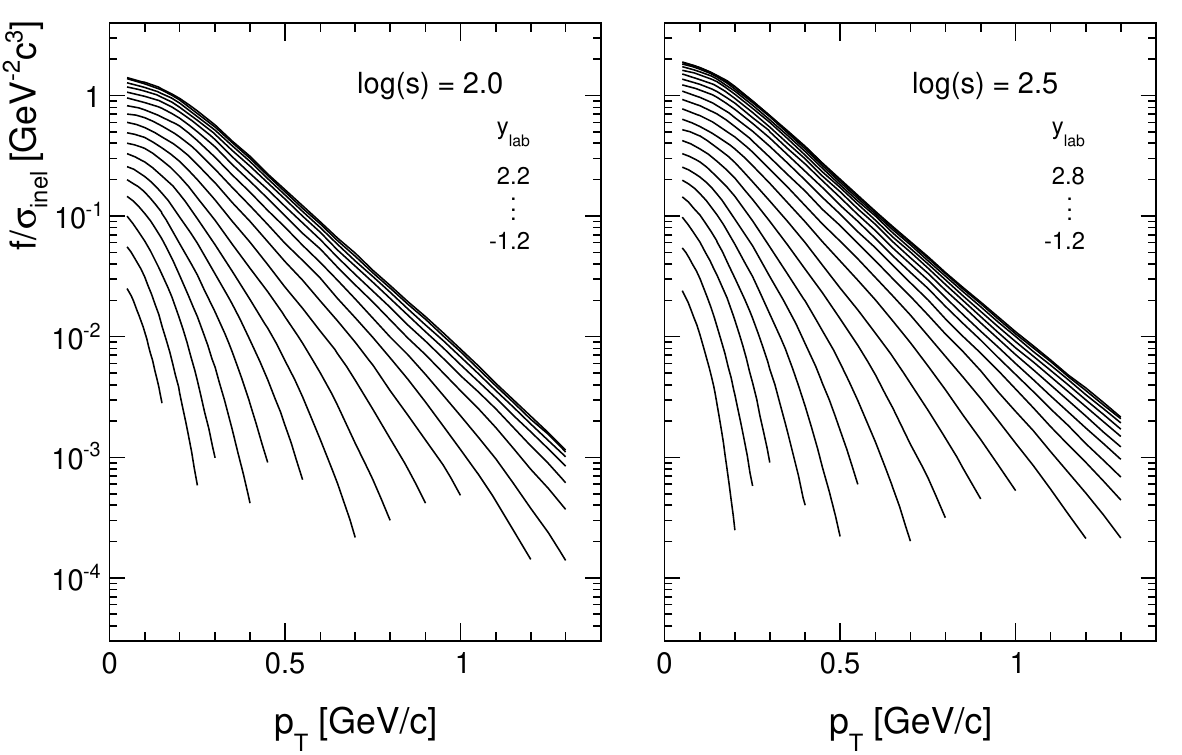} 
 \end{center}
\end{figure}
\begin{figure}[h]
 \begin{center}
   \includegraphics[width=16cm] {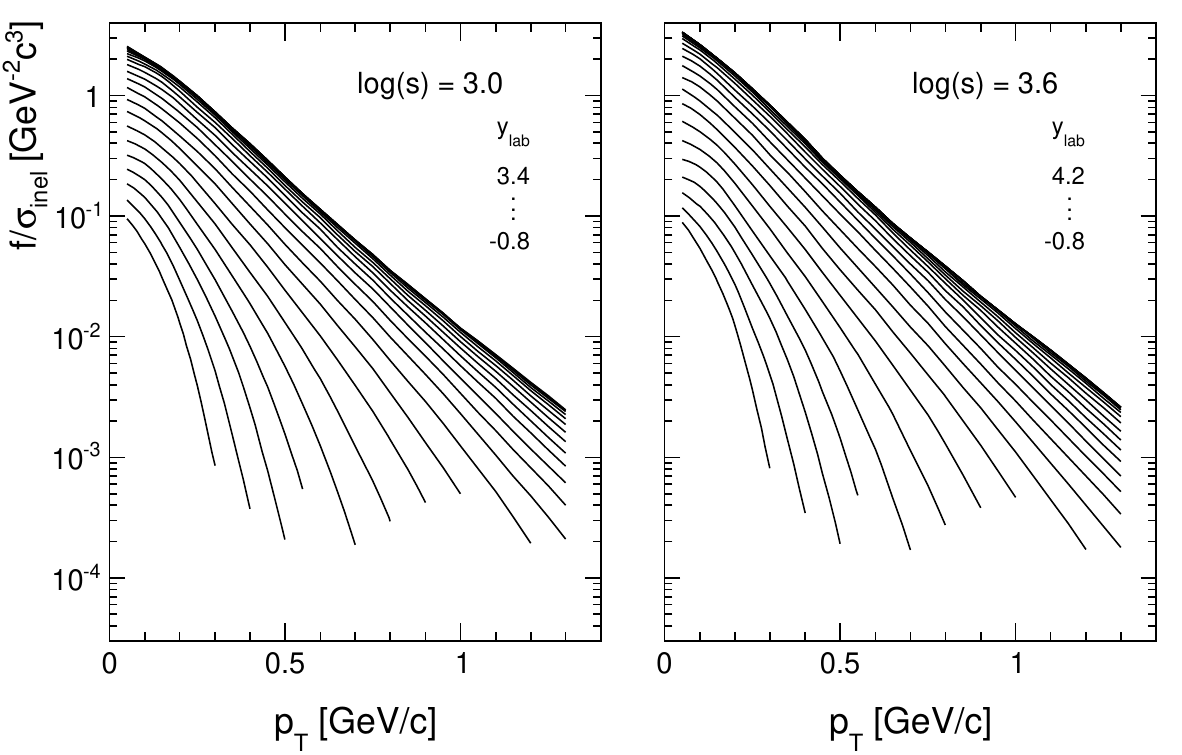} 
	\caption{Interpolated cross sections $f/\sigma_{\textrm{inel}}$ as functions of $p_T$ and $y_{\textrm{lab}}$ for six values of $\log(s)$, from the lower limit corresponding to $\sqrt{s}$~=~3~GeV to the highest ISR energy at $\sqrt{s}$~=~63~GeV}
  \label{fig:interp}
 \end{center}
\end{figure}

%
%
\section{Comparison of the global interpolation to the reference data}
\vspace{3mm}
\label{sec:inter2refdata}

This section will compare the multitude of data points obtained by the 20 reference experiments with the global interpolation in a quantitative way showing, at each energy, the cross sections as functions of $y_{\textrm{lab}}$ and $p_T$ as well as the residual distributions of the respective data points with respect to the interpolation. This includes the mean value and standard deviation in units of $\Delta/\sigma$ where $\Delta$ is the deviation from the interpolation and $\sigma$ gives the statistical error of each point. The following figures will, for clarity, show two experiments per page with, whenever indicated, some remarks.

%
%
\subsection{Bubble chamber and NA49 data}
\vspace{3mm}
\label{sec:interbubble}

\begin{figure}[h]
 \begin{center}
   \begin{turn}{\rotAngle}
   \begin{minipage}{\capw}
	\caption{Cross sections $f/\sigma_{\textrm{inel}}$ as functions of $y_{\textrm{lab}}$ and $p_T$ at $p_{\textrm{beam}}$ of 6.6 and 12~GeV/c. The data at 6.6~GeV/c come from a hitherto unexploited PhD work at LBL \cite{gellert} and give an important and precise reference near the lower range of the $\log(s)$ scale considered here. The data at 12~GeV/c are part of the precision data of Blobel et al. \cite{blobel} with an -- for bubble chamber work -- unprecedented number of 75k reconstructed $\pi^-$}
  	\label{fig:int2data6.6_12}
  	\end{minipage}
  \end{turn}
   \includegraphics[width=\figw,angle=\rotAngle] {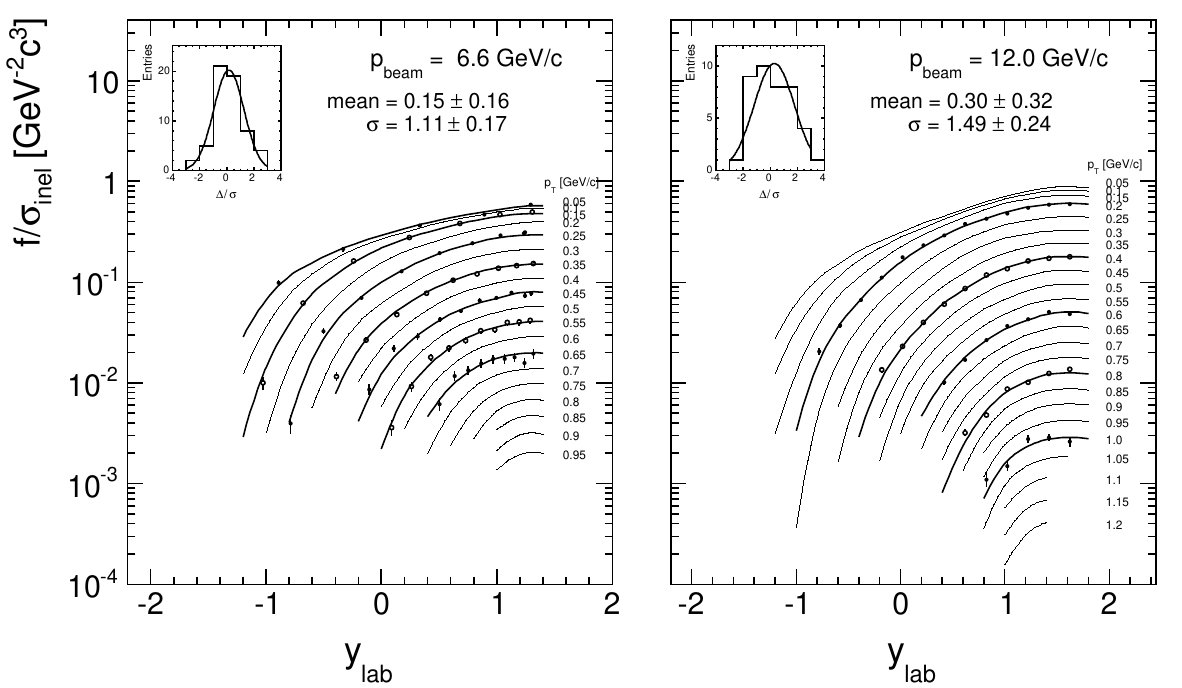} 
 \end{center}
\end{figure}

\begin{figure}[h]
 \begin{center}
   \begin{turn}{\rotAngle}
   \begin{minipage}{\capw}
   	\caption{Cross sections $f/\sigma_{\textrm{inel}}$ as functions of $y_{\textrm{lab}}$ and $p_T$ at $p_{\textrm{beam}}$ of 12.9 and 18~GeV/c. These data from D.B.Smith et al. \cite{smith} are published in Landolt-B\"ornstein \cite{landolt} at 5 beam energies. Although with less statistics than the preceding experiments they give important information in the low-$p_T$ region establishing the deformation of the $y_{\textrm{lab}}$ distributions at $p_T$~=~0.05~GeV/c around $y_{\textrm{lab}}$~=~0.5}
  	\label{fig:int2data12.9_18}
	\end{minipage}
	\end{turn}
   \includegraphics[width=\figw,angle=\rotAngle] {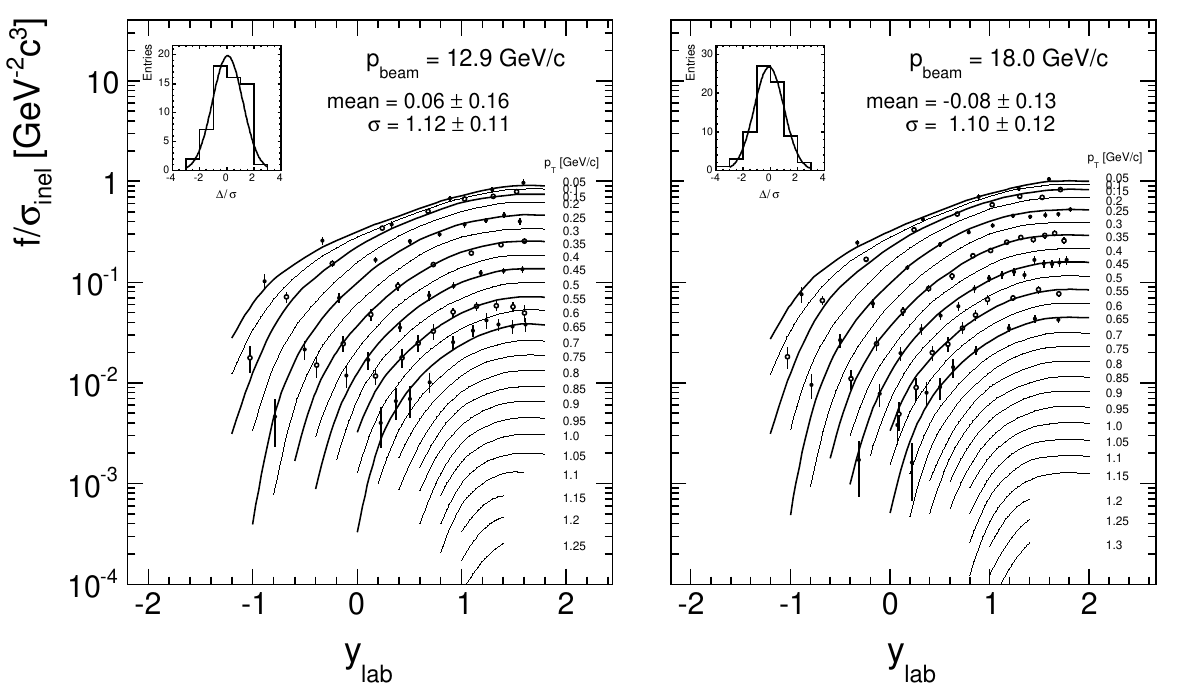} 
 \end{center}
\end{figure}

\begin{figure}[h]
 \begin{center}
   \begin{turn}{\rotAngle}
   \begin{minipage}{\capw}
	\caption{Cross sections $f/\sigma_{\textrm{inel}}$ as functions of $y_{\textrm{lab}}$ and $p_T$ at $p_{\textrm{beam}}$ of 19.2 and 21~GeV/c. The data at 19.2~GeV/c are from B{\o}ggild et al. \cite{boggild} as published in \cite{landolt}, and D.B.Smith et al. \cite{smith} at 21~GeV/c}
  	\label{fig:int2data19_21}
	\end{minipage}
	\end{turn}
  \includegraphics[width=\figw,angle=\rotAngle] {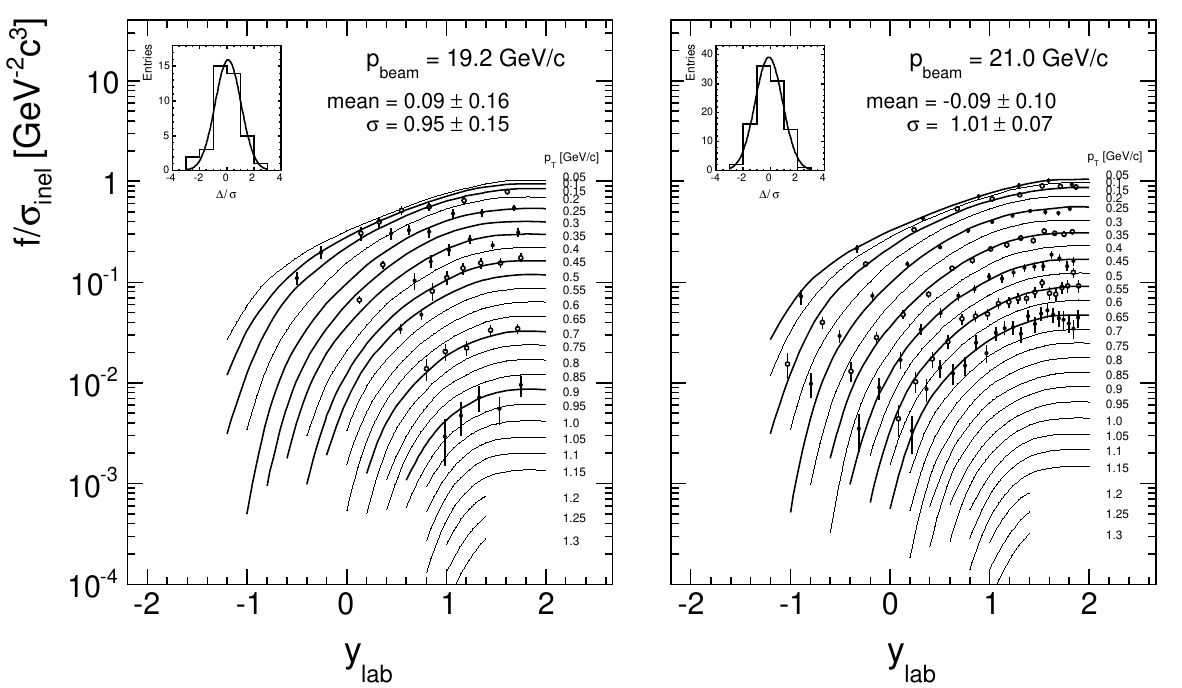} 
 \end{center}
\end{figure}

\begin{figure}[h]
 \begin{center}
   \begin{turn}{\rotAngle}
   \begin{minipage}{\capw}
	\caption{Cross sections $f/\sigma_{\textrm{inel}}$ as functions of $y_{\textrm{lab}}$ and $p_T$ at $p_{\textrm{beam}}$ of 24~GeV/c. Blobel et al. \cite{blobel} have reconstructed 65k $\pi^-$ and the data from Smith \cite{smith} demonstrate the importance of having two independent data sets with a different coverage of phase space}
  	\label{fig:int2data24_24}
	 \end{minipage}
	 \end{turn}
   \includegraphics[width=\figw,angle=\rotAngle] {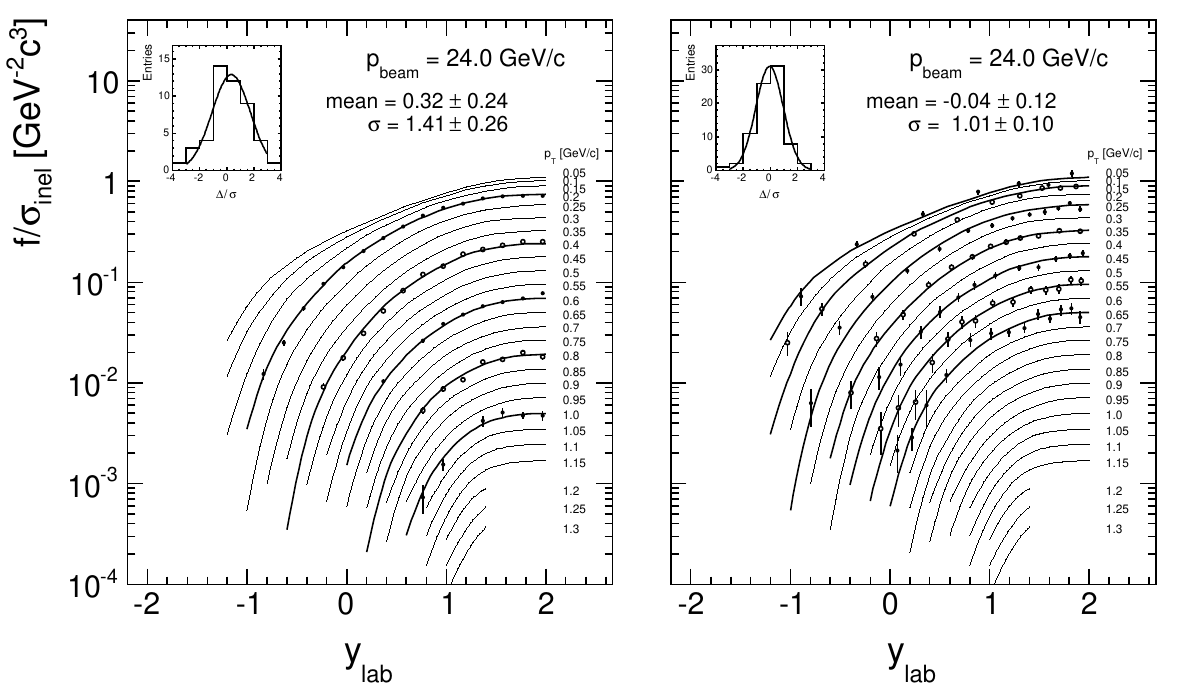} 
 \end{center}
\end{figure}

\begin{figure}[h]
 \begin{center}
   \begin{turn}{\rotAngle}
   \begin{minipage}{\capw}
	\caption{Cross sections $f/\sigma_{\textrm{inel}}$ as functions of $y_{\textrm{lab}}$ and $p_T$ at $p_{\textrm{beam}}$ of 28.4 and 28.5~GeV/c. Again the data of Smith \cite{smith} and Sims \cite{sims} cover the same $p_{\textrm{beam}}$ and comply perfectly with each other. The case of Sims \cite{sims} is special in the sense that the data come from a large exposure of 80k events but only 3 values of $p_T$ are given in the publication. The rest is hidden in a parametrization as functions of $p_T$ (double Gaussian), $y$ (exponential and Gaussian), and $x_F$ (exponential and Gaussian). If calculating the cross sections in one of the parametrizations one obtains the result shown in Fig.~\ref{fig:int2data28_32}}
  	\label{fig:int2data28_28}
	 \end{minipage}
	 \end{turn}
   \includegraphics[width=22.5cm,angle=\rotAngle] {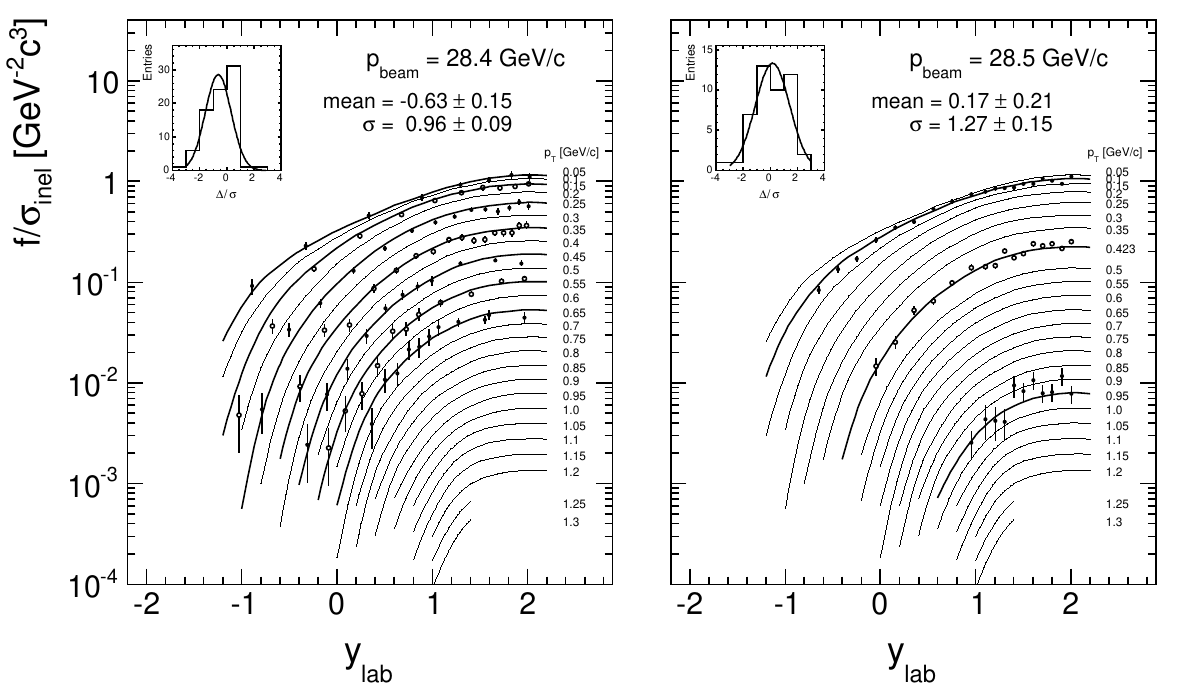} 
 \end{center}
\end{figure}

\begin{figure}[h]
 	\begin{center}
   	\begin{turn}{\rotAngle}
   	\begin{minipage}{\capw}
	 		\caption{Cross section values obtained from the algebraic representation \cite{sims} as a function of $y_{\textrm{lab}}$ and $p_T$ (broken lines). Full lines: global interpolation at $p_{\textrm{beam}}$~=~28.5~GeV/c. This comparison demonstrates the problem of describing data sets with simple algebraic fits. The resulting cross sections do not give justice to the precise, high-statistics data originally obtained. Second panel: data of Zabrodin et al. \cite{zabrodin} at $p_{\textrm{beam}}$ of 32 GeV/c at the Serpukhov accelerator}
   		\label{fig:int2data28_32}
	 	\end{minipage}
		\end{turn}
   	\includegraphics[width=\figw,angle=\rotAngle] {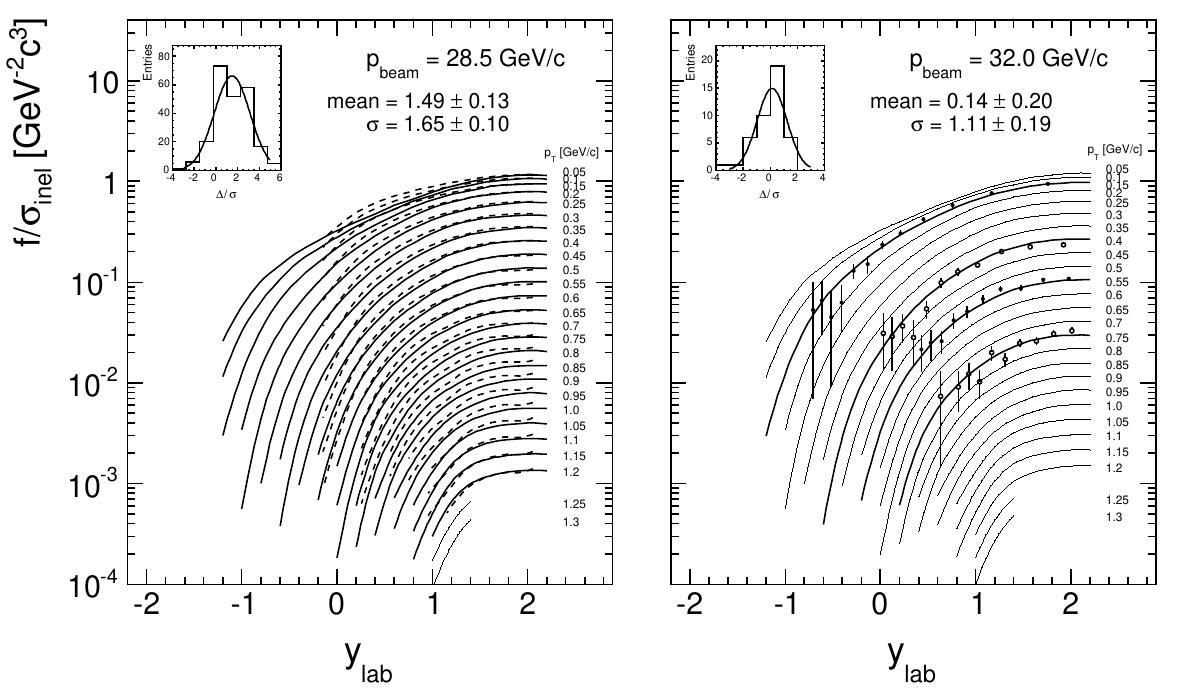} 
 	\end{center}
\end{figure}

\begin{figure}[h]
 \begin{center}
   \begin{turn}{\rotAngle}
   \begin{minipage}{\capw}
	\caption{Cross sections $f/\sigma_{\textrm{inel}}$ as functions of $y_{\textrm{lab}}$ and $p_T$ at $p_{\textrm{beam}}$ of 69~GeV/c, Ammosov et al. \cite{ammosov} and 102~GeV/c, Bromberg et al. \cite{bromberg}}
  	\label{fig:int2data69_102}
	\end{minipage}
	\end{turn}
   \includegraphics[width=\figw,angle=\rotAngle] {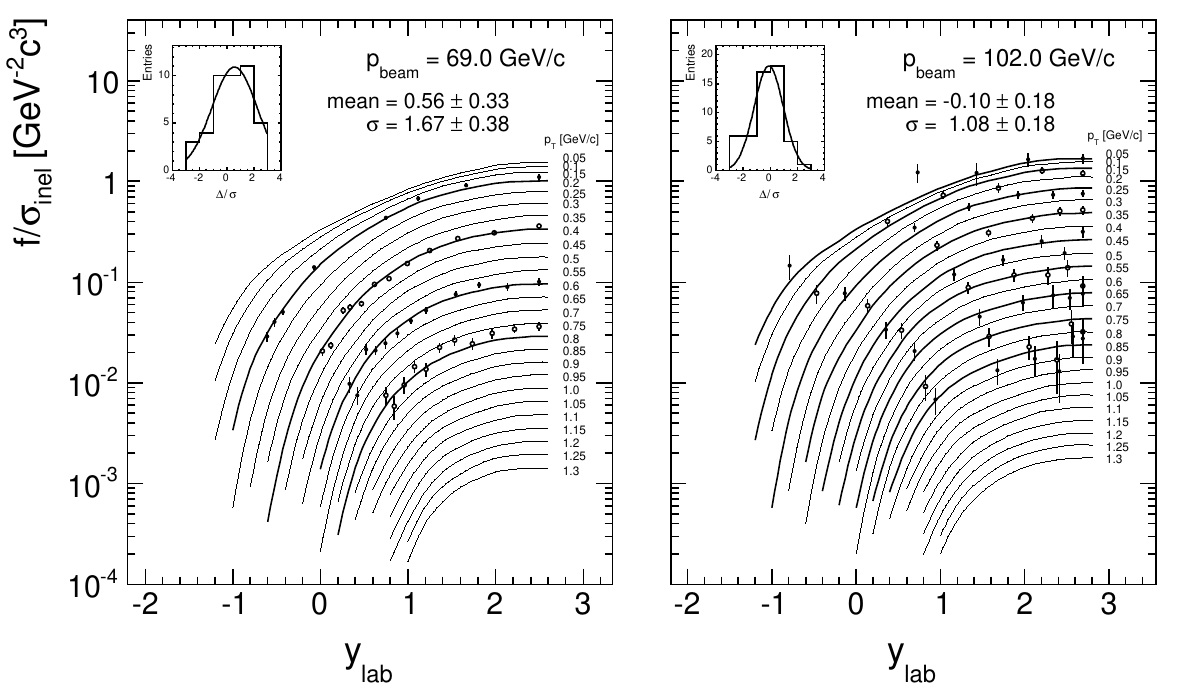} 
 \end{center}
\end{figure}

\begin{figure}[h]
 \begin{center}
   \begin{turn}{\rotAngle}
   \begin{minipage}{\capw}
	\caption{Cross sections $f/\sigma_{\textrm{inel}}$ as functions of $y_{\textrm{lab}}$ and $p_T$ at $p_{\textrm{beam}}$~=~158~GeV/c, Alt et al. \cite{pp_pion} and 400~GeV/c, Bromberg et al. \cite{bromberg}}
  	\label{fig:int2data158_400}
	\end{minipage}
	\end{turn}
   \includegraphics[width=\figw,angle=\rotAngle] {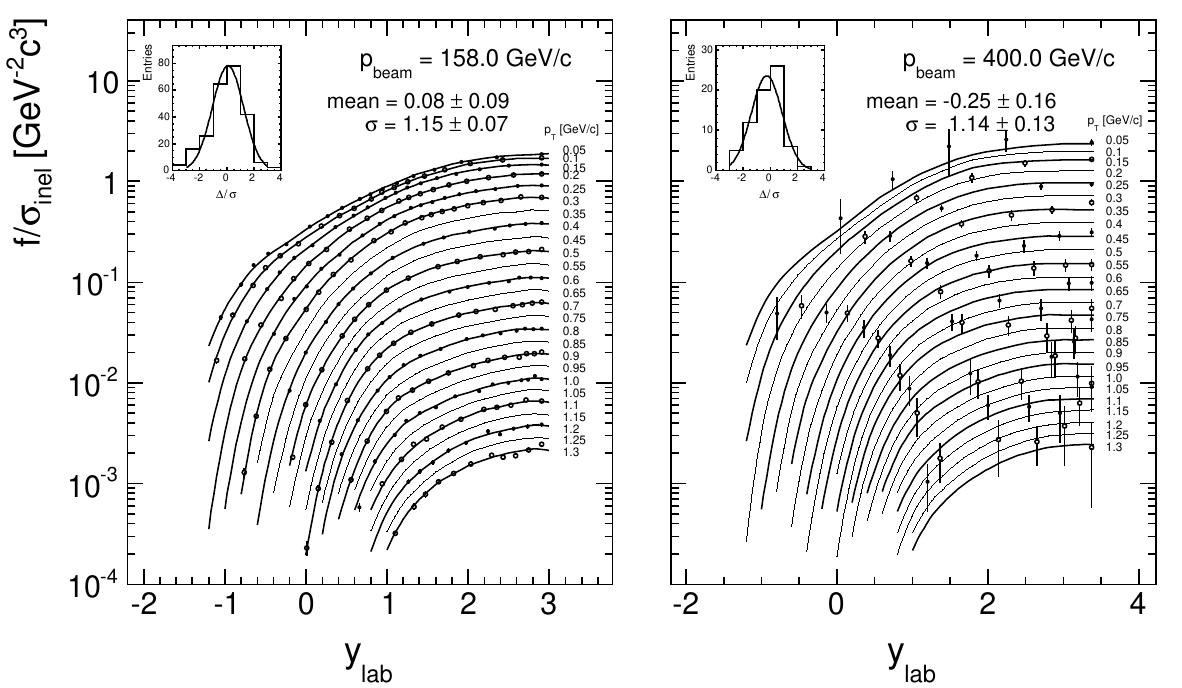} 
 \end{center}
\end{figure}

\newpage
%
%
\subsection{ISR data}
\vspace{3mm}
\label{sec:interisr}

\begin{figure}[h]
 \begin{center}
   \includegraphics[width=12cm] {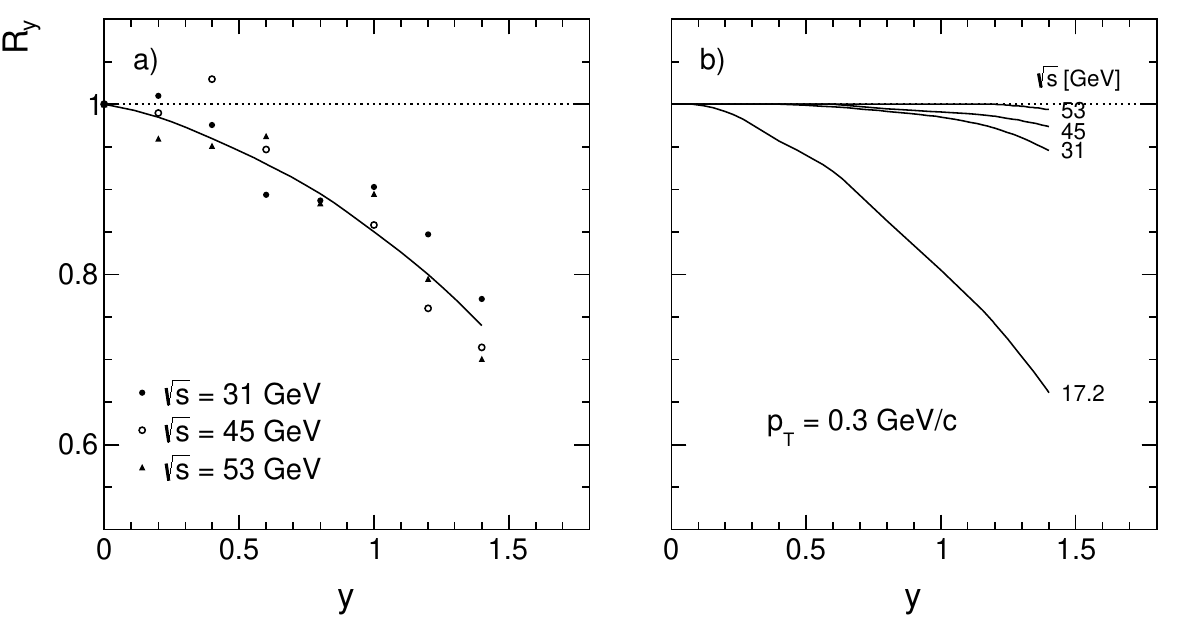} 
	\caption{a) Ratio $R_y$~=~$f(y)/f(0)$ of the forward cross sections \cite{alper} to the central ones as a function of rapidity, averaged over $p_T$, for three ISR energies between 31 and 53~GeV. The full line is given to guide the eye; b) The $y$ dependence of $R_y$ for the general interpolation at $p_T$~=~0.3~GeV/c for the same ISR energies including the NA49 result at $\sqrt{s}$~=~17.2~GeV}
  \label{fig:alpery}
 \end{center}
\end{figure}

As the ISR data have not been corrected for feed-down, Sect.~\ref{sec:isr_data}, the data comparison has to be performed with respect to the general interpolation including feed-down contributions, Sect.~\ref{sec:interpolation}. The four independent experiments, \cite{albrow,alper,capi,guettler,albrow45,albrow53,alper2,capi2}, do not cover the phase space in a continuous fashion but define  a range of high $y_{\textrm{lab}}$ or central rapidity \cite{alper,guettler,alper2}, intermediate $y_{\textrm{lab}}$ below 2 \cite{capi,capi2} and the forward rapidity region down to $y_{\textrm{lab}}$~=~-0.5 \cite{albrow,albrow45,albrow53}. The corresponding general interpolation has therefore to rely on the fact that the normalization of the cross sections is precise to a percent level, Sect.~\ref{sec:isr_data}, and that the data benefit from an overlap with the bubble chamber and NA49 data. In this respect the lack of coverage in rapidity between $y \sim$~0 and $y \sim$~2 needs special attention. While in fact \cite{alper} gives data up to rapidity 1.4 there is evidently an experimental problem with these results as shown in Fig.~\ref{fig:alpery}.

There is a systematic, $s$-independent, drop of the measured cross sections which contradicts the expected development of the rapidity plateau width, Fig.~\ref{fig:alpery}b), with cms energy. Instead the results come close to the rapidity distribution at $\sqrt{s}$~=~17.2~GeV.

\begin{figure}[h]
 \begin{center}
   \includegraphics[width=11.5cm] {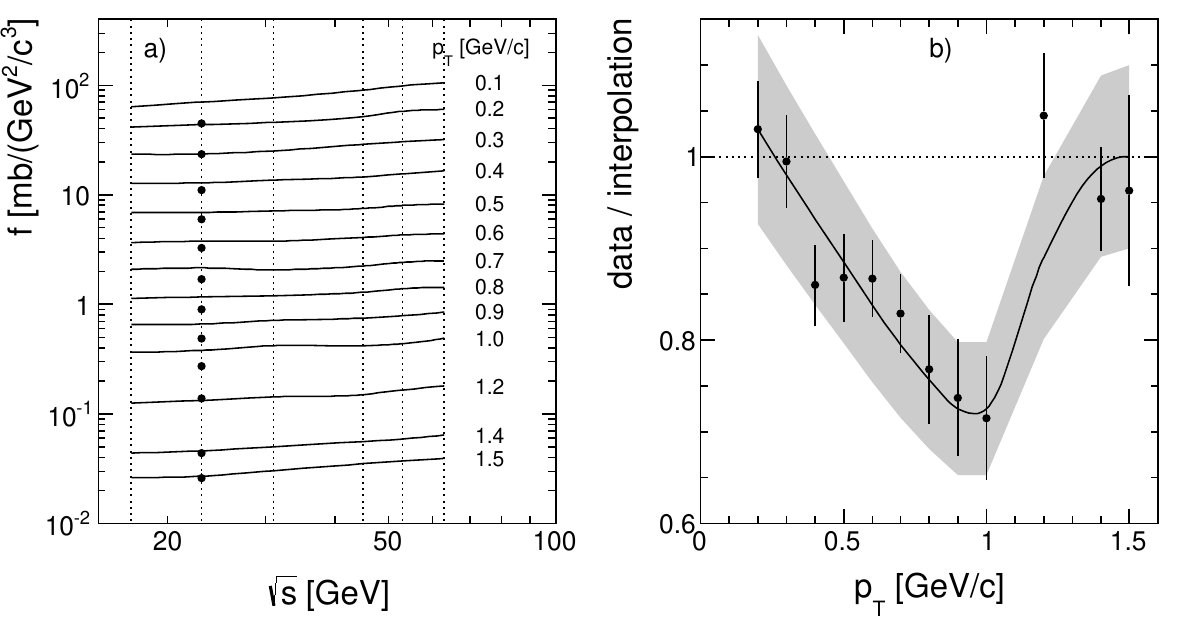} 
	\caption{a) General interpolation at $y$~=~0 as a function of $\log(s)$ and $p_T$ including the data points given by \cite{alper} at $\sqrt{s}$~=~23~GeV; b) Ratio between data and interpolation as a function of $p_T$. The systematic uncertainties are presented as a shaded area}
  \label{fig:alpers}
 \end{center}
\end{figure}

A further problem is apparent in the data \cite{alper} at $\sqrt{s}$~=~23~GeV. Here the cross sections show a sharp drop in the range 0.4~$< p_T <$~1~GeV/c with respect to the general interpolation with deviations up to 30\% as presented in Fig.~\ref{fig:alpers}.

Similar deviations are also observed for kaons \cite{pp_kaon} and protons. The seven data points concerned have been eliminated from the interpolation.

The following Figs.~\ref{fig:int2dataisr23} to \ref{fig:int2dataisr53_63} show the comparison of the ISR data with the global interpolation as mentioned above without feed-down correction.

\begin{figure}[h]
 \begin{center}
   \includegraphics[width=11cm] {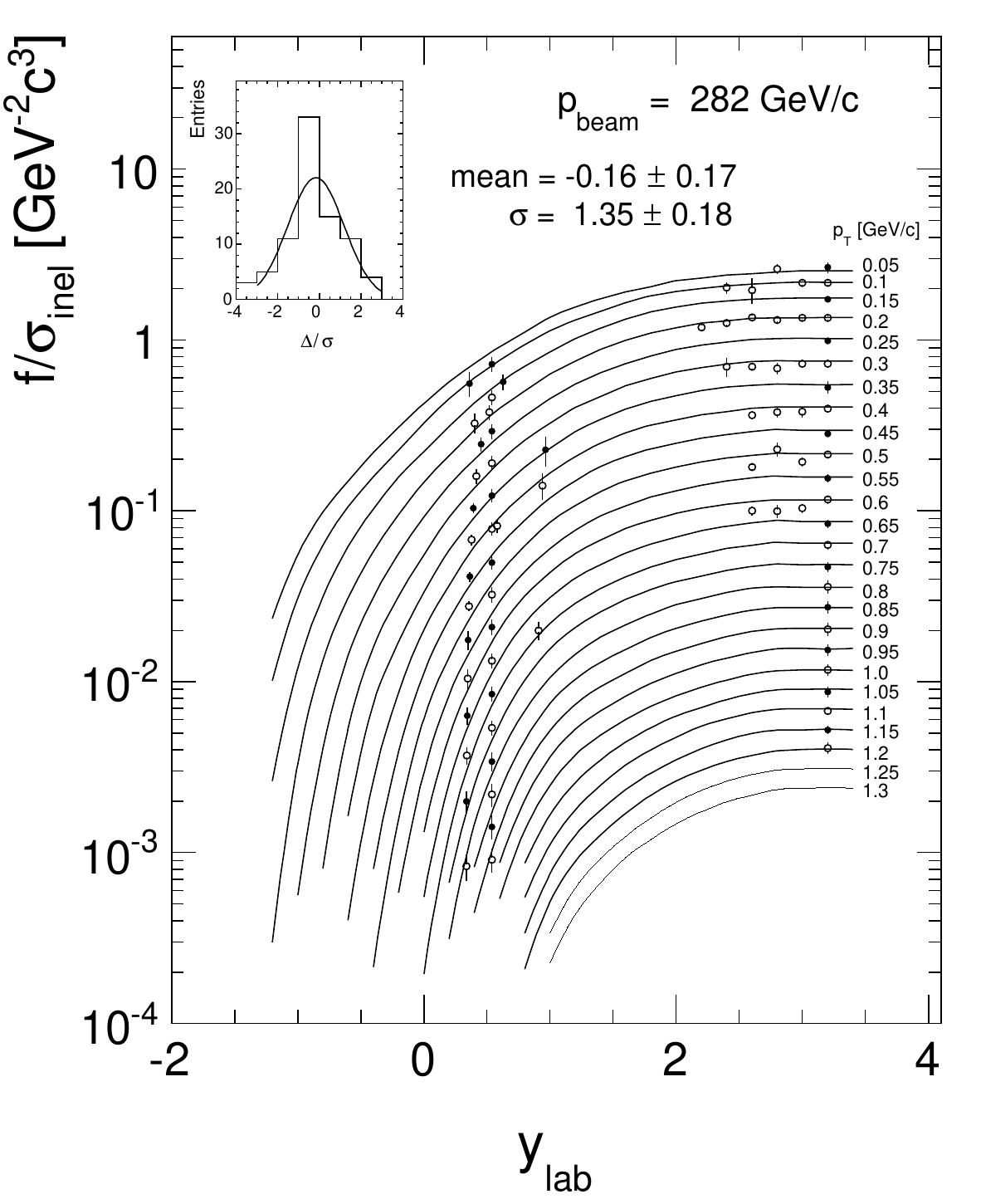} 
	\caption{Cross sections $f/\sigma_{\textrm{inel}}$ as functions of $y_{\textrm{lab}}$ and $p_T$ at $\sqrt{s}$ of 23~GeV corresponding to a beam momentum of 282~GeV/c}
  	\label{fig:int2dataisr23}
 \end{center}
\end{figure}

\begin{figure}[h]
 \begin{center}
   \begin{turn}{\rotAngle}
   \begin{minipage}{\capw}
	\caption{Cross sections $f/\sigma_{\textrm{inel}}$ as functions of $y_{\textrm{lab}}$ and $p_T$ at $\sqrt{s}$ of 31 and 45~GeV corresponding to beam momenta of 505 and 1078~GeV/c}
  	\label{fig:int2dataisr31_45}
	\end{minipage}
	\end{turn}
   \includegraphics[width=\figw,angle=\rotAngle] {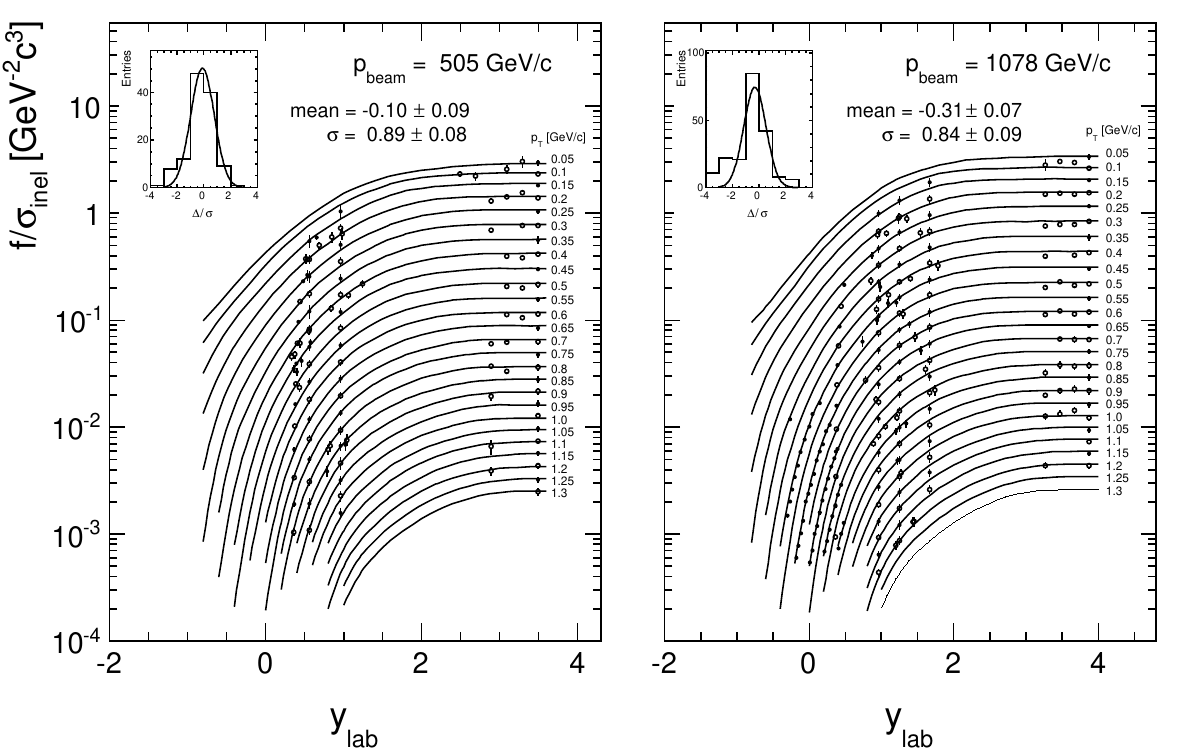} 
 \end{center}
\end{figure}

\begin{figure}[h]
 \begin{center}
   \begin{turn}{\rotAngle}
   \begin{minipage}{\capw}
	\caption{Cross sections $f/\sigma_{\textrm{inel}}$ as functions of $y_{\textrm{lab}}$ and $p_T$ at $\sqrt{s}$ of 53 and 63~GeV corresponding to beam momenta of 1507 and 2114~GeV/c}
  	\label{fig:int2dataisr53_63}
	\end{minipage}
	\end{turn}
   \includegraphics[width=\figw,angle=\rotAngle] {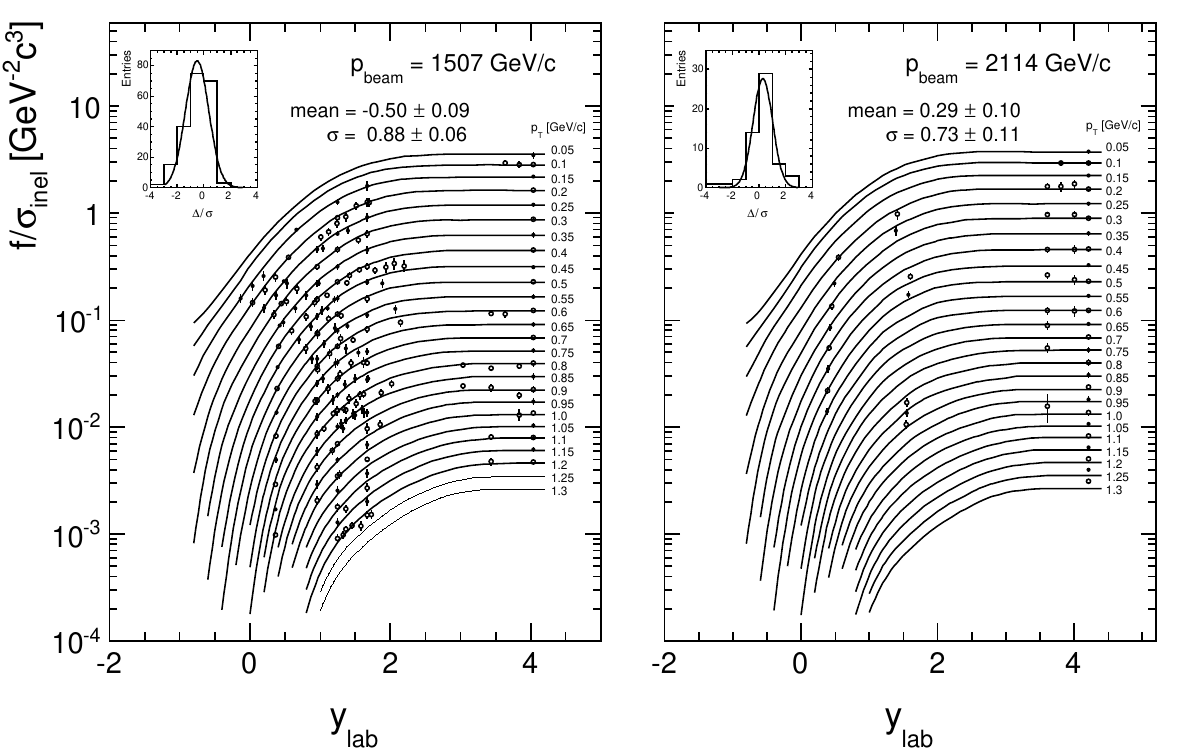} 
 \end{center}
\end{figure}

Although the phase space coverage of the data might seem rather restricted it should be remembered that the lower energies, Fig.~\ref{fig:int2dataisr23} and \ref{fig:int2dataisr31_45}a, bracket the reference data at 158 and 400~GeV/c beam momentum (Fig.~\ref{fig:int2data158_400}) which strongly constrains the overall interpolation. The data at 1078 and 1507~GeV/c beam momentum (Figs.~\ref{fig:int2dataisr31_45}b) and \ref{fig:int2dataisr53_63}) on the other hand feature a more extensive coverage especially in forward direction. In this context it should be stressed again that the ISR data are unique both in phase space extent and in mutual consistency and precision as compared to the more recent results from higher energy colliders discussed in Sect.~\ref{sec:colliders} below.

%
%
\subsection{Reliability and overall precision of the global interpolation scheme for the reference data}
\vspace{3mm}
\label{sec:ref_precision}

\begin{figure}[h]
 \begin{center}
   \includegraphics[width=16cm] {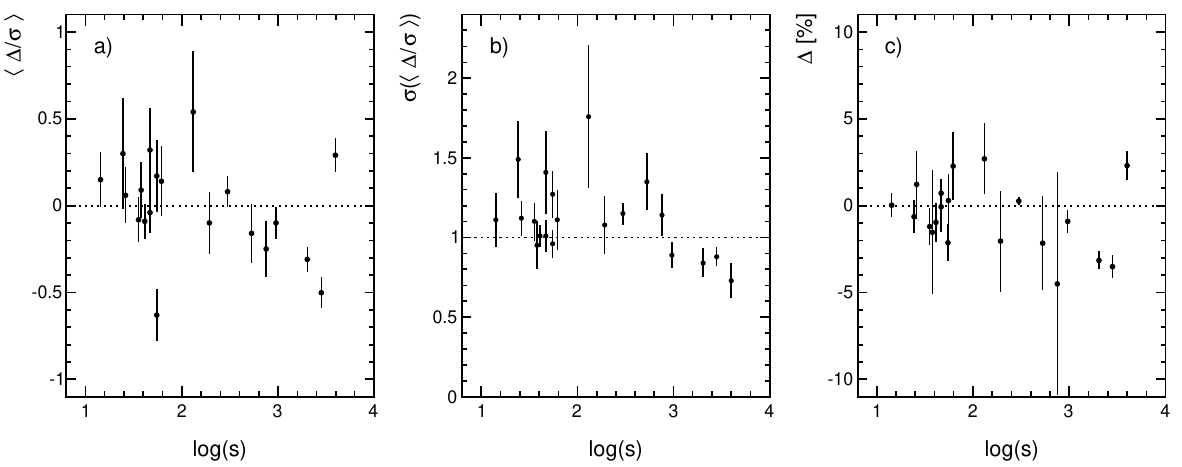} 
   \caption{a) mean deviation $\langle \Delta/\sigma \rangle$ from the global interpolation per experiment as a function of $\log(s)$, b) variance of the $\Delta/\sigma$ distributions as a function of $\log(s)$, c) mean value of the percent deviation between the experimental results and the global interpolation as a function of $\log(s)$}
  	\label{fig:summary_ref}
 \end{center}
\end{figure}

Three main features of the comparison of the experimental results \cite{gellert}--\cite{capi}, Tab.~\ref{tab:pp_pim_data}, with the global interpolation scheme discussed above are presented in Fig.~\ref{fig:summary_ref} as a function of $\log(s)$. The first two panels show that the data comply with the expectation as far as the normalized residuals $\Delta/\sigma$ are concerned. Both the averages $\langle \Delta/\sigma \rangle$, Fig.~\ref{fig:summary_ref}a, and their variances, Fig.~\ref{fig:summary_ref}b, comply within errors with the expectation values 0 and 1. The mean deviations proper, shown in percent in Fig.~\ref{fig:summary_ref}c, are compatible with zero within an error margin of less than $\pm$5\%.

The fact that the percent deviations stay below $\pm$5\% is  to be considered as an important result of this study, keeping in mind that the data for none of the reference experiments had to be re-normalized or modified.

%
%
\section{Comparison of the global interpolation to the spectrometer data (Tab.~\ref{tab:pp_pim_spect})}
\vspace{3mm}
\label{sec:countdata}

As general remark concerns the absence of feed-down corrections for all the spectrometer experiments. The comparison has therefore to be made to the global interpolation including feed-down. A further remark concerns the normalization problem. As already shown in Fig.~\ref{fig:factor} there is, in sharp contrast to the reference data, a very wide range of normalization factors to be applied to the measured cross sections in order to bring them into consistency with the reference data. The fact that these factors have a mean value at 1 implies that indeed there is no systematic trend eventually putting into doubt the normalization of the reference sample. In the following some remarks concerning each experiment will be mandatory.

%
%
\subsection{The data of Melissinos et al. \cite{melissinos} at 3.67~GeV/c beam momentum}
\vspace{3mm}
\label{sec:melissinos}

\begin{figure}[h]
 	\begin{center}
   	\includegraphics[width=14cm] {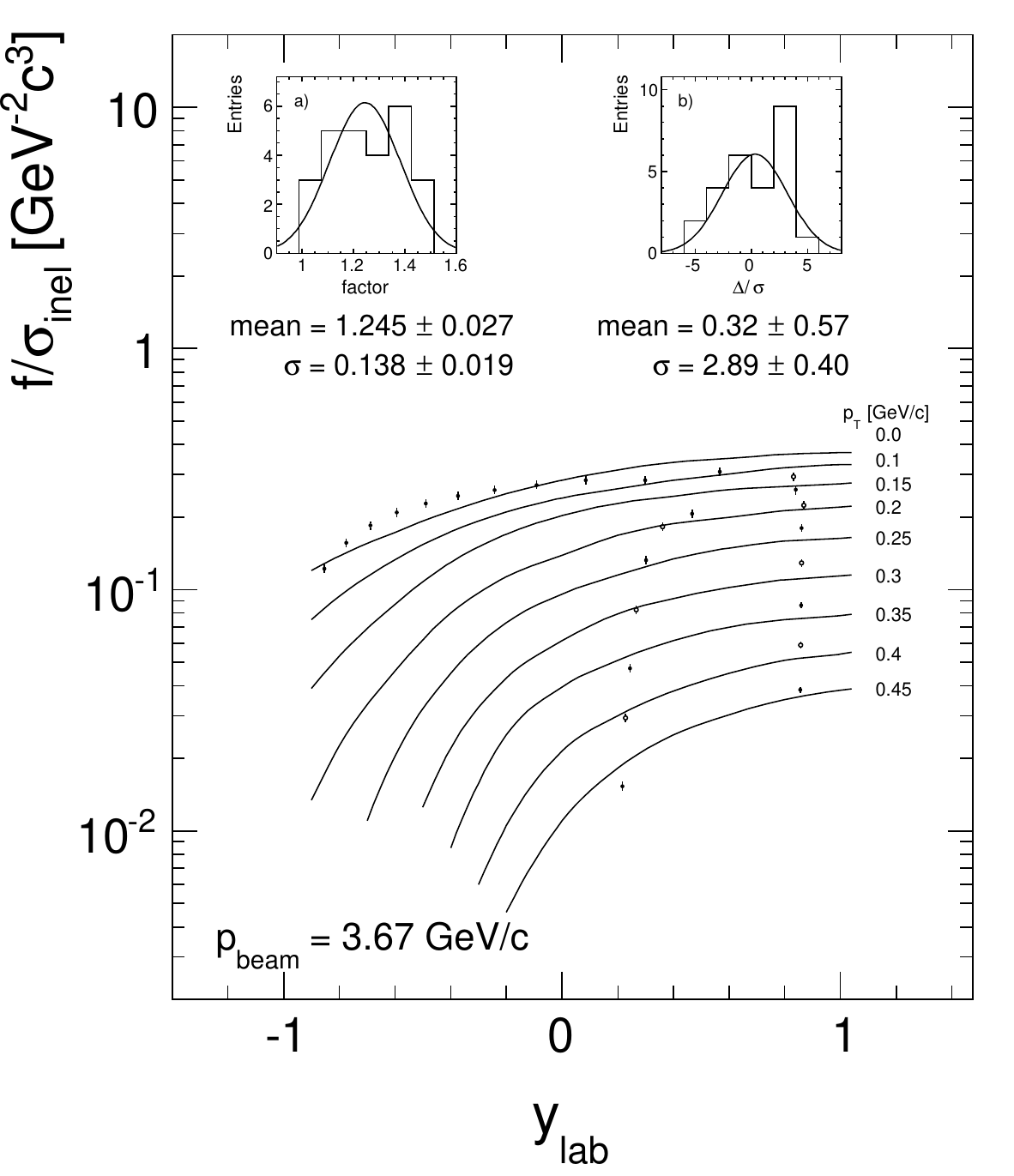} 
	 		\caption{Cross sections $f/\sigma_{\textrm{inel}}$ as functions of $y_{\textrm{lab}}$ and $p_T$ at $\sqrt{s}$~=~2.98~GeV \cite{melissinos}}
  		\label{fig:melissinos}
 	\end{center}
\end{figure}

This is at the same time the earliest (1962) and the lowest-energy experiment in the present comparison. Data were obtained at four lab angles of 0, 17, 32 and 45 degrees. As the latter two angles correspond to cms rapidities of about $\pm$0.17, the corresponding data have been averaged. The data at 0 degrees are, together with \cite{dekkers}, the only existing measurements at $p_T$~=~0~GeV/c. Fig.~\ref{fig:melissinos} shows the resulting comparison to the global interpolation. The distribution of the point-by-point deviations $\Delta$, inset a), is offset from zero by +24\% which is by a factor of 1.5 above the the normalization error given by the authors.

Correcting for this offset the residual distribution, inset b), is centred at zero, however with a broad standard deviation of about 3 units. This indicates further systematic effects (target length, nuclear absorption and multiple scattering in the detector material) as compared to the typical statistical error of 5\% per data point. The results are nevertheless important in order to define the cross sections at the lower edge of the cms energy scale used in the present study, albeit with a somewhat increased systematic error.

%
%
\subsection{The data of Akerlof et al. \cite{akerlof} at 12.5~GeV/c beam momentum}
\vspace{3mm}
\label{sec:akerlof}

\begin{figure}[h]
 	\begin{center}
   	\includegraphics[width=14cm] {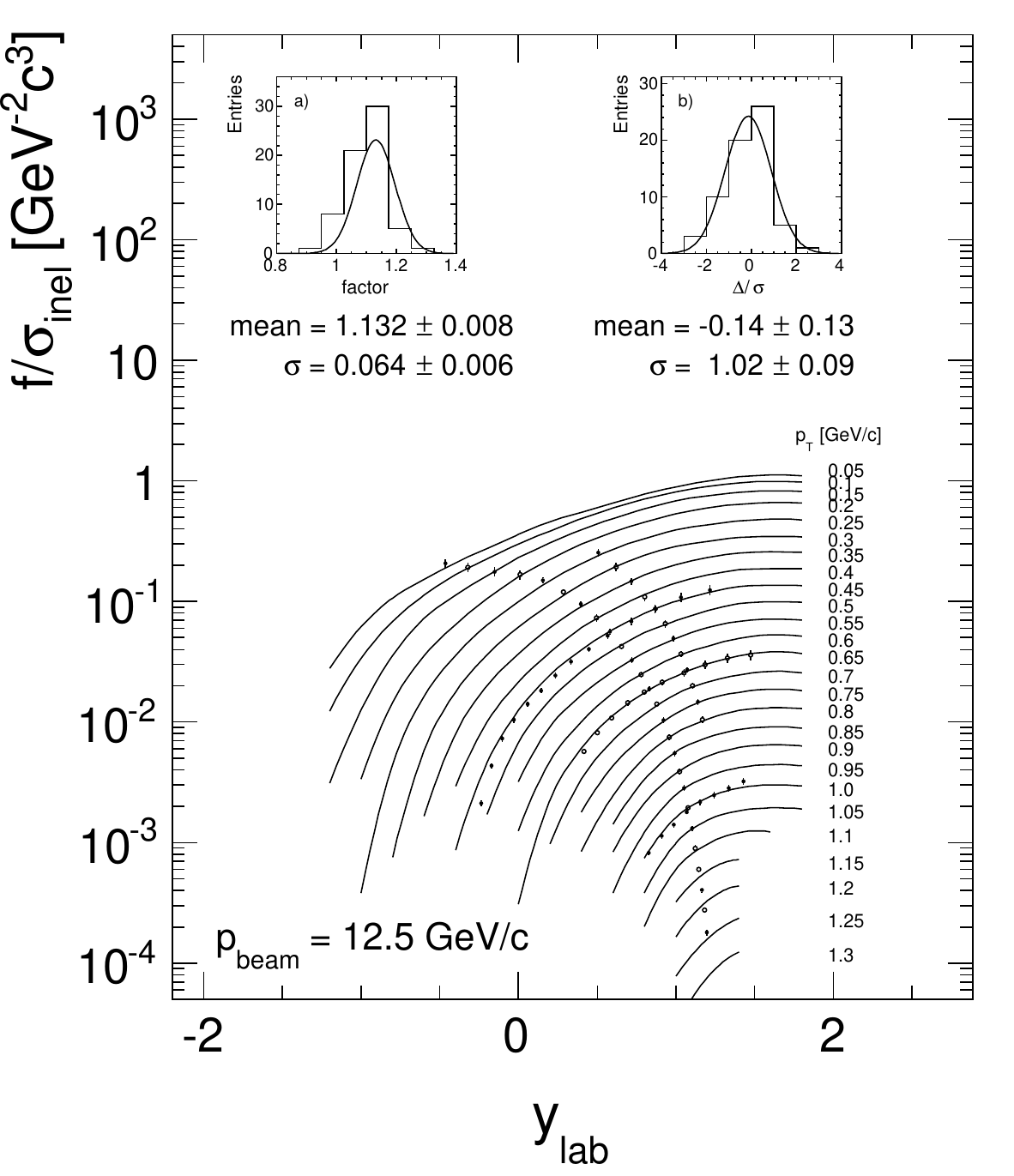} 
	 			\caption{Cross sections $f/\sigma_{\textrm{inel}}$ as functions of $y_{\textrm{lab}}$ and $p_T$ at $\sqrt{s}$~=~5.0~GeV \cite{akerlof}}
  			\label{fig:akerlof}
 	\end{center}
\end{figure}

The experiment gives 70 data points at two values of constant $p_L^{\textrm{cm}}$ and three values of constant $p_T$ which have to be interpolated to the binning scheme in $p_T$ used in this paper.The result is shown in Fig.~\ref{fig:akerlof}.

The distribution of the differences to the global interpolation, insert a) in Fig.~\ref{fig:akerlof}, shows a mean relative factor of 1.1 corresponding to an offset of about +10\% indicating a normalization error which is a factor of two above the value given by the authors. After correcting for this offset the residual distribution is centred at zero with a standard deviation of 1 indicating a rather perfect agreement with the global interpolation.

%
%
\subsection{The data of Dekkers et al. \cite{dekkers} at 18.8~GeV/c beam momentum}
\vspace{3mm}
\label{sec:dekkers18}

\begin{figure}[h]
 	\begin{center}
   	\includegraphics[width=14cm] {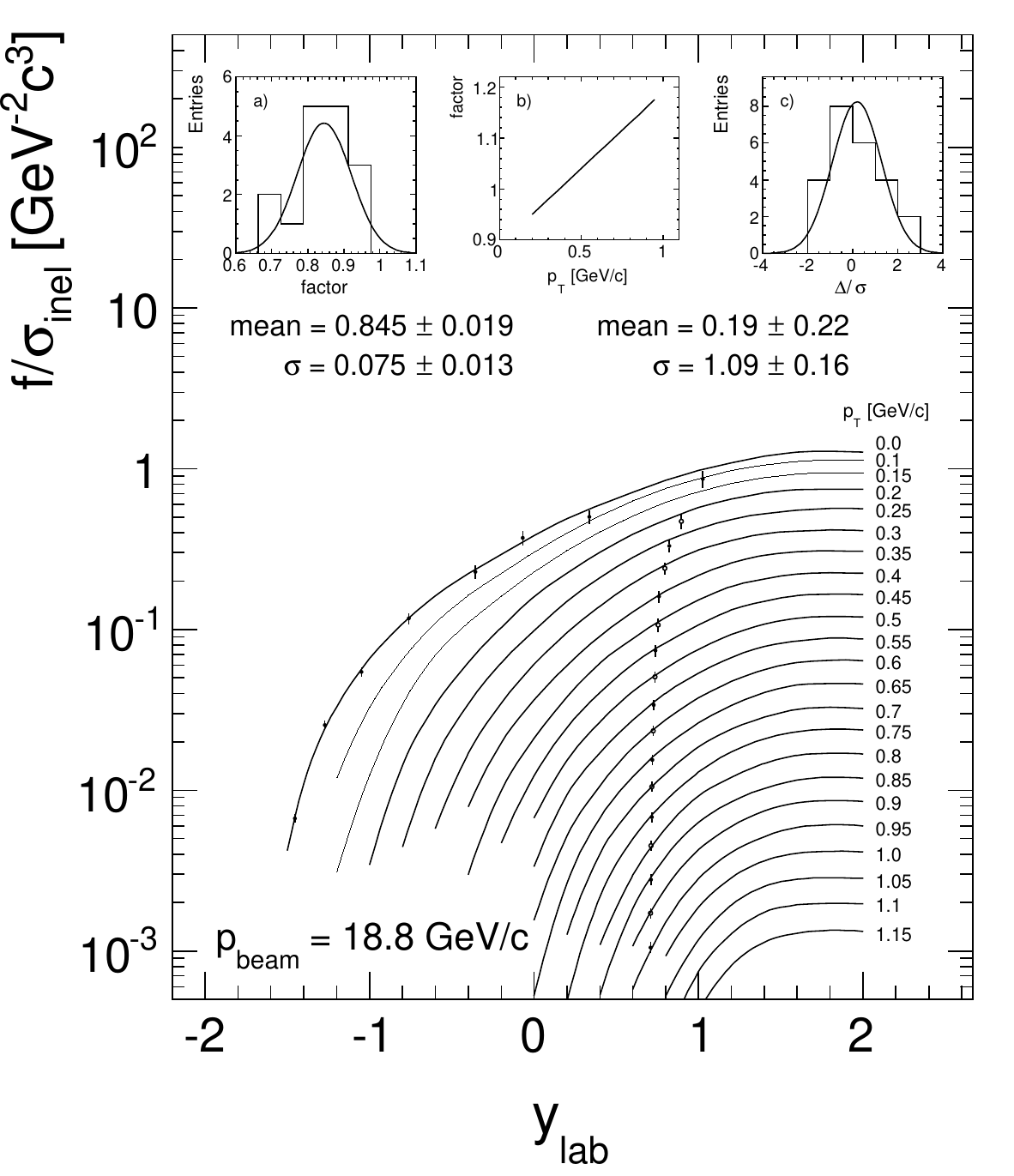} 
	 	\caption{Cross sections $f/\sigma_{\textrm{inel}}$ as functions of $p_T$ and $y_{\textrm{lab}}$ at $\sqrt{s}$~=~6.08~GeV \cite{dekkers}}
  	\label{fig:dekkers18}
 	\end{center}
\end{figure}

The experiment gives 8 data points at lab angle 0~mrad and 7 data points at 100~mrad as a function of $p_{\textrm{lab}}$. The latter data may be interpolated to the standard $p_T$ values between 0.05 and 0.95~GeV/c. Whereas the 0~mrad data comply well with the global interpolation, the 100~mrad data show an important upward deviation with a broad distribution centred around a factor of 0.85, inset a) of Fig.~\ref{fig:dekkers18}. This factor depends strongly on $p_T$ in an approximately linear fashion, inset b). Correcting for this dependence, the residual distribution is centred at 0 with variance 1, inset c) complying well with the global interpolation.

%
%
\subsection{The data of Allaby et al. \cite{allaby1} at 19.2~GeV/c beam momentum}
\vspace{3mm}
\label{sec:allaby19}

\begin{figure}[h]
 	\begin{center}
   	\includegraphics[width=14cm] {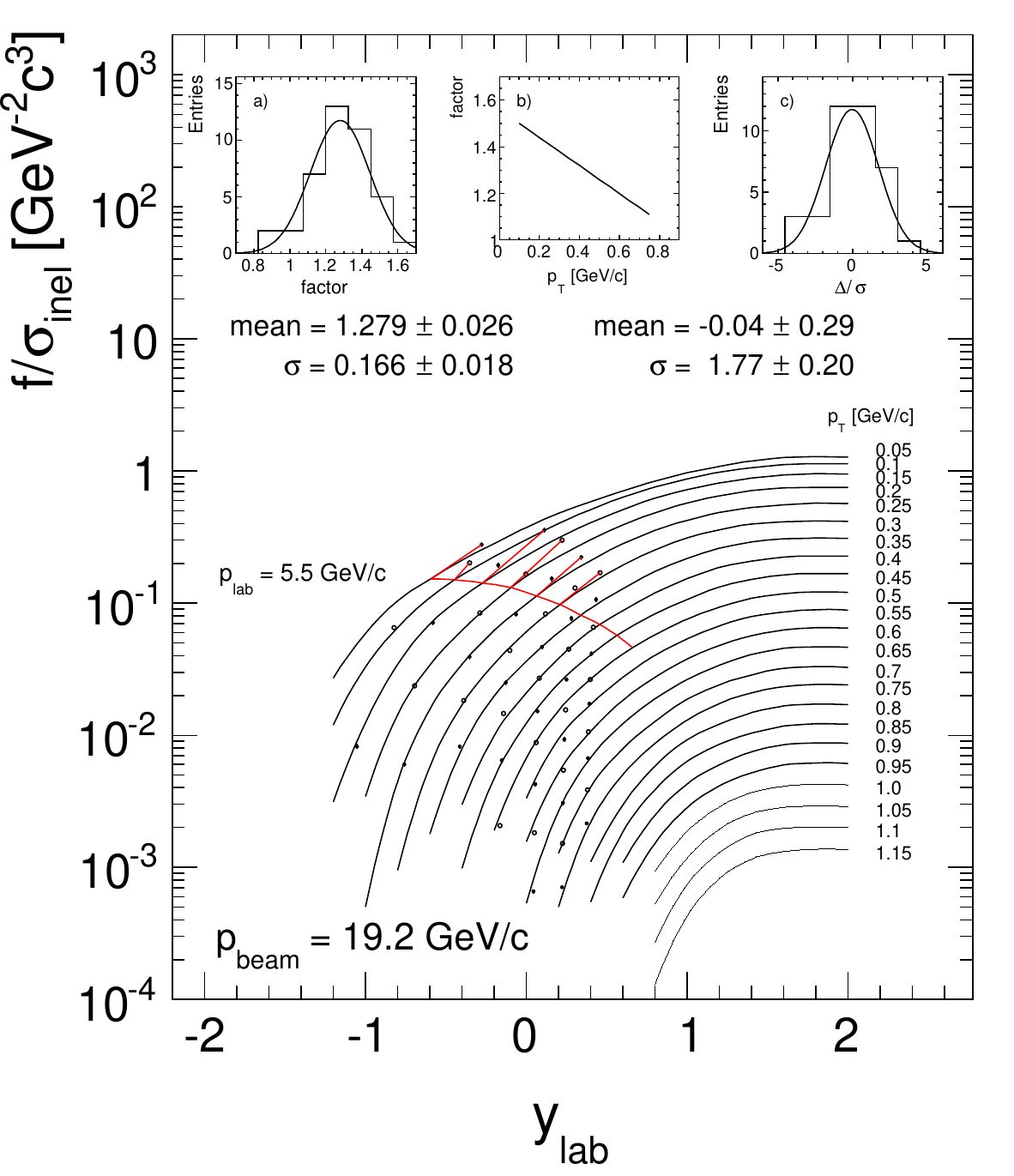} 
	 	\caption{Cross sections $f/\sigma_{\textrm{inel}}$ as functions of $p_T$ and $y_{\textrm{lab}}$ at $\sqrt{s}$~=~6.14~GeV}
  	\label{fig:allaby19}
 	\end{center}
\end{figure}

87 data points have been measured at 6 values of $\Theta_{\textrm{lab}}$ between 12.5 and 70~mrad as a function of $p_{\textrm{lab}}$. After interpolation in $p_T$ the resulting $y_{\textrm{lab}}$ distributions are shown in Fig.~\ref{fig:allaby19}.

A complicated pattern of deviations becomes visible. First of all a large mean positive offset of a factor of 1.3 is apparent, inset a). The standard deviation of this offset is with 15\% five times larger than the statistical error. There are two additional effects to be taken into account. Firstly there is a strong additional upward deviation for the measurements below $p_{\textrm{lab}}$~=~5.5~GeV/c, see the line in Fig.~\ref{fig:allaby19}. The corresponding data are eliminated from the comparison. Secondly there is a strong dependence of the deviations on $p_T$, inset b) where the factor varies between 1.5 and 1.1 over the measured range. Correcting for this second-order effect which is opposite to the one seen in Sect.~\ref{sec:dekkers18}, the residual distribution, inset c), is centred at zero but with a variance which is indicating, with a value of 1.77, further sizeable systematic effects.

%
%
\subsection{The data of Dekkers et al. \cite{dekkers} at 23.1 GeV/c beam momentum}
\vspace{3mm}
\label{sec:dekkers23}

\begin{figure}[h]
 	\begin{center}
   	\includegraphics[width=13cm] {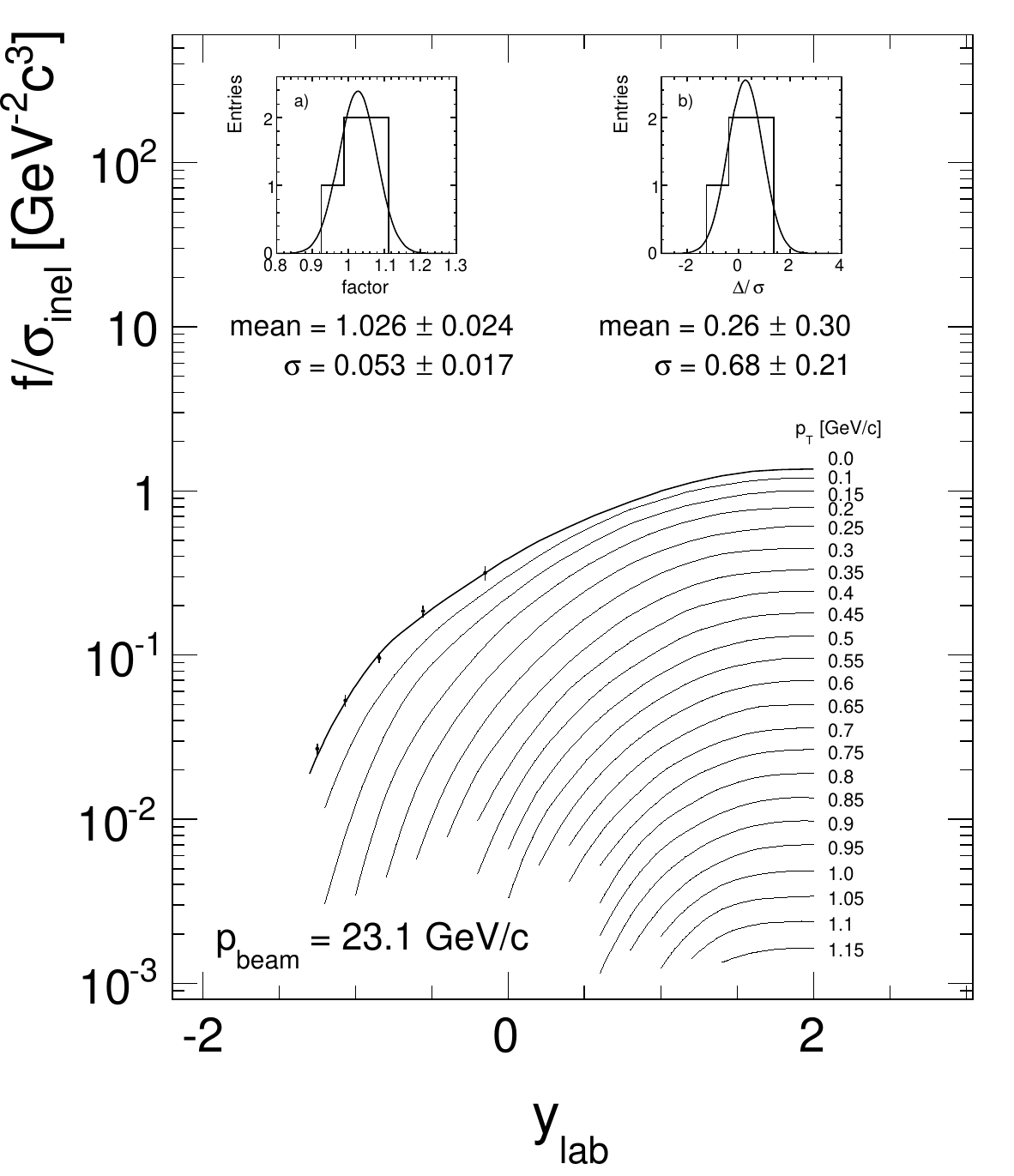} 
	 	\caption{Cross sections $f/\sigma_{\textrm{inel}}$ at $p_T$~=~0~GeV/c as a function of $y_{\textrm{lab}}$ at $\sqrt{s}$~=~6.72~GeV}
  	\label{fig:dekkers23}
 	\end{center}
\end{figure}

The experiment presented under Sect.~\ref{sec:dekkers18} gives also five data points at $p_T$~=~0~GeV/c as a function of $p_{\textrm{lab}}$ as shown in Fig.~\ref{fig:dekkers23}.

Again the data comply well with the global interpolation without a discernible offset in view of the systematic errors of about 6--10\%.


%
%
\subsection{The data of Amaldi et al. \cite{allaby2} at 24.0 GeV/c beam momentum}
\vspace{3mm}
\label{sec:allaby24}

\begin{figure}[h]
 	\begin{center}
   	\includegraphics[width=14cm] {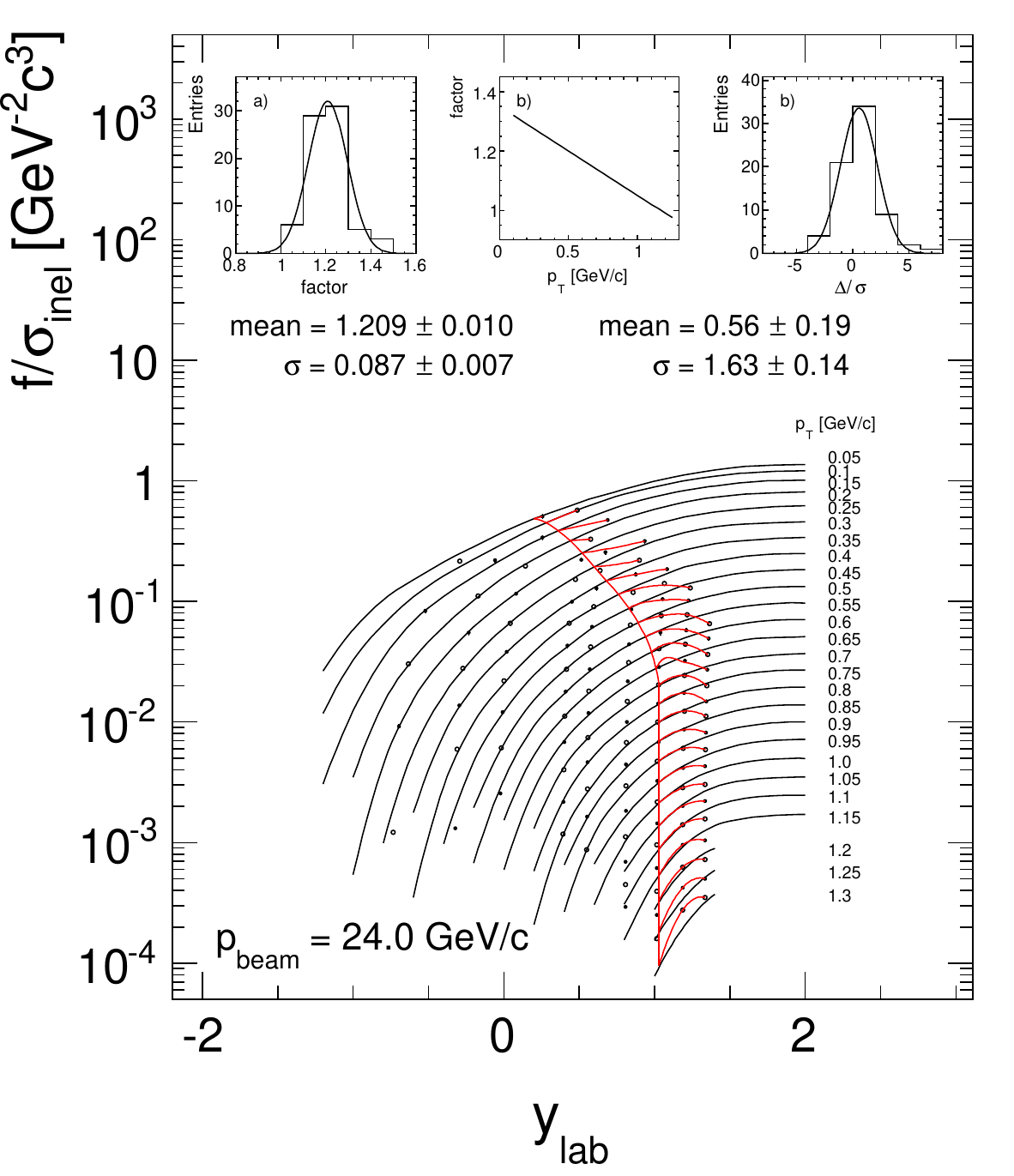} 
	 	\caption{Cross sections $f/\sigma_{\textrm{inel}}$ as functions of $p_T$ and $y_{\textrm{lab}}$ at $\sqrt{s}$~=~6.84~GeV}
  	\label{fig:allaby24}
 	\end{center}
\end{figure}

This is an extension of \cite{allaby1} to 24~GeV/c beam momentum and to lab angles up to 147~mrad. Again a complex scheme of normalization problems and further systematic deviations becomes visible in Fig.~\ref{fig:allaby24}.

Similar to \cite{allaby1} an offset factor of 1.20  is apparent, inset a). In addition there are large downward deviations basically for the lab angles above 87~mrad (line in Fig.47) which lead to unphysical local maxima in the rapidity region around 0.7. These deviations reach -40\% at the largest lab angle. Eliminating this critical region, a strong $p_T$ dependence similar to Sect.~\ref{sec:allaby19} is visible, inset b). Applying the corresponding correction the residual distribution, insert c), is centred at zero. The standard deviation of 1.63 indicates however further sizeable systematic effects. It should be mentioned here that similar inconsistencies have been demonstrated in the study of the cms energy dependence of charged kaon production \cite{pp_kaon}.

%
%
\subsection{The data of Beier et al. \cite{beier} at 24.0~GeV/c beam momentum}
\vspace{3mm}
\label{sec:beier}

\begin{figure}[h]
 	\begin{center}
   	\includegraphics[width=14cm] {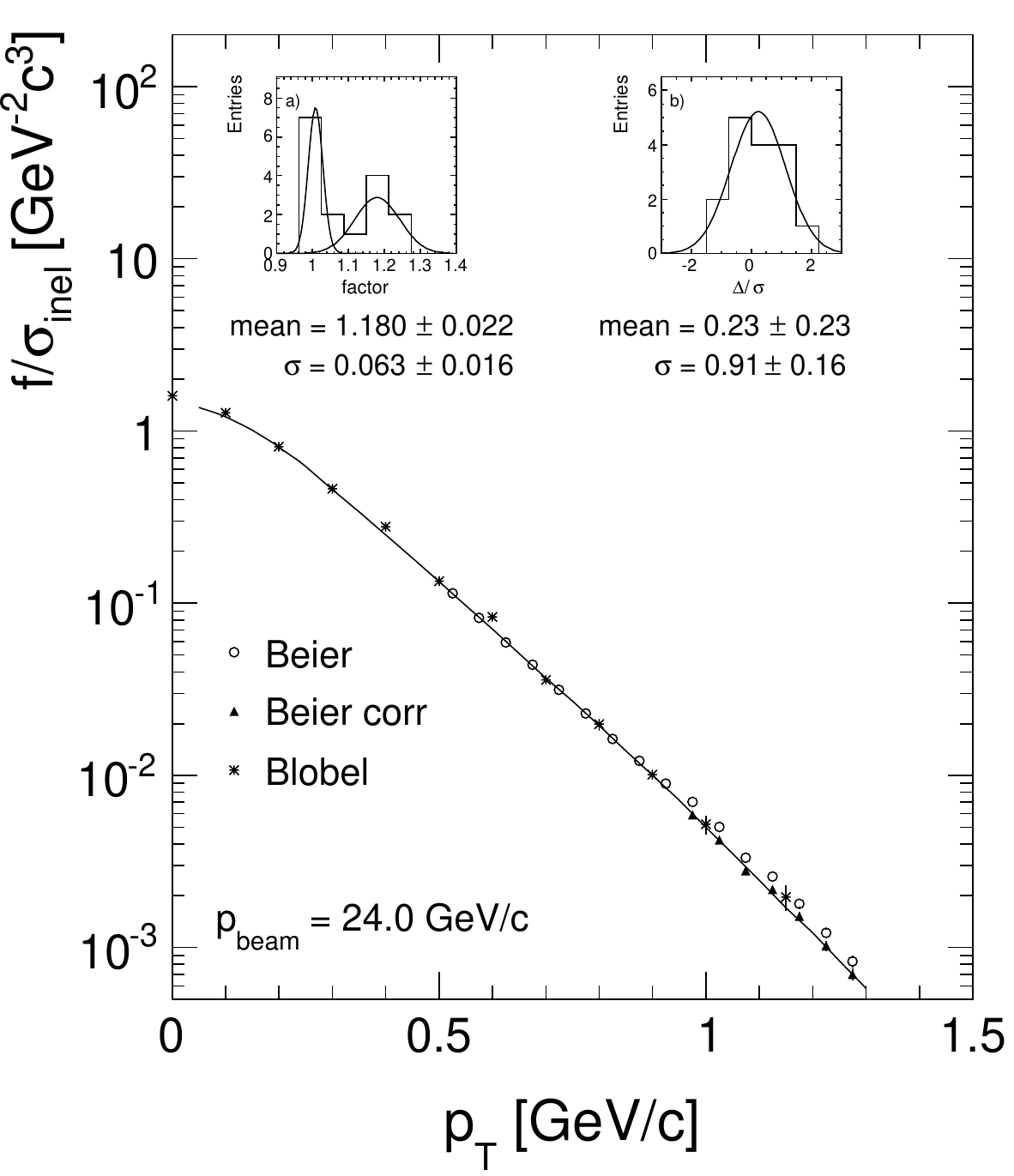} 
	 	\caption{Cross sections $f/\sigma_{\textrm{inel}}$ as a function of $p_T$ at rapidity 0 and $\sqrt{s}$~=~6.84~GeV}
  	\label{fig:beier}
 	\end{center}
\end{figure}

16 cross sections at rapidity zero are given in the range from $p_T$~=~0.525 to 1.375~GeV/c, Fig.~\ref{fig:beier}. The factors with respect to the global interpolation, inset a), show two distinct groups. Up to 0.925~GeV/c there is good agreement, whereas the data above this value group around a factor of 1.18. On the other hand the data of Blobel et al. \cite{blobel} follow the interpolation well in this region after being increased by the feed-down component. After correcting for this (unexplained) break in the data the residual distribution, inset b), is well centred at 0 with an rms corresponding to 1.

%
%
\subsection{The data of Anderson et al. \cite{landolt,anderson} at 29.7~GeV/c beam momentum}
\vspace{3mm}
\label{sec:anderson}

\begin{figure}[h]
 	\begin{center}
 	  \includegraphics[width=14cm] {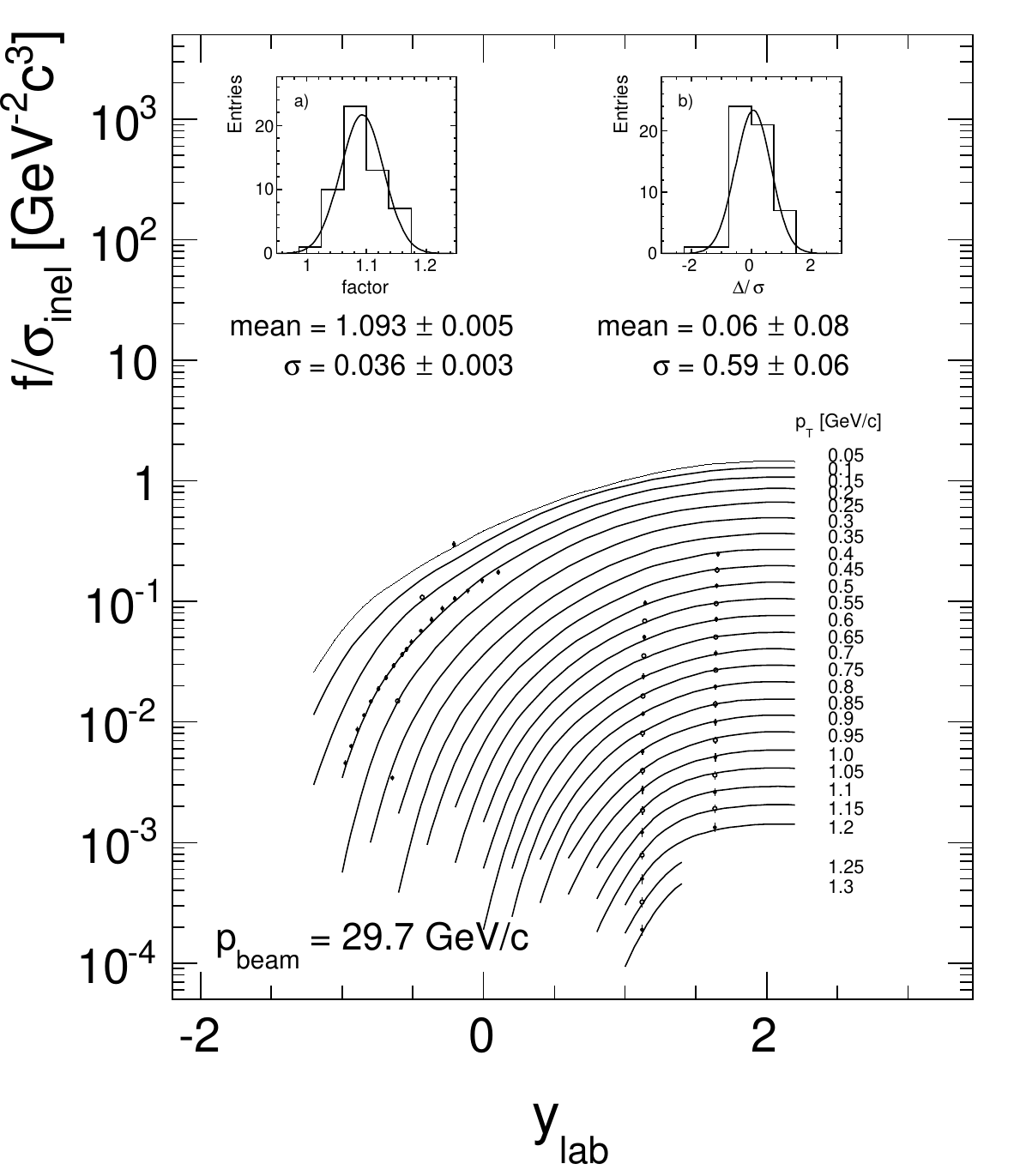} 
	 	\caption{Cross sections $f/\sigma_{\textrm{inel}}$ as functions of $p_T$ and $y_{\textrm{lab}}$ at $\sqrt{s}$~=~7.58~GeV}
  	\label{fig:anderson}
 	\end{center}
\end{figure}

Cross sections at three lab angles of 15, 96 and 160~mrad as well as a momentum distribution at $p_T$~=~0.2~GeV/c have been measured, Fig.~\ref{fig:anderson}. The relative factor to the global interpolation is centred at 1.09, inset a) to be compared to a statistical error of about 5\%. Correcting for this offset the residual distribution is well centred with an rms compatible with 1, inset b). This gives a good example of a precision spectrometer experiment to support and verify the reference data especially in the forward rapidity and larger $p_T$ regions. This is  especially valuable in  comparison with the neighbouring experiment \cite{allaby2}, Sect.~\ref{sec:allaby24} which has been performed at about the same time.

%
%
\subsection{The data of Abramov et al. \cite{abramov} at 70~GeV/c beam momentum}
\vspace{3mm}
\label{sec:abramov}

\begin{figure}[h]
 	\begin{center}
   	\includegraphics[width=14cm] {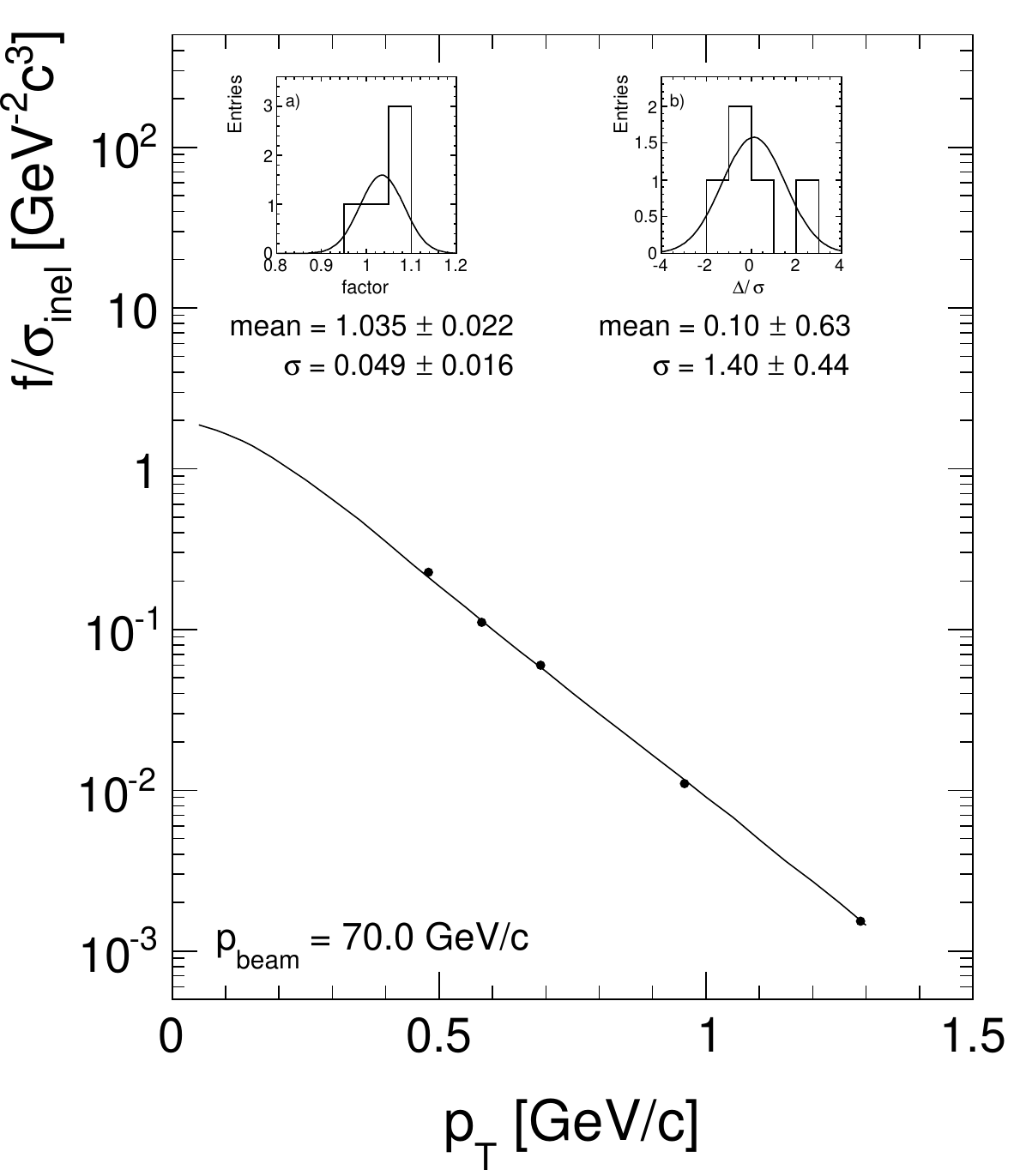} 
	 	\caption{Cross sections $f/\sigma_{\textrm{inel}}$ as a function of $p_T$ at rapidity 0 and $\sqrt{s}$~=~11.5~GeV}
  	\label{fig:abramov}
 	\end{center}
\end{figure}

These data cover, at rapidity zero, a large range of $p_T$ values up to 5~GeV/c of which five points fall into the $p_T$ range of this study, Fig.~\ref{fig:abramov}.

In view of the large normalization uncertainty of 20\% the data comply well with the global interpolation, inset a), with a slight offset of 3.5\%. After application of a small correction of 0.965 the residual distribution is well centred and has an rms of 1.40 with respect to the average statistical error of 3.5\%, inset b).

%
%
\subsection{The data of Brenner et al. \cite{brenner} at 100 and 175~GeV/c beam momentum}
\vspace{3mm}
\label{sec:brenner}

\begin{figure}[h]
 \begin{center}
   \begin{turn}{\rotAngle}
   \begin{minipage}{\capw}
	\caption{Cross sections $f/\sigma_{\textrm{inel}}$ \cite{brenner} at 100 and 175~GeV/c beam momentum as functions of $p_T$ and $y_{\textrm{lab}}$ at $\sqrt{s}$~=~13.8 and 18.2~GeV}
  	\label{fig:brenner}
	\end{minipage}
	\end{turn}
   \includegraphics[width=\figw,angle=\rotAngle] {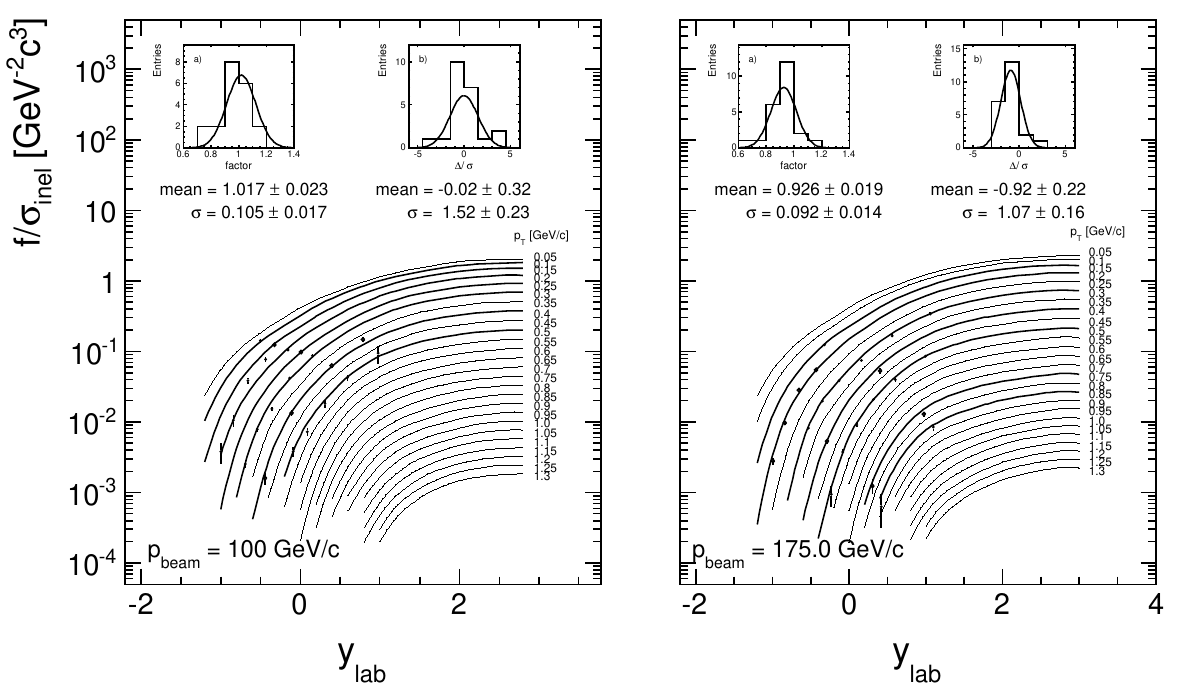} 
 \end{center}
\end{figure}

This experiment provides 25 and 23 cross sections at 100 and 175~GeV/c beam momentum, respectively with statistical errors of about 5\% ranging up to 50\% in some cases, Fig.~\ref{fig:brenner}.

These results are of particular interest as they bracket the NA49 data in cms energy, albeit confined to the forward region at $x_F \geq$~0.3. In both cases the deviations from the global interpolation are small with only -1\% and -8.5\% respectively (panels a) compared to the claimed normalization error of 7\%. After renormalization the residual distributions are well centred (panels b) with rms values which comply with the statistical errors within one (175~GeV/c) and two (100~GeV/c) standard deviations.

%
%
\subsection{The data of Johnson et al. \cite{johnson} at 100, 200 and 400~GeV/c beam momentum}
\vspace{3mm}
\label{sec:johnson}

\begin{figure}[h]
 \begin{center}
   \includegraphics[width=13.6cm] {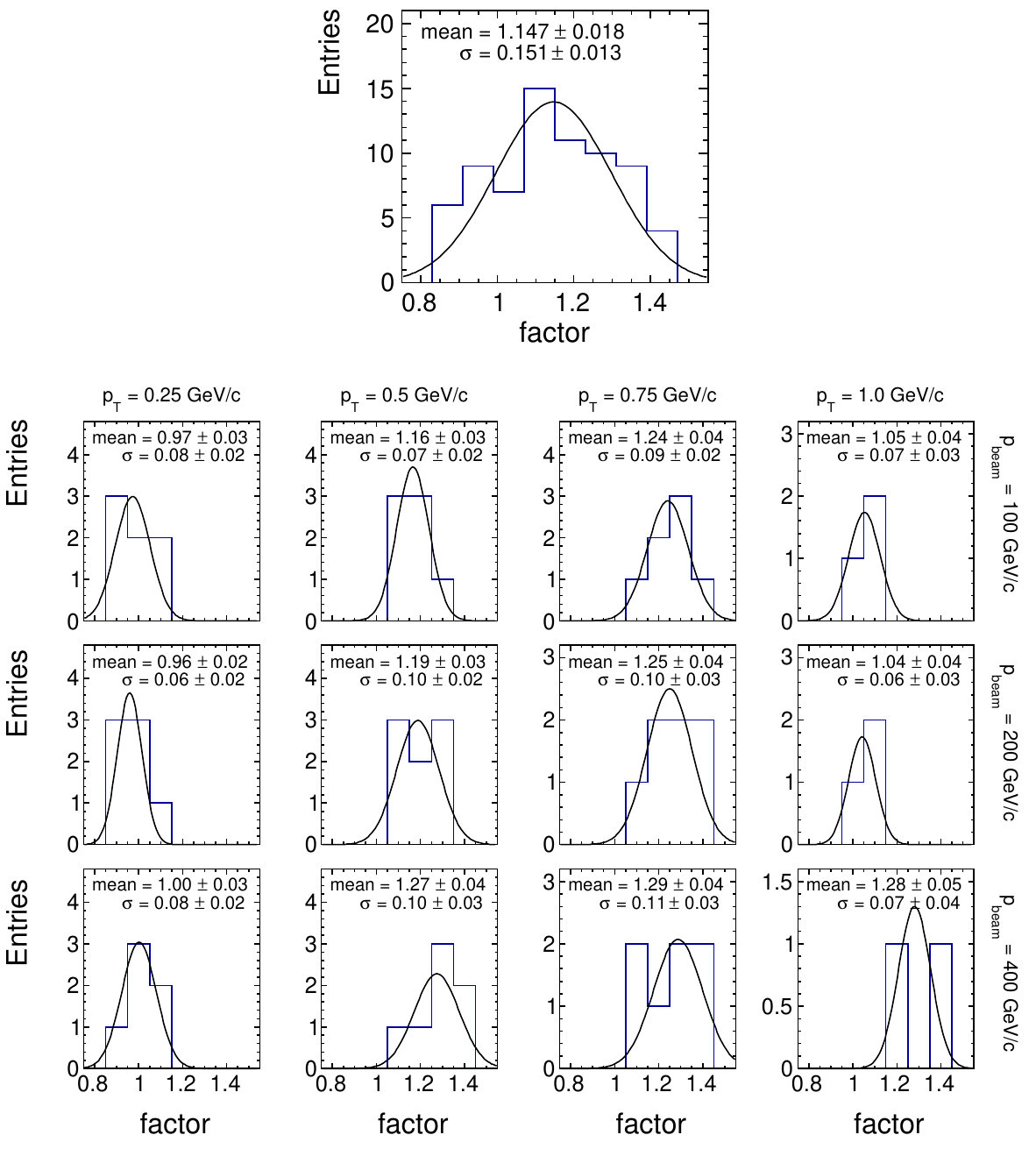} 
	\caption{Distributions of difference to the global interpolation for all data points and separately for the four $p_T$ values and the three beam momenta}
  	\label{fig:johnson1}
 \end{center}
\end{figure}

The experiment gives about 80 cross sections over a wide range of rapidities at the three beam momenta of 100, 200 and 400~GeV/c for the $p_T$ values of 0.25, 0.5, 0.75 and 1~GeV/c. With statistical errors in the range of typically 3\% to 4\% these data are of considerable interest as they span the region from SPS to ISR data. The distribution of the difference factors with the global interpolation shows however a broad distribution with a mean value of 14.7\% and a variance of 15.1\%, Fig.~\ref{fig:johnson1}, which is considerably above the given systematic error of 7\%.

A detailed inspection of the differences reveals a systematic dependence on $p_T$ rather than rapidity or beam momentum as shown in Fig.~\ref{fig:johnson1} with mean deviations which are in general positive, ranging from -4 to 29\%.

\begin{figure}[h]
 \begin{center}
   \begin{turn}{\rotAngle}
   \begin{minipage}{\capw}
	\caption{Corrected cross sections $f/\sigma_{\textrm{inel}}$ \cite{johnson} for the three interaction energies including in the insets the residual distributions}
  	\label{fig:johnson3}
	\end{minipage}
	\end{turn}
   \includegraphics[width=\figw,angle=\rotAngle] {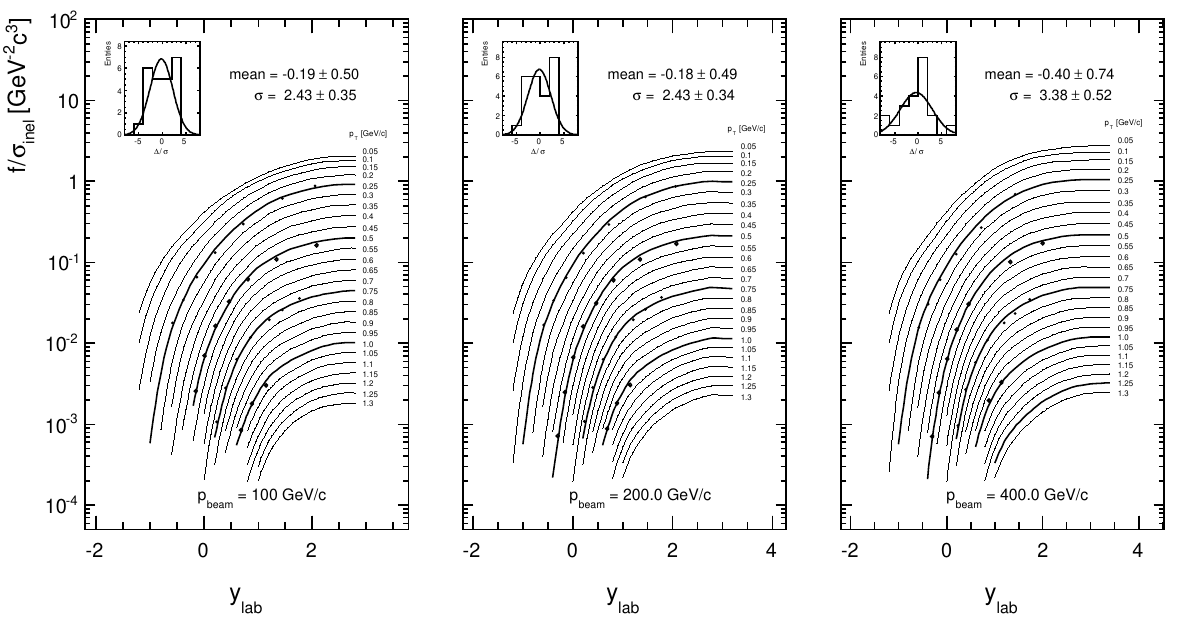} 
 \end{center}
\end{figure}

Correcting for the respective normalization factors the $y_{\textrm{lab}}$ distributions are shown in Fig.\ref{fig:johnson3}. The residual distributions are well centred but the sizeable standard deviations indicate further systematic effects.

The data corrections described above are an example of a complex pattern of systematic effects which go beyond a simple overall normalization error. The connected manipulations are nevertheless useful to show the agreement between spectrometer experiments and reference data in the important transition from SPS/Fermilab energies to the ISR especially in the forward and intermediate rapidity ranges.

%
%
\subsection{The data of Antreasyan et al. \cite{antreasyan} at 200, 300 and 400 GeV/c beam momentm}
\vspace{3mm}
\label{sec:antreasyan}

\begin{figure}[h]
 	\begin{center}
   	\includegraphics[width=14cm] {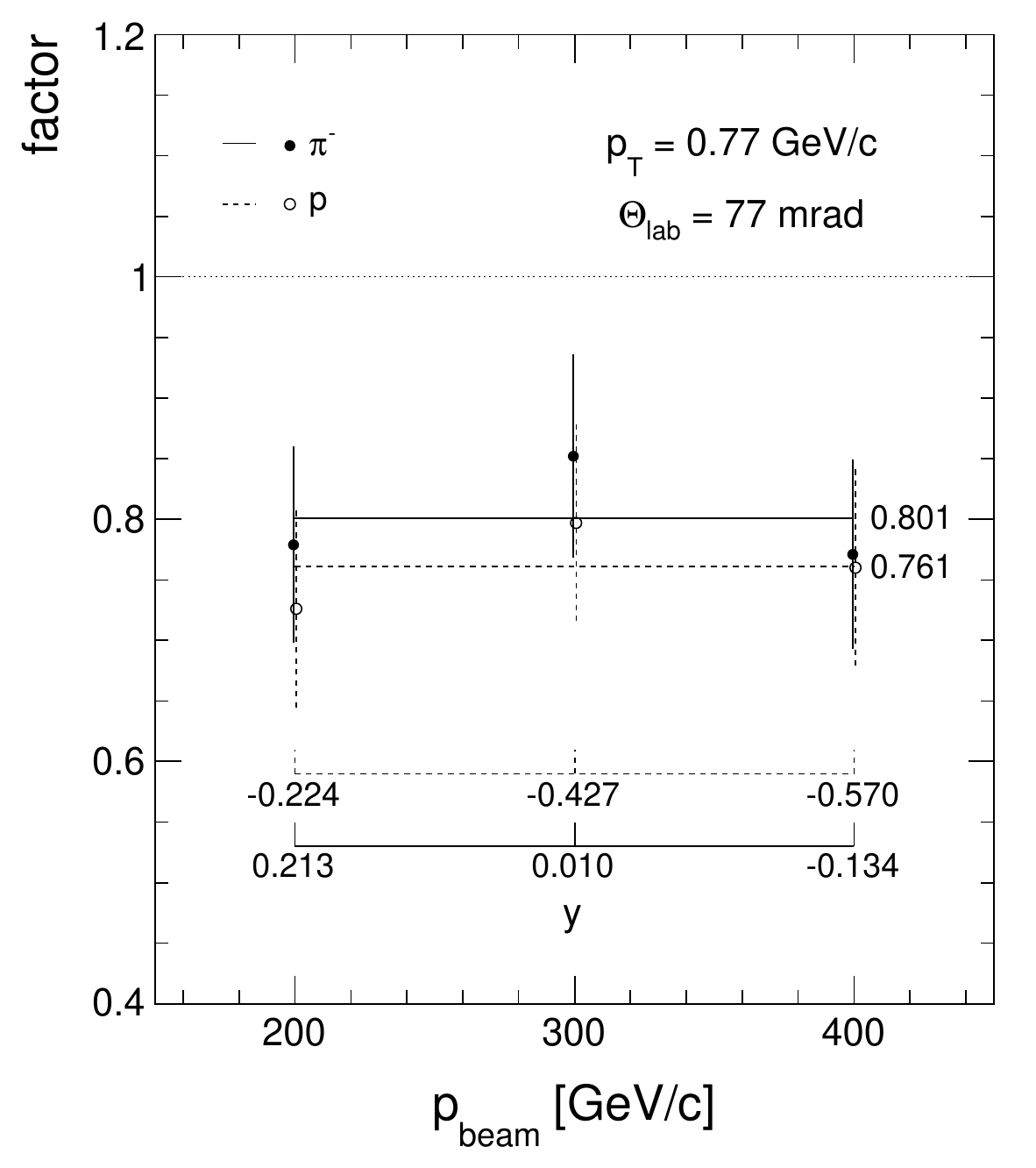} 
	 	\caption{Ratio of data from \cite{antreasyan} and global interpolation at $p_T$~=~0.77~GeV/c at 200, 300 and 400~GeV/c beam momentum indicating the corresponding rapidity values}
  	\label{fig:antreasyan}
 	\end{center}
\end{figure}

Only four data points of this celebrated experiment fall within the $p_T$ range of this paper: at $p_T$~=~0.77~GeV/c for $p_{\textrm{lab}}$~=~200, 300 and 400~GeV/c, and at $p_T$~=~1.16~GeV/c only for 400~GeV/c beam momentum. As the measurements were done at a constant lab angle of 77~mrad, the corresponding rapidity values vary between 0.21 and -0.13. Fig.~\ref{fig:antreasyan} shows the three data points at $p_T$~=~0.77~GeV/c with respect to the reference data.

An average downward shift by a factor 0.8 corresponding to -25\% is apparent. This shift complies with the one already found for protons \cite{pp_proton} also indicated in Fig.~\ref{fig:antreasyan}.

%
%
\subsection{Summary of the results from spectrometer experiments}
\vspace{3mm}
\label{sec:summary}

The following features may be extracted from the discussion of the 12 experiments in Sect.~\ref{sec:melissinos} to \ref{sec:antreasyan}:

\begin{enumerate}
 	\item The main reason for the deviations from the global interpolation is given by the normalization. In fact the mean factors between data and interpolation span a wide range from 0.75 to 1.30 with an average only slightly below 1, Fig.~\ref{fig:factors}.

	\begin{figure}[h]
 		\begin{center}
   		\includegraphics[width=9.5cm] {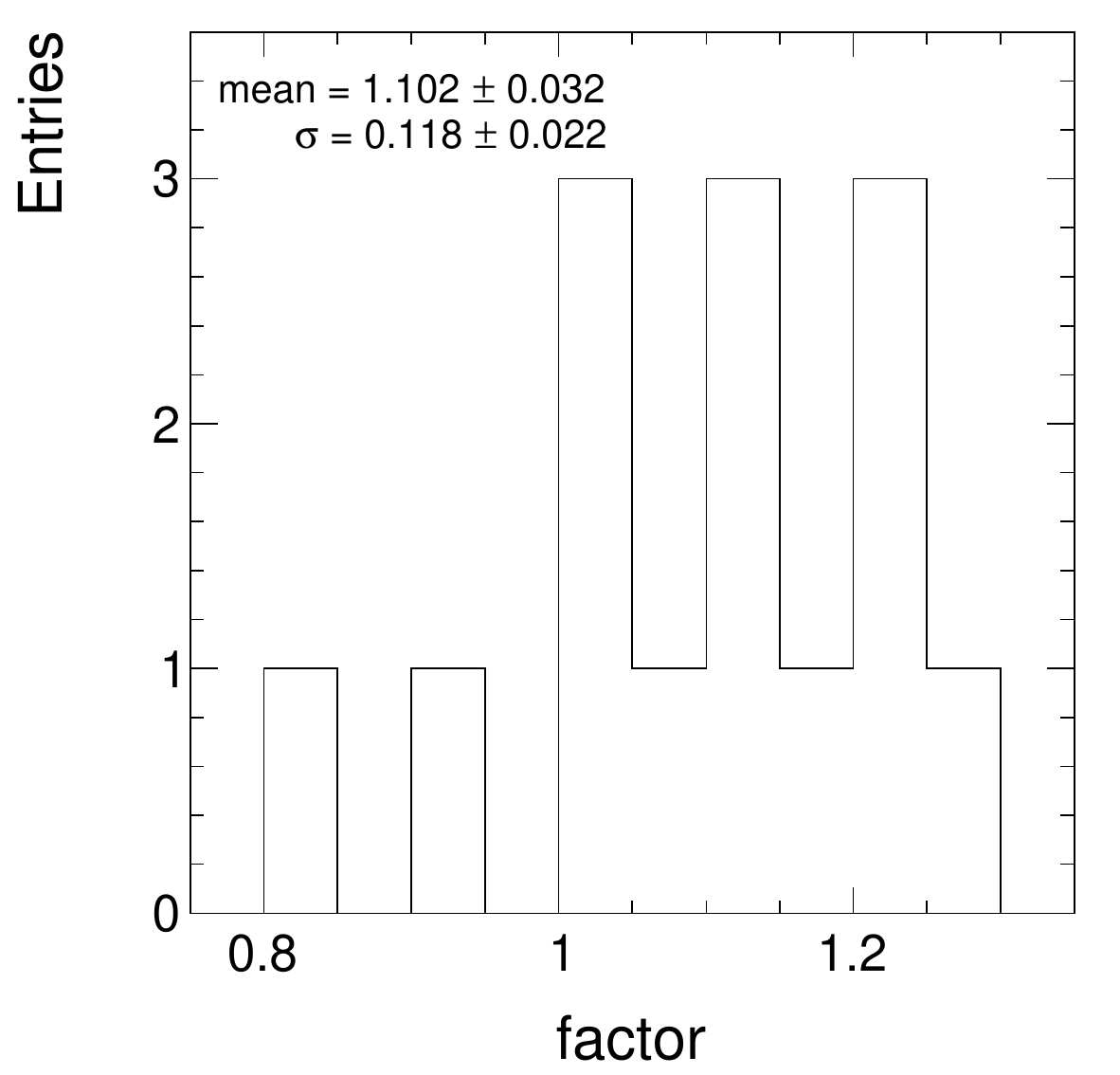} 
	 		\caption{Distribution of factors between data and global interpolation for the spectrometer experiments Sect.~\ref{sec:melissinos} to \ref{sec:antreasyan}}
  		\label{fig:factors}
 		\end{center}
	\end{figure}

	This means that there is no indication of an overall shift with respect to the reference data which do not need re-normalization. It also means that there is a general tendency to under-estimate the normalization errors.
	\item For about half the experiments there are additional systematic aberrations \cite{melissinos,dekkers,allaby1,allaby2,beier,johnson}. This has to do with the fact that the experiments generally do not cover a major fraction of phase space in one go but only see very limited areas one at a time with the danger of varying corrections and time-dependent instability.
	\item The detailed scrutiny of each experiment indicates that a precision evaluation of the behaviour of inclusive cross sections is not possible without the presence of reliable reference data allowing for the judgement and correction of systematic deviations.
	\item After correction for the observed systematic effects the $\log(s)$ dependence of the characteristic quantities $\langle\Delta/\sigma\rangle$ and $\sigma(\langle\Delta/\sigma\rangle)$ as well as the mean deviations in percent are shown in Fig.~\ref{fig:precision} to be compared to Fig.~\ref{fig:summary_ref} for the reference experiments.
\end{enumerate}

\begin{figure}[h]
 	\begin{center}
   	\includegraphics[width=16.2cm] {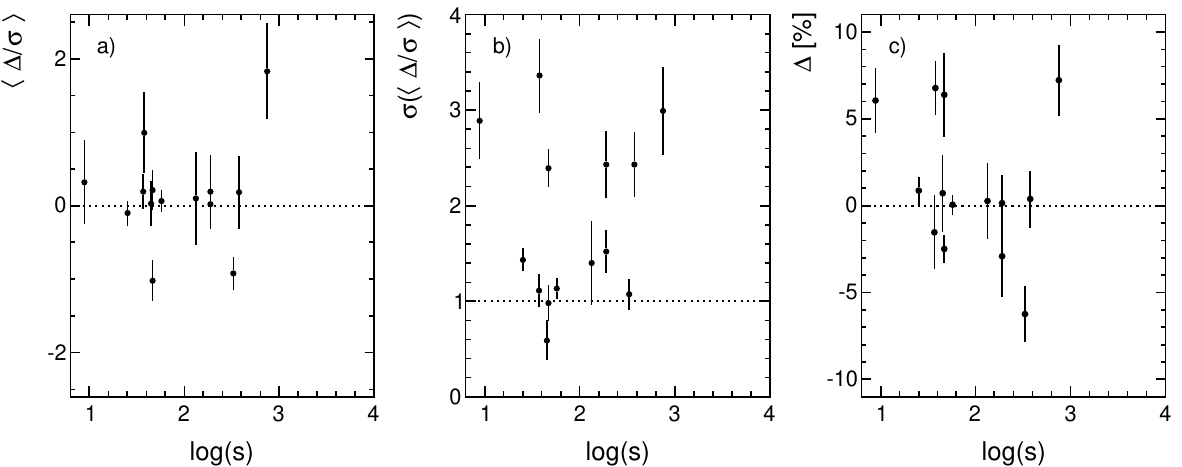} 
	 	\caption{a) Distribution of re-normalization factors applied for the data Figs.~\ref{fig:melissinos}--\ref{fig:johnson3}, b) mean normalized residuals after re-normalization, c) the corresponding variances as a function of $\log(s)$}
  	\label{fig:precision}
 	\end{center}
\end{figure}

Although the mean residuals are centred around the expectation value 0 (Fig.~\ref{fig:precision}a) the variances tend to be sizeably above the expected value 1 for about half the experiments (Fig.~\ref{fig:precision}b). These larger fluctuations indicate further systematic effects in the data which are not taken out completely by the applied re-scaling. The mean deviations in percent, Fig.~\ref{fig:precision}c, are nevertheless in most cases close to zero with some exceptions around up to $\pm$6\%.

%
%
\section{Discussion of the NA61 results}
\vspace{3mm}
\label{sec:na61}
%
%
\subsection{Data without active particle identification \cite{abgrall} (Tab.~\ref{tab:na61})}
\vspace{3mm}
\label{sec:na61_hmin}

NA61 has measured inclusive cross sections at 20, 31, 40, 80 and 158~GeV/c at the CERN SPS, using basically the NA49 detector \cite{nim}. The results \cite{abgrall} were obtained without the use of the available particle identification just reconstructing negative tracks (h$^-$), and using available microscopic models in order to subtract the K$^-$ and $\overline{\textrm{p}}$ yields .The results are binned in the $p_T$ range from 0.026 to 0.575~GeV/c in steps of 0.05 GeV/c,from 0.65 to 0.95~GeV/c in steps of 0.1 GeV/c and from 1.125 to 1.375~GeV/c in two steps of 0.25~GeV/c. The rapidity range extends from 0.1 to 2.9 (3.5) units at 20 (158) GeV/c beam momentum, in steps of 0.2. This yields about 200 to 250 data points per energy. The number of negative pions ranges from 23 k at 20~GeV/c to 500 k at 158~GeV/c beam momentum (Tab.~\ref{tab:na61}).

\begin{figure*}[h]
 	\begin{center}
   	\begin{turn}{\rotAngle}
   	   		\begin{minipage}{\capw}
	 			\caption{Cross sections $f/\sigma_{\textrm{inel}}$ \cite{abgrall} as functions of $p_T$ and rapidity at 20 and 158~GeV/c beam momentum. The inset shows the distribution of normalized residuals}
  			\label{fig:na61_interp}
			\end{minipage}
		\end{turn}
   	\begin{turn}{\rotAngle}
		\begin{minipage}{\capw}
		\hspace{-9mm}
   	\includegraphics[width=\figwh] {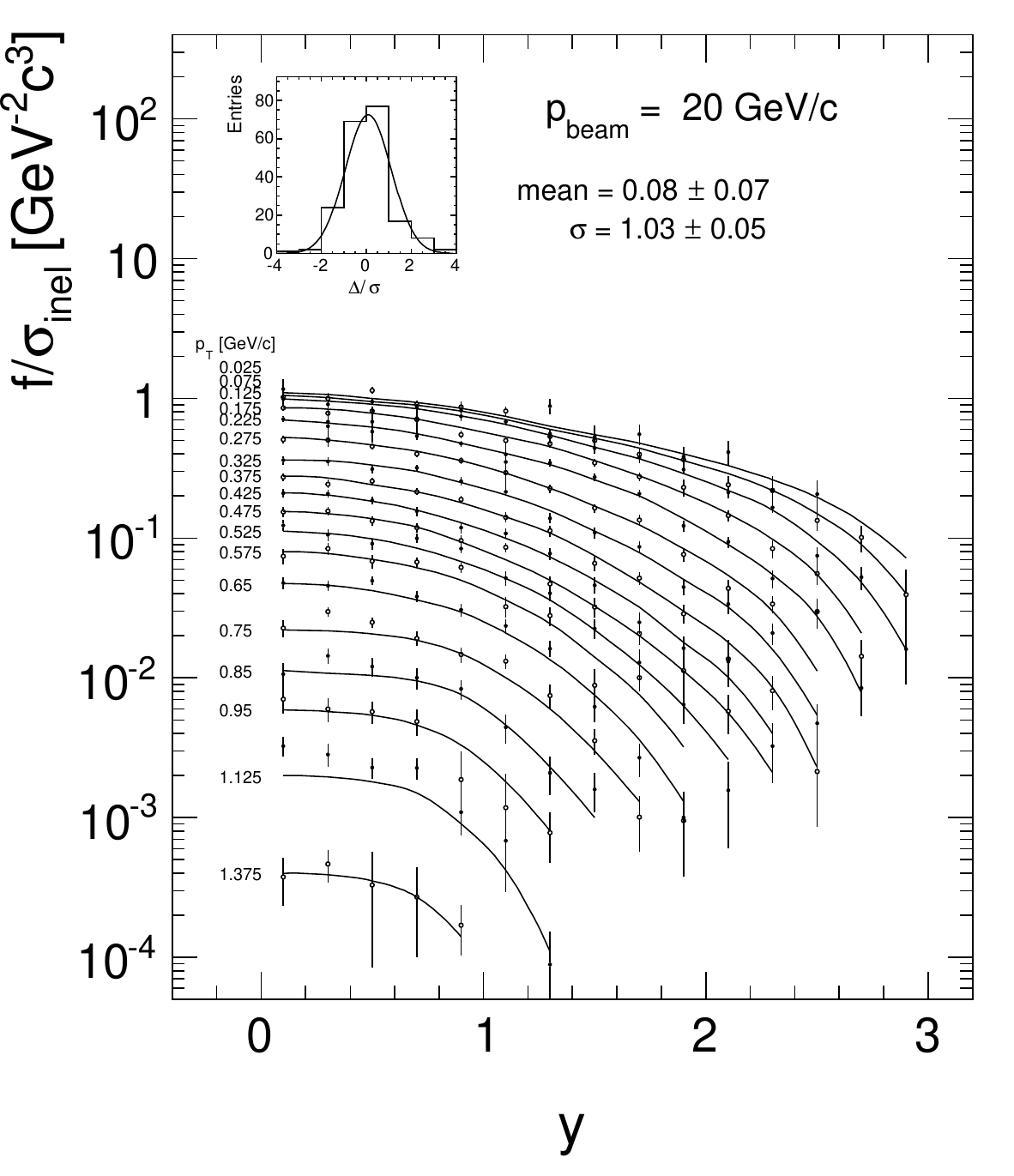} 
		\hspace{5mm}
   	\includegraphics[width=\figwh] {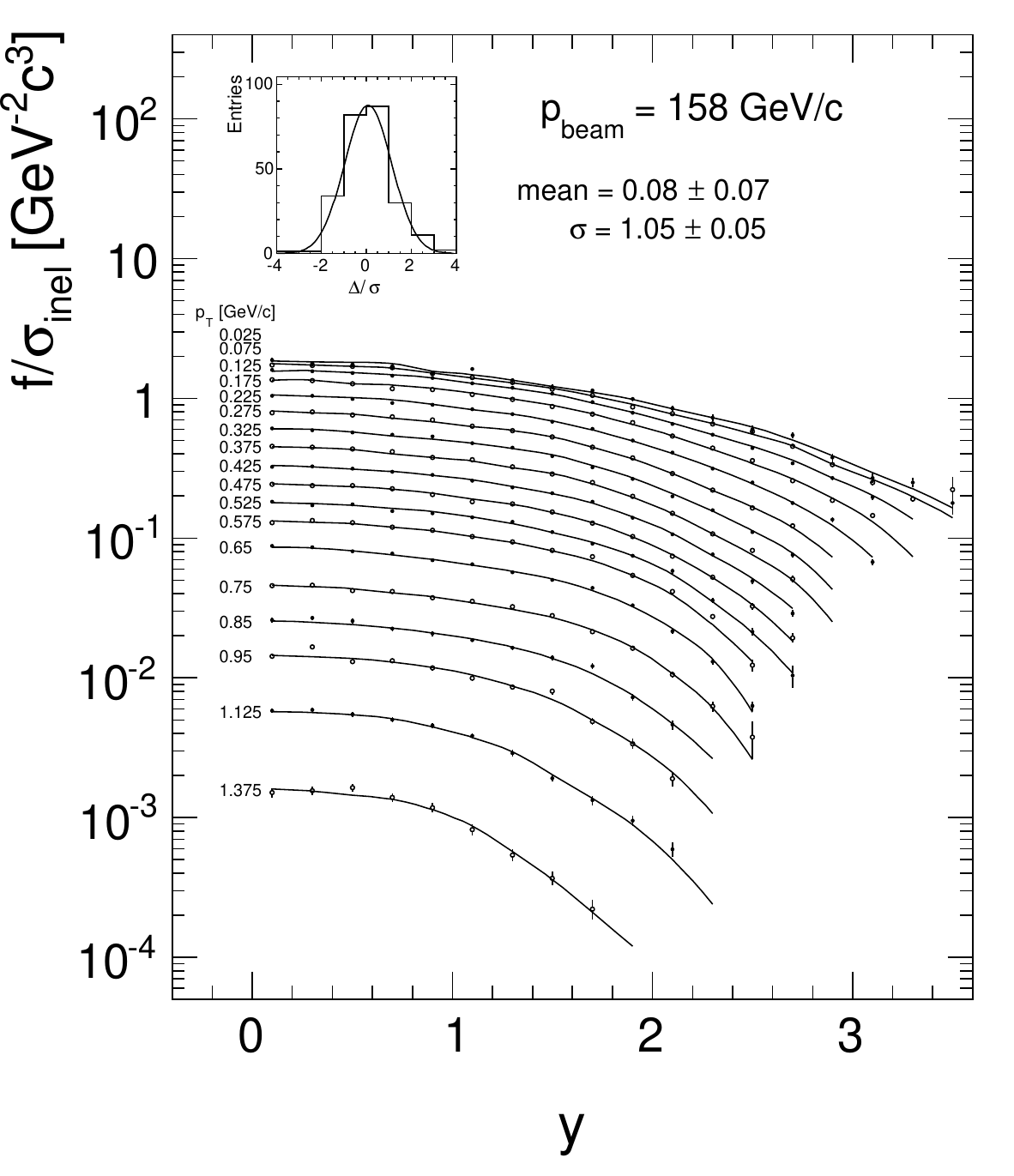} 
		\end{minipage}
		\end{turn}
 	\end{center}
\end{figure*}

As the rapidity bins and most of the $p_T$ bins do not coincide with the ones used for the global interpolation (Sect.~\ref{sec:data_grid}), the data have to be interpolated in both variables. This interpolation is obtained by eye-ball fits (Sect.~\ref{sec:data_interp}) simultaneously in transverse momentum, rapidity and interaction energy thus obtaining a reduction of the statistical errors which reach up to 30\% in the high $p_T$ and high rapidity regions. The interpolation is presented in Fig.~\ref{fig:na61_interp} for the two beam energies 20 and 158~GeV/c including in the insets the distributions of the reduced residuals $\Delta/\sigma$.

\begin{figure*}[h]
 	\begin{center}
 	\vspace{0.1mm}
   	\begin{turn}{\rotAngle}
   		\begin{minipage}{\capsh}
				\hspace{\capsh}
   		\end{minipage}
   		\begin{minipage}{11cm}
	 			\caption{Interpolated cross sections $f/\sigma_{\textrm{inel}}$ \cite{abgrall} as functions of $p_T$ and $y_{\textrm{lab}}$ at 20~GeV/c beam momentum (broken lines) compared to the global interpolation (full lines) in steps of 0.1~GeV/c in $p_T$. The insets show the distribution of the factors relative to the general interpolation for all $p_T$ values and for $p_T >$~0.7~GeV/c separately}
  			\label{fig:na61_comp20}
			\end{minipage}
		\end{turn}
   	\includegraphics[width=10.8cm,angle=\rotAngle] {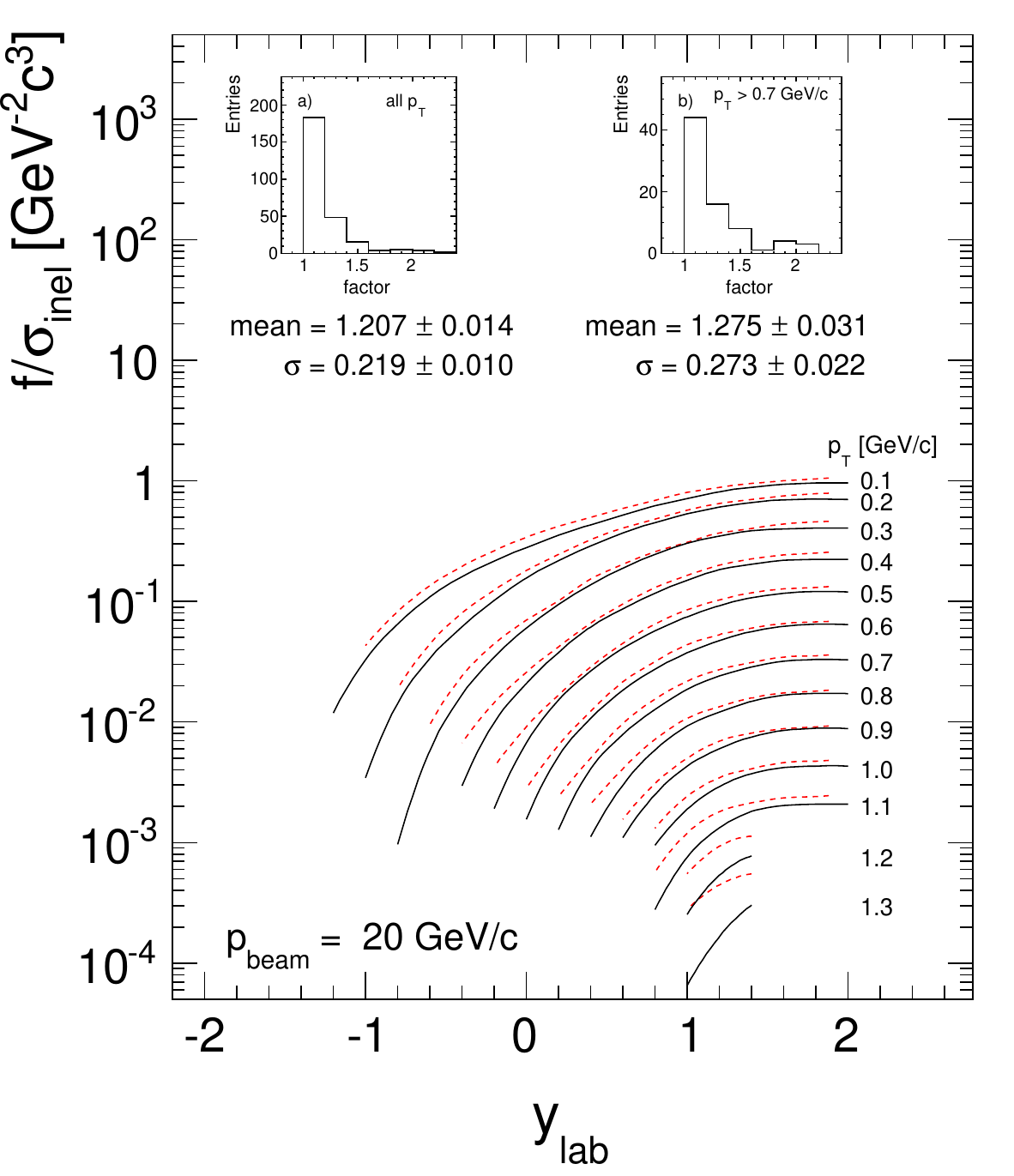} 
   	\begin{turn}{\rotAngle}
   		\begin{minipage}{\capsh}
				\hspace{\capsh}
   		\end{minipage}
   		\begin{minipage}{11cm}
	 	\caption{Interpolated cross sections $f/\sigma_{\textrm{inel}}$ \cite{abgrall} as functions of $p_T$ and $y_{\textrm{lab}}$ at 31~GeV/c beam momentum (broken lines) compared to the global interpolation (full lines) in steps of 0.1~GeV/c in $p_T$. The insets show the distribution of the factors relative to the general interpolation for all $p_T$ values and for $p_T >$~0.7~GeV/c separately}
  	\label{fig:na61_comp31}
			\end{minipage}
		\end{turn}
   	\includegraphics[width=10.8cm,angle=\rotAngle] {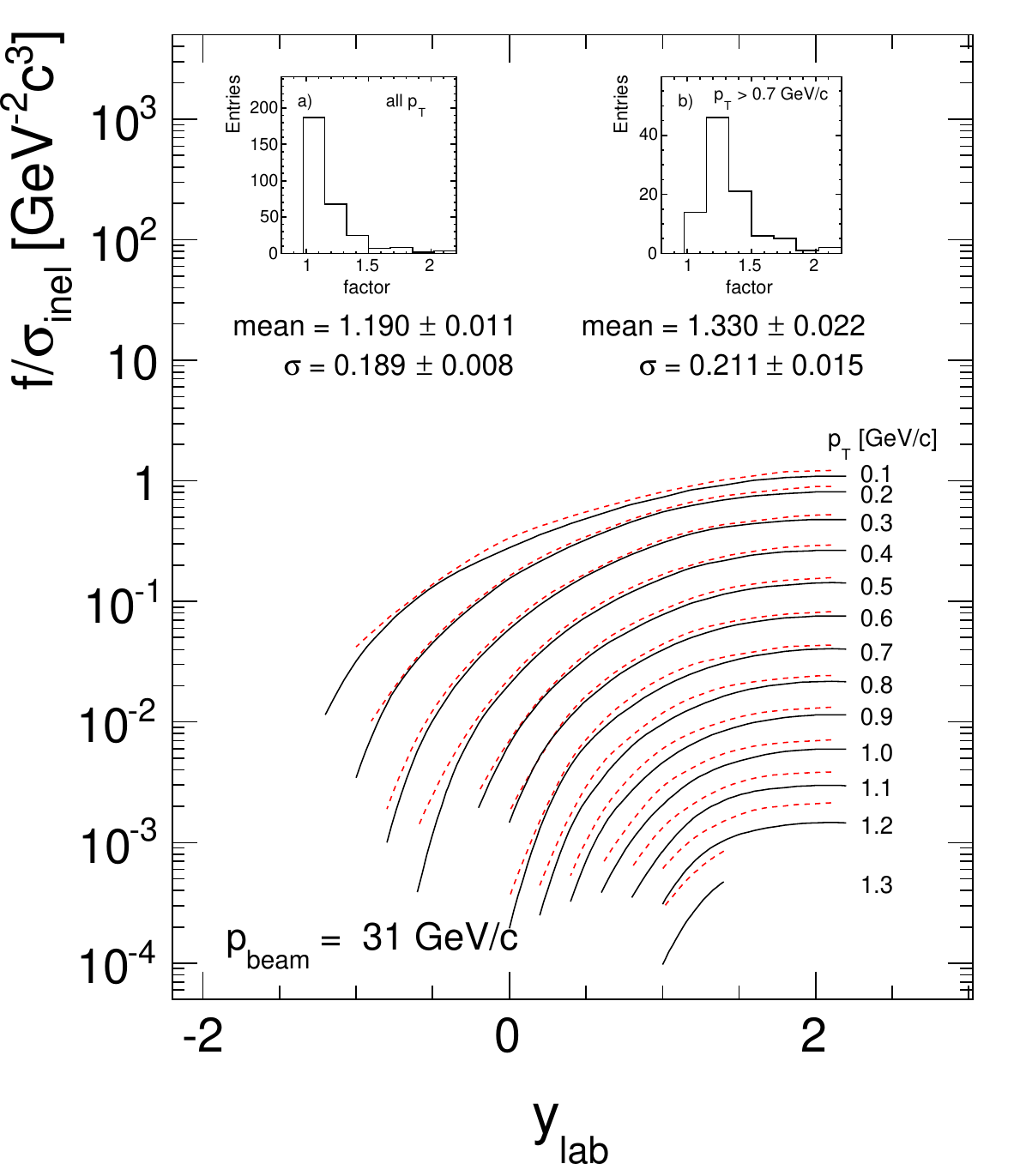} 
 	\end{center}
\end{figure*}

\begin{figure*}[h]
 	\begin{center}
 	\vspace{0.1mm}
   	\begin{turn}{\rotAngle}
   		\begin{minipage}{\capsh}
				\hspace{\capsh}
   		\end{minipage}
   		\begin{minipage}{11cm}
	 			\caption{Interpolated cross sections $f/\sigma_{\textrm{inel}}$ \cite{abgrall} as functions of $p_T$ and $y_{\textrm{lab}}$ at 40~GeV/c beam momentum (broken lines) compared to the global interpolation (full lines) in steps of 0.1~GeV/c in $p_T$. The insets show the distribution of the factors relative to the general interpolation for all $p_T$ values and for $p_T >$~0.7~GeV/c separately}
  			\label{fig:na61_comp40}
			\end{minipage}
		\end{turn}
   	\includegraphics[width=10.8cm,angle=\rotAngle] {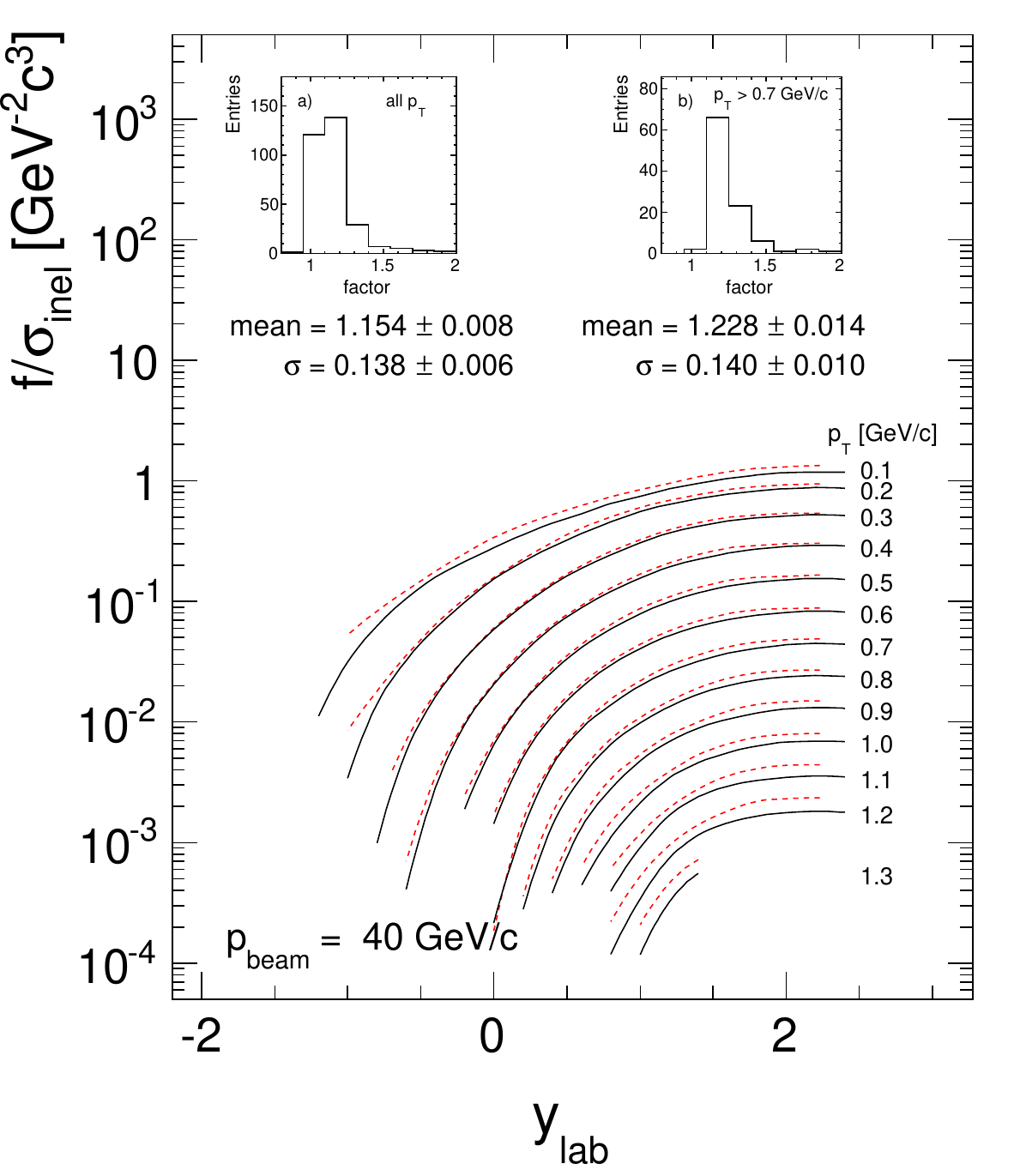} 
   	\begin{turn}{\rotAngle}
   		\begin{minipage}{\capsh}
				\hspace{\capsh}
   		\end{minipage}
   		\begin{minipage}{11cm}
	 	\caption{Interpolated cross sections $f/\sigma_{\textrm{inel}}$ \cite{abgrall} as functions of $p_T$ and $y_{\textrm{lab}}$ at 80~GeV/c beam momentum (broken lines) compared to the global interpolation (full lines) in steps of 0.1~GeV/c in $p_T$. The insets show the distribution of the factors relative to the general interpolation for all $p_T$ values and for $p_T >$~0.7~GeV/c separately}
  	\label{fig:na61_comp80}
			\end{minipage}
		\end{turn}
   	\includegraphics[width=10.8cm,angle=\rotAngle] {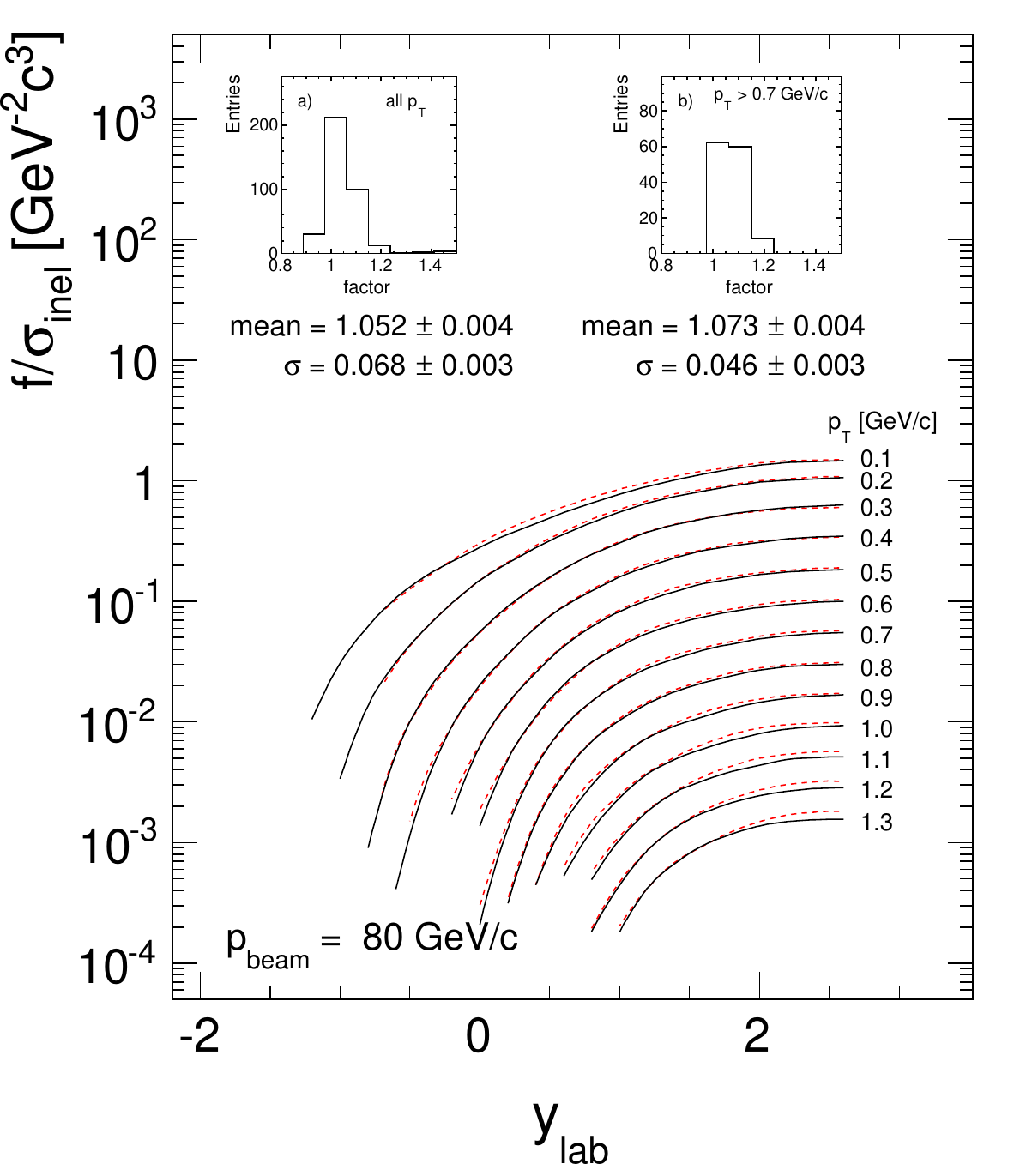} 
 	\end{center}
\end{figure*}

\begin{figure}[h]
 	\begin{center}
   	\includegraphics[width=12cm] {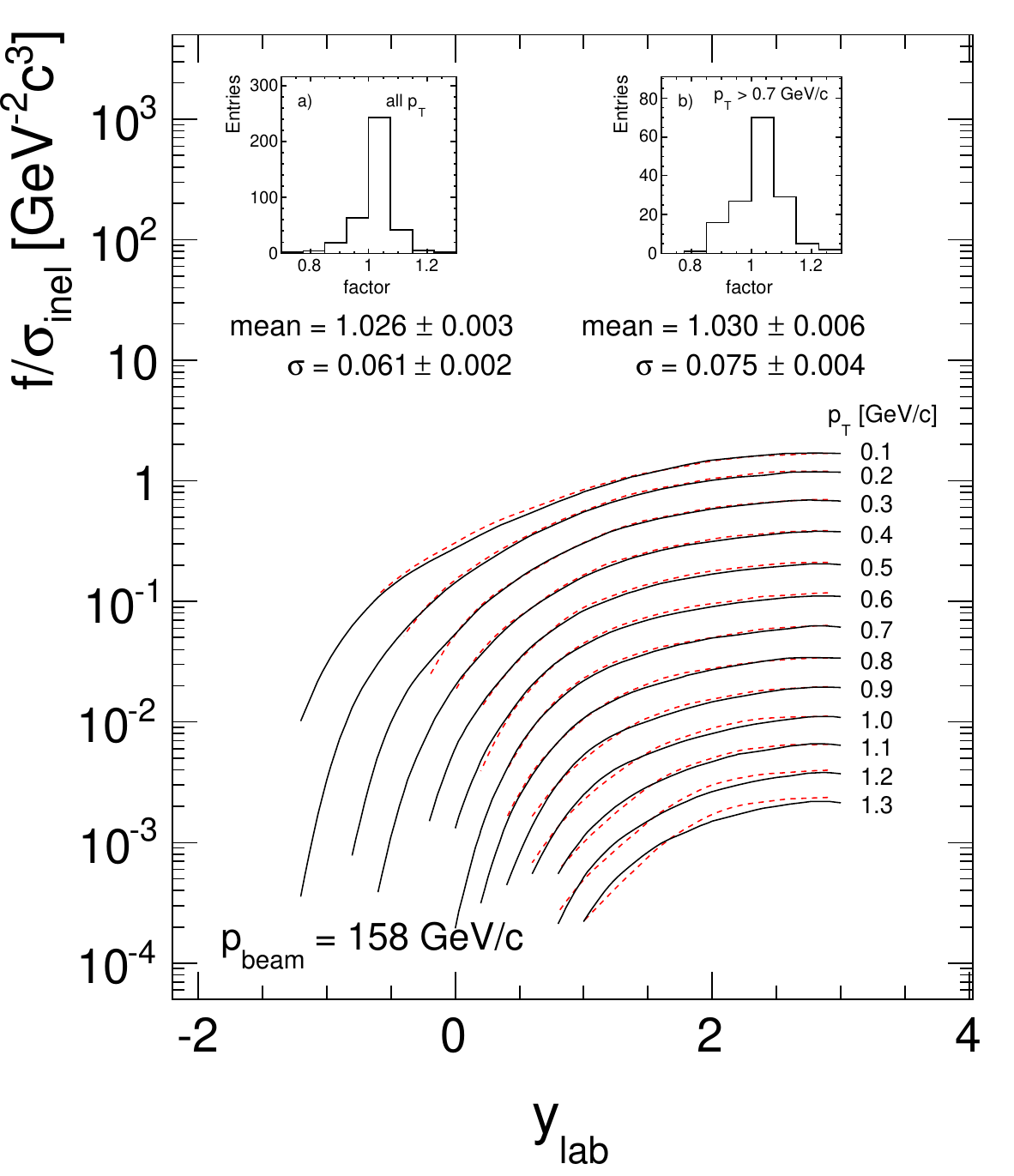} 
	 	\caption{Interpolated cross sections $f/\sigma_{\textrm{inel}}$ \cite{abgrall} as functions of $p_T$ and $y_{\textrm{lab}}$ at 158~GeV/c beam momentum (broken lines) compared to the global interpolation (full lines) in steps of 0.1~GeV/c in $p_T$. The insets show the distribution of the factors relative to the general interpolation for all $p_T$ values and for $p_T >$~0.7~GeV/c separately}
  	\label{fig:na61_comp158}
 	\end{center}
\end{figure}

The normalized residual distributions show the expected Gaussian distributions centred at zero with an rms of one unit within errors. These distributions may now be interpolated in $p_T$ at fixed rapidity to the standard $p_T$ binning and ultimately plotted as functions of $y_{\textrm{lab}}$. At this stage a comparison with the global interpolation becomes possible as shown in Figs.~\ref{fig:na61_comp20} to \ref{fig:na61_comp158}. This comparison is shown in steps of 0.1~GeV/c in transverse momentum as a function of $y_{\textrm{lab}}$ with insets giving the distribution of the relative difference with respect to the global interpolation.

This comparison reveals a complex pattern of systematic deviations. The NA61 data are systematically higher than the global interpolation with an important dependence on beam momentum. The difference distributions show a strong increase with decreasing interaction energy and develop long tails which reach up to more than 100\% at the lower energies. This increase is centred at higher transverse momenta as shown in the insets for $p_T >$~0.7~GeV/c. Only at $p_{\textrm{beam}}$~=~158~GeV/c where the data are directly comparable to the NA49 results \cite{pp_pion} the differences stay below the $\pm$20\% level.

\begin{figure}[h]
 	\begin{center}
   	\includegraphics[width=16cm] {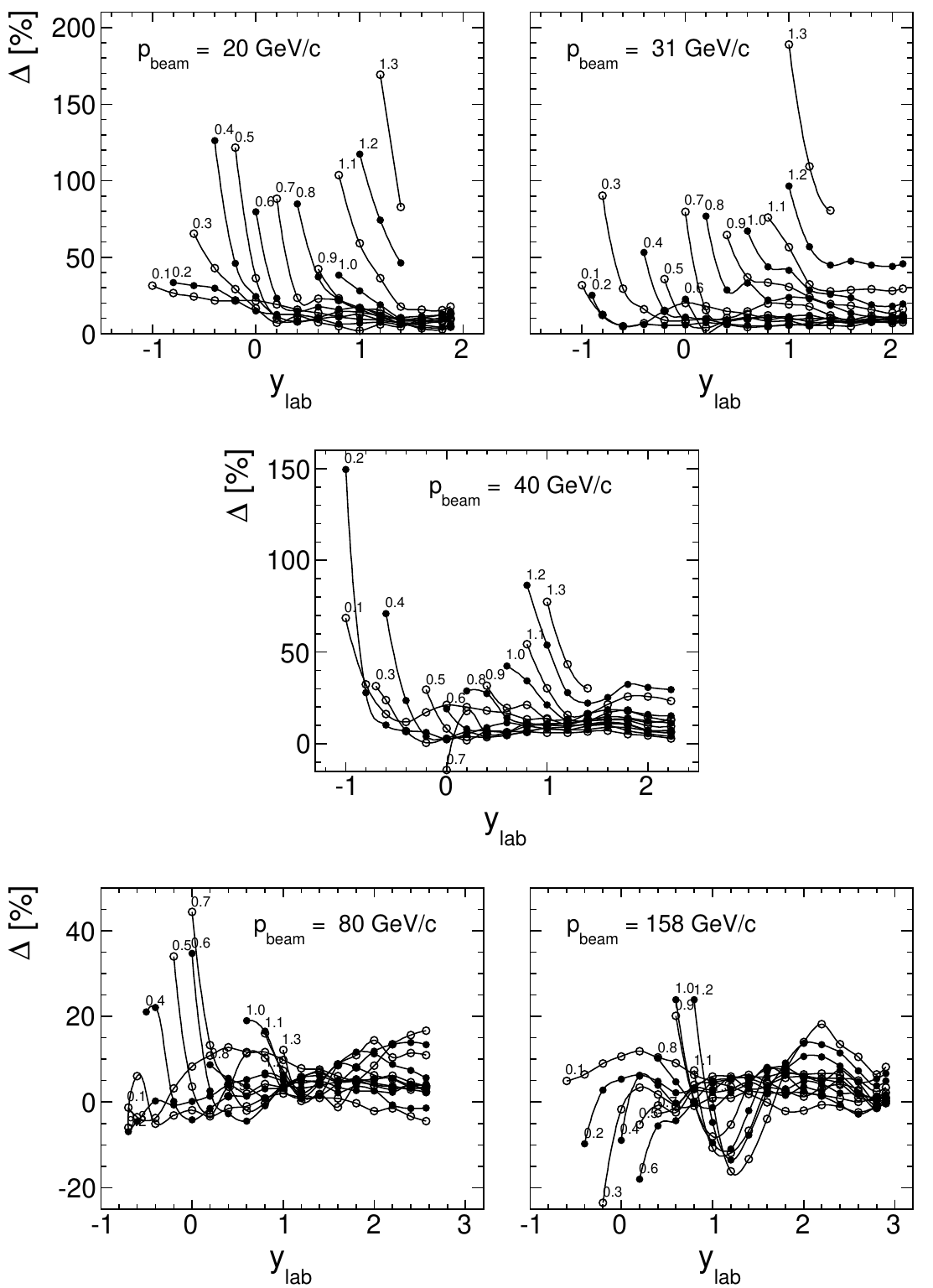} 
	 	\caption{Percent differences between NA61 data and the global interpolation as a function of $y_{\textrm{lab}}$ for fixed $p_T$ at all beam momenta}
  	\label{fig:na61dev}
 	\end{center}
\end{figure}

A detailed inspection of this inter-dependence is presented in Fig.~\ref{fig:na61dev} where the percent deviations are shown as functions of $y_{\textrm{lab}}$ for fixed $p_T$.

\begin{figure}[h]
 	\begin{center}
   	\includegraphics[width=16cm] {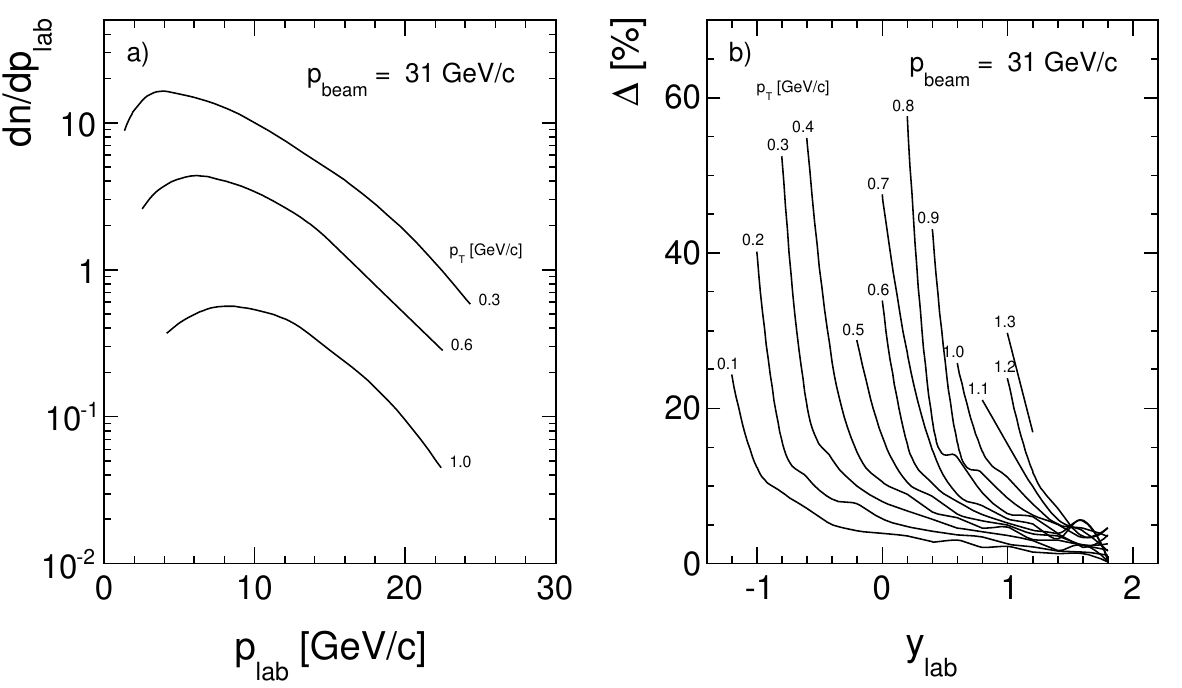} 
	 	\caption{a) $\pi^-$ yields $dn/dp_{\textrm{lab}}$ as a function of $p_{\textrm{lab}}$ for beam momenta of 31~GeV/c and transverse momenta of 0.3, 0.6 and 1~GeV/c; b) predicted changes in percent of the cross sections for 3\% upward modification of the momentum scale starting from the global interpolation, as a function of $y_{\textrm{lab}}$ at fixed $p_T$, for $p_{\textrm{beam}}$~=~31~GeV/c}
  	\label{fig:plabscale}
 	\end{center}
\end{figure}

There is only one possible experimental effect that might explain the large deviations: the momentum scale. The field map of the NA49 detector has been established for the highest beam momentum available at the CERN SPS for Heavy Ion interactions which is 158~GeV/c. If the central solenoidal fields are scaled down with decreasing beam momentum without re-establishing detailed field maps, the change of the corresponding stray fields will decrease the field integral such that the momenta are over-estimated. As the yield distributions at fixed transverse momentum are strongly non-linear in $p_{\textrm{lab}}$, a momentum error will lead to a yield deviation which typically increases with increasing lab momentum. This is evident from the distributions shown in Fig.~\ref{fig:plabscale}a) at $p_{\textrm{beam}}$~=~31~GeV/c for a few values of $p_T$.

The application of a constant shift in lab momentum of +3\% at $p_{\textrm{beam}}$~=~31~GeV/c  indeed reproduces the main features of the observed deviations (Fig.~\ref{fig:na61dev}) as presented in Fig.~\ref{fig:plabscale}b).

\begin{figure*}[h]
 	\begin{center}
 	\vspace{0.1mm}
   	\begin{turn}{\rotAngle}
   		\begin{minipage}{\capsh}
				\hspace{\capsh}
   		\end{minipage}
   		\begin{minipage}{\capwh}
	 			\caption{Cross sections $f/\sigma_{\textrm{inel}}$ as functions of $p_T$ and $y_{\textrm{lab}}$ for $y_{\textrm{lab}}$~=~0.0. The full lines represent the global interpolation, the data points the NA61 results. The broken lines are eyeball fits to these data demonstrating the smooth $s$-dependence of the deviations}
  			\label{fig:na61_ylab0}
			\end{minipage}
		\end{turn}
   	\includegraphics[width=\figwh,angle=\rotAngle] {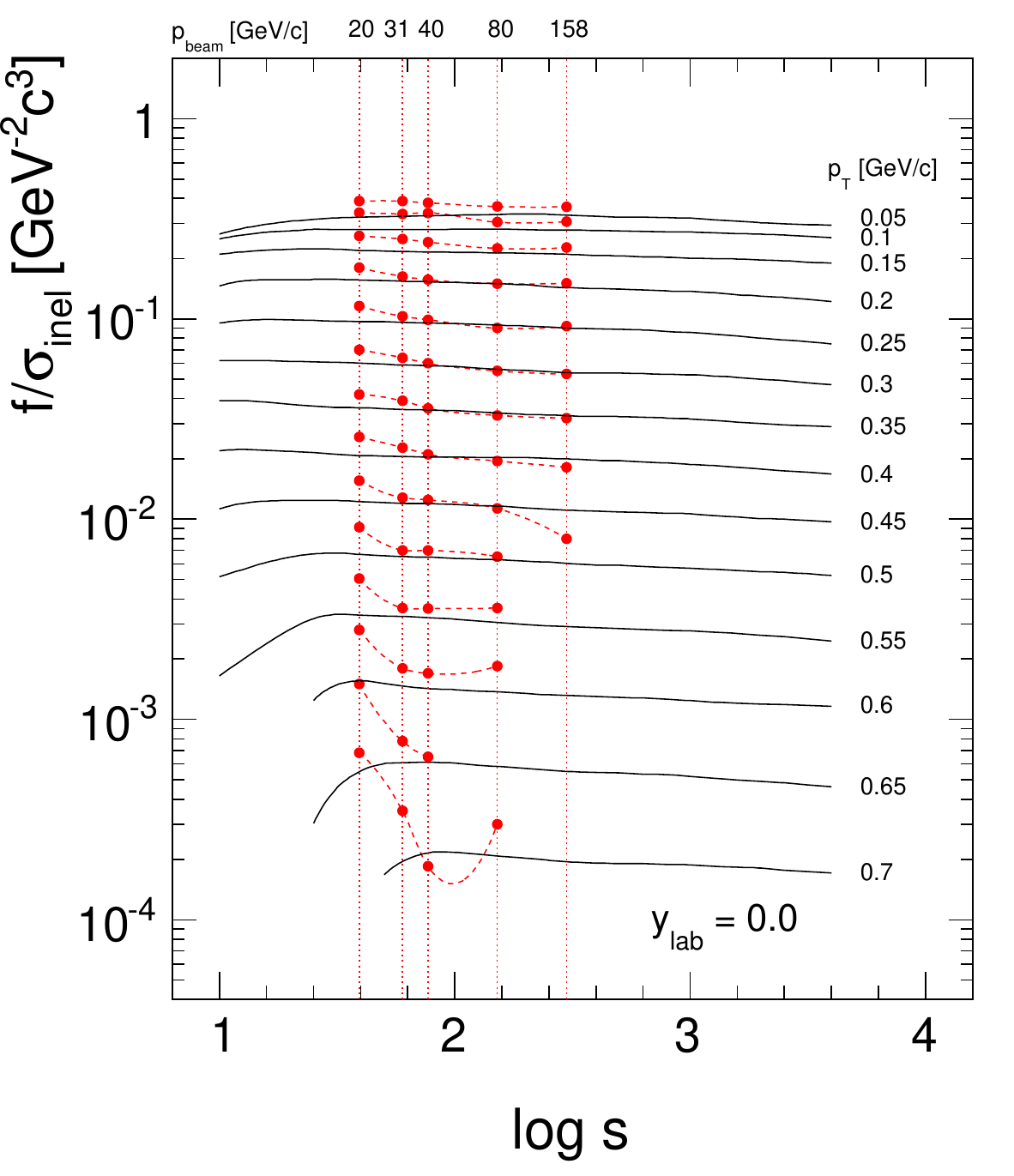} 
   	\begin{turn}{\rotAngle}
   		\begin{minipage}{\capsh}
				\hspace{\capsh}
   		\end{minipage}
   		\begin{minipage}{\capwh}
	 			\caption{Cross sections $f/\sigma_{\textrm{inel}}$ as functions of $p_T$ and $y_{\textrm{lab}}$ for $y_{\textrm{lab}}$~=~1.0. The full lines represent the global interpolation, the data points the NA61 results. The broken lines are eyeball fits to these data demonstrating the smooth $s$-dependence of the deviations}
  			\label{fig:na61_ylab1}
			\end{minipage}
		\end{turn}
   	\includegraphics[width=\figwh,angle=\rotAngle] {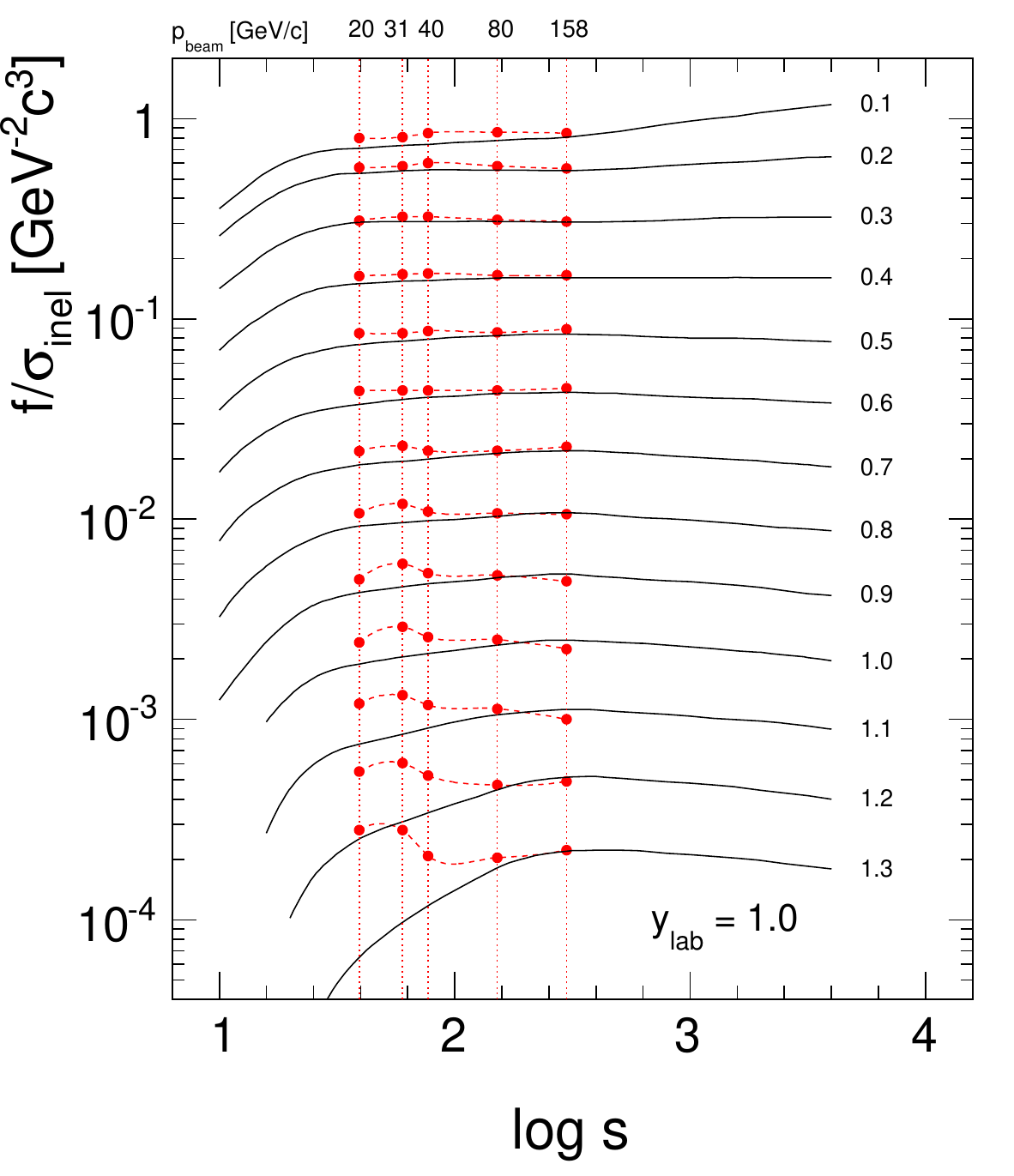} 
 	\end{center}
\end{figure*}

\begin{figure*}[h]
 	\begin{center}
 	\vspace{0.1mm}
   	\begin{turn}{\rotAngle}
   		\begin{minipage}{\capsh}
				\hspace{\capsh}
   		\end{minipage}
   		\begin{minipage}{\capwh}
	 			\caption{Cross sections $f/\sigma_{\textrm{inel}}$ as functions of $p_T$ and $y_{\textrm{lab}}$ for $y_{\textrm{lab}}$~=~1.4. The full lines represent the global interpolation, the data points the NA61 results. The broken lines are eyeball fits to these data demonstrating the smooth $s$-dependence of the deviations}
  			\label{fig:na61_ylab14}
			\end{minipage}
		\end{turn}
   	\includegraphics[width=\figwh,angle=\rotAngle] {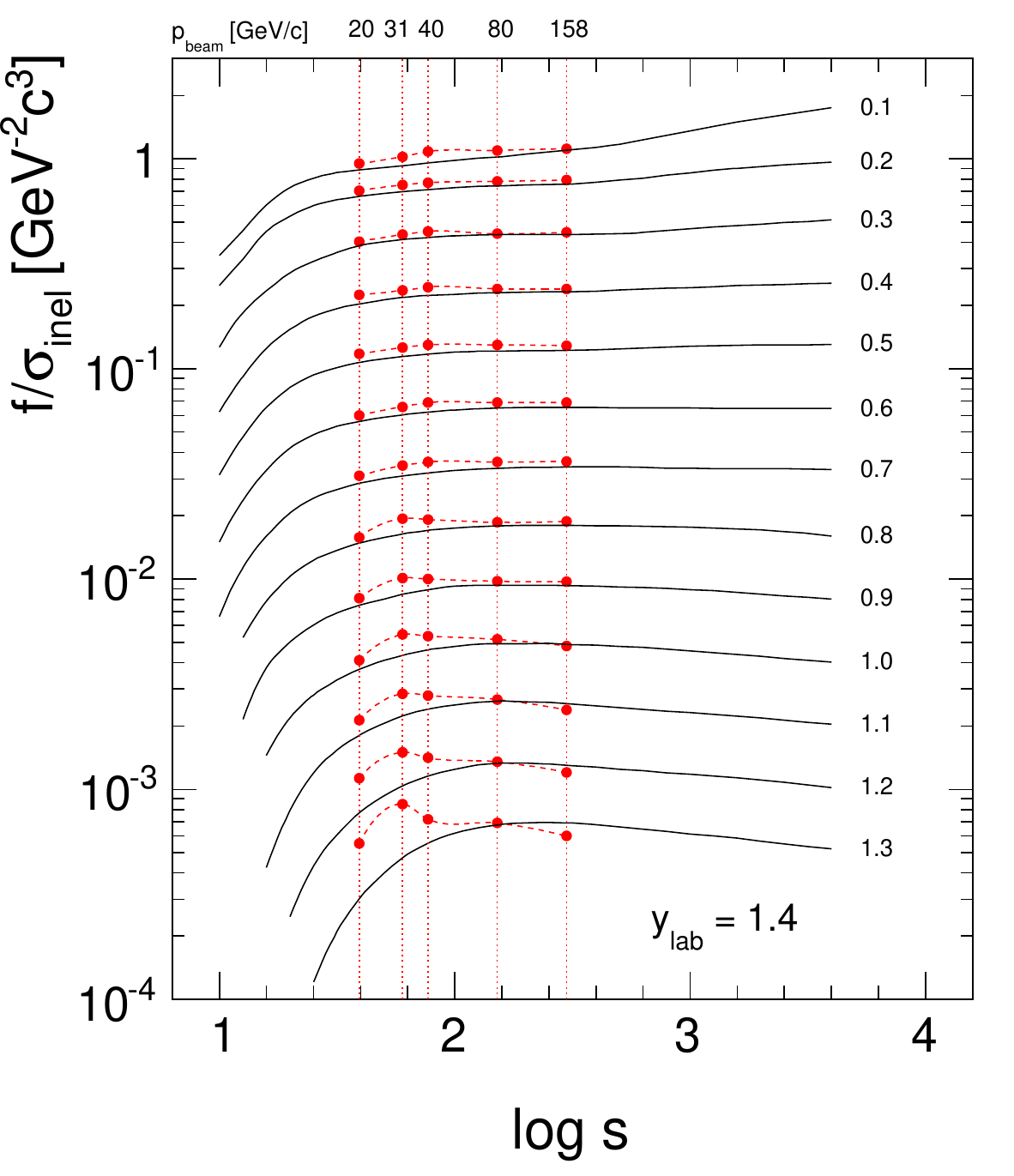} 
   	\begin{turn}{\rotAngle}
   		\begin{minipage}{\capsh}
				\hspace{\capsh}
   		\end{minipage}
   		\begin{minipage}{\capwh}
	 			\caption{Cross sections $f/\sigma_{\textrm{inel}}$ as functions of $p_T$ and $y_{\textrm{lab}}$ for $y_{\textrm{lab}}$~=~2.0. The full lines represent the global interpolation, the data points the NA61 results. The broken lines are eyeball fits to these data demonstrating the smooth $s$-dependence of the deviations}
  			\label{fig:na61_ylab2}
			\end{minipage}
		\end{turn}
   	\includegraphics[width=\figwh,angle=\rotAngle] {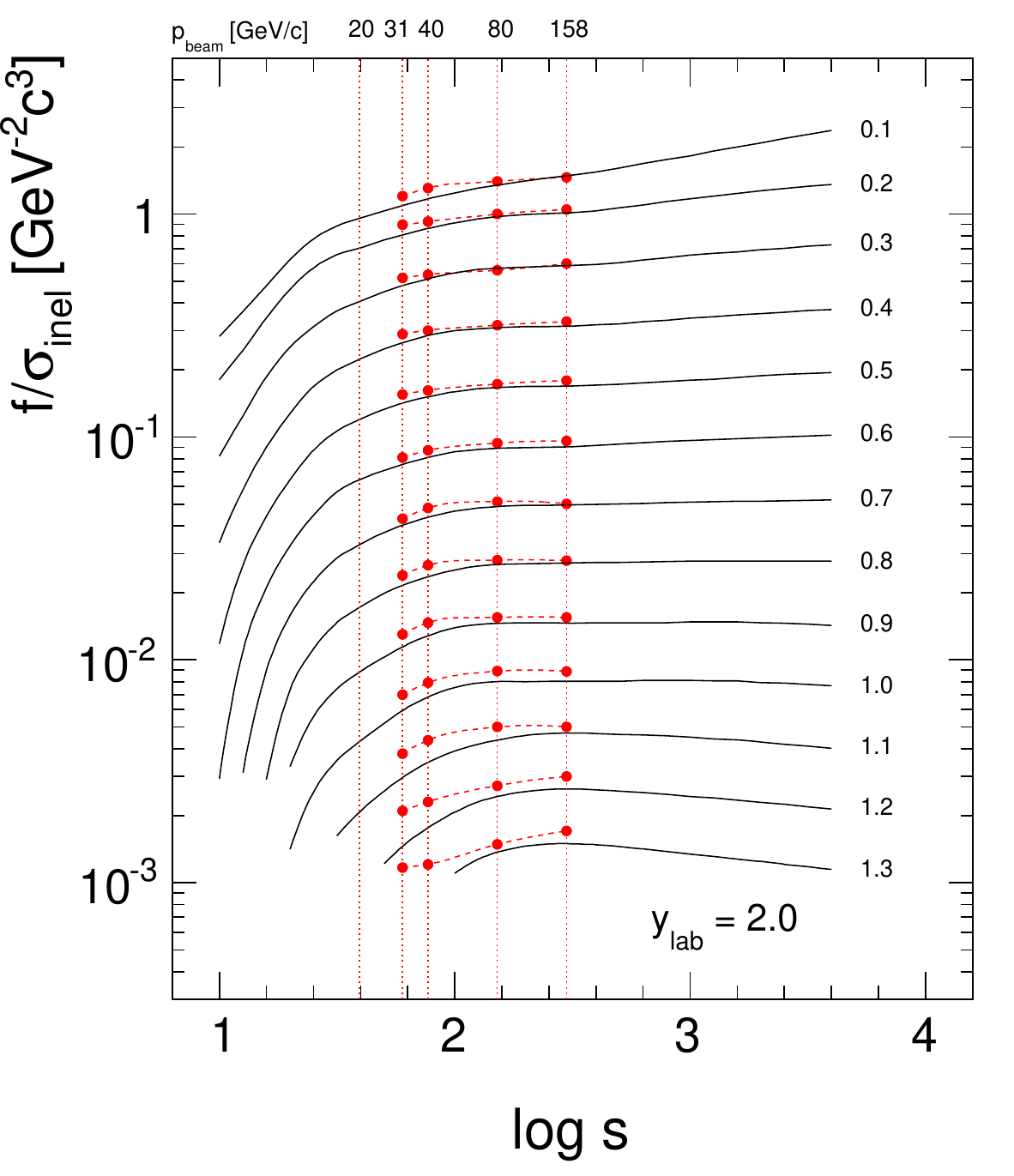} 
 	\end{center}
\end{figure*}

The systematic increase of the deviations with decreasing interaction energy is apparent from Figs.~\ref{fig:na61_ylab0}--\ref{fig:na61_ylab2} where the cross sections are compared to the global interpolation as functions of $\log(s)$ and $p_T$ for four values of $y_{\textrm{lab}}$.

The evidence presented in the preceding Figures leads to several conclusions:

\begin{enumerate}
	\item The NA61 data show deviations from the general interpolation on a scale which goes beyond anything that has been seen in the detailed data analysis of the 30 experiments presented in Sect.~\ref{sec:countdata} above.
	\item The authors claim systematic uncertainties on the level of 5--6 percent.
	\item Although these errors exceed the ones given by the reference experiments by factors of two to three they are completely insufficient to explain deviations which reach more than 100 percent.
	\item Neither modifications of the applied corrections nor the uncertainties involved in the necessary interpolations may reach the observed levels of aberrations.
	\item The deviations have systematic and smooth dependencies on the three kinematical variables $p_T$, $y_{\textrm{lab}}$ and $\log(s)$, in particular visible in the $\log(s)$ scale, Figs.~\ref{fig:na61_ylab0}--\ref{fig:na61_ylab2}.
\end{enumerate}

%
%
\subsection{NA61 data with particle identification \cite{aduszkiewicz}}
\vspace{3mm}
\label{sec:na61_pid}

\begin{figure}[h]
 	\begin{center}
   	\begin{minipage}{10cm}
   		\hspace{-15mm}
   		\includegraphics[width=9.7cm] {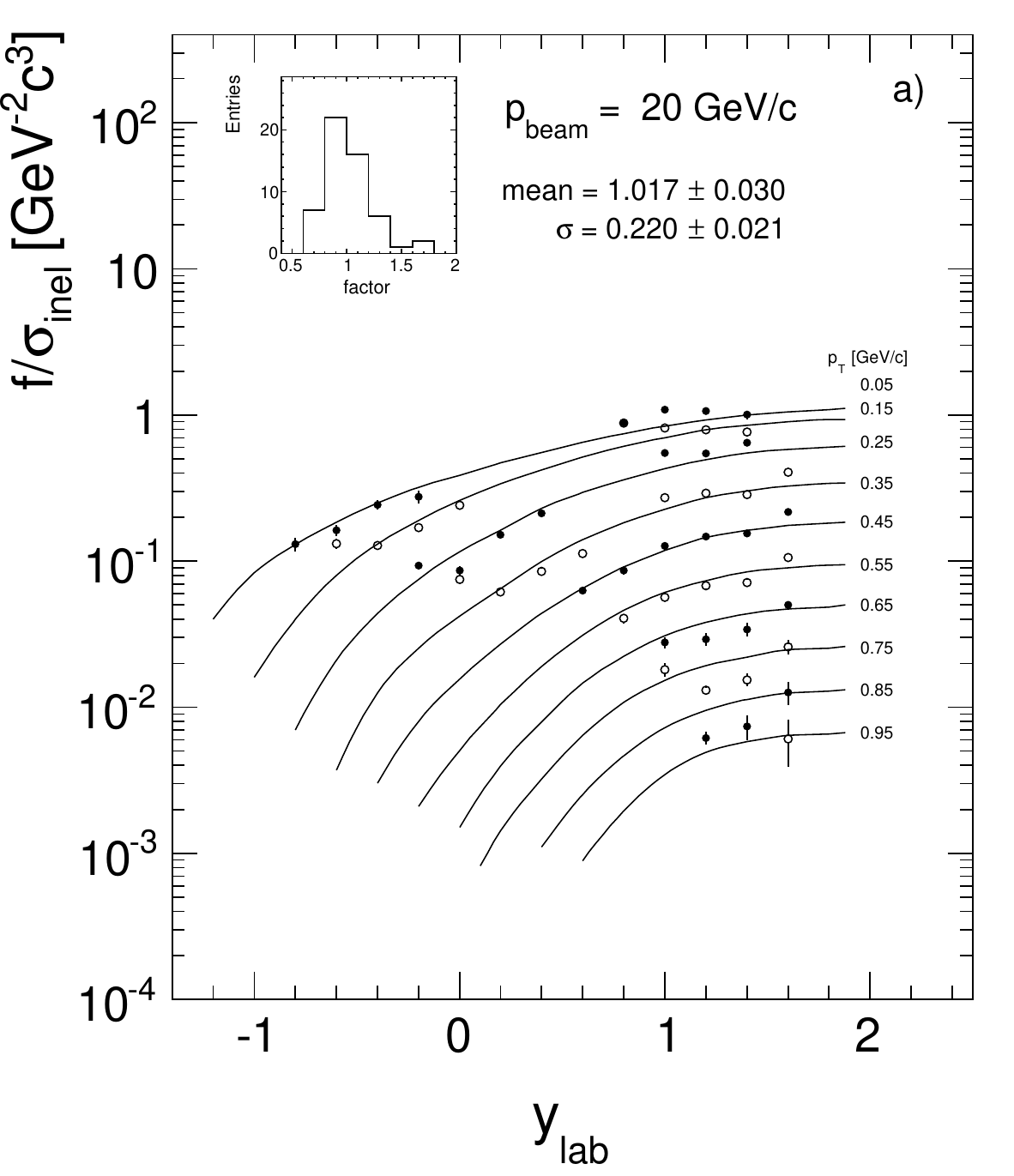} 
		\end{minipage}
   	\begin{minipage}{5.cm}
   		\hspace{-10mm}
   		\includegraphics[width=5.cm] {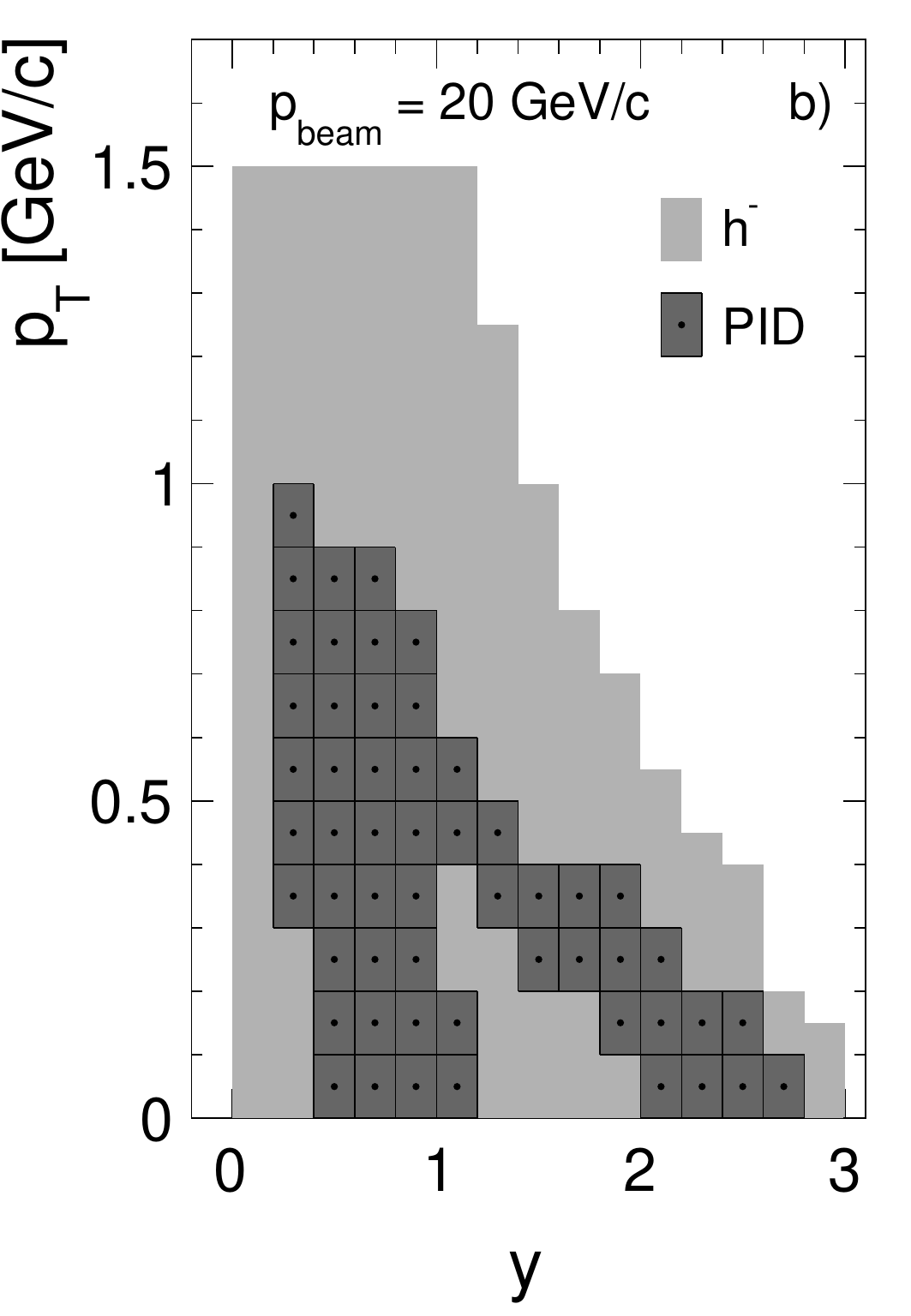} 
		\end{minipage}
	 	\caption{a) Cross sections $f/\sigma_{\textrm{inel}}$ \cite{aduszkiewicz} with particle identification as functions of $p_T$ and $y_{\textrm{lab}}$ compared to the interpolation of the NA61 h$^-$ data (full lines) for 20~GeV/c beam momentum; b) coverage in the $y$/$p_T$. The h- acceptance is shown in comparison as the light grey area}
  	\label{fig:na61pid20}
 	\end{center}
\end{figure}

\begin{figure*}[h]
 	\begin{center}
 	\vspace{0.1mm}
   	\begin{turn}{\rotAngle}
   		\begin{minipage}{\capsh}
				\hspace{\capsh}
   		\end{minipage}
   		\begin{minipage}{\capwh}
	 			\caption{Cross sections $f/\sigma_{\textrm{inel}}$ \cite{aduszkiewicz} with particle identification as functions of $p_T$ and $y_{\textrm{lab}}$ compared to the interpolation of the NA61 h$^-$ data (full lines) for 31~GeV/c beam momentum}
  			\label{fig:na61pid31}
			\end{minipage}
		\end{turn}
   	\includegraphics[width=\figwh,angle=\rotAngle] {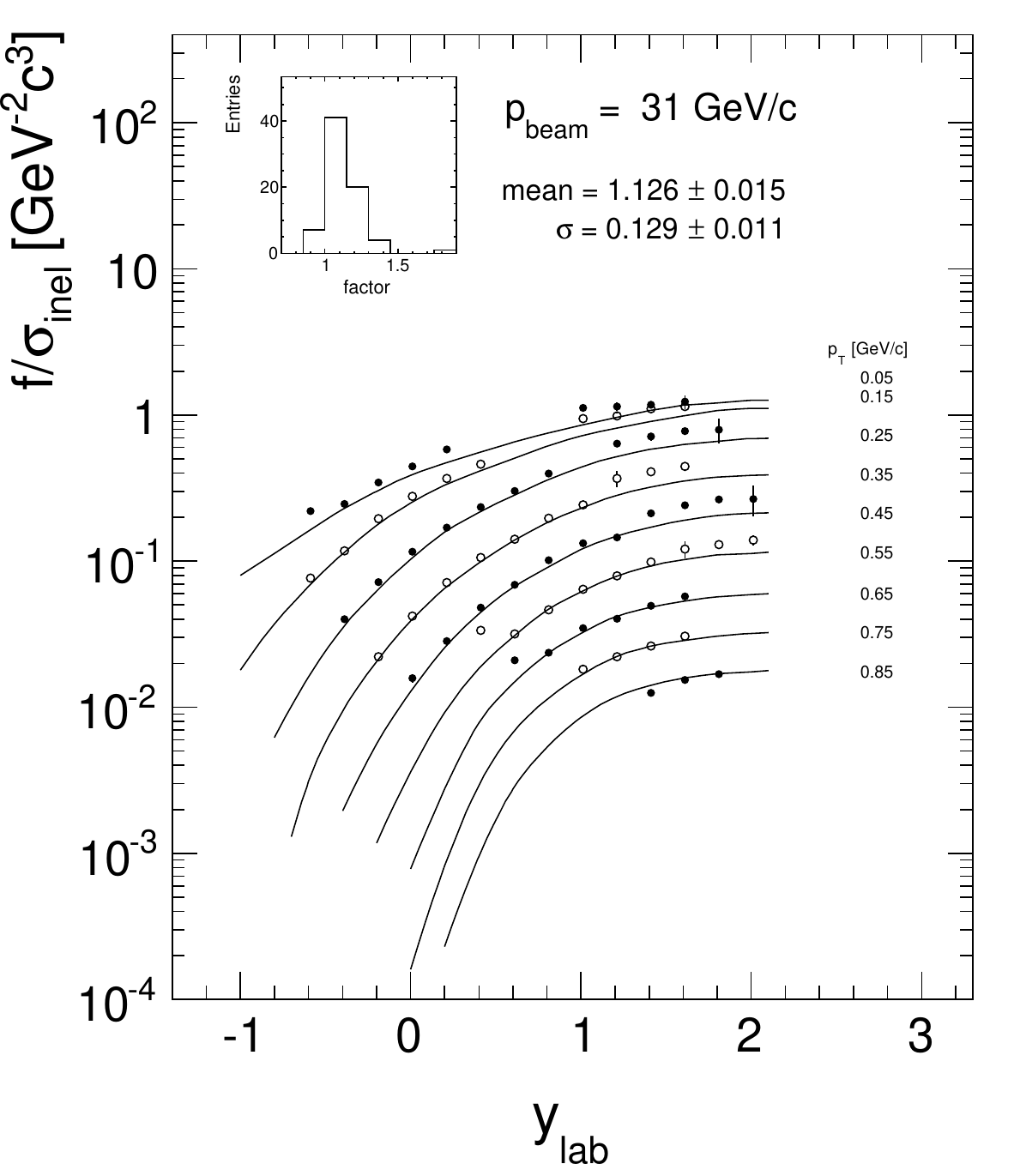} 
   	\begin{turn}{\rotAngle}
   		\begin{minipage}{\capsh}
				\hspace{\capsh}
   		\end{minipage}
   		\begin{minipage}{\capwh}
	 			\caption{Cross sections $f/\sigma_{\textrm{inel}}$ \cite{aduszkiewicz} with particle identification as functions of $p_T$ and $y_{\textrm{lab}}$ compared to the interpolation of the NA61 h$^-$ data (full lines) for 40~GeV/c beam momentum}
  			\label{fig:na61pid40}
			\end{minipage}
		\end{turn}
   	\includegraphics[width=\figwh,angle=\rotAngle] {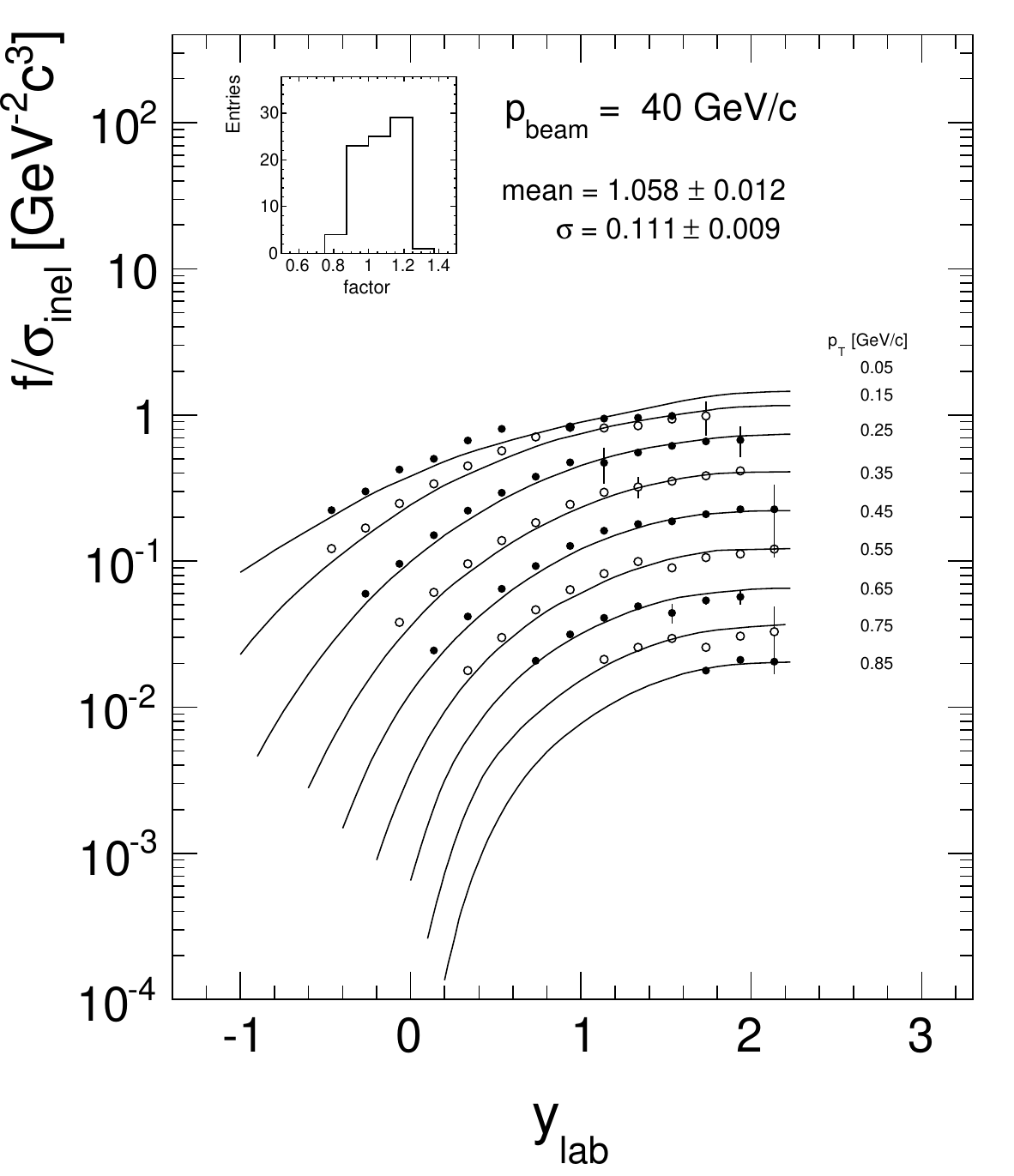} 
 	\end{center}
\end{figure*}

\begin{figure*}[h]
 	\begin{center}
 	\vspace{0.1mm}
   	\begin{turn}{\rotAngle}
   		\begin{minipage}{\capsh}
				\hspace{\capsh}
   		\end{minipage}
   		\begin{minipage}{\capwh}
	 			\caption{Cross sections $f/\sigma_{\textrm{inel}}$ \cite{aduszkiewicz} with particle identification as functions of $p_T$ and $y_{\textrm{lab}}$ compared to the interpolation of the NA61 h$^-$ data (full lines) for 80~GeV/c beam momentum}
  			\label{fig:na61pid80}
			\end{minipage}
		\end{turn}
   	\includegraphics[width=\figwh,angle=\rotAngle] {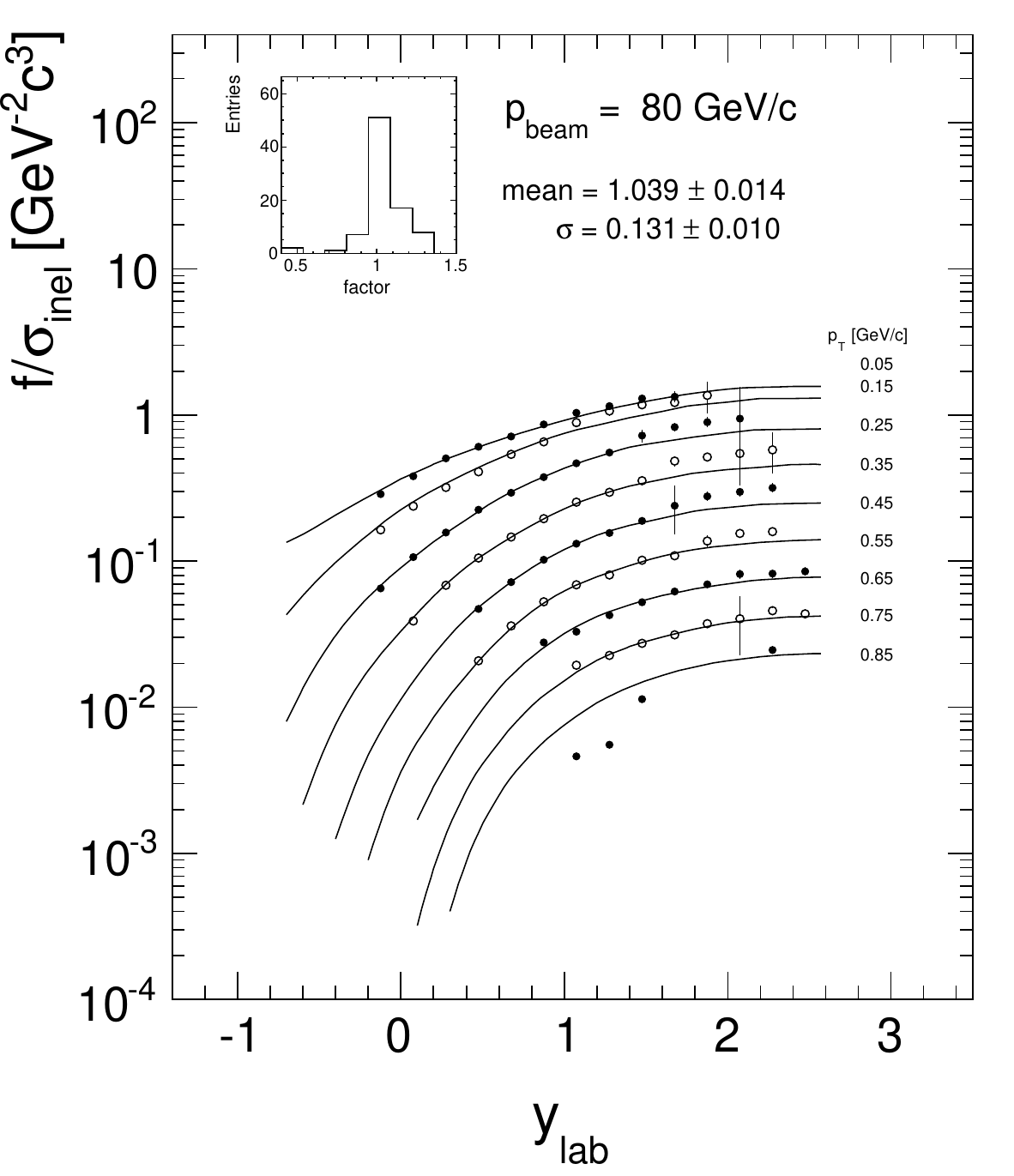} 
   	\begin{turn}{\rotAngle}
   		\begin{minipage}{\capsh}
				\hspace{\capsh}
   		\end{minipage}
   		\begin{minipage}{\capwh}
	 			\caption{Cross sections $f/\sigma_{\textrm{inel}}$ \cite{aduszkiewicz} with particle identification as functions of $p_T$ and $y_{\textrm{lab}}$ compared to the interpolation of the NA61 h$^-$ data (full lines) for 158~GeV/c beam momentum}
  			\label{fig:na61pid158}
			\end{minipage}
		\end{turn}
   	\includegraphics[width=\figwh,angle=\rotAngle] {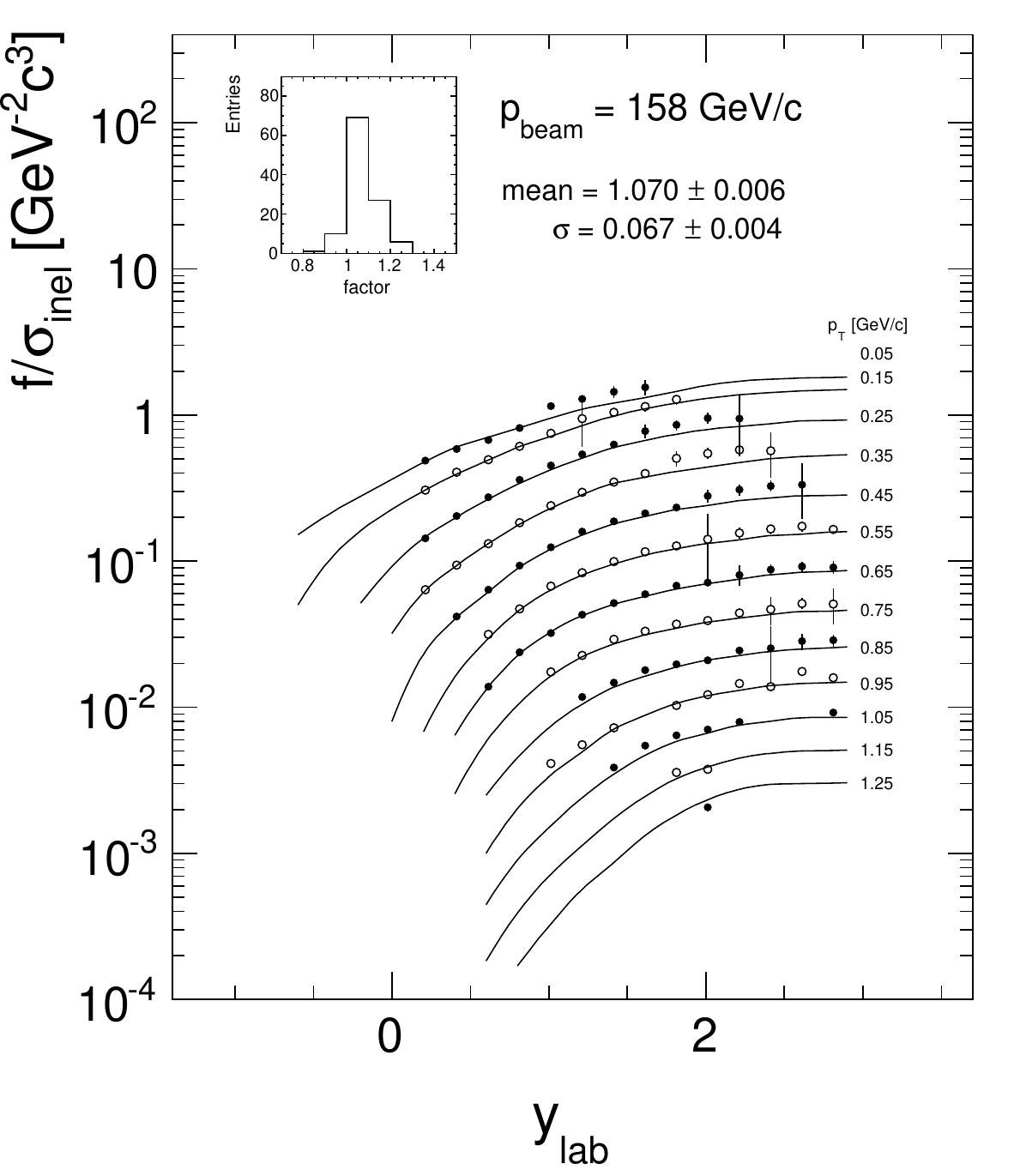} 
 	\end{center}
\end{figure*}

Reference \cite{aduszkiewicz} uses the same raw data as \cite{abgrall}. The bin size in $p_T$ is increased to 0.1~GeV/c centred at values corresponding to the global interpolation but limited, with the exception of $p_{\textrm{beam}}$~=~158~GeV/c, to a range below 0.95~GeV/c strongly decreasing with increasing rapidity. The total number of bins is thus reduced by factors between 2 and 4. As this $p_T$ binning corresponds to the one used for the reference data and for the interpolation of the h$^-$ data, Figs.~\ref{fig:na61_comp20}--\ref{fig:na61_comp158}, a direct comparison between the two NA61 data sets becomes possible. This is shown in Figs.~\ref{fig:na61pid20}a),  \ref{fig:na61pid31}--\ref{fig:na61pid158}. Note that the full lines in these figures describe the NA61 h$^-$ data.

 A first remark concerning the results of \cite{aduszkiewicz} is the very drastic reduction of the phase space coverage compared to the h$^-$ data \cite{abgrall} as given by the full lines. This is obviously due to the limited range of the identification methods (energy loss and energy loss plus time-of-flight (TOF)). This leads at low beam momentum to a split between the low and high
rapidity regions. There is as well a general lack of data at low $p_T$ and low rapidity. Rather large statistical errors in the $dE/dx$ + TOF area indicate drastic fiducial cuts imposed in addition to the ones applying to the h$^-$ tracking. This is borne out by the coverage in the $y$/$p_T$ plane shown in Fig.~\ref{fig:na61pid20}b).

A second remark concerns the sizeable deviations from the h$^-$ results \cite{abgrall} given as the distribution of relative factors in the insets of Figs.~\ref{fig:na61pid20}--\ref{fig:na61pid158}. The possible result of particle identification beyond the subtraction of the K$^-$ and $\overline{\textrm{p}}$ contributions derived from model predictions (Sect.~\ref{sec:na61_hmin}) could at most make up a few percent at low and medium beam momenta. The observed relative factors between the two data sets span a wide range from 0.7 to 1.6 that cannot possibly be explained by differences in particle identification.

A third remark concerns the deviations from the global interpolation of the reference data. The differences shown in Figs.~\ref{fig:na61pid20}--\ref{fig:na61pid158} apply to the internal comparison of the NA61 results. It should be remembered that the h$^-$ results of NA61 are already way above the reference data. The systematics of these deviations has therefore to be added to the results of the internal comparison.

Although there is a direct overlap of measured bins in the $p_T$ range from 0.65 to 0.95~GeV/c no attempt has been made by the authors to compare the two data sets at least in this restricted area.

%
%
\subsection{Conclusion concerning the NA61 results}
\vspace{3mm}
\label{sec:na61_conclusion}

The NA61 experiment is using the NA49 detector \cite{nim}. This is a state-of-the-art detector system deploying two large superconducting magnets and four large Time Projection Chambers with a total drift volume of more than 40 m$^3$. In the context of this paper the NA49 results \cite{pp_pion} at 158~GeV/c beam momentum make part of the reference experiments (Tab.~\ref{tab:pp_pim_data}) which form the basis for a precise description of $\pi^-$ yields as functions of the three inclusive variables, Sects.~\ref{sec:inc_var} and \ref{sec:dep_ener}.

The NA61 experiment has aimed at extending the range of beam momenta from 158~GeV/c down to 20~GeV/c in four steps. This is in principle a welcome effort given the high statistical and systematic precision achievable with this detector as demonstrated by NA49.

In the range of beam momenta mentioned above there are, however, available results from no less than 25 preceding experiments, Tabs.~\ref{tab:pp_pim_data} and \ref{tab:pp_pim_spect}, which date back by up to 50~years, as shown in Fig.~\ref{fig:year}. These results have been analysed in the preceding Sections where it has been demonstrated that a consistent global interpolation of all these results may be achieved with an overall systematic precision of about 2\%.

In contrast the NA61 results show large deviations from this reference as demonstrated in Sects.~\ref{sec:na61_hmin} and \ref{sec:na61_pid} above. In fact the statistical errors are just comparable and the systematic uncertainties larger by factors of 2 to 3 than those of the preceding work as far as the reference experiments are concerned.

%
%
 \section{Data in the range of p+p colliders from RHIC to LHC energies}
\vspace{3mm}
\label{sec:colliders}

The extension of the interaction energy beyond the ISR region meets with two important constraints:

\begin{itemize}
	\item The event selection is generally based on double arm triggering in restricted phase space areas. It therefore does not fulfill the criterion of inclusiveness over the full inelastic cross section.
	\item The phase space coverage for identified hadrons is drastically reduced to a small area around central rapidity thus excluding the complete forward regions from the inclusive data sample, with only one exception \cite{arsene,yang}.
\end{itemize}

The data comparison described in the preceding sections has to rely on the fact that the data are obtained with an event selection covering the complete inelastic interaction cross section. Even small deviations from this condition lead to complex corrections as a function of the inclusive variables. These corrections are non-calculable and have to be obtained experimentally. An example is given by the NA49 experiment where the trigger cross section deviates from the total inelastic one by 14\%, Sect.~\ref{sec:na49_data}. The corresponding corrections have to be obtained by an extrapolation method yielding a $p_T$ and $x_F$ dependent pattern reaching up to 8\% in the forward hemisphere \cite{pp_pion}.

Double-arm triggering as applied by the RHIC and LHC experiments is typically sensitive to about 50--80\% of the inelastic cross section. This trigger condition rejects most of the single and double diffractive events. Strong deviations from the inclusive results described in the preceding sections have therefore to be expected.

In addition, the phase space coverage is reduced to typically less than one unit of rapidity which means that already at RHIC (200~GeV) the corresponding $x_F$ range is less that $\pm$0.02 at $p_T$~=~1.3~GeV/c. At this energy, the BRAHMS forward spectrometer offers data up to 3.5 units of rapidity. As shown in Sect.~\ref{sec:brahms_forward} even this extended $y$ range just covers the transition from the central rapidity plateau to the fragmentation area.

Only the PHENIX experiment at RHIC obtained measurements at 62.4~GeV and central rapidity thus providing a single $p_T$ distribution for direct comparison with the results at the highest ISR energy presented in Sect.~\ref{sec:countdata} above. Beyond this data the list of collider experiments in Tab.~\ref{tab:rhic} contains results from only 5 collaborations at cms energies between 200~GeV and 13~TeV.

%
%
\subsection{The data of Adare et al.\cite{adare} at \boldmath $\sqrt{s}$~=~62.4~GeV}
\vspace{3mm}
\label{sec:phenix62}

\begin{figure}[h]
 	\begin{center}
   	\includegraphics[width=8.6cm] {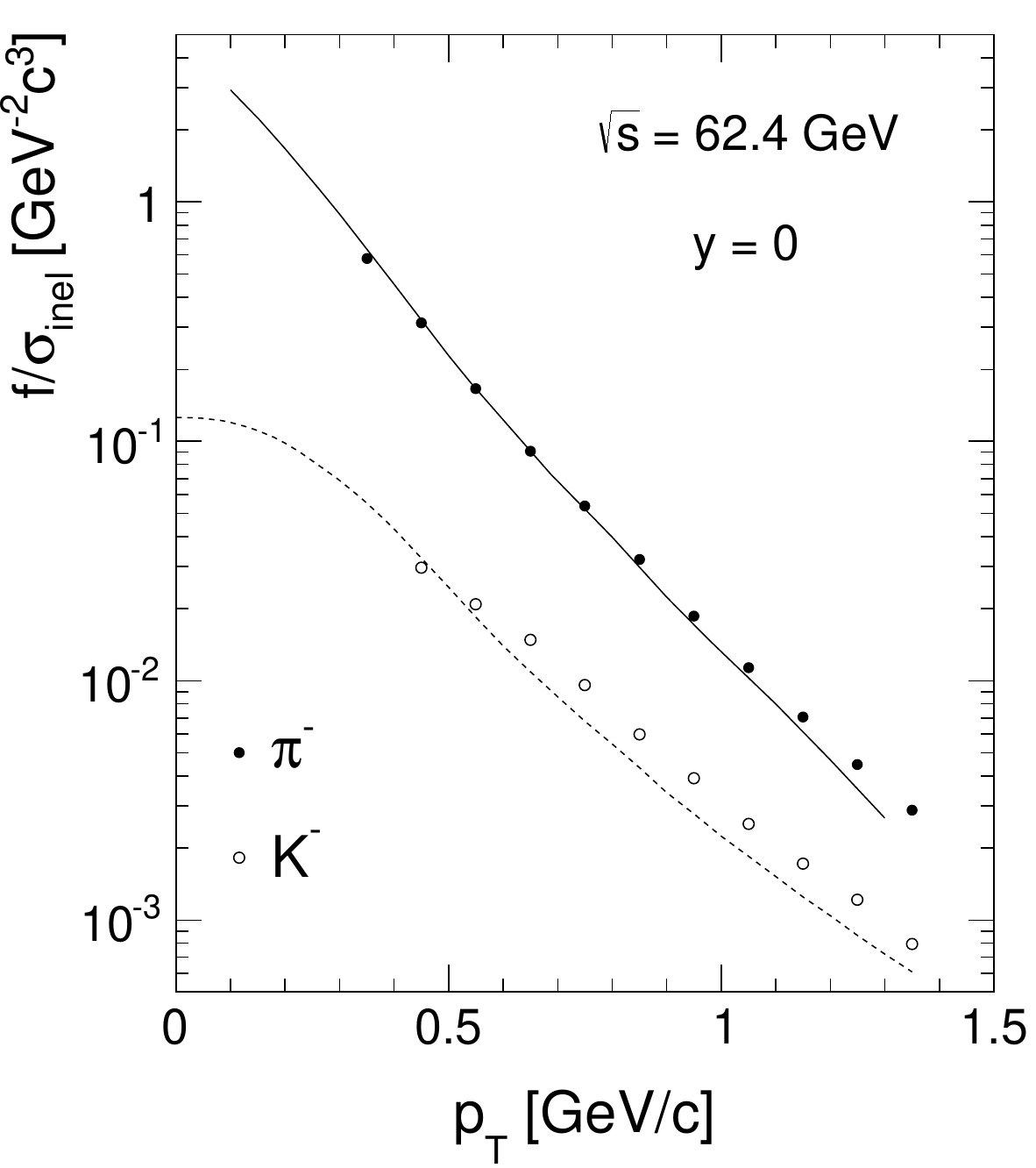} 
	 	\caption{Cross sections $f/\sigma_{\textrm{inel}}$ at $\sqrt{s}$~=~64.2~GeV \cite{adare} as a function of $p_T$ at $y$~=~0, full circles for $\pi^-$, open circles for K$^-$. The global interpolation at this energy is given by the full line for $\pi^-$, the ISR data for K$^-$ \cite{pp_kaon} by the broken line}
  	\label{fig:phenix62}
 	\end{center}
\end{figure}

Double differential cross sections are given in a pseudorapidity range of $\pm$0.35 around central rapidity. The $\pi^-$ data are not feed-down corrected and are thus directly comparable to the ISR data at the same energy. In a $p_T$ range from 0.35 to the upper limit of 1.35~GeV/c as addressed in this paper 11 data points are provided as shown in Fig.~\ref{fig:phenix62}.

\begin{figure}[h]
 	\begin{center}
   	\includegraphics[width=16cm] {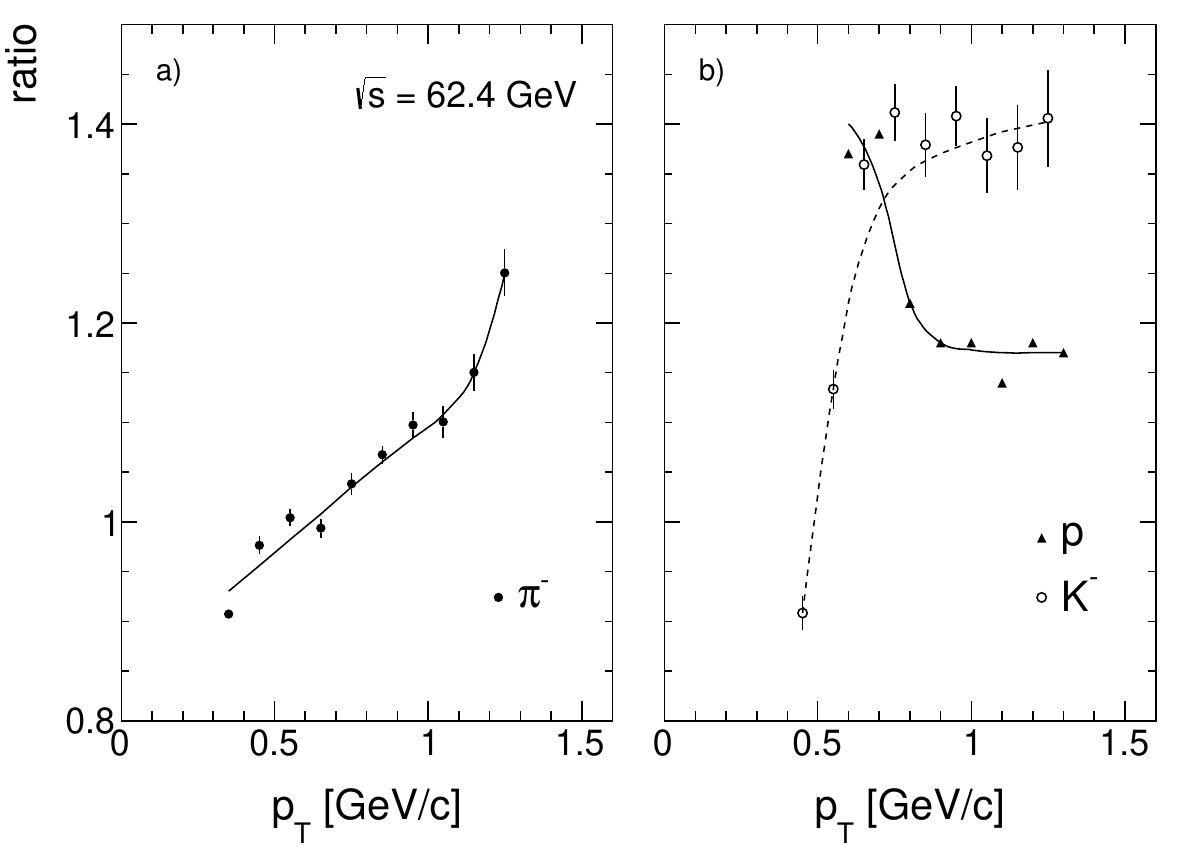} 
	 	\caption{Cross section ratios between the RHIC data and the ISR results as a function of $p_T$ for a) $\pi^-$ (full circles and full line), b) K$^-$ (open circles and broken line) and p (full triangles and full line). The lines are given to guide the eye}
  	\label{fig:phenix62isr}
 	\end{center}
\end{figure}

It is interesting to also compare the K$^-$ and p data \cite{adare} to the corresponding ISR data \cite{pp_kaon}. For all particle species there are important deviations as a function of $p_T$ as shown in Fig.~\ref{fig:phenix62isr}.

The strong decrease of the ratios towards the lower $p_T$ limits at 0.35~GeV/c for $\pi^-$ and 0.45~GeV/c for K$^-$ indicates apparative effects in the PHENIX data. The increase towards higher $p_T$ for $\pi^-$ and K$^-$ together with a decrease for p might be related to the trigger bias effect mentioned above. Indeed the trigger cross section -- asking for a coincidence between upstream and downstream counters -- corresponds to only 13.7~mb compared to the total inelastic cross section of 35.8~mb at this energy. This seems to exclude preferentially peripheral collisions in favour of more central interactions enhancing both the pion and kaon yields. This comparison alerts to the fact that not only the hadron yields but also particle ratios and integrated quantities like $\langle p_T \rangle$ and $dn/dy$ will be strongly affected compared to the reference data.

%
%
\subsection{RHIC data at \boldmath $\sqrt{s}$~=~200~GeV and central rapidity}
\vspace{3mm}
\label{sec:rhic200}

\begin{figure}[h]
 	\begin{center}
   	\includegraphics[width=15cm] {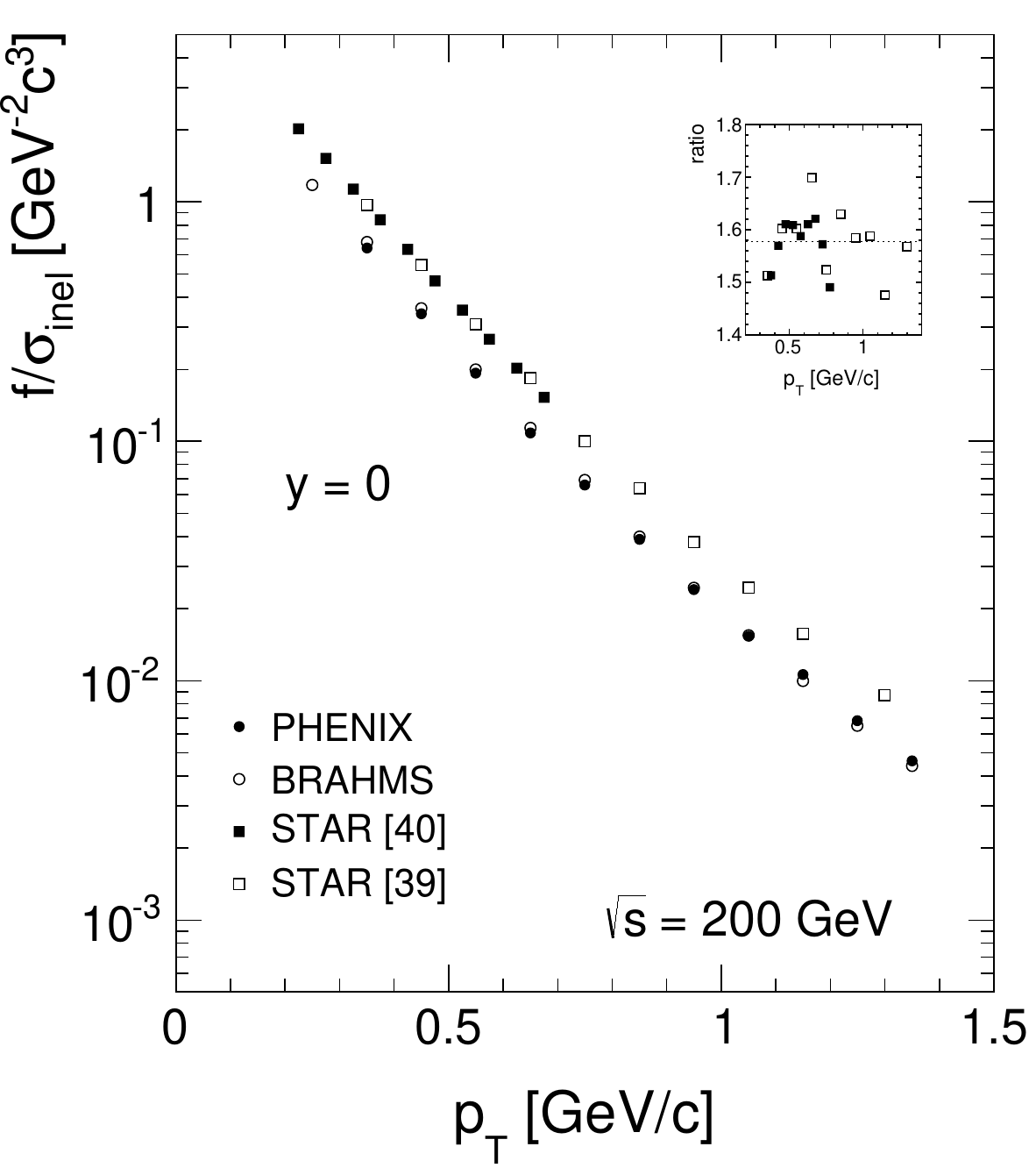} 
	 	\caption{Cross sections $f/\sigma_{\textrm{inel}}$ at $\sqrt{s}$~=~200~GeV as a function of $p_T$ at $y$~=~0. Full circles PHENIX \cite{adare}, open circles BRAHMS \cite{yang} and full squares STAR \cite{abelev,adams}. The inset shows the ratio between STAR and PHENIX data as a function of $p_T$}
  	\label{fig:rhic200}
 	\end{center}
\end{figure}

Three experiments give $\pi^-$ data at central rapidity: BRAHMS \cite{yang}, PHENIX \cite{adare} and STAR \cite{abelev}. Since the STAR data are feed-down corrected the cross sections have been enhanced by the feed-down percentage given in \cite{abelev,adams}. The resulting differential cross sections are presented in Fig.~\ref{fig:rhic200} as a function of $p_T$.

Whereas the data from BRAHMS and PHENIX are equal to within about 10\%, the STAR data are higher by about a factor of 1.6. A similar large discrepancy has been observed for kaons \cite{pp_kaon}. As both the STAR and the BRAHMS data are given as densities whereas PHENIX gives invariant cross sections, the PHENIX data have been divided by the inelastic cross section of 41~mb.

\begin{figure}[h]
 	\begin{center}
   	\includegraphics[width=15cm] {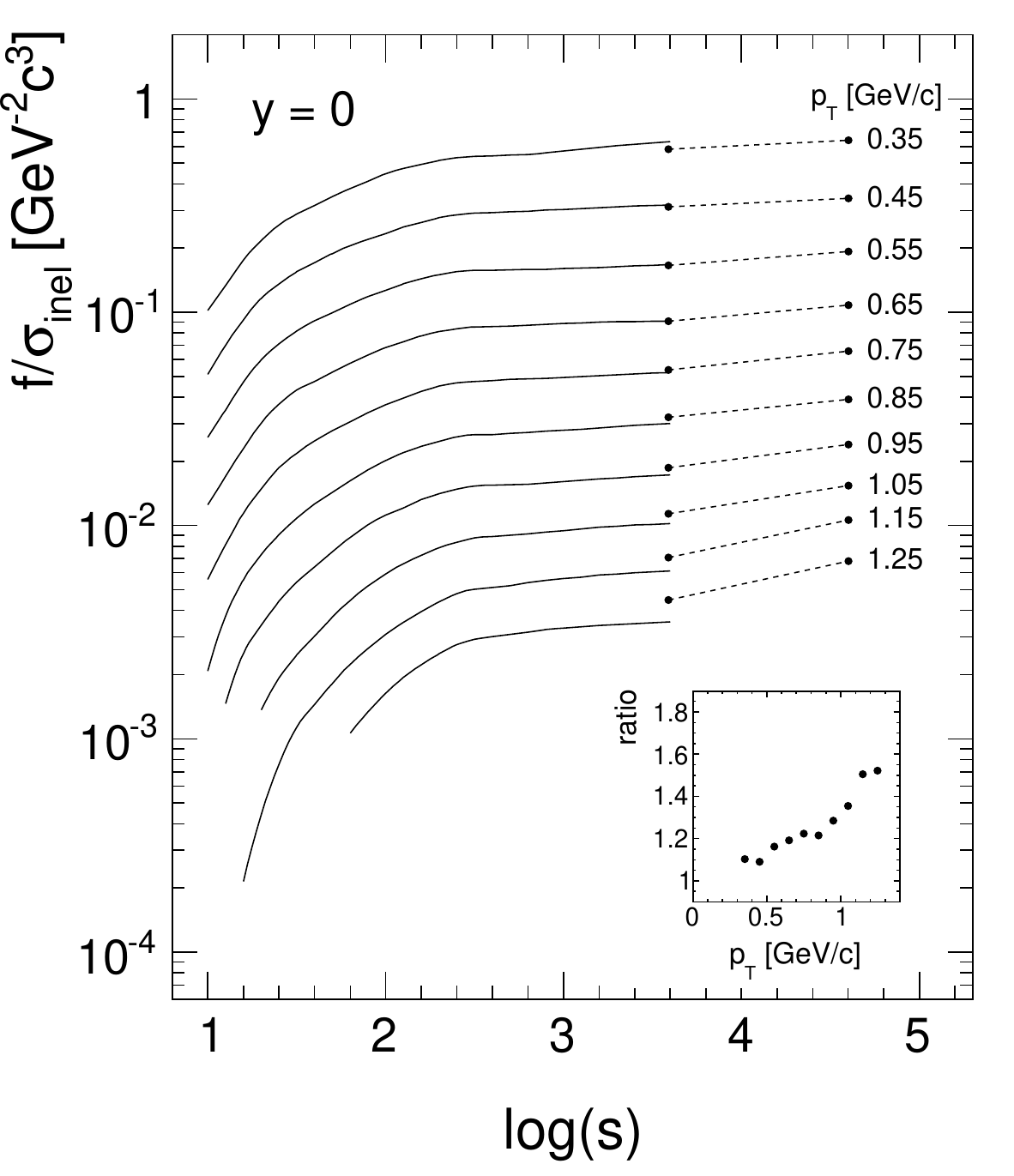} 
	 	\caption{Cross section $f/\sigma_{\textrm{inel}}$ as a function of $\log(s)$ up to RHIC energy for several values of $p_T$. Full lines global interpolation of the reference data, dashed lines RHIC data. The inset shows the ratio between the PHENIX data at 200~GeV and 63~GeV}
  	\label{fig:rhic200logs}
 	\end{center}
\end{figure}

The increase of the cross sections at central rapidity between $\sqrt{s}$~=~200 and 63~GeV shows a behaviour which is incompatible with the $s$-dependence as measured at the lower energies up to the ISR. This is presented in Fig.~\ref{fig:rhic200logs}.

The evolution with $p_T$ shows a similarity with the ratio between the PHENIX and ISR data at fixed $s$~=~63~GeV, Fig.~\ref{fig:phenix62isr}a. This indicates a problem with the RHIC data concerning the definition of "inclusive" physics probably connected with the trigger cross sections which are far below the total inelastic one. This trend will be discussed further in the discussion of the LHC data, Sect.~\ref{sec:lhc}.

%
%
\subsection{BRAHMS data at forward rapidity \cite{arsene2,yang} for \boldmath $\sqrt{s}$=200 GeV }
\vspace{3mm}
\label{sec:brahms_forward}

\begin{figure}[h]
 	\begin{center}
   	\includegraphics[width=14.cm] {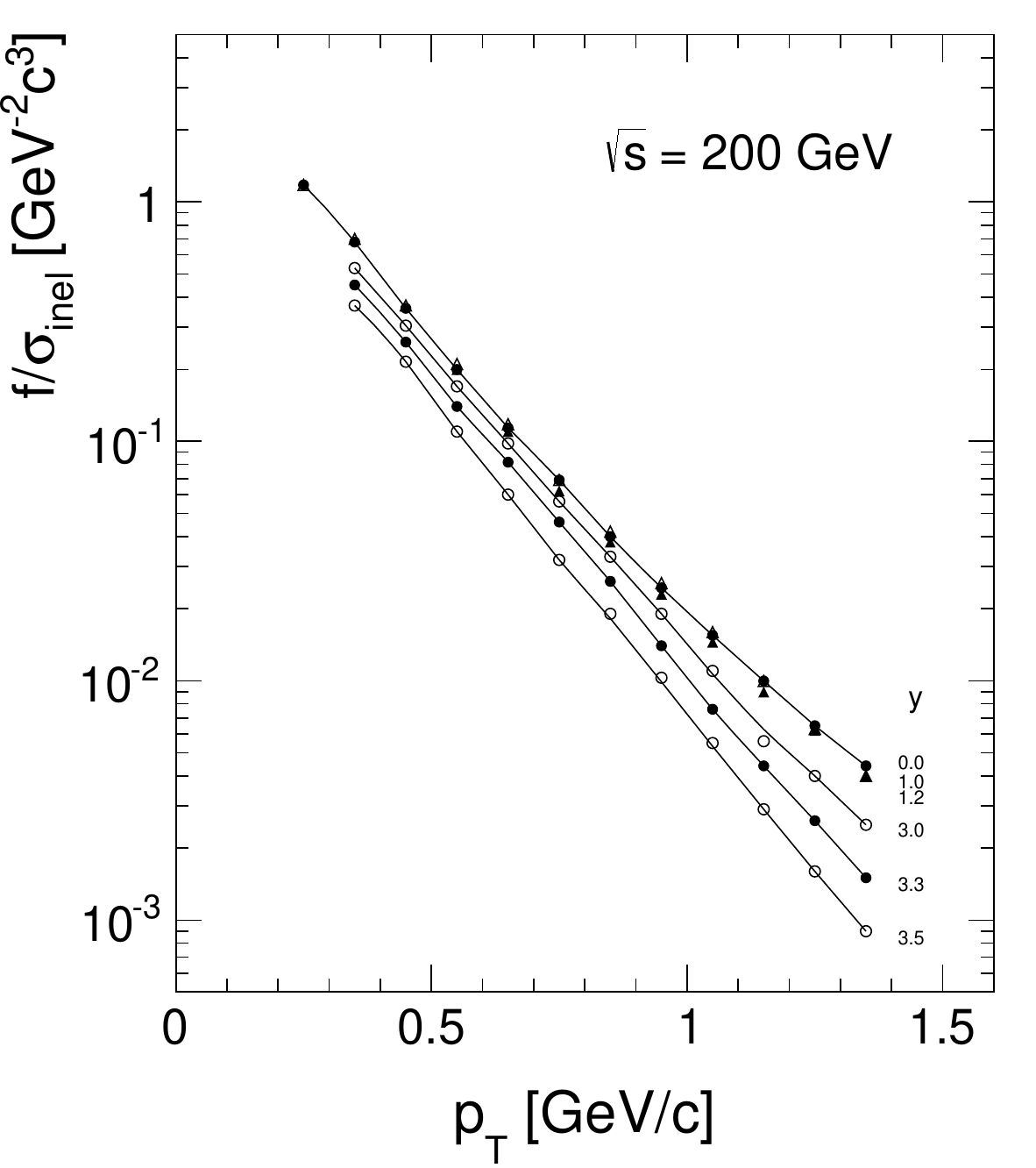} 
	 	\caption{Cross sections $f/\sigma_{\textrm{inel}}$ at $\sqrt{s}$~=~200~GeV as a function of $p_T$ for rapidities 0, 1, 1.2, 3.0, 3.3 and 3.5 from BRAHMS \cite{arsene2,yang}}
  	\label{fig:brahms200}
 	\end{center}
\end{figure}

BRAHMS is the only RHIC experiment offering results in the rapidity range up to 3.5 units. Fig.~\ref{fig:brahms200} shows the cross sections as a function of $p_T$ for six values of rapidity from 0 to 3.5.

\begin{figure}[h]
 	\begin{center}
   	\includegraphics[width=13.5cm] {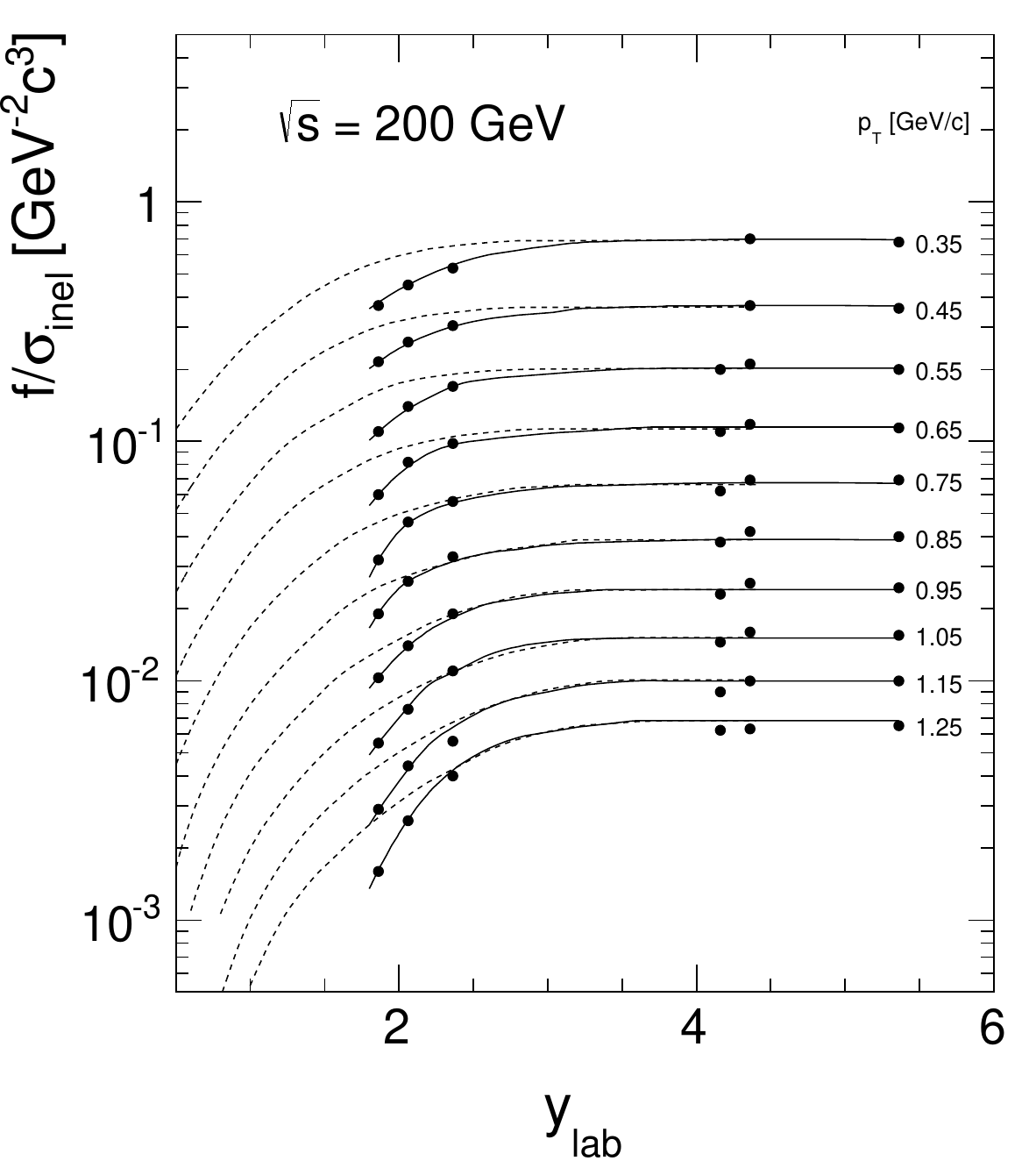} 
	 	\caption{Cross sections $f/\sigma_{\textrm{inel}}$ at $\sqrt{s}$~=~200~GeV as a function of $y_{\textrm{lab}}$ for several values of $p_T$, interpolated by the full lines. In addition the corresponding distributions at the highest ISR energy ($y_{\textrm{beam}}$~=~3.6) are presented as broken lines after re-normalization at $y$~=~0, Fig.~\ref{fig:rhic200logs}}
  	\label{fig:brahms2isr}
 	\end{center}
\end{figure}

The cross sections of Fig.~\ref{fig:brahms200} are re-plotted as a function of $y_{\textrm{lab}}$ in Fig.~\ref{fig:brahms2isr} (full lines) for several $p_T$ values.

As will be shown in the discussion on scaling, Sect.~\ref{sec:scaling}, the lower energy data are nearly energy-independent in the fragmentation region ($y_{\textrm{lab}} <$~2). These  $y_{\textrm{lab}}$ distributions at $y_{\textrm{beam}}$~=~3.6 are shown in Fig.~\ref{fig:brahms2isr} as broken lines after re-normalization to the BRAHMS data at $y$~=~0.

Evidently the RHIC data deviate progressively downwards from the expected shape of the $y_{\textrm{lab}}$ distributions with a suppression factor of about 0.69 at $y_{\textrm{lab}}$~=~1.86 or $y$~=~3.5. It is unclear whether this might be an apparative effect or the consequence of a change in the trigger bias effect between central and forward rapidity.

%
%
\subsection{Data from the LHC}
\vspace{3mm}
\label{sec:lhc}

Invariant yields of identified particles in the $p_T$ range studied here are provided by the ALICE and LHC experiments at five cms energies between 0.9 and 13~TeV, Tab.~\ref{tab:rhic}. The data are all obtained at central rapidity within a range of less than 1 unit. They are usually presented as hadron densities per event

\begin{equation}
	\frac{1}{N_{\textrm{ev}}}\frac{d^2N}{dp_Tdy}
\end{equation}

This may be transformed into the invariant densities

\begin{equation}
	\frac{1}{N_{\textrm{ev}} 2 \pi p_T}\frac{d^2N}{dp_Tdy}
\end{equation}
which is comparable to the densities

\begin{equation}
	\frac{f}{\sigma_{\textrm{inel}}}
\end{equation}
used for the detailed data comparison in the preceding sections. A principal problem is however given by the limited efficiency of the double arm trigger systems of the LHC experiments described in the introduction to Sect.~\ref{sec:colliders}. Systematic deviations between any reasonable extrapolation of the lower-energy data and the LHC results have therefore to be expected.

%
%
\subsubsection{Results from ALICE}
\vspace{3mm}
\label{sec:lhc_alice}

\begin{figure}[h]
 	\begin{center}
   	\includegraphics[width=13.5cm] {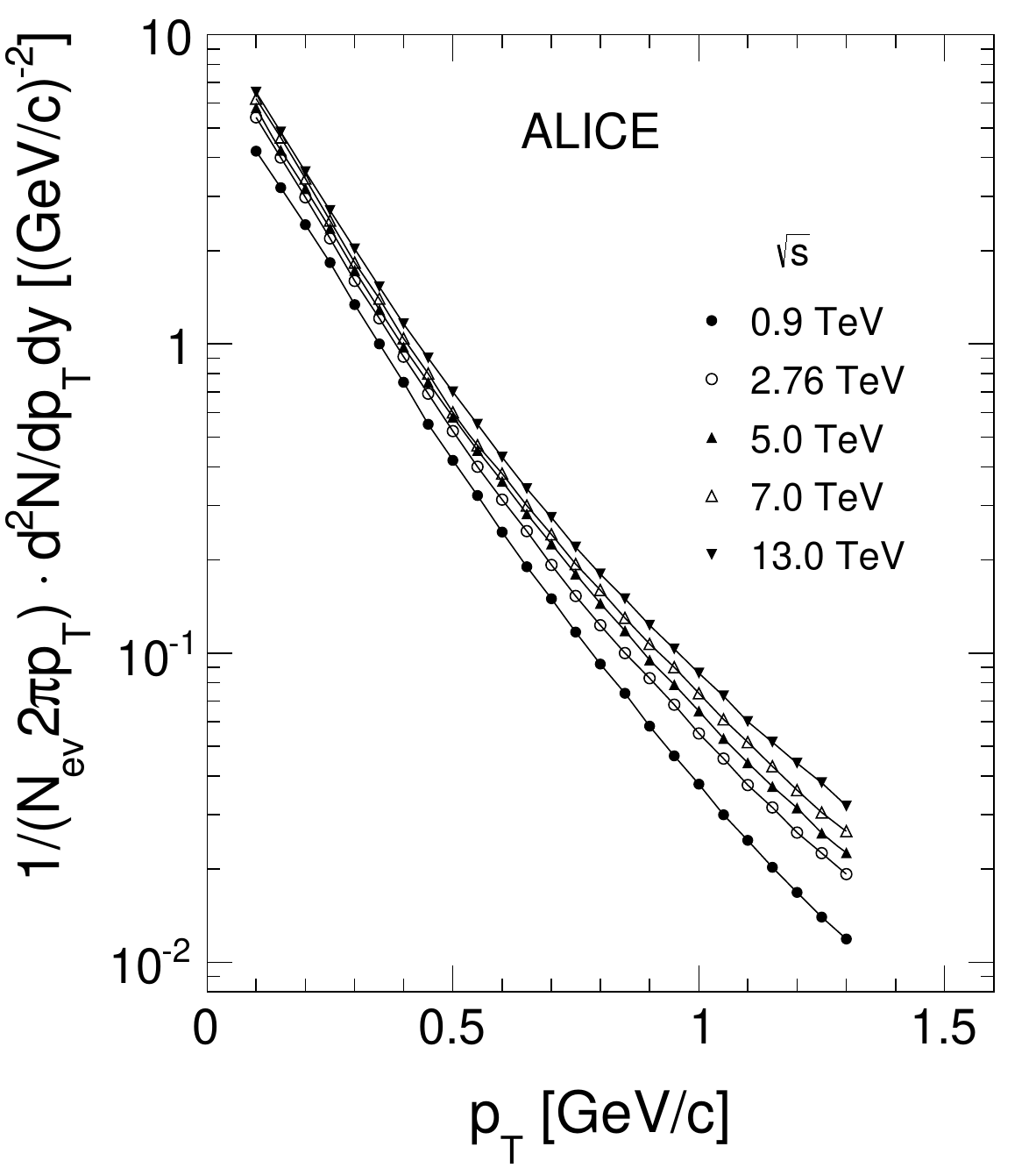} 
	 	\caption{Cross sections $f/\sigma_{\textrm{inel}}$ from ALICE at $\sqrt{s}$~=~0.9~TeV, 2.76~TeV, 7~TeV and 13~TeV as a function of $p_T$}
  	\label{fig:alice}
 	\end{center}
\end{figure}

The ALICE results \cite{aamodt,abelev2,acharya,adam,acharya2} have been re-evaluated as invariant densities and interpolated to the $p_T$ grid used in this paper, $p_T$ from 0.1 to 1.3~GeV/c in steps of 0.05~GeV/c. This is possible as the lowest measured $p_T$ values are at 0.11~GeV/c. The results are shown in Fig.~\ref{fig:alice}.

The given statistical errors are generally below 1\% whereas the systematic errors are typically on the 5--6\% level. The $p_T$ dependence continues down to $p_T$~=~0.1~GeV/c practically exponentially without an indication of a flattening due to the physical constraint of approaching $p_T$~=~0~GeV/c with tangent zero.

%
%
\subsubsection{Results from CMS}
\vspace{3mm}
\label{sec:lhc_cms}

\begin{figure}[h]
 	\begin{center}
   	\includegraphics[width=14cm] {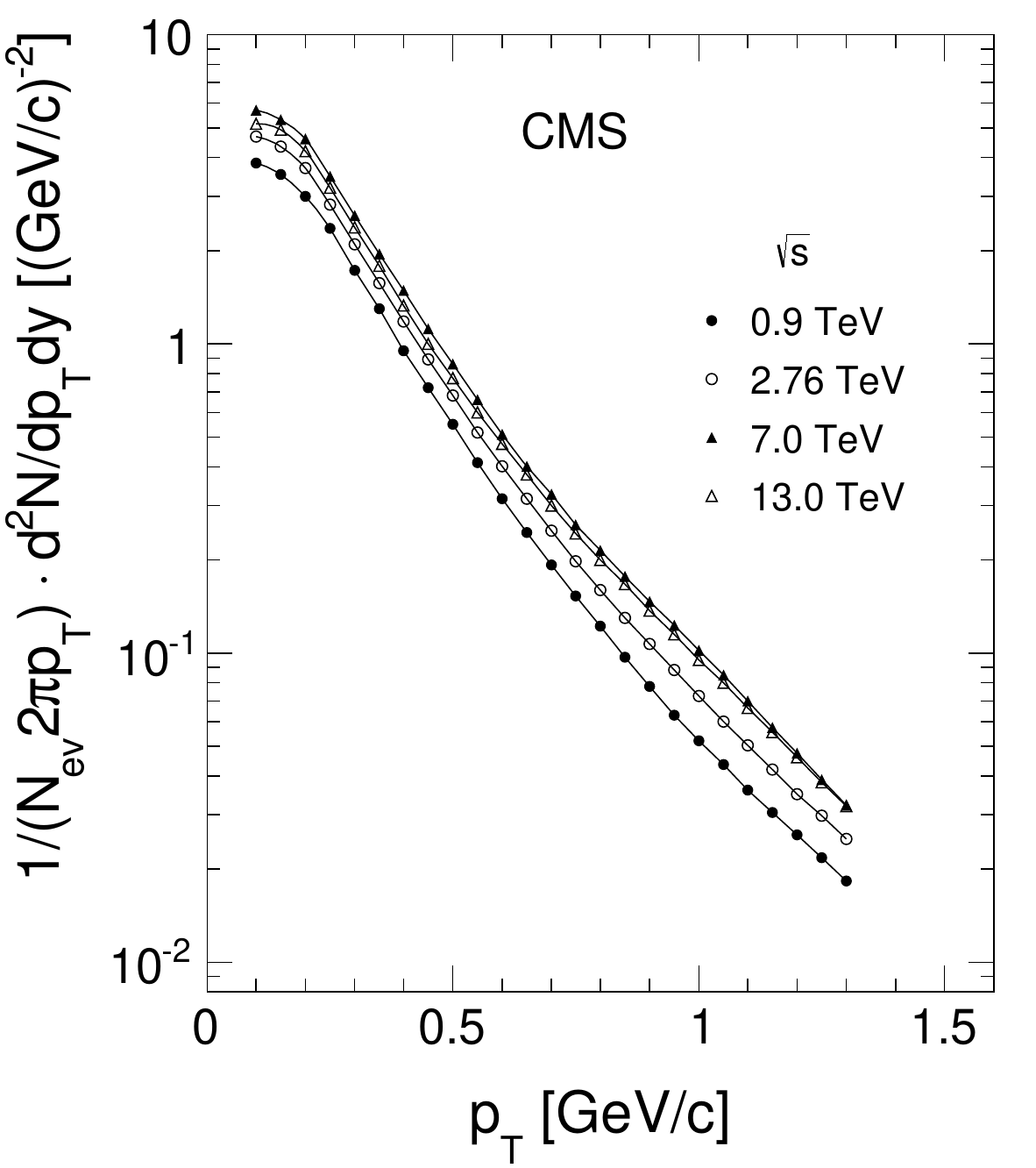} 
	 	\caption{Cross sections $f/\sigma_{\textrm{inel}}$ from CMS at $\sqrt{s}$~=~0.9~TeV, 2.76~TeV, 7~TeV and 13~TeV as a function of $p_T$}
  	\label{fig:cms}
 	\end{center}
\end{figure}

The CMS results \cite{chatrchyan,sirunyan} have been re-evaluated the same way as the ALICE values. Here the smallest measured $p_T$ is 0.125~GeV/c. The invariant densities are presented in Fig.~\ref{fig:cms} as a function of $p_T$.

A flattening in the approach to $p_T$~=~0~GeV/c is clearly visible below $p_T \sim$~0.25~GeV/c. It is somewhat surprising to see the data at $\sqrt{s}$~=~13~TeV below the ones at 7~TeV over the full $p_T$ range. As in the case of ALICE the statistical errors are on the level below 1\% with an estimated systematic error of 5--6\%.

%
%
\subsubsection{Comparison of the LHC results}
\vspace{3mm}
\label{sec:lhc_comp}

\begin{figure}[h]
 	\begin{center}
   	\includegraphics[width=14cm] {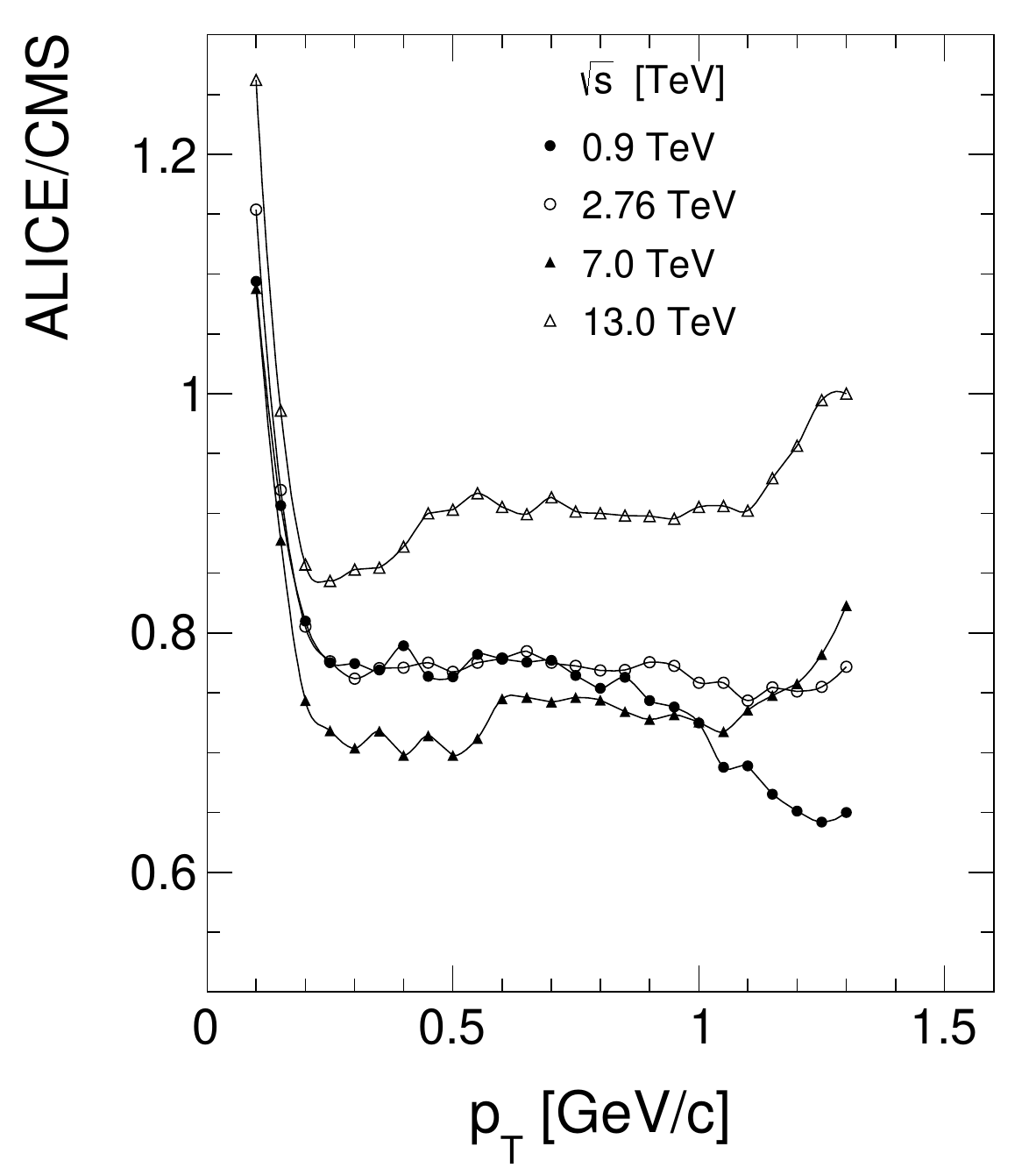} 
	 	\caption{Cross section ratios between ALICE and CMS at $\sqrt{s}$~=~0.9~TeV, 2.76~TeV, 7~TeV and 13~TeV as a function of $p_T$}
  	\label{fig:alice2cms}
 	\end{center}
\end{figure}

The ratio of the $\pi^-$ densities published by ALICE and CMS is given in Fig.~\ref{fig:alice2cms}.

The ratio shows a strong increase of up to 1.2 at $p_T$ below 0.2~GeV/c and flattens out to values between 0.7 and 0.8 at higher $p_T$, with the exception of the data at 13~TeV where the CMS data show a strong internal inconsistency. This situation is reminiscent of the discussion of the spectrometer results in Sect.~\ref{sec:countdata} above. There the main problem has been the absolute normalization whereas the experiments definitely triggered on the total inelastic cross section. At the LHC the large differences which are far outside the given systematic errors seem to be rather due to the trigger schemes which see only some fraction of the inelastic cross section.

\begin{figure}[h]
 \vspace{5mm}
 	\begin{center}
   	\includegraphics[width=16cm] {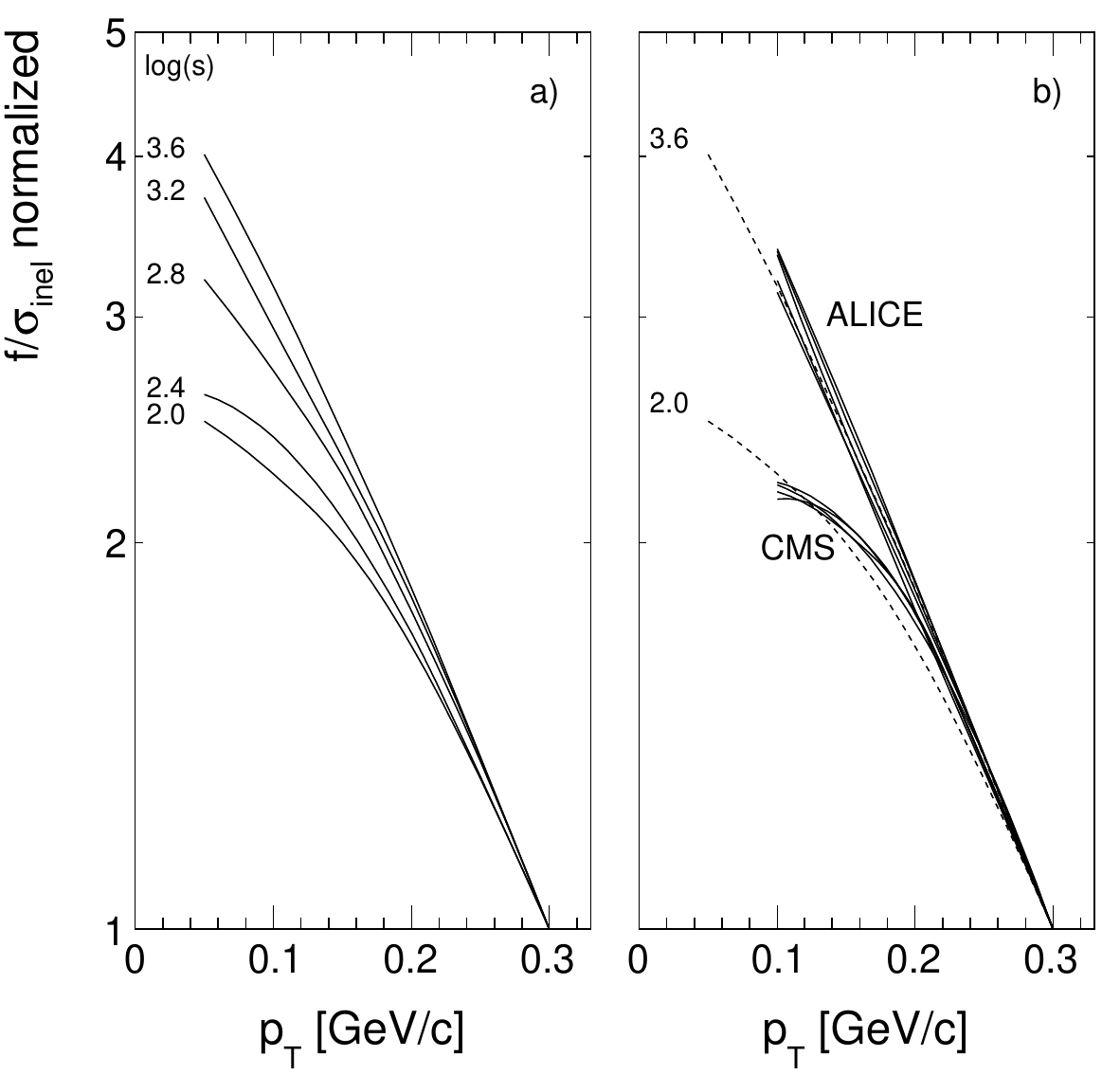} 
	 	\caption{a) Approach of the invariant $\pi^-$ density towards $p_T$~=~0~GeV/c from $\sqrt{s}$~=~10~ GeV to 63~GeV/c. The data are normalized to 1 at $p_T$~=~0.3~GeV/c to clearly bring out the change of the $p_T$ dependence with interaction energy. b) The same for the LHC data from ALICE \cite{aamodt,abelev2,acharya,adam,acharya2} and CMS \cite{chatrchyan,sirunyan}}
  	\label{fig:ptzero}
 	\end{center}
\end{figure}

The discrepancy at low $p_T$ may be resolved by regarding the approach to $p_T$~=~0~GeV/c in the global interpolation of the lower energy data. Here the yields have been measured down to $p_T$~=~0.05~GeV/c by NA49 \cite{pp_pion} and down to $p_T$~=~0.044~GeV/c at the ISR \cite{guettler}. These data show a marked $s$-dependence in the range below $p_T \sim$~0.3~GeV/c. Re-normalizing the data to 1.0 at $p_T$~=~0.3~GeV/c one obtains the trend from $\sqrt{s}$~=~10~GeV to the highest ISR energy as presented in Fig.~\ref{fig:ptzero}.

\begin{figure}[h]
 	\begin{center}
   	\includegraphics[width=16cm] {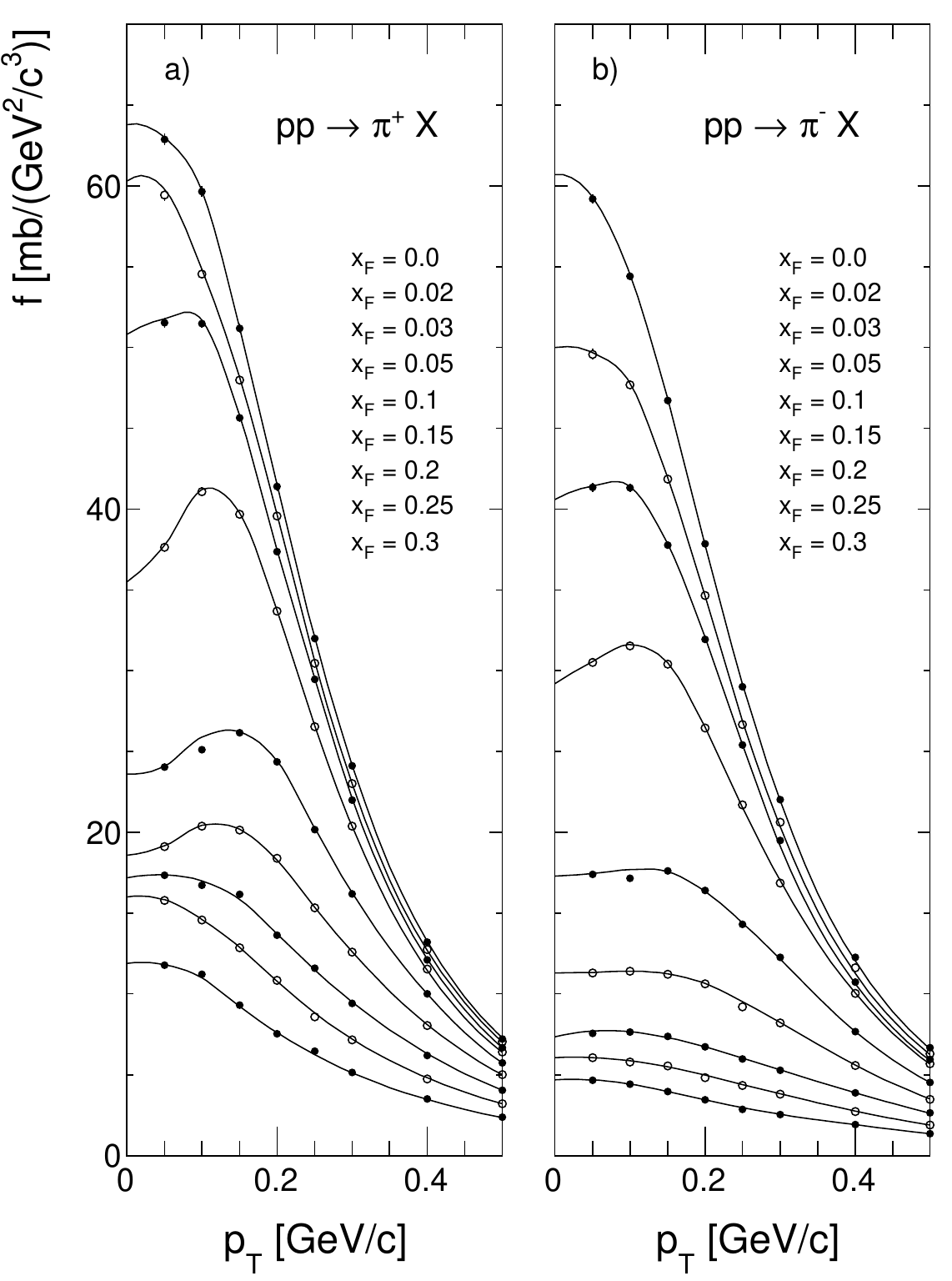} 
	 	\caption{Invariant cross section at low transverse momenta as a function of $p_T$ at fixed $x_F$ for a) $\pi^+$ and b) $\pi^-$ produced in p+p collisions at 158~GeV/c \cite{pp_pion}}
  	\label{fig:pp_lowpt}
 	\end{center}
\end{figure}

The flattening of the $p_T$ distributions at low $p_T$ and low $\log(s)$ has been explained \cite{pp_pion} by the prevalence of the decay of the $\Delta$(1232) resonance in this region. If plotted at fixed $x_F$ the invariant $p_T$ distributions exhibit a typical structure towards $p_T$~=~0~GeV/c, showing a secondary maximum at $p_T \sim$~0.1~GeV/c and $x_F\sim$~0.05 as presented in Fig.~\ref{fig:pp_lowpt} (NA49).

\begin{figure}[h]
 	\begin{center}
   	\includegraphics[width=16cm] {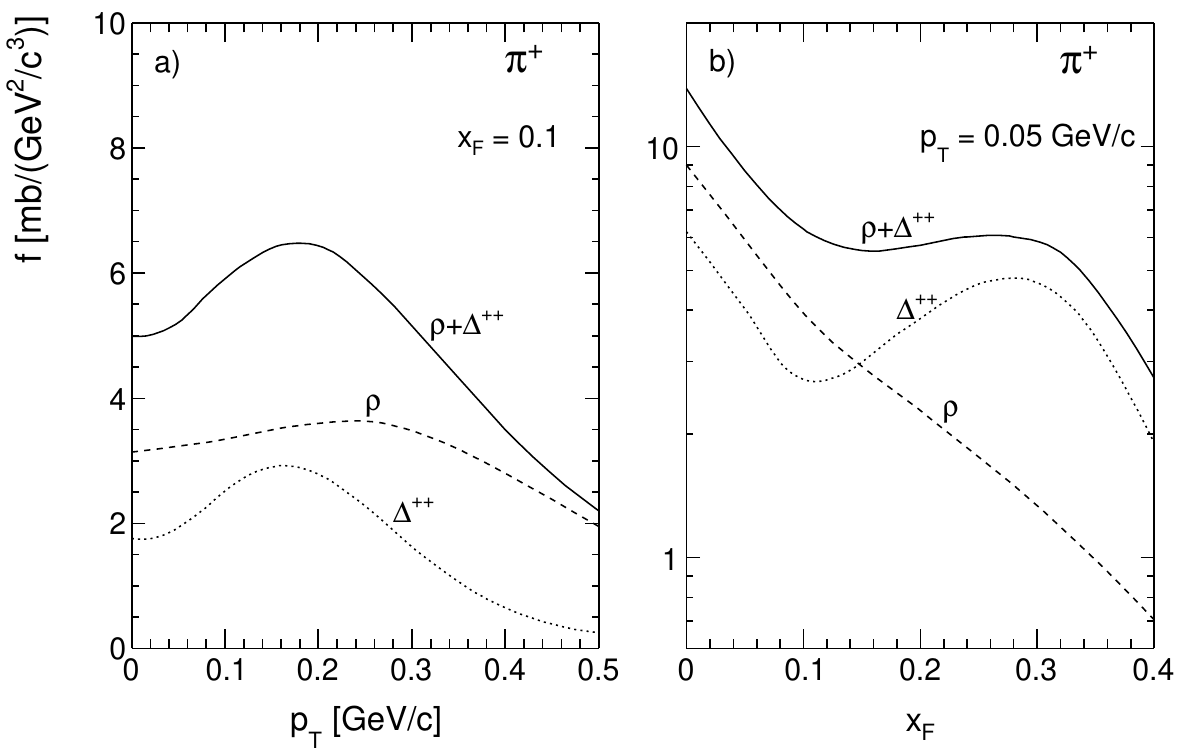} 
	 	\caption{Monte Carlo study of a) $p_T$ distribution at fixed $x_F$, b) $x_F$ distribution at fixed $p_T$ for $\pi^+$ resulting from $\rho$ and $\Delta^{++}$ decay, NA49 \cite{pp_pion}}
  	\label{fig:resonance}
 	\end{center}
\end{figure}

This structure has been quantitatively reproduced by a superposition of $\Delta$(1232) and $\rho$ decay, see Fig.~\ref{fig:resonance}.

The strong isospin dependence of the effect is notable. It originates from the different contributions to $\pi^+$ ($\Delta^{++}$ and $\Delta^+$) and $\pi^-$ ($\Delta^0$ and $\Delta^-$) which have largely different cross sections in p+p collisions. With increasing beam momentum this effect is progressively reduced by the governing decay of higher mass meson and baryon resonances.

The ALICE data follow the trend given at the lower energies quite closely whereas the CMS data correspond in shape rather to the situation at $\sqrt{s}$~=~10~GeV which is definitely unphysical.

%
%
\subsubsection{Dependence on interaction energy}
\vspace{3mm}
\label{sec:lhc_dep_ener}

\begin{figure}[h]
 	\begin{center}
   	\includegraphics[width=11cm] {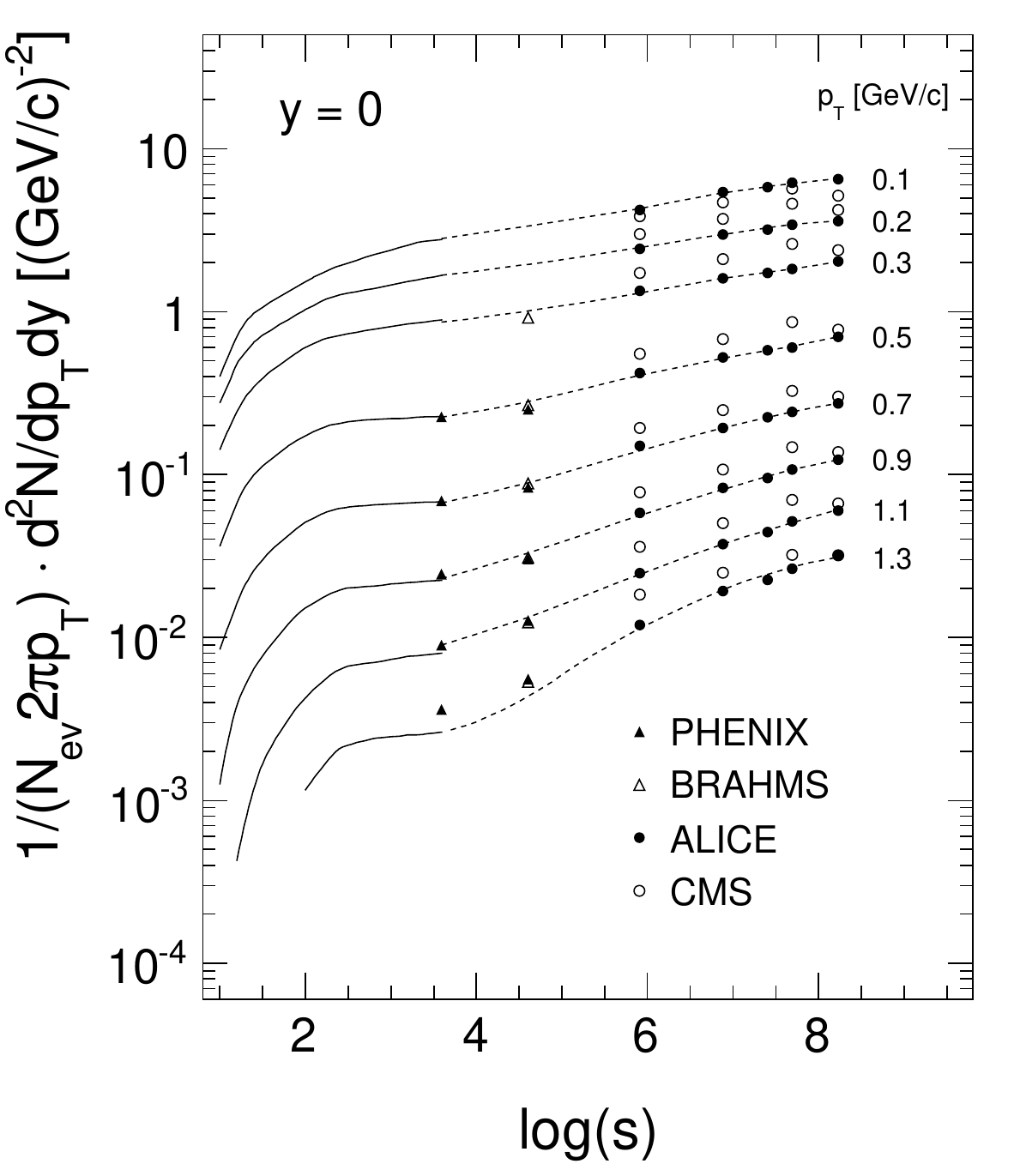} 
	 	\caption{Invariant $\pi^-$ densities $f/\sigma_{\textrm{inel}}$ (low energy data) and $1/(N_{\textrm{ev}} 2 dp_T dy) \cdot d^2N/dp_Tdy$ (LHC data) for several values of $p_T$ as a function of $\log(s)$. All data are feed-down corrected}
  	\label{fig:lhcrhic_sdep}
 	\end{center}
\end{figure}

Further evidence concerning systematic effects comes from the dependence of the $\pi^-$ yields on interaction energy over a wide range of $p_T$. The global interpolation of the low energy data reaching up into ISR energies described in Sects.~\ref{sec:interpolation} and \ref{sec:inter2refdata} shows an evolution of the invariant densities $f/\sigma_{\textrm{inel}}$ as a function of $\log(s)$ which is presented in Fig.~\ref{fig:lhcrhic_sdep} together with results from RHIC and LHC.

\begin{figure}[h]
 	\begin{center}
   	\includegraphics[width=7cm] {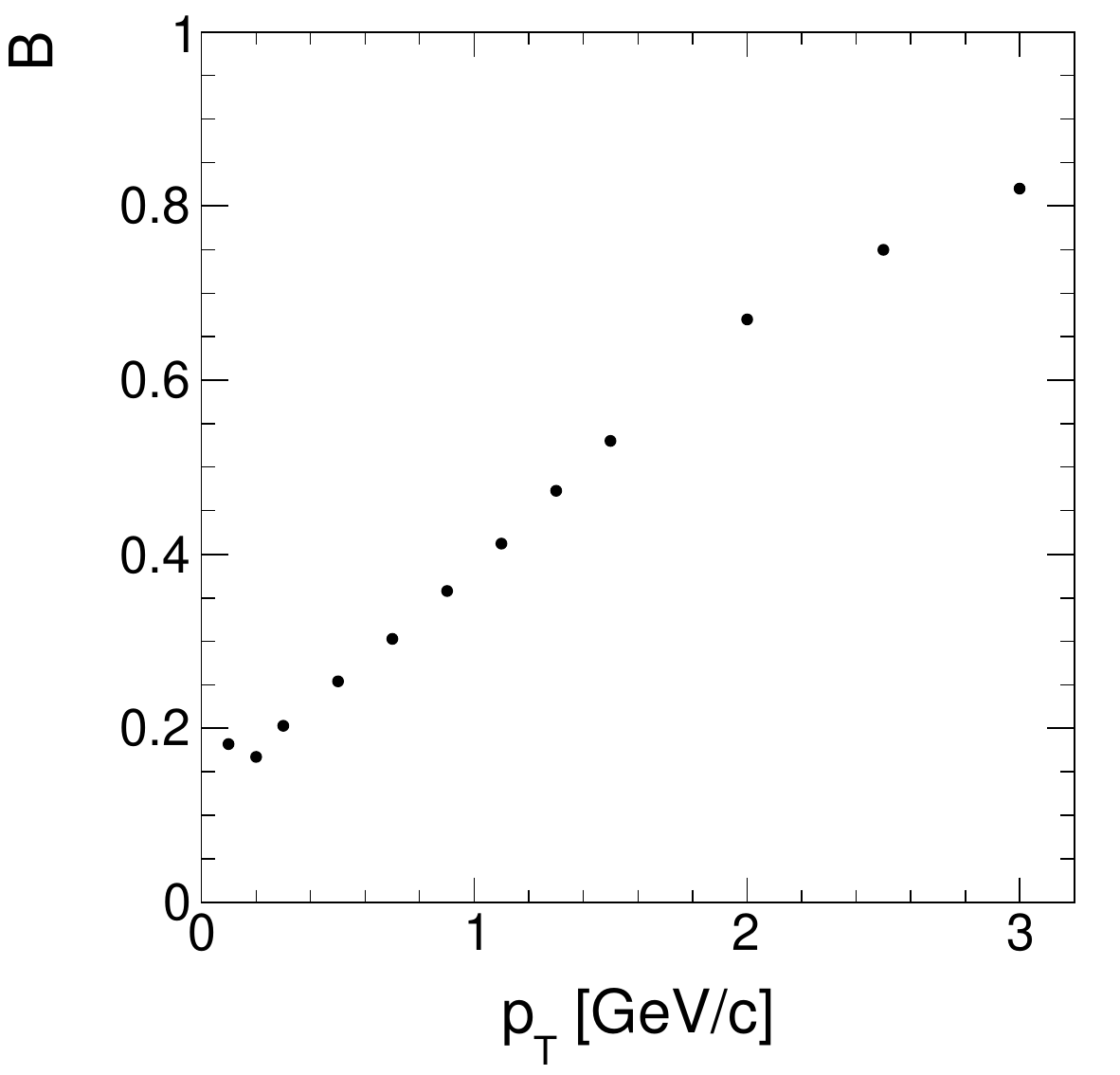} 
	 	\caption{Exponential slope of the $\log(s)$ dependence of the RHIC and LHC data as a function of $p_T$}
  	\label{fig:exp_slope}
 	\end{center}
\end{figure}

The main difference between the lower energy data sets and the RHIC/LHC data is given by the different trigger schemes. Whereas the reference data are referred to the full inelastic cross section the higher energy results do not contain most of the diffractive cross section. As some of the single and double diffraction (considered as "background") is still contained there, the loss of inclusiveness may be estimated to 20--25\%. This loss factor is not to be considered as a constant over the phase space variables, for instance over the $p_T$ dependence shown in Fig.~\ref{fig:lhcrhic_sdep}. There might be regions with only a small variation between diffractive and non-diffractive events, whereas other regions might not contain diffractive components at all. This seems to be borne out in the low-$p_T$ part of Fig.~\ref{fig:lhcrhic_sdep} where a smooth connection between the reference and higher energy data seems present for $p_T$ values below about 0.4~GeV/c. The RHIC and LHC data are characterized by a linear yield dependence on $\log(s)$ of the form

\begin{equation}
    \frac{1}{N_{\textrm{ev}} 2 \pi p_T}\frac{d^2N}{dp_T dy} = A \cdot e^{B\log(s)} = A \cdot s^{0.434 B}
    \label{eq:sdep}
\end{equation}
\vspace{0mm}

\noindent
with the parameter B shown in Fig.~\ref{fig:exp_slope} as a function of $p_T$.

\begin{figure}[h]
 	\begin{center}
   	\includegraphics[width=11cm] {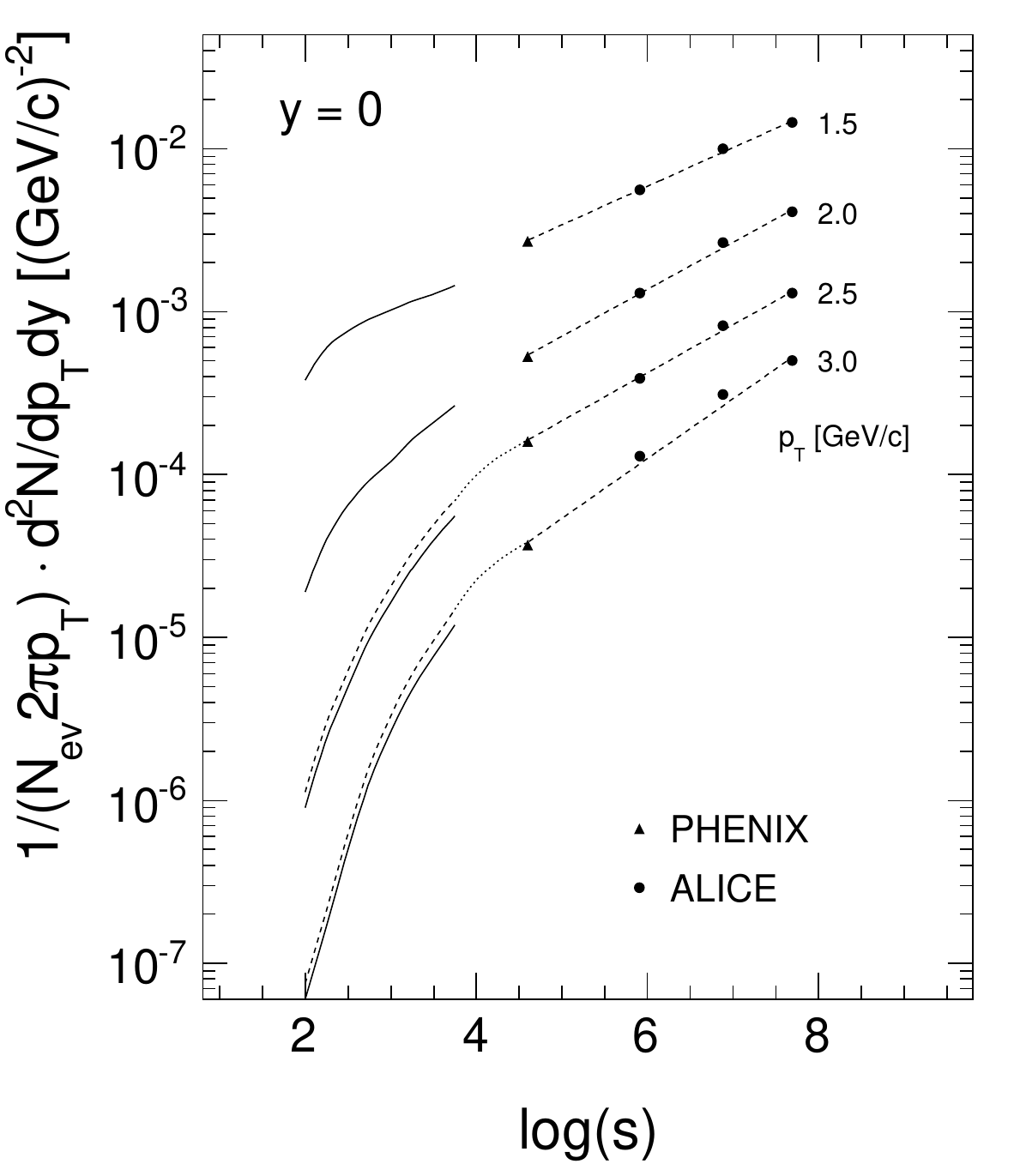} 
	 	\caption{Invariant $\pi^-$ densities $f/\sigma_{\textrm{inel}}$ as a function of $\log(s)$ for $p_T$ values between 1.5 and 3~GeV/c showing the reference data as full lines and data from PHENIX and ALICE as broken lines}
  	\label{fig:highpt}
 	\end{center}
\end{figure}

With $p_T$ increasing beyond 0.3~GeV/c an inconsistency develops. It is interesting to regard this inconsistency for a higher $p_T$ range as shown in Fig.~\ref{fig:highpt} up to $p_T$~=~3~GeV/c.

In fact even at ISR energy the diffractive production of pions with $p_T$ up to 3~GeV/c is strongly suppressed by the presence of a leading baryon which takes of order 80--90\% of the available energy. This is evidenced by the very strong suppression of the yields with increasing $p_T$. It could therefore be assumed that no diffractive component is present in this $p_T$ range. This should entail a suppression of the reference data by about 20\% with respect to the extrapolation from the higher energy region as indicated by the dotted lines in Fig.~\ref{fig:highpt}.

It should be noted that the $p_T$ distribution at $\log(s)$~=~2.1 ($p_{\textrm{beam}}$~=~70~GeV/c) is strictly exponential with the form $f = C e^{-5.9 p_T}$ up to $p_T$~=~4~GeV/c. This is not connected to a thermal behaviour nor to an equivalence to transverse parton fragmentation. It is rather the evolution of shape that evolves with $p_{\textrm{lab}}$ from a steeper, non-exponential form to a flatter, again non-exponential one such that incidentally an exponential behaviour results at $p_{\textrm{beam}}$ around 70~GeV/c.

%
%
\subsubsection{Forward \boldmath $\pi^0$ data from the LHCf experiment}
\vspace{3mm}
\label{sec:lhcf}

LHC data for identified charged hadrons are only available in a very restricted central rapidity range corresponding practically to a delta function in $x_F$ for the $p_T$ range up to 1.3 GeV/c, Fig.~\ref{fig:xdycor}. There are, however, data from LHCf \cite{adriani,adriani2} for $\pi^0$ mesons covering the forward region from about $x_F$~=~0.2 up to the kinematic limit in a $p_T$ range from 0.025 to about 0.6~GeV/c. It is therefore tempting to compare these data to the pion yields at the upper limit of interaction energy available for charged hadrons at the ISR at $\sqrt{s}$~=~63~GeV. This would span a factor of about 100 in $\sqrt{s}$ and thus provide a sensitive experimental cross-check of the energy evolution of the hadronic cross sections, particularly also in view of scaling concepts.

Indeed, invoking Isospin Symmetry and the $\pi^+/\pi^-$ ratio $R_{\pi}$,

\begin{equation}
	\label{eq:pirat}
   R_{\pi}(x_F,p_T) = \frac{f_{\pi^+}(x_F,p_T)}{f_{\pi^-}(x_F,p_T)}
\end{equation}
the $\pi^0$ cross sections may be predicted from the measured $\pi^-$ yields as

\begin{equation}
	\label{eq:f2sig}
  \left(\frac{f}{\sigma_{\textrm{inel}}}\right)_{\pi^0}(x_F,p_T) = \left( \frac{1 + R_{\pi}(x_F,p_T)}{2} \right)  \left(\frac{f}{\sigma_{\textrm{inel}}}\right)_{\pi^-}(x_F,p_T)
\end{equation}

\begin{figure}[h]
 	\begin{center}
   	\includegraphics[width=15cm] {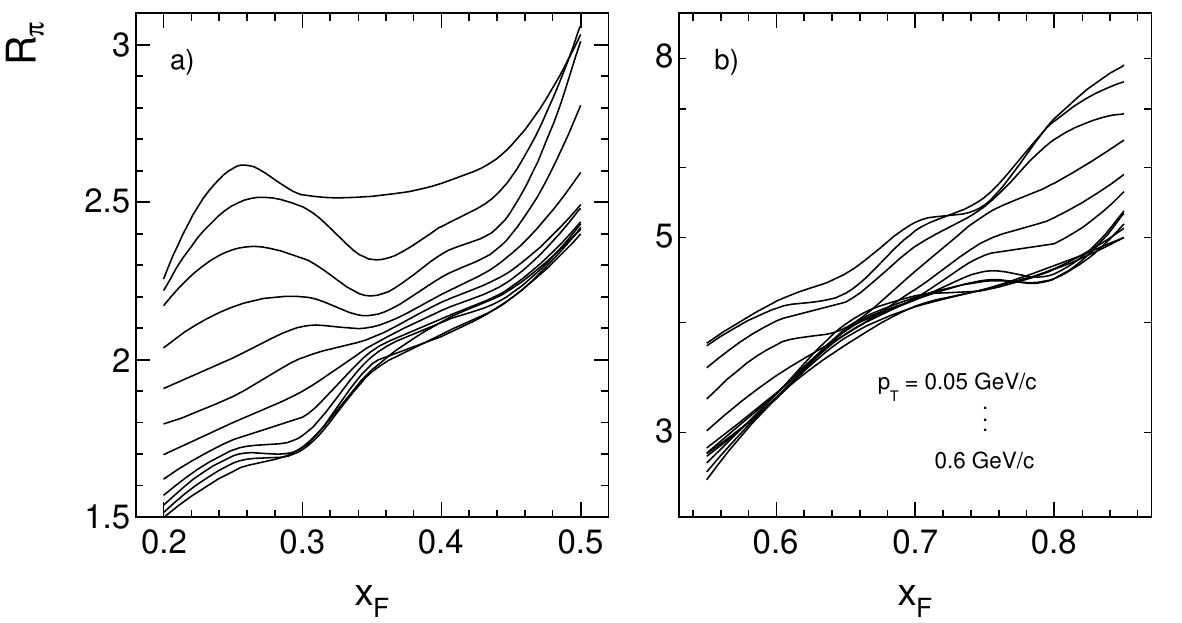} 
	 	\caption{$R_{\pi}$ ratio as a function of $x_F$ for $p_T$ between 0.05 and 0.6~GeV/c. Panel a) $x_F$ from 0.2 to 0.5 in linear scale, panel b) $x_F$ from 0.5 to 0.85 in logarithmic scale}
  	\label{fig:piratio}
 	\end{center}
\end{figure}

$R_{\pi}(x_F,p_T)$ has been measured by NA49 \cite{pp_pion} with a precision of a few percent. It has also been shown \cite{pp_pion} that $R_{\pi}$ is closely $s$-invariant up to ISR energy. The ratio at $\sqrt{s}$~=~17.2~GeV is shown in Fig.~\ref{fig:piratio} as a function of $x_F$ for the $x_F$ and $p_T$ ranges covered by LHCf.

\begin{figure}[h]
 	\begin{center}
   	\includegraphics[width=8cm] {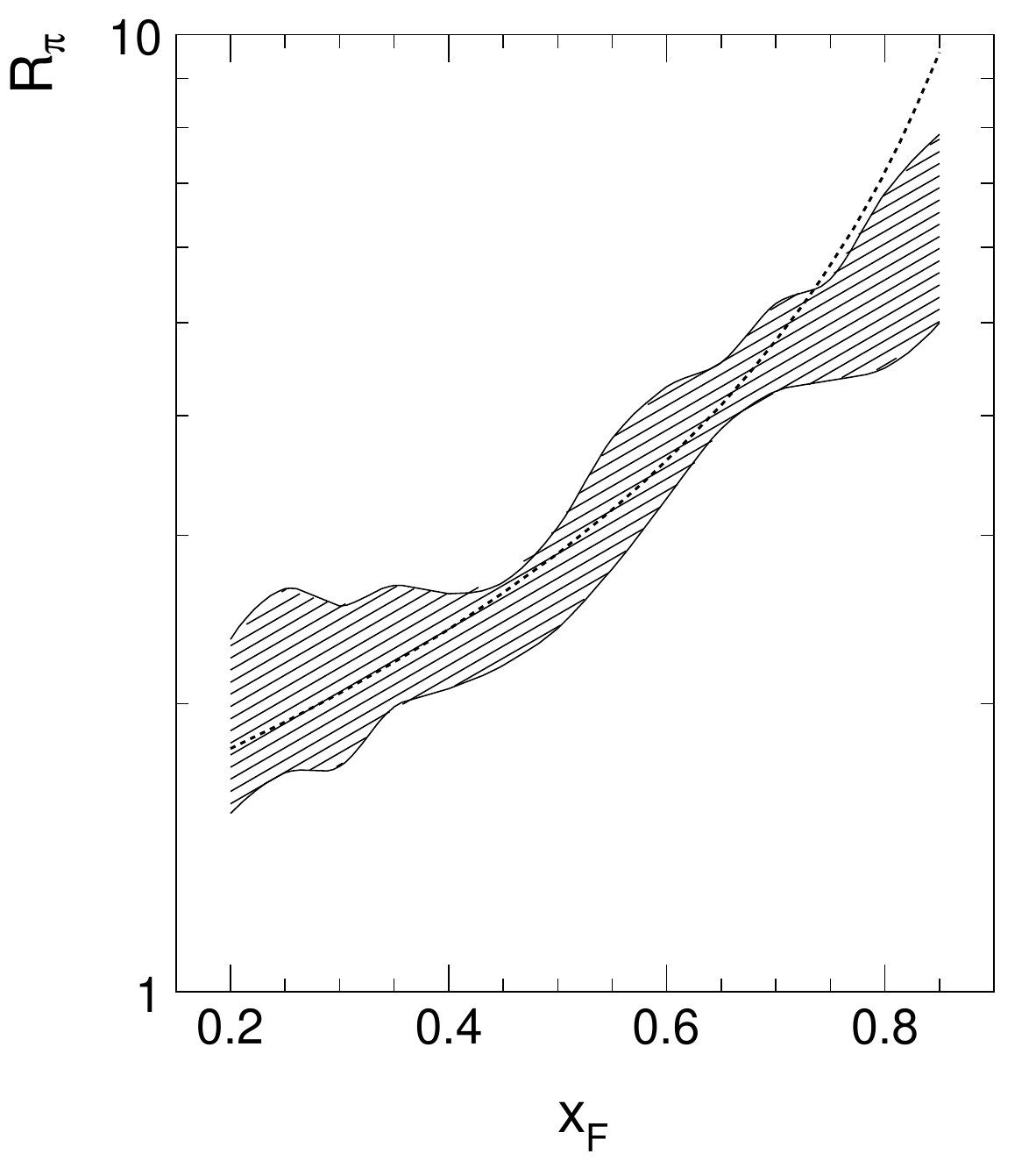} 
	 	\caption{$R_{\textrm{parton}}$ as a function of $x_F$ (dotted line). $R_{\textrm{parton}}$ is adjusted to the mean $p_T$ at $x_F$~=~0.2. The shaded area gives the limits of the experimental $R_{\pi}$ ratio presented in Fig.~\ref{fig:piratio} between $p_T$~=~0.05 and 0.6~GeV/c}
  	\label{fig:partonrat}
 	\end{center}
\end{figure}

It is interesting to compare this evolution with the ratio

\begin{equation}
 R_{\textrm{parton}} = \frac{(1-x_F)^3}{(1-x_F)^4}
\end{equation}
as it has been derived from the structure functions of the leading u and d quarks in the respective $\pi$ mesons as presented in Fig.~\ref{fig:partonrat}.

Of course this integrated ratio, based on the standard invariant and exponential $p_T$ distributions of parton hadronization, does not exhibit any detail of the $p_T$ dependence as it is visible in the experimental results shown in Fig.~\ref{fig:piratio}.

\begin{figure}[h]
 	\begin{center}
   	\includegraphics[width=15.5cm] {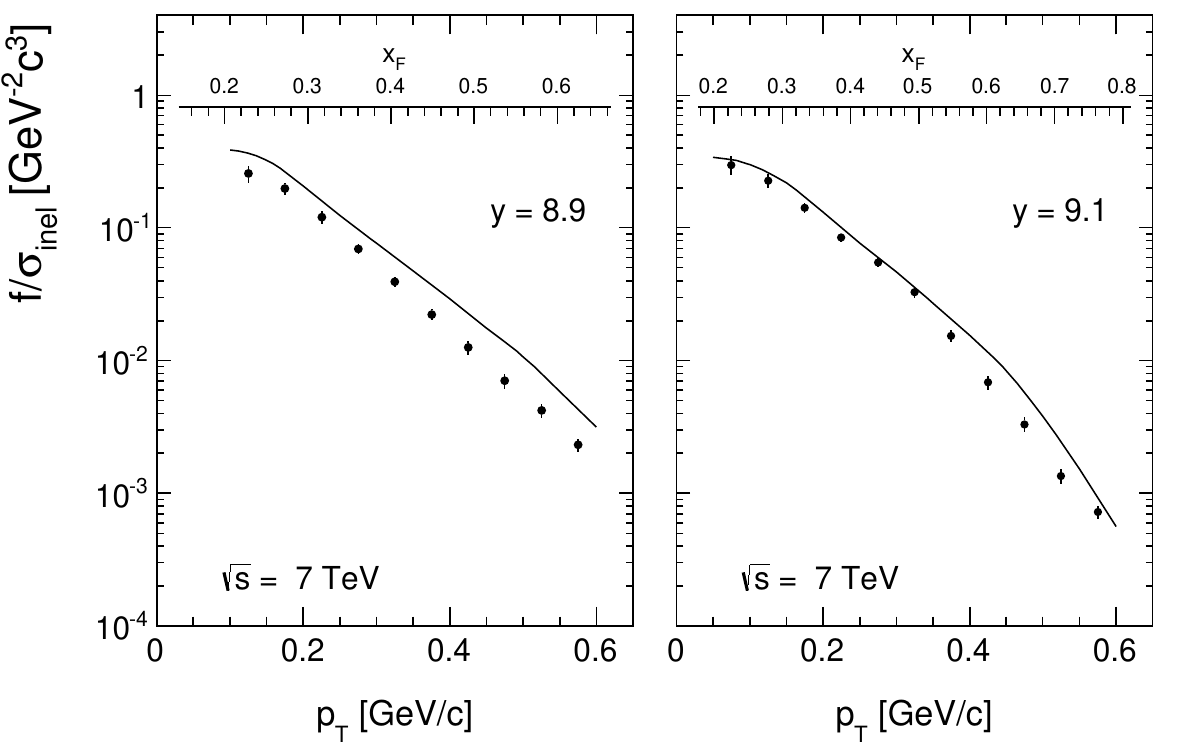} 
	 	\caption{Invariant $\pi^0$ cross sections $f/\sigma_{\textrm{inel}}$ as a function of $p_T$ for rapidities 8.9 and 9.1 at $\sqrt{s}$~=~7~TeV (data points from LHCf) compared to predicted yields at $\sqrt{s}$~=~63 GeV (full lines)}
  	\label{fig:lhcf1}
 	\end{center}
\end{figure}

\begin{figure}[h]
 	\begin{center}
   		\includegraphics[width=15.5cm] {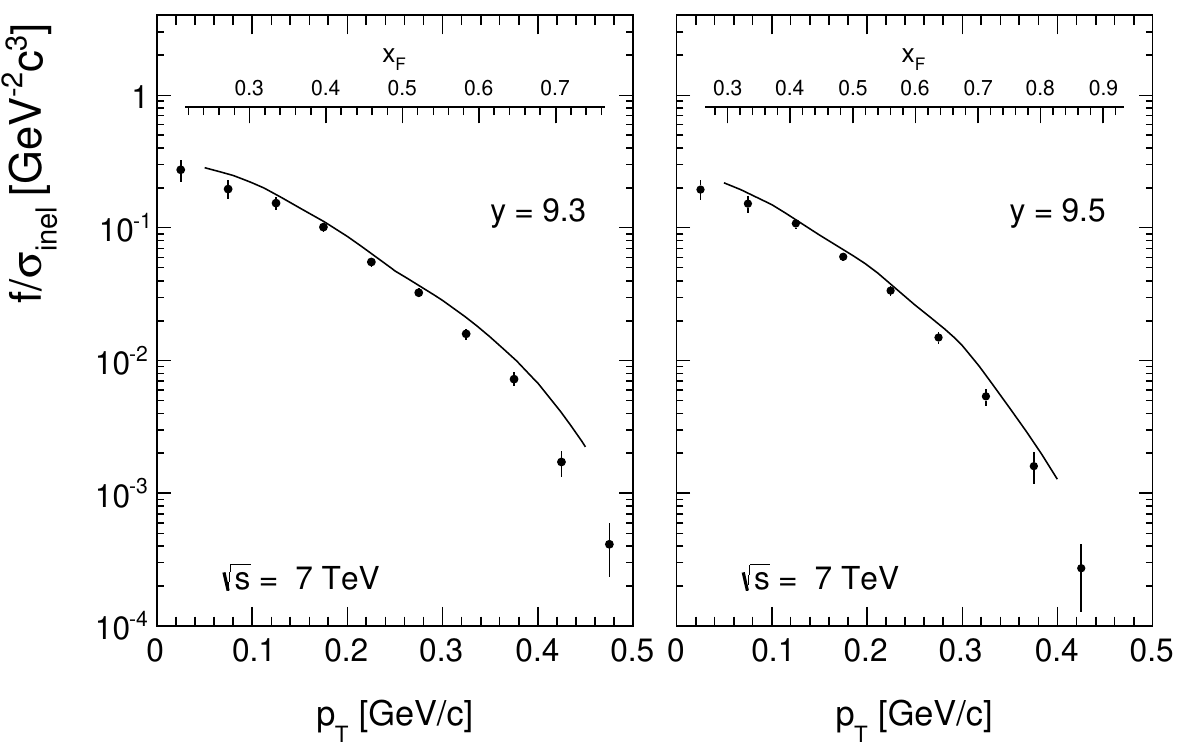} 
	 	\caption{Invariant $\pi^0$ cross sections $f/\sigma_{\textrm{inel}}$ as a function of $p_T$ for rapidities 9.3 and 9.5 at $\sqrt{s}$~=~7~TeV (data points from LHCf) compared to predicted yields at $\sqrt{s}$~=~63 GeV (full lines)}
  		\label{fig:lhcf2}
 	\end{center}
\end{figure}

\begin{figure}[h]
 	\begin{center}
   	\includegraphics[width=15.5cm] {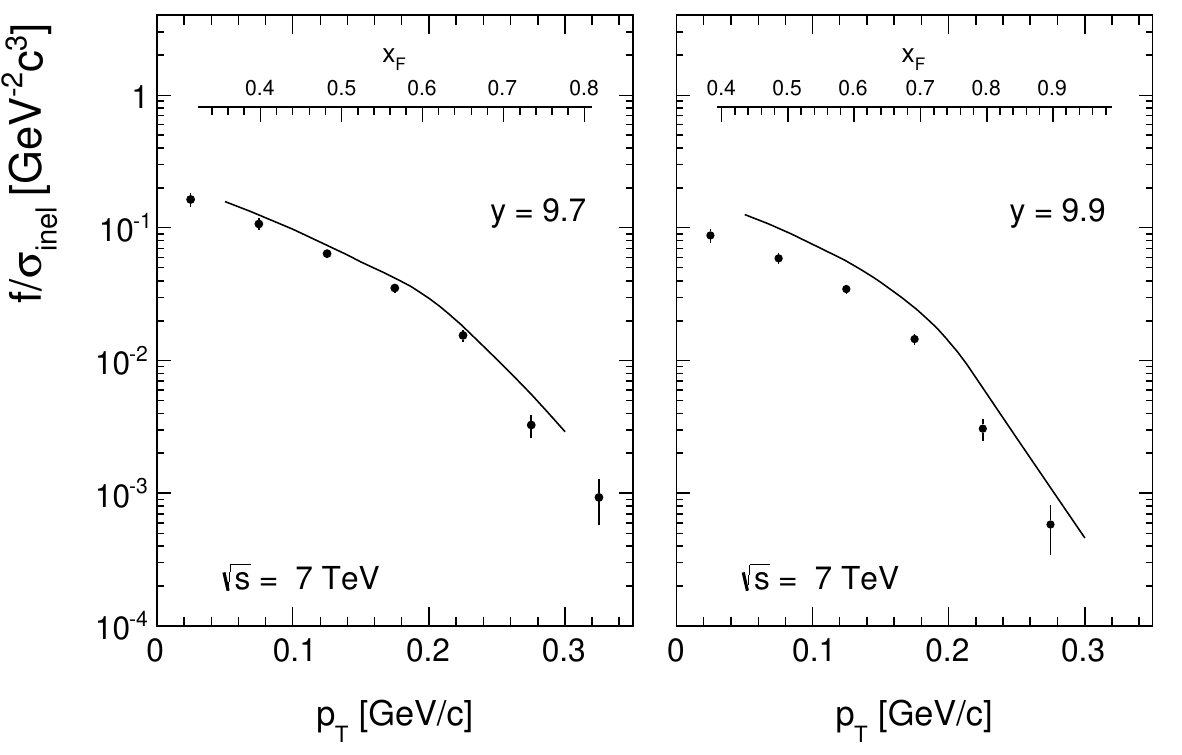} 
	 	\caption{Invariant $\pi^0$ cross sections $f/\sigma_{\textrm{inel}}$ as a function of $p_T$ for rapidities 9.7 and 9.9 at $\sqrt{s}$~=~7~TeV (data points from LHCf) compared to predicted yields at $\sqrt{s}$~=~63 GeV (full lines)}
  	\label{fig:lhcf3}
 	\end{center}
\end{figure}

\begin{figure}[h]
 	\begin{center}
   	\includegraphics[width=9cm] {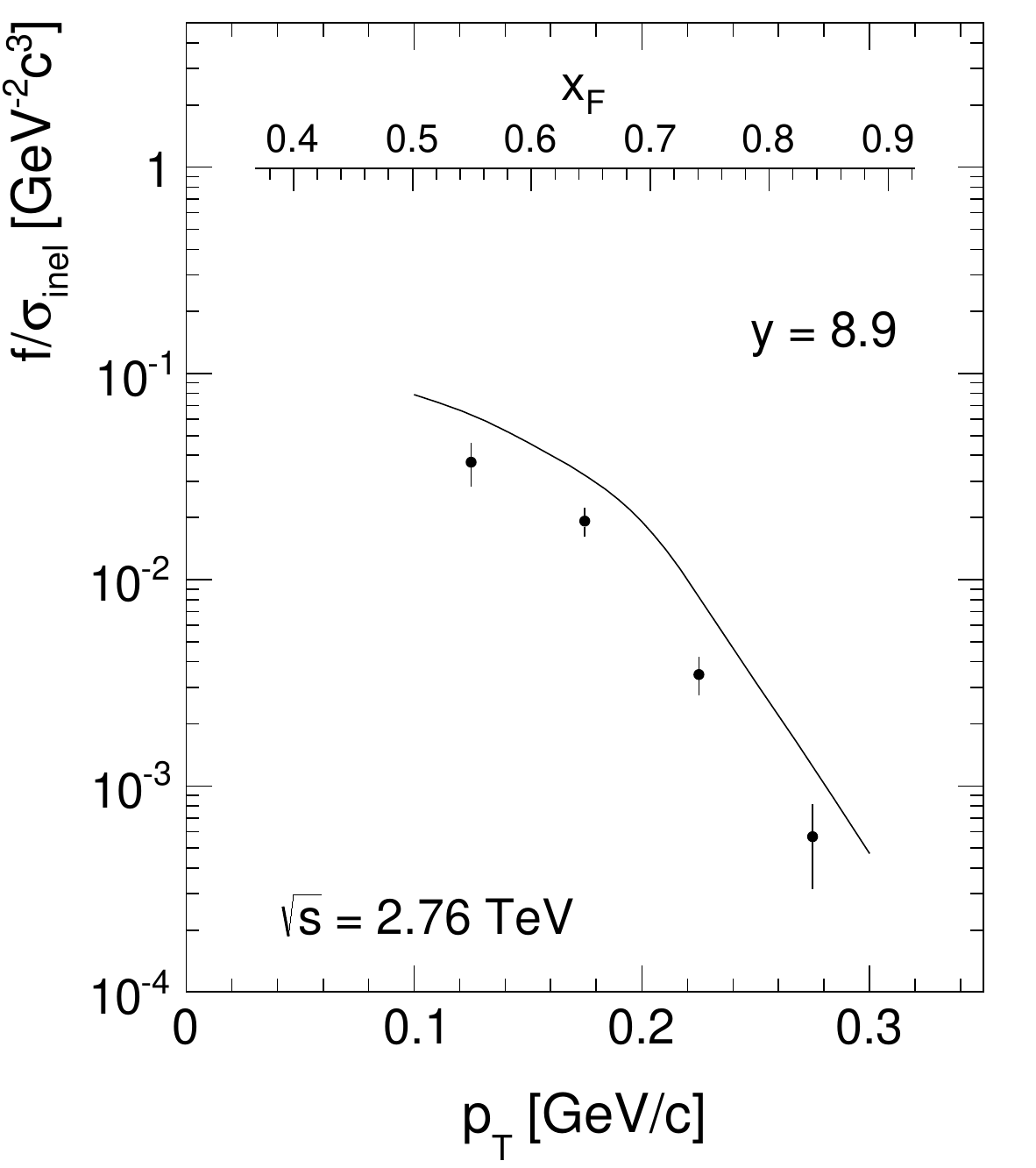} 
	 	\caption{Invariant $\pi^0$ cross sections $f/\sigma_{\textrm{inel}}$ as a function of $p_T$ for rapidity 8.9 at $\sqrt{s}$~=~2.76 TeV (data points from LHCf) compared to predicted $\pi^0$ yields at $\sqrt{s}$~=~63~GeV (full line)}
  	\label{fig:lhcf2.76}
 	\end{center}
\end{figure}

The global interpolation (Sect.~\ref{sec:interpolation}) reaches up to $\log(s)$~=~3.6 corresponding to the highest ISR energy at $\sqrt{s}$~=~63~GeV. The predicted $f/\sigma_{\textrm{inel}}$ for $\pi^0$ at this energy may now be compared to the LHCf data. As the $y_{\textrm{lab}}$ range of the interpolation reaches down to -1, the 6 rapidity values of 8.9, 9.1, 9.3, 9.5, 9.7 and 9.9 are amenable for comparison as presented in Figs.~\ref{fig:lhcf1} to \ref{fig:lhcf3}. At $\sqrt{s}$~=~2.76~TeV only the lowest rapidity value at $y$~=~8.9 falls into the accessible range for comparison as shown in Fig.~\ref{fig:lhcf2.76}.

In addition to the $p_T$ scales the corresponding $x_F$ scales are shown on the abscissa. This demonstrates that it is rather the longitudinal momentum that covers a wide range than the limited $p_T$ region due to the very small cms angle at these high rapidities. It is therefore erroneous to think about $p_T$ distributions as the longitudinal component couples into the yield dependence in a decisive fashion.

Several points are noteworthy in this comparison:

\begin{enumerate}
	\item The shape of the $p_T$ distributions is rather well reproduced by the predicted $\pi^0$ yields at 63~GeV.
	\item All predicted cross sections lie above the LHCf data. This is in sharp contrast with the central rapidity region where all LHC cross sections are above the ISR data with a strong increase as a power of $s$ (\ref{eq:sdep}).
	\item The interesting transition region between $x_F \sim$~0 and 0.2 is not (and will not be) accessible to experiment.
	\item The combined statistical and systematic errors of the LHC data are sizeable and reach values above 20\% at the low and high end of the $p_T$ scale.
\end{enumerate}

\begin{figure}[h]
 	\begin{center}
   	\includegraphics[width=15cm] {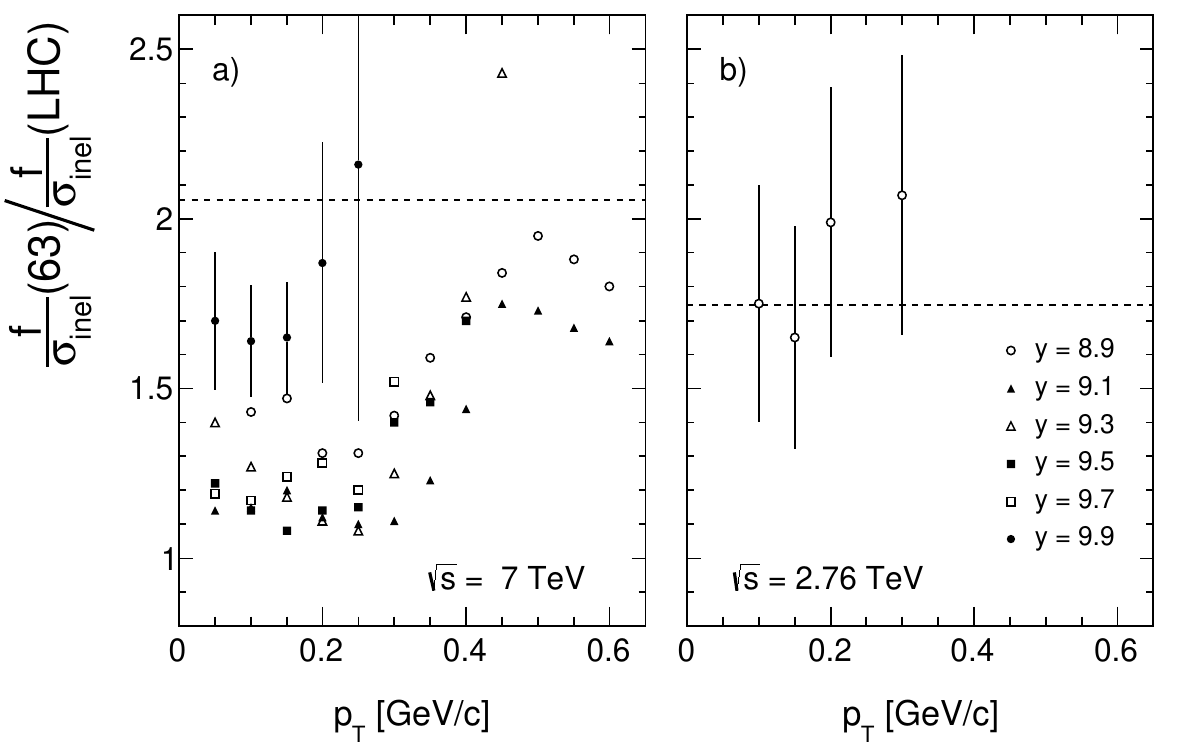} 
	 	\caption{Ratio of invariant cross sections $f/\sigma_{\textrm{inel}}$ between $\sqrt{s}$~=~63~GeV and 7~TeV (panel a) and 2.76~TeV (panel b) as a function of $p_T$ at constant rapidity. The ratios of the corresponding total inelastic cross sections $R_\sigma$ are indicated as broken lines}
  	\label{fig:isrlhcfrat}
 	\end{center}
\end{figure}

In view of the fact that this experimental comparison covers two orders of magnitude in $\sqrt{s}$ as opposed to the fact that the present study of $\pi^-$ yields reaches only over about one order of magnitude from $\sqrt{s}$~=~3 to 63~GeV, the relative closeness of the cross sections at the three energies is rather  impressive. The ratios of the cross sections between 63~GeV and 7~TeV and 2.76~TeV are presented in Fig.~\ref{fig:isrlhcfrat}. The ratios

\begin{equation}
	R_\sigma = \frac{\sigma_{\textrm{inel}}(LHC)}{\sigma_{\textrm{inel}}(63\;GeV)}
\end{equation}
are added as broken lines in the Figure, $R_\sigma$~=~1.75 and 2.05 for 2.76 and 7~TeV respectively.

Notwithstanding the relatively large error margins, these ratios reveal two trends:

\begin{enumerate}
	\item With increasing rapidity the ratios increase and approach the corresponding inelastic cross section ratios at 2.06 at 7~TeV and 1.74 at 2.76~TeV.
	\item A similar trend is visible with increasing $p_T$.
\end{enumerate}

\begin{figure}[h]
 	\begin{center}
    \includegraphics[width=16cm] {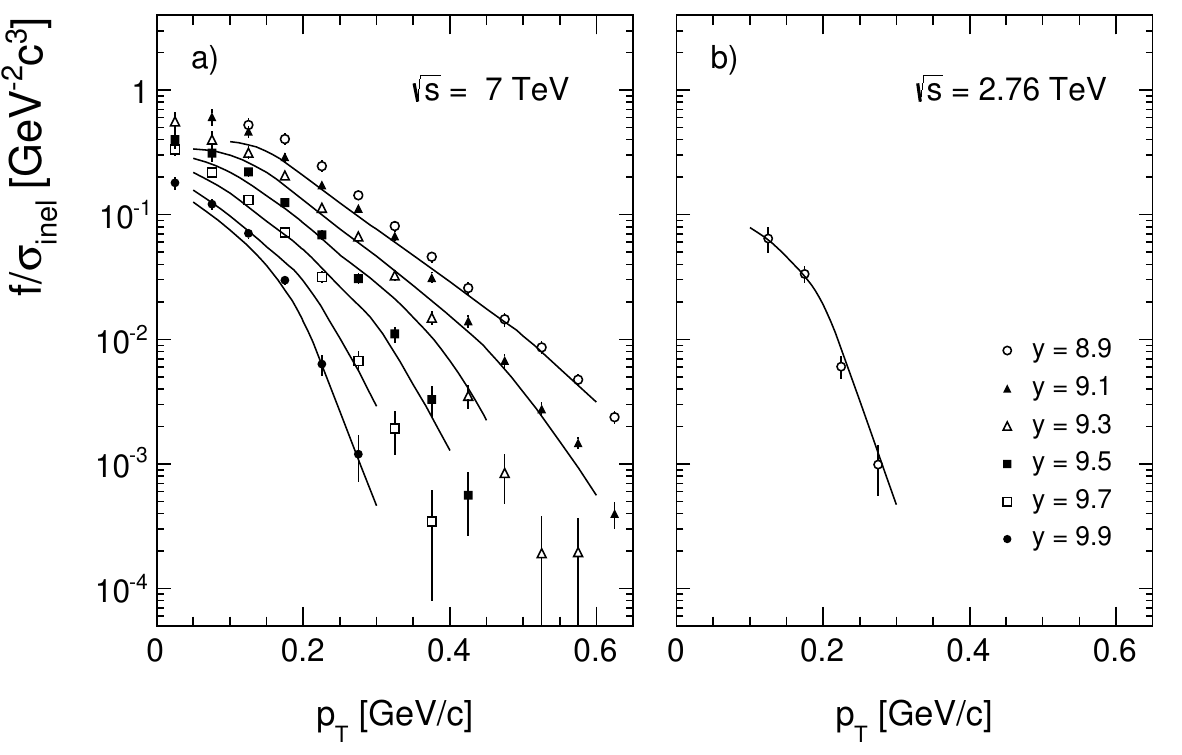} 
	 	\caption{$\pi^0$ cross sections $f/\sigma_{\textrm{inel}}$ as a function of $p_T$ at $\sqrt{s}$~=~7~TeV (panel a) and 2.76~TeV (panel b). Full lines ISR prediction at $\sqrt{s}$~=~63~GeV, data points LHCf data multiplied with the ratio of the inelastic cross sections}
  	\label{fig:isrlhcf}
 	\end{center}
\end{figure}

If multiplied with the cross section ratios of 1.75 and 2.05 at $\sqrt{s}$~=~2.76 and 7~TeV respectively, the LHCf data approach the ISR prediction rather closely at the lowest $y_{\textrm{lab}}$ values available for comparison at $y_{\textrm{lab}}$~=~-0.91 and -0.98 respectively as shown in Fig.~\ref{fig:isrlhcf}.

This seems to correspond to the questions raised in connection with energy scaling in Sect.~\ref{sec:energy_scaling}. Are there first indications that the scaling behaviour might be different in different regions of phase space? Is it the cross section yield per inelastic collision $f/\sigma_{\textrm{inel}}$ or rather the invariant cross section $f$ itself that should be compared? This question will again be invoked in the following section on scaling.

%
%
\section{Scaling}
\vspace{3mm}
\label{sec:scaling}

The concept of "scaling" or, better, "scale invariance", has attracted quite some attention from the early days of particle physics. Indeed, if a quantity in the inclusive sector can be shown to be independent of one of the inclusive variables, one may hope to obtain further, model-independent information about the underlying production process.

Historically it has been the independence of particle yields on the interaction energy parameter $s$ (\ref{eq:incl}) that has been invoked. As early as 1969 when the first inclusive measurements of hadron yields became available, two conjectures in this sense have been published \cite{feynman}, \cite{benecke}. Both conjectures concern the longitudinal momentum components and both are based on results from lepton-nucleon scattering but come to rather different conclusions.

Reference \cite{benecke} evokes "limiting fragmentation" of the target or projectile nucleons in their respective Lorentz frames as a function of longitudinal momentum in the high energy limit.

Reference \cite{feynman} deals with the high energy limit of particle production as a function of the cms longitudinal momentum if referred to the maximum available momentum (\ref{eq:xf_def}) and has obtained the name tag of "Feynman x" in analogy to "Bjorken x" in deep-inelastic lepton-nucleon scattering.

Both conjectures make statements concerning the forward/backward regions of phase space and rely on the notion of "longitudinal phase space" (\ref{eq:pLpT}) where the transverse momentum distributions are sharply (exponentially) limited in contrast to the longitudinal ones which are, for $x_F \gtrsim$~0.3, approximately of the form $\frac{1}{x_F}(1-x_F)^n$ where the exponent $n$ depends on particle type, $n \sim$~4 for $\pi^-$.

Both hypothesis are rather vague as far as numerical predictions of hadronic yields are concerned. They will be confronted in the following sub-sections with the complete set of $\pi^-$ data treated in the preceding sections.

%
%
\subsection{The Hypothesis of Limiting Fragmentation}
\vspace{3mm}
\label{sec:scaling_limiting}

The approach starts from the excitation of the target and projectile nucleons into (unspecified) masses $M > m_p$ and their subsequent decay

\begin{equation}
	\label{eq:mdec}
	M \rightarrow n + 1
\end{equation}
where $n$ concerns secondaries and "1" specifies a final state nucleon thus ensuring baryon number conservation. More specifically the following decays of $M$ are considered:

\begin{equation}
\label{eq:mdecmult}
	M \rightarrow p+\pi,\;p+2\pi,\;p+3\pi,\;...,\;p+n\pi
\end{equation}

In this sense the hypothesis is similar to preceding ideas like the "fireball" \cite{adair}, "isobar" \cite{lindenbaum} or "diffraction dissociation" \cite{good} models. It is specific in the sense that it makes a precise prediction as to the behaviour of particle densities in the target and projectile laboratory systems in the high energy limit. For this purpose a connection to the scaling limit of inelastic electron-proton scattering is established where it is shown that the excitation of target or projectile into a mass $M$ approaches a limit at high energy. This connection assumes of course the independence of target nucleon excitation on the type of projectile particle ("factorization"), one of the many conjectures contained in \cite{benecke}.

In the target rest frame with the coordinates $p_T$ and $p_L^{\textrm{lab}}$, the kinematic limit is described by elastic scattering defining a parabola in $p_T$ and $p_L^{\textrm{lab}}$ centred at $p_L^{\textrm{lab}}$~=~0~GeV/c for protons and negative $p_L^{\textrm{lab}}$ for lighter secondaries. For pions the region of interest is defined  by $p_L^{\textrm{lab}} <$~0~GeV/c \cite{benecke}. At the time of publication, only very scarce experimental data in this area (which corresponds to the very forward region in the experiments) were available, essentially for the two beam momenta of 19 and 30~GeV/c. This led to the statement that the expected limit was already reached at the higher beam momentum, a rather bold claim also in view of the normalization problems of the data (see Sect.~\ref{sec:countdata}).

Given the global data interpolation elaborated above over the wide range of interaction energies from 3~$< \sqrt{s} <$~63~GeV or 1~$< \log{s} <$~3.6, a new sensible test of the hypothesis of limiting fragmentation may be performed. As at the time of publication \cite{benecke} the interaction cross sections were assumed to be constant, the known energy dependence of the inelastic cross section represents an additional challenge to any scaling approach, see Sect.~\ref{sec:energy_scaling}. In the following Figs.~\ref{fig:pllab01}--\ref{fig:pllab05} therefore both the quantities $f/\sigma_{\textrm{inel}}$ and $d^2\sigma/(dp_L^{\textrm{lab}}dp_T)$ are presented although the variation of $\sigma_{\textrm{inel}}$ over the given $s$ range is only about 20\%.

\begin{figure}[h]
 \begin{center}
   \begin{turn}{\rotAngle}
   \begin{minipage}{\capw}
	\caption{$\pi^-$ cross sections as a function of $p_L^{\textrm{lab}}$ for $p_T$~=~0.1~GeV/c and $\log(s)$ from 1 to 3.6, a) $d^2\sigma/(dp_L^{\textrm{lab}}dp_T)$, b) $f/\sigma_{\textrm{inel}}$}
  	\label{fig:pllab01}
	\end{minipage}
	\end{turn}
   \includegraphics[width=\figw,angle=\rotAngle] {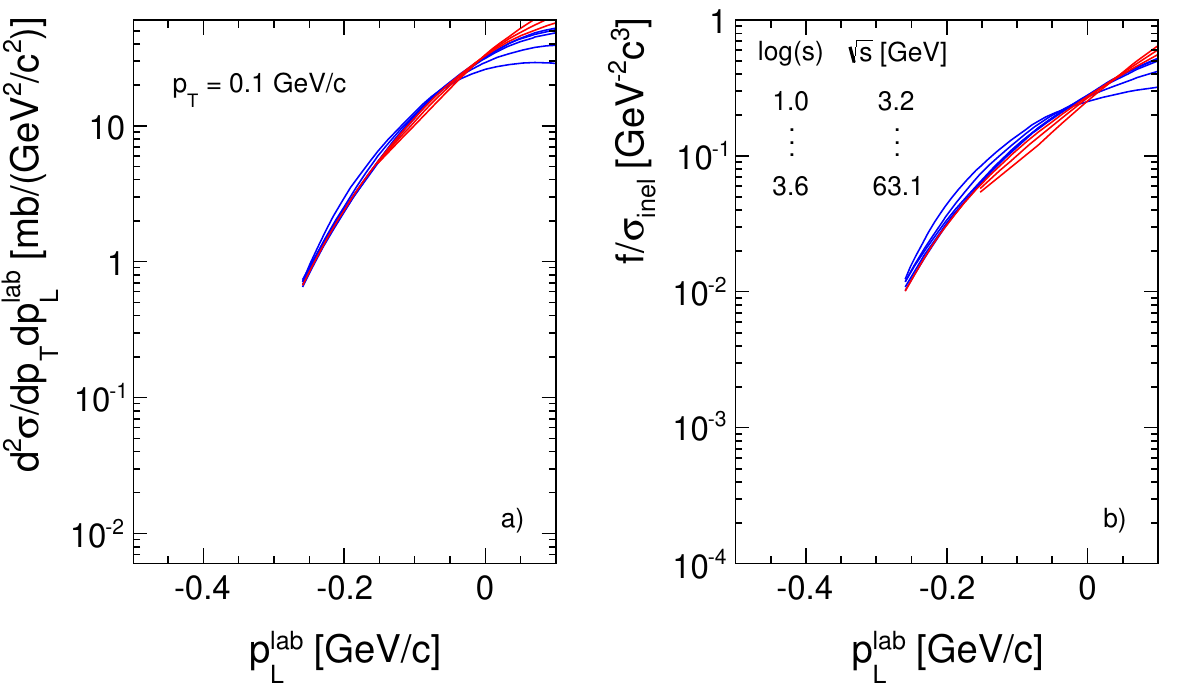} 
 \end{center}
\end{figure}

\begin{figure}[h]
 \begin{center}
   \begin{turn}{\rotAngle}
   \begin{minipage}{\capw}
	\caption{$\pi^-$ cross sections as a function of $p_L^{\textrm{lab}}$ for $p_T$~=~0.2~GeV/c and $\log(s)$ from 1 to 3.6, a) $d^2\sigma/(dp_L^{\textrm{lab}}dp_T)$, b) $f/\sigma_{\textrm{inel}}$}
  	\label{fig:pllab02}
	\end{minipage}
	\end{turn}
   \includegraphics[width=\figw,angle=\rotAngle] {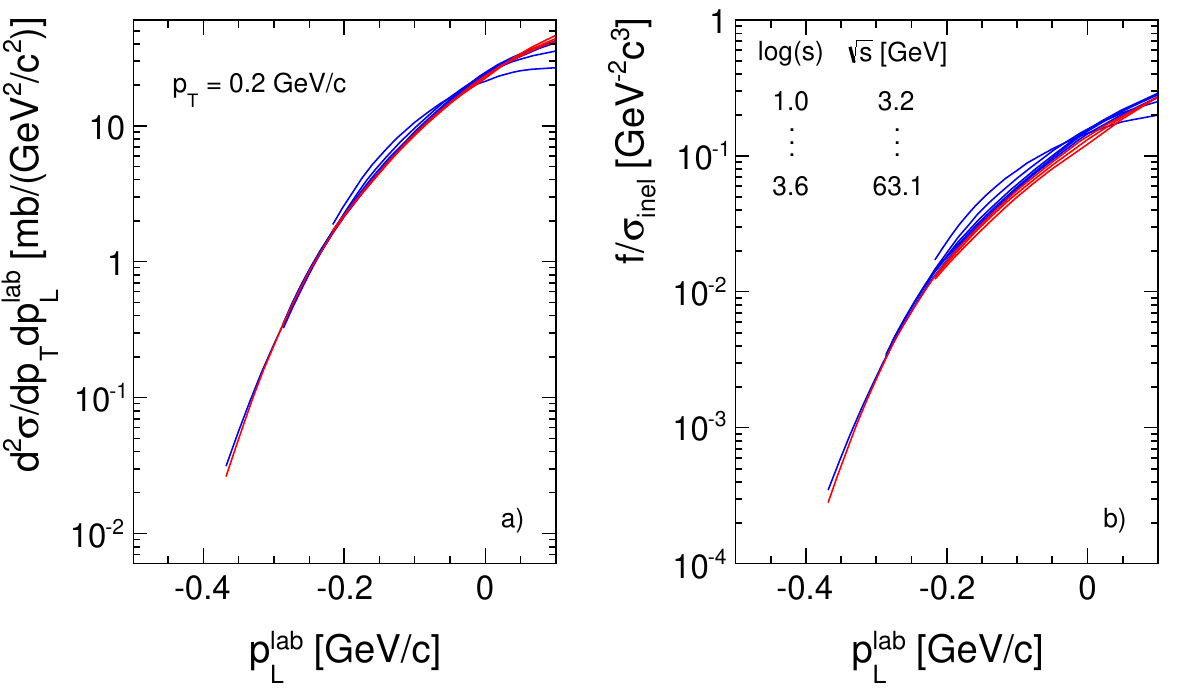} 
 \end{center}
\end{figure}

\begin{figure}[h]
 \begin{center}
   \begin{turn}{\rotAngle}
   \begin{minipage}{\capw}
	\caption{$\pi^-$ cross sections as a function of $p_L^{\textrm{lab}}$ for $p_T$~=~0.3~GeV/c and $\log(s)$ from 1 to 3.6, a) $d^2\sigma/(dp_L^{\textrm{lab}}dp_T)$, b) $f/\sigma_{\textrm{inel}}$}
  	\label{fig:pllab03}
	\end{minipage}
	\end{turn}
   \includegraphics[width=\figw,angle=\rotAngle] {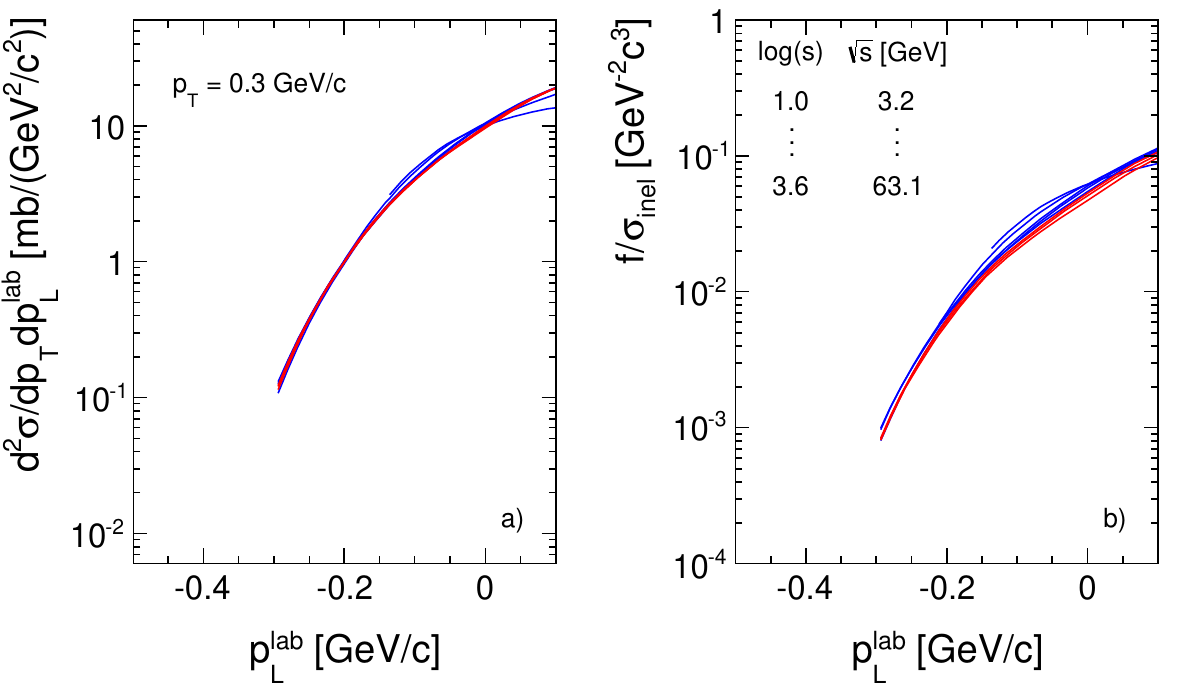} 
 \end{center}
\end{figure}

\begin{figure}[h]
 \begin{center}
   \begin{turn}{\rotAngle}
   \begin{minipage}{\capw}
	\caption{$\pi^-$ cross sections as a function of $p_L^{\textrm{lab}}$ for $p_T$~=~0.4~GeV/c and $\log(s)$ from 1 to 3.6, a) $d^2\sigma/(dp_L^{\textrm{lab}}dp_T)$, b) $f/\sigma_{\textrm{inel}}$}
  	\label{fig:pllab04}
	\end{minipage}
	\end{turn}
   \includegraphics[width=\figw,angle=\rotAngle] {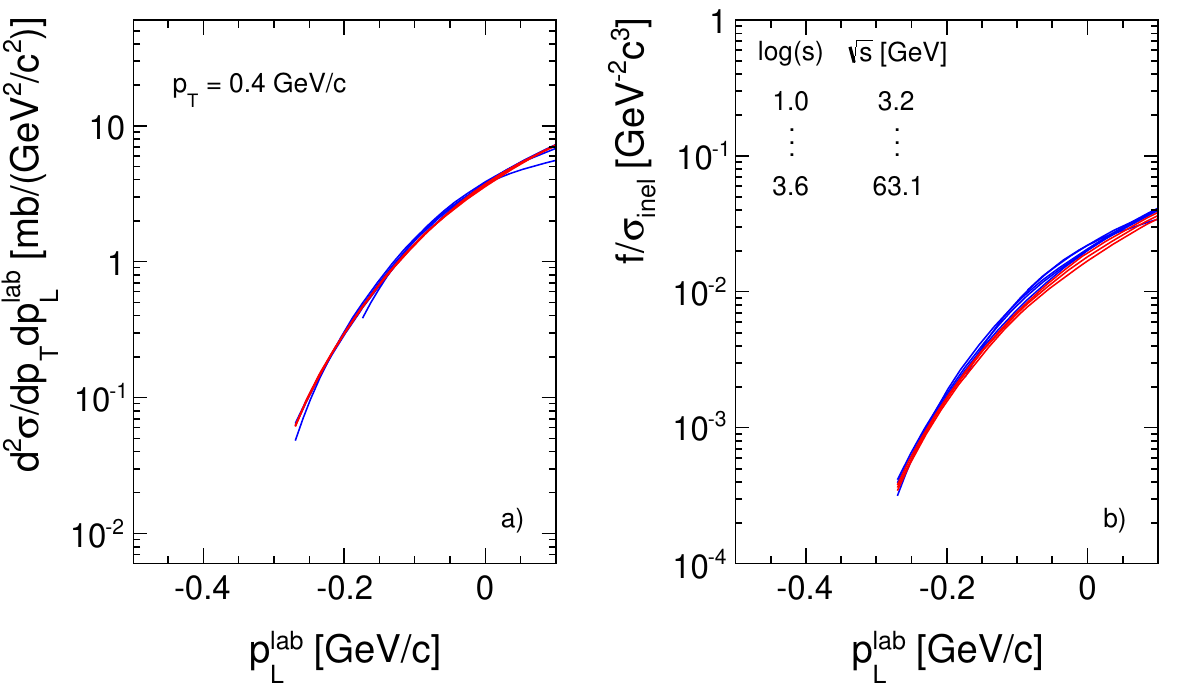} 
 \end{center}
\end{figure}

\begin{figure}[h]
 \begin{center}
   \begin{turn}{\rotAngle}
   \begin{minipage}{\capw}
	\caption{$\pi^-$ cross sections as a function of $p_L^{\textrm{lab}}$ for $p_T$~=~0.5~GeV/c and $\log(s)$ from 1 to 3.6, a) $d^2\sigma/(dp_L^{\textrm{lab}}dp_T)$, b) $f/\sigma_{\textrm{inel}}$}
  	\label{fig:pllab05}
	\end{minipage}
	\end{turn}
   \includegraphics[width=\figw,angle=\rotAngle] {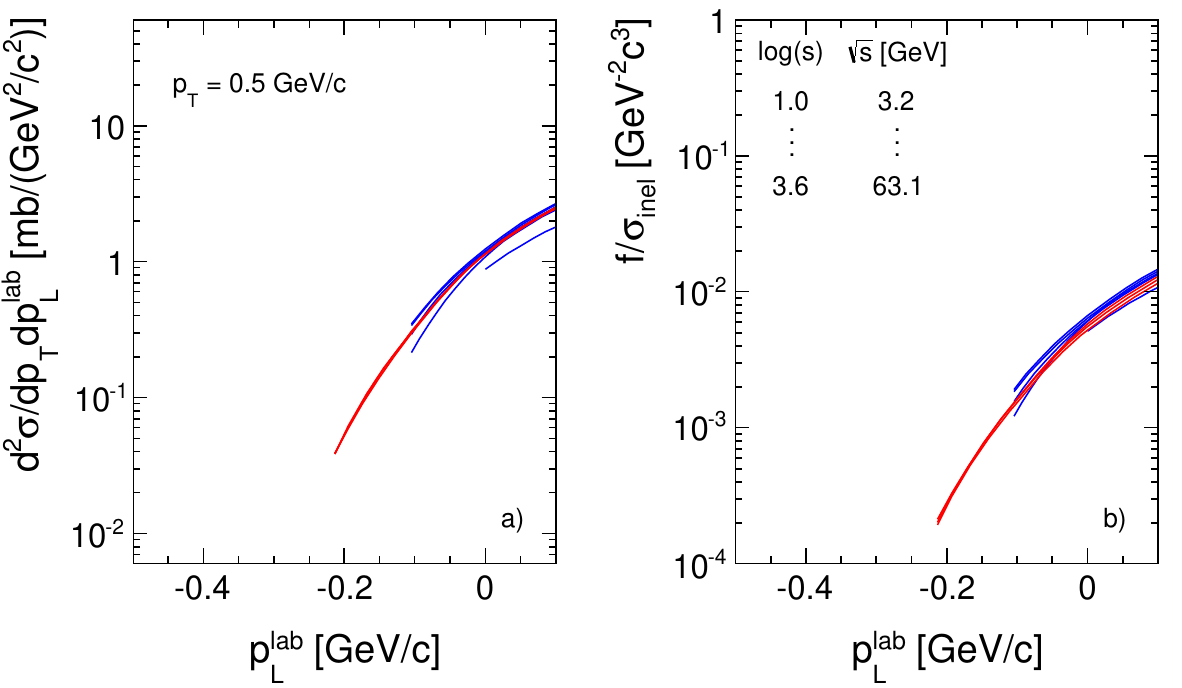} 
 \end{center}
\end{figure}

The closeness of the $d^2\sigma/(dp_L^{\textrm{lab}}dp_T)$ distributions over the full available range of interaction energies is striking (panels a). This provides strong evidence for the hypothesis of Limiting Fragmentation. The normalized cross sections $f/\sigma_{\textrm{inel}}$ (panels b) on the other hand spread considerably more with $\log(s)$. This indicates a preference for constant invariant cross sections in this specific area of phase space rather than constant hadronization over the full increasing interaction area.

\begin{figure}[h]
 \begin{center}
   \begin{turn}{\rotAngle}
   \begin{minipage}{\capw}
	\caption{$\pi^0$ cross sections as a function of $p_L^{\textrm{lab}}$ for $p_T$~=~0.1~GeV/c for NA49 and ISR (predictions) and LHC energies, a) $d^2\sigma/(dp_L^{\textrm{lab}}dp_T)$, b) $f/\sigma_{\textrm{inel}}$}
  	\label{fig:pllhc01}
	\end{minipage}
	\end{turn}
   \includegraphics[width=\figw,angle=\rotAngle] {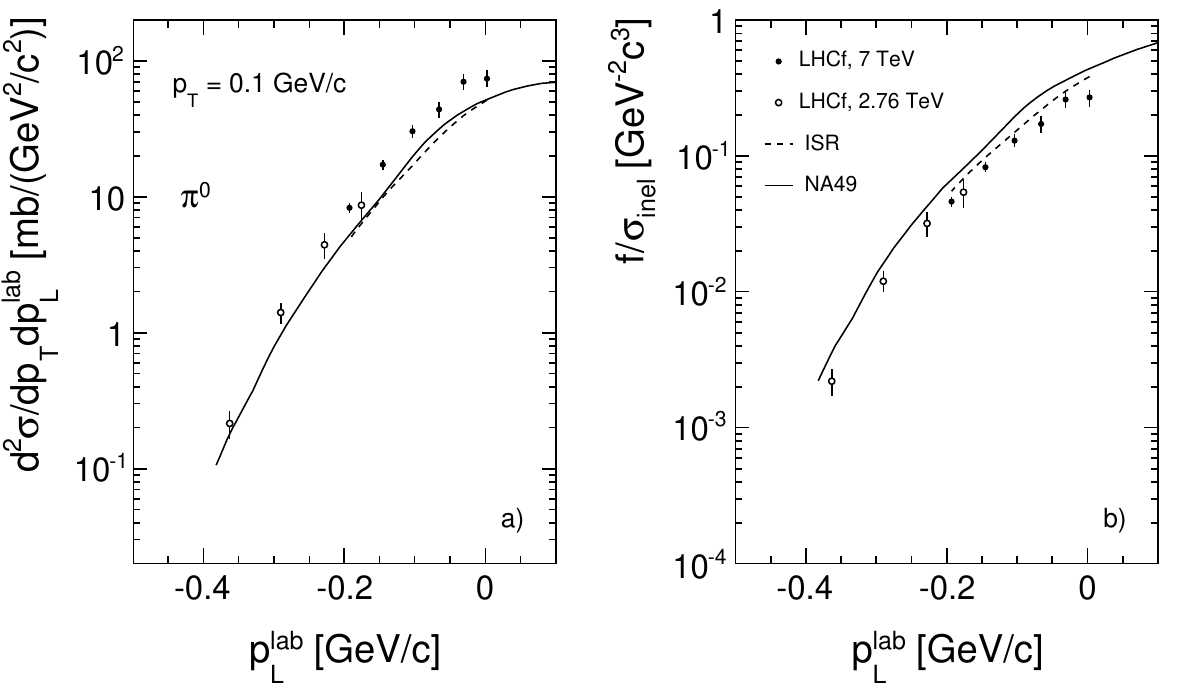} 
 \end{center}
\end{figure}

\begin{figure}[h]
 \begin{center}
   \begin{turn}{\rotAngle}
   \begin{minipage}{\capw}
	\caption{$\pi^0$ cross sections as a function of $p_L^{\textrm{lab}}$ for $p_T$~=~0.2~GeV/c for NA49 and ISR (predictions) and LHC energies, a) $d^2\sigma/(dp_L^{\textrm{lab}}dp_T)$, b) $f/\sigma_{\textrm{inel}}$}
  	\label{fig:pllhc02}
	\end{minipage}
	\end{turn}
   \includegraphics[width=\figw,angle=\rotAngle] {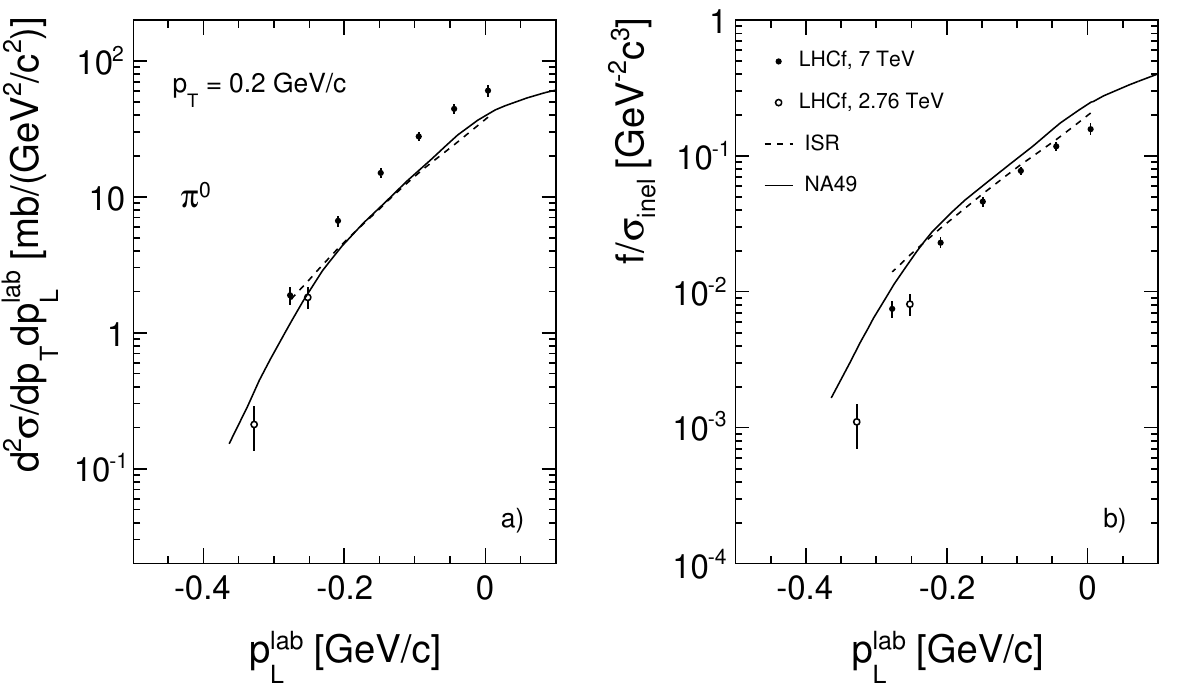} 
 \end{center}
\end{figure}

\begin{figure}[h]
 \begin{center}
   \begin{turn}{\rotAngle}
   \begin{minipage}{\capw}
	\caption{$\pi^0$ cross sections as a function of $p_L^{\textrm{lab}}$ for $p_T$~=~0.3~GeV/c for NA49 and ISR (predictions) and LHC energies, a) $d^2\sigma/(dp_L^{\textrm{lab}}dp_T)$, b) $f/\sigma_{\textrm{inel}}$}
  	\label{fig:pllhc03}
	\end{minipage}
	\end{turn}
   \includegraphics[width=\figw,angle=\rotAngle] {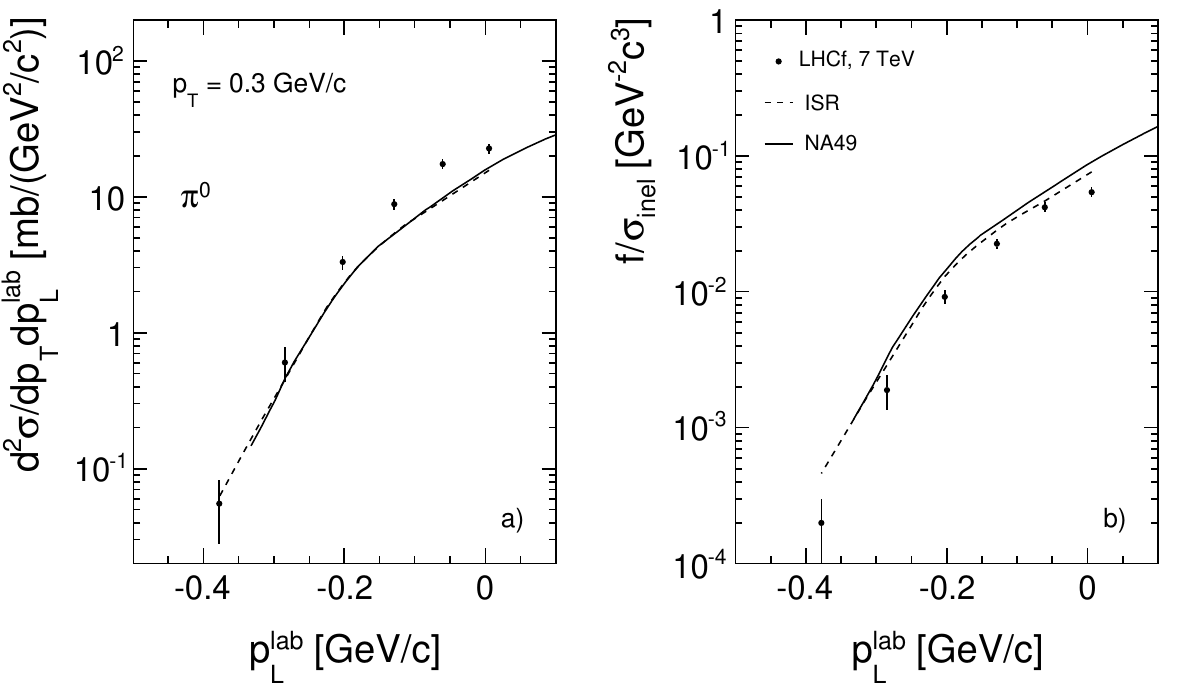} 
 \end{center}
\end{figure}

\begin{figure}[h]
 \begin{center}
   \begin{turn}{\rotAngle}
   \begin{minipage}{\capw}
	\caption{$\pi^0$ cross sections as a function of $p_L^{\textrm{lab}}$ for $p_T$~=~0.4~GeV/c for NA49 and ISR (predictions) and LHC energies, a) $d^2\sigma/(dp_L^{\textrm{lab}}dp_T)$, b) $f/\sigma_{\textrm{inel}}$}
  	\label{fig:pllhc04}
	\end{minipage}
	\end{turn}
   \includegraphics[width=\figw,angle=\rotAngle] {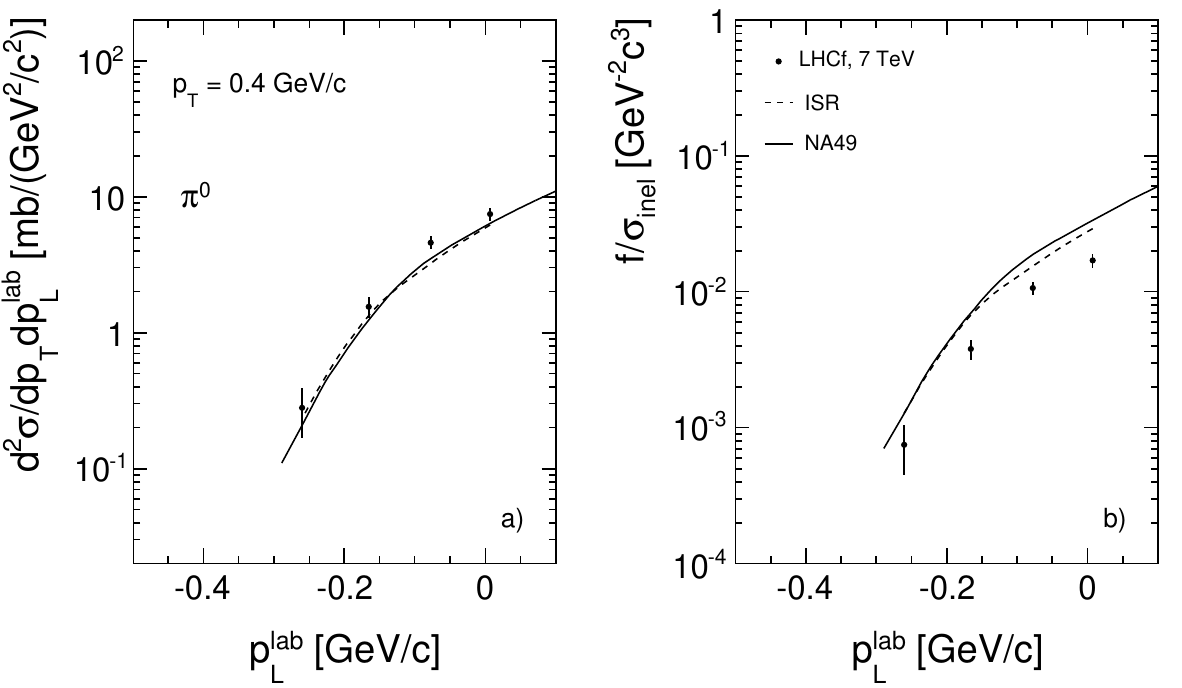} 
 \end{center}
\end{figure}

\begin{figure}[h]
 \begin{center}
   \begin{turn}{\rotAngle}
   \begin{minipage}{\capw}
	\caption{$\pi^0$ cross sections as a function of $p_L^{\textrm{lab}}$ for $p_T$~=~0.5~GeV/c for NA49 and ISR (predictions) and LHC energies, a) $d^2\sigma/(dp_L^{\textrm{lab}}dp_T)$, b) $f/\sigma_{\textrm{inel}}$}
  	\label{fig:pllhc05}
	\end{minipage}
	\end{turn}
   \includegraphics[width=\figw,angle=\rotAngle] {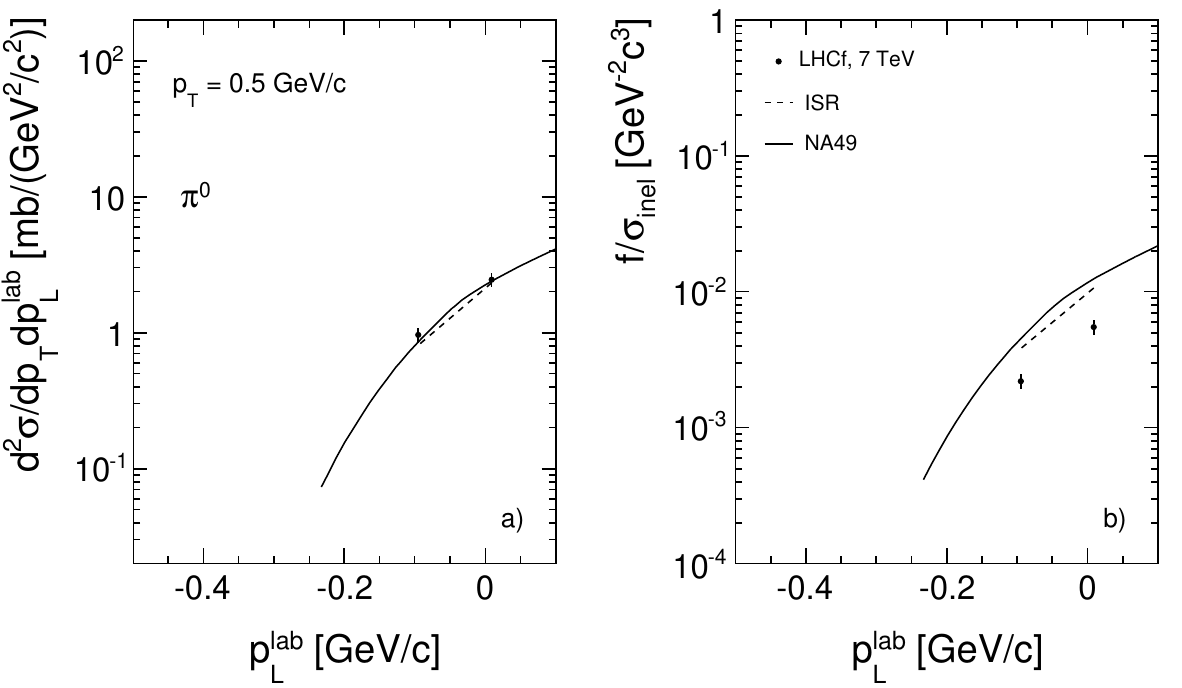} 
 \end{center}
\end{figure}

A further, more stringent test may be provided by the $\pi^0$ data of LHCf (Sect.~\ref{sec:lhcf}). Here the predicted $\pi^0$ cross sections (\ref{eq:pirat}), (\ref{eq:f2sig}) at the NA49 and ISR energies can be compared to the LHCf data. This corresponds to a range of more than 2 orders of magnitude in $\sqrt{s}$. The results are presented in Figs.~\ref{fig:pllhc01}--\ref{fig:pllhc05}.

As expected from the discussion of the $p_T$ distributions for different large rapidities in Sect.~\ref{sec:lhcf}, the $d^2\sigma/(dp_L^{\textrm{lab}}dp_T)$ distributions are surprisingly $s$-independent for $p_T \gtrsim$~0.2~GeV/c up to 7~TeV and for all $p_T$ at 2.76~TeV. The convergence with increasing $\sqrt{s}$ improves with decreasing $p_L^{\textrm{lab}}$ and $p_T$, see Fig.~\ref{fig:isrlhcfrat}. On the other hand, the $f/\sigma_{\textrm{inel}}$ distributions split very considerably in these regions.

At this point some general remarks are in place. Pions are in fact rather special items to be regarded at negative $p_L^{\textrm{lab}}$ as it is rather difficult to transport them there. Their yields are orders of magnitude below the proton ones at low $p_L^{\textrm{lab}}$. Indeed a widespread belief would have it that pions from resonance decay ((\ref{eq:mdec}) and (\ref{eq:mdecmult})) are all centred at low $x_F$ and low $p_T$. This is true for $\Lambda$ decay, Figs.~\ref{fig:lambdaS} and \ref{fig:lam2pi} where the pion yields are effectively cut off at $p_T \sim$~0.3~GeV/c and $x_F \sim$~0.3. The $\Lambda$ is a low-$Q$ resonance in its p+$\pi$ decay with a mass distribution of zero width. In contrast, already the K$^0_S$ decay with its modest $Q$ of 200~MeV and its symmetric 2$\pi$ decay yields pions at large $x_F$ and considerable $p_T$, see Fig.~\ref{fig:kazeros}. For strong decays, the wide mass distributions with the very long Breit-Wigner tails deconfine the decay pions to cover the full phase space. This is demonstrated in the Sect.~\ref{sec:res_mass_spec} dealing with resonance decay contributions. Another point is the decay chain p+n$\pi$ (\ref{eq:mdecmult}). In a two body decay the center of the pion decay ellipse is placed with respect to the protons at

\begin{equation}
	\label{eq:pi2p}
	\langle x_F^\pi \rangle \sim \frac{m_\pi}{m_p} x_F(p)
\end{equation}
such that a diffractively produced baryon resonance yields pions with an average $x_F \sim$~0.15 only. This value moves to smaller $x_F$ for multi-pion decays.

As a consequence several general remarks may be made as to the production of pions at negative $p_L^{\textrm{lab}}$ concerning:

\begin{enumerate}[label=(\roman*)]
	\item \label{i:i} Two body decays of high-$Q$ resonances.
	\item \label{i:ii} In the decay p+$\pi$ backward decay of the proton in the overall cms.
	\item Strong resonance decays with important relativistic Breit-Wigner tails.
	\item Decays which are symmetric in the decay particle masses.
	\item Contribution of both baryonic and mesonic resonances.
	\item With increasing interaction energy the rate of low multiplicity decays -- especially two-body decays -- is strongly reduced thus de-populating the region of negative $p_L^{\textrm{lab}}$.
\end{enumerate}

The authors of \cite{benecke} have clearly pointed out points \ref{i:i} and \ref{i:ii} claiming that at higher $n$ (\ref{eq:mdecmult}) the pions will move towards the central rapidity area. As far as resonance mass distributions are concerned they see only a modest "widening of the mesa-like structures" in phase space corresponding to the decay process. Mesonic resonances are not foreseen, nor are decays with symmetric decay particle masses.

%
%
\subsection{The hypothesis of Feynman scaling}
\vspace{3mm}
\label{sec:feynman_scaling}

In his original suggestion \cite{feynman} Feynman took as well as \cite{benecke} reference to deep inelastic lepton-proton scattering, however in a completely different context. In fact the existence of scaling longitudinal "structure functions" of proton constituents in this process had just been established \cite{bjorken} with $x_{Bj}$ (Bjorken $x$) as characteristic variable giving the momentum fraction of a parton relative to the proton momentum.

The conjecture of Feynman suggested a similar behaviour of secondary hadrons in p+p interactions if plotted in the fractional longitudinal momentum variable $x_F = 2p_L/\sqrt{s}$ (\ref{eq:xf_def}) thus creating an analogy with the partonic structure of the proton. In 1972 Berman, Bjorken and Kogut \cite{berman} predicted in fact partonic scattering in large momentum transfer hadronic interactions, first with electromagnetic coupling but already foreseeing strong coupling by "gluons" with correspondingly much higher cross sections.

The rapid evolution of these ideas into a full-blown theory of the strong interaction (Quantum Chromodynamics, QCD) cannot be followed here. Unfortunately the application of QCD is limited, due to the variable strong coupling constant $\alpha_s$, to the perturbative sector of QCD. This was followed up by Field and Feynman in 1977 \cite{field} for e+e collisions and by Feynman, Field and Fox in 1978 \cite{feynman2} for the production of hadrons at high $p_T$ in hadronic interactions. A crucial problem in all applications of parton dynamics is the fact that for the prediction of final state hadronic yields one is invariably confronted with non-perturbative QCD in the final stage of parton hadronization.

For soft hadronic interactions as they are studied in this paper there are basically two approaches. Either higher-order QCD is pushed to the limit ("next-to-next-to leading order") in order to make predictions at lower and lower $p_T$. Today it is believed that this is possible down to about 1~GeV/c. Or one assumes colour exchange as the source of hadronic interactions which leads to the breakup of both participating hadrons into quarks and, in the case of baryons, diquarks thus connecting the target and projectile systems by "strings" fragmenting into the final state secondaries.

This paper does not attempt to refer to any of the practically unlimited number of "string" models which are today available. These are rather characterized by a large number of adjustable parameters and very limited predictive power. If for the quark end of the strings experimental data from leptonic interactions may be invoked, the fictitious diquark systems which have by the way to ensure baryon number conservation have no referable analogue in different sectors of particle physics. It is a question why the baryon number should be contained in a partonic subsystem of the nucleon.

In addition, the point-like interactions assumed to describe the hadronic collisions do not carry any connection to the fact that hadrons are extended objects where the impact parameter is a decisive variable for any collision. This is apparent in the sector of peripheral interactions \cite{increase_rim1} ("diffraction") which is not really describable by string models.

In the following the energy dependence of longitudinal $\pi^-$ distributions will be shown.

%
%
\subsubsection{\boldmath $y_{\textrm{lab}}$ distributions}
\vspace{3mm}
\label{sec:scaling_ylab}

It should be remembered that $y_{\textrm{lab}}$ is equivalent to $x_F$ (Fig.~\ref{fig:xycor}) in the fragmentation zone for $\sqrt{s} \gtrsim$~10~GeV and $x_F \gtrsim$~0.2. The corresponding cross sections are readily available from the global interpolation, Sect.~\ref{sec:interpolation}. Both the invariant cross sections $f(y_{\textrm{lab}},p_T)$ and the yields per inelastic event, $f(y_{\textrm{lab}},p_T)/\sigma_{\textrm{inel}}$ will be shown.

\begin{figure}[h]
 	\begin{center}
    \includegraphics[width=16cm] {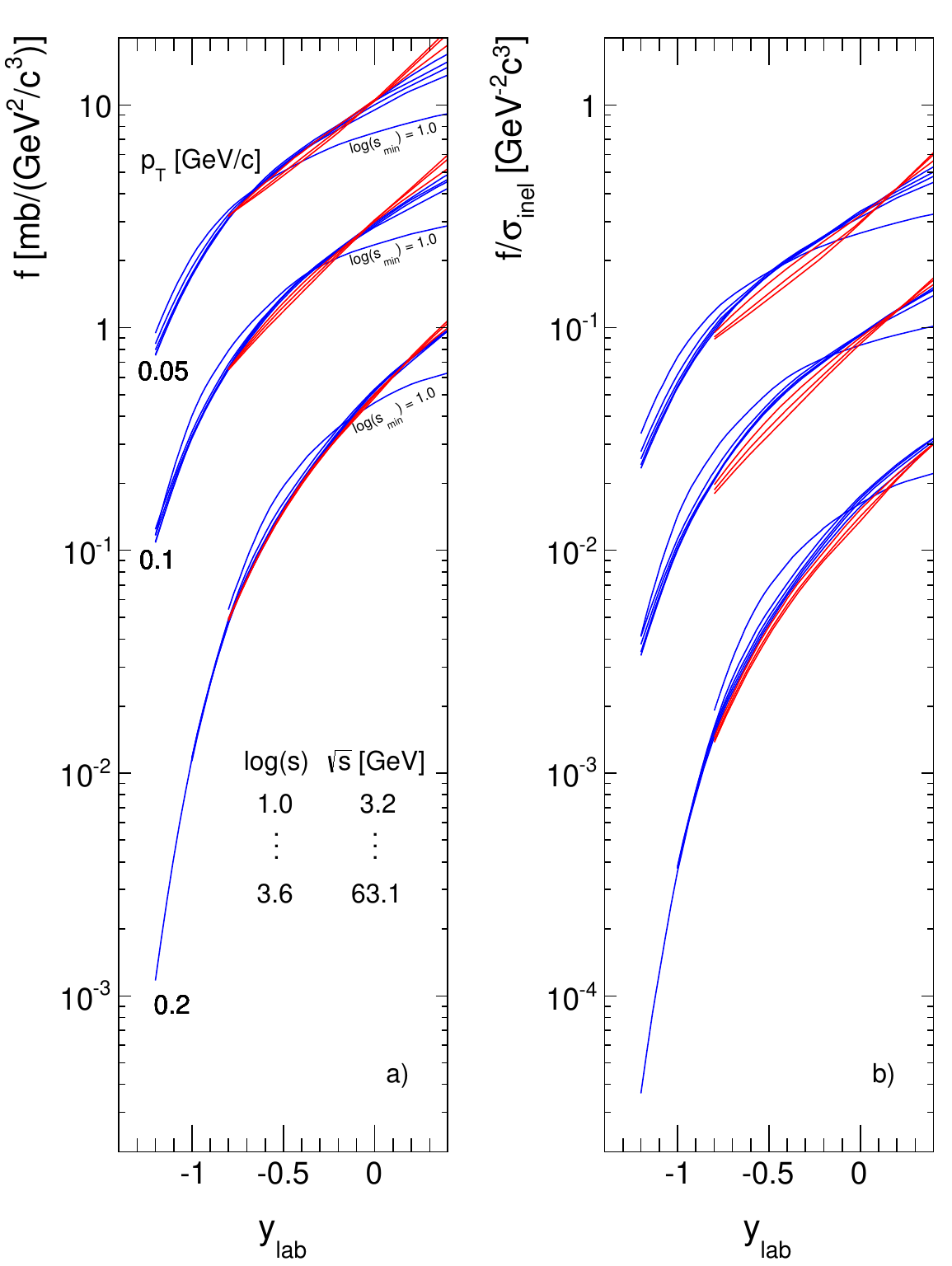} 
	 	\caption{$\pi^-$ cross sections as a function of $y_{\textrm{lab}}$ for $p_T$~=~0.05, 0.1, 0.2~GeV/c and different $\log(s)$ values in the range 1~$< \log{s} <$~3.6; a) $f(y_{\textrm{lab}},p_T)$, b) $f(y_{\textrm{lab}},p_T)/\sigma_{\textrm{inel}}$. The distributions for different $p_T$ values are successively scaled down by factor of 1/3 for better separation}
  	\label{fig:ylabdist}
 	\end{center}
\end{figure}

\begin{figure}[h]
 	\begin{center}
    \includegraphics[width=16cm] {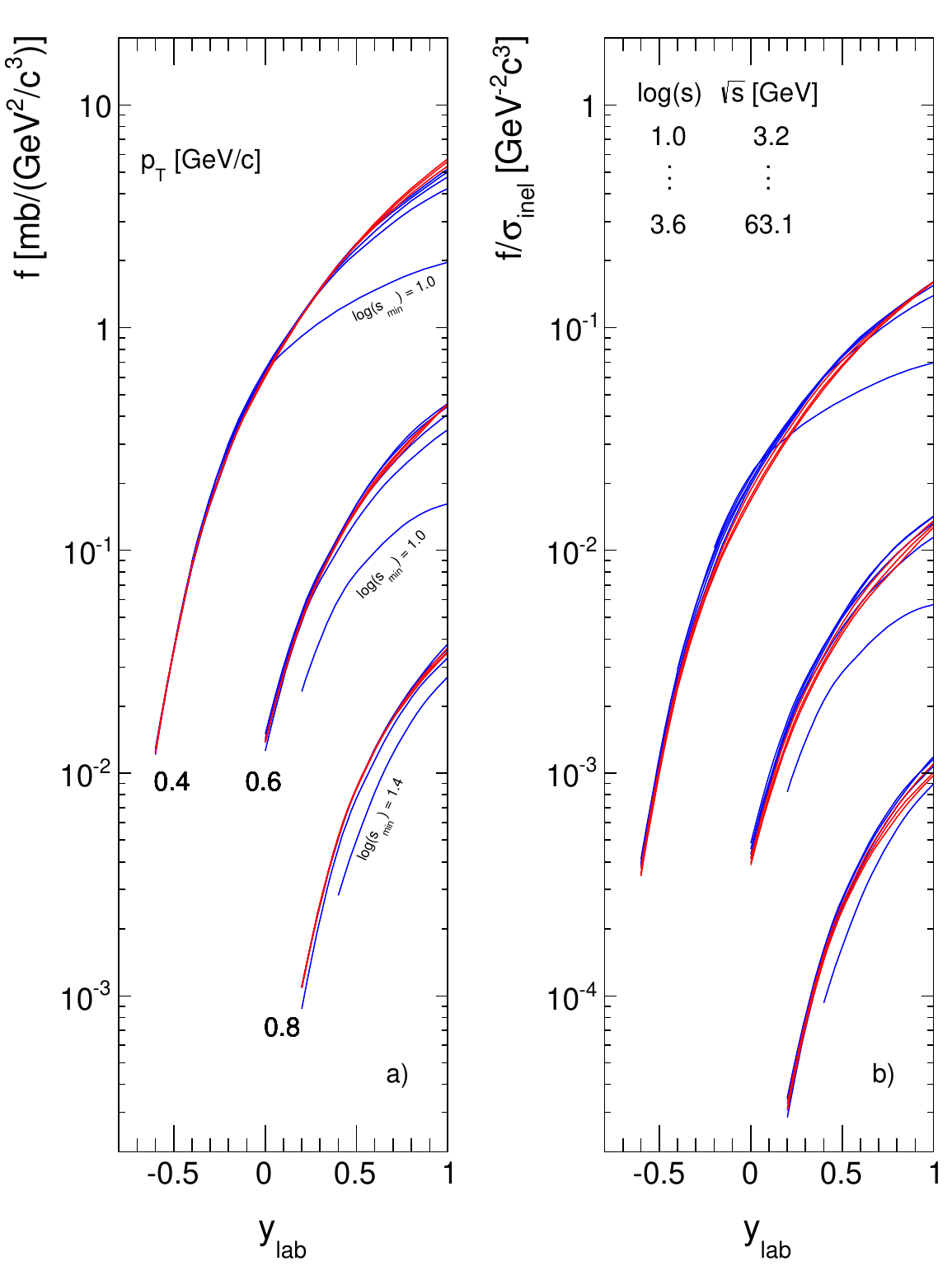} 
	 	\caption{$\pi^-$ cross sections as a function of $y_{\textrm{lab}}$ for $p_T$~=~0.4, 0.6, 0.8~GeV/c and different $\log(s)$ values in the range 1~$< \log{s} <$~3.6; a) $f(y_{\textrm{lab}},p_T)$, b) $f(y_{\textrm{lab}},p_T)/\sigma_{\textrm{inel}}$. The distributions for different $p_T$ values are successively scaled down by factor of 1/3 for better separation}
  	\label{fig:ylabdist1}
 	\end{center}
\end{figure}

\begin{figure}[h]
 	\begin{center}
    \includegraphics[width=16cm] {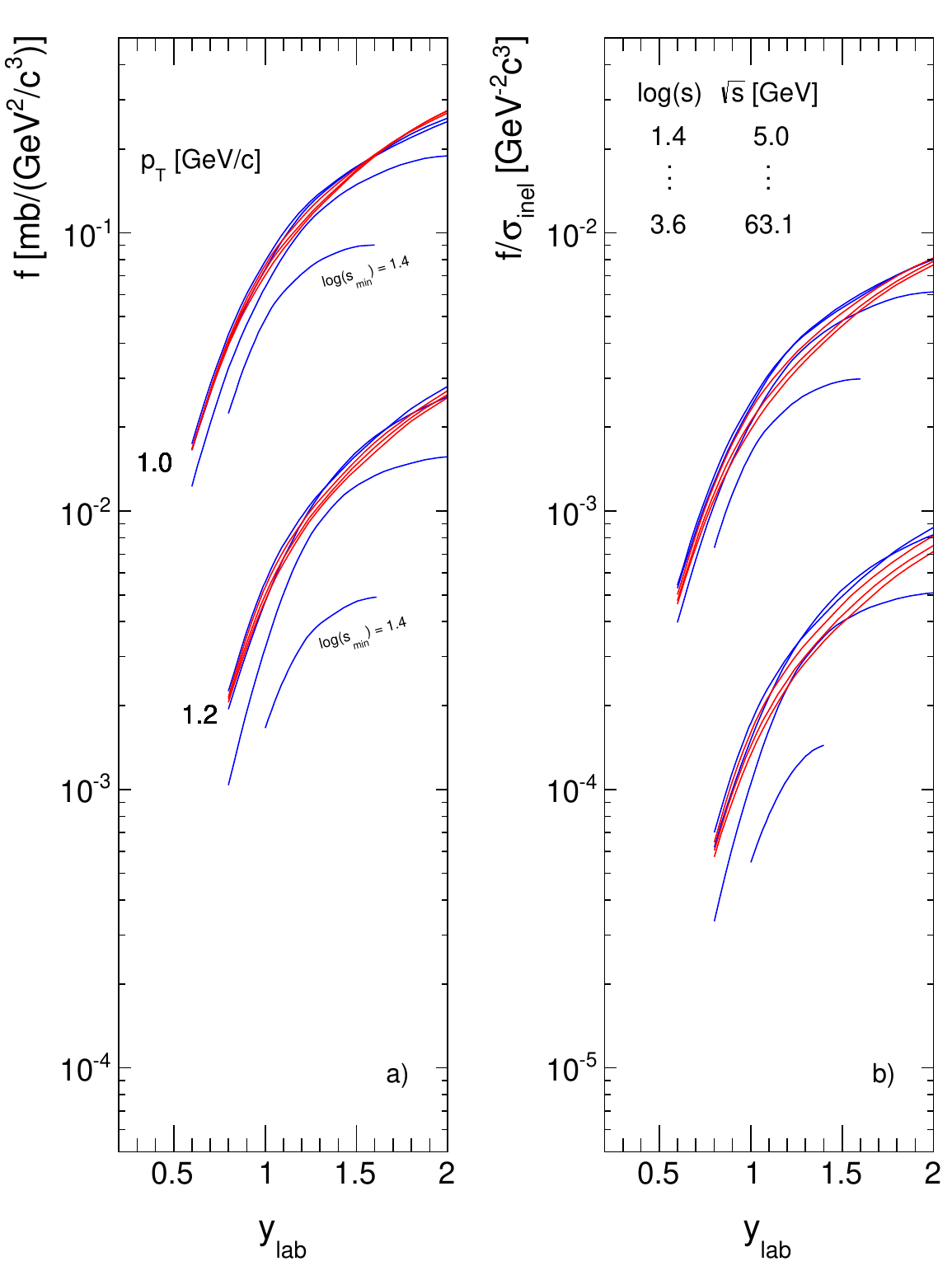} 
	 	\caption{$\pi^-$ cross sections as a function of $y_{\textrm{lab}}$ for $p_T$~=~1.0, 1.2~GeV/c and different $\log(s)$ values in the range 1~$< \log{s} <$~3.6; a) $f(y_{\textrm{lab}},p_T)$, b) $f(y_{\textrm{lab}},p_T)/\sigma_{\textrm{inel}}$. The distributions for different $p_T$ values are successively scaled down by factor of 1/3 for better separation}
  	\label{fig:ylabdist2}
 	\end{center}
\end{figure}

\begin{figure}[h]
 	\begin{center}
    \includegraphics[width=16cm] {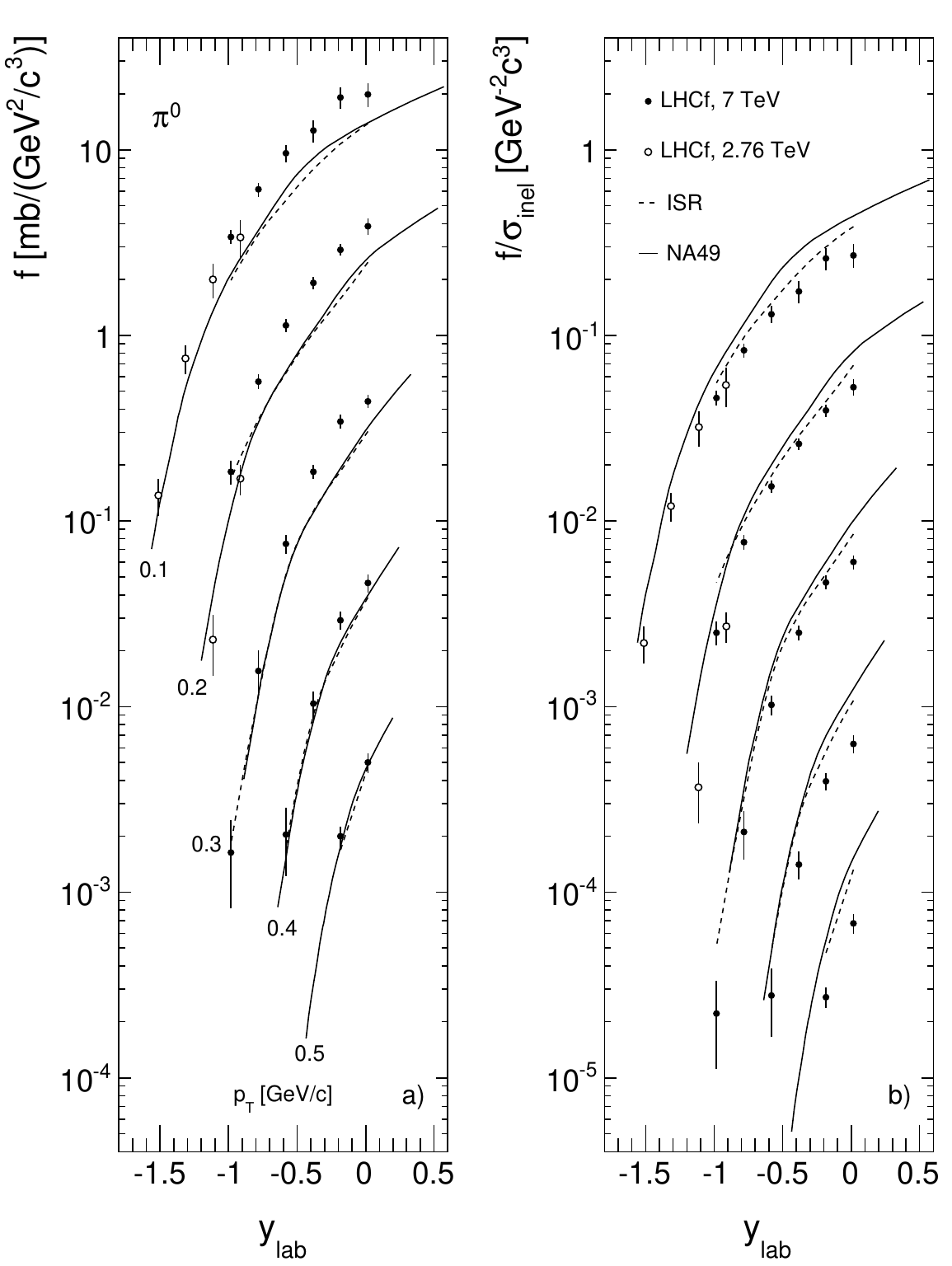} 
	 	\caption{$\pi^0$ cross sections as a function of $p_L^{\textrm{lab}}$ for $p_T$~=~0.1, 0.2, 0.3, 0.4, 0.5~GeV/c for NA49 and ISR (predictions) and LHC energies; a) $f(y_{\textrm{lab}},p_T)$, b) $f(y_{\textrm{lab}},p_T)/\sigma_{\textrm{inel}}$. The distributions for different $p_T$ values are successively scaled down by factor of 1/3 for better separation}
  	\label{fig:lhcfylab}
 	\end{center}
\end{figure}

There is rather precise scaling over a wide range of $y_{\textrm{lab}}$ and $p_T$ values. In the higher $p_T$ range there are of course deviations which are just reflecting energy-momentum conservation at the lower interaction energies. And it is again plotting the invariant cross sections $f(y_{\textrm{lab}},p_T)$ rather than the yield per inelastic event which gives narrower $s$-dependences. Extending the $s$ range up to LHC energies the $\pi^0$ distributions (Sect.~\ref{sec:lhcf}) show corresponding performance for the invariant cross sections as presented in Figs.~\ref{fig:ylabdist} -- \ref{fig:ylabdist2} as a function of $y_{\textrm{lab}}$ both for $f(y_{\textrm{lab}},p_T)$ and for $f(y_{\textrm{lab}},p_T)/\sigma_{\textrm{inel}}$. The lowest $\log(s)$ value in each plot is given on the lowest-lying curve. Several points are noteworthy in this context:

\begin{enumerate}[label=(\roman*)]
  	\item Again as in Sect.~\ref{sec:scaling} the invariant cross sections $f(y_{\textrm{lab}},p_T)$ show a more precise scaling behaviour than the yields per inelastic event $f(y_{\textrm{lab}},p_T)/\sigma_{\textrm{inel}}$.
	\item For medium $p_T$ values the spread is only on the few percent level.
	\item The distributions split up, depending on $p_T$, at different $y_{\textrm{lab}}$ values in the approach to the non-scaling central region (Sect.~\ref{sec:non-scaling}).
  	\item With increasing $p_T$ the distributions at the lowest $log{s}$ values indicated in the Figures move down. This is an effect of energy-momentum conservation.
\end{enumerate}

The corresponding distributions for the $\pi^0$ data from LHCf \cite{adriani2} are shown in comparison with the $\pi^0$ predictions from NA49 and ISR(63~GeV) in Fig.~\ref{fig:lhcfylab} for the invariant cross sections $f(y_{\textrm{lab}},p_T)$ and $f(y_{\textrm{lab}},p_T)/\sigma_{\textrm{inel}}$. The approach to scaling over this very extended $\sqrt{s}$ range with increasing $p_T$ and decreasing $y_{\textrm{lab}}$ (Figs.~\ref{fig:isrlhcfrat} and \ref{fig:isrlhcf}) is clearly evident.

%
%
\subsubsection{\boldmath $x_F$ and $x_F'$ distributions}
\vspace{3mm}
\label{sec:scaling_xf}

The global interpolation in ($y_{\textrm{lab}}$, $p_T$) (Sect.~\ref{sec:data_interp}) has been interpolated to constant $x_F$ values for a grid of 24 $x_F$ and the standard 26 $p_T$ values. As the distributions $f(x_F,p_T)/\sigma_{\textrm{inel}}$ are rather steep at low $x_F$ the following bins have been used:

\begin{table}[h]
\begin{tabular}{ccccccccccccc}
$x_F$ = &0,   & 0.01,& 0.02,& 0.03,& 0.04,& 0.05,& 0.075,& 0.1,& 0.125,& 0.15,& 0.2,& 0.25, \\
        & 0.3, & 0.35,& 0.4, & 0.45,& 0.5, & 0.55, & 0.6,& 0.65,& 0.7,& 0.75,& 0.8,& 0.85  \\
\end{tabular}
\nonumber
\end{table}

\begin{figure}[h]
 	\begin{center}
    \includegraphics[width=16cm] {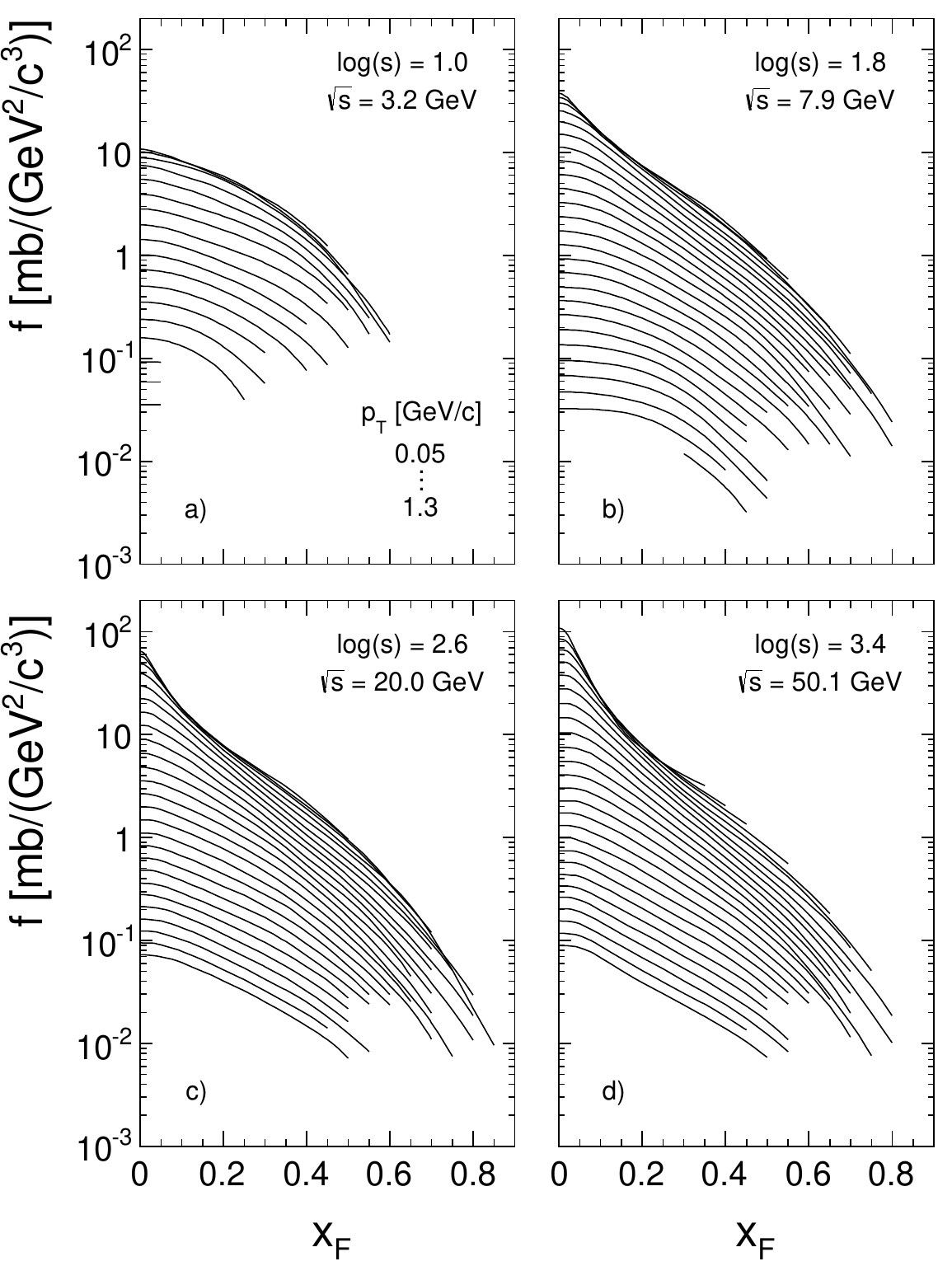} 
	 	\caption{$f(x_F,p_T)$ as a function of $x_F$ for the four interaction energies $\log{s)}$~=~1, 1.8, 2.6, 3.4 or $\sqrt{s}$~=~3.2, 7.9, 20.0 and 50.1~GeV. The full lines give the global interpolation for $p_T$ values between 0.05 and 1.3~GeV/c}
  	\label{fig:xfdist}
 	\end{center}
\end{figure}

An overall impression of the $x_F$ distributions is given by Fig.~\ref{fig:xfdist} for four $\log(s)$ values where the $x_F$ definition, see (\ref{eq:xf_def}):

\begin{equation}
	\label{eq:xf1}
  x_F = \frac{2p_L}{\sqrt{s}}
\end{equation}

\noindent
is used. As $x_F$ is defined in the region 0~$< x_F <$~1, this allows a unified overview over the complete phase space at each energy.

Three distinct zones are distinguishable:

\begin{enumerate}[label=(\roman*)]
	\item The central area at $x_F$~=~0 where an increase of the cross section by one order of magnitude over the range 3.6~$< \sqrt{s} <$~50~GeV is visible.
	\item The forward or "fragmentation" area at $x_F \gtrsim$~0.2 where the cross section is about $s$-invariant.
	\item An intermediate area at 0.05~$< x_F <$~0.2.
\end{enumerate}

 The central and intermediate zones will be treated in the subsequent Sect.~\ref{sec:non-scaling}. Several features of the distributions shown in Fig.~\ref{fig:xfdist} are noteworthy:

\begin{enumerate}[label=(\alph*)]
	\item The cross sections at $p_T <$~0.2~GeV/c are hardly distinguishable in the intermediate and forward regions.
	\item At the lowest value at $\log(s)$~=~1.0 there are no data for $p_T >$~0.75~GeV/c.
	\item At this energy the distributions are shrinking in their $x_F$ range with increasing $p_T$.
	\item For $p_T \gtrsim$~1~GeV/c there is a clear change in the shape of the $p_T$ distributions.
\end{enumerate}

\begin{figure}[h]
 	\begin{center}
    \includegraphics[width=16cm] {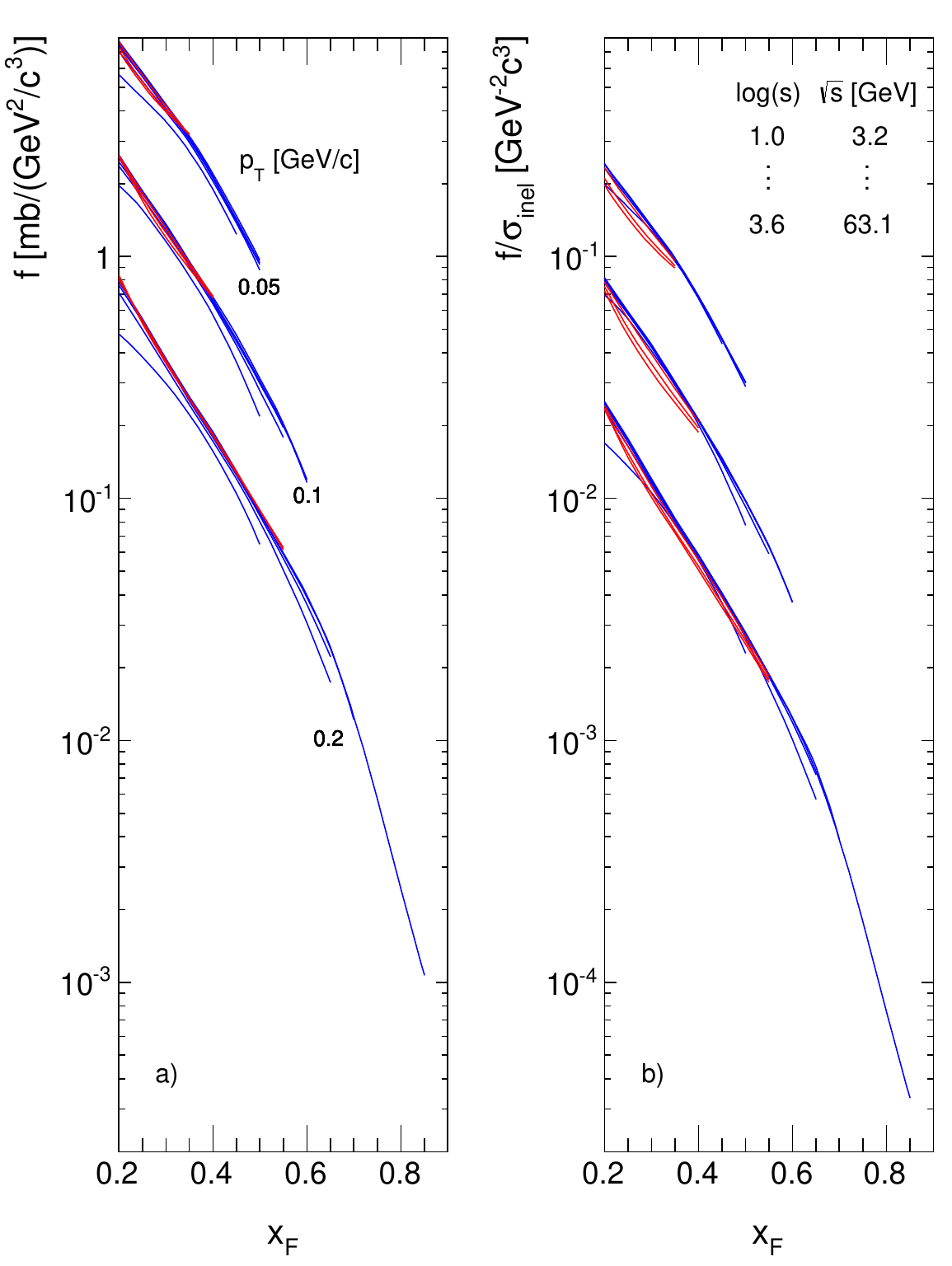} 
	 	\caption{a) $f(x_F,p_T)$ as a function of $x_F$ for $p_T$~=~0.05, 0.1 and 0.2~GeV/c for the global interpolation at $\log(s)$ values from 1.0 to 3.6. The lines at subsequent $p_T$ values are scaled down by 1/3 for better separation. b) $f(x_F,p_T)/\sigma_{\textrm{inel}}$}
  	\label{fig:xf1}
 	\end{center}
\end{figure}

\begin{figure}[h]
 	\begin{center}
    \includegraphics[width=16cm] {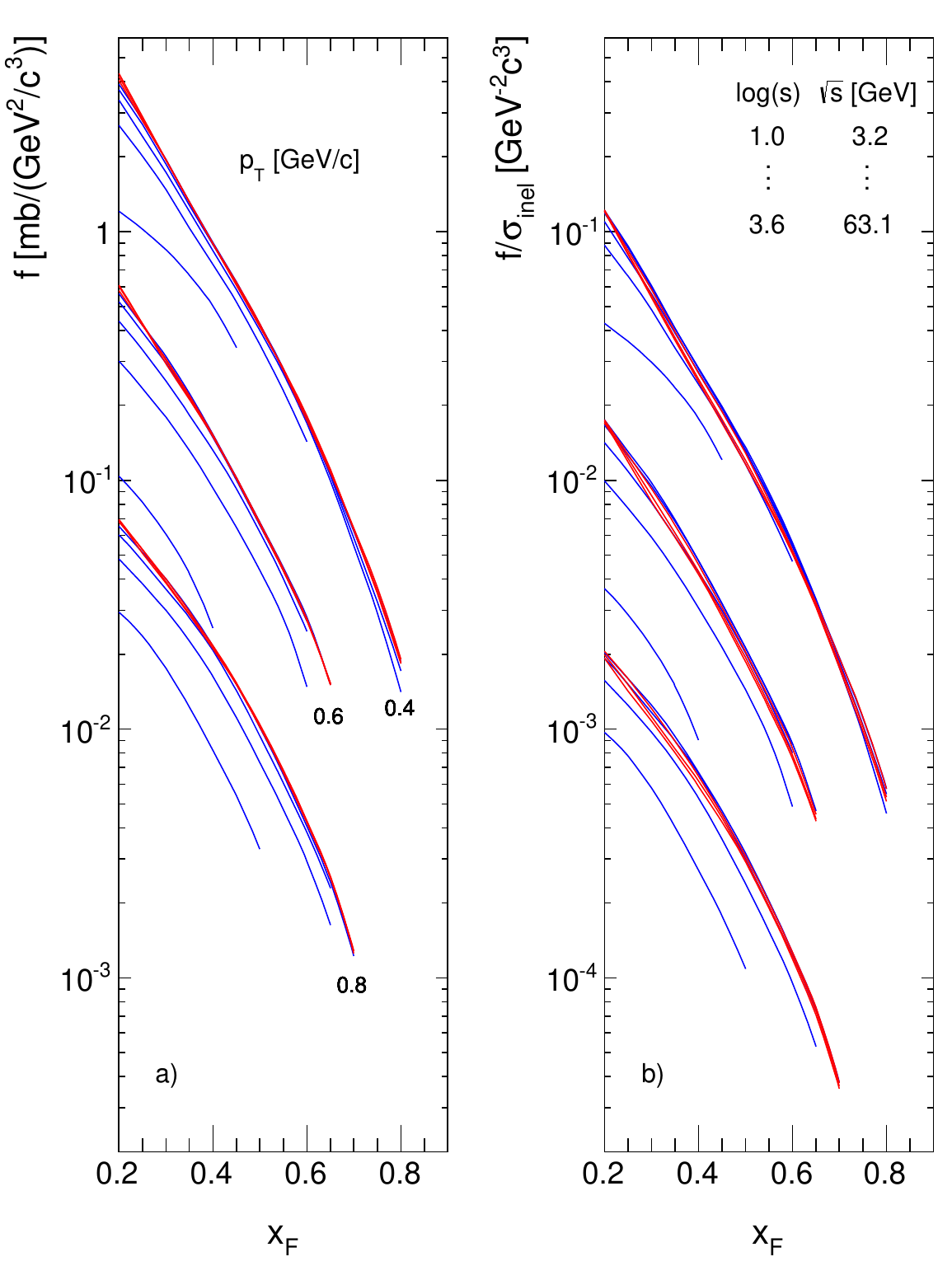} 
	 	\caption{a) $f(x_F,p_T)$ as a function of $x_F$ for $p_T$~=~0.4, 0.6 and 0.8~GeV/c for the global interpolation at $\log(s)$ values from 1.0 to 3.6. The lines at subsequent $p_T$ values are scaled down by 1/3 for better separation. b) $f(x_F,p_T)/\sigma_{\textrm{inel}}$}
  	\label{fig:xf2}
 	\end{center}
\end{figure}

\begin{figure}[h]
 	\begin{center}
    \includegraphics[width=16cm] {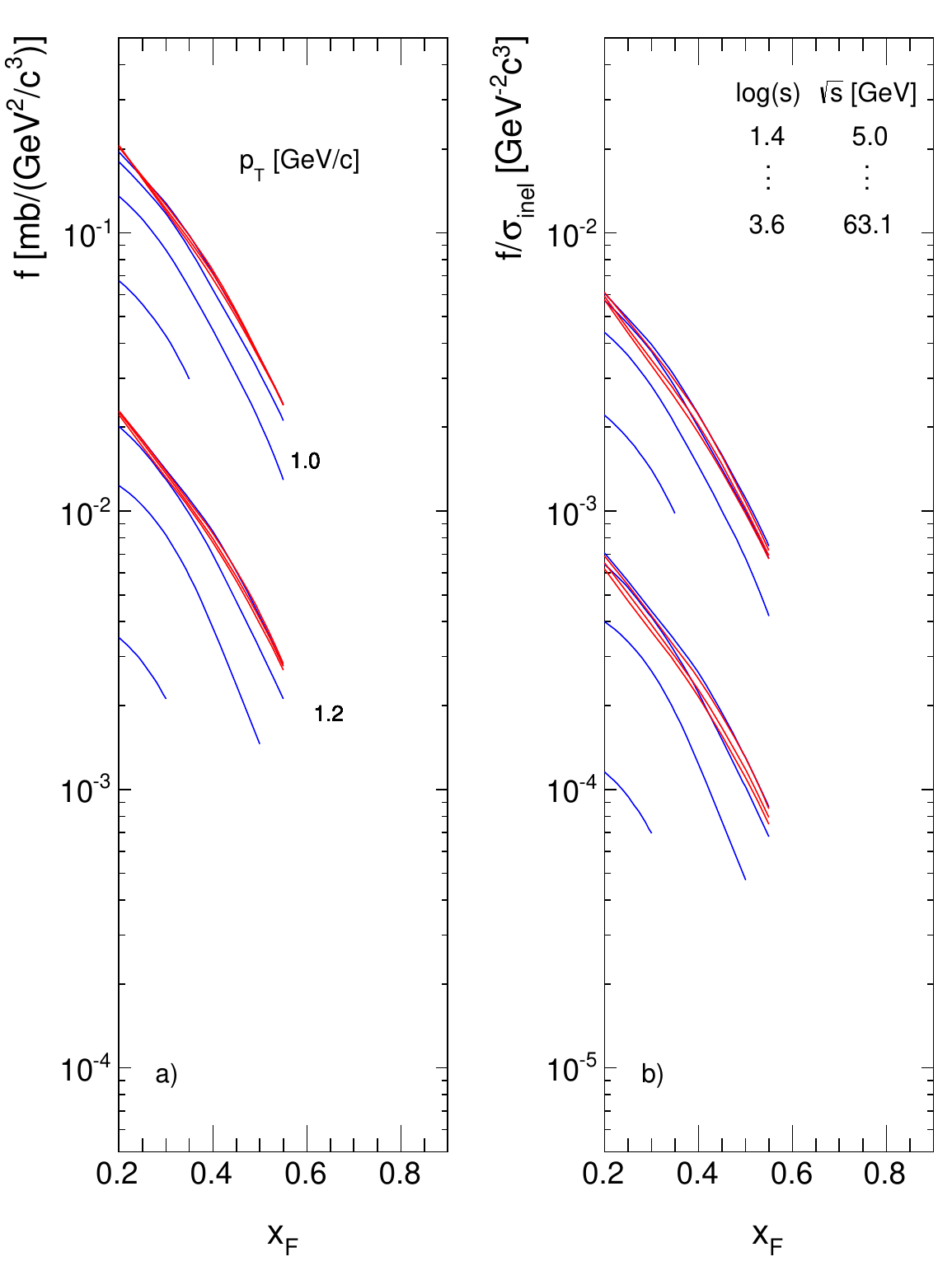} 
	 	\caption{a) $f(x_F,p_T)$ as a function of $x_F$ for $p_T$~=~1.0 and 1.2~GeV/c for the global interpolation at $\log(s)$ values from 1.0 to 3.6. The lines at subsequent $p_T$ values are scaled down by 1/3 for better separation. b) $f(x_F,p_T)/\sigma_{\textrm{inel}}$}
  	\label{fig:xf3}
 	\end{center}
\end{figure}

A more detailed view of this situation is presented in Figs.~\ref{fig:xf1} to \ref{fig:xf3} where the invariant cross sections $f(x_F,p_T)$ and $f(x_F,p_T)/\sigma_{\textrm{inel}}$ are shown as a function of $x_F$ for seven values of $\log(s)$ between 1 and 3.4. For the higher range of $\log(s)$ there is again a smaller spread of the yields for $f(x_F,p_T)$ than for $f(x_F,p_T)/\sigma_{\textrm{inel}}$. However, for $\sqrt{s}$ below about 12~GeV ($\log(s)\lesssim$~1.8) there is an explicit non-scaling effect that reduces the yields especially at higher $p_T$ values. In addition, the $x_F$ range in the low-$s$ range is confined to successively smaller $x_F$ ranges with increasing $p_T$.

This has to do with the definition of $x_F$ (\ref{eq:xf1}) being in principle only valid in an infinite momentum frame -- as is the definition of $x_{Bj}$ for partons. For finite energies, the maximum final state momentum $p_L{\textrm{max}}$ is $s$-dependent and unequal to $\sqrt{s}$/2 as energy-momentum, charge and baryon number conservation have to be taken into account. This leads, for final state protons, to the value

\begin{equation}
	p_L^{\textrm{max}} (p_T=0) = \frac{1}{2} \sqrt{s-4m_p^2} \: \neq \: \frac{1}{2} \sqrt{s}
\end{equation}

\begin{figure}[h]
 	\begin{center}
    \includegraphics[width=14cm] {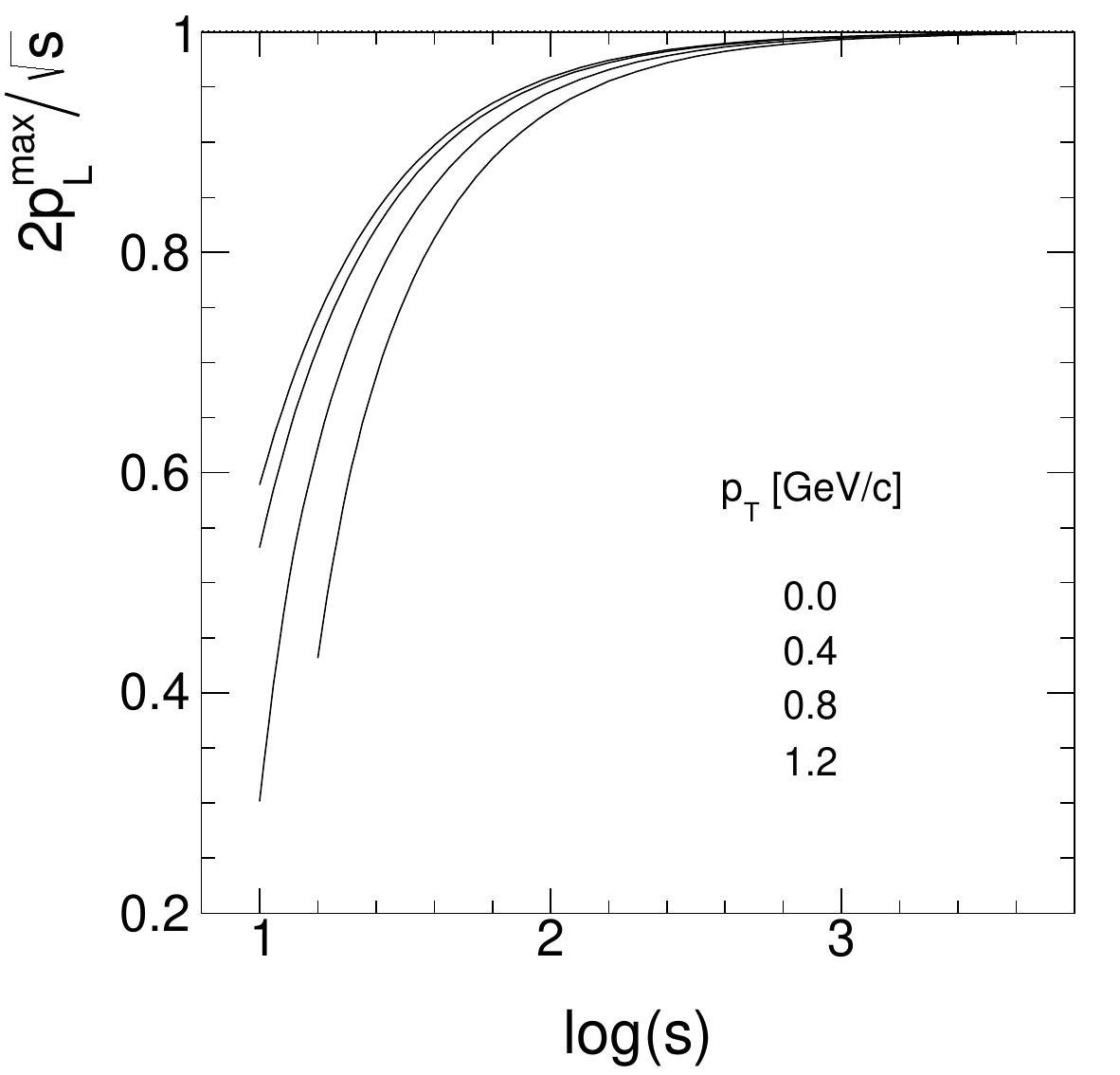} 
	 	\caption{The ratio $x_F'$/$x_F$ as a function of $\log(s)$ for different $p_T$ values}
  	\label{fig:xfprime2xf}
 	\end{center}
\end{figure}

For $\pi^-$ production, there have to be, in addition to the $\pi^-$ mass, 2 nucleons for baryon number conservation and at least one $\pi^+$ for charge conservation in the final state,

\begin{equation}
	p_L^{\textrm{max}} (p_T=0) = \frac{1}{2\sqrt{s}} \sqrt{s-(2m_{\pi}+2m_p)^2} \;\sqrt{s-4m_p^2}
\end{equation}

\noindent
see \cite{allaby2} for further explanations. The definition of $x_F$ should  therefore be replaced by

\begin{equation}
	x_F' = \frac{p_L}{p_L^{\textrm{max}}}
\end{equation}

\noindent
see also (\ref{eq:xfprime}). This leads to sizeable deviations from definition (\ref{eq:xf1}) at $\log(s) <$~3 and subsequently to a dependence on $p_T$ as shown in Fig.~\ref{fig:xfprime2xf}.

This traces the corresponding deviations from the correlation between $x_F$ and $y_{\textrm{lab}}$ shown in Fig.~\ref{fig:xycor}. A tolerable correctable deviation is only reached for the SPS energy range $\sqrt{s} \gtrsim$~15~GeV. In fact the values $p_L^{\textrm{max}}(s,p_T)$ are corresponding to a scale $y_{\textrm{lab}}^{\textrm{min}}(s,p_T)$,

\begin{equation}
  y_{\textrm{lab}}^{\textrm{min}} = y_{\textrm{beam}} - \frac{1}{2}ln\frac{ E + p_L^{\textrm{max}} }{ E - p_L^{\textrm{max}} }
\end{equation}

\noindent
with

\begin{equation}
	E = \sqrt{ (p_L^{\textrm{max}})^2 + p_T^2 + m_\pi^2}
\end{equation}

\begin{figure}[h]
 	\begin{center}
    \includegraphics[width=11.5cm] {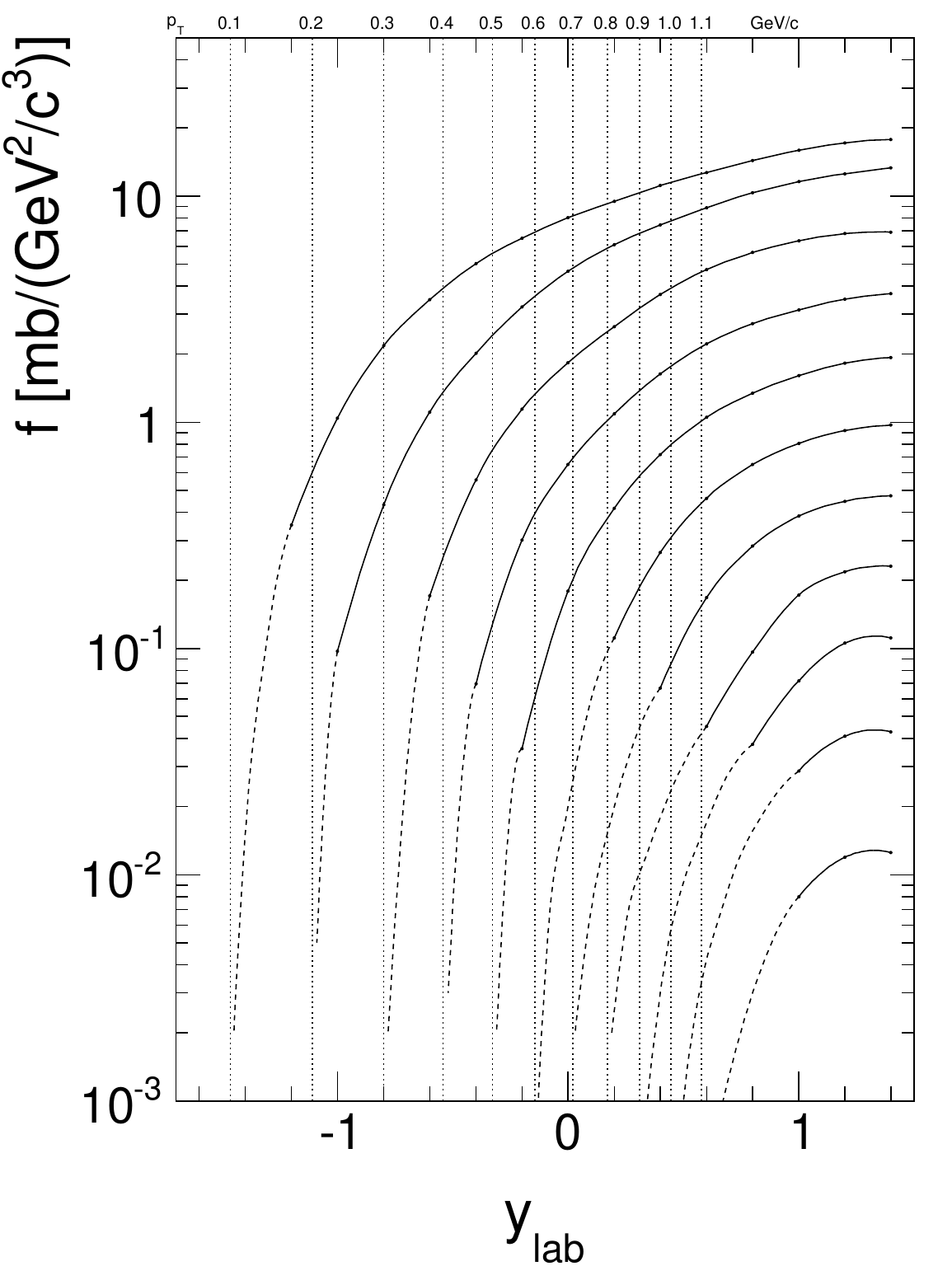} 
	 	\caption{$f(y_{\textrm{lab}},p_T)$ as a function of $y_{\textrm{lab}}$ for $p_T$ values between 0.1 and 1.1~GeV/c in steps of 0.1~GeV/c and $\log(s)$~=~1.2. For each $p_T$ the corresponding $y_{\textrm{lab}}^{\textrm{min}}$ is indicated as a vertical line}
  	\label{fig:kin_lim}
 	\end{center}
\end{figure}

This is quantified in Fig.~\ref{fig:kin_lim} where the invariant cross section $f(y_{\textrm{lab}},p_T)$ is presented as a function of $y_{\textrm{lab}}$ for the relatively low interaction energy $\sqrt{s}$~=~4.0~GeV or $\log(s)$~=~1.2. For each $p_T$ the corresponding $y_{\textrm{lab}}^{\textrm{min}}$ value is indicated as a vertical line on the abscissa.

The approach of the measured cross sections to the limiting $y_{\textrm{lab}}$ values as imposed by the conservation laws is clearly visible.

\begin{figure}[h]
 	\begin{center}
    \includegraphics[width=16cm] {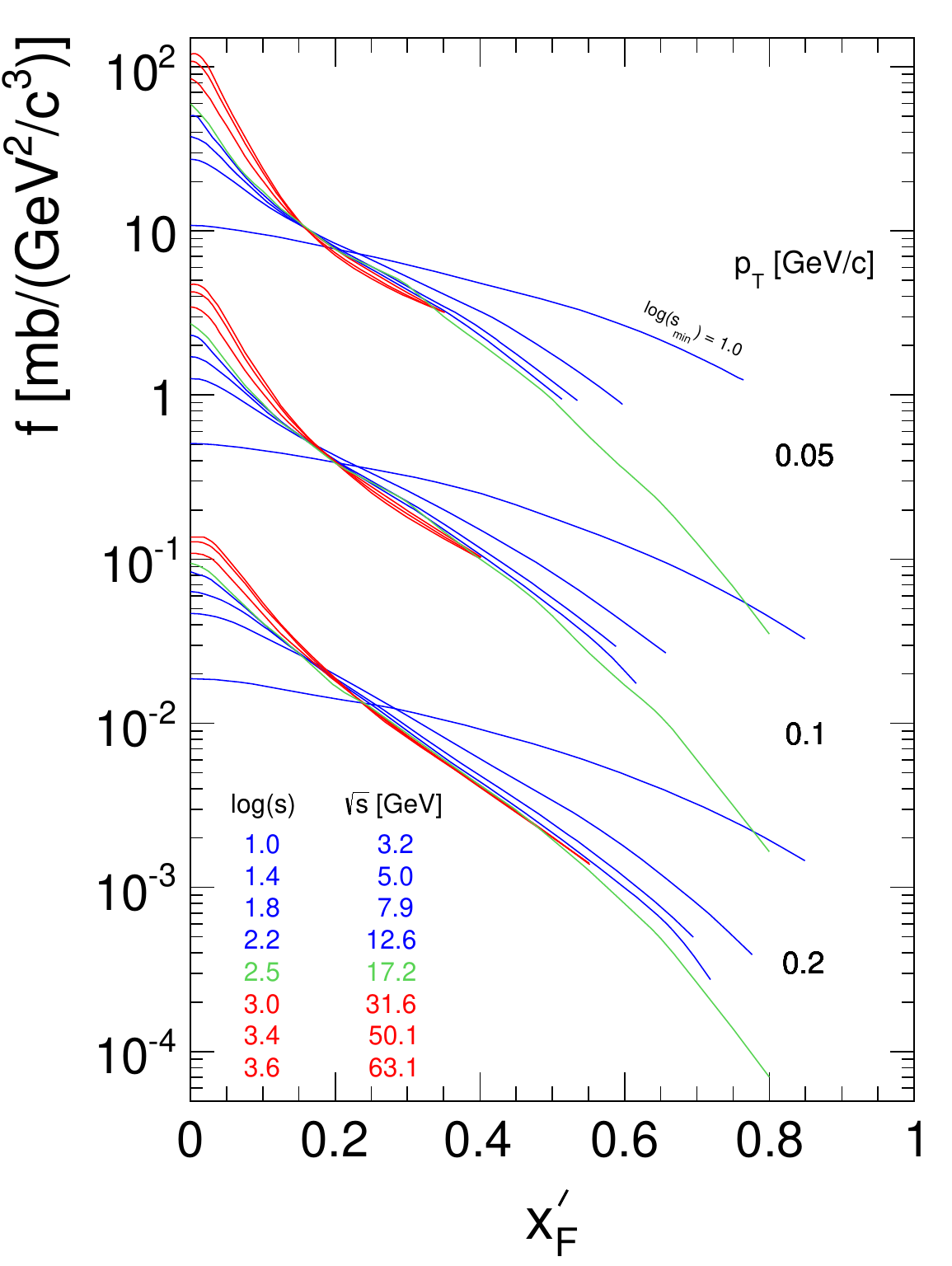} 
	 	\caption{$f(x_F',p_T)$ as a function of $x_F'$ for $p_T$~=~0.05, 0.1 and 0.2~GeV/c for the global interpolation at $\log{s}$ values from 1.0 to 3.6. The plots at subsequent $p_T$ values are scaled down by 1/20 for better separation. The NA49 interpolation at $\log(s)$~=~2.48 is shown as a green line}
  	\label{fig:xfprime1}
 	\end{center}
\end{figure}

\begin{figure}[h]
 	\begin{center}
    \includegraphics[width=16cm] {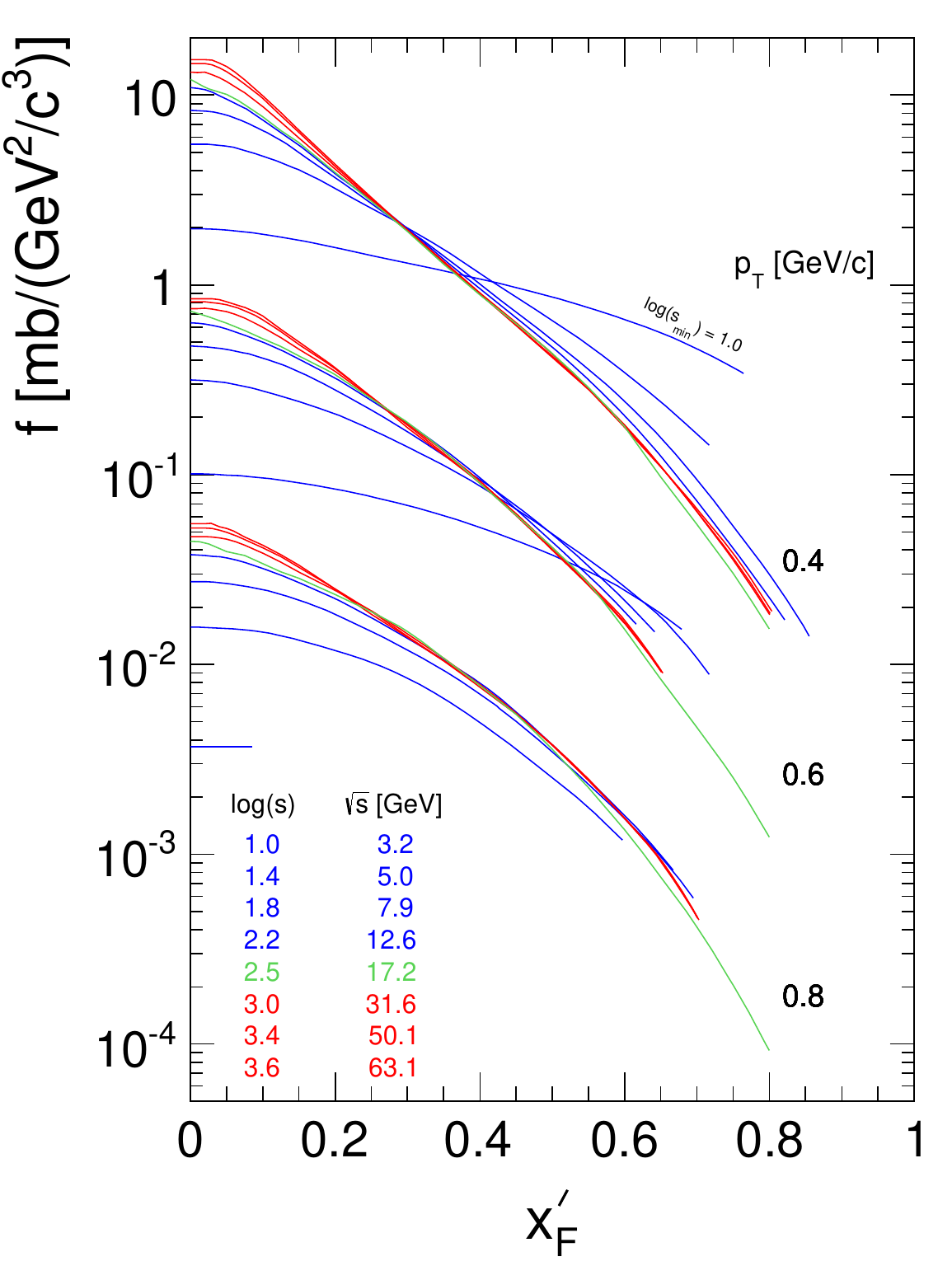} 
	 	\caption{$f(x_F',p_T)$ as a function of $x_F'$ for $p_T$~=~0.4, 0.6 and 0.8~GeV/c for the global interpolation at $\log(s)$ values from 1.0 to 3.6. The plots at subsequent $p_T$ values are scaled down by 1/5 for better separation. The NA49 interpolation at $\log(s)$~=~2.48 is shown as a green line}
  	\label{fig:xfprime2}
 	\end{center}
\end{figure}

\begin{figure}[h]
 	\begin{center}
    \includegraphics[width=16cm] {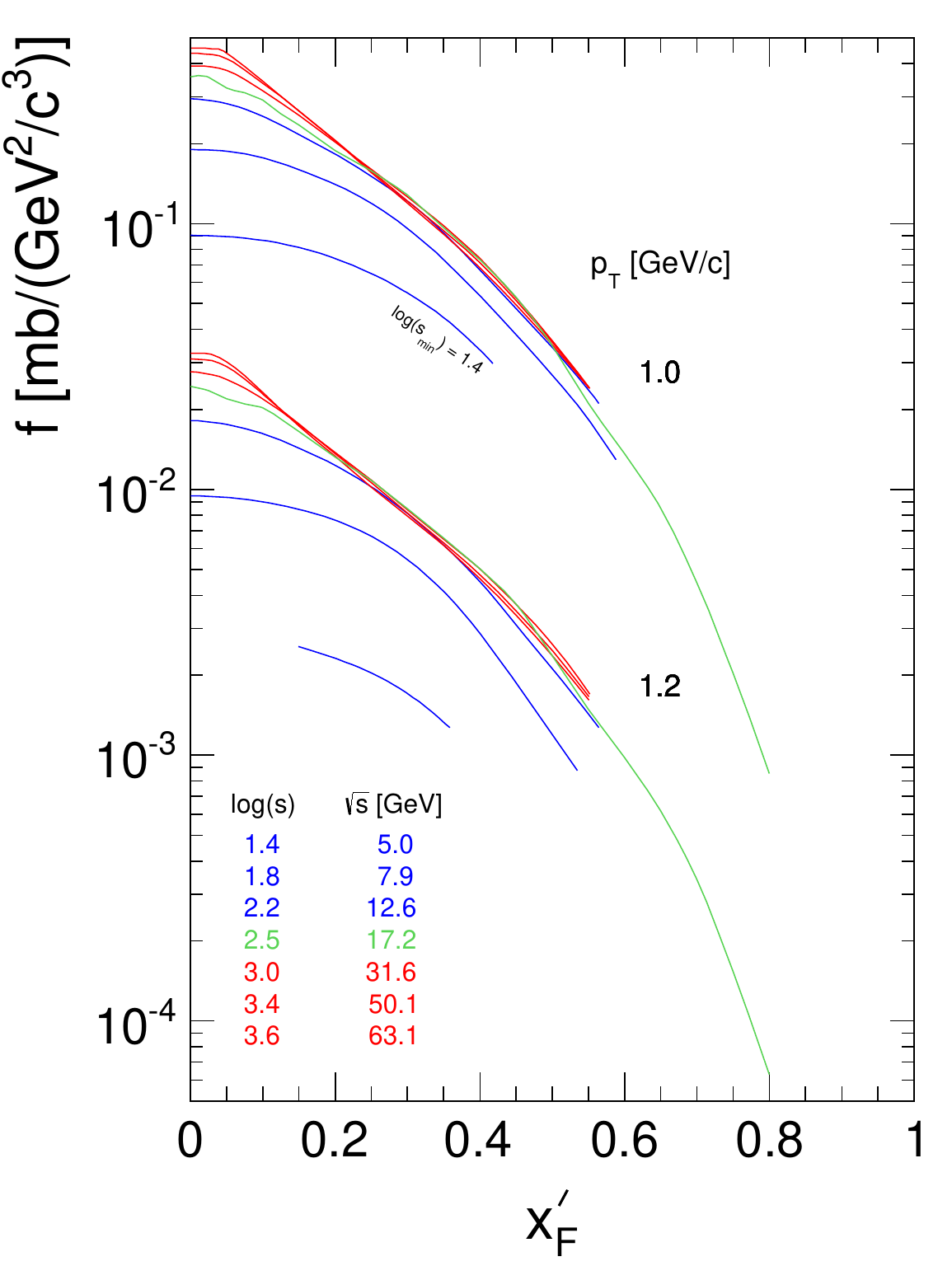} 
	 	\caption{$f(x_F',p_T)$ as a function of $x_F'$ for $p_T$~=~1.0 and 1.2~GeV/c for the global interpolation at $\log(s)$ values from 1.0 to 3.6. The plots at subsequent $p_T$ values are scaled down by 1/5 for better separation. The NA49 interpolation at $\log(s)$~=~2.48 is shown as a green line}
  	\label{fig:xfprime3}
 	\end{center}
\end{figure}

The invariant cross sections $f(x_F',p_T)$ are shown in Figs.~\ref{fig:xfprime1} to \ref{fig:xfprime3} as a function of $x_F'$ for 8 values of $p_T$.

\begin{figure}[h]
 	\begin{center}
   \includegraphics[width=13cm] {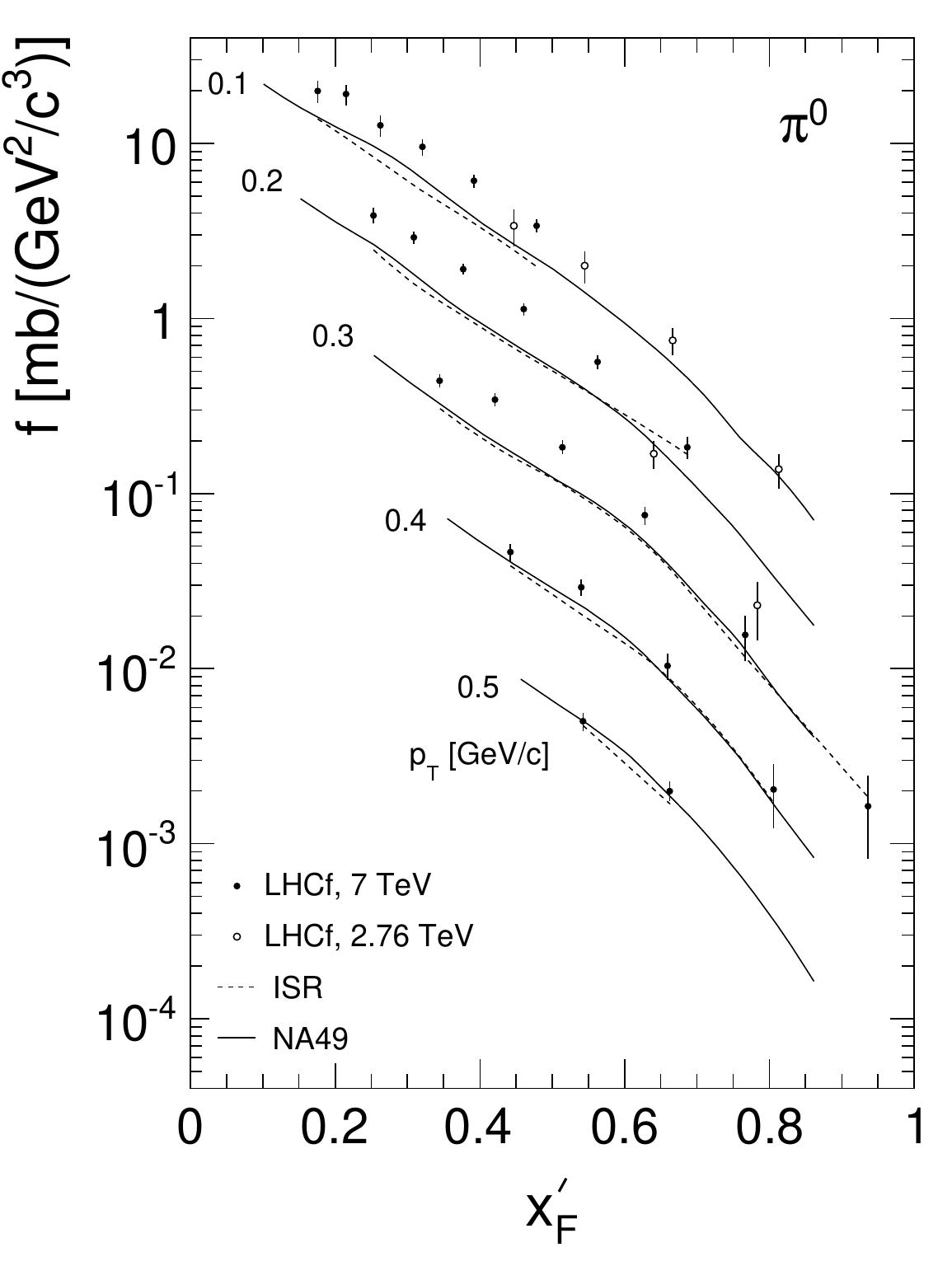} 
	 	\caption{Invariant $\pi^0$ cross sections $f(x_F',p_T)$ as a function of $x_F'$ for $p_T$ values between 0.1 and 0.5~GeV/c}
  	\label{fig:pi0xfprime}
 	\end{center}
\end{figure}

The $\pi^0$ data of LHCf \cite{adriani2} are presented in Fig.~\ref{fig:pi0xfprime} as a function of $x_F'$ for different $p_T$ values. As already evident from the $y_{\textrm{lab}}$ distributions (Fig.~\ref{fig:lhcfylab}), scaling in $x_F'$ is approached for the higher $x_F'$ and $p_T$ regions if plotting the invariant cross sections $f(x_F',p_T)$.

For the lower $x_F'$ and $p_T$ regions the data exceed the NA49 and ISR results by factors of up to 1.7 corresponding to about 73\% of the ratio of the inelastic cross sections. This indicates that the scaling data come from a constant "core" area of the increasing proton size with a transition region approaching a major fraction of the total interaction zone at lower $x_F'$ and $p_T$.

%
%
\subsection{The non-scaling central region}
\vspace{3mm}
\label{sec:non-scaling}

The $x_F'$ distributions of Figs.~\ref{fig:xfprime1} to \ref{fig:xfprime3} which are extended to the complete kinematical range clearly demonstrate a non-scaling region at low $x_F'$ with a sharply defined transition to approximate scaling located between $x_F \sim$~0.16 at low $p_T$ approaching $x_F' \sim$~0.3 in the higher $p_T$ range (always with the exception of the lowest $\log(s)$ values). This corresponds to a "rising rapidity plateau" \cite{cernisr1,cernisr2,cernisr3} and has led to the assumption of a special source of central particle production ("pionization") and in consequence to an undue concentration of interest in the central rapidity area in contrast to the forward/backward production regions. This is specially flagrant for Heavy Ion collisions where a central "hot" zone is postulated as a "Quark-Gluon Plasma".

\begin{figure}[h]
 	\begin{center}
    \includegraphics[width=11.4cm] {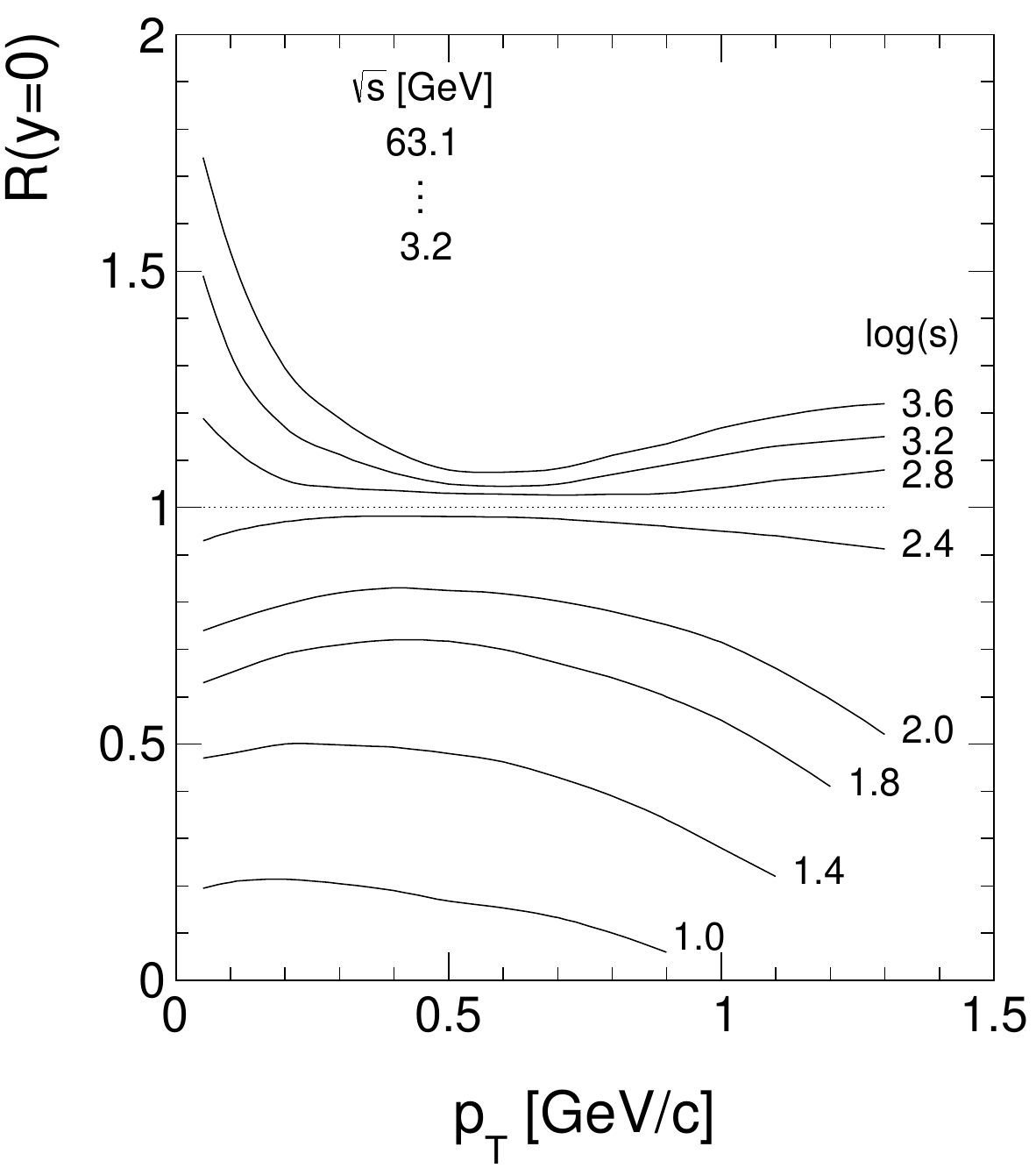} 
	 	\caption{$R(y=0,s,p_T)$ as a function of $p_T$ for the reference data, $\log(s)$ from 1.0 to 3.6}
  	\label{fig:rrefy0}
 	\end{center}
\end{figure}

\begin{figure}[h]
 	\begin{center}
    \includegraphics[width=15cm] {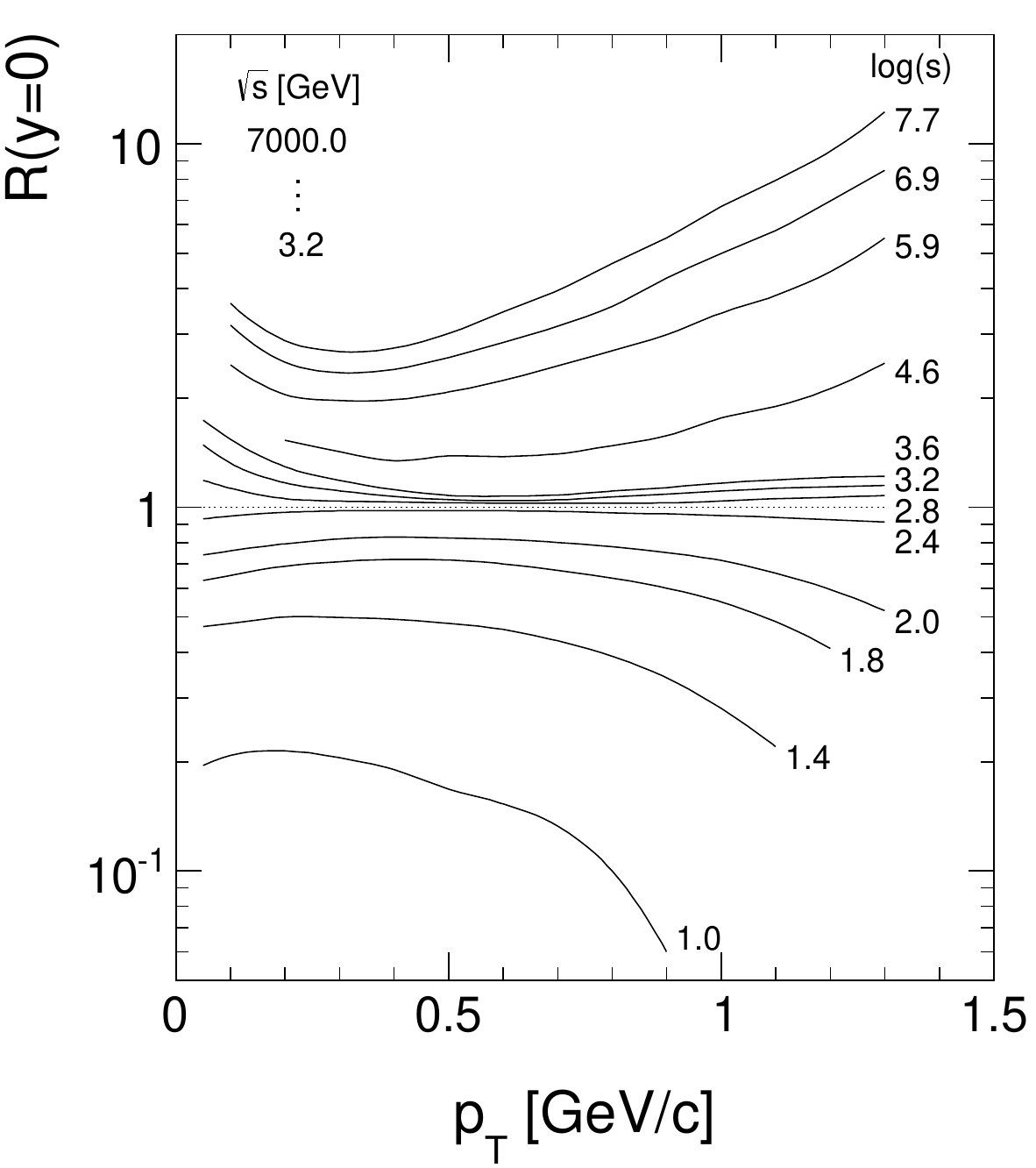} 
	 	\caption{$R(y=0,s,p_T)$ as a function of $p_T$ for $\sqrt{s}$ values from 3~GeV to 7~TeV}
  	\label{fig:rcoly0}
 	\end{center}
\end{figure}

A detailed inspection of the low-$x_F'$ region allows however for a more differentiated picture especially if the $p_T$ dependence is taken into account. This is demonstrated in Figs.~\ref{fig:rrefy0} and \ref{fig:rcoly0} where the evolution of the $s$-dependence with respect to the NA49 data is shown using the cross section ratio at $y$~=~0:

\begin{equation}
	R(y=0,s,p_T) = \left(\frac{f}{\sigma_{\textrm{inel}}}\right)(s,p_T)\Bigg/\left(\frac{f}{\sigma_{\textrm{inel}}}\right)(NA49,p_T)
	\label{eq:rat_y0}
\end{equation}

$R(y=0,s,p_T)$ is plotted as a function of $p_T$ for the $\log(s)$ range of the reference data in Fig.~\ref{fig:rrefy0}.

Two main contributions to the evolution with increasing $s$ are clearly evident;

\begin{enumerate}[label=(\alph*)]
	\item An enhancement at low $p_T$ in the range $p_T <$~0.3~GeV/c.
	\item A steady increase of yields up to the limit of $p_T$ at 1.3~GeV/c.
\end{enumerate}

These trends continue up to LHC energies, Fig.~\ref{fig:rcoly0}, with a strong evolution of the higher $p_T$ region which exceeds the low $p_T$ enhancement above RHIC energies (notice the log scale).

\begin{figure}[h]
 	\begin{center}
    \includegraphics[width=8.8cm] {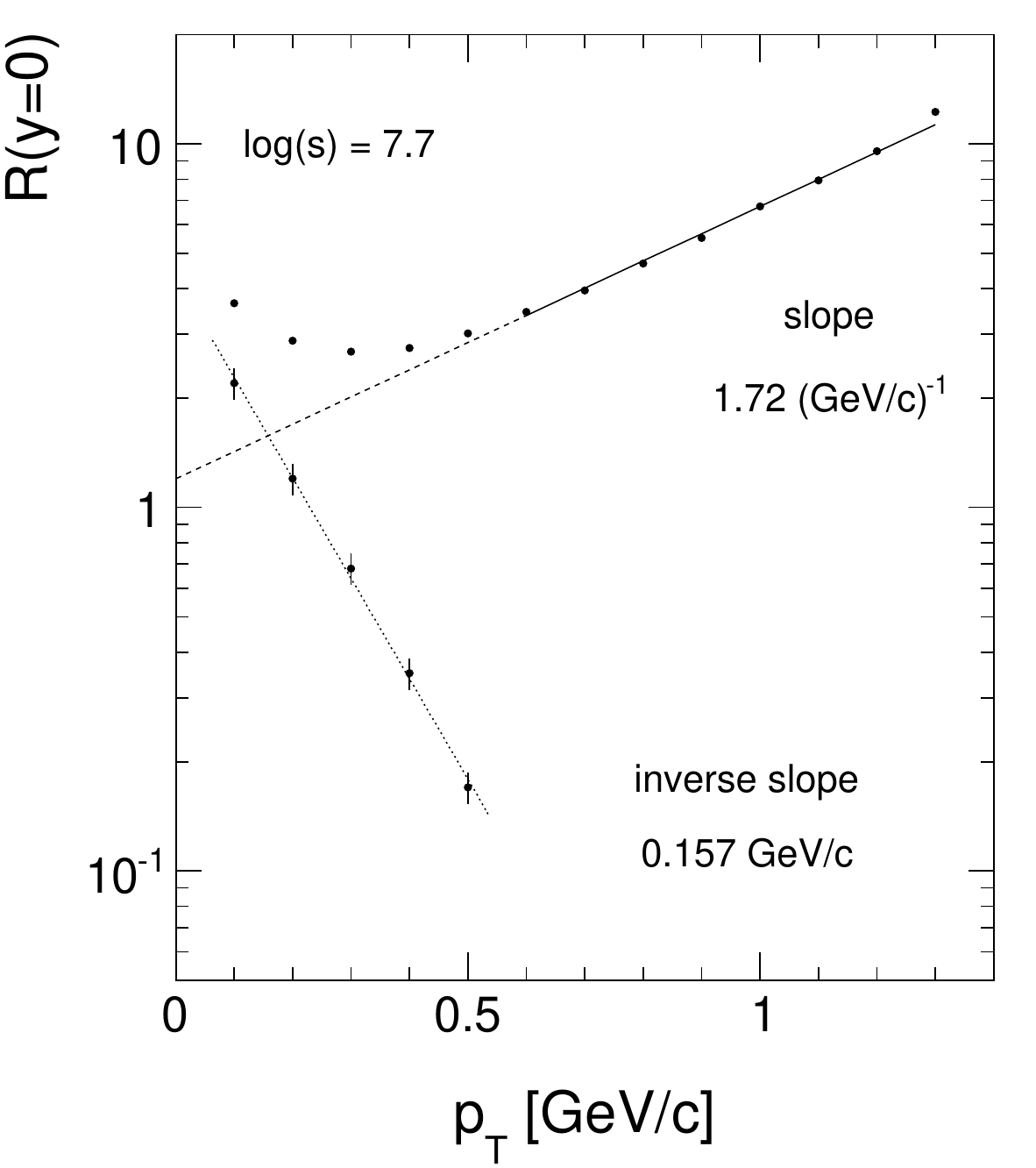} 
	 	\caption{$R(y=0,s,p_T)$ as a function of $p_T$ for $\log(s)$~=~7.7 with exponential fit to $p_T >$~0.6~GeV/c, extrapolation to low $p_T$ (broken line) and subtracted ratio (dotted line)}
  	\label{fig:r7.7}
 	\end{center}
\end{figure}

\begin{figure}[h]
 	\begin{center}
    \includegraphics[width=8.8cm] {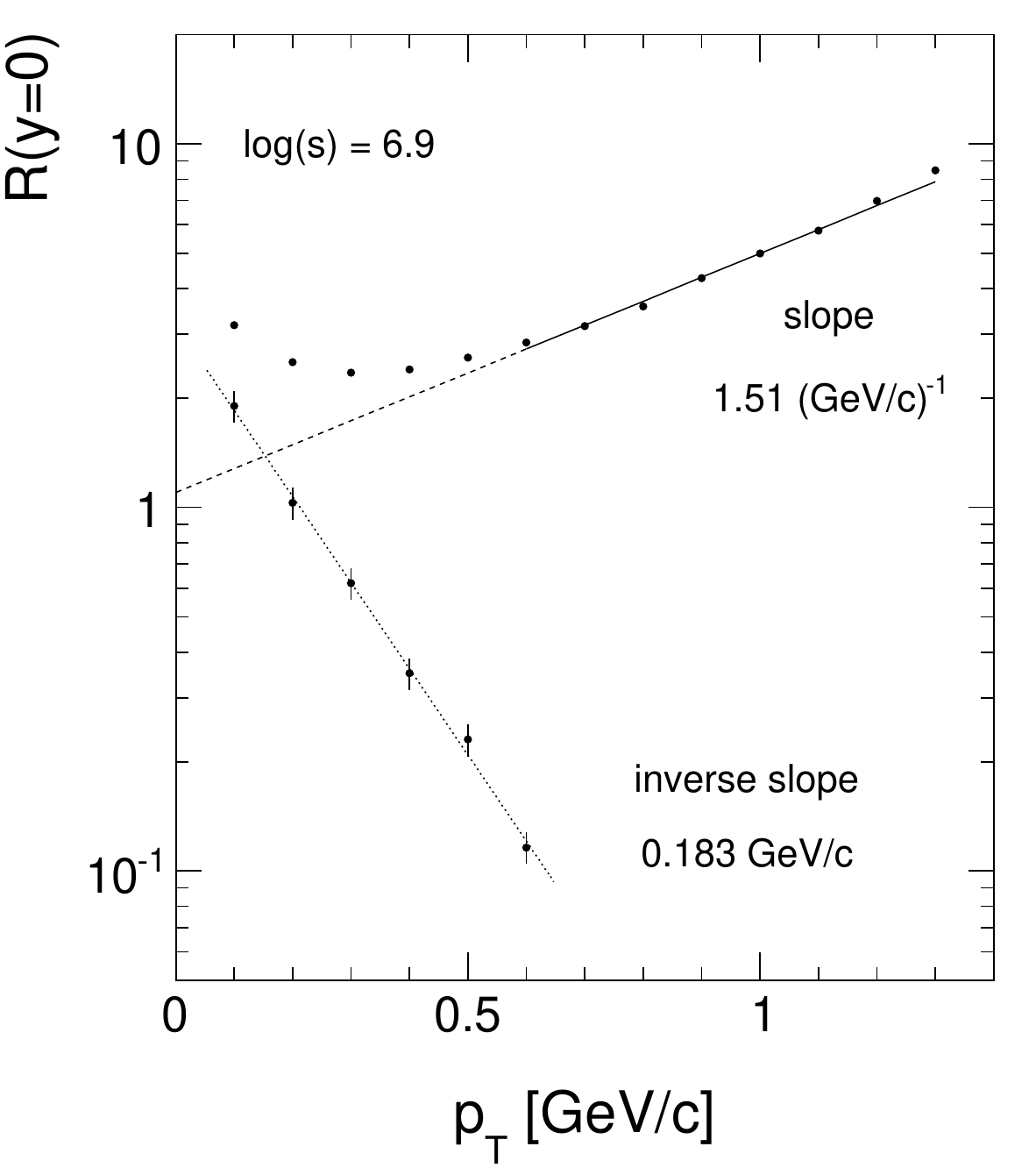} 
	 	\caption{$R(y=0,s,p_T)$ as a function of $p_T$ for $\log(s)$~=~6.9 with exponential fit to $p_T >$~0.6~GeV/c, extrapolation to low $p_T$ (broken line) and subtracted ratio (dotted line)}
  	\label{fig:r6.9}
 	\end{center}
\end{figure}

\begin{figure}[h]
 	\begin{center}
    \includegraphics[width=8.8cm] {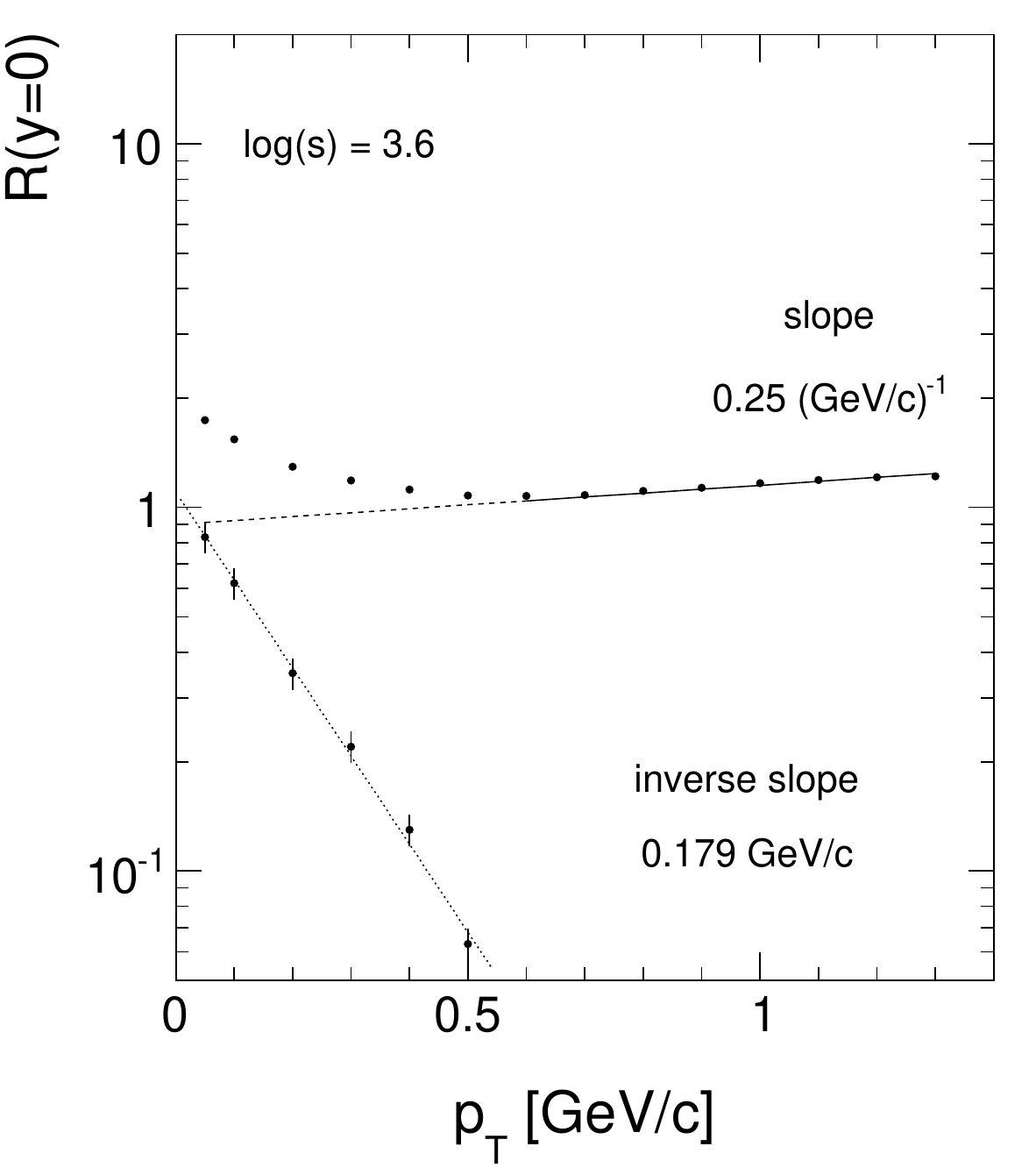} 
	 	\caption{$R(y=0,s,p_T)$ as a function of $p_T$ for $\log(s)$~=~3.6 with exponential fit to $p_T >$~0.6~GeV/c, extrapolation to low $p_T$ (broken line) and subtracted ratio (dotted line)}
  	\label{fig:r63}
 	\end{center}
\end{figure}

The approximately exponential increase of the cross section ratio above $p_T \sim$~0.6~GeV/c may be used to separate the two components by extrapolating to low $p_T$ as shown in Figs.~\ref{fig:r7.7} to \ref{fig:r63}.

\begin{figure}[h]
 	\begin{center}
    \includegraphics[width=15cm] {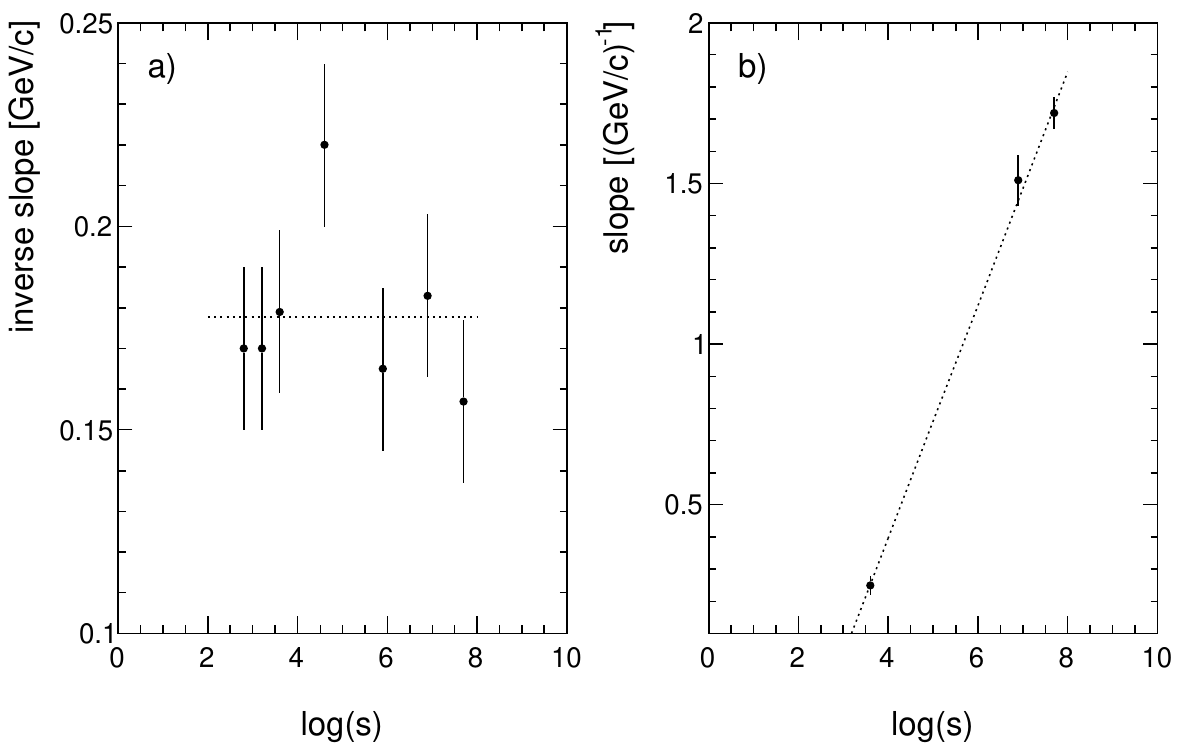} 
    	\caption{Exponential fits to the cross section ratio (\ref{eq:rat_y0}) as given in Figs.~\ref{fig:r7.7} to \ref{fig:r63} a) inverse slopes for the low $p_T$ region $p_T <$~0.5~GeV/c and b) slopes for the higher $p_T$ region at $p_T >$~0.6~GeV/c}
  	\label{fig:col_inverse}
 	\end{center}
\end{figure}

The subtraction of the fit to higher $p_T$ yields exponential distributions with inverse slope parameters as shown in Fig.~\ref{fig:col_inverse}

\begin{figure}[h]
 	\begin{center}
    \includegraphics[width=16cm] {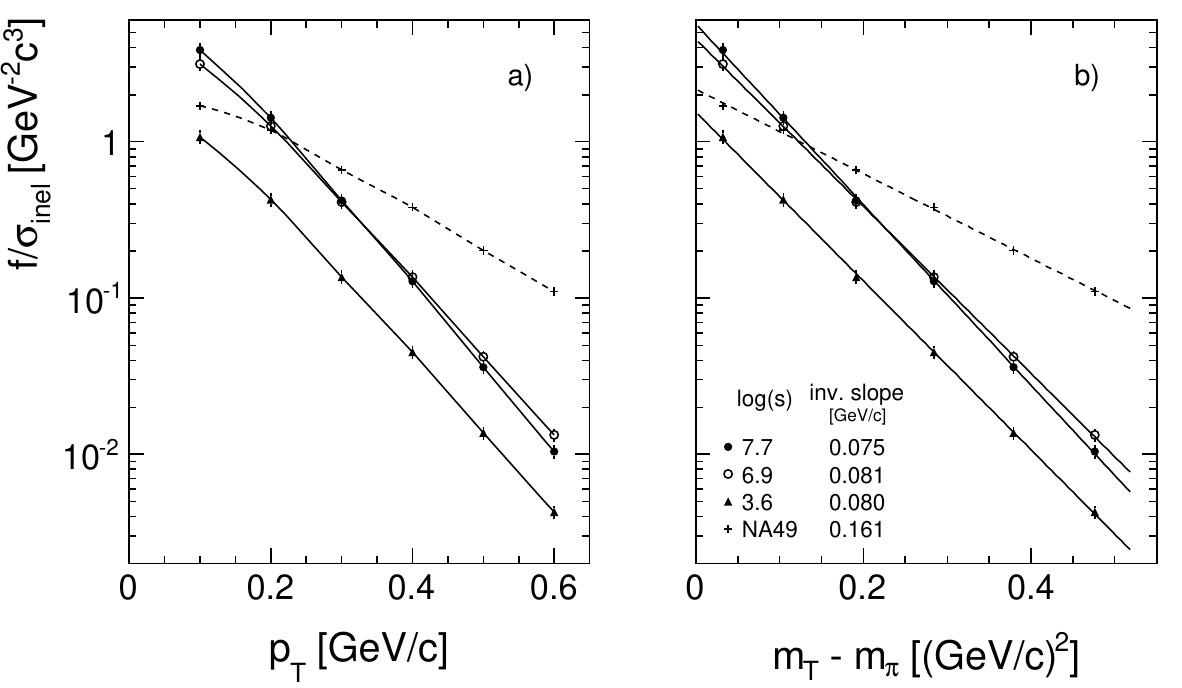} 
	 	\caption{a) $f/\sigma_{\textrm{inel}}(y=0,s,p_T)$ as a function of $p_T$ for different values of $\log(s)$ compared to the invariant cross section of NA49; b) corresponding distribution as a function of $m_T-m$}
  	\label{fig:ptmtdist}
 	\end{center}
\end{figure}

The extracted exponential behaviour of $R(y=0)$ in the low-$p_T$ region may be converted into invariant cross sections using the $p_T$ distribution from NA49 at $y$~=~0 with its inverse slope of 0.17~GeV/c, Fig.~\ref{fig:ptmtdist}a, and finally into the $m_T-m$ distribution of Fig.~\ref{fig:ptmtdist}b.

The inverse slope or "temperature" of the $m_T$ distribution in the low-$p_T$ region is with 0.08~GeV/c about half the value of the inclusive data. In fact it is located about half-way between the "temperature" of the feed-down pions from K$^0_S$ and $\Lambda$ decay, Fig.~\ref{fig:mtslopes}. This gives a strong indication concerning the origin of pions in this part of the central rapidity region. In fact it is the decay of excited baryons \cite{benecke} which gives preferentially low-$x_F$ pions due to the small mass ratio $m_\pi/m_p$ (\ref{eq:pi2p}). There are many channels open with small $Q$-values to pions like

\begin{equation}
	\begin{split}
 		\Sigma(1385) &\rightarrow \Lambda + \pi \\
                 &\rightarrow \Sigma  + \pi
   \end{split}
\end{equation}

(in a consistent picture these strange resonances ought to be subtracted from the inclusive pion sample like $\Lambda$, $\Sigma$ and K$^0$ decays. They are, however, as strong decays on-vertex).

\begin{equation}
	\Xi(1320) \rightarrow \Lambda + \pi
\end{equation}

\noindent
as well as heavy flavour baryons in the RHIC/LHC energy region. Concerning "normal" baryonic resonances there are many channels open with small $Q$-values like

\begin{equation}
	\begin{split}
		N^*(1440) &\rightarrow \Delta + \pi \\
  	N^*(1520) &\rightarrow p + \pi + \pi \\
             &\rightarrow \Delta + \pi  \\
             &\rightarrow p + \rho
  \end{split}
\end{equation}

\noindent
and of course higher resonances.

\begin{figure}[h]
 	\begin{center}
   	\includegraphics[width=16cm] {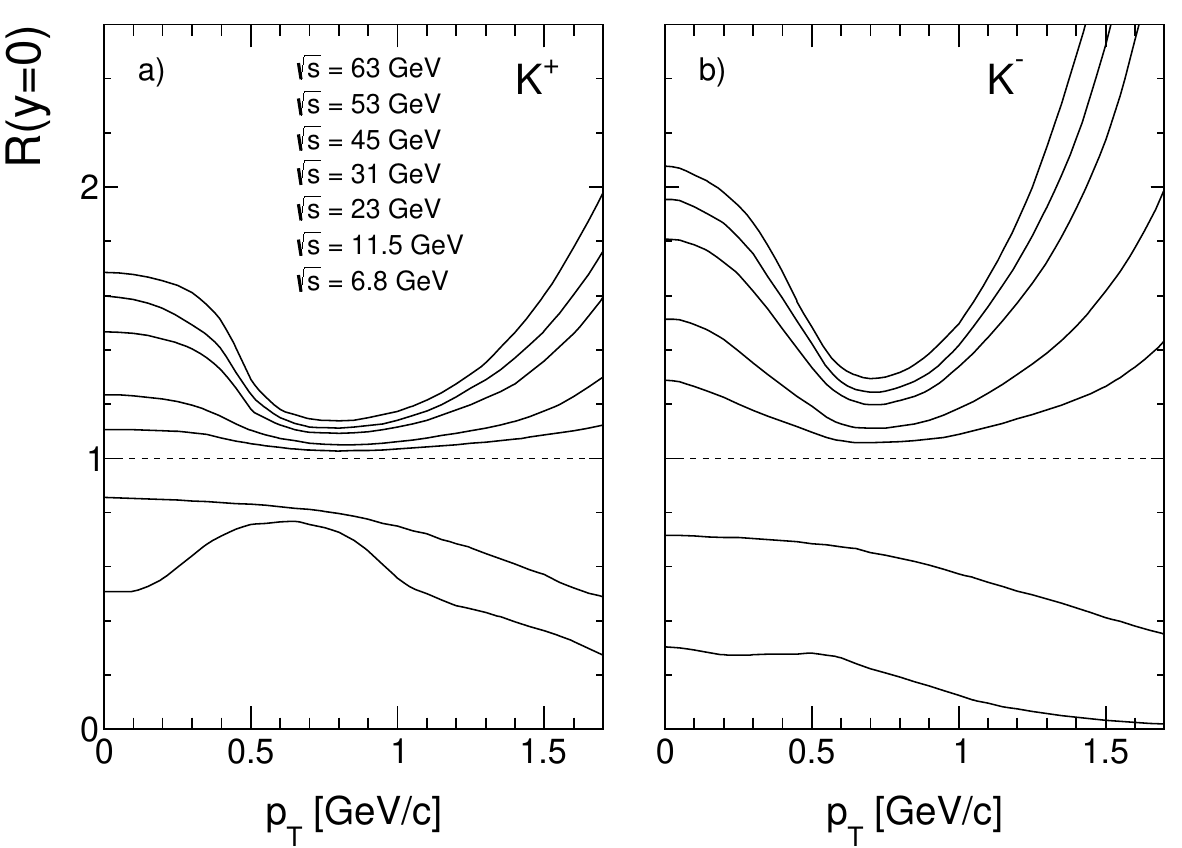} 
	 	\caption{$R(y=0,s,p_T)$ for charged kaons \cite{pp_kaon} as a function of $p_T$ for $\sqrt{s}$ values between 6.8 and 63~GeV}
  	\label{fig:ratka}
 	\end{center}
\end{figure}

In this context it is interesting to look at comparable results for charged kaons \cite{pp_kaon} as shown in Fig.~\ref{fig:ratka} for the ratio $R(y=0,s,p_T)$ which are directly comparable to the $\pi^-$ data (Fig.~\ref{fig:rrefy0}).

Also in this case the low-$p_T$ enhancement has been connected to low-$Q$ resonance decay, namely

\begin{equation}
	\begin{split}
		\Phi(1020)    &\rightarrow K^+ + K^- \\
		\Lambda(1520) &\rightarrow p + K^-
  \end{split}
\end{equation}

The quantitative results depend on the kaon charge due to the relatively large difference between the inclusive K$^+$ and K$^-$ yields in face of the charge symmetric $\Phi$ decay and the fact that $\Lambda(1520)$ exclusively feeds into K$^-$.

\begin{figure}[h]
 	\begin{center}
   	\includegraphics[width=16cm] {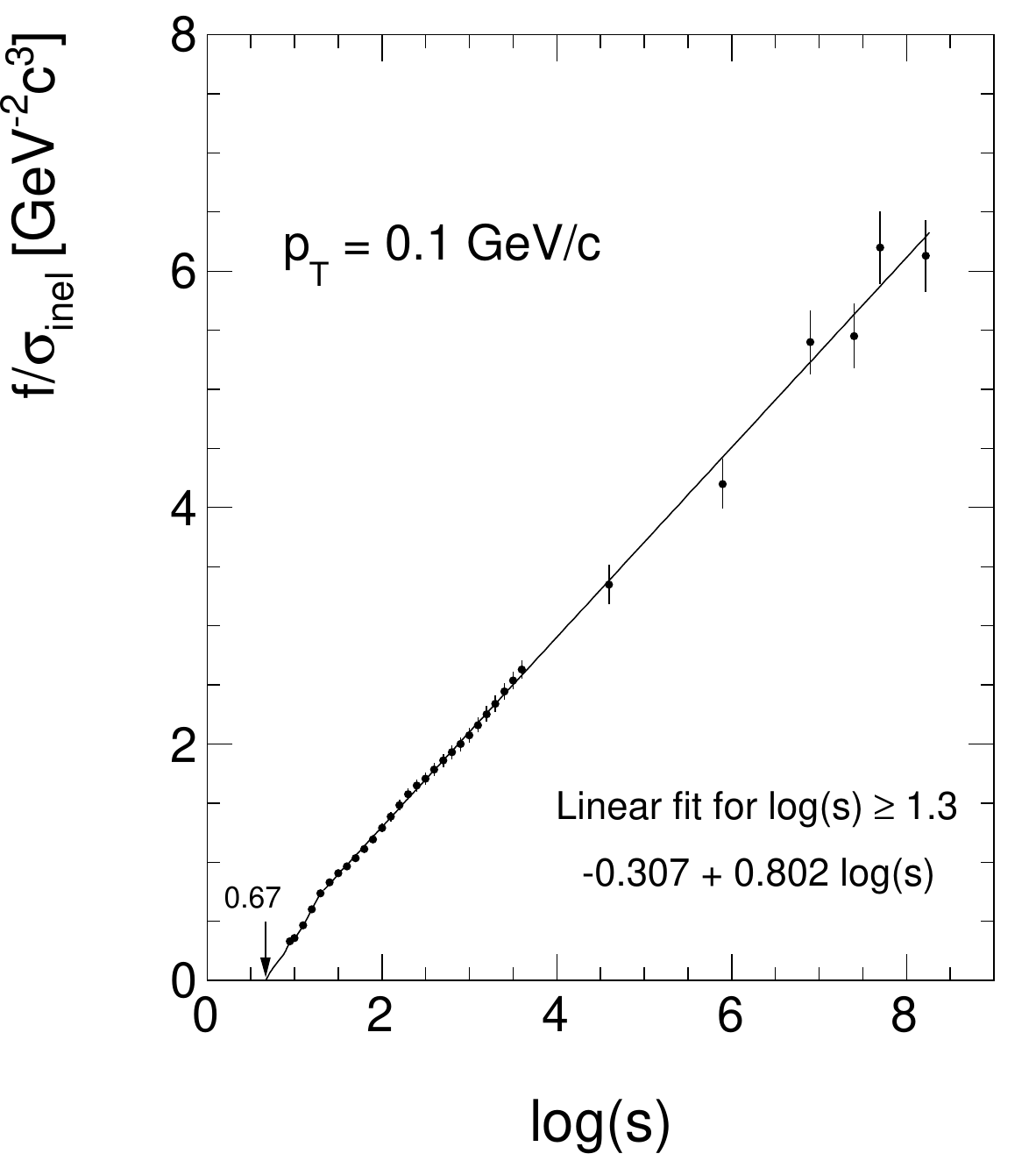} 
	 	\caption{$f/\sigma_{\textrm{inel}}$ at $p_T$~=~0.1~GeV/c as a function of $\log(s)$}
  	\label{fig:fspt01}
 	\end{center}
\end{figure}

The energy dependence of the invariant cross section at low $p_T$ is shown in Fig.~\ref{fig:fspt01} for $p_T$~=~0.1~GeV/c as a function of $\log(s)$.

The yield rises sharply from the threshold at $\log(s)$~=~0.67 or $\sqrt{s}$~=~2.2~GeV. It reaches an almost linear $\log(s)$ dependence above $\log(s)$~=~1.3 of the form

\begin{equation}
	f/\sigma_{\textrm{inel}}(\log{s}) \sim -0.307 + 0.802\log{s}
\end{equation}

\noindent
with small ondulating deviations between $\log{s} \sim$~1.25 and 3.2. The yields at RHIC and LHC energy have negligible statistical errors but sizeable systematic uncertainties on the level of 5\%.

\begin{figure}[h]
 	\begin{center}
   	\includegraphics[width=10cm] {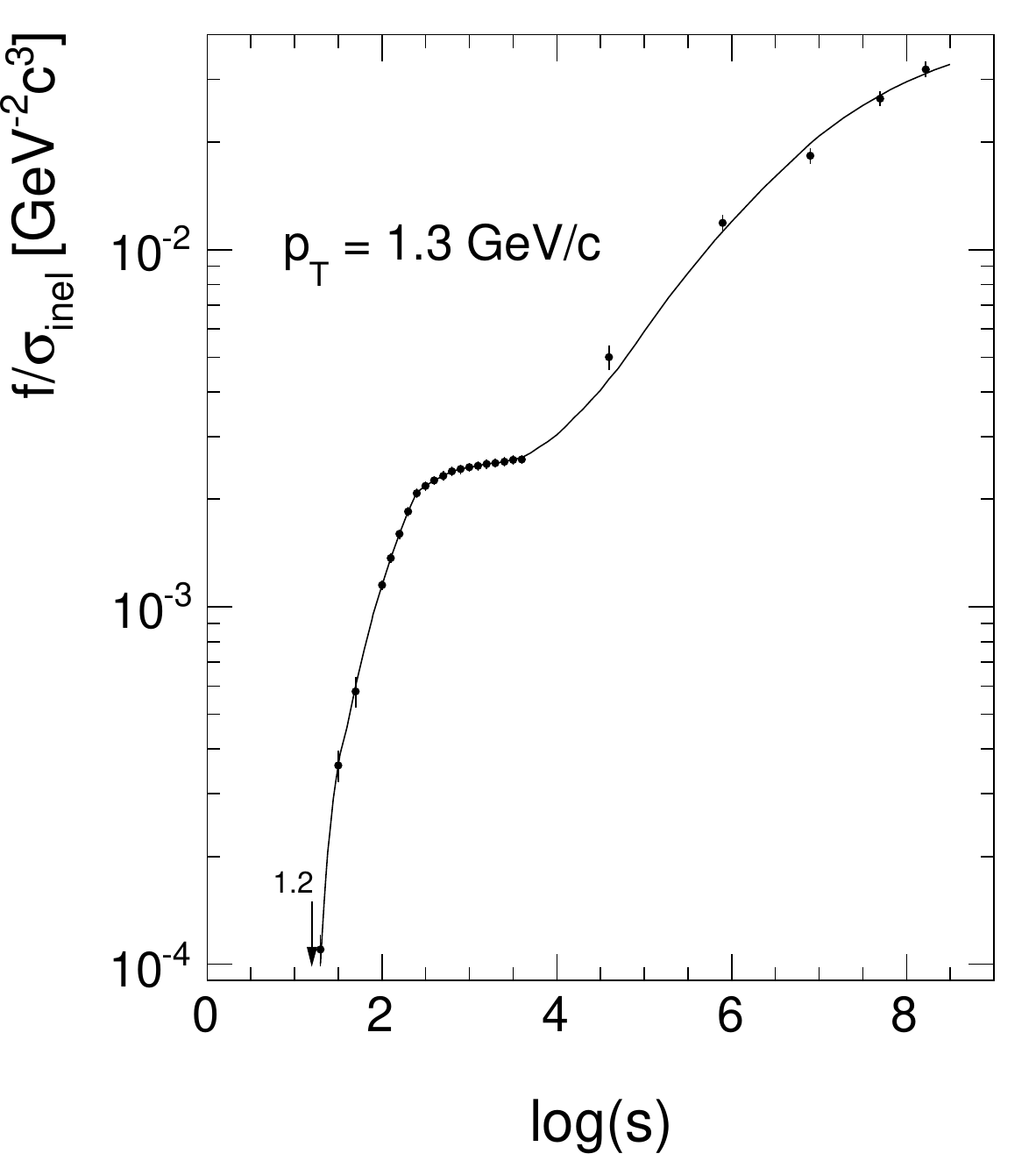} 
	 	\caption{$f/\sigma_{\textrm{inel}}$ at $p_T$~=~1.3~GeV/c as a function of $\log(s)$}
  	\label{fig:fspt13}
 	\end{center}
\end{figure}

For the higher $p_T$ region a more complex energy dependence is evident. This is presented for $p_T$~=~1.3~GeV/c in Fig.~\ref{fig:fspt13}.

The invariant cross section again rises sharply from the threshold at $\log(s)$~=~1.2 or $\sqrt{s}$~=~4.0~GeV to a first plateau at $\sqrt{s} \sim$~40~GeV with a following non-linear increase up to $\sqrt{s}$~=~13~TeV. It will be shown in Sect.~\ref{sec:resonance_highpt} that resonance decay governs the $\pi^-$ yields up to a transverse momentum of at least 3~GeV/c. In fact the total yields of hadronic resonances show a strongly non-linear energy dependence (see for instance Fig.~\ref{fig:feedtot}) with a fast increase from threshold followed by a flattening off towards higher energies.

%
%
\subsection{Pionization and factorization}
\vspace{3mm}
\label{sec:pionization}

The increase of central particle density in the ISR energy range came as a surprise \cite{cernisr1,cernisr2,cernisr3}, as cosmic ray data had indicated, on the contrary, a depletion. In consequence, a specific central production process ("pionization") was postulated as opposed to the independent hadronization of the target and projectile hemispheres with a central overlap. This independent fragmentation could also be described as "factorization" \cite{benecke} in the sense that target (or projectile) particle densities would be independent of the type of projectile (or target) present in the collision. This would mean that experimentally for the inclusive reactions

\begin{align}
  p + p                   &\rightarrow \pi^- + X \\
  \langle \pi \rangle + p &\rightarrow \pi^- + X \\
  \gamma + p              &\rightarrow \pi^- + X
\end{align}

\noindent
the cross sections would be equal in the target hemisphere. The shape and extent of the central overlap region would be accessible via different experimental signatures such as long-range two-particle and multiplicity correlations, charge ratios or the extent of net baryon production. Also hadron-nucleus interactions may be used,

\begin{equation}
 p + A \rightarrow \pi^- + X
\end{equation}

\noindent
for the same purpose although the hadronization itself is of course different for both the target and projectile in this case.

\begin{figure}[h]
 	\begin{center}
   	\includegraphics[width=16cm] {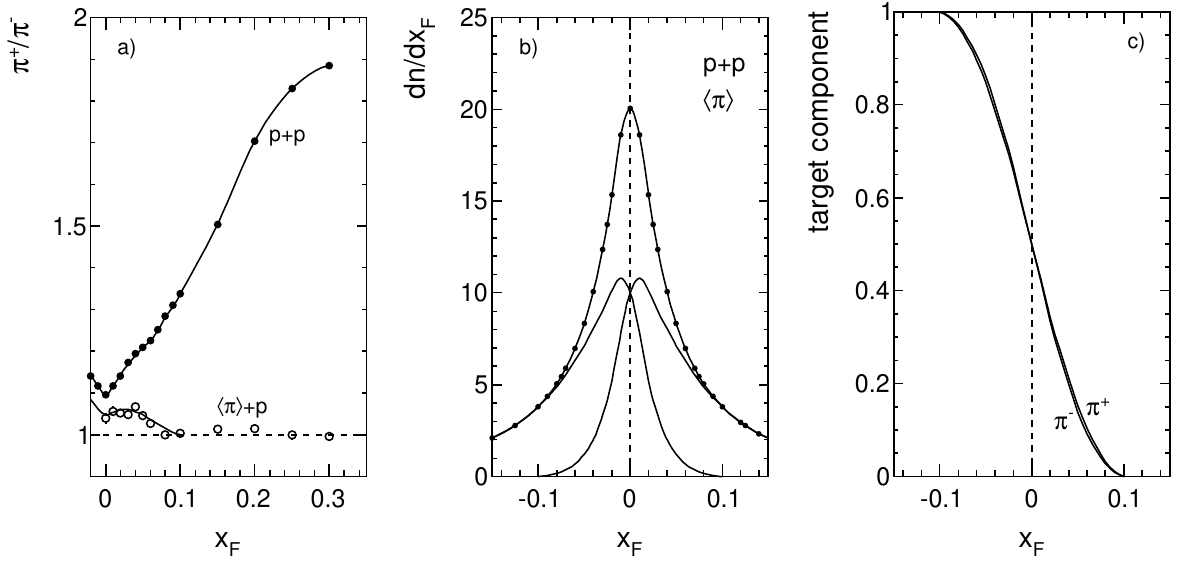} 
	 	\caption{a) $\pi^+$/$\pi^-$ ratio in p+p and $\langle \pi \rangle$+p collisions; b) target and projectile components of charge-averaged pion production and c) target component with respect to the total yield of $\pi^+$ and $\pi^-$ in p+p interactions}
  	\label{fig:pc_pions}
 	\end{center}
\end{figure}

\begin{figure}[h]
 	\begin{center}
   	\includegraphics[width=13cm] {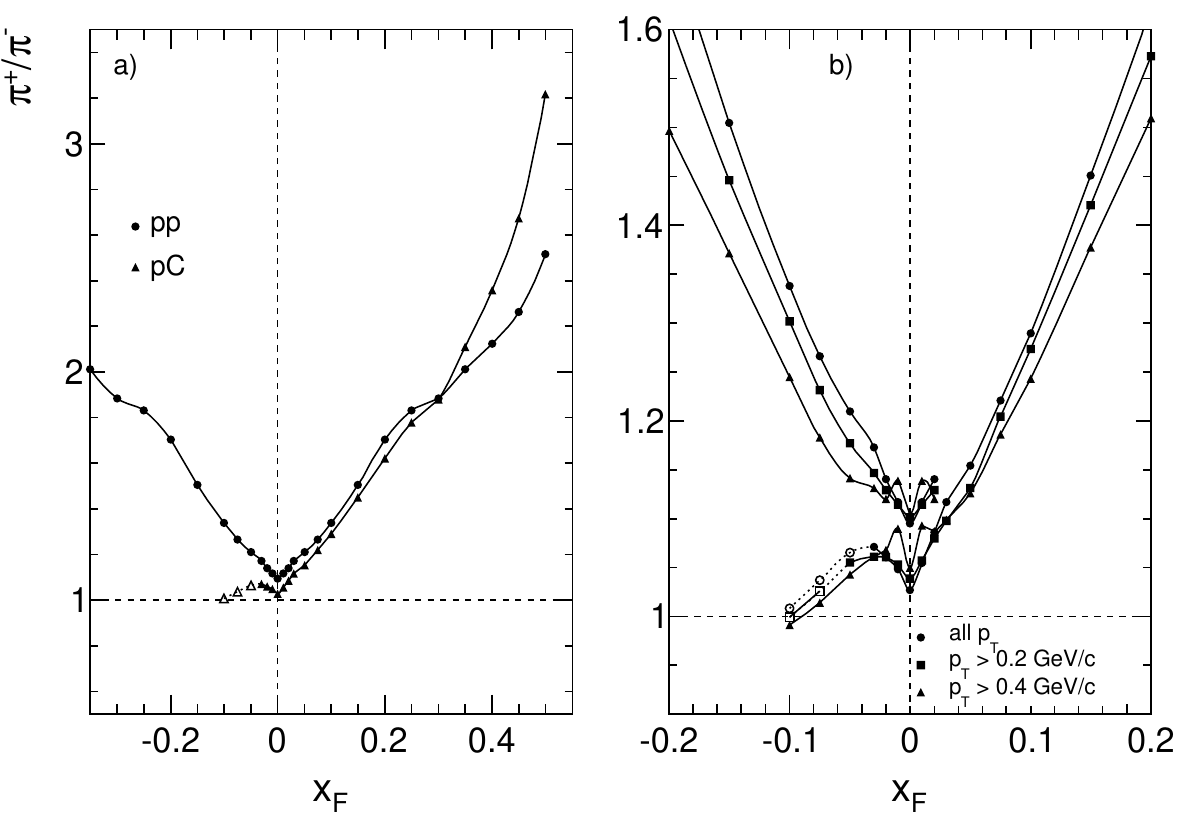} 
	 	\caption{$p_T$ integrated $\pi^+$/$\pi^-$ ratio as a function of $x_F$ in p+p and p+C: a) without $p_T$ cut off and b) with $p_T$ cut off included as the data points in p+p collisions are plotted only up to $x_F$~=~0.02. The data points concerned by an extrapolation of the low $p_T$ region are given as open symbols}
  	\label{fig:pc_target}
 	\end{center}
\end{figure}

\begin{figure}[h]
 	\begin{center}
   	\includegraphics[width=16cm] {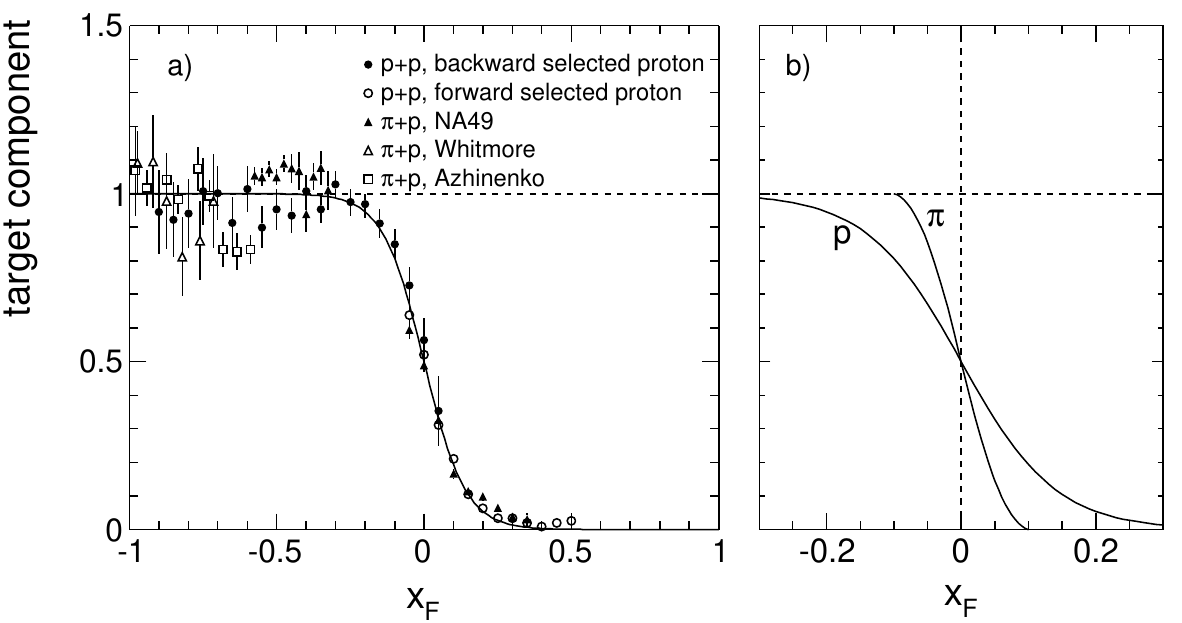} 
	 	\caption{a) Target component for net protons measured with different methods and b) target components of $\langle \pi \rangle$ and net protons}
  	\label{fig:pc_survey}
 	\end{center}
\end{figure}

A very detailed study of factorisation and overlap functions has been conducted in \cite{pc_discuss,pc_proton}. Some examples are shown in Figs.~\ref{fig:pc_pions} to \ref{fig:pc_survey} concerning the feedover-distributions for pions and for net protons. extracted from p+p, $\langle \pi \rangle$ + p and p + C collisions.

The fact that the $\pi^+/\pi^-$ ratio has to be unity for the projectile contribution in $\langle \pi \rangle$ + p interactions is experimentally  verified in Fig.~\ref{fig:pc_pions}a such that only the target component will contribute to the measured $\pi^+/\pi^-$ ratio. The symmetric target and projectile distributions in p + p reactions, Fig.~\ref{fig:pc_pions}b, allow for the direct extraction of the target feed-over, Fig.~\ref{fig:pc_pions}c both in shape and range in $x_F$. A similar argument may be used in p + C interactions where the isoscalar Carbon nucleus has a $\pi^+/\pi^-$ ratio equal to unity (now in the target hemisphere) as presented in Fig.~\ref{fig:pc_target}.
Note that this argument holds since there is no pion exchange at SPS energies as demonstrated in \cite{pc_survey}.

Concerning protons, $\langle \pi \rangle$ + p interactions should not have any net protons in the projectile hemisphere. In p + p collisions, a complementary method consists of fixing net baryon number by selecting protons either forward or backward at $x_F$ values outside the range of pair produced protons \cite{deszo}. These methods allow for a completely model-independent way of measuring the feed-over distributions for net protons as shown in Fig.~\ref{fig:pc_survey}.

This study shows that there is an important difference between the feed-over distributions of pions and net protons, Fig.~\ref{fig:pc_survey}b. In both cases the $x_F$ range of feed-over is well outside the accessible $x_F$ range of high energy colliders.

Note that also a test of factorization becomes possible between $\pi$ + p and p + p reactions as shown in Fig.~\ref{fig:pc_survey} in backward direction below $x_F \sim$~-0.5 where proton identification is possible in bubble chambers ($p_{\textrm{lab}}^{\textrm{p}} <$~1.3~GeV/c). In addition, \cite{pp_proton} has shown factorization for baryon production between p + p and leptoproduction at HERA \cite{chekanov1,chekanov2}.

In the context of the present argumentation it is important to realize that it has been experimentally verified that no central "pionization" region is present in addition to the (factorizing) target and projectile contributions as already hypothesized in \cite{benecke}. This is also true for p + C interactions.

%
%
\subsection{Summary on scaling}
\vspace{3mm}
\label{sec:summary_scaling}

The critical comparison of the complete set of existing data and the resulting establishment of a global interpolation scheme with a systematic uncertainty below the 5\% level allow for a confrontation of certain physics hypotheses with the experimental reality over the full energy range available today. This has been attempted here with special emphasis on energy dependence or, rather, the absence thereof which is usually labelled as "scaling".

Several aspects are of importance here:

\begin{enumerate}[label=(\alph*)]
	\item The complete coverage of not only the energy range but also of a maximum of the available phase space is absolutely mandatory.
	\item Even on the single-particle inclusive level the complexity of the experimental results precludes any simplistic description of a physics situation which is not amenable to theoretical predictions.
	\item The increase of the total inelastic cross section by a factor of three over the available energy range imposes a very careful scrutiny of the geometrical aspects in terms of dependencies on the impact parameter and its consequences on different sectors of phase space, especially at LHC energies.
	\item In view of this fact any description in terms of point-like interactions as they are for instance imposed by parton dynamics is problematic.
\end{enumerate}

Two hypotheses which were phrased rather early on in the development of particle physics (1969) have been confronted with the data in detail.

The first hypothesis concerns "limiting fragmentation" \cite{benecke}.

\begin{enumerate}[label=(\alph*)]
	\setcounter{enumi}{4}
	\item This hypothesis was for the first time tested over the full energy range (Sect.~\ref{sec:scaling_limiting}).
	\item The predicted energy invariance is verified within tight limits if plotting the invariant cross section as a function of $p_L^{\textrm{lab}}$.
	\item On the other hand if using the cross section per inelastic event ($f/\sigma_{\textrm{inel}}$) sizeable deviations from $p_L^{\textrm{lab}}$ scaling are observed.
	\item This is a first indication of a yield dependence on the production point in the overlapping hadronic disks.
\end{enumerate}

Ref. \cite{benecke} is an astounding document as it contains a number of conjectures that were not testable at the time of publication. To be mentioned here is hadronic factorization and the absence of central pionization (Sect.~\ref{sec:summary_scaling}). The inclusion of baryon  number conservation as a basic ingredient as well as resonance production and decay are further assets.

\begin{enumerate}[label=(\alph*)]
	\setcounter{enumi}{8}
	\item The study of $y_{\textrm{lab}}$ distributions (Sect.~\ref{sec:scaling_ylab}) is complementary to $p_L^{\textrm{lab}}$ scaling and shows similar results, again in preference of invariant cross sections as compared to yields per inelastic event.
\end{enumerate}

The second hypothesis, "Feynman scaling", is inspired by the partonic structure of the colliding hadrons as it is measured in deep inelastic lepton scattering.

\begin{enumerate}[label=(\alph*)]
	\setcounter{enumi}{9}
	\item In consequence the conjecture postulates scaling in longitudinal momentum if referred to the maximum available momentum (Sect.~\ref{sec:scaling_xf}).
	\item In the very definition of this relative variable it is necessary to take full account of energy and baryon number conservation.
	\item Nevertheless there are strong deviations from scaling at all energies below the ISR range.
	\item An inherent weakness of this hypothesis is the baryonic sector. In fact there are no informations to be gained on baryon distributions and baryon number conservation from deep inelastic lepton scattering. Hypothetical "di-quark" states have therefore to be introduced which carry unknown fragmentation functions.
	\item The sector of diffraction which is characterized by large impact parameters \cite{increase_rim1} is in no way describable in this approach.
\end{enumerate}

\begin{figure}[h]
 	\begin{center}
   		\includegraphics[width=13cm] {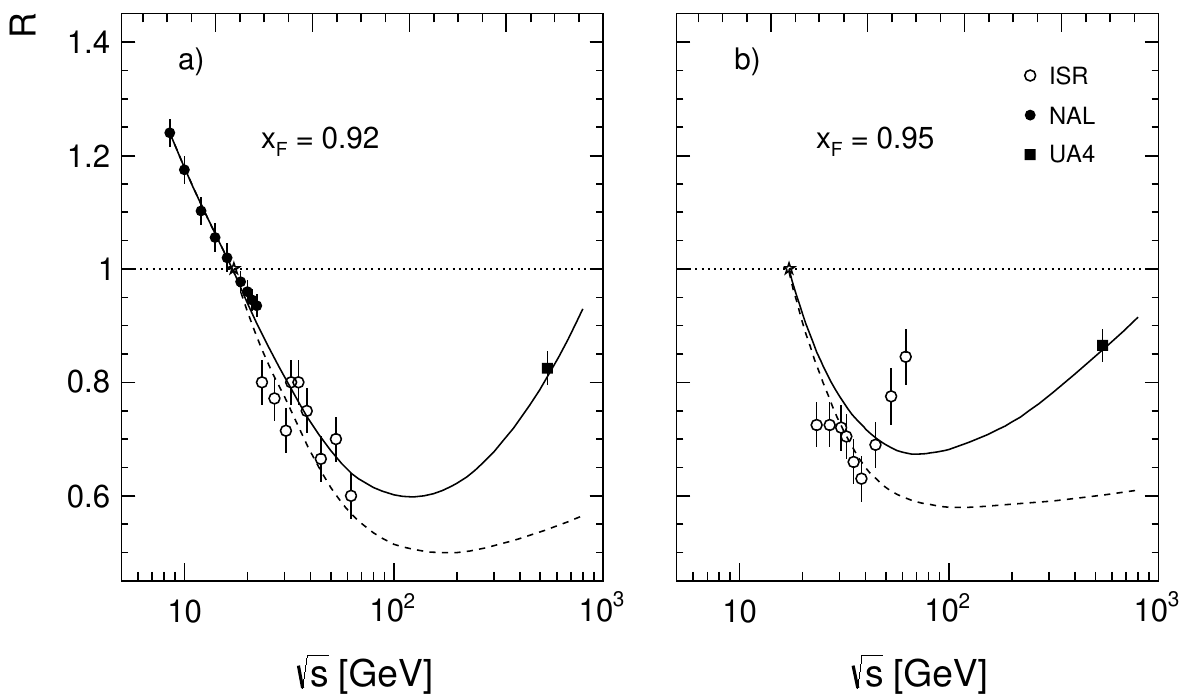} 
	 	\caption{Ratio of proton cross sections $R$ relative to NA49 data as a function of $\sqrt{s}$ in the diffractive sector at $x_F$~=~0.92 and 0.95 including data from UA4 \cite{bozzo}. The full lines correspond to the invariant cross sections, the broken lines to the cross sections per inelastic collision}
  	\label{fig:isr_ua4_prot}
 	\end{center}
\end{figure}

In this context of impact-parameter dependencies it is worth while looking as a side remark at diffractive proton production which would stem \cite{increase_rim1,increase_rim2} from the expanding rim of the interaction region. One would therefore, contrary to the $\pi^-$ production, expect cross sections to scale in the yield per inelastic collision rather than the invariant cross section itself which should increase with the interaction area. This has been shown in \cite{bozzo} reproduced here as Fig.~\ref{fig:isr_ua4_prot}.

The UA4 experiment at the CERN p$\overline{\textrm{p}}$ collider published data \cite{bozzo} only at $\sqrt{s}$~=~530~GeV. It would be of considerable interest to have data also up to the highest LHC energies.

%
%
\section{Transverse momentum}
\vspace{3mm}
\label{sec:transverse}

The available data concerning double differential inclusive $\pi^-$ cross sections present -- given sufficient precision on the experimental
level -- an overall phenomenology of great complexity defying attempts
at simple interpretations or algebraic approximations. It is however
the largely asymmetric occupation of the phase space in longitudinal
and transverse direction, the "longitudinal phase space" (\ref{eq:pLpT}), which seems
to offer a natural approach to some underlying physics. The longitudinal
part has been analysed in the preceding Sect.~\ref{sec:scaling} in terms of energy dependence and "scaling" with only marginal success. The transverse
part shows a steep $p_T$ dependence that has been from early on tentatively approximated by an exponential form

\begin{equation}
	\frac{f}{\sigma( x_F', p_T, s )} = A( x_F', s )e^{-B( x_F', s ) p_T}
\end{equation}

\begin{figure}[h]
 	\begin{center}
   		\includegraphics[width=16cm] {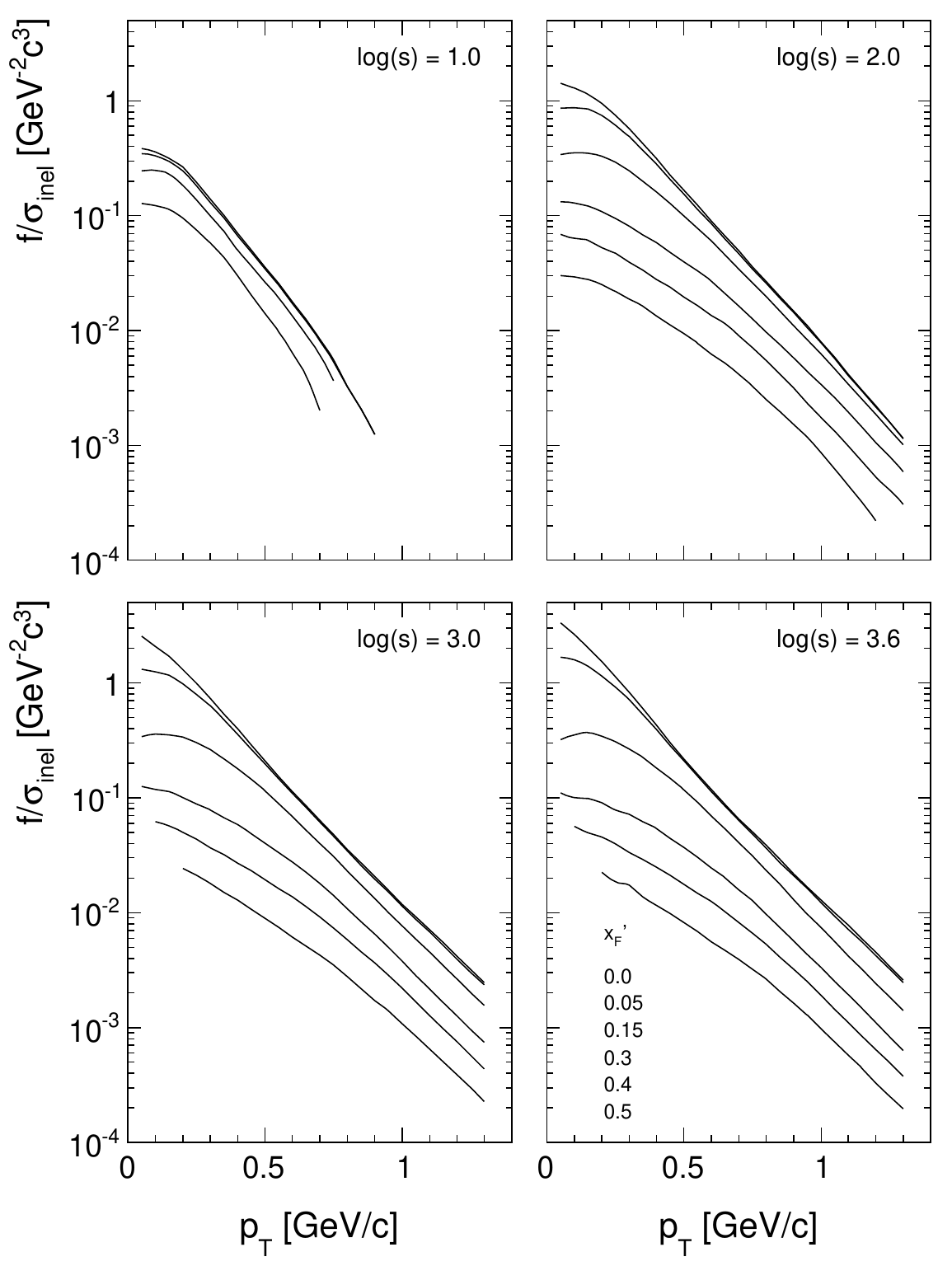} 
	 	\caption{Invariant $\pi^-$ cross section $f/\sigma_{\textrm{inel}}$ as a function of $p_T$ for several $x_F'$ and $\log(s)$ values}
  		\label{fig:ptinv_dist}
 	\end{center}
\end{figure}

Of course this simplistic form is only defendable as a zero-order approximation to the rather complex overall $p_T$ dependence that follows from the global interpolation (Sect.~\ref{sec:interpolation}). This is shown in
Fig.~\ref{fig:ptinv_dist} where the invariant cross section is presented as a function of $p_T$ and several values of $x_F'$ and $\log(s)$.

In fact there is only a very small region at $\sqrt{s} \sim$~11~GeV where
an exponential shape is realized at $x_F' \sim$~0, in this area even up to large $p_T$ values of $\sim$~4~GeV/c \cite{abramov}. Always of course exception made for small $p_T$ where the invariant cross section has to approach $p_T$~=~0~GeV/c with
tangent zero. Towards higher interaction energies, where only data at central rapidity are available, the again very complex $p_T$ and $\log(s)$
dependence has already been shown in Fig.~\ref{fig:rcoly0}.

Notwithstanding this complex experimental situation, rather general
attempts at understanding the transverse phenomenology by way of statistical or thermal "models" have been and still are enjoying
widespread interest. These attempts are characterized by only one
or a few parameters besides the particle mass. They will be confronted
with the results of resonance decay in the following Sections.

%
%
\subsection{General considerations concerning the choice of co-ordinates}
\vspace{3mm}
\label{sec:transverse_coord}

\begin{figure}[h]
	\begin{center}
		\includegraphics[width=16cm] {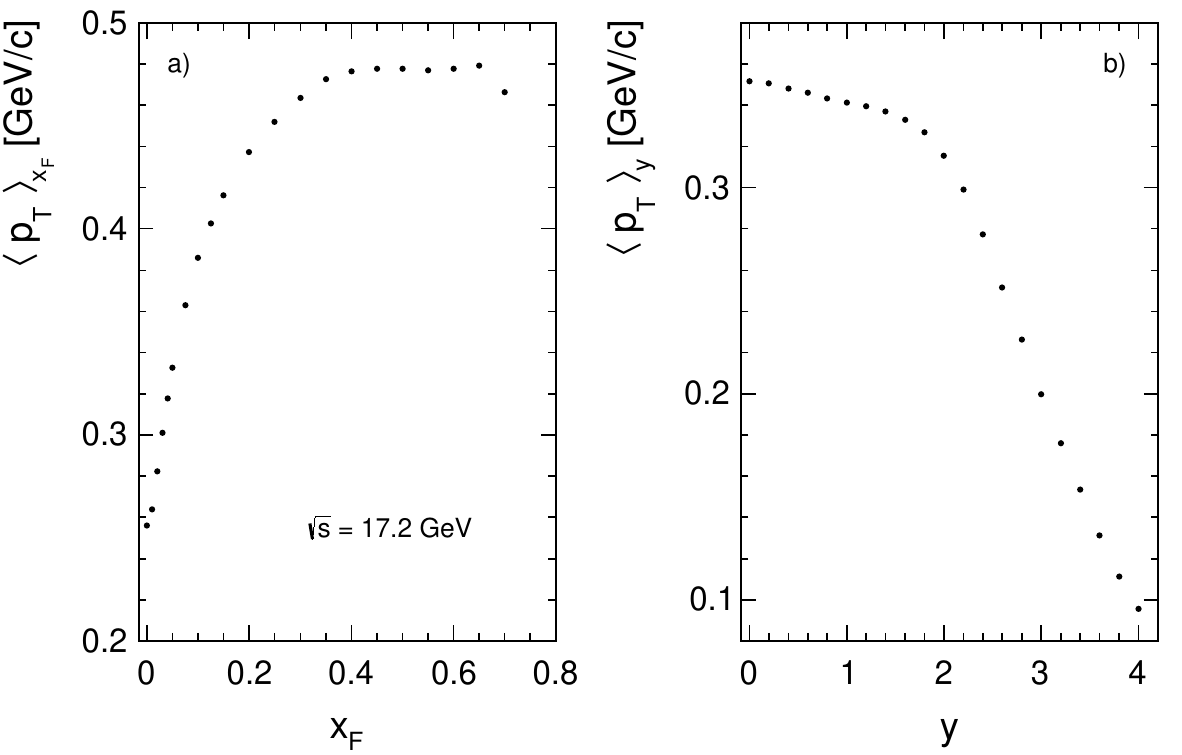} 
		\caption{Mean transverse momentum $\langle p_T \rangle$ as a function of
a) $x_F$ and b) rapidity integrated in the limit 0~$< p_T <$~1.3~GeV/c}
		\label{fig:meanptna49}
	\end{center}
\end{figure}

The in-depth analysis of transverse momentum phenomena needs
first of all some clarification, in particular concerning the
coordinate systems to be used. Unlike the longitudinal momentum
distributions discussed in the preceding Section in different
reference systems, the use of orthogonal coordinates is mandatory. This is evident in Fig.~\ref{fig:meanptna49} where the mean transverse momentum distributions are shown as functions of $x_F$ and rapidity for the
NA49 data at $\sqrt{s}$~=~17.2~GeV.

\begin{figure}[h]
	\begin{center}
		\includegraphics[width=15cm] {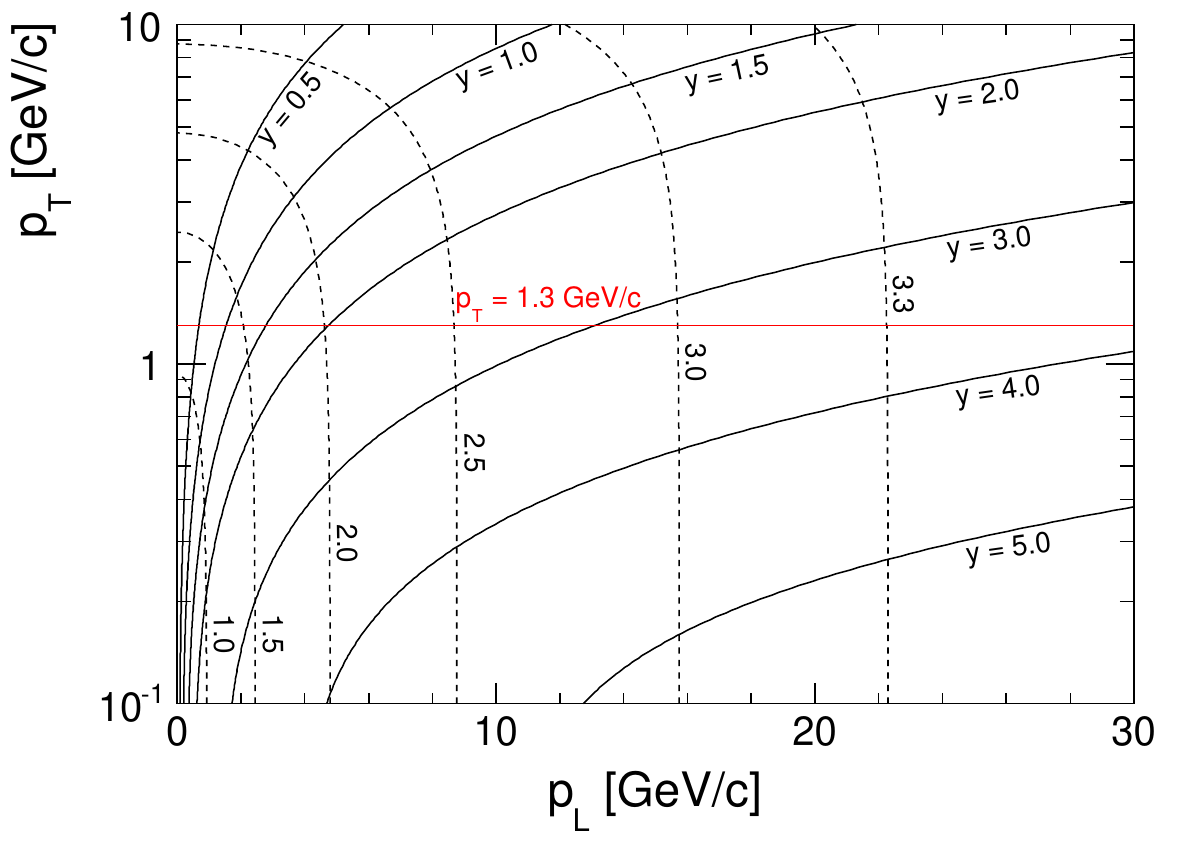} 
		\caption{Kinematical relation between the orthogonal longitudinal and transverse momentum variables $p_L$ and $p_T$ and lines of constant rapidity (full lines). For different interaction energies $\log(s)$ the kinematic limits are indicated (dashed lines)}
		\label{fig:longvstrans}
	\end{center}
\end{figure}

The completely different functional behaviour is explained by the fact that the rapidity variable is not orthogonal in $p_L$ and $p_T$ but links both variables such that, for pions, the cross sections are closely following, for constant rapidity, fixed cms angles thus folding the $p_T$ distributions with the strongly decreasing longitudinal yields as presented in Fig.~\ref{fig:longvstrans}.

For the NA49 energy the kinematical limit is reached for the limiting $p_T$ value of this study of 1.3~GeV/c at $y$~=~2.56 at which point the pion yield has to vanish. Nothing could be more erroneous than to interpret the decreasing $\langle p_T \rangle$ as a function of $y$ (Fig.~\ref{fig:meanptna49}b) as a reduction of "hadronic temperature" in forward direction. On the contrary, there is a strong increase of $\langle p_T \rangle$ with increasing $p_L$ or, as shown in Fig.~\ref{fig:meanptna49}a), with $x_F$ -- an effect known since decades as "seagull effect" \cite{bardadin,apeldoorn}.

In connection with the determination of the mean transverse momentum the pion densities as functions of $x_F$ and rapidity are used:

\begin{align}
	\frac{d^2n}{dx_Fdp_T} &= \frac{p_T}{E}f(x_F,p_T) \\
    \frac{d^2n}{dydp_T} &= p_T f(y,p_T)
\end{align}

\begin{figure}[h]
	\begin{center}
		\includegraphics[width=13cm] {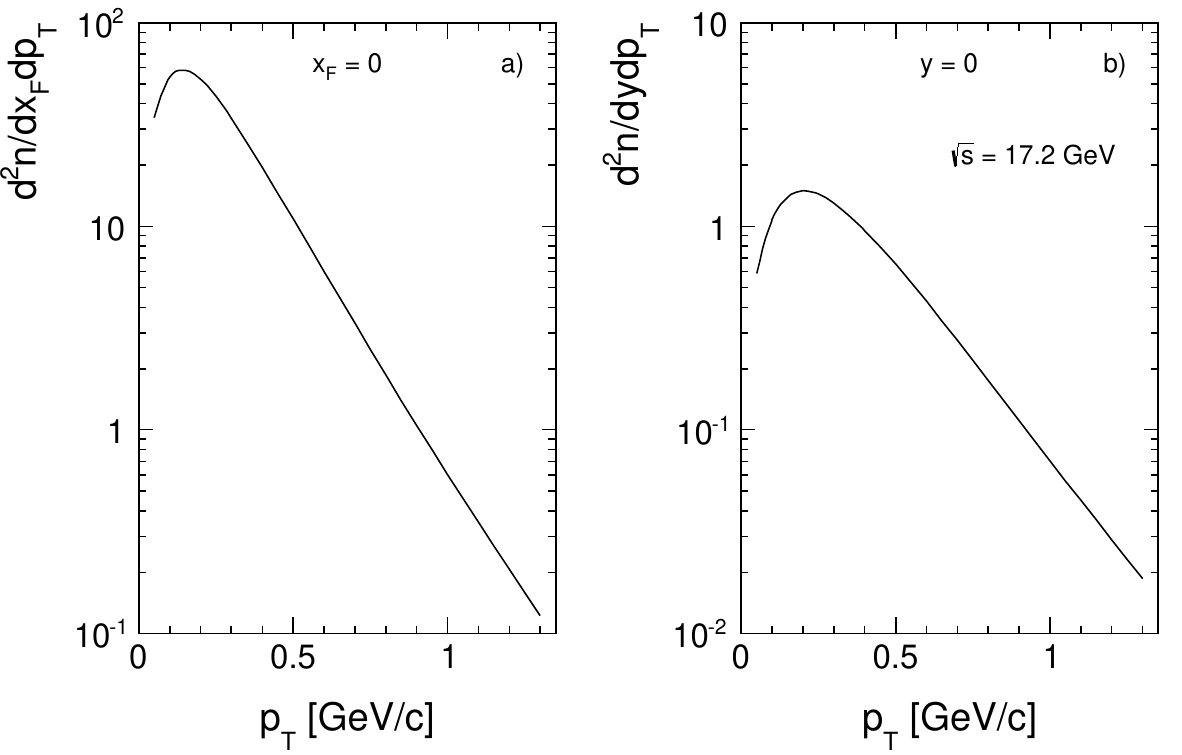} 
		\caption{a) Pion densities $d^2n/dx_Fdp_T$ for $x_F$~=~0 and b) $d^2n/dydp_T$ for $y$~=~0 as a function of $p_T$ at $\sqrt{s}$~=~17.2~GeV/c}
		\label{fig:xfvsydist}
	\end{center}
\end{figure}

These two pion densities have a different shape as shown in Fig.~\ref{fig:xfvsydist}.

This results in $\langle p_T \rangle$ values as

\begin{align}
\label{eq:mpt_xf}
\langle p_T \rangle_{x_F} &= \frac{\displaystyle \int p_T \,\frac{dn}{dx_Fdp_T} \,dp_T}
{\displaystyle \vphantom{\sum_{a}}\int \frac{dn}{dx_Fdp_T}\, dp_T} = \frac{\displaystyle \int \frac{p_T^2}{E} \,f\, dp_T}
{\displaystyle \int \frac{p_T}{E} \,f \,dp_T} \\
\label{eq:mpt_y}
\langle p_T \rangle_{y}   &= \frac{\displaystyle \vphantom{\sum^{a}}\int p_T \,\frac{dn}{dydp_T}\, dp_T}
{\displaystyle \vphantom{\sum_{a}}\int \frac{dn}{dydp_T}\, dp_T} = \frac{\displaystyle \int p_T^2 \,f \,dp_T}
{\displaystyle \int p_T\, f\, dp_T}
\end{align}

The resulting $\langle p_T \rangle$ values differ appreciably by 0.1~GeV/c for $\pi$, 0.06~GeV/c for K and 0.03~GeV/c for p($\overline{\textrm{p}}$). This difference depends on the particle mass as presented in Tab.~\ref{tab:mpt0all}.

 \begin{table}[h]
 	\renewcommand{\tabcolsep}{1.0pc}
 	\renewcommand{\arraystretch}{1.}
 	\begin{center}
 		\begin{tabular}{ccc}
			\hline
	         & $\langle p_T \rangle_{x_F}$ [GeV/c] &  $\langle p_T \rangle_{y}$ [GeV/c]\\ \hline
	        $\pi^+$                 &   0.258   &  0.352  \\
	        $\pi^-$                 &   0.256   &  0.351  \\
	        K$^+$                   &   0.406   &  0.466  \\
	        K$^-$                   &   0.395   &  0.451  \\
			K$^0_S$                 &   0.402   &  0.460  \\
	        p                       &   0.495   &  0.528  \\
	        $\overline{\textrm{p}}$ &   0.477   &  0.507  \\
			\hline
		\end{tabular}
	\end{center}
	\caption{$\langle p_T \rangle_{x_F}$ for $x_F$~=~0 and $\langle p_T \rangle_{y}$ for $y$~=~0 at $\sqrt{s}$~=~17.2~GeV of feed-down subtracted data for different particle species}
	\label{tab:mpt0all}
\end{table}

\begin{figure}[h]
	\begin{center}
		\includegraphics[width=10cm] {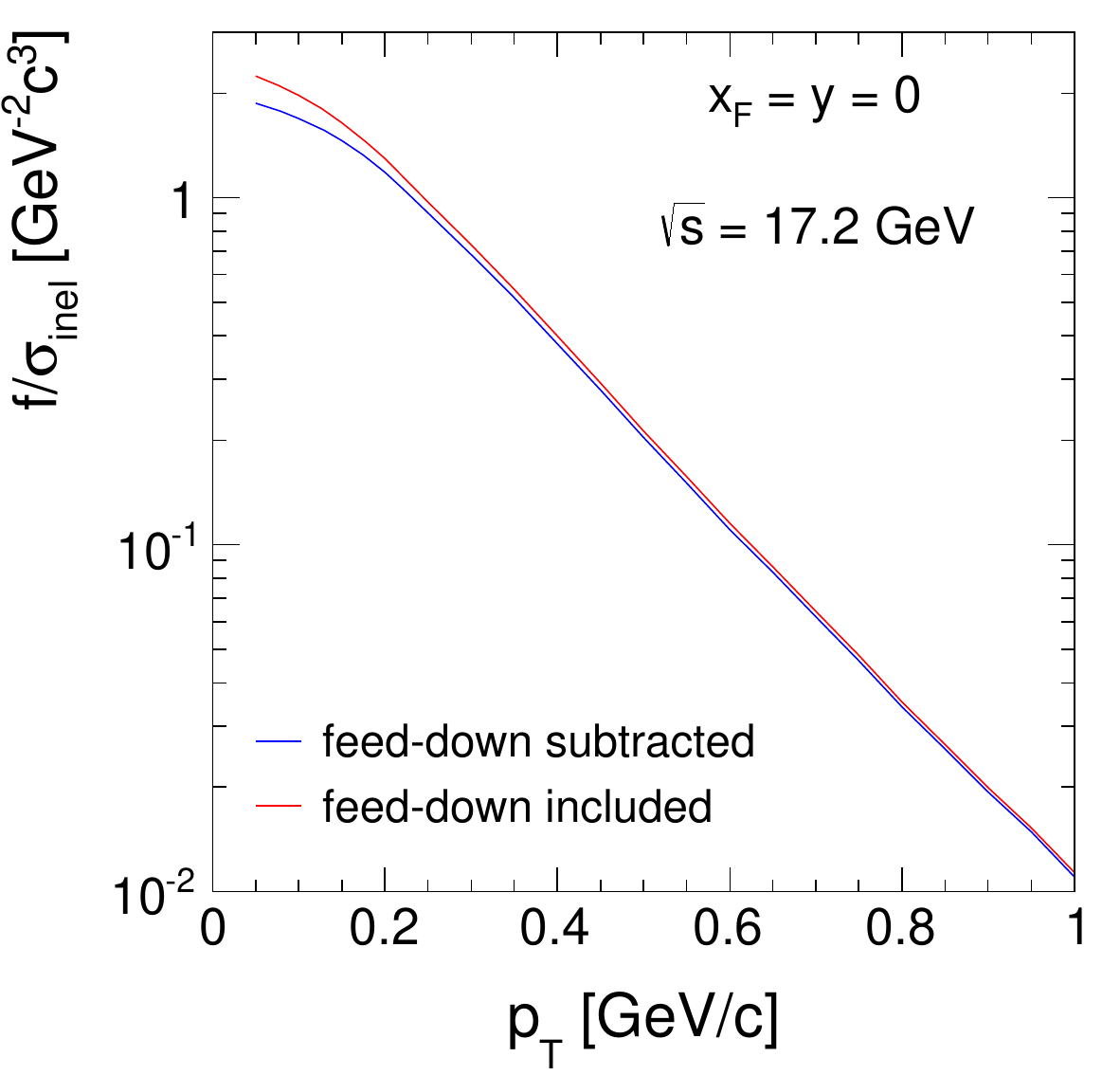} 
		\caption{$f/\sigma_{\textrm{inel}}$ for $\pi^-$ as a function of $p_T$ at $x_F$~=~$y$~=~0 with and without feed-down subtraction}
		\label{fig:feedsubadd}
	\end{center}
\end{figure}

A last general remark concerns the influence of the feed-down correction on the determination of mean $p_T$. As shown in Fig.~\ref{fig:feedsubadd} at $x_F$~=~$y$~=~0 the $p_T$ distributions are different with and without feed-down correction, see Sect.~\ref{sec:fd_corr}.

The corresponding $\langle p_T \rangle_{x_F}$ and $\langle p_T \rangle_{y}$ values are different as given in Tab.~\ref{tab:mptpim}.

\begin{table}[h]
	\renewcommand{\tabcolsep}{1.0pc}
	\renewcommand{\arraystretch}{1.3}
	\small
	\begin{center}
		\begin{tabular}{ccc}
			\hline
			& $\langle p_T \rangle_{x_F}$ [GeV/c] &  $\langle p_T \rangle_{y}$ [GeV/c] \\ \hline
			feed-down subtracted   &  0.256   &  0.351  \\
			feed-down included     &  0.249   &  0.343  \\
			\hline
		\end{tabular}
	\end{center}
	\caption{$\langle p_T \rangle_{x_F}$ for $x_F$~=~0 and $\langle p_T \rangle_{y}$ for $y$~=~0 at $\sqrt{s}$~=~17.2~GeV of feed-down subtracted and feed-down included data}
	\label{tab:mptpim}
\end{table}

%
%
\subsection{Statistical and thermodynamical models}
\vspace{3mm}
\label{sec:thermo}

Transverse momentum distributions, if plotted for constant $s$ and different $y_{\textrm{lab}}$ (Fig.~\ref{fig:interp}) or different $x_F'$ (Fig.~\ref{fig:ptinv_dist}) or at central rapidity for different s (Fig.~\ref{fig:rcoly0}), exhibit an extraordinary variety of shapes that seems to defy any attempt at a simple algebraic description covering the full phase space. In the preceding Sections, mostly concentrated on longitudinal momentum dependencies, the transverse component was rather characterized by its rapid, quasi-exponential and quasi invariant, decrease ("longitudinal phase space") without attempts to describe the detailed shape which is, of course, not calculable in non-perturbative QCD.

On the other hand, early on in the 1950's, rather general statistical models \cite{fermi,belenskij} based on thermodynamical analogous were developed and applied to hadronic interactions and particle production.

This approach has been extended to predictions concerning momentum distributions, principally also the transverse component, in the mid-1960's \cite{hagedorn1} and quantified in subsequent publications \cite{hagedorn2}, \cite{hagedorn}.

The evolution of interaction energies through the ISR and RHIC regions up to the LHC necessitated, however, important modifications and extensions concerning the shapes of the transverse distributions. This was attempted by adding at least one further parameter to the original thermodynamic fits, or taking reference to another statistical mechanical formulation \cite{tsallis,bueyuekkilic}.

To date the original claim of thermodynamical behaviour of hadronic production, in particular concerning the notion of "hadronic temperature" and, in connection with a "limiting temperature", the purported existence of a hadronic phase transition, is still widely proclaimed. In the following sections, a critical view at this problematics will be attempted for the sector of elementary hadronic interactions.

%
%
\subsubsection{Hagedorn's Statistical Bootstrap Model (SBM)}
\vspace{3mm}
\label{sec:hagedorn}

The "bootstrap" yields an exponentially increasing mass spectrum of "fireballs" or "resonances" whose inverse slope defines a limiting "temperature" -- hence the inherent possibility of a hadronic phase transition should this limiting temperature ever be exceeded. The momentum distributions of fireballs, resonances and their decay products are postulated to obey the characteristics of an ideal gas in equilibrium with only longitudinal motion. This leads, taking account of the different particle masses via the transverse mass $m_T$ (\ref{eq:mt}) to the prediction of exponential $m_T$ distributions with a unique inverse slope defined by the hadronic temperature $T$. This means that the SBM model should describe the transverse momentum distributions by a single parameter, independent of interaction energy and particle type.

\begin{figure}[h]
	\begin{center}
		\includegraphics[width=15.5cm] {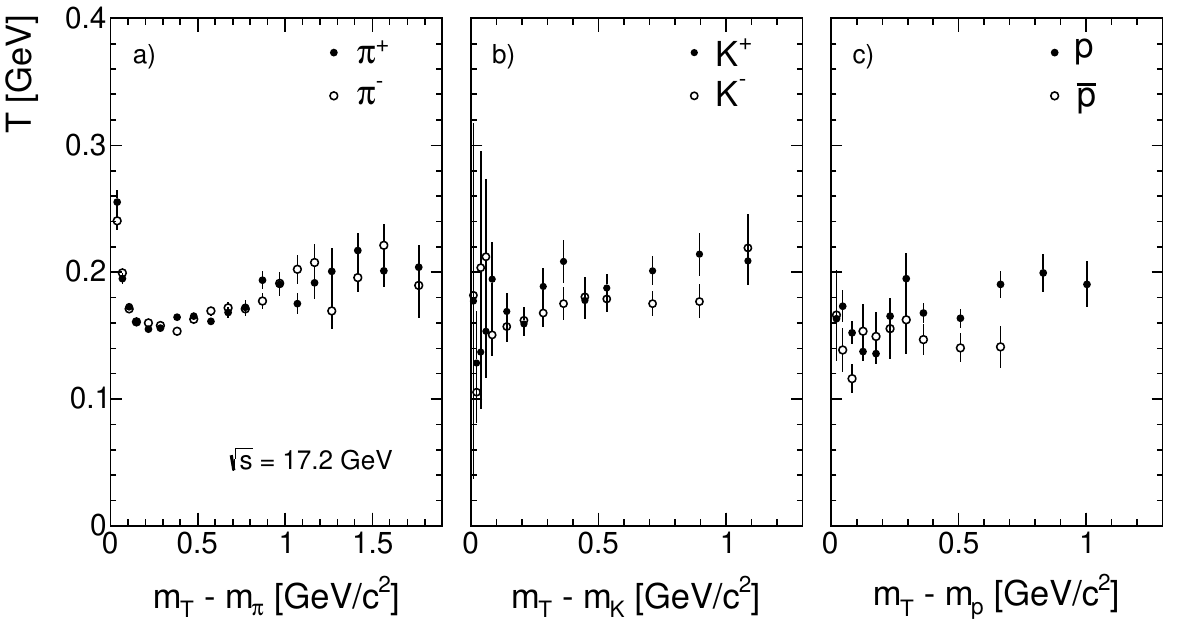} 
		\caption{Inverse slopes of the transverse mass distributions as a function of $m_T-m$ for a) pions \cite{pp_pion}, b) kaons \cite{pp_kaon} and c) protons and anti-protons \cite{pp_proton} at $\sqrt{s}$~=~17.2~GeV}
		\label{fig:invna49}
	\end{center}
\end{figure}

A first confrontation with the experimental reality is given in Fig.~\ref{fig:invna49} where $m_T-m$ distributions for pions, kaons and protons are shown for the NA49 data at $\sqrt{s}$~=~17.2~GeV and central rapidity.

Although these distributions have comparable shapes they are far from showing a constant inverse slope. In fact minima are reached at 160~MeV for $m_T-m$ between about 150 and 350~MeV (pions), at 150~MeV for 50~MeV (kaons), at 140~MeV for 150~MeV (protons) and at 130~MeV for 100~MeV (anti-protons). The strong increase below these minima is due to the flat approach to $p_T$~=~0~GeV/c (Fig.~\ref{fig:pp_lowpt}). All plots increase to temperatures of about 200~MeV above the minima at the $p_T$ limit of this study. This value is far above the "limiting temperature" given by Hagedorn \cite{hagedorn3} as 160$\pm$10~MeV.

\begin{figure}[h]
	\begin{center}
		\includegraphics[width=14cm] {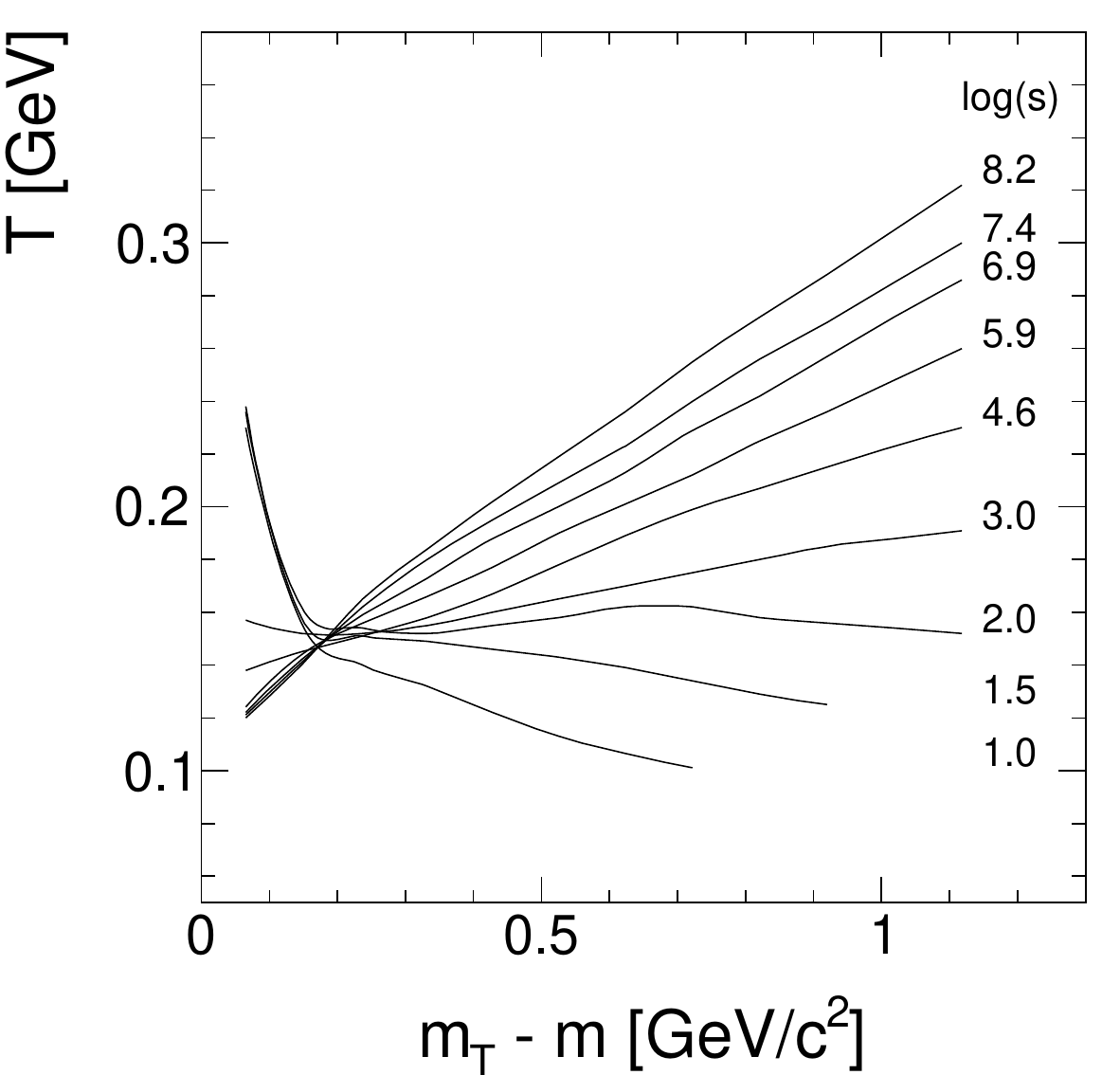} 
		\caption{Inverse slopes of the transverse mass distributions as a function of $m_T-m$ for $\pi^-$ from $\log(s)$~=~1.0 to 8.0}
		\label{fig:invtransv}
	\end{center}
\end{figure}

The situation gets more involved when using the global interpolation and LHC results over the complete energy region as shown in Fig.~\ref{fig:invtransv}.

\begin{figure}[h]
	\begin{center}
		\includegraphics[width=13cm] {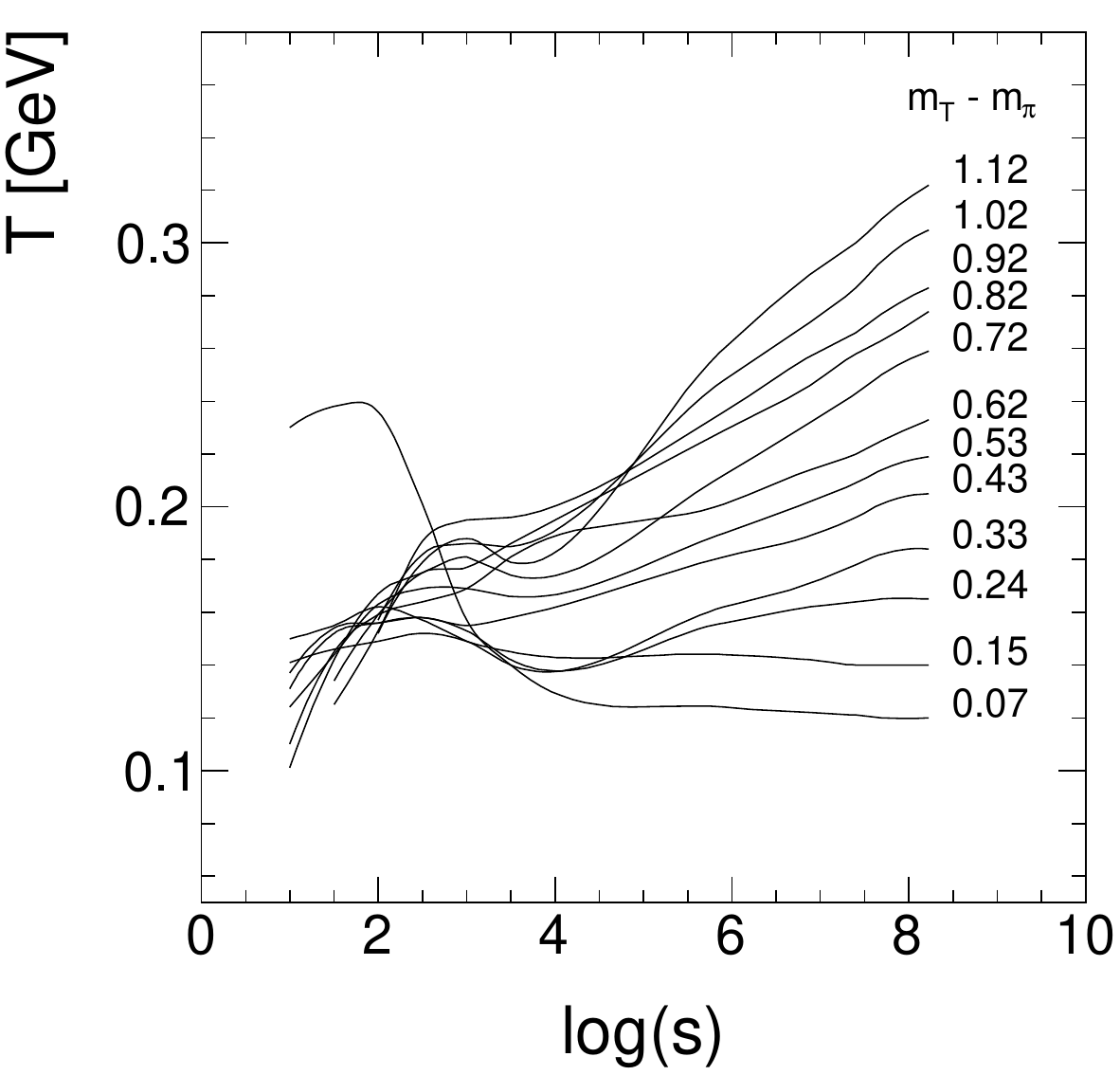} 
		\caption{Inverse slopes $T$ for constant $m_T-m$ as a function of $\log(s)$}
		\label{fig:invT}
	\end{center}
\end{figure}

Only the region of $\sqrt{s} \sim$~10~GeV (Serpukhov) exhibits a rather constant inverse slope which corresponds to the exponential $p_T$ distribution in this energy range (Figs.~\ref{fig:abramov} and \ref{fig:ptinv_dist}). Further information is given in Fig.~\ref{fig:invT} where $T$ is presented for constant values of $m_T-m$ as a function of $\log(s)$.

\begin{figure}[h]
	\begin{center}
		\includegraphics[width=14cm] {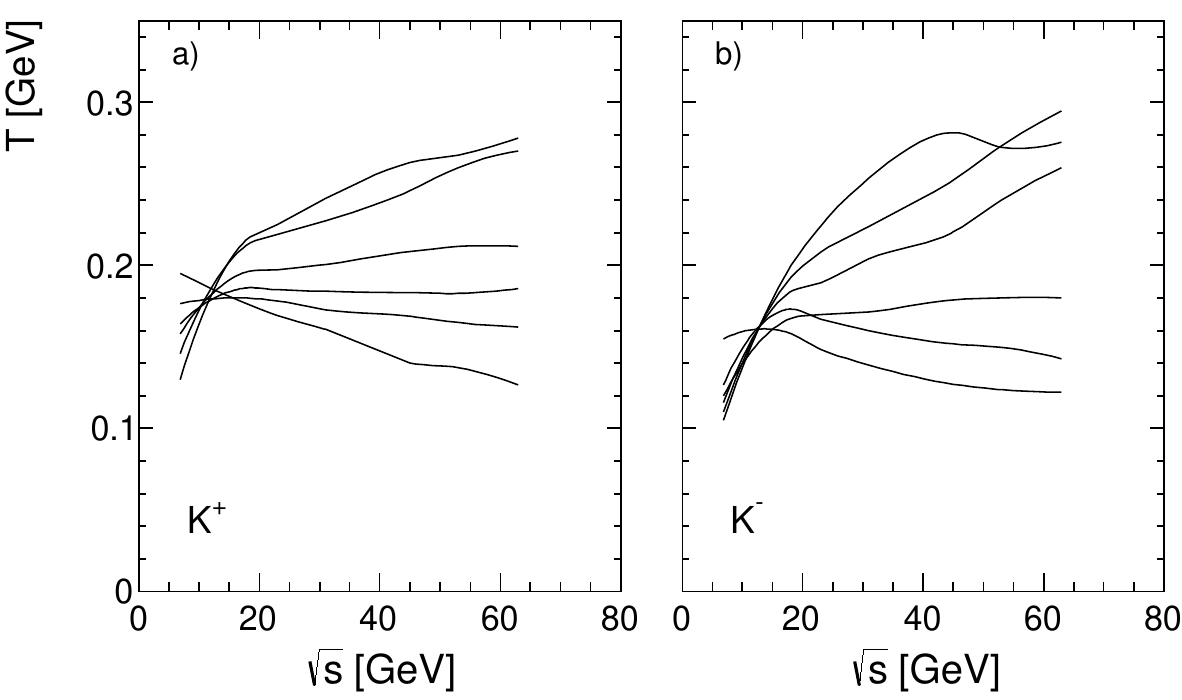} 
		\caption{Inverse slopes $T$ for constant $m_T-m$ as a function of $\sqrt{s}$ for a) K$^+$ and b) K$^-$}
		\label{fig:invTka}
	\end{center}
\end{figure}

Here again only at $\log(s)$~=~2 there is a reasonably small range of inverse slopes around 160~MeV. Below and above this energy there is a wide spread of temperatures from 100 to 300~MeV. A similar result has been obtained for kaons \cite{pp_kaon} as shown in Fig.~\ref{fig:invTka}.

The increase of inverse slopes at low $m_T-m$ (Fig.~\ref{fig:invna49}) may be attributed to the fact that a purely exponential approach becomes unphysical at $m_T-m$~=~$p_T$~=~0 since the invariant cross sections have to flatten out such that the axis at $p_T$~=~0~GeV/c is crossed horizontally. The increase beyond $m_T-m \sim$~0.4~GeV/c$^2$ poses a more principle problem as it continues to $p_T$ values much beyond the limit imposed in this paper. Originally this was attributed to occasional thermodynamic turbulence, to local temperature fluctuations or even to the "escape" of hadrons from a plasma state. An explanation seemed to be offered by QCD in the perturbative sector which allowed the description of the "high $p_T$" results at ISR and the SPS $\overline{\textrm{p}}$p collider as basically a power-law behaviour. Hence a second parameter beyond the limiting temperature was introduced \cite{hagedorn2} in this sense. A problem is however posed by the fact that perturbative QCD is certainly not applicable in a $p_T$ range of 0.4 to 1.3~GeV/c. In addition it will be shown in Sect.~\ref{sec:resonances} that particle yields are saturated at least up to $p_T$~=~3~GeV/c by resonance decay. For a practical application of statistical thermodynamics in an adapted two-parameter form see Sect.~\ref{sec:tsallis}.

There are other inconsistencies inherent to the SBM approach. The "fireballs" are supposed to have a sequence cascading from higher to lower mass states. In the thermodynamic sense each fireball should follow a Boltzmann-type thermal transverse distribution such that already in a second generation chain m$^*$ $\rightarrow$ m$^{**}$ $\rightarrow$ final state hadrons these hadrons would have a non-thermal $m_T$ distribution. This has lead to the introduction of an "effective temperature" that is far above the thermal one up to cascading masses m$^*$ of 5~GeV and more \cite{hagedorn2}. Finally, only the very last step to final hadrons is taken into account with m$^*$ having no transverse momentum. Furthermore, even in this approximation, two-body decays pose a grave problem as they would yield specific lines in $p_T$. In order to describe these cases which are common in hadron spectroscopy, $m^*$ has to have at least a thermal $m_T$ distribution. Even then the resulting final state hadron distributions are non-thermal. It is supposed that two-body decays only happen at the very end of the cascading chain and would not have incidence on the overall $m_T$ distributions. However for most of the heavier hadronic decays the multibody final states regroup into effective two-body decays. This is apparent both for baryonic and mesonic resonances, for instance

\begin{equation}
	\begin{split}
          N^*(2050) \rightarrow& N+3\pi       \\
				    &\rightarrow N+\eta       \\
					&\rightarrow N+\omega     \\
					\rightarrow& N+2\pi       \\
					&\rightarrow N+\rho       \\
					&\rightarrow \Delta+\pi    \\
					&\rightarrow N+2\pi       \\
\textrm{or}         &                         \\
		f_2(2300) 	\rightarrow& \phi+\phi     \\
					&\rightarrow K+\overline{K} + K+\overline{K}
	\end{split}
\end{equation}

\noindent
which contain effective two-body decays with large branching fractions.

It is also claimed \cite{hagedorn2} that already for three-body decays the secondary distributions would be thermal which is not the case, see Sects.~\ref{sec:feed_ddk0l} and \ref{sec:mtslopes}.

%
%
\subsubsection{Application of Tsallis statistics}
\vspace{3mm}
\label{sec:tsallis}

The use of the Tsallis form of statistical thermodynamics \cite{tsallis} to hadronic production \cite{bueyuekkilic} is not uncontested \cite{parvan}.

Nevertheless it offers a two-parameter fit similar to extension of the purely thermal fits given in \cite{hagedorn2} of the form

\begin{equation}
	\label{eq:tsallis}
	f = C m_T\left( 1 + (q-1)\frac{m_T}{T} \right)^{-\frac{1}{q-1}}
\end{equation}

\begin{figure}[h]
	\begin{center}
		\includegraphics[width=10cm] {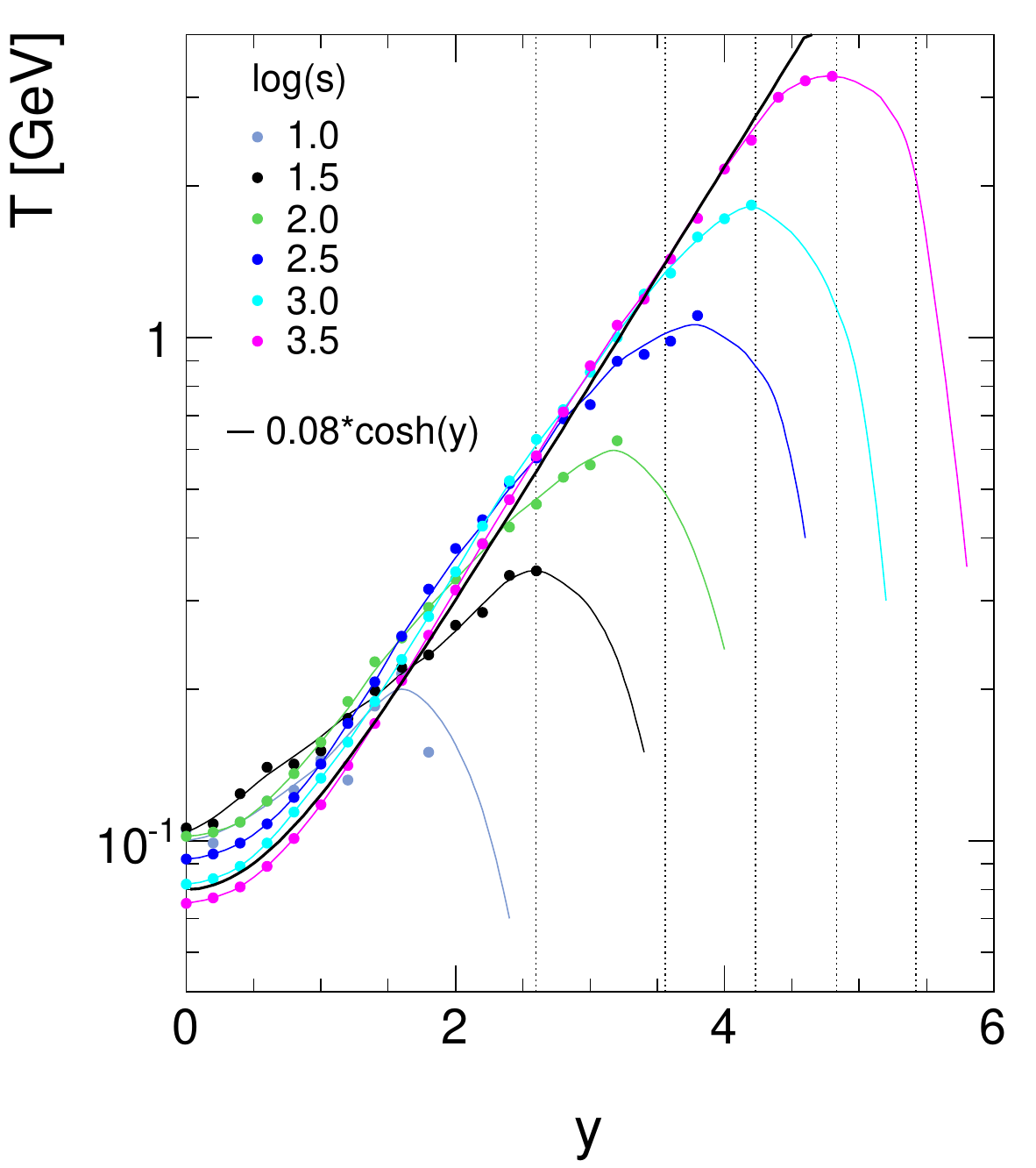} 
		\caption{Tsallis fit parameters $T$ and $q$ (\ref{eq:tsallis}) for $\log(s)$ from 1 to 3.5 as a function of rapidity}
		\label{fig:tsallis}
	\end{center}
\end{figure}

Fits of this type are widely used in the energy region from RHIC to LHC. In the lower energy range, $\log(s)$~=~1.0--3.5, using the general interpolation of $\pi^-$ results, the resulting fits with the parameters $T$ and $q$ are given in Fig.~\ref{fig:tsallis}.

A wide spread of the fit parameters is obtained which are not described by a common behaviour. In particular at the lower interaction energies the approach to the kinematic limit has to be taken into account leading to a progressive shrinking of the $p_T$ range (see also Fig.~\ref{fig:longvstrans}). A line corresponding to a $p_T$ range of 0.2~GeV/c is indicated in Fig.~\ref{fig:tsallis}.

\begin{figure}[h]
	\begin{center}
		\includegraphics[width=11.6cm] {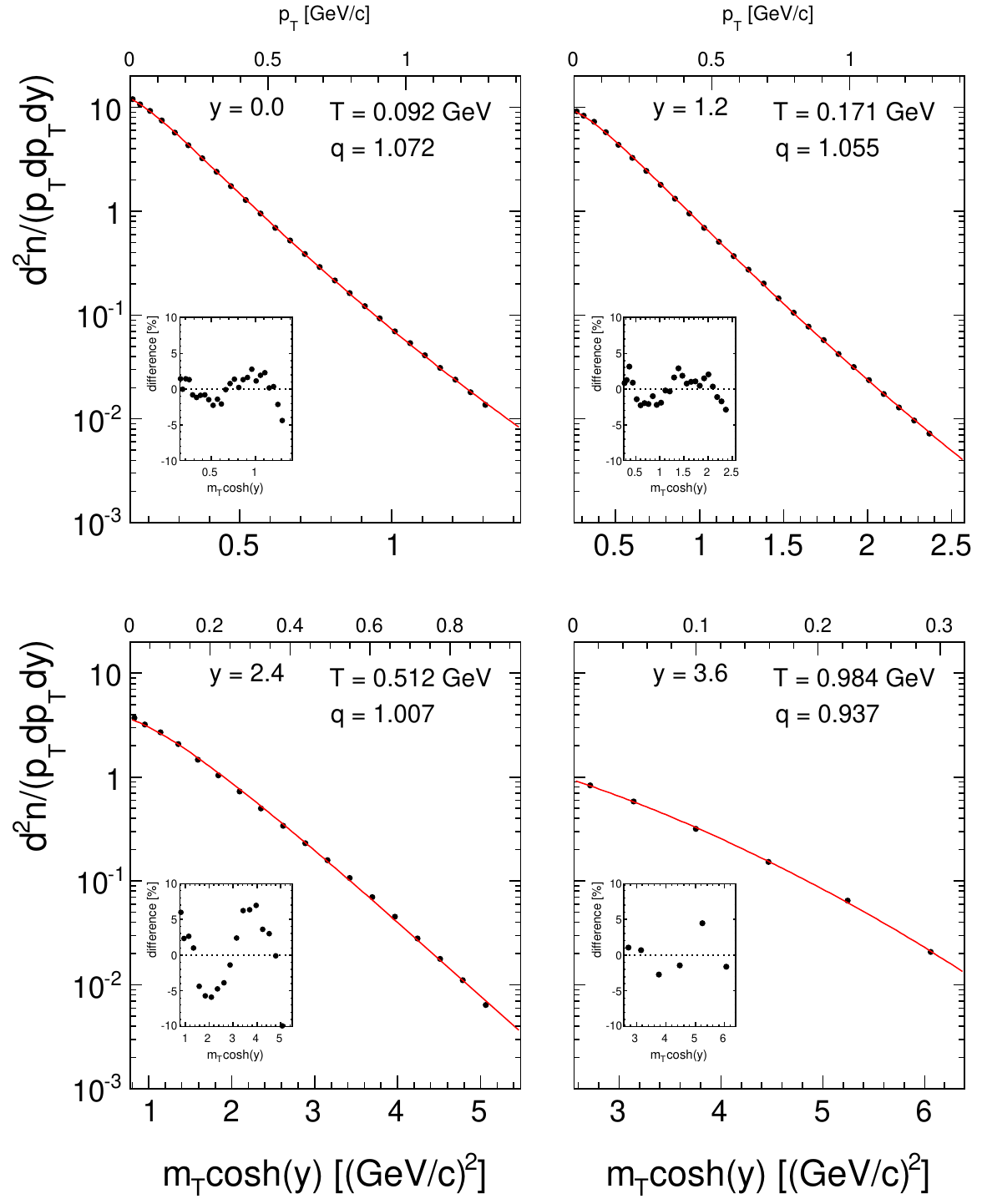} 
		\caption{Tsallis fits of $\pi^-$ cross sections for rapidities between 0 and 3.6 at $\log{s})$~=~2.5. The abscissa shows both the quantity $m_T*\cosh(y)$ and $p_T$. For each rapidity the fit parameters $T$ and $q$ as well as the residual distributions of the fits are shown}
		\label{fig:tsallis_fits}
	\end{center}
\end{figure}

Further indications of the quality of the fits are given in Fig.~\ref{fig:tsallis_fits} for $\log(s)$~=~2.5 (NA49 energy range). The cross sections $d^2n/p_Tdp_Tdy$ are given for rapitities from 0 to 3.6 together with the residual distributions as a function of both $m_T*\cosh(y)$ and $p_T$ indicating the strong decrease of the $p_T$ range at forward rapidities.

%
%
\section{\boldmath $p_T$ integrated distributions and total $\pi^-$ yields}
\vspace{3mm}
\label{sec:integrated}

Following the discussion of the double differential results in the preceding chapters integrated distributions up to the determination of the resulting total $\pi^-$ yields will be presented here. There are several constraints and boundary conditions to be considered:

\begin{enumerate}[label=(\roman*)]
	\item The global interpolation is limited in $p_T$ to an upper limit of 1.3~GeV/c. The influence of this limit on the integration will be quantified by extending the range to 1.9~GeV/c using an exponential extrapolation.
	\item There is an upper limit also in $x_F$ which depends on $\sqrt{s}$ and is imposed by the limit of published cross sections at about 1~$\mu$b. Also here an extrapolation to higher $x_F$ will be used to quantify the influence on the integral.
	\item Results both with and without feed-down subtraction will be presented in order to quantify this important contribution.
	\item Due to the limited phase space coverage of the high-energy colliders the results only reach up to ISR energy at $\log(s)$~=~3.6.
\end{enumerate}

%
%
\subsection{\boldmath $dn/dx_F$ distributions}
\vspace{3mm}
\label{sec:dndx}

\begin{figure}[h]
	\begin{center}
		\includegraphics[width=15.cm] {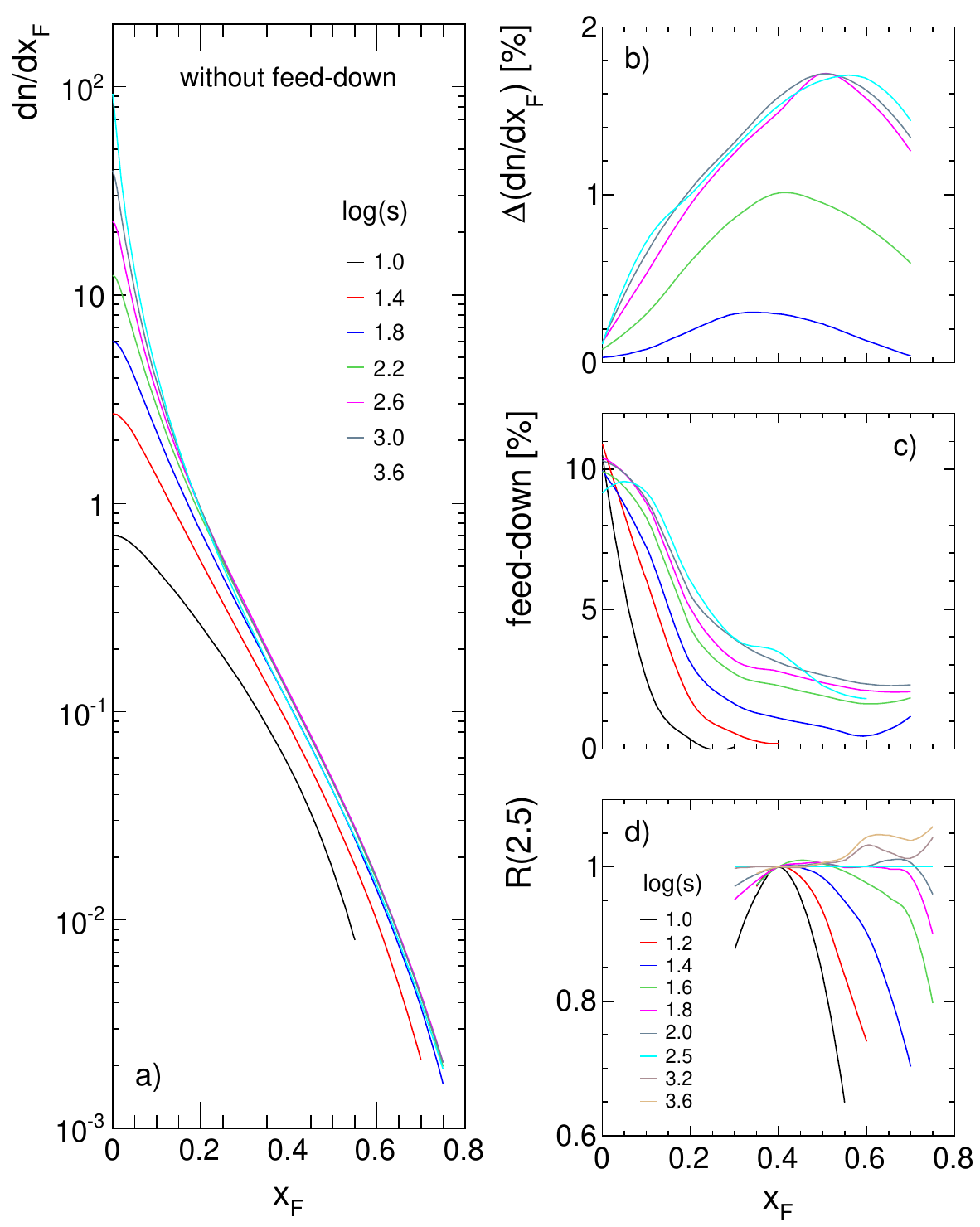} 
		\caption{a) $dn/dx_F$ distributions as function of $x_F$ for several $\log(s)$ values from 1 to 3.6 using the $p_T$ integration up to 1.3~GeV/c, b) percent increase of $dn/dx_F$ extending the $p_T$ range from 1.3 to 1.9~GeV/c, c) percent increase of $dn/dx_F$ by adding the feed-down contribution, d) variation of $dn/dx_F$ in the high-$x_F$ region relative to $\log(s)$~=~2.5 normalizing all distributions to unity at $x_F$~=~0.4}
		\label{fig:dndxf}
	\end{center}
\end{figure}

$dn/dx_F$ distributions are shown in Fig.~\ref{fig:dndxf} as a function of $x_F$ for several values of $\log(s)$ from the lowest available value to the highest ISR energy for feed-down subtracted data.

The distributions span five orders of magnitude in $dn/dx_F$ and feature a strong $\log(s)$ dependence for $x_F \lesssim$~0.2 followed by a region of shape similarity up to the highest $x_F$ at around 0.75. The $p_T$ integration is performed up to the limit of the global interpolation at $p_T$~=~1.3~GeV/c. The increase of $dn/dx_F$ when extending the upper $p_T$ limit to 1.9~GeV/c is shown in panel b). The increase is limited to 2\% in the upper range of $\log(s)$. Panel c) presents the \% increase of $dn/dx_F$ when adding the feed-down contribution. The percentage increases steadily with $\log(s)$ up to a maximum
of about 10\% at low $x_F$ followed by a strong decrease towards high $x_F$. Panel d) addresses the shape similarity of the $dn/dx_F$ distributions in the high-$x_F$ region. The distributions are normalized to unity at $x_F$~=~0.4 and their ratio to a reference at $\log(s)$~=~2.5 is plotted.

This last panel allows a rather precise check of longitudinal scaling. This concept has been treated in Sect.~\ref{sec:scaling} on the double differential level of cross sections. With respect to the reference line at $\log(s)$~=~2.5 or $\sqrt{s}$~=~17.8~GeV a steady evolution of the $\pi^-$ density is evident. The line is approached from below with increasing energy and a short plateau between 0.5 and 0.7 in $x_F$ and between 2.4 and 2.8 in $\log(s)$ is followed by a further increase at ISR energy. This phenomenology does not correspond to a manifestation of the down-quark structure function as postulated  by some models \cite{ochs,das}. In this context it is regrettable that the far forward region is not attainable at the high energy colliders above the ISR energy range.

%
%
\subsection{\boldmath $y$ and $y_{\textrm{lab}}$ distributions}
\vspace{3mm}
\label{sec:dndy}

\begin{figure}[h]
	\begin{center}
		\includegraphics[width=14.cm] {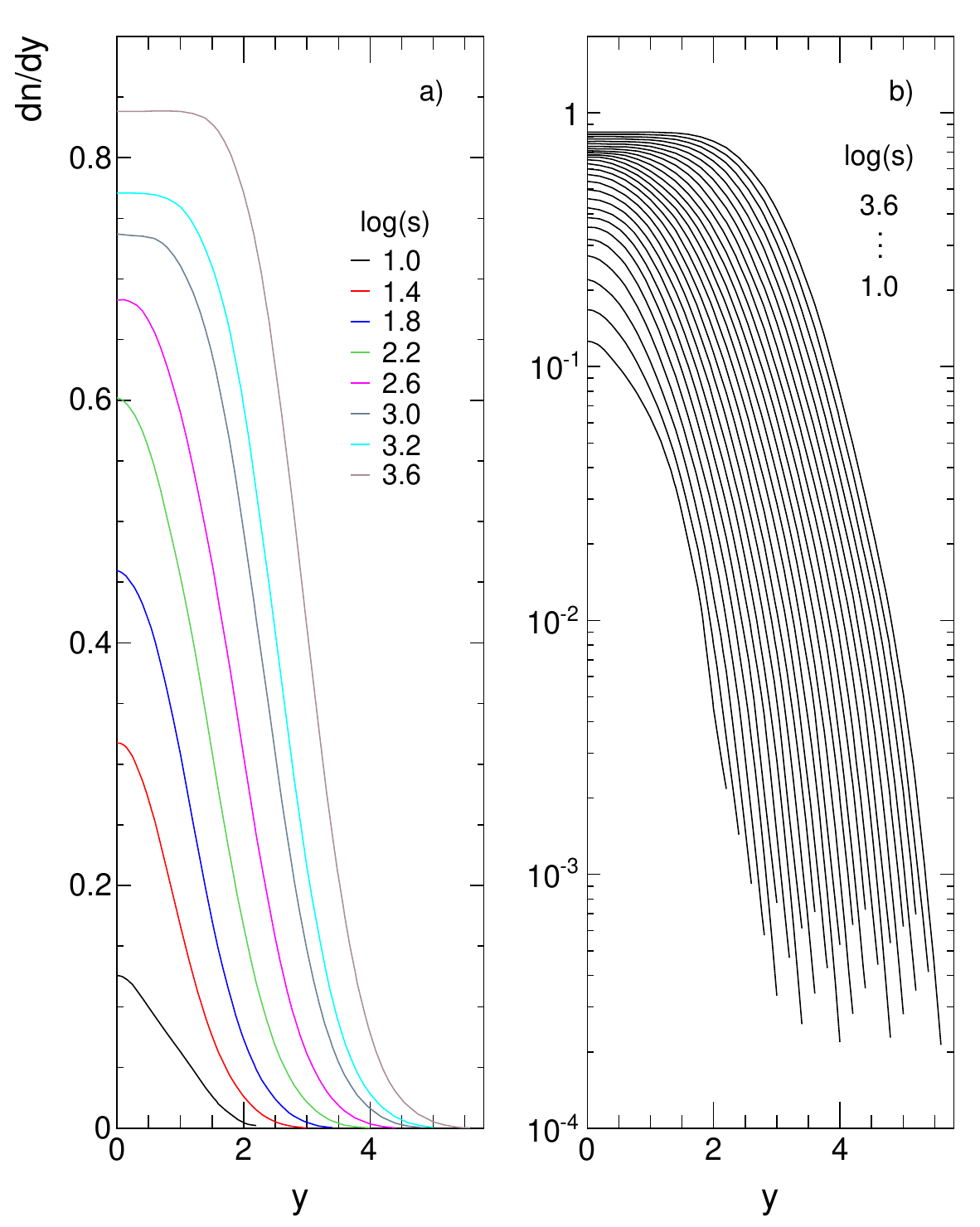} 
		\caption{$dn/dy$ distributions as functions of $y$ for several $\log(s)$ values from 1 to 3.6 using the $p_T$ integration up to 1.3~GeV/c a) with a linear scale on the ordinate  b) with a logarithmic scale for a dense coverage in $\log(s)$ from 1 to 3.6 in steps of 0.1}
		\label{fig:dndy}
	\end{center}
\end{figure}

\begin{figure}[h]
	\begin{center}
		\includegraphics[width=14.8cm] {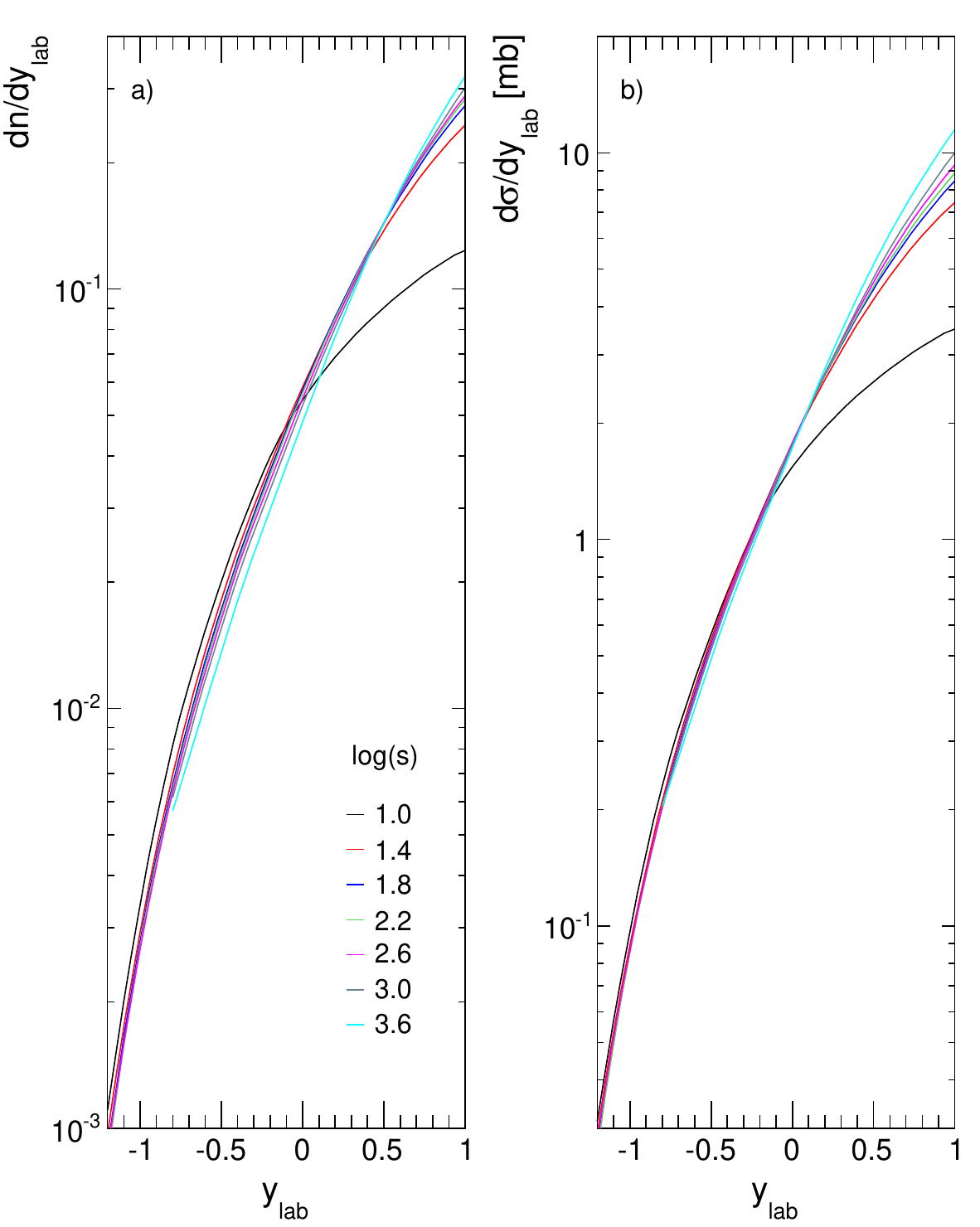} 
		\caption{$y_{\textrm{lab}}$ distributions of a) the rapidity density per inelastic event and b) the integrated invariant cross section as a function of $y_{\textrm{lab}}$ for seven values of $\log(s)$}
		\label{fig:dndylab}
	\end{center}
\end{figure}

$p_T$ integrated rapidity distributions are shown in Figs.~\ref{fig:dndy} and \ref{fig:dndylab} for a subset of interaction energies ranging from $\sqrt{s}$~=~3~GeV to the highest ISR energy at 63~GeV. The linear plot of Fig.~\ref{fig:dndy} brings out the well-known features of such distributions: a strong increase of the central $\pi^-$ density with interaction energy developing into a "rapidity plateau" followed by a steep decrease into the forward direction.

The presence of an increasing rapidity plateau fostered ideas of a "central" production mechanism as opposed to a "fragmentation region" in forward direction with little if any energy dependence.

The study of the central rapidity region using double differential cross sections in Sect.~\ref{sec:non-scaling} shows however a more complex situation. In fact there are two components contributing to the increasing particle yield: a first part at low $p_T$ typical of low-$Q$ resonance decay as it is also seen in the feed-down from weak decays, Sect.~\ref{sec:fd_corr}, followed by a steady increase towards higher $p_T$ again due to the decay of resonances with increasing mass hence high $Q$ decay.

The steady broadening of the $y$ distributions with $\log(s)$ is a trivial consequence of the logarithmic increase of the extent of the longitudinal phase space with energy (\ref{eq:ybeam}). The decrease of the cross section towards higher rapidity is closely conformal as shown in Fig.~\ref{fig:dndy}b) for a dense sequence of $\log(s)$ values. Hence the ideas of "energy scaling" in this region.

The scaling hypothesis is further detailed by replacing the rapidity variable by $y_{\textrm{lab}} = y_{\textrm{beam}} - y$ (\ref{eq:ylab}) thus taking account of the increasing phase space and eliminating one trivial component. This is shown in Fig.~\ref{fig:dndylab} which in addition presents two normalizations of the yield: In panel a) the density per inelastic event,

\begin{equation}
	\frac{dn}{dy} = \frac{\pi}{\sigma_{\textrm{inel}}} \int f dp_T^2
\end{equation}

\noindent
and panel b) the integrated invariant cross section

\begin{equation}
	\frac{d\sigma}{dy} = \pi \int f dp_T^2
\end{equation}

As already evident in the double differential case, Sect.~\ref{sec:feynman_scaling}, the integrated invariant cross section shows less $s$-dependence than the rapidity density. This is another evidence for the necessity to properly address the variation of the inelastic cross section as a function of interaction energy and, connected to this, the question of the impact parameter dependence of a given inclusive phenomenology.

%
%
\subsection{Total \boldmath $\pi^-$ yields}
\vspace{3mm}
\label{sec:totalpim}

The $p_T$ integrated distributions discussed above may now be integrated over $x_F$ or $y$ in order to determine the total $\pi^-$ yield as a function of $\log(s)$. The resulting numerical values are given in Table~\ref{tab:totalpim} both with and without feed-down contribution. The extrapolation from the $p_T$ limit at 1.3~GeV/c to 1.9~GeV/c as well as the extrapolation from $x_F$~=~0.75 to $x_F$~=~0.95 contribute each less than 0.4\% to the yield.

\begin{table}[h]
	\renewcommand{\tabcolsep}{0.5pc}
	\renewcommand{\arraystretch}{1.15}
	\footnotesize
	\begin{center}
		\begin{tabular}{l|llllllllll}
			\hline
	$\log(s)$ & 1.0  &  1.1  &  1.2  &  1.3  &  1.4  &  1.5  &  1.6  &  1.7  &  1.8  &  1.9 \\ \hline
	with FD		&0.2649 &0.3574 &0.4736 &0.6077 &0.7473 &0.8844 &1.0214 &1.1659 &1.3229 &1.4920 \\
	without FD	&0.2581 &0.3446 &0.4527 &0.5767 &0.7049 &0.8305 &0.9557 &1.0874 &1.2306 &1.3847 \\ \hline \hline
	$\log(s)$ & 2.0  &  2.1   & 2.2  &  2.3 &   2.4  &  2.5  &  2.6   & 2.7  &  2.8  &  2.9 \\ \hline
	with FD		&1.6710 &1.8572 &2.0479 &2.2422 &2.4395 &2.6417 &2.8496 &3.0685 &3.2986 &3.5403 \\
	without FD	&1.5474 &1.7161 &1.8883 &2.0631 &2.2406 &2.4238 &2.6125 &2.8113 &3.0209 &3.2419 \\ \hline \hline
	$\log(s)$  &  3.0   & 3.1 &   3.2  &  3.3  &  3.4  &  3.5  &  3.6 &&& \\ \hline
	with FD		&3.7895 &4.0504 &4.3200 &4.5963 &4.8778 &5.1648 &5.4547 &&& \\
	without FD	&3.4700 &3.7094 &3.9578 &4.2137 &4.4757 &4.7440 &5.0158 &&& \\ \hline
		\end{tabular}
		\caption{Total $\pi^-$ yield as a function of $\log(s)$ with and without feed-down contribution}
		\label{tab:totalpim}
	\end{center}
\end{table}

\begin{figure}[h]
	\begin{center}
		\includegraphics[width=8cm] {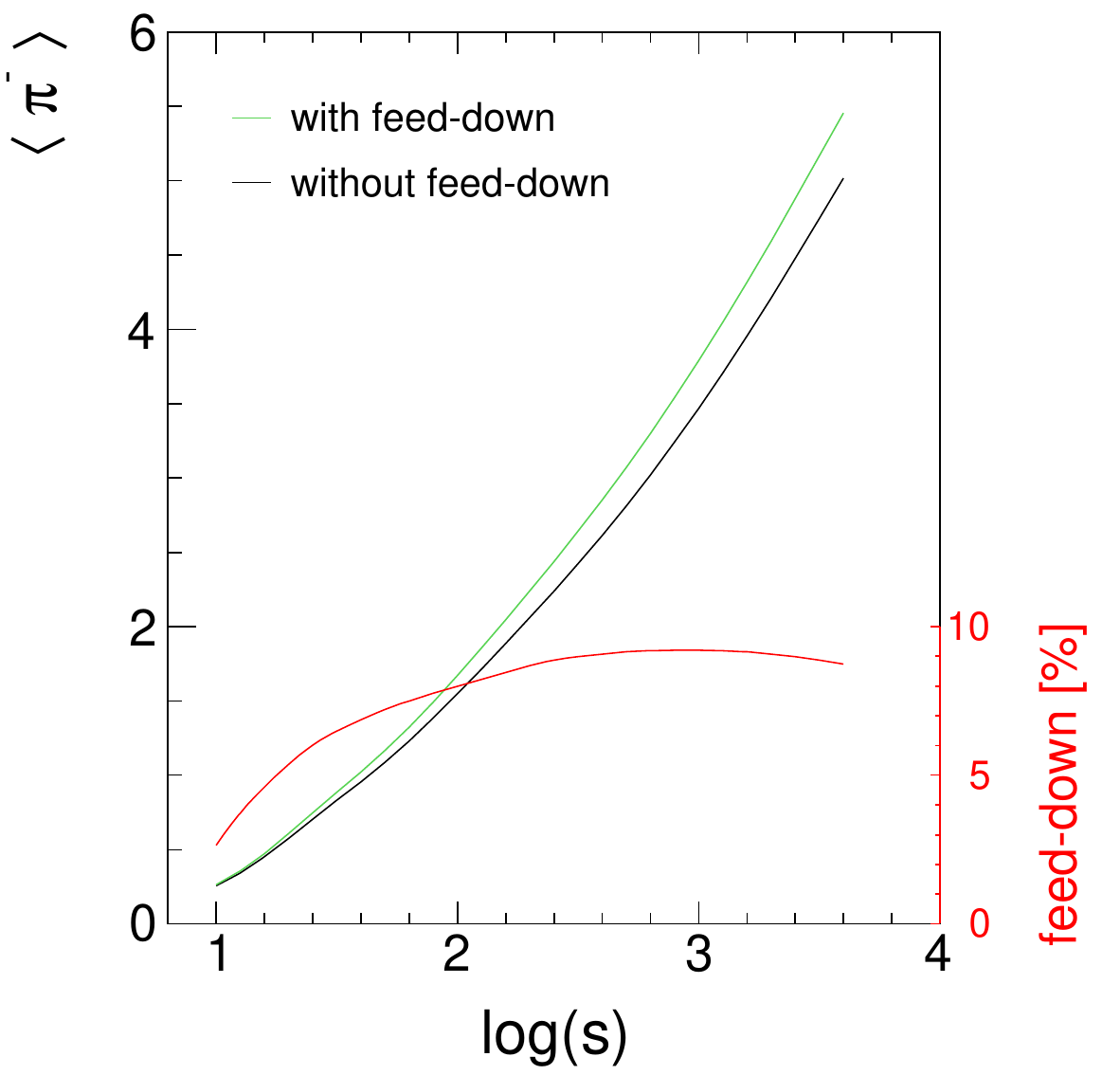} 
		\caption{Total $\pi^-$ yields as a function of $\log(s)$ with and without feed-down contribution (left scale) together with the feed-down component in percent (right scale)}
		\label{fig:totalpim}
	\end{center}
\end{figure}

These values are shown in Fig.~\ref{fig:totalpim} as a function of $\log(s)$ together with the contribution of the feed-down component in percent.

%
%
\section{Resonances and their decay}
\vspace{3mm}
\label{sec:resonances}

It has been known since decades that final state hadrons are decay products of resonances \cite{grassler,jancso,whitmore} to a fraction which ranged, at the time, from 80 to 90\%. To date one may assume that all of these hadrons stem from resonances, in other words there is no "direct" production. Resonances form a cloud of states above the ground state of stable hadrons which is of extraordinary diversity and extent concerning their masses, quantum numbers and decay branching fractions. Their study is thus indispensable for any progress of understanding that goes beyond the realm of purely inclusive single particle physics.

In order to determine the contribution of a given resonance to the final state inclusive particle cross section several quantities have to be known:

\begin{enumerate}
	\item Central mass and width \\
	The resonance width may change from zero for weak decays (Sect.~\ref{sec:fd_corr}) to several hundred MeV for high-mass, strongly decaying states.
	\item Mass distribution \\
	Resonances are in general characterized by asymmetric Breit-Wigner type mass distributions which may extend up to masses far above the nominal value. Their important influence on tail effects in the inclusive cross sections will be evoked below.
	\item Decay branching fractions \\
	For non-exotic decays as they are treated in this paper, branching fractions span a range from a few to almost 100\%. They are generally well measured for mesonic and heavy flavour resonances as they are readily accessible in e+e colliders. For baryon resonances the situation is less favourable. Here in general only a few channels have been isolated with rather large error margins.
	\item Resonance production cross sections over full phase space \\
	Here the production yields have in principle to be known, on a double-differential level, over the full phase space. Given the problems of coverage for the final state hadrons evoked in this paper, it is not surprising that this condition is hardly fulfilled for most resonances. Here single-differential distributions integrated over $p_T$ or $x_F$ will have to do in most cases.
\end{enumerate}

In the following sections, a particular resonance, namely the $\Delta^{++}$(1232), will be used to clarify some details concerning the problems mentioned above. This resonance has been studied by about 30 experiments over a wide range of interaction energies and for different projectiles on proton targets. The $\Delta^{++}$ has an almost 100\% branching fraction into the two-body channel p + $\pi^+$. The asymmetry in the decay particle masses allows, in addition, to elaborate the mass effect on the final state particle distributions.

Some references for p+p interactions are given in \cite{ammosov2,erwin,debrion,brick1,barish,dao,drijard,aguilar,bracinik,kreps}.

These earlier publications concern mostly bubble chamber work and are concentrated in the 1970's, see the discussion in Sect.~\ref{sec:exp_sit}, and Fig.~\ref{fig:year}. Refs.\cite{drijard,aguilar} come from spectrometer experiments at the ISR and SPS.

Experimental information on the $\Delta$(1232) quadruplet at $\sqrt{s}$~=~17.2~GeV is available from two Theses \cite{bracinik,kreps} in the context of the NA49 experiment.

%
%
\subsection{\boldmath $\Delta^{++}$ decay: General considerations}
\vspace{3mm}
\label{sec:deltapp_dec_gen}

The phase space distribution of hadrons from resonance decay is characterized by a two-step process.

The first step is given by the disintegration process in the resonance cms which is defined by the resonance and the decay particle masses. For a two body decay, the secondary particles have equal momentum independent of their respective masses. This disintegration is modulated by a spin dependent decay angular distribution which is in general not uniform.

In a second step a Lorentz transformation from the resonance cms to the overall cms as defined by the colliding initial state hadrons has to be performed. As the resonance has both longitudinal and transverse momentum in this system the transformation is three-dimensional.

There are two complications involved with this straight-forward decay process. Firstly, the resonance has in general not a fixed mass but features a complex asymmetric mass distribution with a long tail towards high masses. Secondly, a three dimensional boost is not easily amenable to simple algebraic formulations. Here it is specifically the resonance transverse momentum which creates uncomfortable complications. It is therefore rather customary to neglect the one or the other complication (or both) in hadron production models. Although it has generally to be admitted that a hadronic collision results in a spectrum of massive objects (often named "fireballs") their decay into the final state hadrons is rarely specified in detail. As an example  the "NOVA" model \cite{jacob} specifies a rather narrow Gaussian-type "fireball" mass spectrum centred at relatively low masses, but it does not allow these objects to have transverse momentum. On the other hand Hagedorn's "Thermal" model \cite{hagedorn1} allows for a fireball spectrum but again neglects their transverse momentum. In addition final state particle production is assumed to be described by Boltzmann radiation from an equilibrium thermodynamic state where mass dependence is introduced at this level only.


In view of this situation $\Delta^{++}$ decay will be treated in the following Sections in some detail by starting with the fixed nominal mass 1.232~GeV/c$^2$ and an $x_F$ distribution without allowing for transverse momentum. The transverse momentum and the Breit-Wigner mass distributions will then be successively introduced.

%
%
\subsection{\boldmath $\Delta^{++}$ resonance: $x_F$ and $p_T^2$ distributions}
\vspace{3mm}
\label{sec:deltapp_res}

\begin{figure}[h]
	\begin{center}
		\includegraphics[width=12.cm] {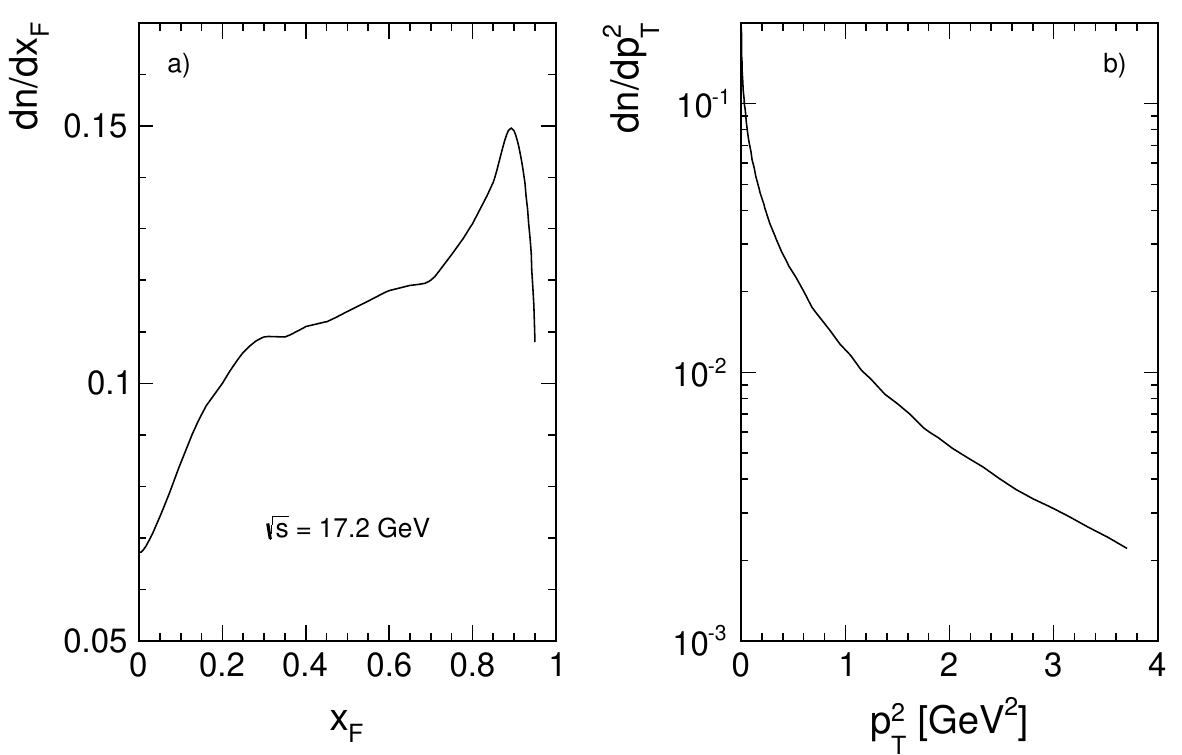} 
		\caption{Single differential $\Delta^{++}$ distributions at $\sqrt{s}$~=~17.2~GeV: a) $dn/dx_F$ as a function of $x_F$, b) $dn/dp_T^2$ as a function of $p_T^2$}
		\label{fig:delta2ppi}
	\end{center}
\end{figure}

From the rich sample of experimental results mentioned above the phase space distribution of the Delta resonance may be extracted, in the present argumentation at the NA49 energy of 17.2~GeV. These distributions are shown in Fig.~\ref{fig:delta2ppi}.

The $dn/dx_F$ distribution integrates to 0.22 $\Delta^{++}$ per inelastic event at $\sqrt{s}$~=~17.2~GeV. This contributes, given the 100\% branching fraction, about 7\% to the total $\pi^+$ yield and 20\% to the total p yield at this energy \cite{pp_pion,pp_proton}.

%
%
\subsection{\boldmath $\Delta^{++}$ decay: Mass as a delta function at 1.232~GeV and no $p_T$ allowed}
\vspace{3mm}
\label{sec:deltapp_dec_mass}

The Lorentz transformation between the resonance cms and the p+p cms is defined by the momentum vector of the resonance and decay particle masses via the energy and momentum values in the resonance cms. For the nominal $\Delta^{++}$ mass the decay momentum is $q$~=~0.227~GeV/c independent of the particle mass $m_{\textrm{dec}}$ which enters only via the energy factor

\begin{equation}
	E_{\textrm{dec}} = \sqrt{ q^2 + m_{\textrm{dec}}^2}
	\label{eq:edec}
\end{equation}

\begin{figure}[h]
	\begin{center}
		\includegraphics[width=9.cm] {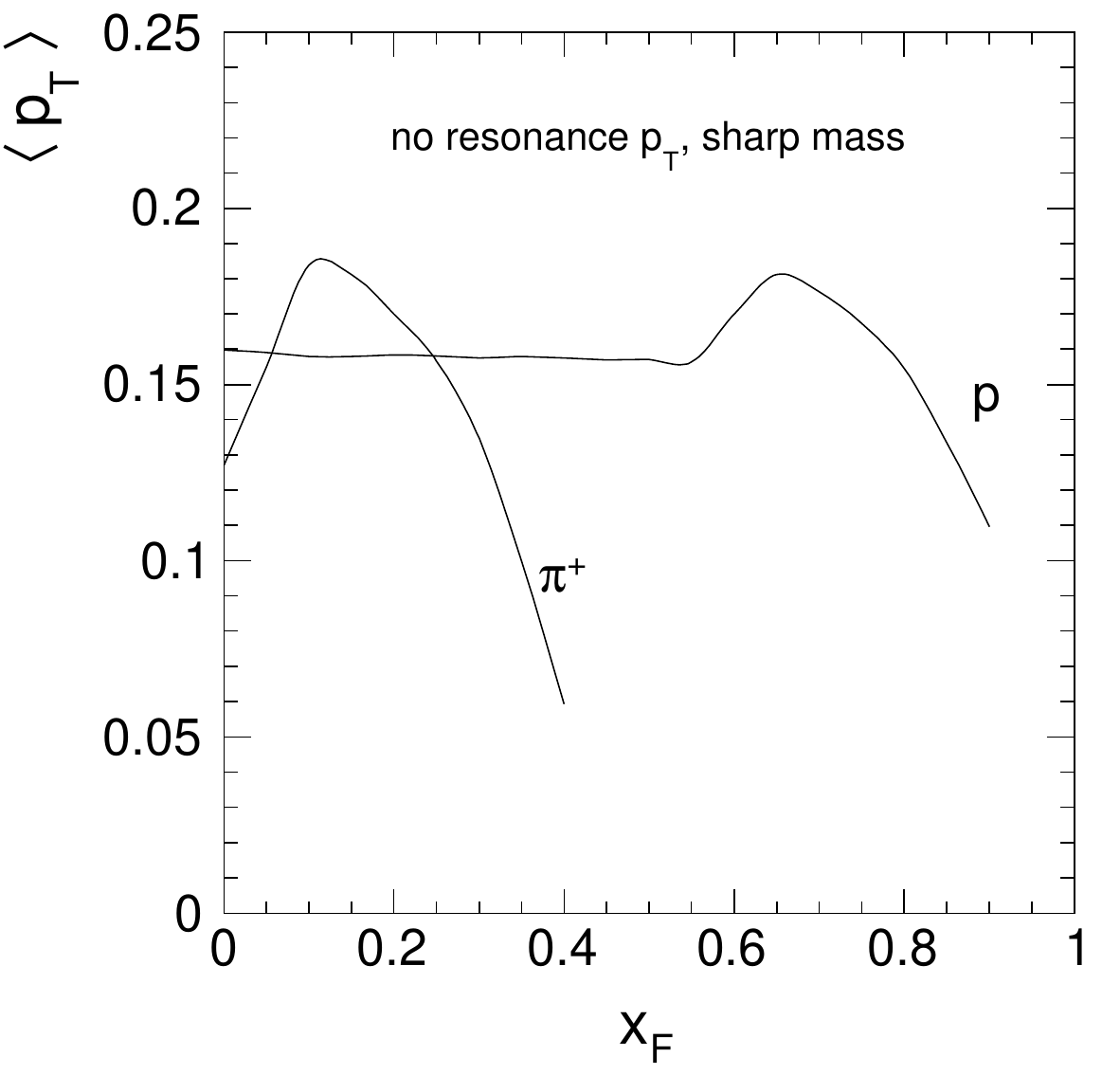} 
		\caption{Mean $p_T$ for decay protons and pions as a function of $x_F$}
		\label{fig:delta_meanpt}
	\end{center}
\end{figure}

Excluding the resonance transverse momentum, the $p_T$ of the decay particles is limited by the decay momentum $q$ to 0.227~GeV/c, independent on particle mass. This results in the mean $p_T$ as a function of $x_F$ as shown in Fig.~\ref{fig:delta_meanpt} for the decay protons and pions.

The Lorentz boost is purely longitudinal and yields, for a given resonance momentum $p_{l,\textrm{res}}$, the Lorentz factors

\begin{align}
	\beta =& \frac{p_{l,\textrm{res}}}{\sqrt{ p_{l,\textrm{res}}^2 + m_{\Delta}^2 }} \label{eq:beta}\\
	\textrm{and} & \nonumber \\
	\gamma =& \frac{1}{\sqrt{ 1 - \beta^2 }} \label{eq:gamma}
\end{align}

\noindent
and hence the decay particle longitudinal momenta between

\begin{align}
	p_{l,\textrm{min}} &= \gamma( -q + \beta E_{\textrm{dec}} )  \label{eq:pmin} \\
	\textrm{and} & \nonumber \\
	p_{l,\textrm{max}} &= \gamma( +q + \beta E_{\textrm{dec}} )  \label{eq:pmax}
\end{align}

\begin{figure}[h]
	\begin{center}
		\includegraphics[width=9cm] {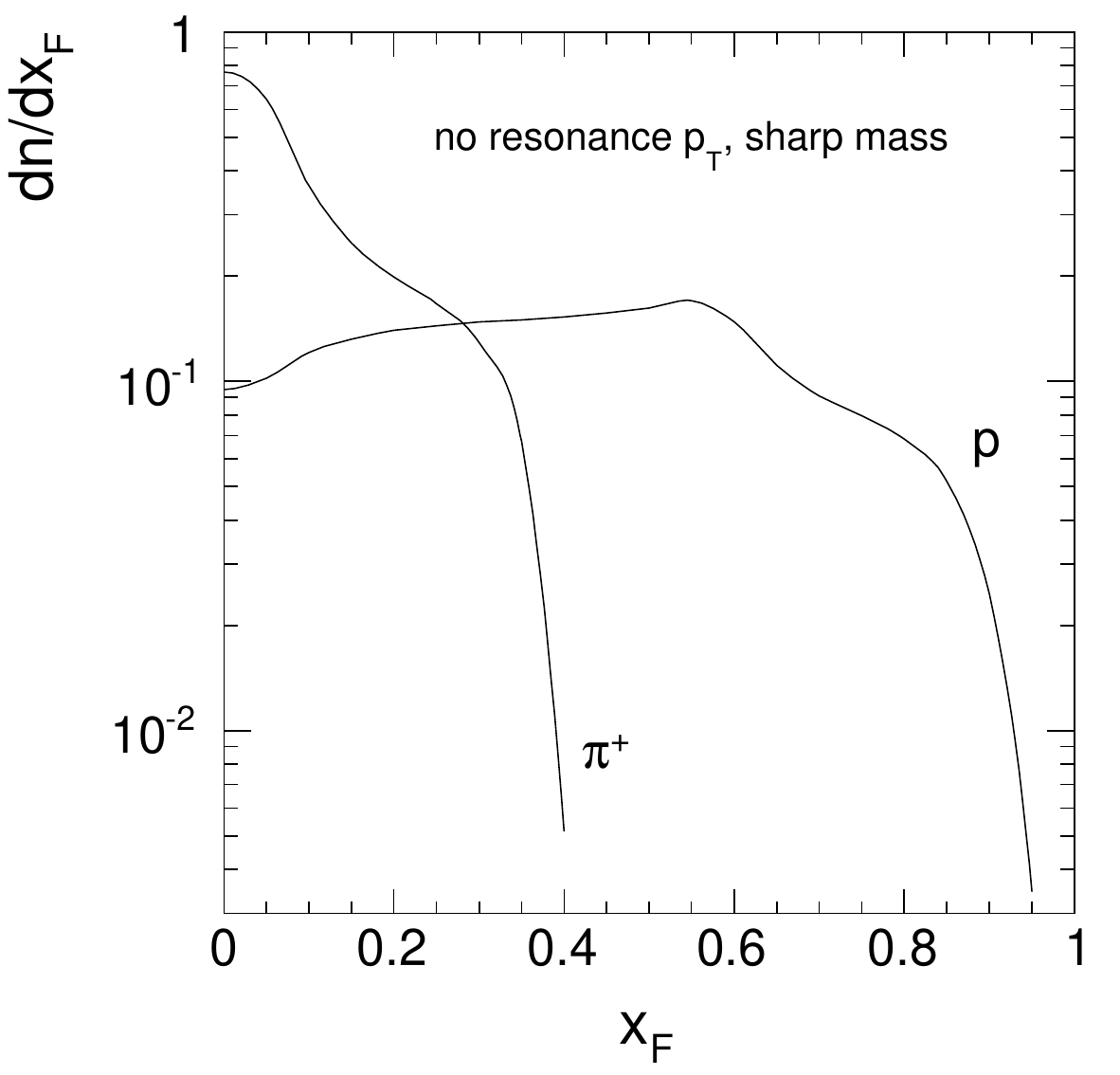} 
		\caption{$dn/dx_F$ for decay protons and pions as a function of $x_F$}
		\label{fig:delta_xfdep}
	\end{center}
\end{figure}

This leads to an explicit mass dependence in the $x_F$ dependence as shown in Fig.~\ref{fig:delta_xfdep} for the $p_T$ integrated quantity $dn/dx_F$.

The neglecting of the resonance transverse momentum and mass distribution is rather common in model calculations, albeit for practical considerations concerning the possibility of algebraic solutions \cite{hagedorn1,jacob}. It leads to the wide-spread belief that low-mass secondaries are all centred at low $x_F$ and low $p_T$ \cite{whitmore}.

%
%
\subsection{\boldmath $\Delta^{++}$ decay: adding resonance transverse momentum}
\vspace{3mm}
\label{sec:deltapp_dec_pt}

\begin{figure}[h]
	\begin{center}
		\includegraphics[width=9.5cm] {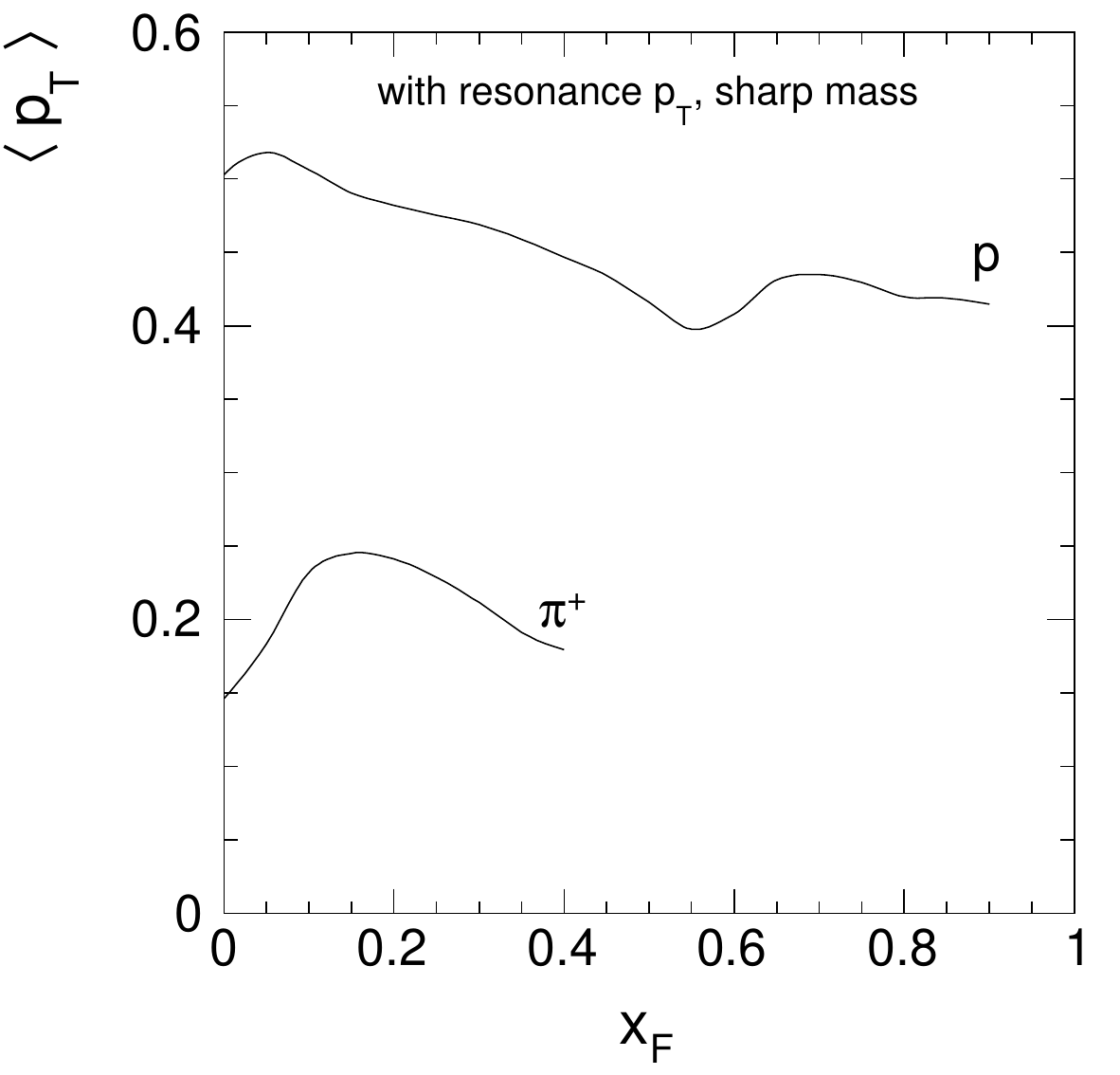} 
		\caption{Mean $p_T$ for decay protons and pions as a function of $x_F$}
		\label{fig:delta_ppimeanpt}
	\end{center}
\end{figure}

\begin{figure}[h]
	\begin{center}
		\includegraphics[width=9.5cm] {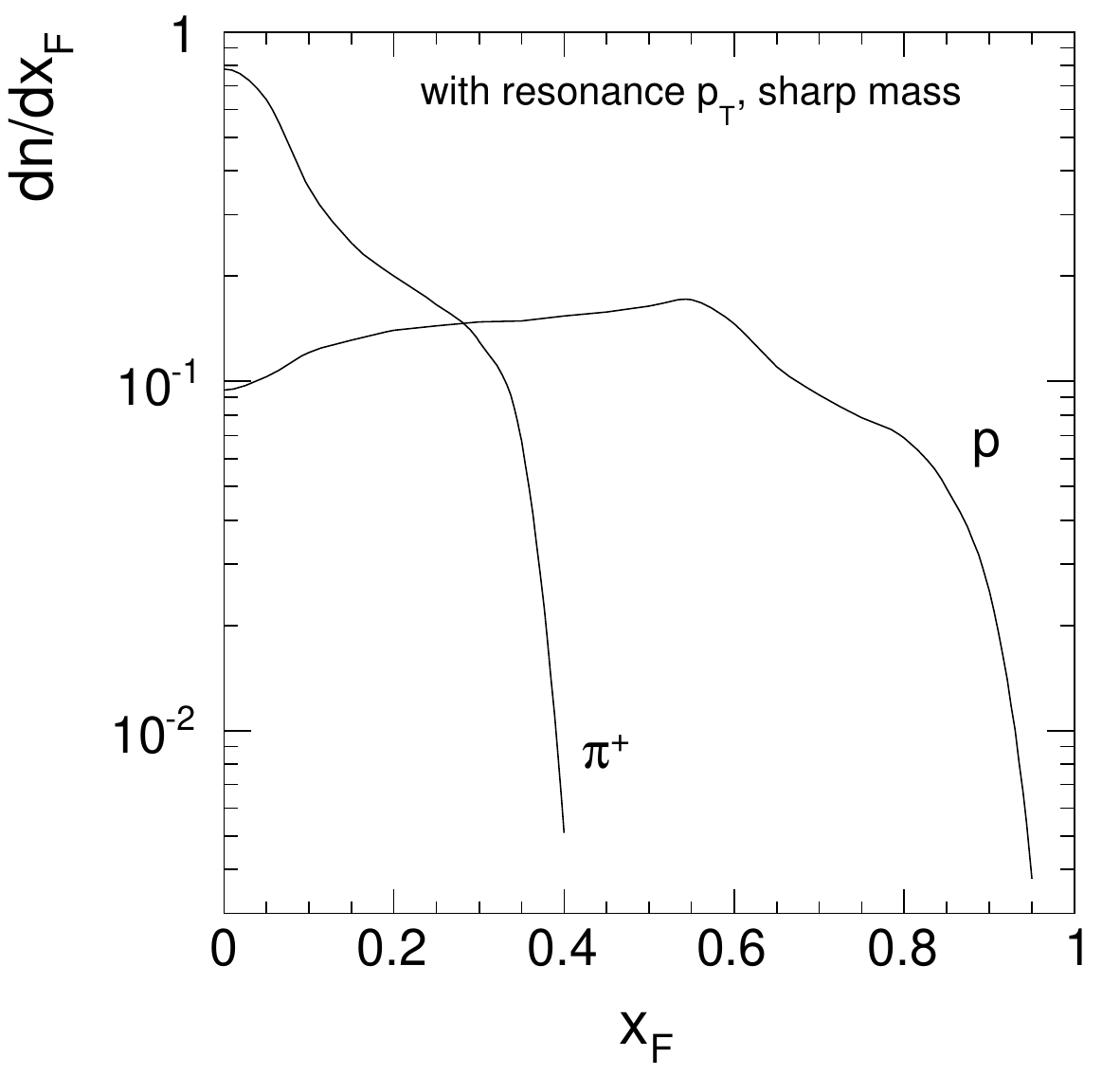} 
		\caption{$dn/dx_F$ for decay protons and pions as a function of $x_F$}
		\label{fig:delta_ppixfdep}
	\end{center}
\end{figure}

In taking account of the measured $p_T$ distribution of the $\Delta^{++}$ the Lorentz transformation gets a transverse component. At $x_F$~=~0. this transverse boost replaces $p_L$ with $p_T$ in Eqs.~\ref{eq:beta} to \ref{eq:pmax} thereby creating a mass dependence also in transverse direction. This is shown in Figs.~\ref{fig:delta_ppimeanpt} and \ref{fig:delta_ppixfdep} for the mean $p_T$ and $dn/dx_F$ as a function of $x_F$ both for the decay pion and proton.

It is to be noted that the increase of $\langle p_T \rangle$ for pions varies from 16\% at $x_F$~=~0 to 300\% at $x_F$~=~0.4 whereas for protons it is much larger at $x_F$~=~0 with 300\% decreasing slightly to 250\% at $x_F$~=~0.7. This originates from the mass dependence of the energy factor also in the transverse Lorentz boost (\ref{eq:edec}) and explains the observation that heavy particles have in general considerably higher $\langle p_T \rangle$ than lighter ones at low $x_F$.

For $dn/dx_F$ there is no noticeable variation both for protons and for pions as compared to Fig.~\ref{fig:delta_xfdep} without resonance $p_T$.

%
%
\subsection{Delta decay: adding the resonance mass distribution}
\vspace{3mm}
\label{sec:deltapp_dec_mt}

The Breit-Wigner (BW) mass distribution of resonances has a decisive influence both on mean $p_T$ and $dn/dx_F$ for pions whereas for protons there are only relatively small modifications. This is again due to the fact that tails in the $Q$ value of the Lorentz boosts (\ref{eq:edec}) are most effective for small decay mass values $m_{\textrm{dec}}$.

Limitations to the shape and mass range of the BW distribution will be discussed in the following Sect.~\ref{sec:res_mass_spec}. In the present explanatory section a BW distribution with linear mass damping up to 3~GeV for the $\Delta^{++}$ will be used.

\begin{figure}[h]
	\begin{center}
		\includegraphics[width=9.5cm] {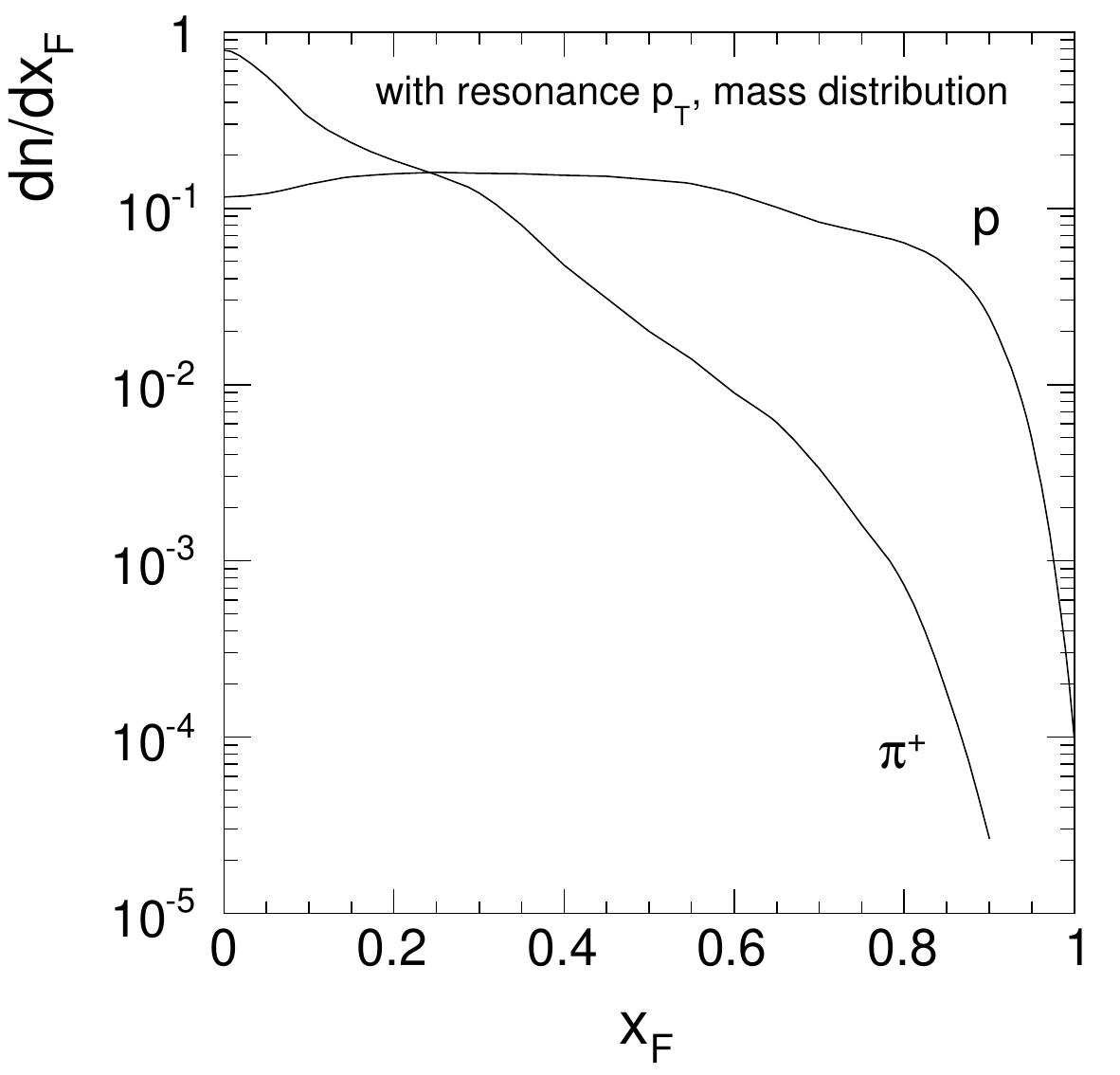} 
		\caption{$dn/dx_F$ for decay protons and pions as a function of $x_F$}
		\label{fig:delta_resxfdep}
	\end{center}
\end{figure}

\begin{figure}[h]
	\begin{center}
		\includegraphics[width=9.5cm] {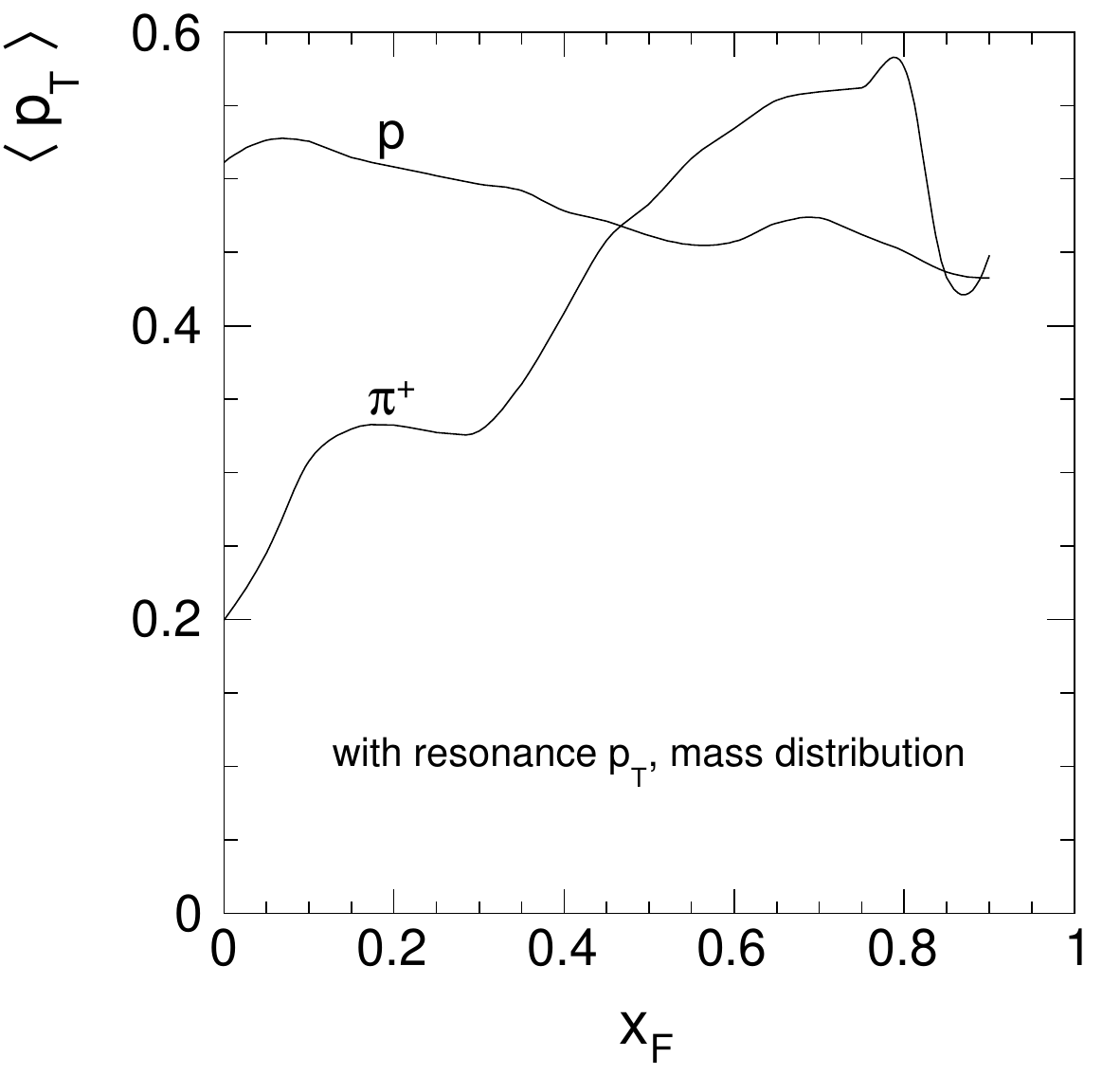} 
		\caption{Mean $p_T$ for decay protons and pions as a function of $x_F$}
		\label{fig:delta_resmeanpt}
	\end{center}
\end{figure}

The most important effect of the resonance mass distribution consists in the fact that the decay pions now cover the complete $x_F$ range and also reach the $\langle p_T \rangle$ values of the protons in the higher $x_F$ region. This is demonstrated in Fig.~\ref{fig:delta_resxfdep} for $dn/dx_F$ and in Fig.~\ref{fig:delta_resmeanpt} for $\langle p_T \rangle$.

The $dn/dx_F$ distribution for protons increases by 20\% at $x_F$~=~0 and decreases by about the same amount at $x_F$~=~0.55 with respect to Fig.~\ref{fig:delta_ppixfdep}. For pions there is a decrease of about 10\% up to $x_F \sim$~0.3. Beyond this limit there is a dramatic increase of the yield such that the cross section of the decay pions reaches a level of about 20\% of the total inclusive $\pi^+$ yield (also shown in Fig.\ref{fig:delta_resxfdep} up to the highest $x_F$ range.

\begin{figure}[h]
	\begin{center}
		\includegraphics[width=9.cm] {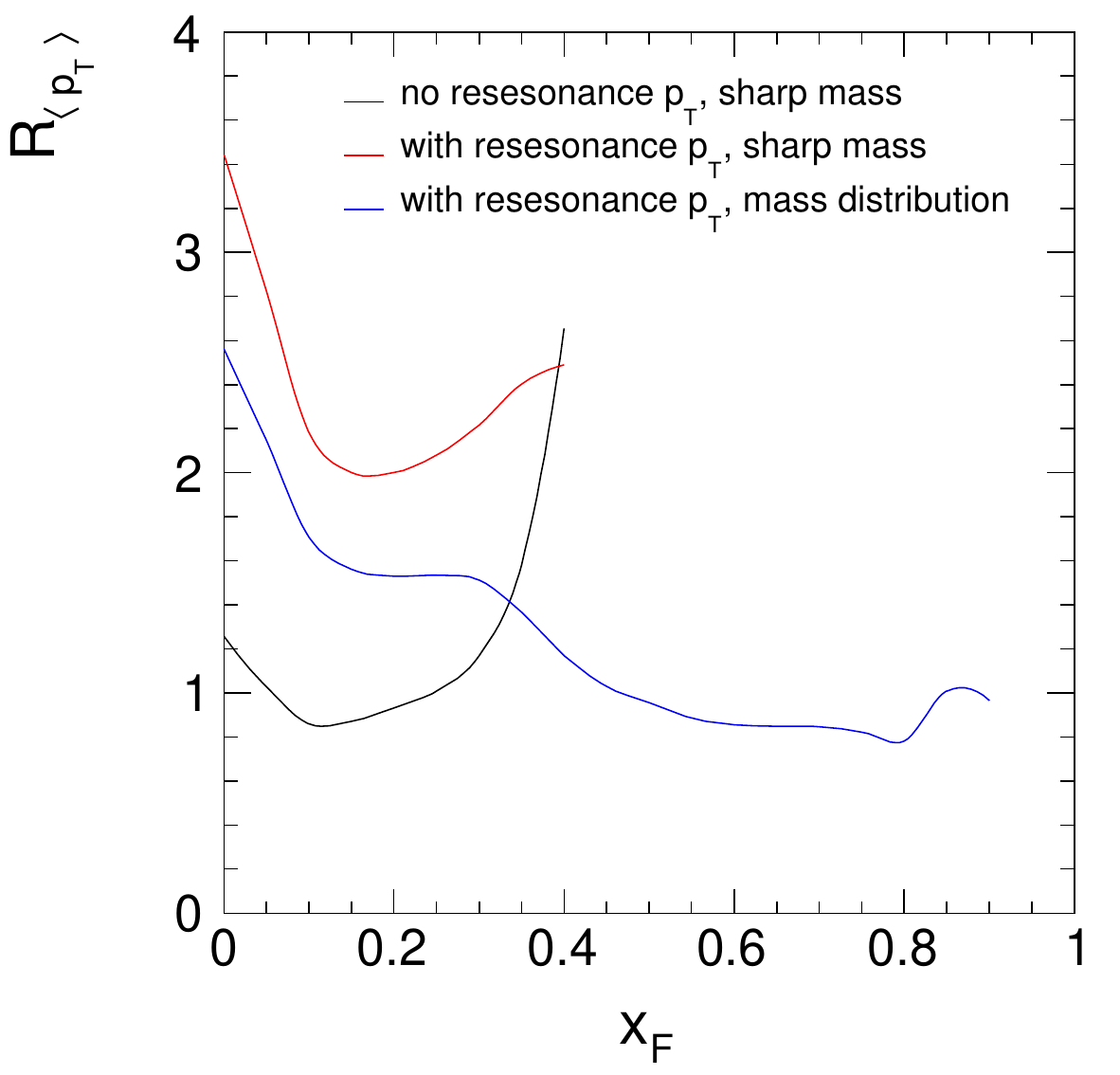} 
		\caption{$R_{\langle p_T \rangle}$ as a function of $x_F$ for the different decay configurations Sect.~\ref{sec:deltapp_dec_mass}--\ref{sec:deltapp_dec_mt}}
		\label{fig:delta_ratio}
	\end{center}
\end{figure}

For the mean $p_T$ a similar picture emerges. There is a strong increase of 30\% already at $x_F$~=~0 compared to Fig.~\ref{fig:delta_ppimeanpt} which continues rising until it crosses and exceeds the proton value at $x_F$ above 0.45. The relative factor between $\langle p_T \rangle$ for protons and pions,

\begin{equation}
	R_{\langle p_T \rangle}(x_F) = \frac{\langle p_T \rangle_{\textrm{prot}}(x_F)}{\langle p_T \rangle_{\textrm{pion}}(x_F)}
\end{equation}

\noindent
is shown in Fig.~\ref{fig:delta_ratio}.

\begin{figure}[h]
	\begin{center}
		\includegraphics[width=12.5cm] {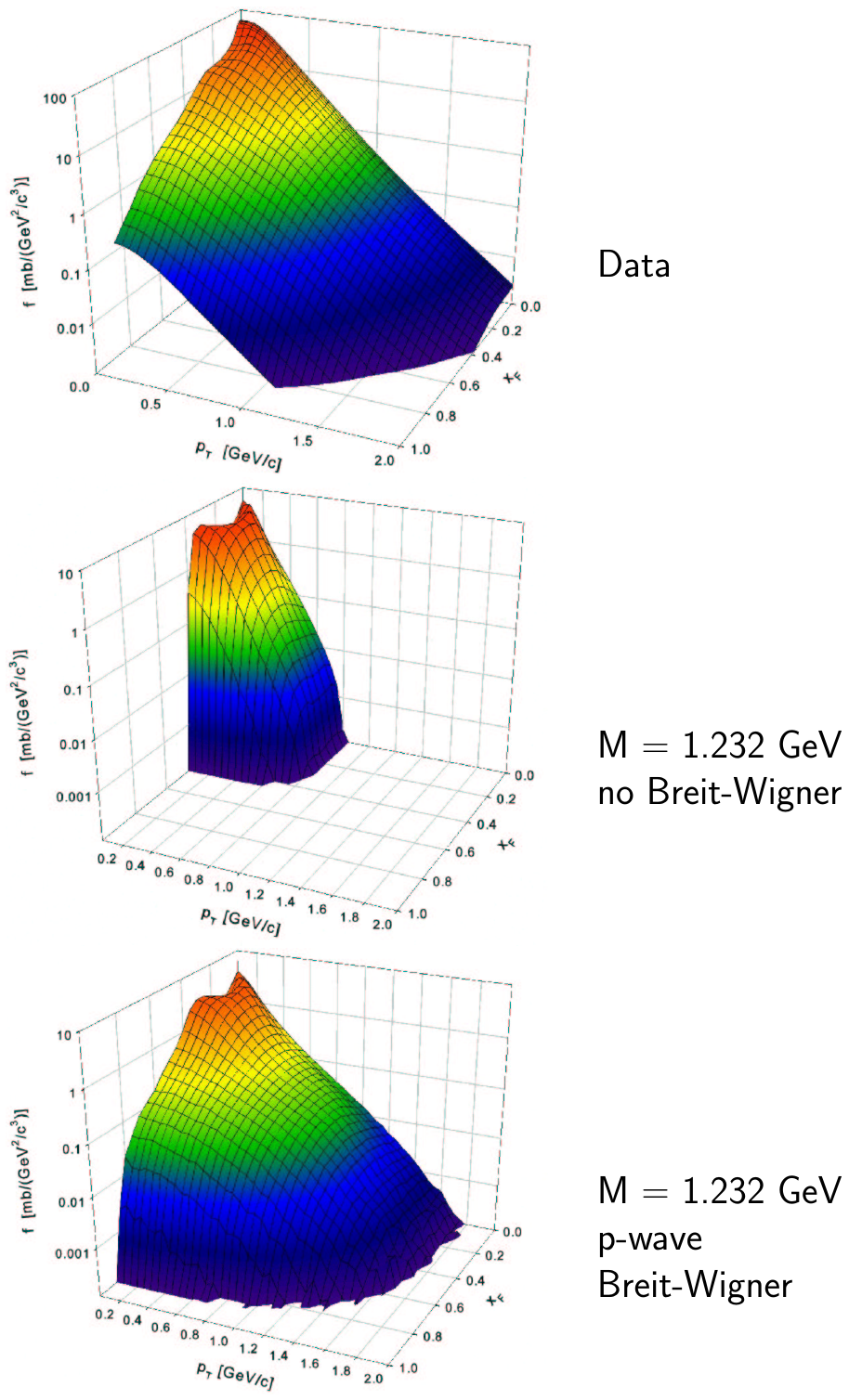} 
		\caption{Invariant pion cross section as a function of $p_T$ and $x_F$ for data \cite{pp_pion} and decay of the baryon resonance $\Delta(1232) \rightarrow N + \pi$ with zero width and Breit-Wigner mass distribution}
		\label{fig:delta_pi_inv}
	\end{center}
\end{figure}

An overview of the double-differential pion cross section $f(x_F,p_T)$ is presented in Fig.~\ref{fig:delta_pi_inv} showing the total inclusive $\pi^+$ yield, the decay pion distribution at fixed resonance mass and its evolution allowing for a Breit-Wigner mass distribution.

From the above discussion it should be clear that a detailed treatment of resonances and of their decay is absolutely mandatory in order to come to an understanding of their contribution to the measured inclusive data. In this sense resonances offer a model-independent step towards a better understanding of non-perturbative hadronic phenomena.

%
%
\section{The resonance mass spectrum}
\vspace{3mm}
\label{sec:res_mass_spec}

The discussion of $\Delta^{++}$ decay in the preceding Sections has demonstrated the necessity of properly taking into account both the transverse momentum and the mass dependencies. Whereas the transverse momentum distribution is experimentally accessible this is in general not true for the mass distribution due to problems in extracting its tails from an important background. Here experimental constraints as well as energy-momentum conservation and cascading decays play a role. This will be discussed in the following Sections.

%
%
\subsection{The unconstrained mass distribution}
\vspace{3mm}
\label{sec:res_uncons_dist}

Performing an energy scan over the mass range of a resonance in a "formation experiment" combined with a partial wave analysis the unbiased Breit-Wigner mass distribution may be obtained:

\begin{equation}
	BW(m) = \frac{m m_0\Gamma(m)}{(m_0^2 - m^2)^2+m_0^2\Gamma^2(m)}
	\label{eq:bw_mass}
\end{equation}

\noindent
where $m_0$ is the resonance mass and $\Gamma(m)$ is the mass and spin dependent width

\begin{equation}
	\Gamma(m) = \Gamma_0 \left(\frac{Q}{Q_0}\right)^{2L+1} \frac{2Q_0^2}{Q^2+Q_0^2}
\end{equation}

\begin{figure}[h]
	\begin{center}
		\includegraphics[width=10cm] {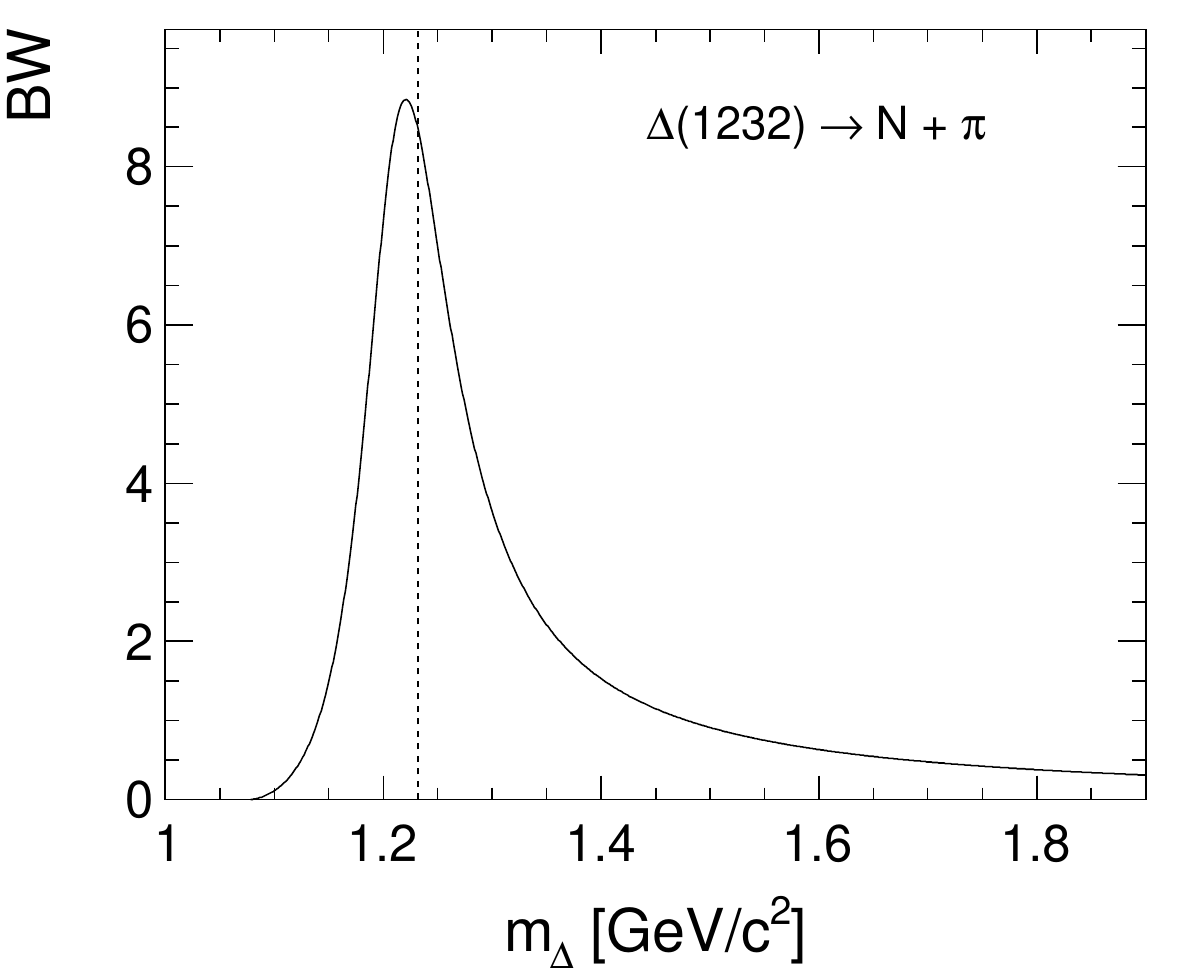} 
		\caption{BW mass spectrum of the $\Delta(1232)$ resonance with full width 118~MeV}
		\label{fig:delta_mass_spectrum}
	\end{center}
\end{figure}

\noindent
$\Gamma_0$ is the central width value, $Q$ and $Q_0$ the momenta of the decay products in the resonance rest frame and $L$ the spin of the resonance. This mass distribution is shown in Fig.~\ref{fig:delta_mass_spectrum} for the decay

\begin{equation}
	\Delta(1232) \rightarrow p + \pi
\end{equation}

It is limited at low mass by the threshold $m_{\textrm{thr}}$~=~$m_p$~+~$m_\pi$ and features a long, unconstrained tail. This tail is subject, for production experiments as they are discussed here, to several constraints presented in the following sections.

%
%
\subsection{Mass limitations due to the available interaction energy}
\vspace{3mm}
\label{sec:res_mass_limit}

\begin{figure}[h]
	\begin{center}
		\includegraphics[width=10cm] {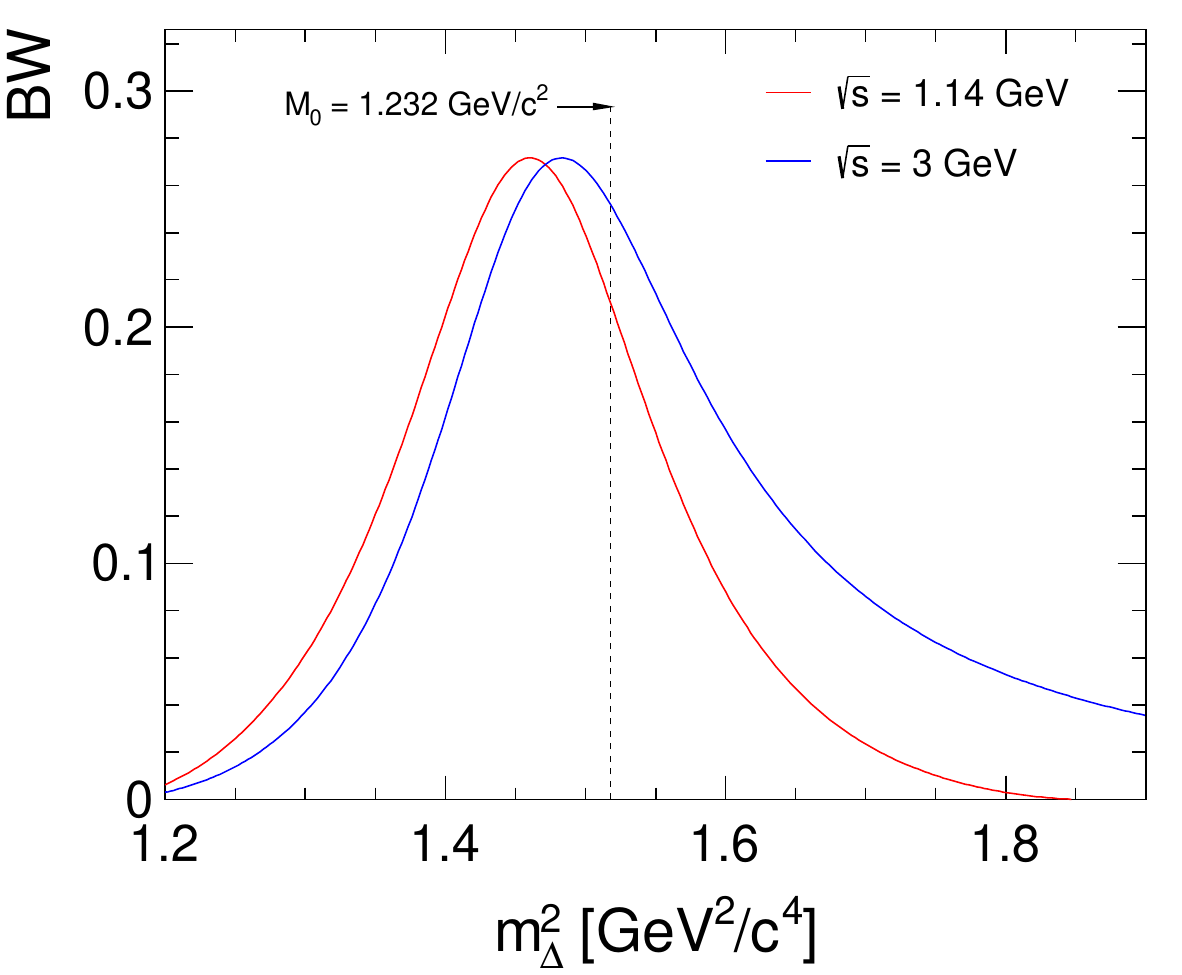} 
		\caption{Relativistic Breit-Wigner mass distributions of the final state $\Delta^{++}$ in the reaction K$^+$~+~p~$\rightarrow$~K$^0$~+~$\pi$~+~p at $\sqrt{s}$~=~1.86 and 2.61~GeV, \cite{jackson}}
		\label{fig:delta_mass}
	\end{center}
\end{figure}

As production experiments are performed at fixed interaction energy there is, by energy-momentum conservation a cut-off mass smoothly approached from below according to the $x_F$ distribution of the resonance. This is demonstrated in the original Jackson paper \cite{jackson} for the interaction K$^+$~+~p~$\rightarrow$~K$^0$~+~$\pi$~+~p at 1.14 and 3~Gev/c incident K$^+$ momentum. As shown in Fig.~\ref{fig:delta_mass} both the resonance shape and the position of the central mass are depending on the interaction energy.

\begin{figure}[h]
	\begin{center}
		\includegraphics[width=10cm] {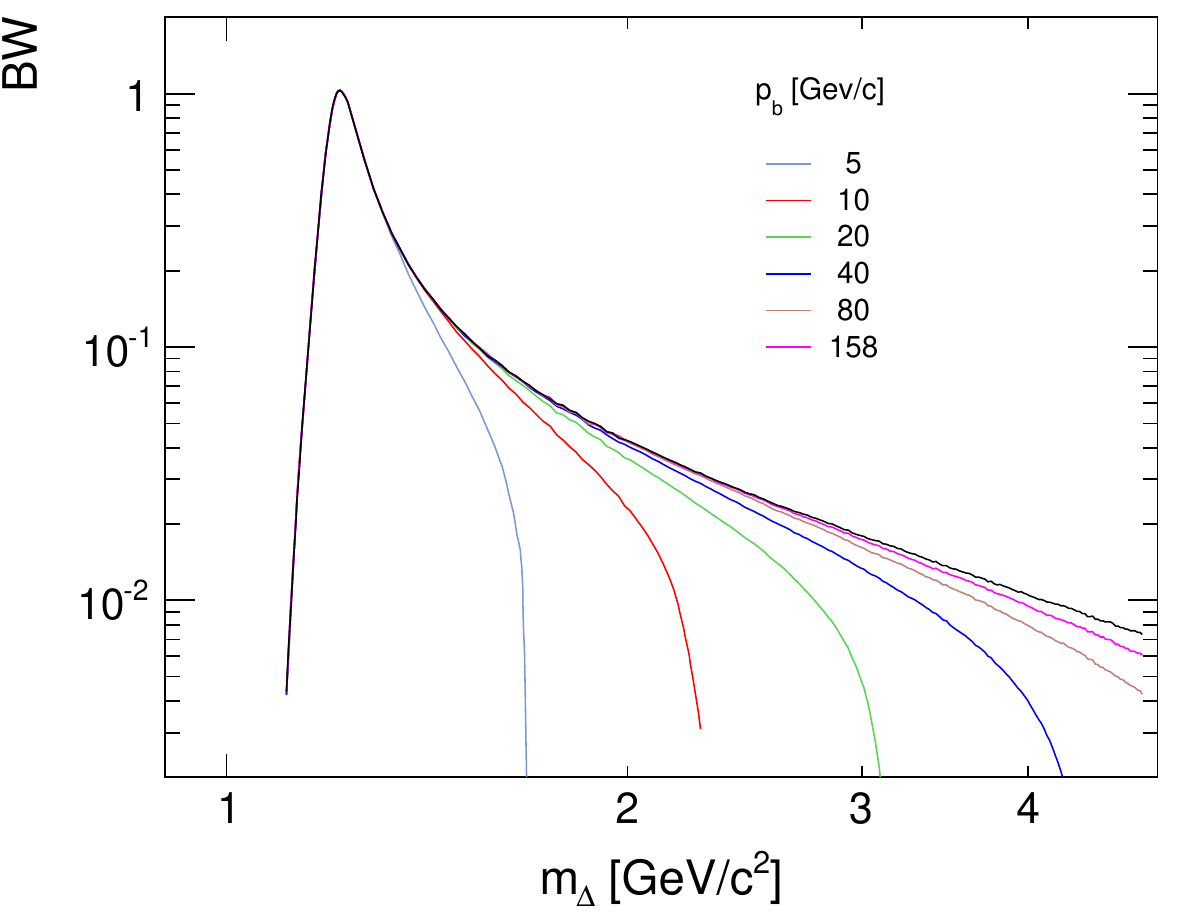} 
		\caption{Relativistic Breit-Wigner mass distributions for final state Delta baryons in p+p interactions at beam momenta between 5 and 158~GeV/c}
		\label{fig:delta_energies}
	\end{center}
\end{figure}

At higher interaction energy the central mass distribution stays largely unchanged but the high mass tails are curtailed by energy-momentum conservation as as shown in Fig.~\ref{fig:delta_energies} for p+p interactions in the range of beam momenta between 5 and 158~GeV/c. At the lowest beam momentum the mass distribution is conformal up to about 1.4~GeV and vanishes at about 1.7~GeV whereas at 158~GeV/c the change is smaller than 20\% even at a mass of 5~GeV.

\begin{figure}[h]
	\begin{center}
		\includegraphics[width=10cm] {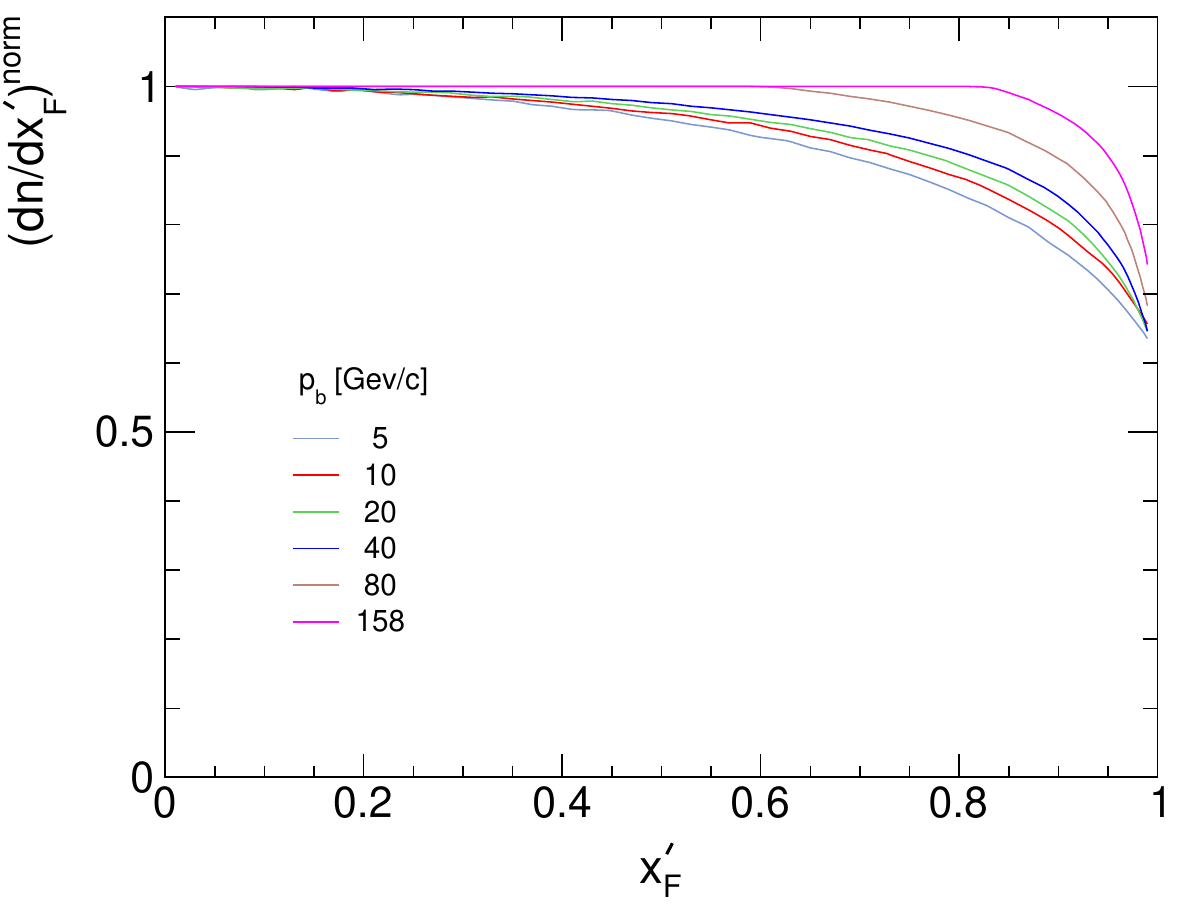} 
		\caption{Suppression of $x_F'$ distributions with respect to the unconstrained case for beam momenta between 5 and 158~GeV/c}
		\label{fig:xf_suppression}
	\end{center}
\end{figure}

Correspondingly the $x_F'$ distributions (Sect~\ref{sec:scaling_xf}) are progressively suppressed above $x_F'\sim$~0.4 as shown in Fig.~\ref{fig:xf_suppression}.

%
%
\subsection{Constraints due to proton momentum cuts}
\vspace{3mm}
\label{sec:res_cons_momcut}

\begin{figure}[h]
	\begin{center}
		\includegraphics[width=10cm] {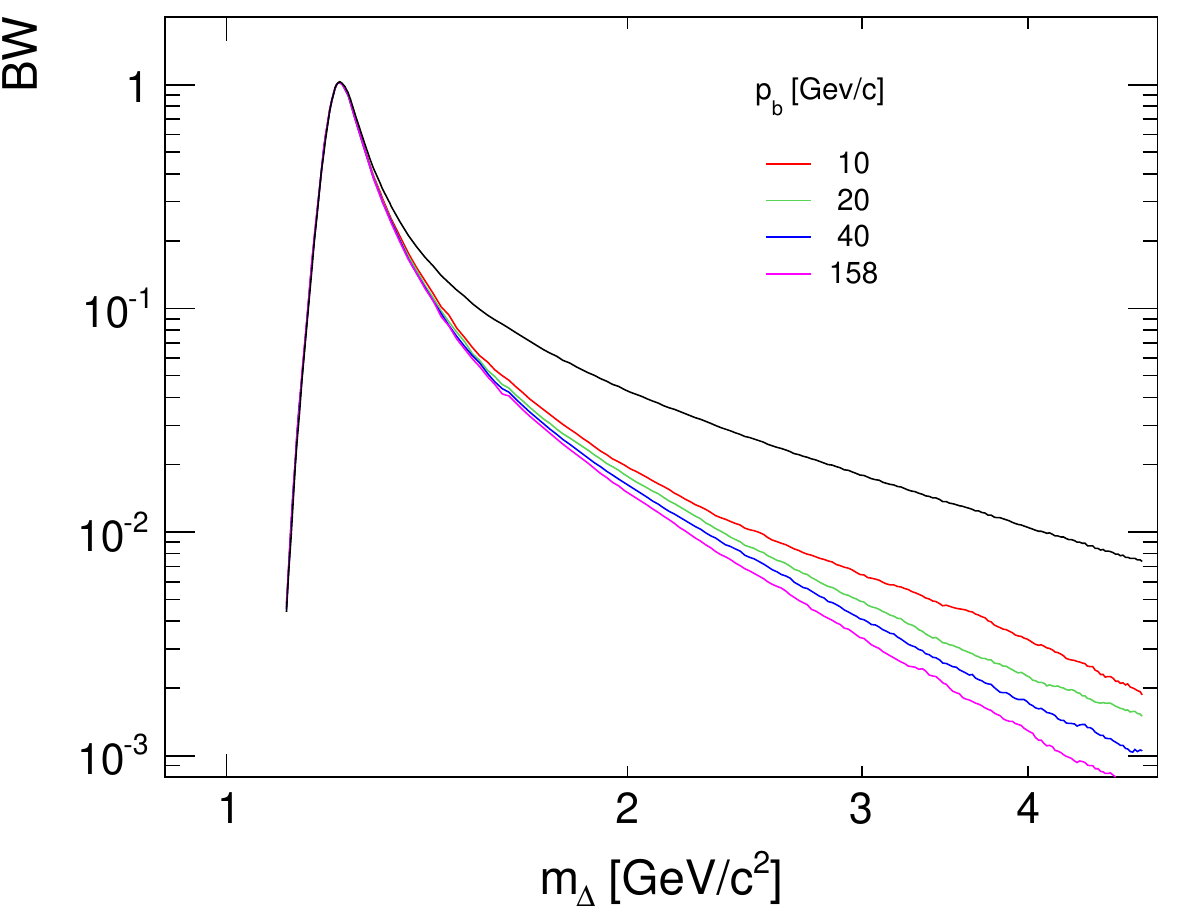} 
		\caption{Measured relativistic Breit-Wigner mass distributions of a $\Delta$(1232) resonance imposing a lab momentum cut on the decay protons at $p_{\textrm{lab}}$~=~1.0~GeV/c for beam momenta between 10 and 158~GeV/c}
		\label{fig:delta_measured}
	\end{center}
\end{figure}

\begin{figure}[h]
	\begin{center}
		\includegraphics[width=10cm] {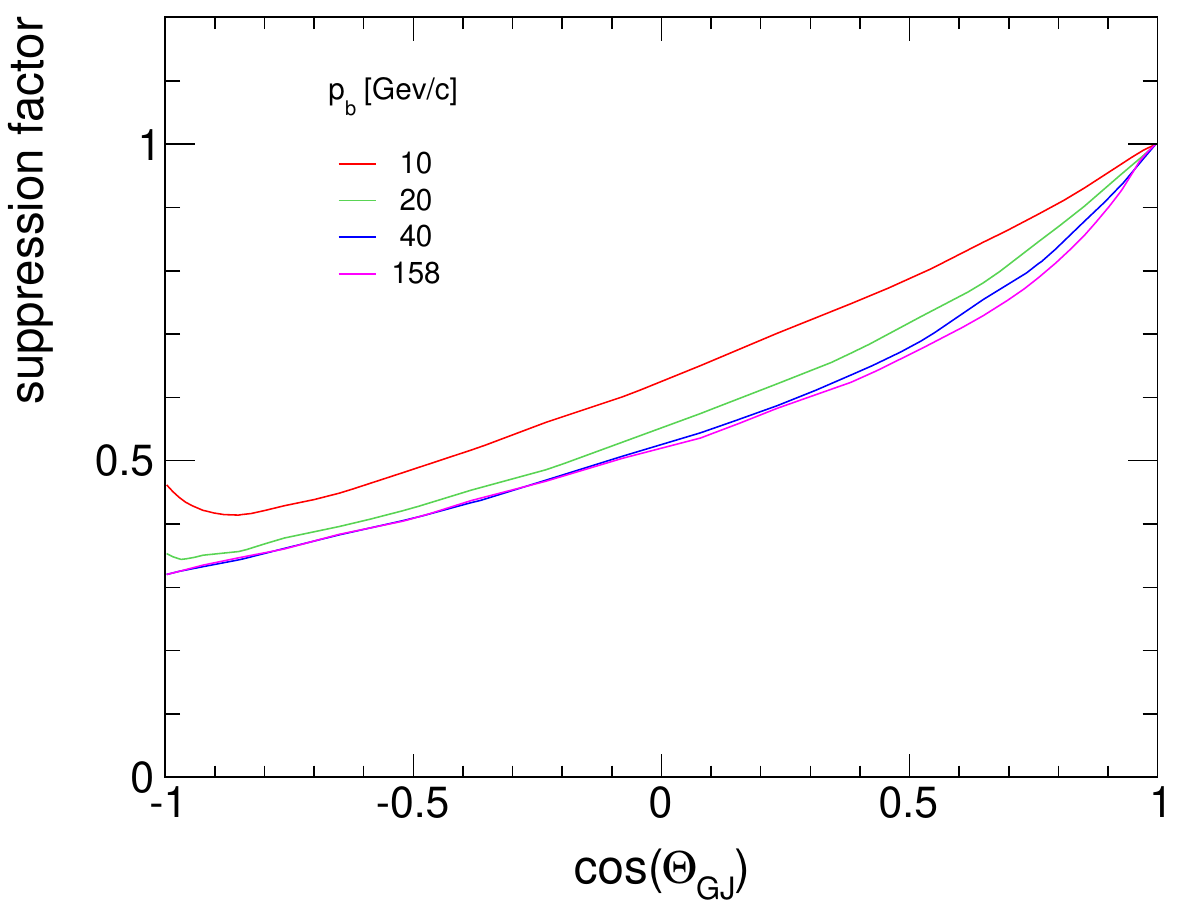} 
		\caption{Suppression factors of the Gottfried-Jackson decay angular distribution in backward direction as a consequence of a proton $p_{\textrm{lab}}$ cut at 1.0~GeV/c as a function of cos($\Theta_{GJ}$) and beam momenta between 10 and 158~GeV/c}
		\label{fig:angle_supp}
	\end{center}
\end{figure}

In bubble chamber experiments the positive identification of protons via bubble density is only feasible for lab momenta below 1.0--1.3~GeV/c. In consequence both the high mass tails of the relativistic Breit-Wigner distribution are curtailed and the decay angular distributions become suppressed in the backward direction. This is demonstrated in Figs.~\ref{fig:delta_measured} for the mass distributions and Fig.~\ref{fig:angle_supp} for the Gottfried-Jackson angle $\Theta_{GJ}$.

The resulting modification of the relativistic Breit-Wigner shape has an opposite energy dependence as the one imposed by energy-momentum conservation, Fig.~\ref{fig:delta_energies}.

The lab momentum cut also influences the acceptance of the decay angle distribution by suppressing backwards decays. This is shown in Fig.~\ref{fig:angle_supp} where the resulting suppression factor is plotted as a function of the Gottfried-Jackson angle.

Evidently the resulting distortions have to be corrected for in the analysis of the experimental data.

%
%
\subsection{Cascading}
\vspace{3mm}
\label{sec:res_cascading}

An important aspect of resonance production and decay is cascading. Given sufficient cms energy, the lower-mass resonances are decay products of higher mass states. This means that they are not directly produced but rather cascade down from heavier objects. It is interesting to estimate the amount of cascading steps prior to the appearance of the final state particles. In the baryonic sector this may be done for final state "net" baryons invoking baryon number conservation. Given the total sum of net baryonic resonance cross sections $\sum \sigma(\textrm{res})$, the number of cascading steps $n_{\textrm{casc}}$ is given by

\begin{equation}
	n_{\textrm{casc}} = \frac{\sum \sigma(res)}{2\sigma_{\textrm{inel}}}
\end{equation}

At SPS energy this may be estimated to $n_{\textrm{casc}} >$~2 \cite{fischer_private} where this value is to be regarded as a lower limit because the number of measured resonances gets rather limited towards high masses.

%
%
\subsubsection{Cascading: Production mechanism of resonances - charge exchange versus Pomeron exchange}
\vspace{3mm}
\label{sec:res_casc_exchange}

\begin{figure}[h]
	\begin{center}
		\includegraphics[width=14cm] {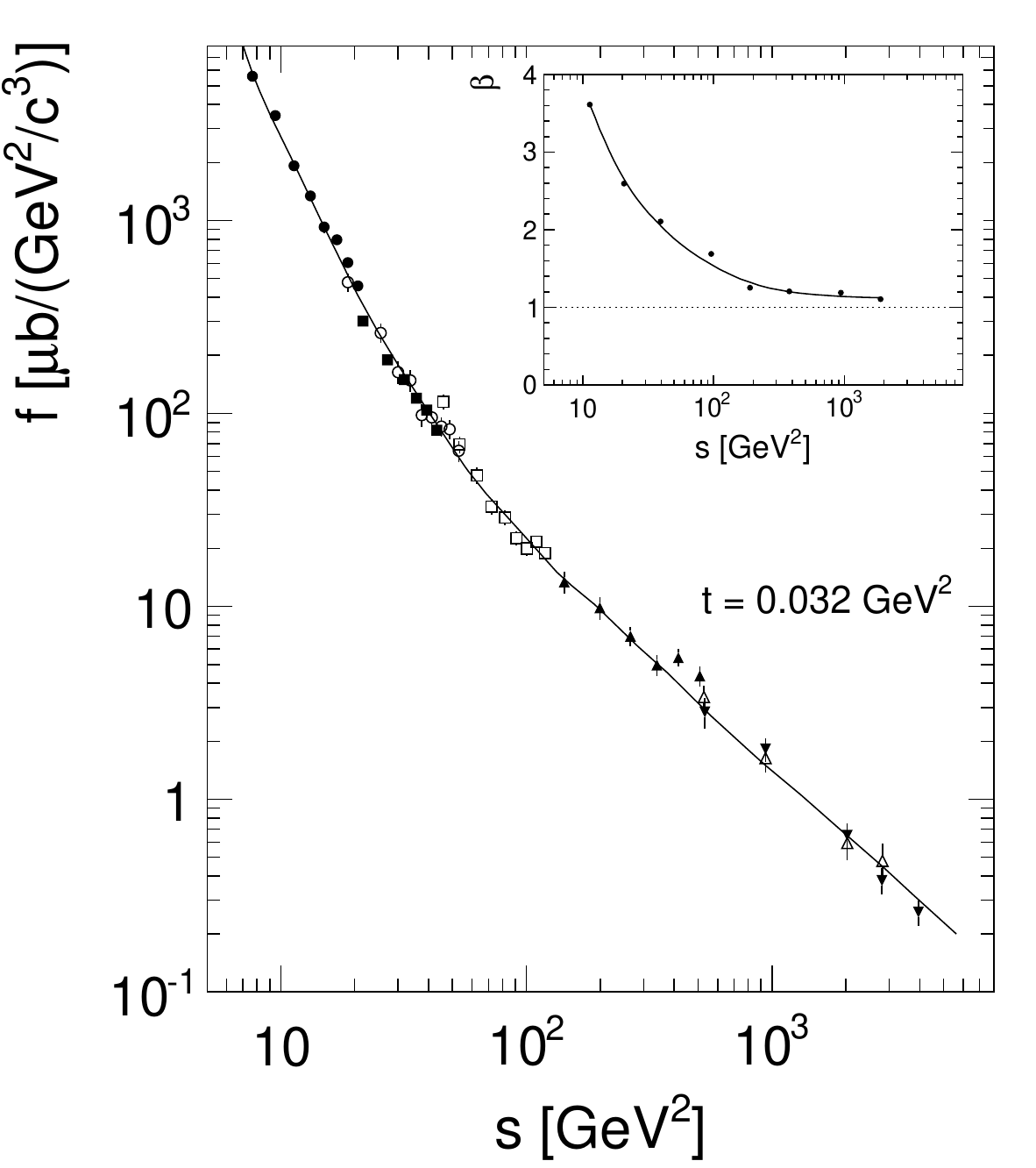} 
		\caption{Invariant cross sections of charge exchange and single and double dissociation in nucleon-nucleon interactions as a function of $s$ at a momentum transfer $t$~=~0.032~GeV$^2$. The	full line represents an interpolation of the data points. The insert gives the local slope $\beta$ in the parametrization $f \sim s^{-\beta}$ as a function of $s$}
		\label{fig:chargeexch}
	\end{center}
\end{figure}

The interplay of charge and Pomeron exchange has been studied for baryons and pions in connection to Regge theory \cite{pc_survey}. For baryons the $s$-dependence of the diffractive production of two-body final states of the type

\begin{align}
	p + p &\rightarrow n + (p + \pi^+) \\
	      &  \nonumber \textrm{or} \\
    p + p &\rightarrow (p + \pi^-) + (p + \pi^+)
\end{align}

\noindent
approaches a slope compatible with Pomeron exchange at SPS energy \cite{increase_rim1} whereas at lower energies Reggeon exchange prevails as shown in Fig.~\ref{fig:chargeexch}.

\begin{figure}[h]
	\begin{center}
		\includegraphics[width=12.cm] {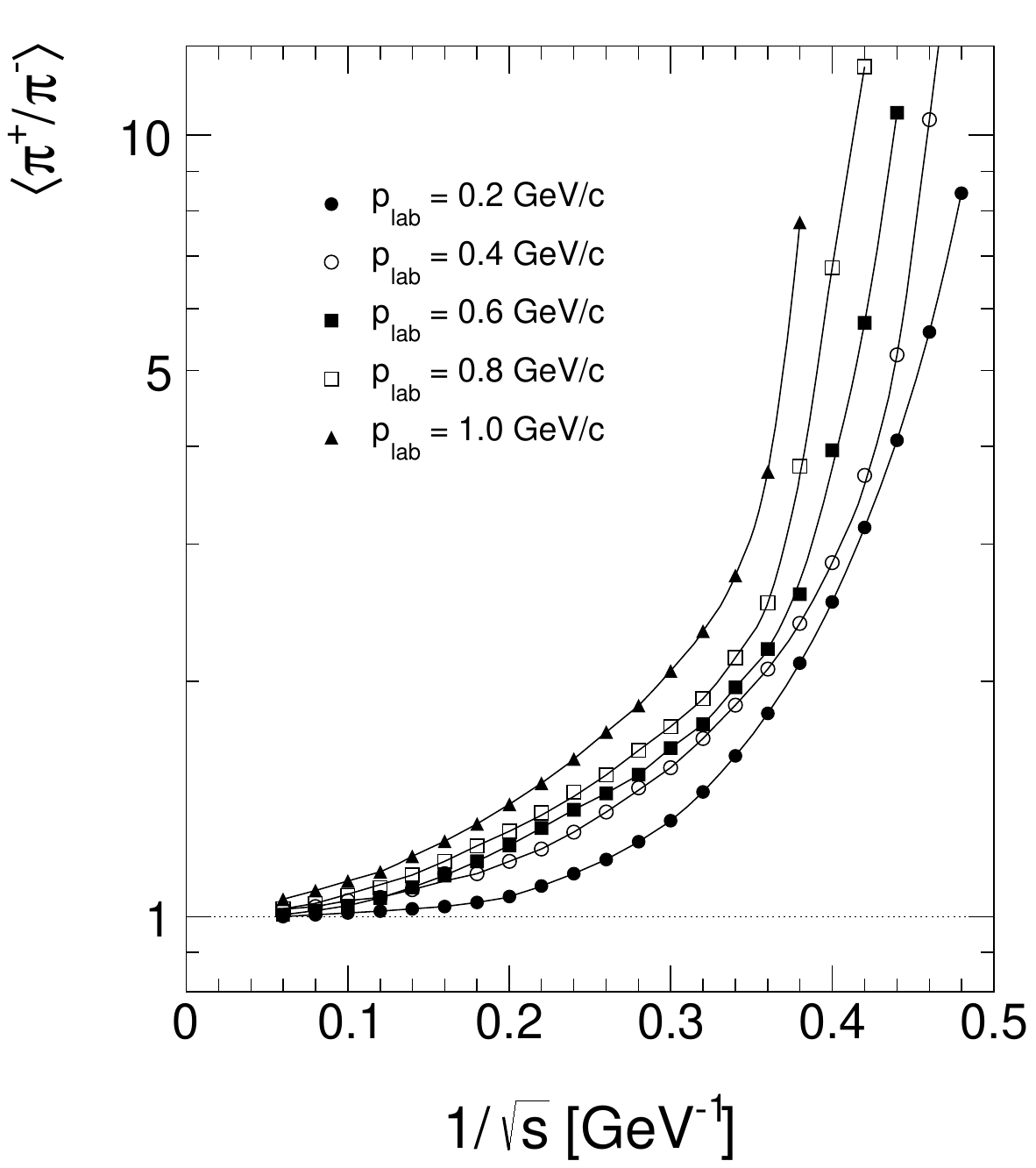} 
		\caption{$\langle \pi^+/\pi^- \rangle$ as a function of $1/\sqrt{s}$ for five values of $p_{\textrm{lab}}$ between 0.2 and 1.0~GeV/c. The full lines are hand interpolations through the data points for p+C interactions in the target frame \cite{pc_survey}}
		\label{fig:p2n_meanang}
	\end{center}
\end{figure}

For pions the $\pi^+/\pi^-$ ratio in the target hemisphere of the isoscalar Carbon nucleus in the reaction

\begin{equation}
	p + C \rightarrow \pi^+, \pi^-
\end{equation}

\noindent
may be used for a similar separation of charge and Pomeron exchange \cite{pc_survey}. As shown in Fig.~\ref{fig:p2n_meanang} the $\pi^+/\pi^-$ ratio approaches unity again in the SPS energy region whereas at lower energy it increases to values typical of charge exchange.

It may therefore be concluded that Pomeron exchange prevails over quantum number exchange both for diffractive and central interactions at SPS energies and above. This has the important consequence that N$^*$ production dominates there over $\Delta$ production, see also \cite{increase_rim1}. It is in sharp contrast to most "microscopic" production models where N$^*$ production tends to be neglected with respect to $\Delta$ by isospin counting arguments. In fact $\Delta$ resonances turn up in the final state as decay products of higher mass N$^*$ states with their mass distributions being inscribed in the respective N$^*$ decay distributions. Indeed three body decays of the type

\begin{equation}
	N^* \rightarrow N + \pi + \pi
\end{equation}

\noindent
are characterized by

\begin{equation}
	N + \pi + \pi \rightarrow \Delta + \pi
\end{equation}

\noindent
with large branching fractions \cite{pdg}.

%
%
\subsubsection{Cascading: Consequences for the Breit-Wigner mass distribution}
\vspace{3mm}
\label{sec:res_casc_bw}

In a situation where resonances proper are decay products of higher mass states, a modification of their mass distributions with respect to the unconstrained Breit-Wigner shape, Sect.~\ref{sec:res_uncons_dist}, must be expected. Due to the absence of theoretical predictivity in the soft sector of QCD it is difficult to estimate the effect of cascading in the Breit-Wigner mass tails. A look at particle production with different assumptions concerning the extent of the mass tails may help to at least establish some limits. Taking up the example of $\Delta^{++}$ decay into p and $\pi^+$ the yield of the decay products may be studied for different upper limits of the Breit-Wigner tails using a simple, linear tail suppression of the form

\begin{equation}
	f_{\textrm{supp}}(m) = 1 - \frac{ m - m_{\textrm{0}} }{ m_{\textrm{up}} - m_{\textrm{0}}}
\end{equation}

\noindent
where $f_{\textrm{supp}}(m)$ is the suppression factor applied to the form (\ref{eq:bw_mass}) and $m_{\textrm{up}}$ an upper mass limit at which the Breit-Wigner distribution vanishes.

\begin{figure}[h]
	\begin{center}
		\includegraphics[width=13.cm] {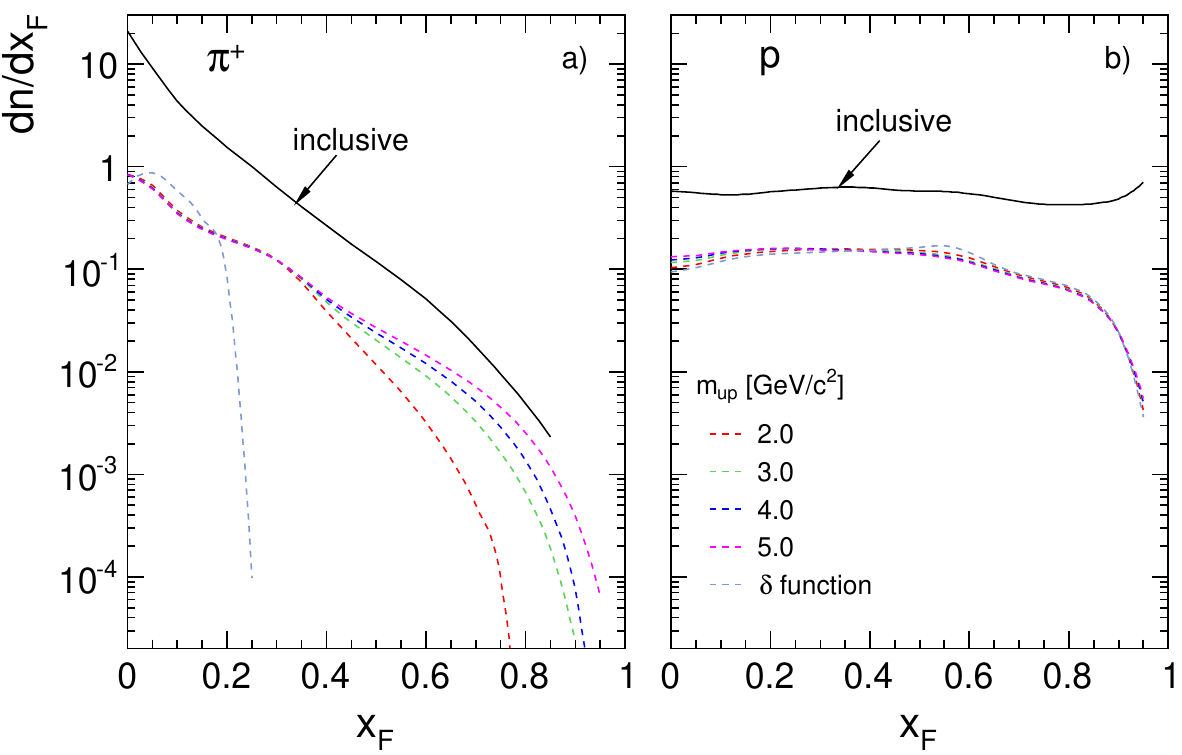} 
		\caption{$dn/dx_F$ distributions from $\Delta^{++}$ decay as a function of $x_F$ for different damping factors $f_{\textrm{supp}}(m)$ at $\sqrt{s}$~=~17.2~GeV, a) for decay pions, b) for decay protons. The respective measured inclusive cross sections are indicated as full lines}
		\label{fig:delta2ppixf}
	\end{center}
\end{figure}

The resulting $dn/dx_F$ distributions for the decay pion and protons are shown in Fig.~\ref{fig:delta2ppixf} for different values of the upper mass limit $m_{\textrm{up}}$ at SPS energy ($\sqrt{s}$~=~17.2~GeV). For comparison the total inclusive yields are also indicated.

\begin{figure}[h]
	\begin{center}
		\includegraphics[width=11cm] {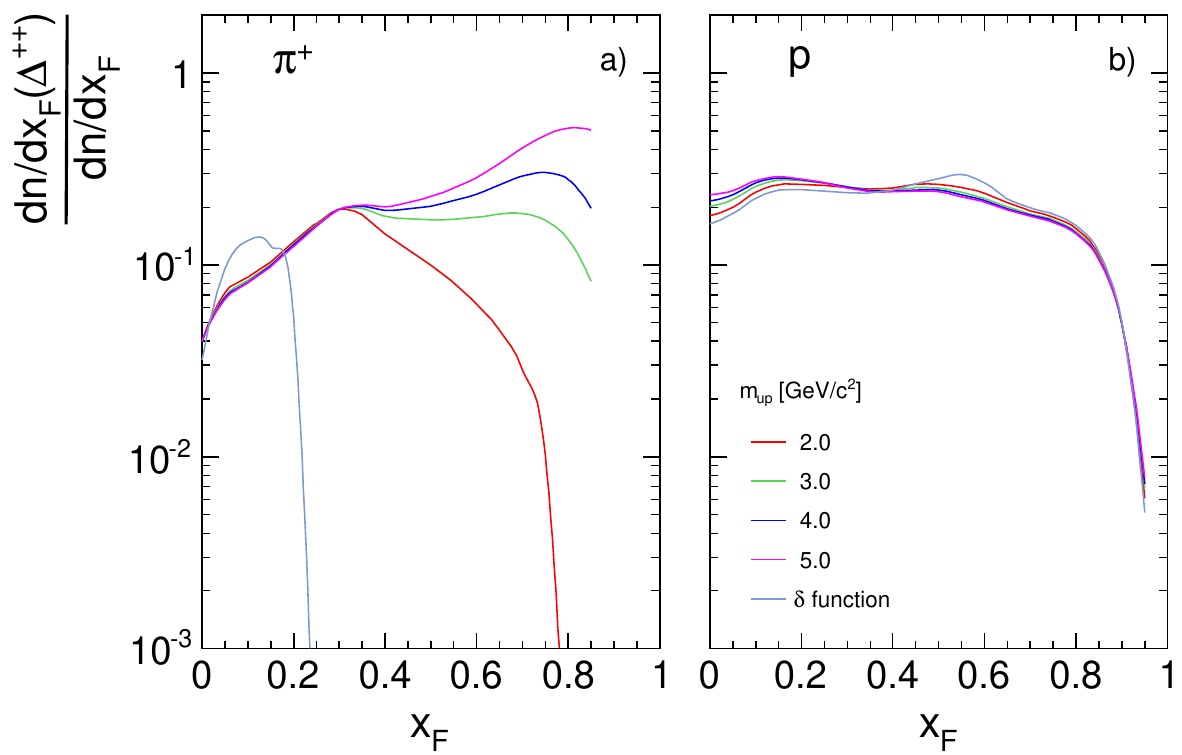} 
		\caption{Decay particle yields relative to the total inclusive cross sections as a function of $x_F$ a) for the decay pions and b) for the decay protons at $\sqrt{s}$~=~17.2~GeV and different damping factors}
		\label{fig:delta2rat}
	\end{center}
\end{figure}

The effect of the resonance mass tail on the decay particle yields is strong for $\pi^+$ above $x_F$~=~0.3 whereas for protons there is a smaller dependence spread over the complete $x_F$ range. This is quantified in Fig.~\ref{fig:delta2rat} where the yield ratios relative to the total inclusive cross section are given.

It is interesting to note that the yield ratios reach more than 20\% for the decay protons over most of the $x_F$ range. For $\pi^+$ the ratios increase from a few percent at low $x_F$ to about 20\% at $x_F$~=~0.3 and then fan out for higher $x_F$ up to unphysical values at high $m_{\textrm{up}}$.

\begin{figure}[h]
	\begin{center}
		\includegraphics[width=15.5cm] {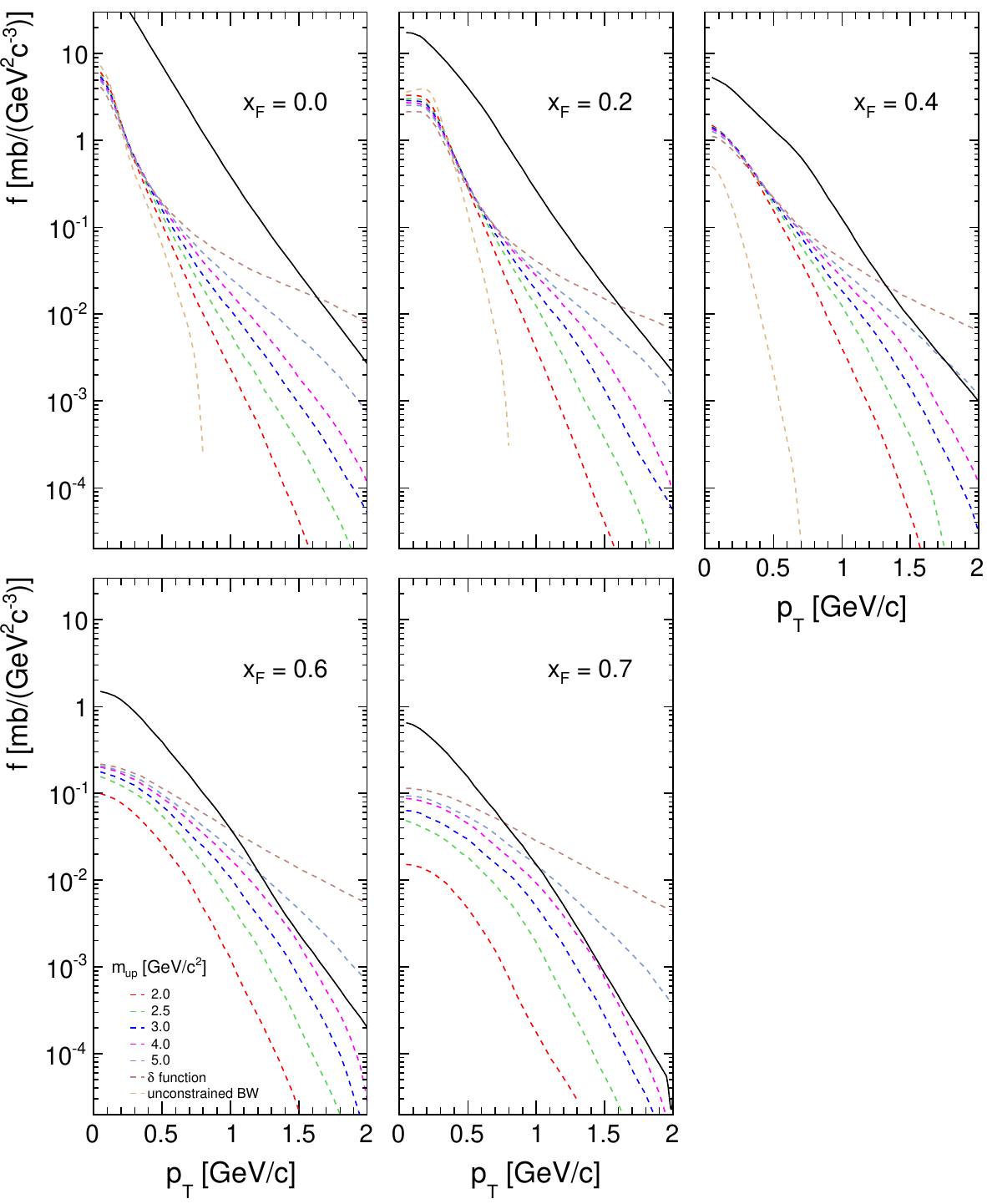} 
		\caption{Invariant decay pion cross sections as a function of $p_T$ for different $x_F$ values and mass cuts at $\sqrt{s}$~=~17.2~GeV. The measured total inclusive cross sections are shown as full lines}
		\label{fig:delta2picuts}
	\end{center}
\end{figure}

Further constraints on the mass cut-off may be obtained by inspecting the double-differential invariant cross sections as functions of $p_T$ for different $x_F$ values shown in Fig.~\ref{fig:delta2picuts} for the decay pions.

\begin{figure}[h]
	\begin{center}
		\includegraphics[width=16.cm] {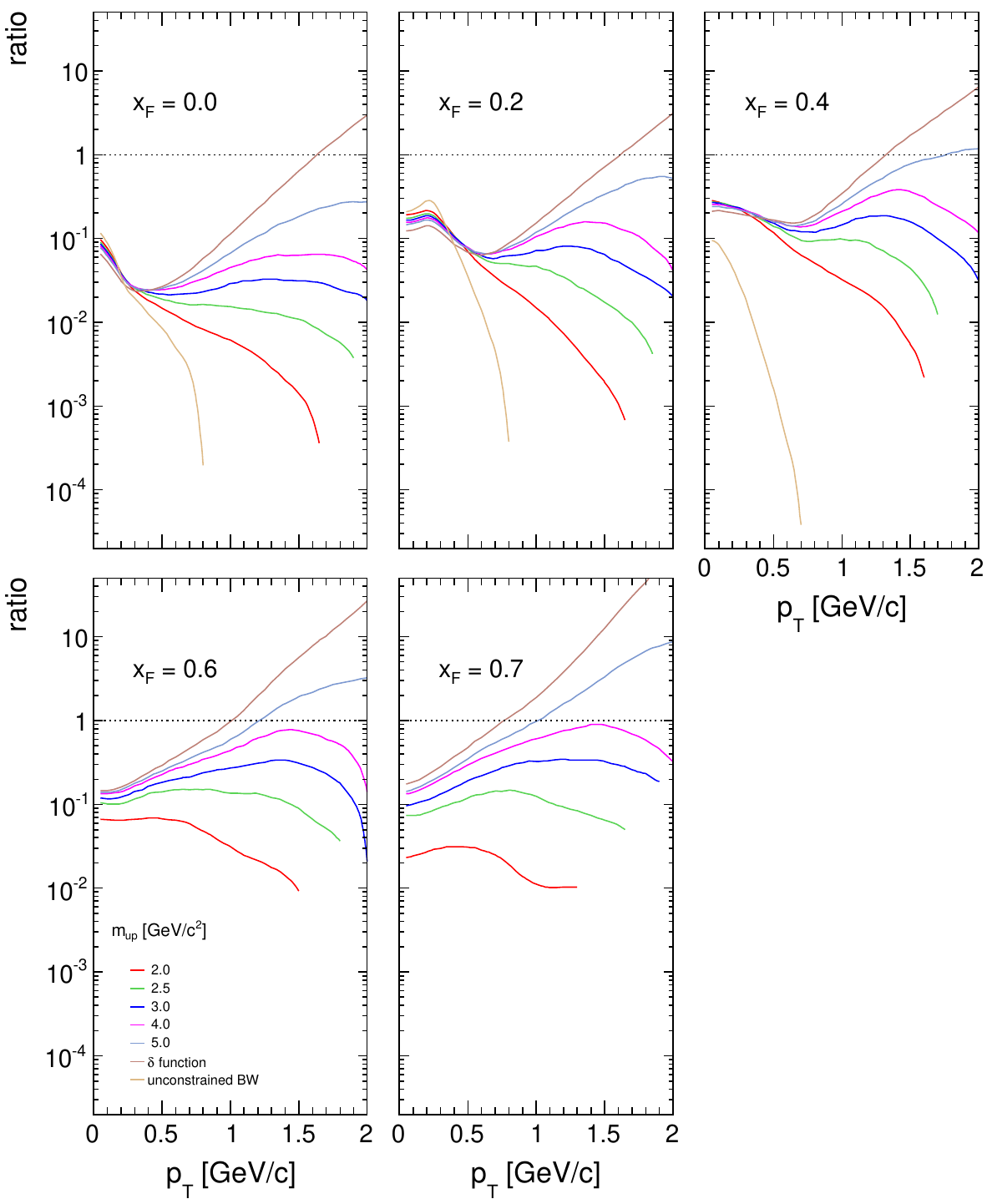} 
		\caption{Yield ratio $R_{\pi^+} = f_{\pi^+}(\Delta^{++})/f_{\textrm{incl}}(\pi^+)$ as a function of $p_T$ for $x_F$ between 0.0 and 0.7 and different mass cuts}
		\label{fig:delta2piratpt}
	\end{center}
\end{figure}

The corresponding yield ratios relative to the measured inclusive cross section are presented in Fig.~\ref{fig:delta2piratpt} as a function of $p_T$ for different mass cuts

The yield ratio exceeds the total inclusive level at $x_F >$~0.4 for $m_{\textrm{up}}$~=~5.0~GeV/c$^2$ and reaches 1.0 for $m_{\textrm{up}}$~=~4.0~GeV/c$^2$ at $x_F >$~0.6. At low $p_T$ the ratio stays consistently on a 10--20\% level, decreases to a flat minimum at $p_T \sim$~0.5~GeV/c and stabilizes again at 10--20\% for higher $x_F$ and $m_{\textrm{up}}$~=~2.5--3~GeV.

\begin{figure}[h]
	\begin{center}
		\includegraphics[width=16.cm] {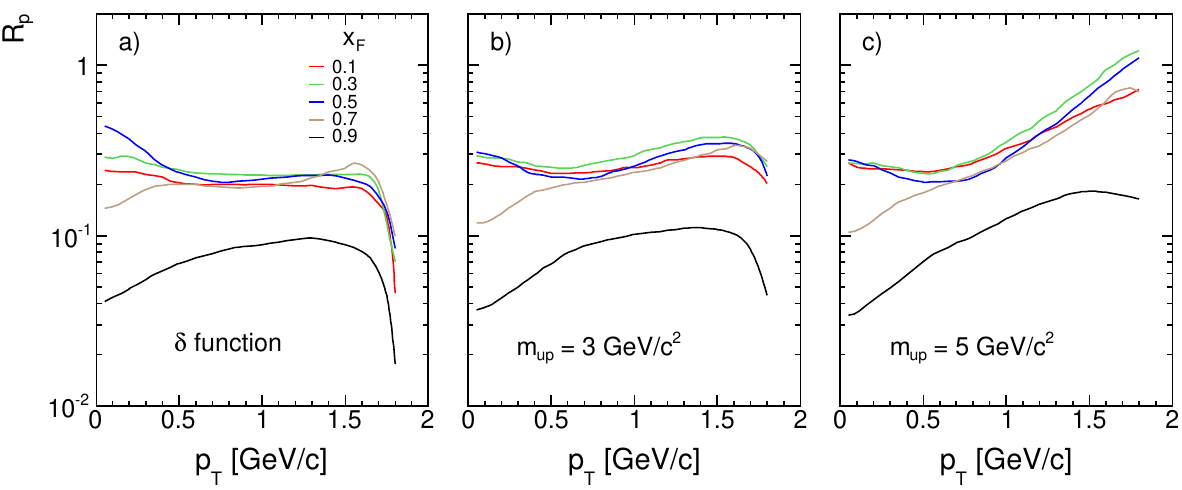} 
		\caption{Yield ratio $R_{\textrm{p}} = f_p(\Delta^{++})/f_{\textrm{incl}}(p)$ as a function of $p_T$ for $x_F$ from 0.1 to 0.9, a) mass delta function,  b) $m_{\textrm{up}}$~=~3.0~GeV/c$^2$ and c) $m_{\textrm{up}}$~=~5.0~GeV/c$^2$}
		\label{fig:delta2pratpt}
	\end{center}
\end{figure}

For the decay protons the influence of the mass cut-off is less important as shown in Fig.~\ref{fig:delta2pratpt} which gives directly the yield ratio $R_{\textrm{p}} = f_p(\Delta^{++})/f_{\textrm{incl}}(p)$ as a function of $p_T$ for different values of $x_F$ and three different mass cuts.

The ratio $R_{\textrm{p}}$ is in general on the level of 20--30\% of the inclusive proton yield increasing in the higher $p_T$ range from 30 to 50\% at high $x_F$ and $m_{\textrm{up}}$ up to 3.0~GeV/c$^2$. At $m_{\textrm{up}}$~=~5.0~GeV/c$^2$ this increase exceeds the limit of 100\% in the medium $x_F$ range. There is a general decrease of $R_{\textrm{p}}$ with values below 10\% towards $x_F$~=~0.9 which is due to the transition to single proton production for highly peripheral collisions.

%
%
\subsubsection{Mean \boldmath $p_T$ and inverse $m_T$ slopes}
\vspace{3mm}
\label{sec:res_inv_slope}

The influence of the resonance mass distribution pervades, in addition to the examples discussed above, all secondary particle distributions. As two further examples the mean $p_T$ and the inverse slopes of the transverse mass distributions ("hadronic temperature") are presented below for $\Delta^{++}$ decay.

\begin{figure}[h]
	\begin{center}
		\includegraphics[width=12.cm] {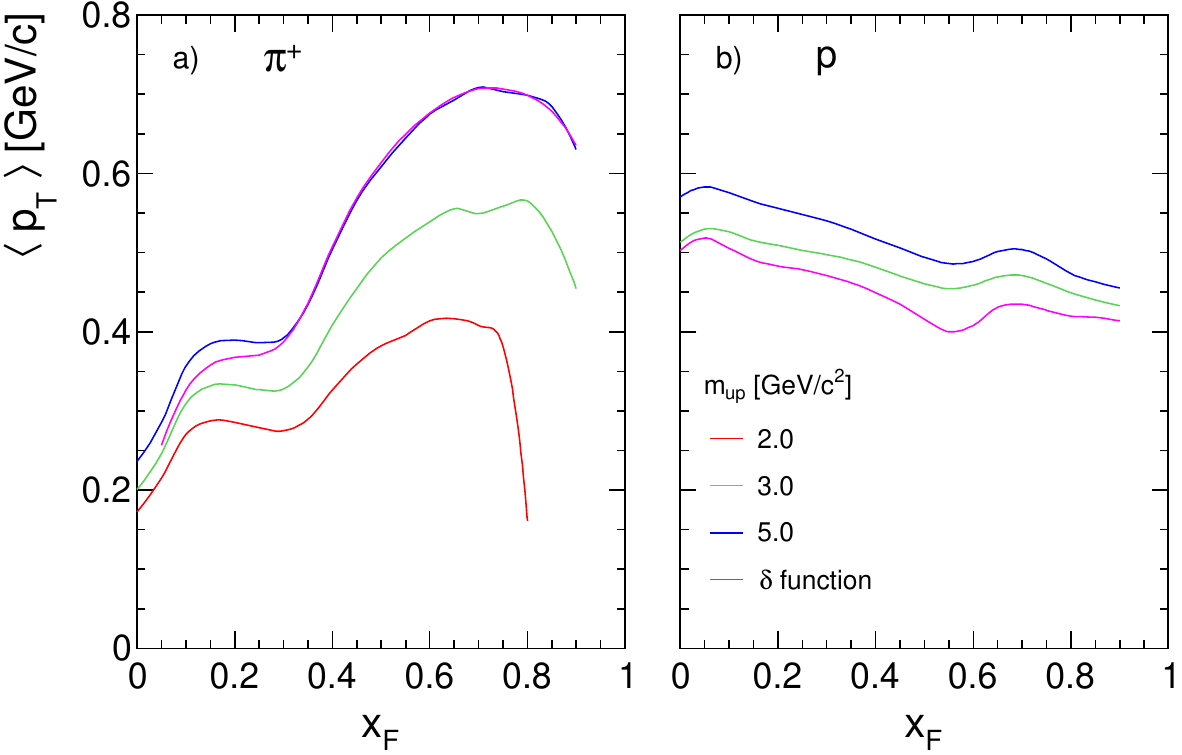} 
		\caption{Mean transverse momentum as a function of $x_F$, a) for the decay pions and b) for the decay protons from the $\Delta^{++}(1232)$ resonance}
		\label{fig:delta2ppi_meanpt}
	\end{center}
\end{figure}

The mean transverse momentum of the decay pions and protons is shown in Fig.~\ref{fig:delta2ppi_meanpt} as a function of $x_F$ for different mass cuts $m_{\textrm{up}}$, a) for the decay pions and b) for the decay protons.

While there is a strong dependence of $\langle p_T \rangle$ on $m_{\textrm{up}}$ for pions in particular for $x_F >$~0.3, the mean $p_T$ of protons is less affected over the full $x_F$ range. As already discussed in Sect.~\ref{sec:deltapp_dec_mt} (Fig.~\ref{fig:delta_resmeanpt}) the sizeable difference of the $\langle p_T \rangle$ values for pions and protons at small $x_F$ vanishes with increasing $x_F$ for $m_{\textrm{up}}$ between 2.0 and 3.0~GeV/c$^2$, while again the extreme assumptions of a delta function in mass or of $m_{\textrm{up}} >$~3~GeV/c$^2$ allow for the limitation of the range of possible resonance mass distributions.

\begin{figure}[h]
	\begin{center}
		\includegraphics[width=12.cm] {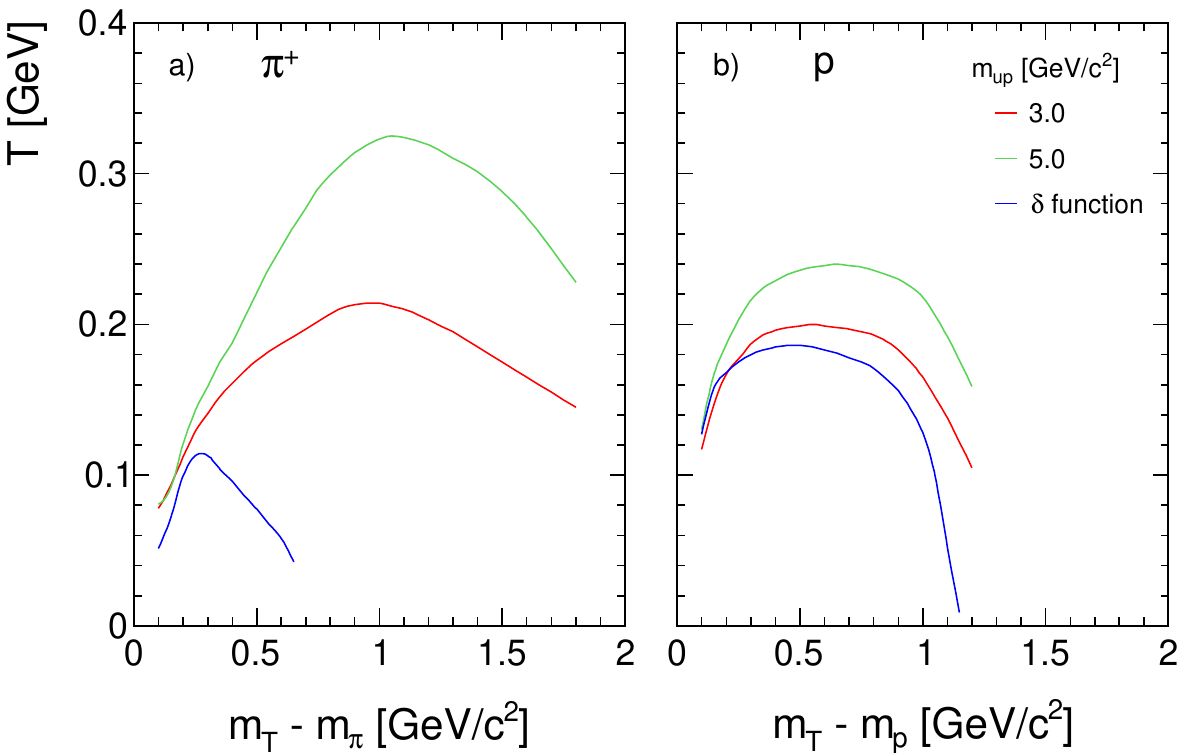} 
		\caption{Inverse slopes $T$ of the $m_T$ distributions a) for decay pions and b) for decay protons from the $\Delta^{++}$ resonance as a function of $x_F$ for different values of $m_{\textrm{up}}$}
		\label{fig:delta_ppi_invslope}
	\end{center}
\end{figure}

A similar conclusion may be drawn from the inverse $m_T$ slopes shown in Fig.~\ref{fig:delta_ppi_invslope}.

While the $T$ values of the decay protons tend to level out for $m_T - m >$~0.3~GeV/c$^2$ after a strong increase for all $m_{\textrm{up}}$, the inverse slopes of the decay pions show, in contrast, a  strong $m_{\textrm{up}}$ dependence in the same $m_T - m$ region contradicting any "thermal" behaviour.

%
%
\subsection{Summary of Sects.~\ref{sec:resonances} and \ref{sec:res_mass_spec} concerning the decay of the \boldmath $\Delta^{++}$ resonance}
\vspace{3mm}
\label{sec:res_summary}

The in-depth discussion of several features of $\Delta^{++}$ decay as presented in Sects.~\ref{sec:resonances} and \ref{sec:res_mass_spec} points out the importance of several resonance parameters beyond the mass, half width and branching fraction. This concerns in particular two important ingredients:

\begin{enumerate}
	\item In order to predict the inclusive distribution of the decay particles the knowledge of the resonance cross section over the full phase space is mandatory
	\item In strong decays a long upper mass tail (Breit-Wigner distribution) exists which depends on several parameters like energy-momentum conservation, experimental cuts and in particular cascading decay from higher mass resonances. The effect of cascading on the mass distribution is as yet not calculable as it makes part of non-perturbative QCD. It may at least be approximated by a detailed study of the phase space distribution of the decay particles with different assumptions on the extent of the upper mass tail. This way a range of possible decay parameters may be established
\end{enumerate}

In the present study of $\Delta^{++}$ decay a damping of the unconstrained Breit-Wigner distribution up to mass values between 2 and 3~GeV/c$^2$ has been shown to be compatible with a wide range of secondary particle distributions.

%
%
\section{Inverse slopes from the decay of different resonances ("Temperature")}
\vspace{3mm}
\label{sec:res_invslope}

\begin{figure}[h]
	\begin{center}
		\includegraphics[width=12.5cm] {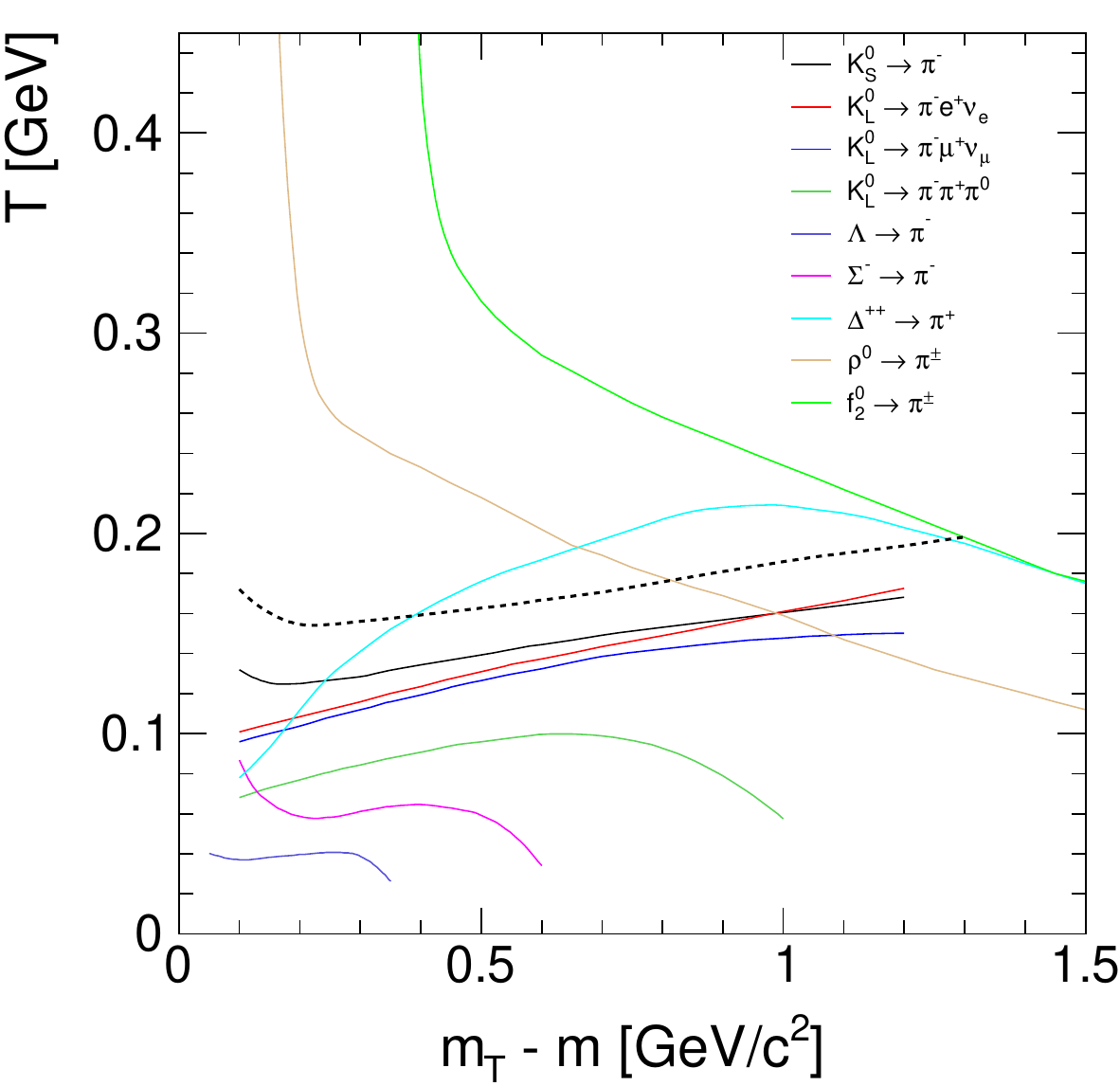} 
		\caption{Inverse slopes $T$ (GeV) as a function of transverse mass at central rapidity and $p_{\textrm{beam}}$~=158~GeV/c for decay pions from different baryonic and mesonic resonances. The measured inclusive value for pions is indicated by a thick dashed line}
		\label{fig:res_invslope}
	\end{center}
\end{figure}

Following the preceding discussion of the influence of the resonance parameters on a number of inclusive quantities of the final state hadrons, the inverse slope parameters of decay pions from a set of different baryonic and mesonic resonances will be discussed. In addition to the $\Delta$ baryon decay (Sect.~\ref{sec:resonances} and \ref{sec:res_mass_spec}) the inverse slopes of $\pi^-$ from the weak decays of K$^0$, $\Lambda$ and $\Sigma^-$ (Sect.~\ref{sec:mtslopes}) and from $\rho^0$(770) and f$_2^0$(1270) are presented in Fig.~\ref{fig:res_invslope}. Here the weakly decaying strange hadrons have zero width whereas for the strong decays a linear damping of the respective Breit-Wigner distributions up to 3~GeV/c$^2$ mass is introduced. The inverse slopes $T$ are given at 158~GeV/c beam momentum and central rapidity.

Several conclusions may be drawn from Fig.~\ref{fig:res_invslope}:

\begin{enumerate}[label=(\roman*)]
	\item Each resonance decay creates its proper inverse slope distribution.
	\item These distributions are far from flat at a given "temperature" as it would be imposed by a Boltzman-type emission from an equilibrium thermal ensemble.
	\item The corresponding inverse slope values are imposed by a set of resonance parameters like the $Q$ value, the central width and the mass extent of the Breit-Wigner distribution as discussed in Sect.~\ref{sec:resonances} and \ref{sec:res_mass_spec} above.
	\item The actual $T$ values show a very large spread from a few tens of MeV up to several hundred MeV.
	\item The inclusively measured $T$ distributions of a particle must be seen as an incidental result of the overlap of many distinct individual contributions.
\end{enumerate}

Given the fact that most if not all final state hadrons stem from resonance decay the very notion of a thermodynamic origin is not tenable.

%
%
\section{Mean transverse momentum}
\vspace{3mm}
\label{sec:meanpt}

%
%
\subsection{\boldmath $\langle p_T \rangle$ for pions, kaons and baryons from NA49}
\vspace{3mm}
\label{sec:meanpt_part}

\begin{figure}[h]
	\begin{center}
		\includegraphics[width=15cm] {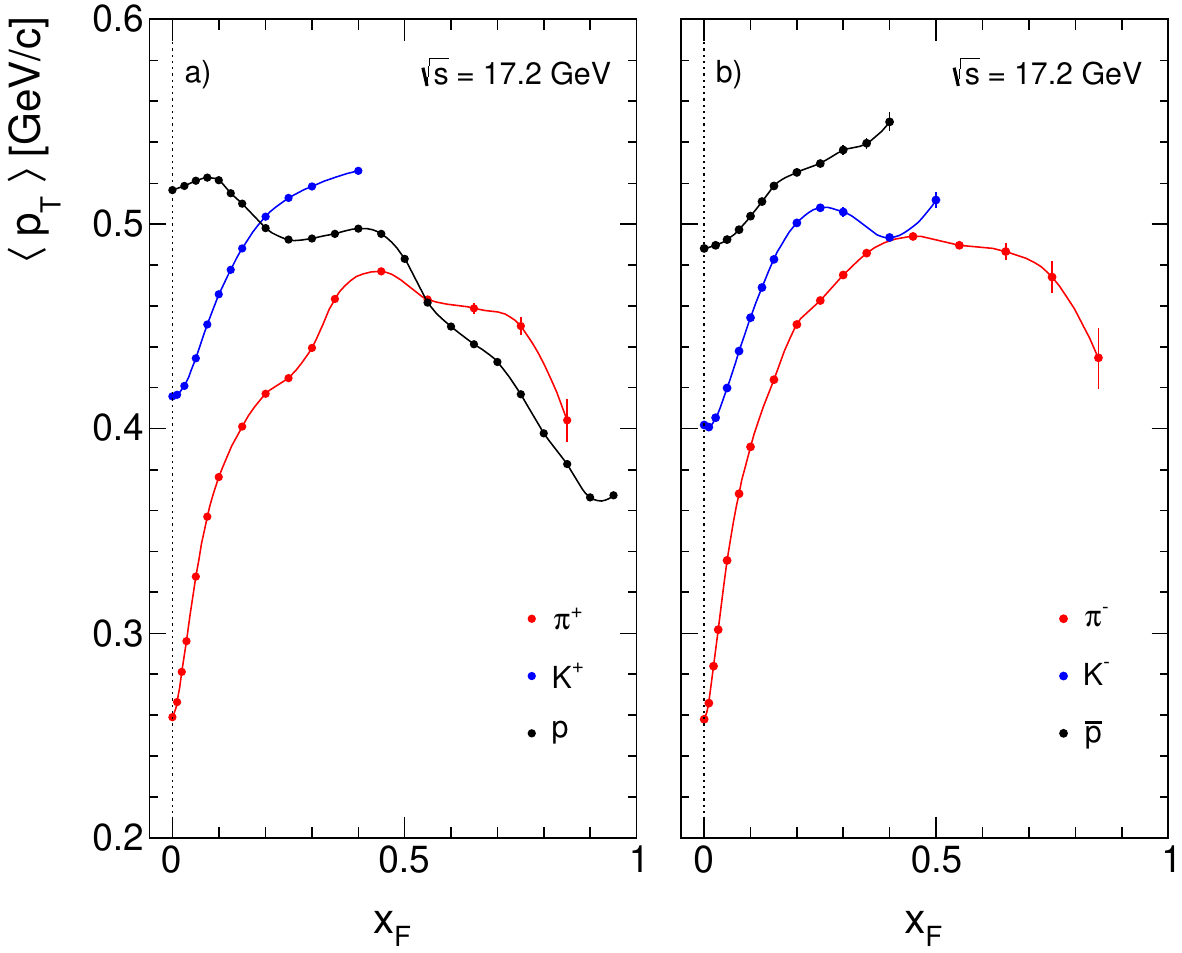} 
		\caption{Mean $p_T$ as a function of $x_F$ a) $\pi^+$, K$^+$, protons and b) $\pi^-$, K$^-$, anti-protons}
		\label{fig:meanpt_part}
	\end{center}
\end{figure}

The NA49 experiment has provided precision data of mean transverse momenta for $\pi^+$, $\pi^-$, K$^+$, K$^-$, protons and anti-protons at $\sqrt{s}$~=~17.2~GeV \cite{pp_pion,pp_kaon,pp_proton}. These data are presented in Fig.~\ref{fig:meanpt_part} as a function of $x_F$.

Several features are noteworthy:

\begin{enumerate}[label=(\arabic*)]
	\item Opposite charges give very similar $\langle p_T \rangle$ to within about 20~MeV/c.
	\item \label{it:2} At low $x_F$ there are substantial differences in $\langle p_T \rangle$ between particle species of about 150~MeV/c between pions and kaons and about 250 MeV/c between pions and baryons.
	\item \label{it:3} These differences tend to vanish with increasing $x_F$ such that at $x_F \sim$~0.4 all particle types yield $\langle p_T \rangle$ values which are equal to within about 50~MeV/c.
	\item \label{it:4} There are two exceptions to this effect: $\langle p_T \rangle$ for anti-protons continues to increase with $x_F$ whereas for protons there is a pronounced decrease for $x_F >$~0.4.
\end{enumerate}

In the framework of resonance decays treated in Sects.~\ref{sec:resonances} and \ref{sec:res_mass_spec} above, the effects \ref{it:2} and \ref{it:3} are explained by the resonance transverse momentum and the resonance mass distribution, respectively, as shown in Figs.~\ref{fig:delta_ppimeanpt} and \ref{fig:delta_resmeanpt}. In this sense the data of Fig.~\ref{fig:meanpt_part} represent a model independent check of the importance of resonance decay for the final state inclusive data. The deviations \ref{it:4} are a consequence of the detailed production mechanism. For protons, there is the approach to diffraction at large $x_F$ with sharper $p_T$ distributions \cite{pp_proton} whereas anti-proton production is characteristic of high-mass isospin triplets of mesonic origin \cite{fischer} with a steeply declining $x_F$ distribution. In this case it is the high mass tail of the respective resonances which is exploited.

In this context it should be stressed that many salient features of inclusive final state hadron distributions follow naturally from resonance decay in contrast to statistical and thermal models which have to take reference to an initial equilibrium state and neither predict the experimental inverse slopes (Sect.~\ref{sec:res_invslope}) nor the dependence of mean $p_T$ on $x_F$.

%
%
\subsection{\boldmath $\langle p_T \rangle$ for $\pi^-$ as a result of the global interpolation for different values of $\log(s)$}
\vspace{3mm}
\label{sec:meanpt_pim_interp}

\begin{figure}[h]
	\begin{center}
		\includegraphics[width=9cm] {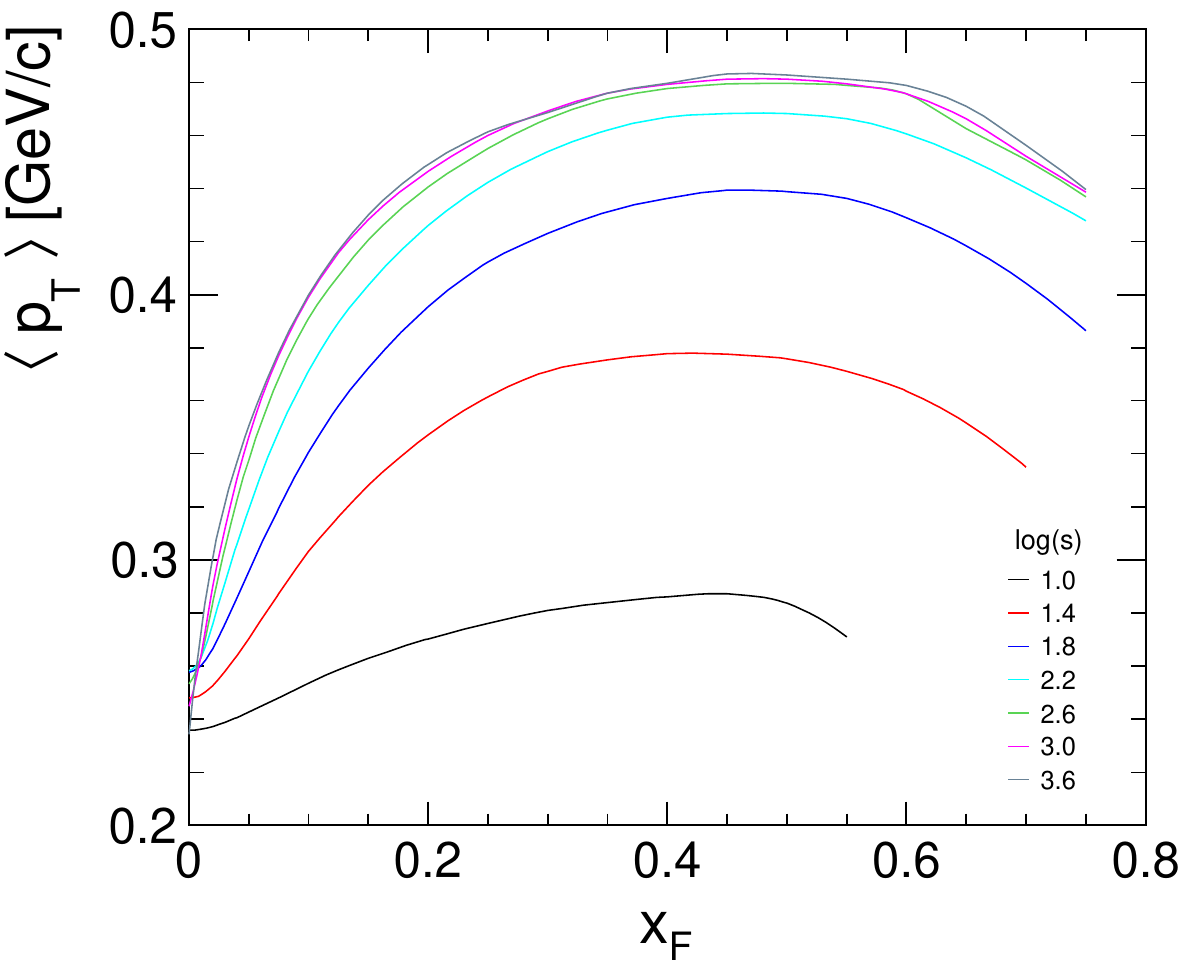} 
		\caption{$\langle p_T \rangle$ of $\pi^-$ as a function of $x_F$ for 7 values of $\log(s)$ from 1 to 3.6}
		\label{fig:logs_meanpt}
	\end{center}
\end{figure}

The interpolation scheme presented in Sect.~\ref{sec:interpolation} allows for the calculation of mean transverse momentum in a range of $\log(s)$ from 1 to 3.6 or $\sqrt{s}$ from 3 to 63~GeV over the full phase space. The result is shown in Fig.~\ref{fig:logs_meanpt} as a function of $x_F$ for $\log(s)$~=~1 to 3.6
in steps of 0.4 for feed-down corrected data.

Fig.~\ref{fig:logs_meanpt} shows a definite progression of the mean $p_T$ by factors which vary from 10\% at $x_F$~=~0 to 70\% at $x_F$~=~0.4 over the range of $\sqrt{s}$ from 3 to 63~GeV.

Several remarks are to be made concerning the strong dependence of $\langle p_T \rangle$ on both the longitudinal momentum and the interaction energy as is evident from Fig.~\ref{fig:logs_meanpt}:

\begin{enumerate}
\item $\langle p_T \rangle$ increases with $x_F$ from $x_F$~=~0 to $x_F \sim$~0.4 where the increase reaches a plateau at all energies.
\item There is a strong dependence on $\sqrt{s}$ which saturates at $\log{s} >$~3.0.
\item There is as well a strong increase with $x_F$ which ranges from about 20\% at $\log(s)$~=~1 to 90\% at $\log(s)$~=~3.6.
\end{enumerate}

\begin{figure}[h]
	\begin{center}
		\includegraphics[width=8.5cm] {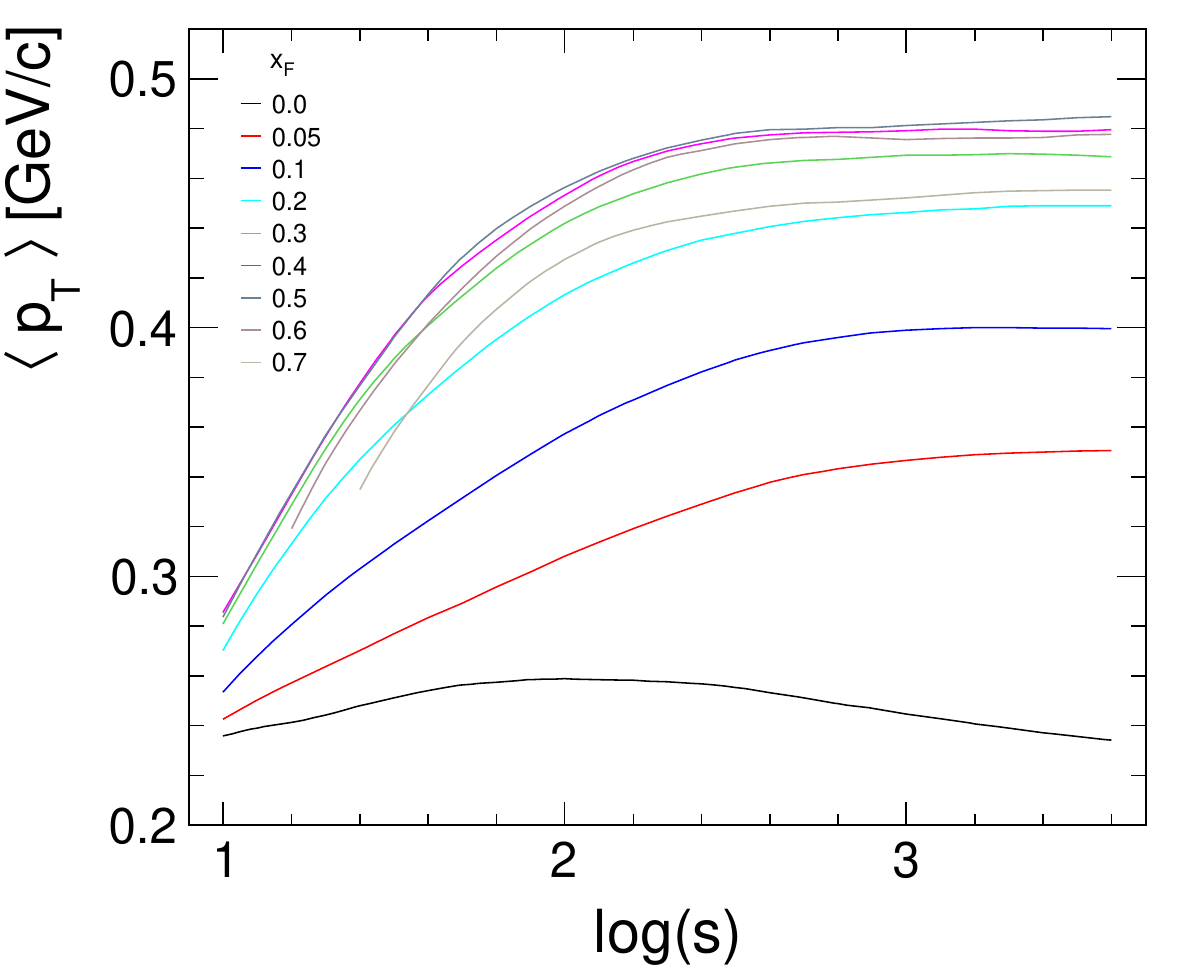} 
		\caption{$\langle p_T \rangle$ of $\pi^-$ as a function of $\log(s)$ for several values of $x_F$}
		\label{fig:logs_meanpt_xf}
	\end{center}
\end{figure}

This situation is visualized in Fig.~\ref{fig:logs_meanpt_xf} which shows the mean $p_T$ as a function of $\log(s)$ for several values of $x_F$.

\begin{figure}[h]
	\begin{center}
		\includegraphics[width=9cm] {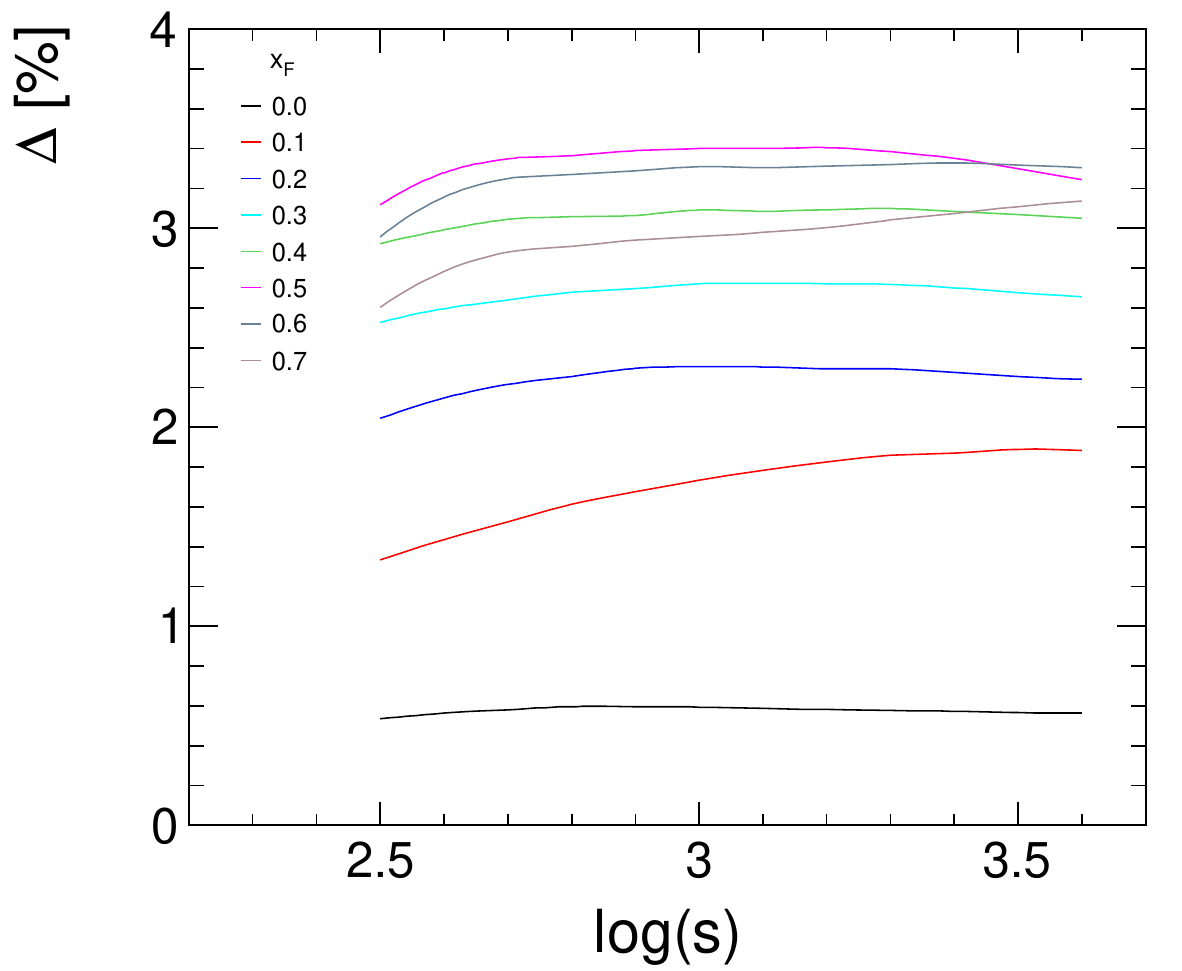} 
		\caption{The increase of $\langle p_T \rangle$ between an upper integration limit at 1.3~GeV/c and a limit at 1.9~GeV/c in \% as a function of $\log(s)$}
		\label{fig:meanpt_increase}
	\end{center}
\end{figure}

The saturation in $\log(s)$ is due to two underlying effects. Firstly the invariant cross sections show only a small $s$-dependence above $\log(s)$~=~2.5, see Fig.~\ref{fig:fspt13} for $p_T$~=~1.3~GeV/c. Secondly the limit at $p_T <$~1.3~GeV/c leaves the tail of the $p_T$ distributions unaccounted for. The effect of this $p_T$ cut is shown in Fig.~\ref{fig:meanpt_increase} where the percentage increase of $\langle p_T \rangle$ from an upper limit at 1.3~GeV/c to an upper limit at 1.9~GeV/c is presented.

There is a steady increase of $\langle p_T \rangle$ between the two integration limits as a function of $\log(s)$ which reaches more than 3.5\% or about 0.017~GeV/c at $\log(s)$~=~3.6.

\begin{figure}[h]
	\begin{center}
		\includegraphics[width=13cm] {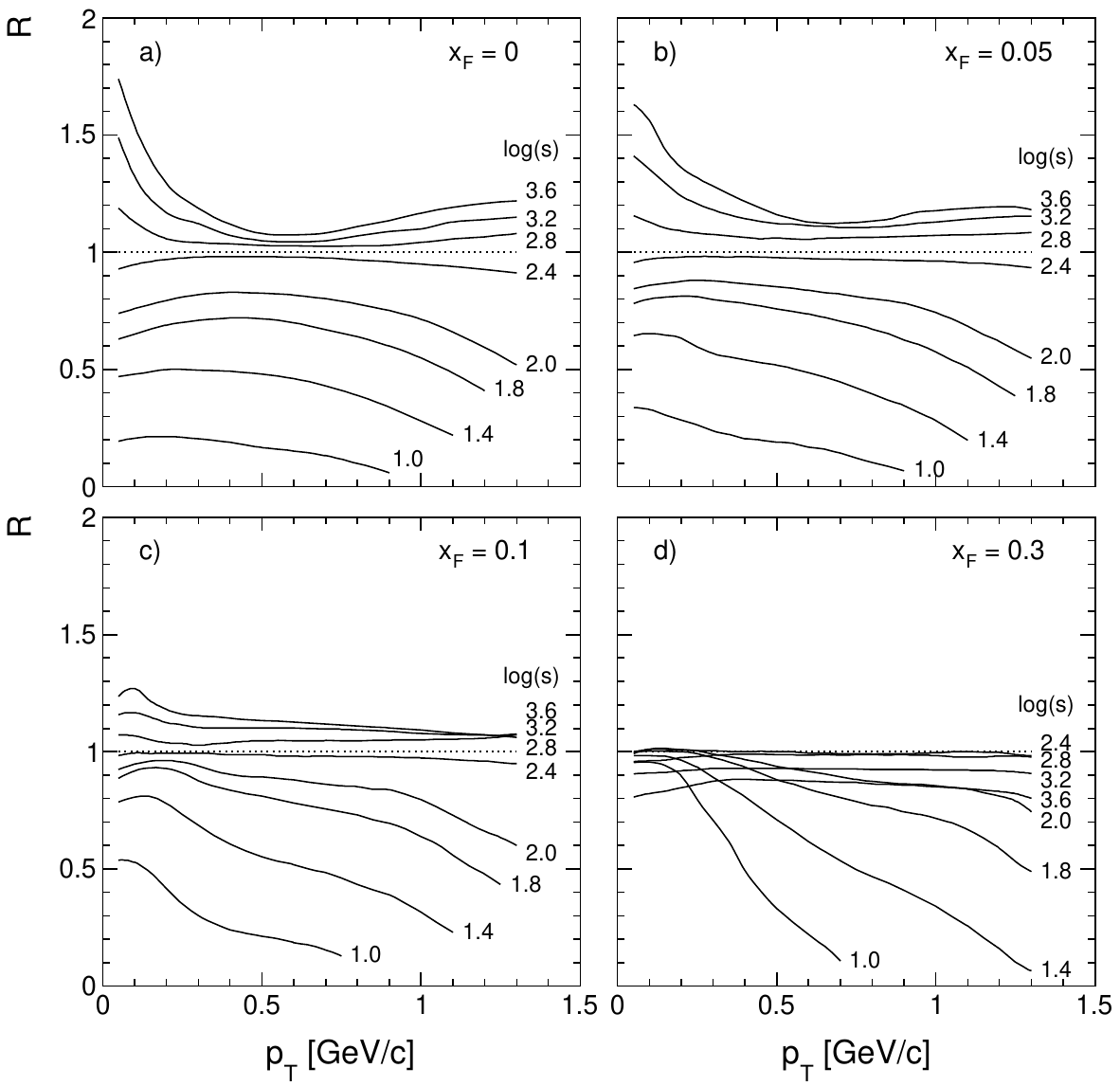} 
		\caption{The ratio $R$ of the invariant $\pi^-$ cross section to the NA49 data (\ref{eq:rat_y0}) as a function of $p_T$ for different values of $\log(s)$ for a) $x_F$~=~0, b) $x_F$~=~0.05, c) $x_F$~=0.1 and d) $x_F$~=~0.3}
		\label{fig:ratio_pt}
	\end{center}
\end{figure}

As far as the $x_F$ dependence is concerned ("seagull effect") two underlying components may be isolated using the detailed $p_T$ dependence of the cross section ratios (\ref{eq:rat_y0}) shown in Figs.~\ref{fig:r7.7}--\ref{fig:ptmtdist} at various $x_F$ values, Fig.~\ref{fig:ratio_pt}.

At central rapidity, a first component at low $p_T$ exhibits an exponential shape with inverse $p_T$ slopes of about 0.18~GeV/c which is independent of the interaction energy, Figs.~\ref{fig:r7.7}--\ref{fig:col_inverse}. This corresponds, if converted into invariant cross sections and plotted against $m_T$ (Fig.~\ref{fig:ptmtdist}), to inverse $m_T$ slopes of about 0.08~GeV/c. The (non-thermal) value is similar to the results for feed-down pions from weakly decaying strange hadrons (Fig.~\ref{fig:mtslopes}) and indicates the presence of further resonant states with low $Q$ values and small width.

A second component corresponds to the higher $p_T$ range and is clearly visible at $p_T >$~0.5~GeV/c. It features an exponential increase with $p_T$ whose inverse slope increases with the interaction energy as shown in Fig.~\ref{fig:col_inverse}b. This slope becomes negative below $\sqrt{s} \sim$~10~GeV. At the base of this behaviour is of course energy-momentum conservation at low interaction energy together with the production of resonances with increasing mass towards higher energies.

In the central region both the exponential decrease at low $p_T$ and the exponential increase at higher $p_T$ tend to compensate such that the mean $p_T$ remains small with little dependence on $\log(s)$.

In forward direction both components vanish as shown in Fig.~\ref{fig:ratio_pt}b) to d) for $x_F$~=~0.3 up to $\sqrt{s}$~=~63~GeV, see also Figs.~\ref{fig:kazeros}, \ref{fig:kazerol} and \ref{fig:lambdaS}. The behaviour towards higher $\sqrt{s}$ will probably never be accessible at present-day colliders.

Here the prevailing effect of resonance decay (Sect.~\ref{sec:beyond_inclusive} and the discussion in Sect.~\ref{sec:resonances}, Figs.~\ref{fig:delta_resmeanpt} and \ref{fig:delta2ppi_meanpt}) leads to the strong increase of $\langle p_T \rangle$ with $x_F$ which is mostly driven by the resonance mass distribution via the Lorentz transformation from the resonance cms to the lab system.

\begin{figure}[h]
	\begin{center}
		\includegraphics[width=8.5cm] {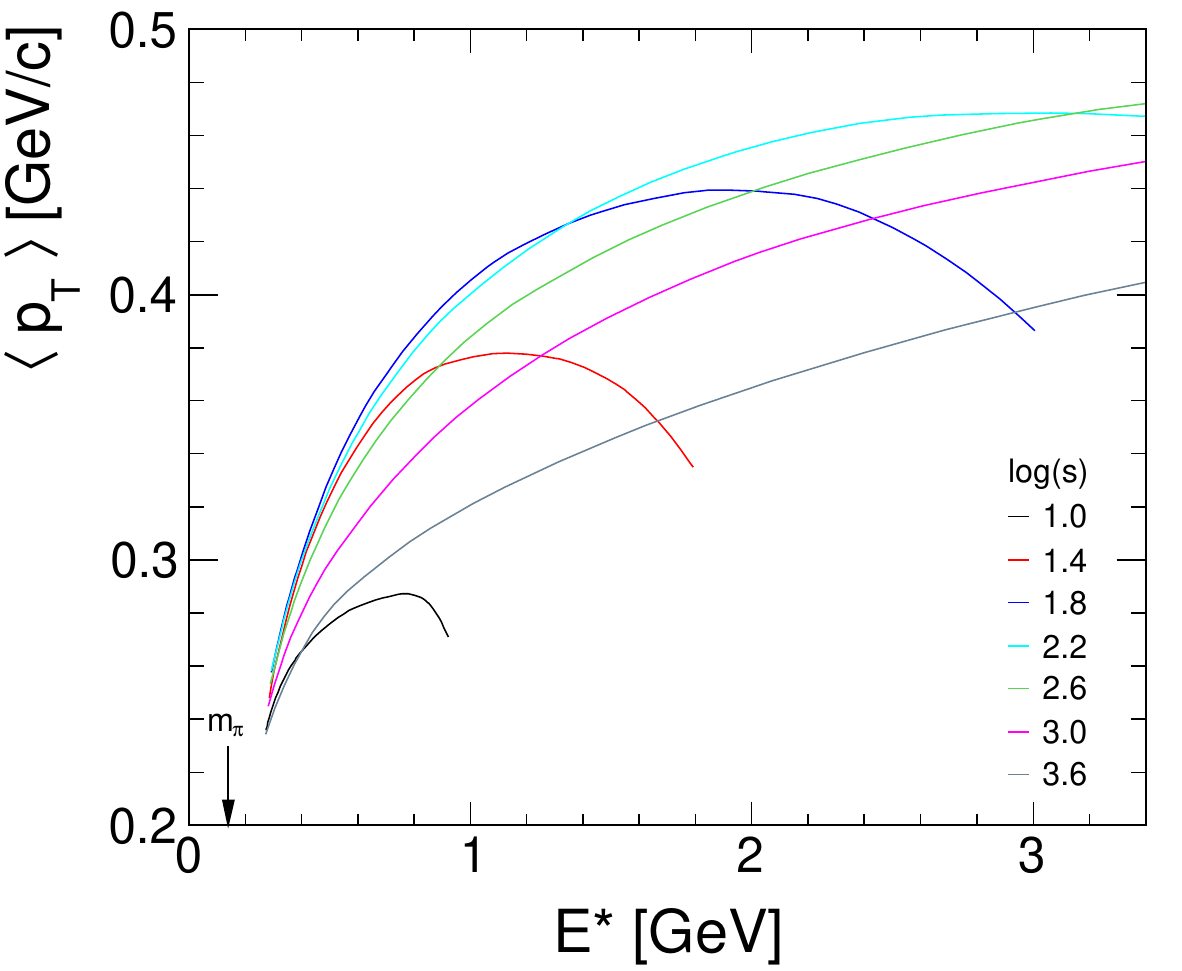} 
		\caption{$\langle p_T \rangle$ of $\pi^-$ as a function of cms energy $E^*$ for seven interaction energies $\sqrt{s}$ from 3.2 to 63~GeV}
		\label{fig:logs_meanpt_ener}
	\end{center}
\end{figure}

A further remark concerns the approach to the plateau in $\langle p_T \rangle$ up from $x_F$~=~0, Fig.~\ref{fig:logs_meanpt}. With increasing interaction energy this approach becomes successively steeper indicating that the relative longitudinal momentum $x_F$ is not the correct variable to describe this phenomenon. In fact there is a factor of 20 in longitudinal momentum for constant $x_F$ between the lowest and the highest interaction energy. The mean transverse momentum is therefore plotted in Fig.~\ref{fig:logs_meanpt_ener} as a function of cms energy $E^*$ rather than $x_F$ in order to cover the increase of $\langle p_T \rangle$ from the lowest longitudinal momentum, $E^*$~=~$m_\pi$ up to $E^*$~=~3~GeV.

The $E^*$ dependence shows a common, $s$-independent increase from $E^*_{\textrm{min}} = m_\pi$ up to $\sqrt{s}$~=~12.6~GeV. The deviations at 3.2, 5, 7.9 and 12.6~GeV above $E^*$~=0.4, 0.8, 1.4 and 2.4~GeV are imposed by energy-momentum conservation which cuts in at about 35\% of the available cms energy $E^*_{max} = \sqrt{s}/2$. The saturation at $\log(s)$ above 2.6, Fig.~\ref{fig:ratio_pt}, transforms into a subsequent lowering of the mean $p_T$ in the $E^*$ scale such that similar $\langle p_T \rangle$ values are reached multiplying $E^*$ by the $\sqrt{s}$ ratio ($x_F$ scaling).

%
%
\section{Physics beyond the inclusive level: contribution from known resonances to the \boldmath $\pi^-$ yield}
\vspace{3mm}
\label{sec:beyond_inclusive}

In the preceding sections some aspects of resonance decay and its influence on different inclusive quantities of the final state particles have been discussed. If a complete set of hadronic resonances decaying into given final state hadrons would be known including detailed yields and phase space distributions, quantum numbers, branching fractions, decay widths and mass distributions, a new level of inclusive physics beyond the single particle inclusive state might be defined. This could ultimately yield a deeper understanding of non-perturbative QCD on a purely experimental level.

Unfortunately the present state of knowledge concerning resonance production and decay does not allow such a complete study. There are however quite a number of experimental results available which may be used to obtain at least a preliminary picture of the salient features of the resulting inclusive quantities.

The following argumentation is based on a set of eight mesonic and five hadronic resonances which has been discussed some years ago \cite{spsc} concerning final state $\pi^-$ production in connection with the NA49 experiment at $\sqrt{s}$~=~17.2~GeV. Although the corresponding results are necessarily incomplete some important conclusions may nevertheless be drawn.

%
%
\subsection{The resonance sample}
\vspace{3mm}
\label{sec:resonance_shape}

The following resonances decaying into $\pi^-$ have been included in the present study:

\begin{table*}[h]
	\renewcommand{\tabcolsep}{0.1pc}
	\renewcommand{\arraystretch}{1.4}
	\begin{center}
		\begin{tabular}{cl@{\hspace{10mm}}c}
			mesonic           &&  baryonic          \\   \cline{1-1} \cline{3-3}
			$\eta$(548)       &&  $\Delta^0$(1232)  \\
			$\rho^0$(770)     &&  $\Delta^-$(1232)  \\
			$\rho^-$(770)     &&  N$^*$(1440)       \\
			$\omega$(782)     &&  N$^*$(1520)       \\
			f$_2$(1270)       &&  N$^*$(1680)       \\
			$\rho_3^0$(1690)  &&  \\
			$\rho_3^-$(1690)  &&  \\
			f$_4$(2050)       &&  \\
		\end{tabular}
	\end{center}
\end{table*}

In order to avoid double counting, only two-body decays into final state hadrons have been included (three-body decays for $\eta$ and $\omega$). This excludes cascading decays like

\begin{align}
 \rho_3 &\rightarrow \omega+\pi \label{eq:rho3}\\
 N^* &\rightarrow \Delta+\pi  \label{eq:nstar}
\end{align}

 The resulting $\pi^-$ yields present therefore a lower limit, where cascading is expected to contribute preferentially to the low $p_T$ and $x_F$ regions.

%
%
\subsection{Resonance yields as functions of \boldmath $x_F$ and $p_T$}
\vspace{3mm}
\label{sec:resonance_yields}

\begin{figure}[h]
	\begin{center}
		\includegraphics[width=12cm] {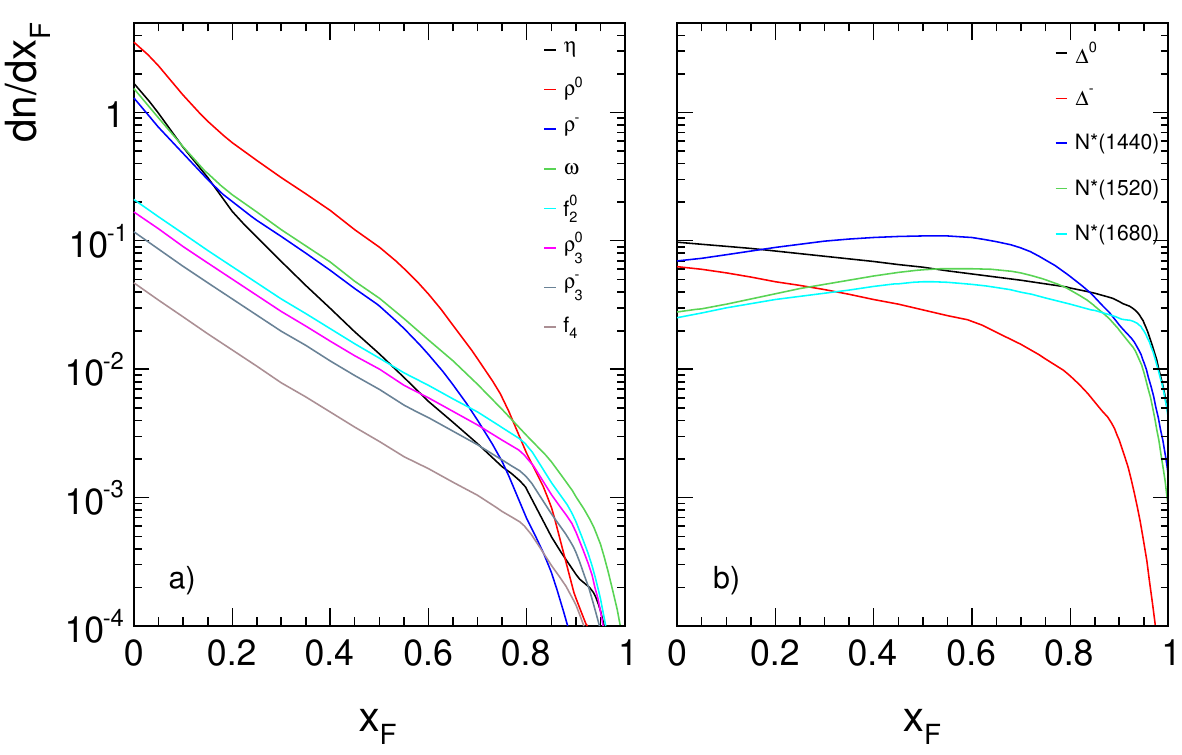} 
		\caption{$dn/dx_F$ distributions as a function of $x_F$, a) for a set of 8 mesonic resonances b) for five baryonic resonances}
		\label{fig:dndxf_resonances}
	\end{center}
\end{figure}

As there are practically no double-differential resonance cross sections available the measured $p_T$ integrated yields $dn/dx_F$ and $dn/dp_T^2$ are used for the generation of complete resonance spectra \cite{drijard,aguilar,blobel2,albrow1,drijard1,suzuki,suzuki1}. $dn/dx_F$ distributions are presented in Fig.~\ref{fig:dndxf_resonances} as a function of $x_F$ for the mesons and baryons defined above, inter- or extrapolated to the NA49 energy, $\sqrt{s}$~=~17.2~GeV.

The $dn/dx_F$ distributions give, when integrated over $x_F$, the total resonance yields that serve as normalization for the generated particle yields.

\begin{figure}[h]
	\begin{center}
		\includegraphics[width=12cm] {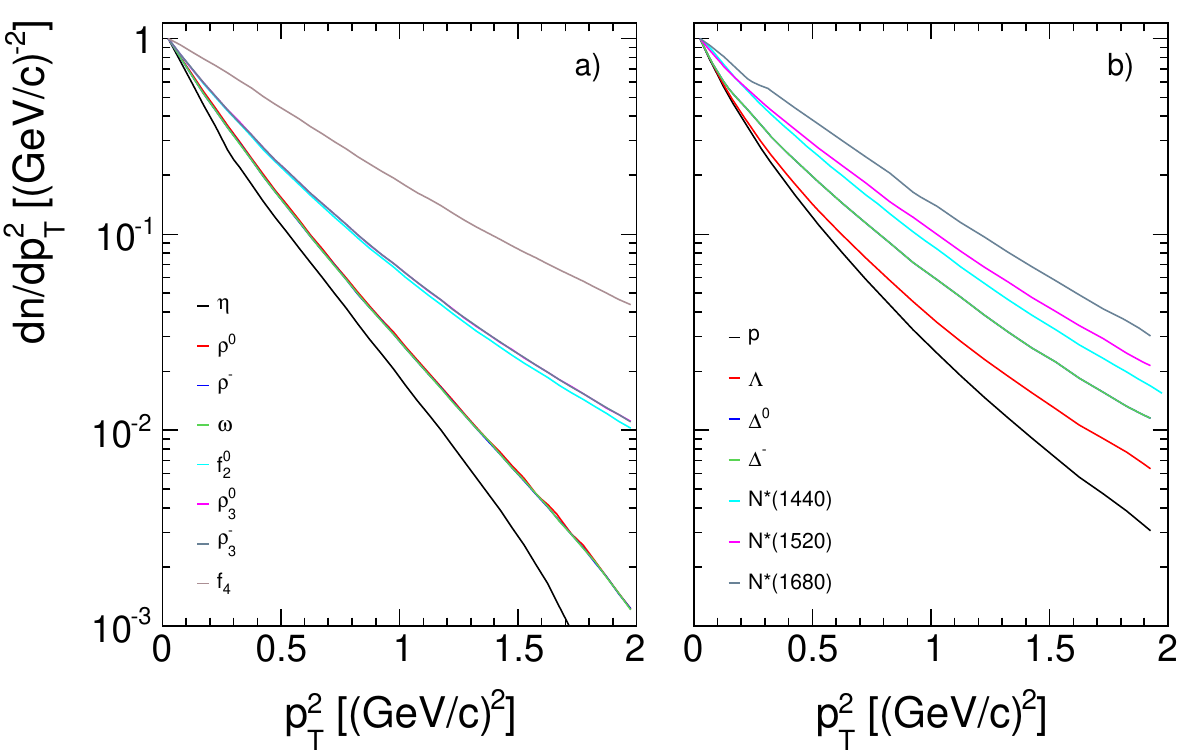} 
		\caption{Normalized $dn/dp_T^2$ distributions as a function of $p_T^2$, a) for mesonic b) for baryonic states}
		\label{fig:dndpt_resonances}
	\end{center}
\end{figure}

$dn/dp_T^2$ distributions are given in Fig.~\ref{fig:dndpt_resonances} as a function of $p_T^2$, here normalized at $p_T^2$~=~0.025~GeV/c$^2$.

For the mesonic states there is an important evolution of the $p_T^2$ slopes with resonance mass. This evolution is smaller for the baryonic states as the baryonic mass differences are smaller. The distribution for protons is also given in Fig.~\ref{fig:dndpt_resonances}. Here the yield difference is a function of $p_T^2$ similar to the same effect between $\Lambda$ and proton as indicated in Fig.~\ref{fig:lam_dd} above.

The integration of the $dn/dx_F$ distributions results in total resonance yields and, given the two-body branching fractions, in the $\pi^-$ cross sections given in Table~\ref{tab:resonance_list}. All yields are given per inelastic event.

\begin{table}[h]
	\renewcommand{\tabcolsep}{1.0pc}
	\renewcommand{\arraystretch}{1.15}
	\begin{center}
		\begin{tabular}{llll}
			\hline
			Resonance         & Total yield & Branching fraction   & $\pi^-$ yield \\ \hline
			$\Delta^0$(1232)  &   0.123     &        0.333         &   0.041   \\
			$\Delta^-$(1232)  &   0.0584    &        1.0           &   0.0584  \\
			N$^*$(1440)       &   0.157     &        0.43          &   0.0675  \\
			N$^*$(1520)       &   0.0848    &        0.40          &   0.0339  \\
			N$^*$(1680)       &   0.0716    &        0.43          &   0.0308  \\ \hline
			$\eta$(548)       &   0.316     &        0.273         &   0.0862  \\
			$\rho^0$(770)     &   0.837     &        1.0           &   0.837   \\
			$\rho^-$(770)     &   0.286     &        1.0           &   0.286   \\
			$\omega$(782)     &   0.333     &        0.907         &   0.302   \\
			f$_2$(1270)       &   0.0709    &        0.62          &   0.0440  \\
			$\rho_3^0$(1690)  &   0.0566    &        0.236         &   0.0134  \\
			$\rho_3^-$(1690)  &   0.0397    &        0.236         &   0.0094  \\
			f$_4$(2050)       &   0.0159    &        0.113         &   0.00180 \\ \hline
			total baryons     &   0.495     & $\pi^-$ from baryons &   0.232   \\
			total mesons      &   1.955     & $\pi^-$ from mesons  &   1.579   \\ \hline
			\multicolumn{3}{l}{total $\pi^-$ from 2-body decays (3-body from $\eta$ and $\omega$)}    &   1.811   \\ \hline
			\multicolumn{3}{l}{total inclusive $\pi^-$ \cite{pp_pion}} &   2.36    \\ \hline
			\multicolumn{3}{l}{fraction from 2-body decays (3-body from $\eta$ and $\omega$)}        &  76.7\%   \\ \hline
		\end{tabular}
		\caption{Resonance yields}
		\label{tab:resonance_list}
	\end{center}
\end{table}

Several consequences follow from the results of Table~\ref{tab:resonance_list}:

\begin{enumerate}
	\item About three quarters of all inclusive $\pi^-$ are coming from 2-body decays of measured baryonic and mesonic resonances.
	\item As stated above this result has to be regarded as a lower limit since higher multiplicity decays are disregarded and clearly not all possible resonant states have been included. This is especially true for strange particle decays which are not included here.
	\item As decay particles from two-body resonance decays are non-thermal in the sense of Hagedorn's statistical bootstrap model \cite{hagedorn2} their overwhelming contribution to the final-state inclusive yields puts grave doubts on the applicability of statistical or thermal models to hadronic interactions.
	\item The biggest contributions in Table~\ref{tab:resonance_list} come from the low-mass mesonic resonances $\eta$, $\rho$ and $\omega$. It is clear from the available branching fractions of higher mass states that these mesons are cascading down from both baryonic and mesonic resonances. Hence the necessity to restrict their contribution to their direct decays into final state hadrons. On the other hand decays of the type $\Delta \rightarrow N^*(1440) + \pi$ or $b_1(1235) \rightarrow \omega + \pi$ yield -- again via two-body decays -- pion contributions which have to be counted into the total inclusive cross sections.
	\item The resonance cross sections given in Table~\ref{tab:resonance_list} span three orders of magnitude and their decay pions another 4 to 5 orders of magnitude. It has therefore to be realized that the relative contribution of different resonances are strongly depending on their $Q$ values, mass and phase space distributions such that the tails of the inclusive cross sections both in $x_F$ and in $p_T$ have a different resonance heritage than their mean values.
\end{enumerate}

%
%
\subsection{Predictions for double-differential cross sections from resonance decay}
\vspace{3mm}
\label{sec:resonance_predict}

\begin{figure}[b]
	\begin{center}
		\includegraphics[width=15.5cm] {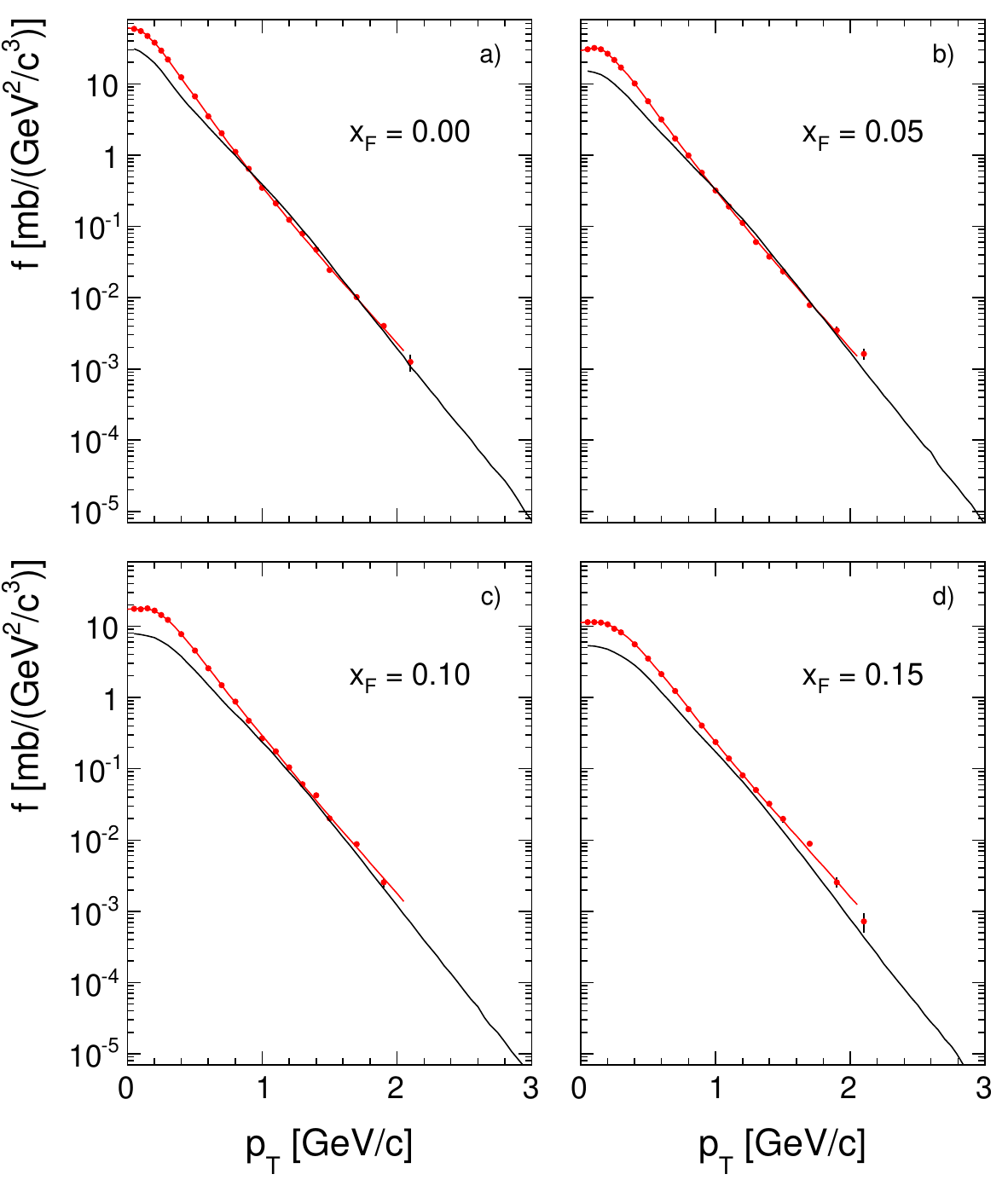} 
		\caption{Predicted invariant $\pi^-$ cross sections from resonance decay as a function of $p_T$ for fixed $x_F$~=~0, 0.05, 0.1 and 0.15. The predictions are given as full lines, the NA49 data as dots and the corresponding interpolations as red lines}
		\label{fig:res_xf0-015}
	\end{center}
\end{figure}

\begin{figure}[h]
	\begin{center}
		\includegraphics[width=15.5cm] {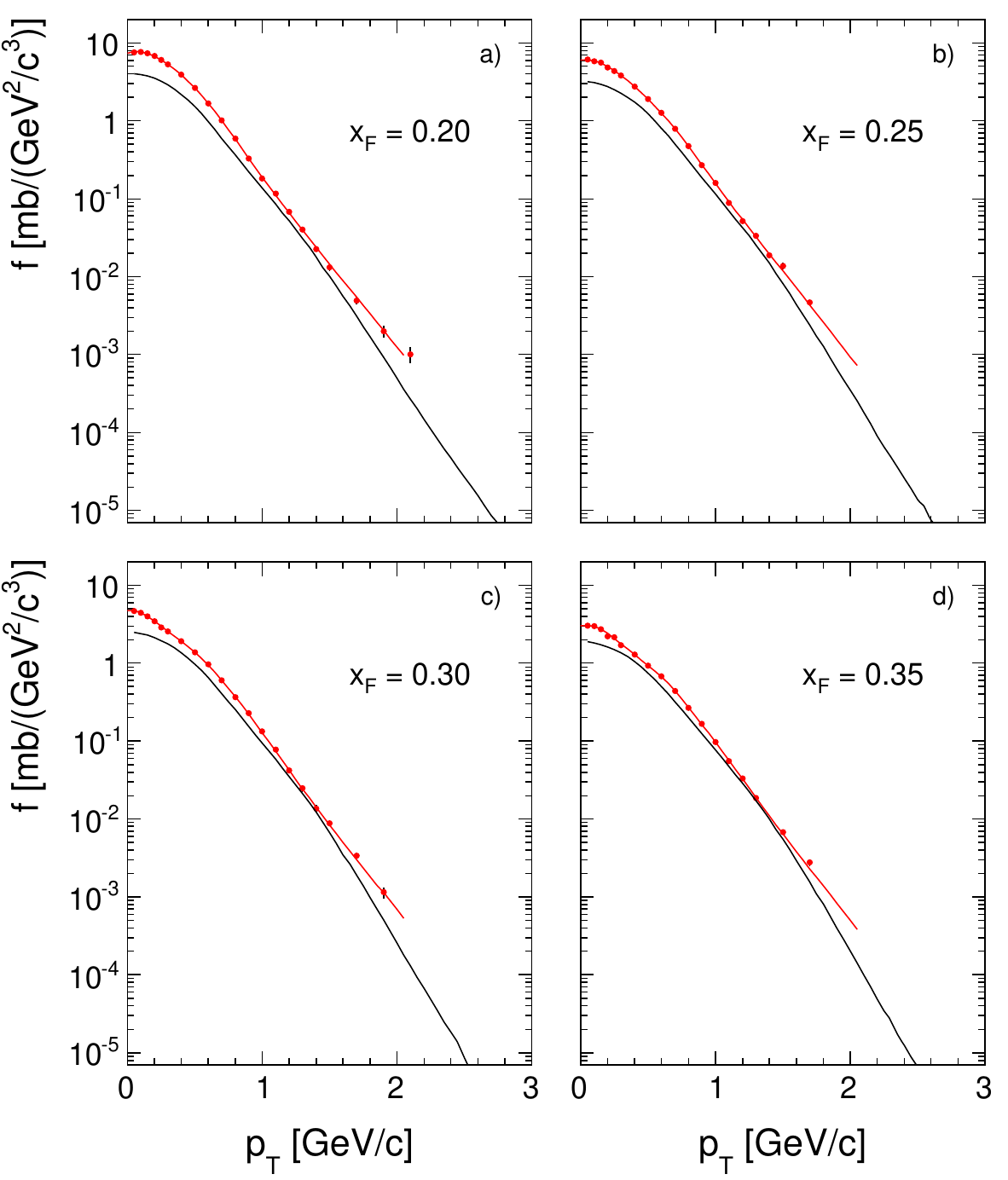} 
		\caption{Predicted invariant $\pi^-$ cross sections from resonance decay as a function of $p_T$ for fixed $x_F$~=~0.2, 0.25, 0.3 and 0.35. The predictions are given as full lines, the NA49 data as dots and the corresponding interpolations as red lines}
		\label{fig:res_xf02-035}
	\end{center}
\end{figure}

\begin{figure}[h]
	\begin{center}
		\includegraphics[width=15.5cm] {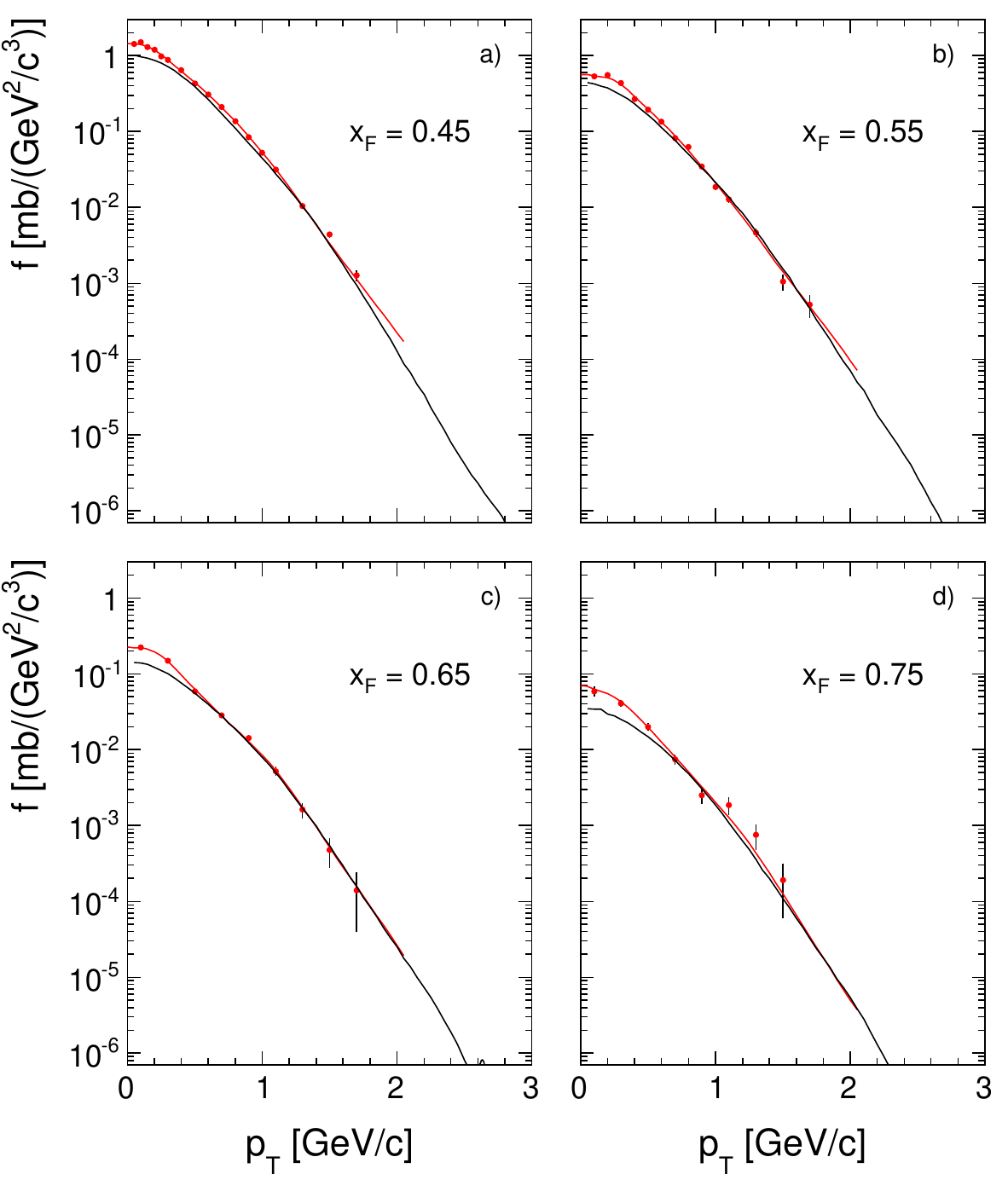} 
		\caption{Predicted invariant $\pi^-$ cross sections from resonance decay as a function of $p_T$ for fixed $x_F$~=~0.45, 0.55, 0.65 and 0.75. The predictions are given as full lines, the NA49 data as dots and the corresponding interpolations as red lines}
		\label{fig:res_xf045-075}
	\end{center}
\end{figure}

Invariant cross sections of $\pi^-$ from the decay of the resonances given in Table~\ref{tab:resonance_list} are given in Figs.~\ref{fig:res_xf0-015} to \ref{fig:res_xf045-075} as a function of $p_T$ for fixed $x_F$. They are directly comparable to the NA49 data \cite{pp_pion}.

Note that the plots are continued to $p_T$ values above the limit of 1.3~GeV/c imposed in this paper as the NA49 data extend partially up to 2~GeV/c. The simulation of the resonance decay results in $\pi^-$ cross sections which are rather exponential at low $x_F$ and $p_T >$~1~GeV/c. The prediction has hence been extended to $p_T$ beyond 2~GeV/C.

Several features are apparent from these figures:

\begin{enumerate}
	\item On first sight there is a surprising reproduction of the main features of the inclusive distributions as far as the dependencies on $p_T$ and $x_F$ are concerned.
	\item At $x_F >$~0.4 the data and their interpolation coincide with the prediction to a precision of a few percent for $p_T >$~0.5~GeV/c.
	\item There is an offset centred around $p_T \sim$~1.2$\pm$0.3~GeV/c which is $x_F$ dependent. Such offset would be expected if the resonance sum of Table~\ref{tab:resonance_list} would lack contributions from further resonances which are of course to be expected.
	\item There is an enhancement below $p_T \sim$~0.8 GeV/c which is strongly $x_F$ dependent.
	\item There is another enhancement at $p_T \gtrsim$~1.3 GeV/c and $x_F <$~0.6.
\end{enumerate}

\begin{figure}[h]
	\begin{center}
		\includegraphics[width=9cm] {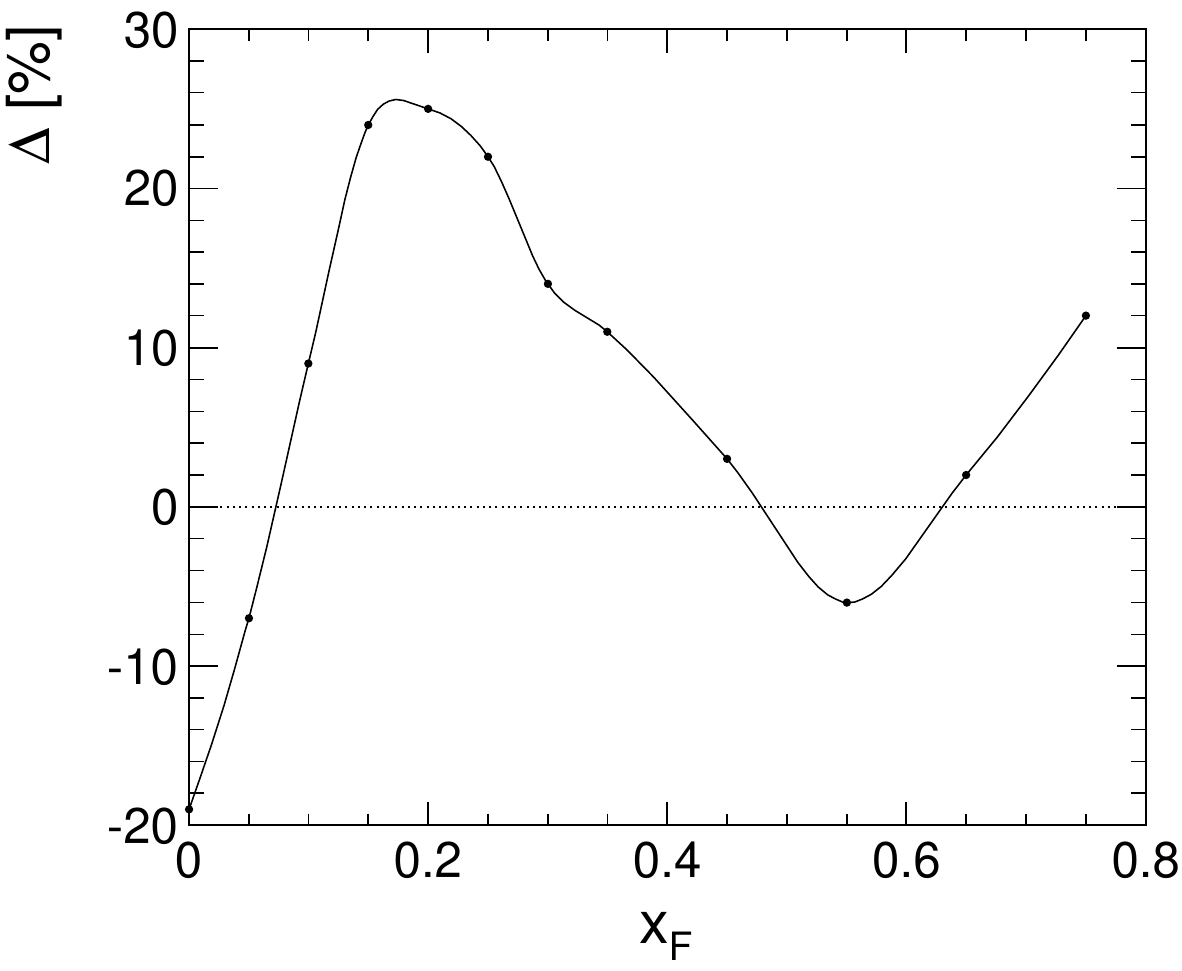} 
		\caption{Offset between data and $\pi^-$ from resonance decay at $p_T$~=~1.2$\pm$0.3~GeV/c in percent as a function of $x_F$}
		\label{fig:res_offset}
	\end{center}
\end{figure}

In order to quantify these effects the offset shown in Fig.~\ref{fig:res_offset} as a function of $x_F$ has been subtracted from the data interpolation.

\begin{figure}[h]
	\begin{center}
		\includegraphics[width=10.5cm] {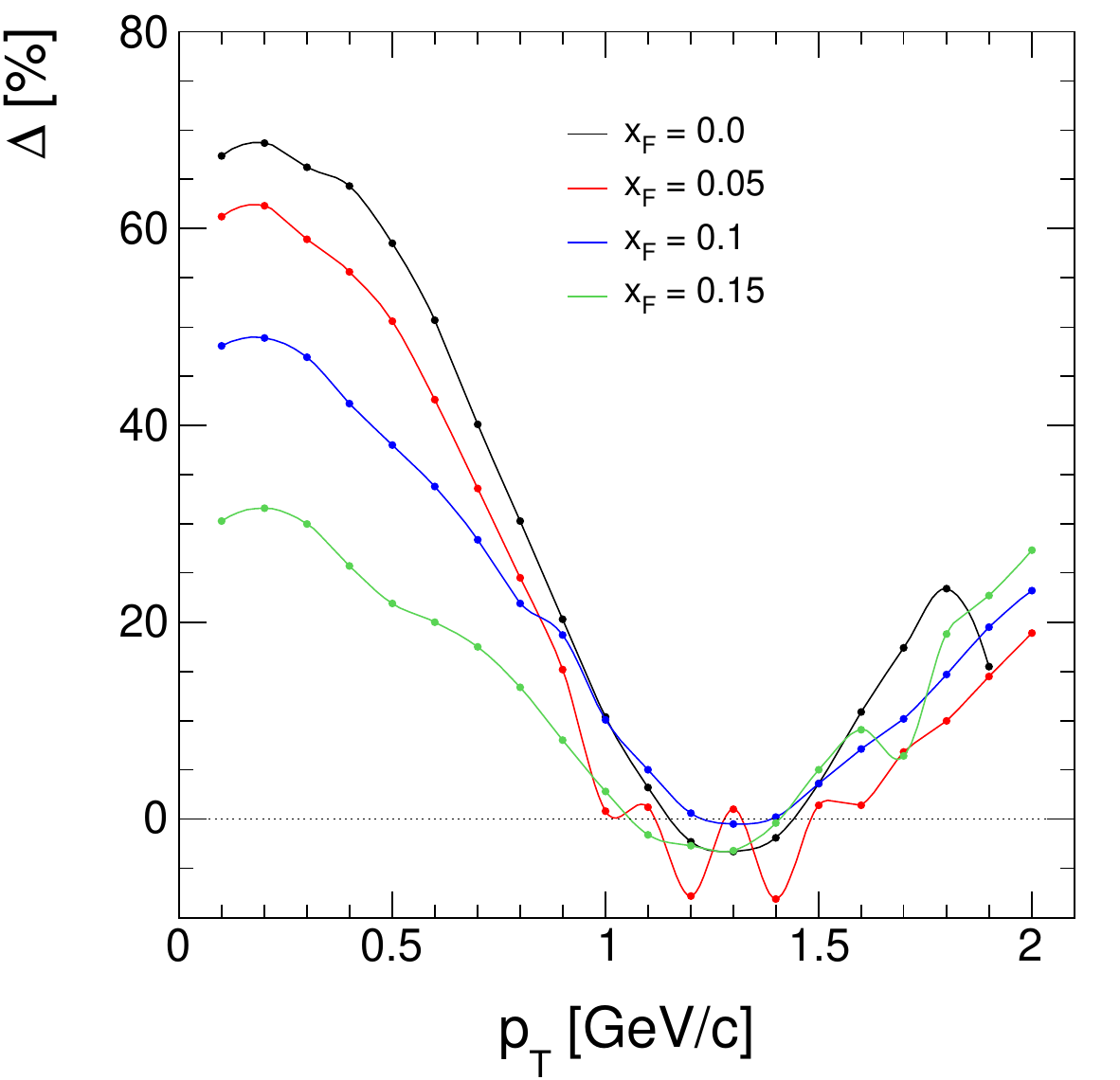} 
		\caption{Percent deviation between data and resonance sum as a function of $p_T$ for $x_F$~=~0, 0.05, 0.1 and 0.15}
		\label{fig:res_perc0-015}
	\end{center}
\end{figure}

\begin{figure}[h]
	\begin{center}
		\includegraphics[width=10.2cm] {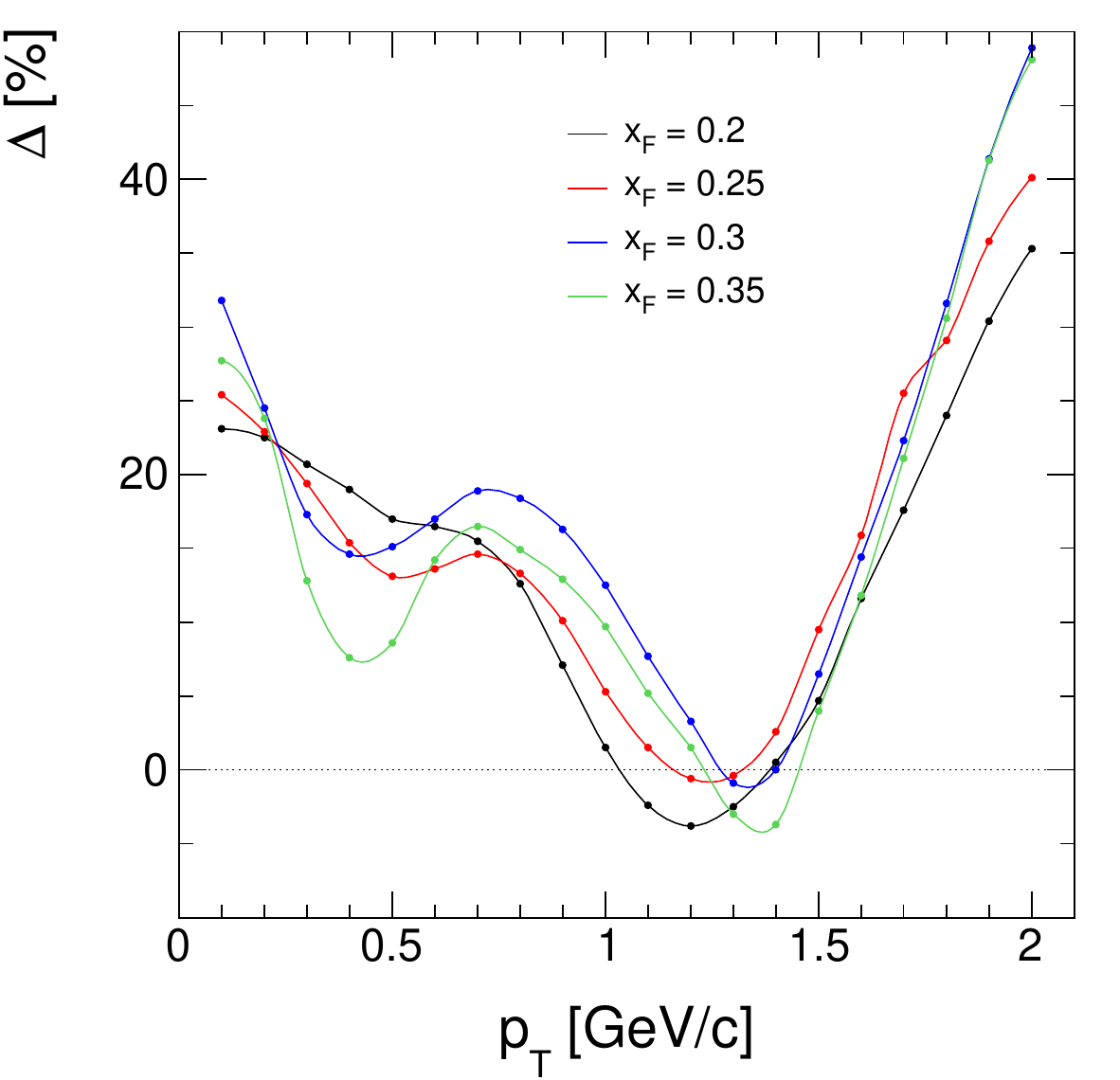} 
		\caption{ Percent deviation between data and resonance sum as a function of $p_T$ for $x_F$~=~0.2, 0.25, 0.3 and 0.35}
		\label{fig:res_perc02-035}
	\end{center}
\end{figure}

\begin{figure}[h]
	\begin{center}
		\includegraphics[width=10.2cm] {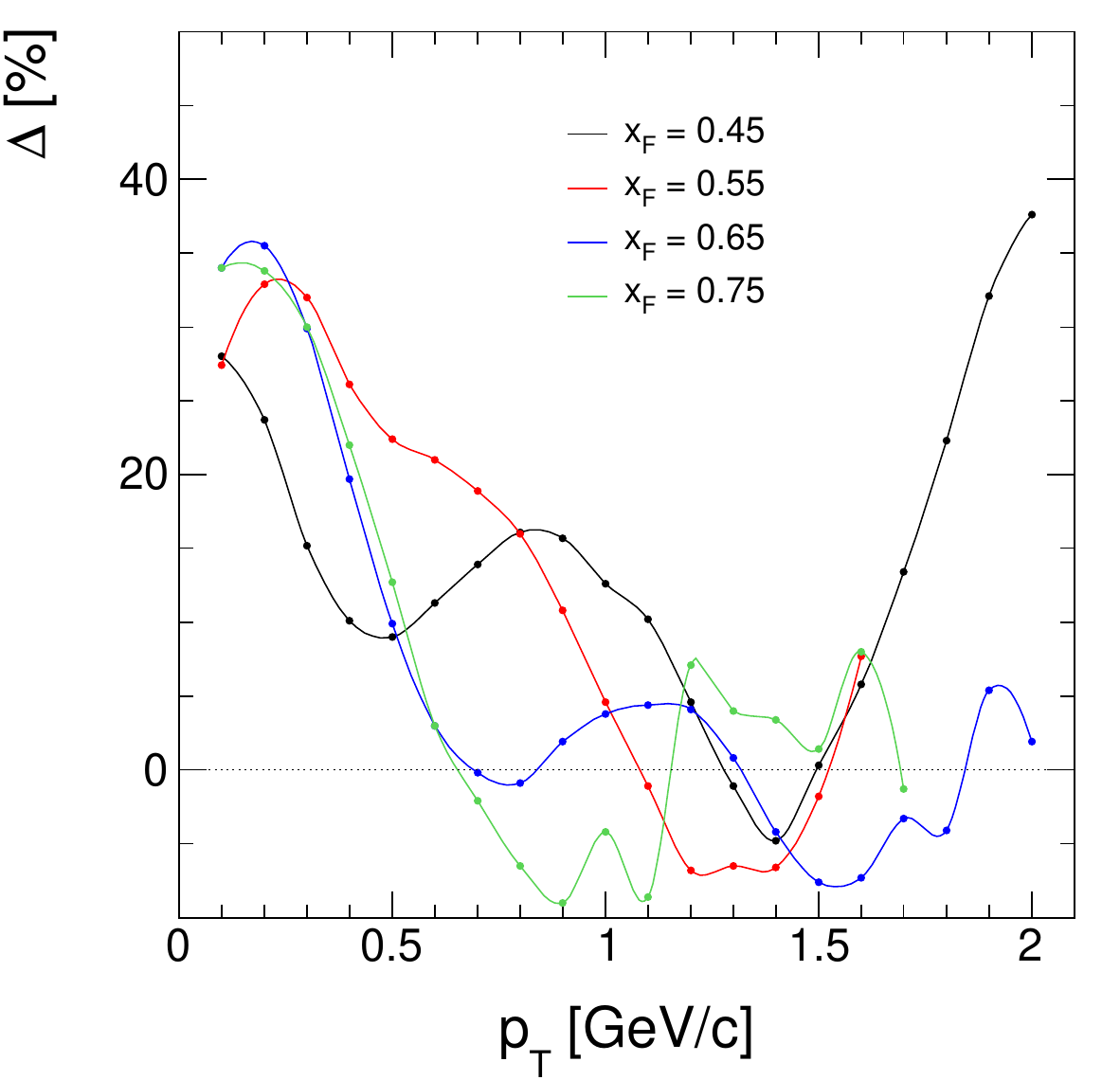} 
		\caption{Percent deviation between data and resonance sum as a function of $p_T$ for $x_F$~=~0.45, 0.55, 0.65 and 0.75}
		\label{fig:res_perc045-075}
	\end{center}
\end{figure}

After this subtraction the percent difference between data and resonance prediction is presented in Figs.~\ref{fig:res_perc0-015}--\ref{fig:res_perc045-075}.

\begin{figure}[h]
	\begin{center}
		\includegraphics[width=9.8cm] {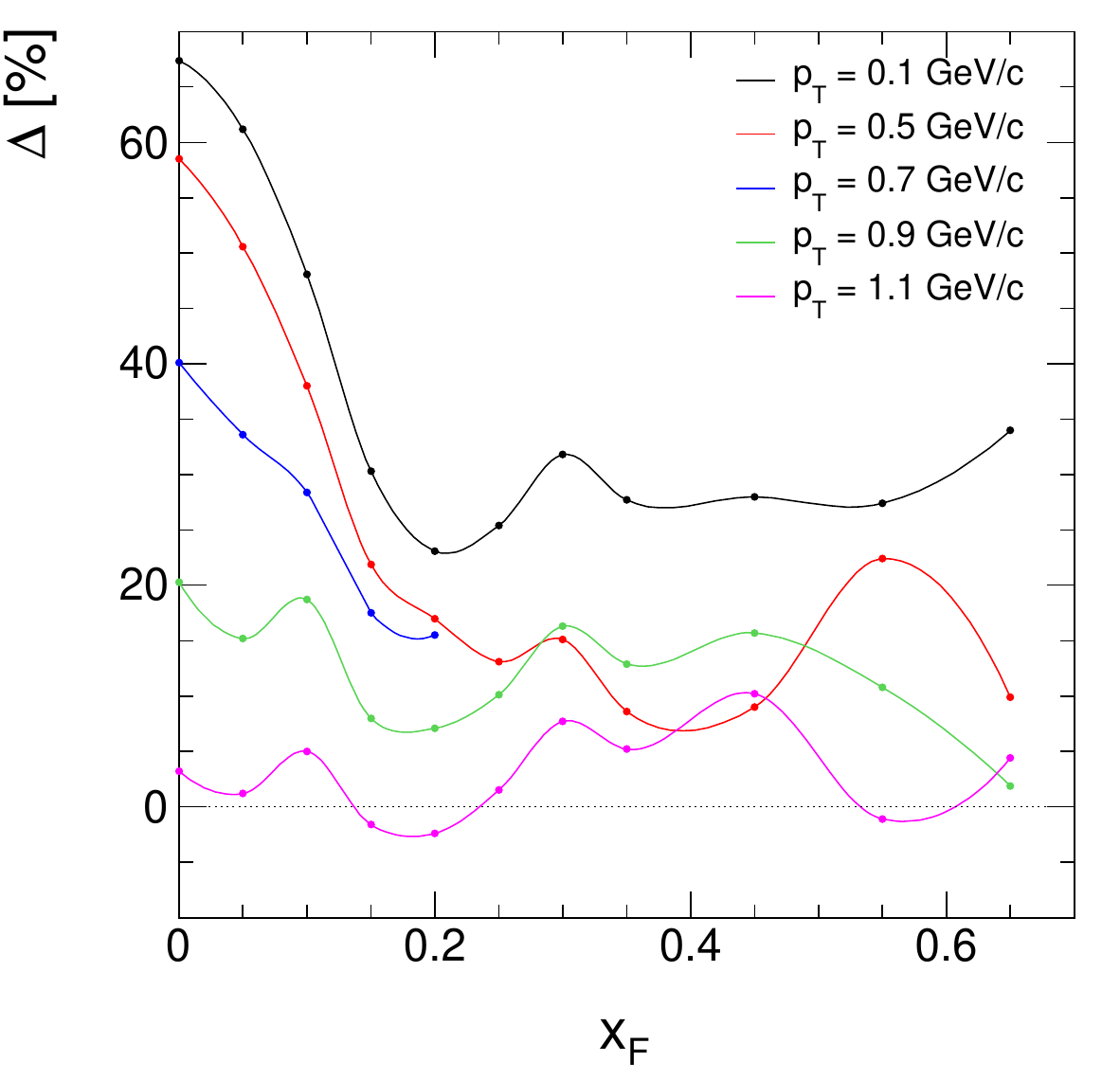} 
		\caption{Percent difference $\Delta$ between data and resonance prediction as a function of $x_F$ for $p_T$~=~0.1, 0.5, 0.7, 0.9 and 1.1~GeV/c}
		\label{fig:res_perc01-11}
	\end{center}
\end{figure}

\begin{figure}[h]
	\begin{center}
		\includegraphics[width=9.8cm] {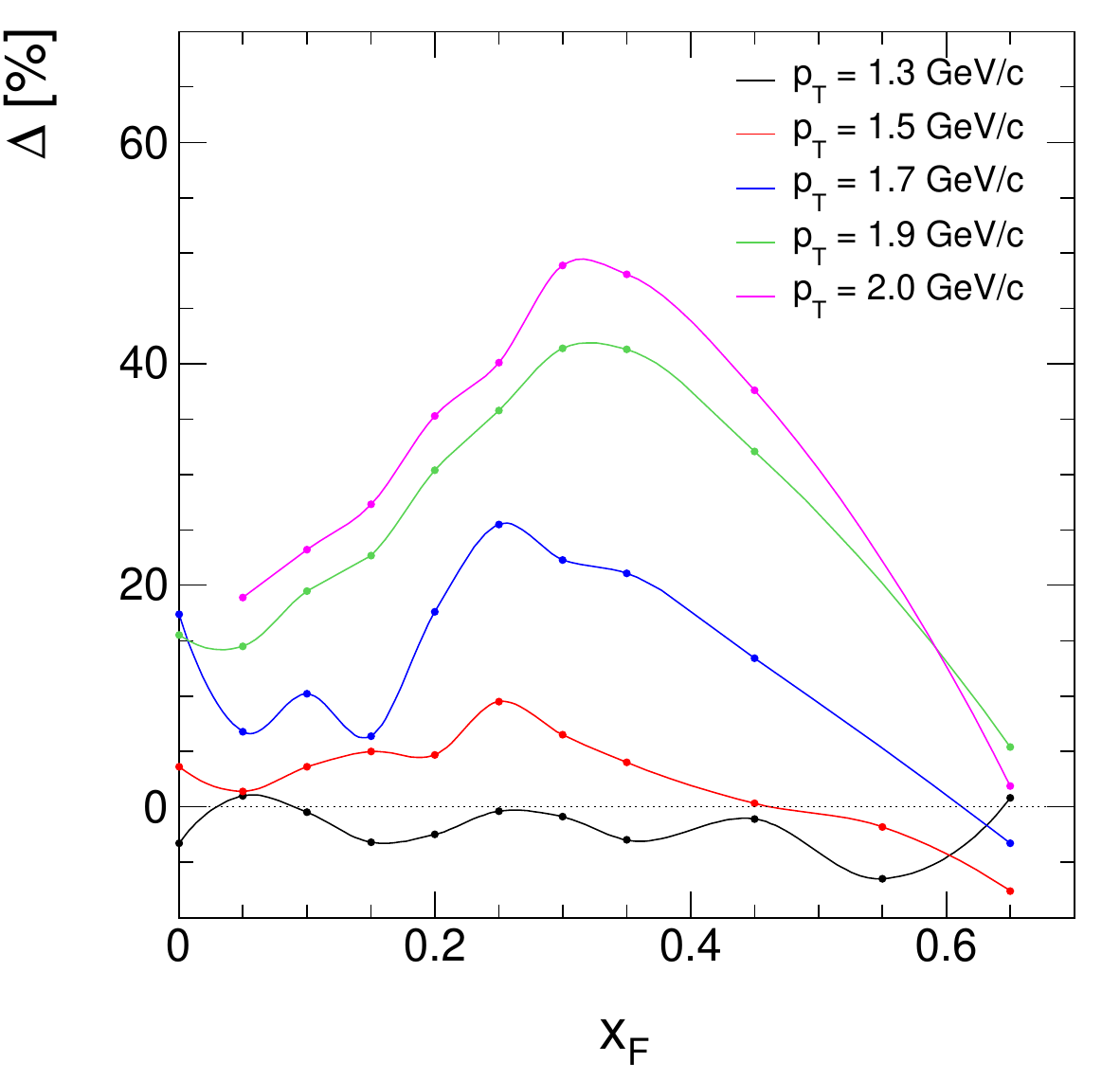} 
		\caption{Percent difference $\Delta$ between data and resonance prediction as a function of $x_F$ for $p_T$~=~1.3, 1.5, 1.7. 1.9 and 2.0~GeV/c}
		\label{fig:res_perc13-20}
	\end{center}
\end{figure}

It is also useful to show the $x_F$ dependence for fixed $p_T$, Figs.~\ref{fig:res_perc01-11} and \ref{fig:res_perc13-20}.

Some remarks are in place with respect to Figs.~\ref{fig:res_perc0-015}--\ref{fig:res_perc13-20}.

\begin{enumerate}
	\item Both in $p_T$ and in $x_F$ a strong decrease of the excess yield is apparent up to about 1~GeV/c in $p_T$ and 0.2 in $x_F$.
	\item This is followed by a systematic increase towards higher $p_T$ and $x_F$.
	\item The increase with $p_T$ is continuing well beyond the actual $p_T$ limit at 2.0 GeV/c.
	\item The increase with $x_F$ shows a maximum at $x_F \sim$~0.3~--~0.4 and decreases strongly towards higher $x_F$.
\end{enumerate}

The low $p_T$/low $x_F$ enhancements are reminiscent of the decay of low-$Q$ resonances without Breit-Wigner tails, see Fig.~\ref{fig:delta_pi_inv}. In this context the resonance decays into final-state hadrons

\begin{equation}
	 R \rightarrow h_1 + h_2 \qquad \textrm{($R$ - resonance, $h_1$ and $h_2$ - final state hadrons)}
	\label{eq:res1}
\end{equation}

\noindent
as discussed above have to be confronted with the cascading decay

\begin{equation}
	R \rightarrow R' + h \qquad \textrm{($R$ and $R'$ - resonances, $h$ - final state hadron)}	\label{eq:res2}
\end{equation}

Some examples for (\ref{eq:res2}) have been evoked in Sect.~\ref{sec:resonance_shape}, (\ref{eq:rho3}) and (\ref{eq:nstar}). Whereas in (\ref{eq:res1}) the Breit-Wigner tail of the decaying resonance comes fully into play, Sect.~\ref{sec:res_mass_spec}, the cascading decays like (\ref{eq:res2}) have been shown to damp the extent of the mass tail, Sect.~\ref{sec:res_cascading}. By energy-momentum conservation, the second resonance in (\ref{eq:res2}) has to be strongly correlated in its mass distribution with the first one: indeed the mass of the second resonance cannot be too different from the first one. This would explain the mass cut-offs necessitated by unitarity arguments shown in Sect.~\ref{sec:res_casc_bw}. The final state hadron will therefore be emitted at low $Q$ with a correlation that keeps the masses of both resonances close together with a result corresponding to Fig.~\ref{fig:delta_ppixfdep}. It would of course be highly desirable to look with high-statistics at cascading reactions like (\ref{eq:res2}) directly assessing mass correlations between $R$ and $R'$

The increase towards high $p_T$ and medium $x_F$ is, on the other hand, rather suggestive of higher mass resonances not contained in the list of Table~\ref{tab:resonance_list}, see for instance Fig.~\ref{fig:delta2ppi_meanpt}.

%
%
\subsection{Integrated quantities: mean \boldmath $p_T$}
\vspace{3mm}
\label{sec:resonance_meanpt}

\begin{figure}[h]
	\begin{center}
		\includegraphics[width=10.6cm] {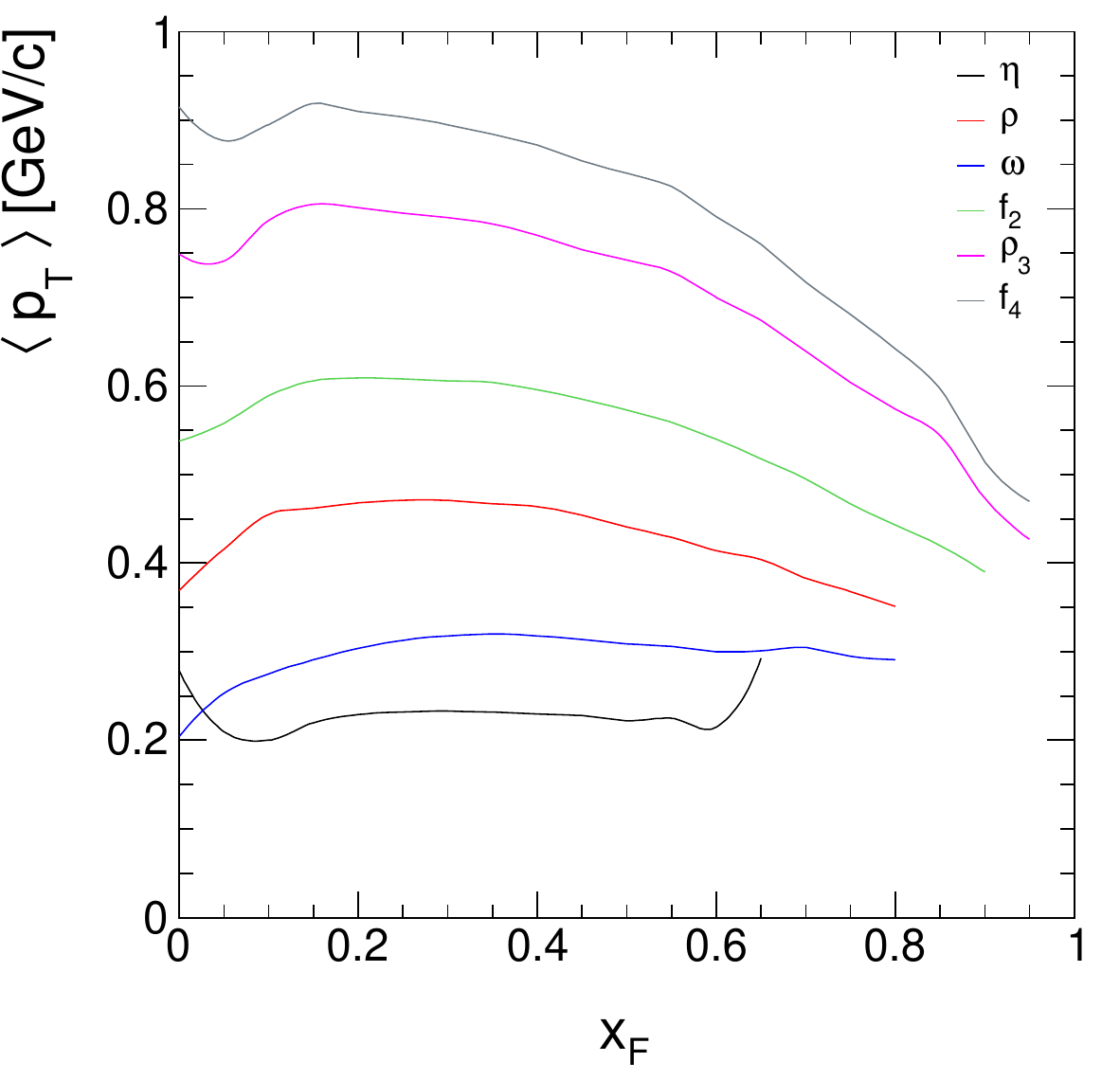} 
		\caption{Mean $p_T$ for $\pi^-$ from the decay of mesonic resonances as a function of $x_F$}
		\label{fig:res_mpt_mesonic}
	\end{center}
\end{figure}

\begin{figure}[h]
	\begin{center}
		\includegraphics[width=10.6cm] {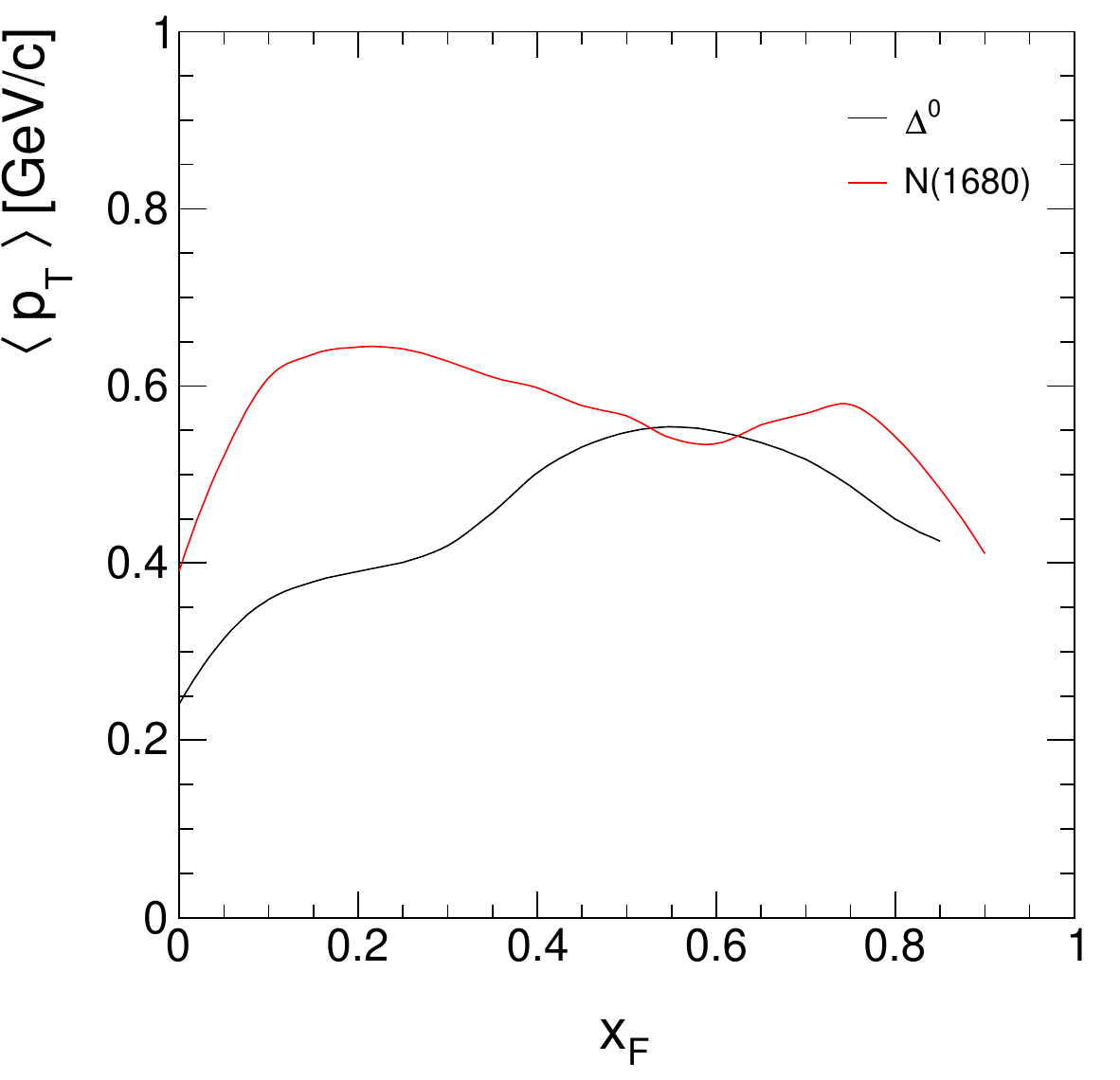} 
		\caption{Mean $p_T$ for $\pi^-$ from the decay of baryonic resonances as a function of $x_F$}
		\label{fig:res_mpt_baryonic}
	\end{center}
\end{figure}

The first moments of the $p_T$ distributions of $\pi^-$ from resonance decay, Table~\ref{tab:resonance_list}, are shown in Figs.~\ref{fig:res_mpt_mesonic} for mesonic and Fig.~\ref{fig:res_mpt_baryonic} for baryonic initial states.

The $\langle p_T \rangle$ values show important differences for the individual parent resonances depending on resonance mass and two- or three-body decays. There is no trace of a common behaviour neither for the shape of the $\langle p_T \rangle$ distributions nor for their absolute values: this corresponds of course to the problems encountered with the Statistical Bootstrap Model, Sect.~\ref{sec:thermo}.

\begin{figure}[h]
	\begin{center}
		\includegraphics[width=10.5cm] {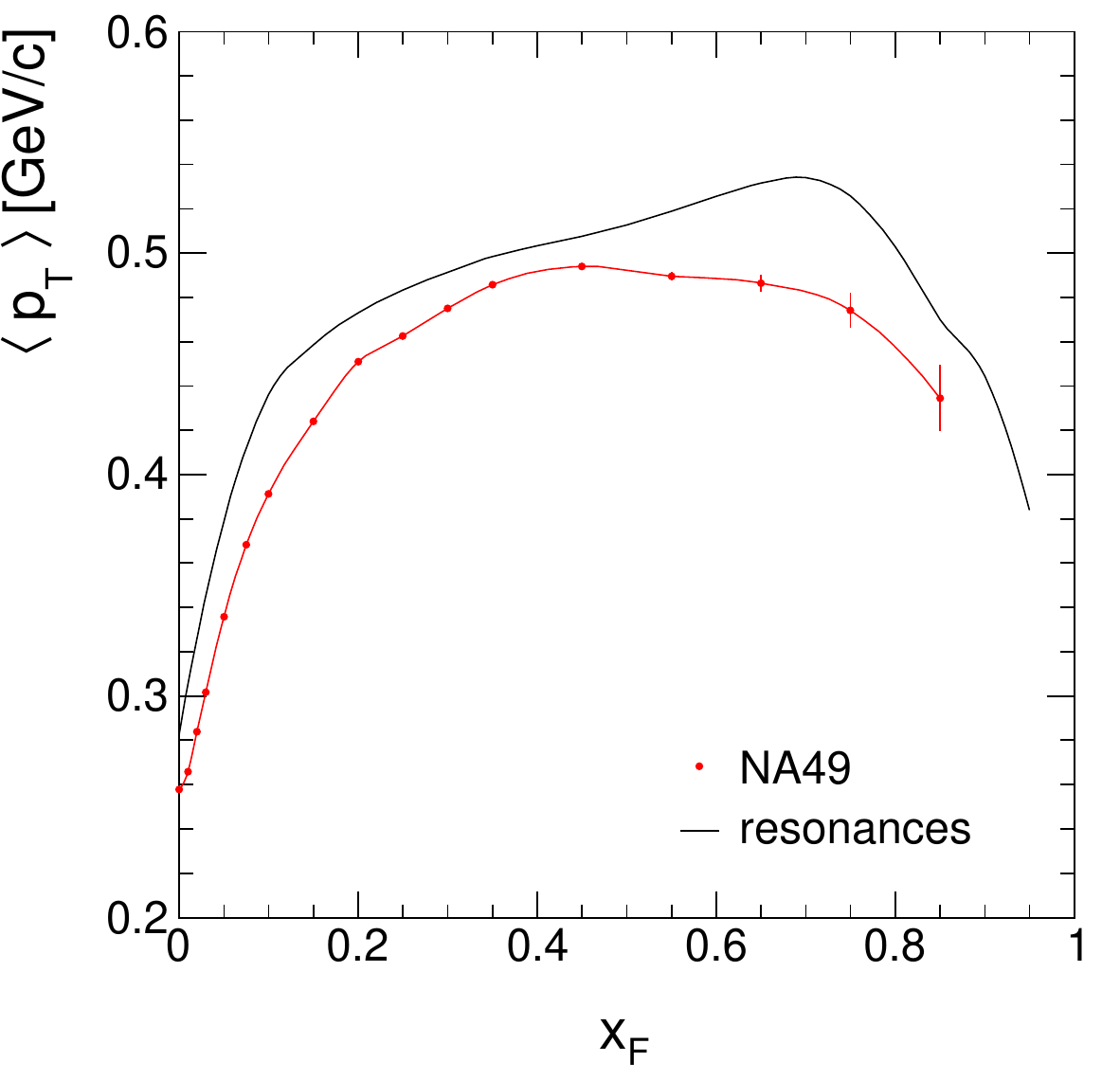} 
		\caption{Mean $p_T$ for $\pi^-$ from the complete resonance sample of Table~\ref{tab:resonance_list} as a function of $x_F$ in comparison to the measured inclusive data from NA49}
		\label{fig:res_mpt_all}
	\end{center}
\end{figure}

The superposition of $\langle p_T \rangle$ as a function of $x_F$ for the complete resonance sample is presented in Fig.~\ref{fig:res_mpt_all} and compared to the $\pi^-$ data from NA49 \cite{pp_pion}. The rather precise reproduction of the main features of the experimental result proves once more the necessity to go beyond the most general inclusive level in order to develop an understanding of the underlying physics.

%
%
\subsection{Integrated quantities: \boldmath $dn/dx_F$ distributions}
\vspace{3mm}
\label{sec:resonance_dndxf}

\begin{figure}[h]
	\begin{center}
		\includegraphics[width=15.5cm] {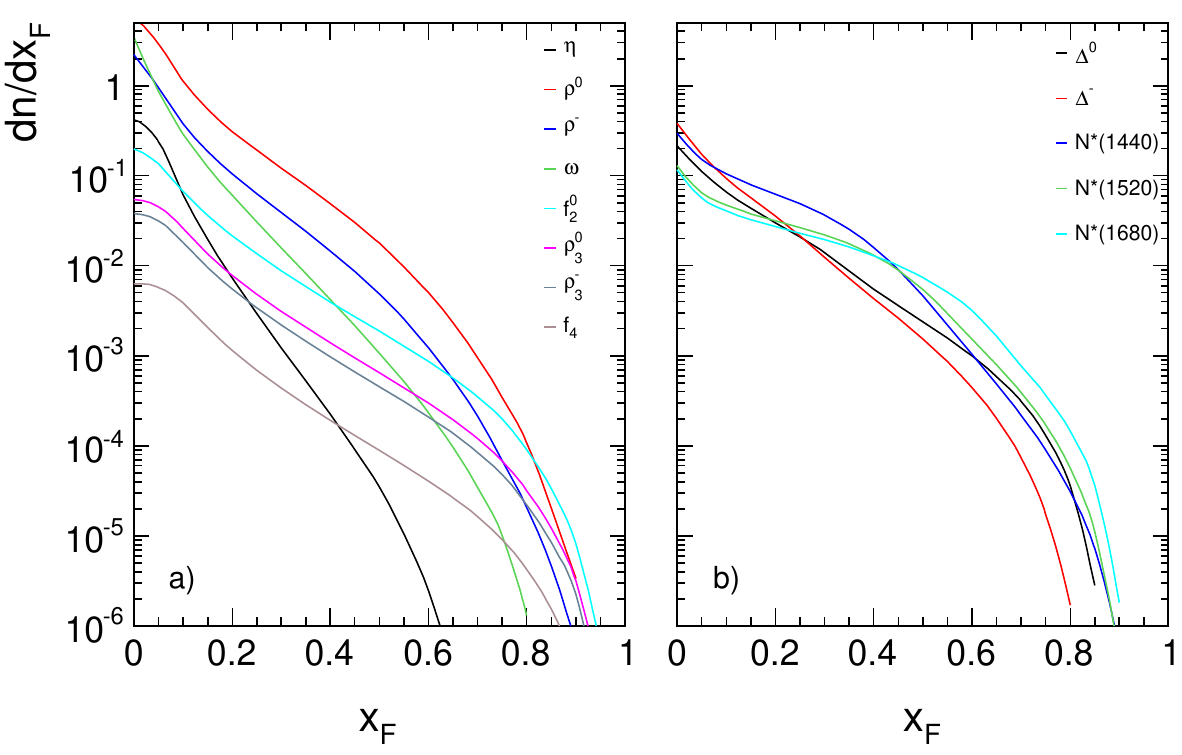} 
		\caption{$dn/dx_F$ distributions for $\pi^-$ from the decay of the resonances used in Table~\ref{tab:resonance_list} as a function of $x_F$}
		\label{fig:res_dndxf_all}
	\end{center}
\end{figure}

The integration over $p_T$ at fixed $x_F$ yields the $dn/dx_F$ distributions of the decay pions shown in Fig.~\ref{fig:res_dndxf_all} for the individual resonances used in Table~\ref{tab:resonance_list}:

\begin{equation}
	dn/dx_F = \pi/\sigma_\textrm{inel} \cdot \sqrt{s}/2 \cdot \int{f/E \cdot dp_T^2
}
	\label{eq:dndxf}
\end{equation}

\begin{figure}[h]
	\begin{center}
		\includegraphics[width=14cm] {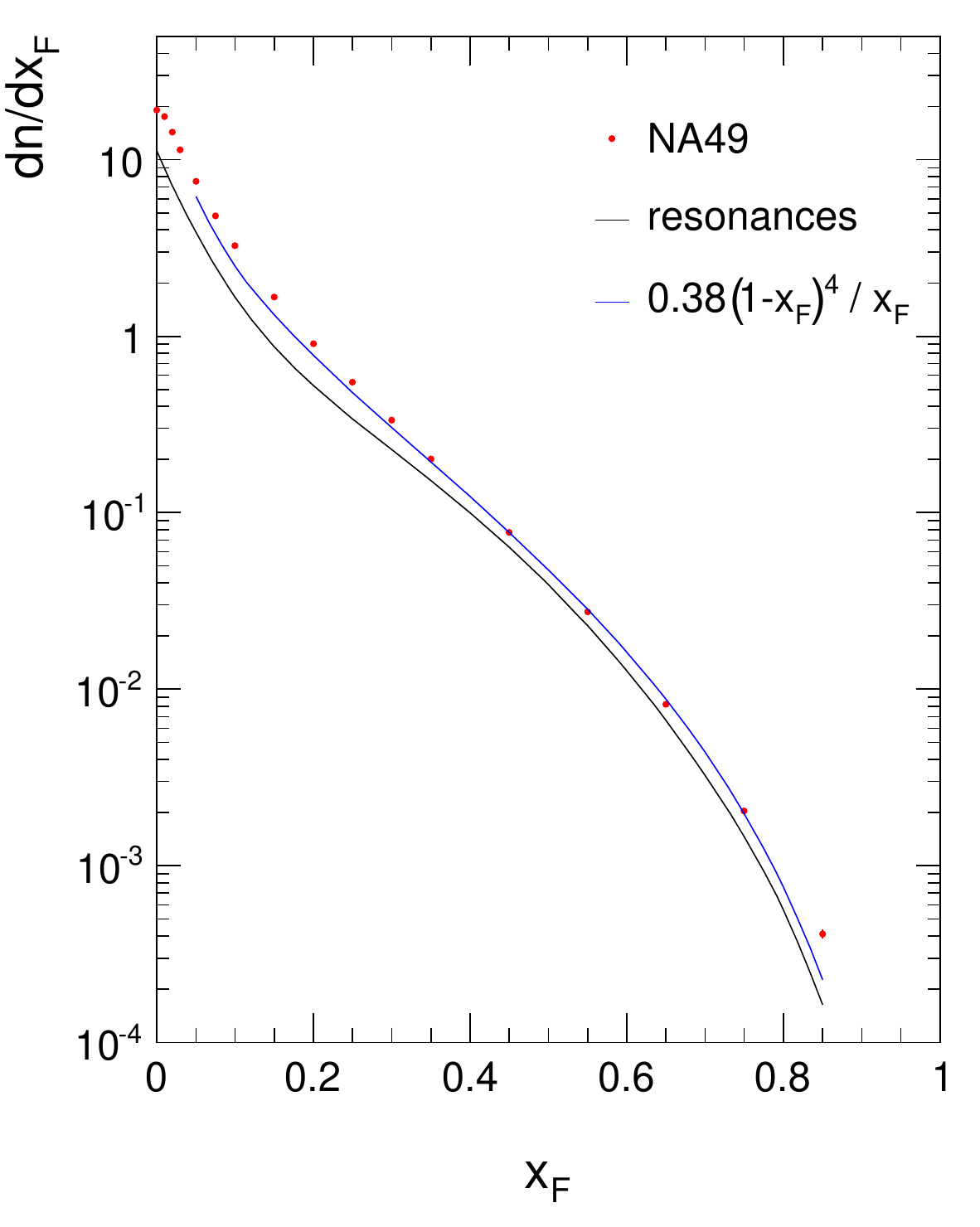} 
		\caption{$dn/dx_F$ of $\pi^-$ as a function of $x_F$ for the inclusive NA49 data and for the decay of the resonances contained in Table~\ref{tab:resonance_list}. A fit to the data suggesting the down-quark structure function is superimposed}
		\label{fig:res_dndxf2down}
	\end{center}
\end{figure}

The $dn/dx_F$ distributions show the expected steepening when compared to the parent particles, Fig.~\ref{fig:dndxf_resonances}. And again, as for $\langle p_T \rangle$, there are marked differences between the different mesonic and baryonic decays. The sum over all individual contributions is shown in Fig.~\ref{fig:res_dndxf2down}, and compared to the NA49 data.

\begin{figure}[h]
	\begin{center}
		\includegraphics[width=12.5cm] {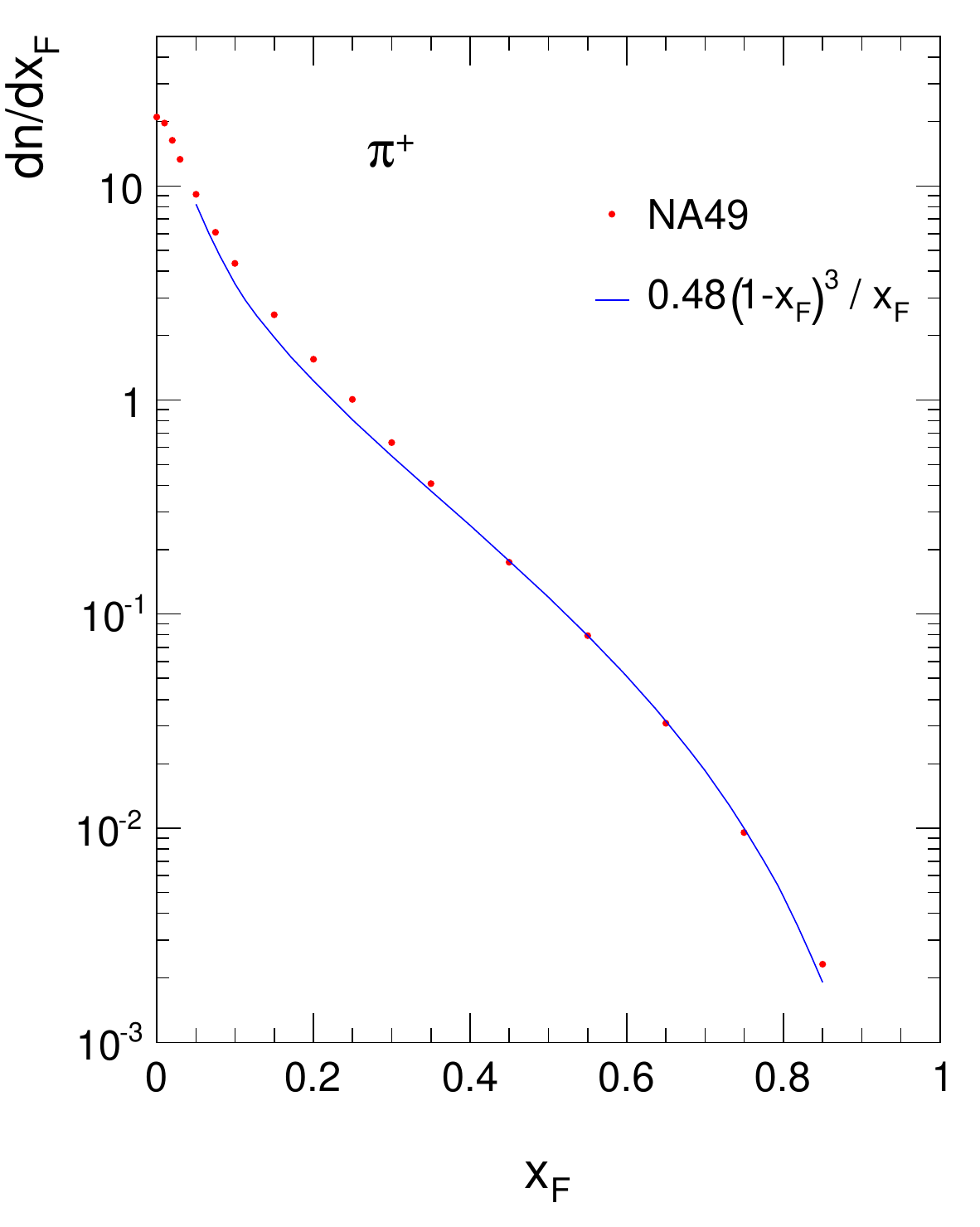} 
		\caption{$dn/dx_F$ of $\pi^+$ as a function of $x_F$ for the inclusive NA49 data. A fit to the data suggesting the up-quark structure function is superimposed}
		\label{fig:res_dndxf2up}
	\end{center}
\end{figure}

Several remarks should be made in this context:

\begin{enumerate}
	\item The distribution from resonance decay tracks the $\pi^-$ data rather precisely for $x_F >$~0.4 with a constant difference of about 25\%.
	\item Towards lower $x_F$ this difference increases to about 60\% at $x_F$~=~0. This is to be expected since the difference between data and resonance decay increases strongly for $x_F <$~0.2, Fig.~\ref{fig:res_perc01-11}.
	\item Both distributions reproduce the shape of the down-quark structure function where it is thought to apply at $x_F >$~0.35 as given by the function fitted to the data in Fig.~\ref{fig:res_dndxf2down}.
	\item The $\pi^+$ data of NA49 \cite{pp_pion} are presented in Fig.~\ref{fig:res_dndxf2up} including a fit corresponding to the up-quark structure function. Again a rather precise correspondence is visible beyond $x_F$~=~0.35.
	\item The interpretation of this shape similarity invoking parton dynamics is rather daring when realizing that the final state pions come -- after several cascading steps -- from parent resonances where each parent creates a  proper pionic $dn/dx_F$ distribution of different shape, Fig.~\ref{fig:res_dndxf_all}.
\end{enumerate}

These examples illustrate the necessity to follow inclusive phenomena beyond the lowest level of simplification thus opening up a new way towards the understanding of the underlying physics processes. Instead, the shape similarities have given rise to "recombination" models \cite{ochs,das} and to "quark counting rules" which were widely proposed and used in the 1970´s.

%
%
\subsection{The approach to "high" \boldmath $p_T$}
\vspace{3mm}
\label{sec:resonance_highpt}

Another example for the use of inclusive data for the introduction of parton dynamics in order to explain an observed phenomenology is the so-called high-$p_T$ sector. About 50 years ago it was observed at the CERN ISR collider that the $p_T$ distributions at central rapidity developed non-exponential tails above about 1~GeV/c which were in contradiction to the scaling hypotheses of the time. This has been discussed in Sect.~\ref{sec:scaling}, see for instance Fig.~\ref{fig:rrefy0}.

At the same time results from deep-inelastic lepton-proton scattering were used to predict the appearance of just this phenomenon at high transverse momentum by hard parton-parton scattering inside the colliding hadrons \cite{berman}. Partonic scattering amplitudes were established using the strong coupling constant to first order and the partonic structure functions available from lepton-proton experiment.

The term "high $p_T$" is rather ill-defined. As the early data were very limited in their $p_T$ range the tendency was to push the application of parton-parton scattering down to the very limit of momentum transfer compatible with perturbative QCD. This led to some problems with the reproduction of the observed yields at $p_T \lesssim$~2 GeV/c which were partially solved by the introduction of a parton transverse momentum $k_T$ \cite{feynman2,della}. Transverse momentum is however not a priory contained in the definition of partonic structure functions.

In a later stage higher-order QCD graphs including gluon radiation were introduced thus opening up a source of transverse momentum. This approach was used to compare to the data even down to $p_T \sim$~1.5~GeV/c.

\begin{figure}[h]
	\begin{center}
		\includegraphics[width=12cm] {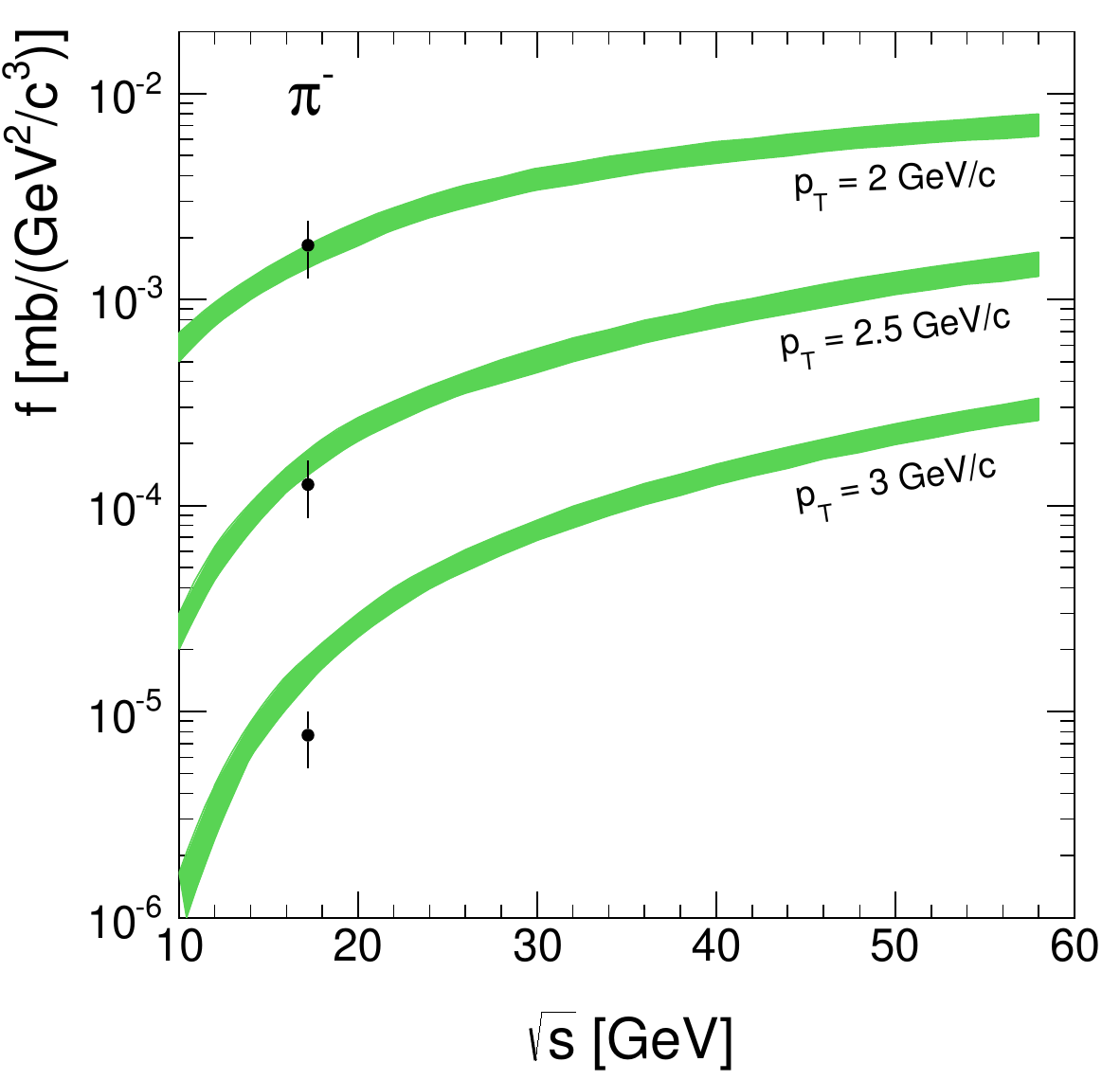} 
		\caption{Invariant $\pi^-$ cross sections at $x_F$~=~0.0 as a function of $\sqrt{s}$ indicated as bands with a width of $\pm$15\% at $p_T$~=~2.0, 2.5 and 3.0~GeV/c. The predicted yields from resonance decay are given as points at $\sqrt{s}$~=~17.2~GeV at the same $p_T$ values with estimated errors of 30\%}
		\label{fig:res_na49_pim}
	\end{center}
\end{figure}

In connection with the present discussion it is however absolutely mandatory to regard resonance decay as an eventual source of "high $p_T$" hadrons. The predicted $\pi^-$ yields from the two-body decay of the resonance sample given in Table~\ref{tab:resonance_list} have therefore been extended up to $p_T$ values of 3~GeV/c which should be safely contained in the regime of perturbative QCD. The corresponding cross sections (at $\sqrt{s}$~=~17.2~GeV) are visible in Figs.~\ref{fig:res_xf0-015} to \ref{fig:res_xf045-075}. As the data from NA49 at this energy are limited to $p_T <$~2~GeV/c, available data from the ISR and Serpukhov up to $p_T$~=~3~GeV/c have been combined in Fig.~\ref{fig:res_na49_pim}.

It is apparent from Fig.~\ref{fig:res_na49_pim} that the predicted cross sections are well within an error margin of $\pm$15\% from the measured yields for $p_T$~=~2 and 2.5~GeV/c and fall below by about 40\% at $p_T$~=~3~GeV/c. Here it should be remembered that the resonance sample of Table~\ref{tab:resonance_list} is not complete and is lacking high-mass baryons and mesons as well as heavy flavours which would contribute substantially, notwithstanding their low cross sections and small branching fractions, see the study of K$^-$ from D mesons in \cite{pp_kaon}.

In general the application of perturbative QCD for the prediction of inclusive processes has to be seen very critically. This concerns also one of the generally undoubted "successes" of QCD, the production of high mass lepton pairs via the Drell-Yan effect. It has been shown 30 years ago \cite{geist} that the production of dilepton states may be well described by semi-leptonic decays of charm-anticharm and beauty-antibeauty mesons, including characteristic inclusive quantities like mass and $s$-dependence, $x_F$ and helicity angle distributions as well as average transverse momentum.

%
%
\section{Conclusions}
\vspace{3mm}
\label{sec:conclusion}

The conclusions will follow the three main sections of this paper as specified in the introduction:

\begin{enumerate}
	\item Critical review of all available data and establishment of a general interpolation scheme including results from the high-energy proton colliders.
	\item Confronting the interpolation with various physics hypotheses both in longitudinal and transverse direction concerned with energy dependences and eventual manifestations of partonic effects.
	\item Extension of the study beyond the purely inclusive level by regarding resonance production and decay in its decisive influence on virtually all inclusive phenomena.
\end{enumerate}

%
%
\subsection{Data evaluation}
\vspace{3mm}
\label{sec:conc_data_eval}

In a first step the inclusiveness of the existing data sets with respect to the treatment of $\pi^-$ feed-down from weak decays is established by isolating two classes of experiments with and without feed-down correction. Corresponding corrections over the full phase space are worked out. In a second step a subgroup of experiments yielding internally consistent cross sections is defined as "reference data". In a third step a three-dimensional interpolation scheme in the co-ordinates interaction energy, longitudinal and transverse momentum is developed  using these data. In a forth step this interpolation is confronted with the remaining "spectrometer" data which are shown to essentially necessitate corrections connected to normalization problems. In this context data sets which fall far out of all other results in terms of systematical deviations are eliminated. A detailed statistical analysis shows that the interpolation has an unprecedented systematic precision on the 5\% level from $\sqrt{s} \sim$~3~GeV up to the highest ISR energy. The interpolation is made available both with and without feed-down correction in the coordinate triplets $y_{\textrm{lab}}$, $p_T$, $\log(s)$ and $x_F$, $p_T$ and $\log(s)$.

The situation at the high-energy colliders RHIC and LHC is less favourable as the phase-space coverage of the data is reduced to the very central or very forward directions and characterized by rather substantial discrepancies between experiments. Nevertheless an attempt is made to use these data up to $\sqrt{s}$~=~13~TeV at the LHC.

%
%
\subsection{Confronting the data with different physics hypotheses}
\vspace{3mm}
\label{sec:conc_phys_hypotheses}

Considering the wide range in interaction energy a first rather fundamental question is concerned with the data re-normalization. The inelastic cross section grows by about a factor of three from the lowest to the highest $\sqrt{s}$ available. Recent conjectures seem to indicate that this increase involves mostly the peripheral regions of the hadrons as opposed to a constant, central component. Therefore both the invariant cross section proper and the cross section divided by the inelastic cross sections are shown: due to the high precision of the interpolation, differences between the two normalizations become visible already at ISR energy. Some interesting effects up to the LHC range are exploited for $\pi^0$ production.

As far as longitudinal momentum dependencies are concerned the rather vague connection to parton dynamics by introducing the "scaling" variable $x_F$ does not yield satisfactory results when regarded over the complete $\sqrt{s}$ range and confronted with data of sufficient precision.

On the other hand the hypothesis of "limiting fragmentation" proposed at a time where very few and uncertain data at low energy were available, shows surprising predictivity even up to LHC energy. This approach has the advantage to contain built-in baryon number conservation and the definition of a cascading chain of high-mass excited states including proper treatment of the involved Lorentz-transformations.

On the level of transverse momentum there used to be a general idea of exponential damping corresponding to the early establishment of the "longitudinal phase space". A strong predictivity has been introduced via the notion of "statistical thermodynamics". This model postulates unified exponential distributions in "transverse mass" where the inverse slope defines a "temperature" that should be independent of interaction energy and particle mass and smaller than a limiting value. In connection with this a hadronic phase transition makes part of the thermodynamic treatment.

The in-depth study of the available data over the full energy range shows that this simplistic picture is not tenable. Being applicable essentially at central rapidity, the model misses as an essential part the strong increase of the mean transverse momentum as a function of $x_F$ known as "seagull effect" which is followed here with precision over the full range of interaction energies.

%
%
\subsection{Resonance decay as a first step beyond the single-particle inclusive level}
\vspace{3mm}
\label{sec:conc_res}

Whoever looks at inclusive data is met with clear indications of resonance decay contributing to the measured observables. This is put in this paper to a quantitative test by introducing a group of eight mesonic and five baryonic resonances whose production characteristics are sufficiently well known in order to study the impact of their decay on the establishment of the single-inclusive quantities studied in the first parts of this paper. The resonance masses range from $\eta^0$(550) to $f_4$(2050) for the mesons and from $\Delta$(1232) to $N^*$(1680) for the baryons. In order to avoid double counting only two-body decays are included.
As resonance production is relatively well measured in the energy range of the CERN SPS the study is carried out at $\sqrt{s}$~=~17.2~GeV corresponding to the NA49 experiment.

As a first result it turns out that 77\% of all inclusive $\pi^-$ are accounted for by resonance two or three-body ($\eta$ and $\omega$) decays -- which is a lower limit as the list of resonances is incomplete. This result is not new as the importance of resonance decay has been claimed decades ago. It is new to the point that both mesonic and baryonic resonances are involved and their decays traced in parallel.

In a first step the basic kinematic features of resonance decay are pointed out in particular concerning the resonance transverse momentum and -- for strong decays -- the resonance mass distribution. For this aim the well-measured $\Delta^{++}$(1232) is used to also bring out the effect of asymmetric masses.

In a second step the decay contribution to most of the inclusive distributions presented in this paper are quantified.

On the level of double-differential cross sections a surprisingly exact reproduction of the essential features of the $p_T$ and $x_F$ distributions is achieved. The deviations of up to 20\% from the data are studied in their $p_T$ and $x_F$ dependence and brought into relation to further expected decay features like cascading decays.

Concerning single-differential, integrated quantities it may be stated that their features are reproduced with surprising precision.This is true both for mean $p_T$ and its $x_F$ dependence and for $dn/dx_F$ as a function of $x_F$.

The study is extended to $p_T$ values up to 3~GeV/c at central rapidity. It is shown that resonance decay reproduces the measured cross sections also in this sector which has been ascribed to hard parton scattering as an agreed consensus invoking perturbative QCD over the past decades.

As a general consequence it becomes clear that the term of "soft hadronic physics" must be redefined as far as the application of parton dynamics and "perturbative QCD" is concerned.

%
%
\section{Future studies: experimental and accelerator constraints}
\vspace{3mm}
\label{sec:future_studies}

In this paper a considerable effort has been made to bring into perspective the experimental and phenomenological information concerning one single final state particle in soft proton-proton interactions and this by covering the full available range of collision energy, the complete production phase space and, of course, particle identification.

Looking back at this effort, the authors have to confess to a feeling of frustration. Not only would this study have to be repeated for the full range of secondary particles and types of collision concerning different hadron beams on both protons and neutron targets. In addition the nuclear sector should not be disregarded as it presents a vast laboratory for the study of multiple hadronic collisions. The actual status of the field gives rise to a number of questions:

\begin{itemize}
	\item How is it possible that non-perturbative QCD, as the by far biggest sector of the standard model, has been more or less completely abandoned by experiment as well as theory?
	\item How is it possible that in spite of the non-deniable progress made both in detector and in accelerator technology there are still experimental results produced which do not come up in quality and precision to work done five or six decades ago?
	\item How is it possible that the problem of parton dynamics and its purported contribution to final state inclusive results is still treated as a non-touchable certainty where experimental results, when obtained with proper precision and reliability, point just to the opposite?
	\item How is it possible that -- extending the range of hadronic interactions to the nuclear sector -- a possible phase transition to a QCD plasma state is considered as existing when no indisputable experimental proof has ever been put forward?
	\item Why no real attempts have been made to have a look beyond the purely inclusive level in order to open up another source of quantitative understanding underlying the observed phenomena on the single inclusive level?
\end{itemize}

For the time being it is apparent that it is on the experimental level that further progress is relying as the actual ideas based on statistical thermodynamics on the one hand and parton dynamics on the other hand seem to be so deep rooted that no real movement towards putting the one or the other to a decisive test are visible. This in a situation where progress on the experimental side should allow a decisive improvement if certain constraints both concerning detector design and construction as well as accelerator systems are properly taken into account. Part of these constraints have hopefully become apparent in the present paper.

\begin{enumerate}
	\item Phase space coverage \\
New detector systems should be designed such that the full production phase space is covered uniformly from the far backward to the far forward direction with precision tracking. It has been a mistake to concentrate since decades more or less exclusively on the very central part of the collision.
	\item Particle identification \\
Particle identification should cover the full phase space including calorimetry for photon/electron and neutral hadron detection.
	\item Systematic uncertainties \\
The systematic uncertainties should be on the percent level. This is the real problem -- as has become apparent in the above comparison of the "classic" experiments using bubble chambers with more or less all "modern" approaches using different kinds of spectrometer layouts or so-called "detector facilities".
	\item Versatility of the strong interaction \\
The tremendous versatility of the strong interaction in terms of initial state conditions defining the collision itself as well as the choice of secondary particles should be fully exploited.
	\item Order of approach \\
Research should start with the so-called "elementary" collisions using the full spectrum of possible beam particles on proton targets. This should also contain lepton beams which are sometimes called "noble probes" as compared to "dirty hadrons". A definite high precision comparison is still lacking.
	\item The nuclear sector \\
This sector should be worked up successively from light to heavy nuclei. Here the deuteron offers a (non-trivial) "neutron" target. Again all kinds of projectile hadrons should be used.
	\item Impact parameter \\
Hadron-nucleus collision should be studied with a precise impact parameter measurement defining the number of intranuclear projectile collisions which gives access to a new sector of hadronic physics.
	\item For all the above points, full attention should be given to the precision measurement of hadronic resonances.
	\item Nucleus-nucleus collisions \\
The interest here should not be exclusively concentrated on "Heavy Ion" collisions. Full attention should also be paid to peripheral interactions.
\end{enumerate}

This list is not aimed at "discovery". In the past too much attention has been paid to the perturbative sector concentrating the attention upon more and more remote sectors hence necessitating extreme accelerator luminosities whereas the soft regime is accessible with convenient interaction rates.

From the above (non-complete) list of experimental constraints and tasks several consequences follow immediately:

\begin{enumerate}[label=(\roman*)]
	\item Proton colliders cannot fulfil the wide spectrum of possibilities, neither the homogeneous coverage of production phase space nor the versatility of projectile/target combinations.
	\item Fixed-target operation is on the contrary well suited to cover the complete range of tasks. For a first generation of experiments, the interaction energy may be kept conveniently low in order to cope with high precision tracking and the necessities of complete particle identification. \newline
\end{enumerate}

In view of this the CERN SPS complex offers an available, at present under-exploited, environment.

\appendix

\section{Availability of numerical information}
\vspace{3mm}
\label{sec:appendix}

The NA49 working group on proton-proton and proton-nucleus interactions has created a web-page
  \href{https://spshadrons.web.cern.ch}{https://spshadrons.web.cern.ch}, where salient numerical results as well as documents may be addressed. For the present paper a number of quantitative results may be found under the header \href{https://spshadrons.web.cern.ch/pp_piminus.html}{$\pi^-$ in p+p} . Here the following quantities are listed:

\begin{itemize}
	\item Invariant cross sections normalized to the inelastic cross section concerning the global interpolation in the three dimensions $p_T$, $\log{s}$) and $y_{\textrm{lab}}$ with and without feed-down in both csv and text files, $p_T$, $\log{s}$ and $x_F$ with and without feed-down, $p_T$, $\log{s}$ and cms rapidity with and without feed-down. Each data set covers the binning scheme with about 10$^4$ entries.
	\item $p_T$ integrated yields as functions of $\log{s}$ and $x_F$ with and without feed-down, $p_T$ integrated yields as functions of $\log{s}$ and cms rapidity
	\item Total $\pi^-$ yields as a function of $\log{s}$ with and without feed-down.
\end{itemize}

Similar detailed information may be found on this web-page for the preceding papers as well:

\begin{tabular}{llll}
	p+p & $\rightarrow$ &  \cite{pp_pion,pp_kaon,pp_proton}    &   \href{https://spshadrons.web.cern.ch/ppdata.html}{p+p data}  \\
	p+C & $\rightarrow$ &  \cite{pc_paper,pc_discuss,pc_proton}   &   \href{https://spshadrons.web.cern.ch/pCdata.html}{p+C data}  \\
	p+p & $\rightarrow$ & \cite{pp_kaon} &   \href{https://spshadrons.web.cern.ch/pp_kaon_sdep.html}{kaon in p+p -- $s$-dependence}  \\
\end{tabular}

\end{document}

%% file: table.tex
\begin{tabular}{|cccccccccccccc|}
\hline
\multicolumn{14}{|c|}{$f/\sigma_{\textrm{inel}}(y_{\textrm{lab}},p_T,\log(s)) \quad \log(s)$ =  1.0} \\ \hline
$p_T \backslash y_{\textrm{lab}}$ & -1.2 & -1.0 & -0.8 & -0.6 & -0.4 & -0.2 &  0.0 &  0.2 &  0.4 &  0.6 &  0.8 &  1.0 &  1.2\\ \hline
0.05 &  0.03349 &  0.07367 &  0.11890 &  0.15890 &  0.19710 &  0.23420 &  0.26550 &  0.29560 &  0.32360 &  0.34560 &  0.36470 &  0.37980 &  0.38100 \\ 0.10 &  0.01242 &  0.04270 &  0.08365 &  0.12910 &  0.17960 &  0.21890 &  0.25080 &  0.27930 &  0.30420 &  0.32580 &  0.34330 &  0.35580 &  0.35680 \\ 0.20 &  &  &  0.01725 &  0.04508 &  0.07851 &  0.11330 &  0.14640 &  0.17760 &  0.19930 &  0.22280 &  0.24430 &  0.25980 &  0.25830 \\ 0.30 &  &  &  &  &  0.02093 &  0.04209 &  0.06197 &  0.08002 &  0.09579 &  0.11080 &  0.12800 &  0.14160 &  0.13940 \\ 0.40 &  &  &  &  &  &  0.01028 &  0.02200 &  0.03236 &  0.04242 &  0.05212 &  0.06182 &  0.06958 &  0.06907 \\ 0.50 &  &  &  &  &  &  &  0.00515 &  0.01102 &  0.01810 &  0.02453 &  0.03081 &  0.03505 &  0.03535 \\ 0.60 &  &  &  &  &  &  &  &  0.00247 &  0.00636 &  0.01042 &  0.01459 &  0.01715 &  0.01754 \\ 0.70 &  &  &  &  &  &  &  &  &  0.00118 &  0.00337 &  0.00611 &  0.00776 &  0.00797 \\ 0.80 &  &  &  &  &  &  &  &  &  &  &  &  0.00326 &  0.00333 \\ 0.90 &  &  &  &  &  &  &  &  &  &  &  &  0.00125 &  0.00125 \\ \hline
\end{tabular}
\begin{tabular}{|ccccccccccccccccc|}
\hline
\multicolumn{17}{|c|}{$f/\sigma_{\textrm{inel}}(y_{\textrm{lab}},p_T,\log(s)) \quad \log(s)$ =  1.5} \\ \hline
$p_T \backslash y_{\textrm{lab}}$ & -1.2 & -1.0 & -0.8 & -0.6 & -0.4 & -0.2 &  0.0 &  0.2 &  0.4 &  0.6 &  0.8 &  1.0 &  1.2 &  1.4 &  1.6 &  1.8\\ \hline
0.05 &  0.02732 &  0.05975 &  0.10440 &  0.15040 &  0.20400 &  0.25810 &  0.31970 &  0.38890 &  0.46200 &  0.55500 &  0.65820 &  0.74760 &  0.83060 &  0.91280 &  0.97260 &  0.98160 \\ 0.10 &  0.01218 &  0.03302 &  0.06601 &  0.11010 &  0.16290 &  0.21830 &  0.27940 &  0.34990 &  0.42880 &  0.51380 &  0.60720 &  0.70320 &  0.79380 &  0.85860 &  0.89820 &  0.90650 \\ 0.20 &  &  0.00341 &  0.01433 &  0.03452 &  0.06270 &  0.10280 &  0.15750 &  0.21920 &  0.28920 &  0.36760 &  0.44220 &  0.52670 &  0.58600 &  0.63650 &  0.66490 &  0.66660 \\ 0.30 &  &  &  0.00089 &  0.00576 &  0.01726 &  0.03502 &  0.06059 &  0.09403 &  0.13700 &  0.18920 &  0.24310 &  0.29400 &  0.32980 &  0.35740 &  0.37160 &  0.37440 \\ 0.40 &  &  &  &  &  0.00302 &  0.00992 &  0.02108 &  0.03772 &  0.06062 &  0.08539 &  0.11500 &  0.14650 &  0.17440 &  0.19370 &  0.20190 &  0.20350 \\ 0.50 &  &  &  &  &  &  0.00184 &  0.00678 &  0.01529 &  0.02629 &  0.04024 &  0.05563 &  0.07214 &  0.08815 &  0.10130 &  0.10860 &  0.11010 \\ 0.60 &  &  &  &  &  &  &  0.00149 &  0.00502 &  0.01036 &  0.01785 &  0.02656 &  0.03610 &  0.04514 &  0.05322 &  0.05825 &  0.05882 \\ 0.70 &  &  &  &  &  &  &  &  0.00119 &  0.00379 &  0.00747 &  0.01228 &  0.01788 &  0.02285 &  0.02668 &  0.02950 &  0.02992 \\ 0.80 &  &  &  &  &  &  &  &  &  0.00100 &  0.00278 &  0.00554 &  0.00881 &  0.01140 &  0.01370 &  0.01511 &  0.01524 \\ 0.90 &  &  &  &  &  &  &  &  &  &  0.00098 &  0.00236 &  0.00407 &  0.00566 &  0.00690 &  0.00759 &  0.00764 \\ 1.00 &  &  &  &  &  &  &  &  &  &  &  0.00087 &  0.00179 &  0.00265 &  0.00330 &  0.00362 &  0.00366 \\ 1.10 &  &  &  &  &  &  &  &  &  &  &  0.00024 &  0.00070 &  0.00115 &  0.00154 &  0.00165 &  0.00166 \\ 1.20 &  &  &  &  &  &  &  &  &  &  &  &  0.00021 &  0.00044 &  0.00061 &  &  \\ 1.30 &  &  &  &  &  &  &  &  &  &  &  &  0.00005 &  0.00012 &  0.00021 &  &  \\ \hline
\end{tabular}
\begin{tabular}{|ccccccccccccccccccc|}
\hline
\multicolumn{19}{|c|}{$f/\sigma_{\textrm{inel}}(y_{\textrm{lab}},p_T,\log(s)) \quad \log(s)$ =  2.0} \\ \hline
$p_T \backslash y_{\textrm{lab}}$ & -1.2 & -1.0 & -0.8 & -0.6 & -0.4 & -0.2 &  0.0 &  0.2 &  0.4 &  0.6 &  0.8 &  1.0 &  1.2 &  1.4 &  1.6 &  1.8 &  2.0 &  2.4\\ \hline
0.05 &  0.02507 &  0.05571 &  0.10000 &  0.14520 &  0.19950 &  0.25790 &  0.32930 &  0.40340 &  0.48950 &  0.59880 &  0.70070 &  0.82190 &  0.95470 &  1.06500 &  1.17100 &  1.27200 &  1.37300 &  1.42000 \\ 0.10 &  0.01089 &  0.03109 &  0.06214 &  0.10520 &  0.15740 &  0.21270 &  0.28040 &  0.36270 &  0.44580 &  0.54080 &  0.65390 &  0.75980 &  0.87540 &  0.98240 &  1.08800 &  1.16800 &  1.24500 &  1.29000 \\ 0.20 &  &  0.00340 &  0.01400 &  0.03200 &  0.05922 &  0.09811 &  0.15210 &  0.21290 &  0.28240 &  0.36380 &  0.45430 &  0.55460 &  0.64980 &  0.72680 &  0.80020 &  0.86730 &  0.91400 &  0.94680 \\ 0.30 &  &  &  0.00099 &  0.00560 &  0.01600 &  0.03312 &  0.05790 &  0.09193 &  0.13620 &  0.18940 &  0.24460 &  0.30700 &  0.37470 &  0.42940 &  0.47590 &  0.51450 &  0.54550 &  0.56970 \\ 0.40 &  &  &  &  0.00042 &  0.00278 &  0.00939 &  0.02039 &  0.03690 &  0.06102 &  0.09025 &  0.12150 &  0.15750 &  0.19450 &  0.22650 &  0.25860 &  0.28280 &  0.29970 &  0.31540 \\ 0.50 &  &  &  &  &  &  0.00185 &  0.00638 &  0.01462 &  0.02581 &  0.04194 &  0.05979 &  0.08095 &  0.10040 &  0.12000 &  0.13800 &  0.14940 &  0.15980 &  0.16650 \\ 0.60 &  &  &  &  &  &  &  0.00141 &  0.00502 &  0.01110 &  0.01974 &  0.03047 &  0.04102 &  0.05326 &  0.06376 &  0.07303 &  0.08017 &  0.08591 &  0.08973 \\ 0.70 &  &  &  &  &  &  &  0.00022 &  0.00131 &  0.00435 &  0.00897 &  0.01435 &  0.02049 &  0.02697 &  0.03287 &  0.03861 &  0.04259 &  0.04646 &  0.04941 \\ 0.80 &  &  &  &  &  &  &  &  0.00030 &  0.00143 &  0.00357 &  0.00670 &  0.00998 &  0.01377 &  0.01746 &  0.02074 &  0.02340 &  0.02537 &  0.02664 \\ 0.90 &  &  &  &  &  &  &  &  &  0.00042 &  0.00139 &  0.00288 &  0.00488 &  0.00708 &  0.00923 &  0.01095 &  0.01256 &  0.01394 &  0.01479 \\ 1.00 &  &  &  &  &  &  &  &  &  &  0.00049 &  0.00116 &  0.00221 &  0.00358 &  0.00478 &  0.00585 &  0.00684 &  0.00753 &  0.00795 \\ 1.10 &  &  &  &  &  &  &  &  &  &  &  0.00044 &  0.00097 &  0.00176 &  0.00252 &  0.00312 &  0.00357 &  0.00390 &  0.00416 \\ 1.20 &  &  &  &  &  &  &  &  &  &  &  0.00014 &  0.00038 &  0.00080 &  0.00125 &  0.00162 &  0.00190 &  0.00208 &  0.00222 \\ 1.30 &  &  &  &  &  &  &  &  &  &  &  &  0.00014 &  0.00037 &  0.00062 &  0.00084 &  0.00101 &  0.00110 &  0.00116 \\ \hline
\end{tabular}
\begin{tabular}{|ccccccccccccccccccc|}
\hline
\multicolumn{19}{|c|}{$f/\sigma_{\textrm{inel}}(y_{\textrm{lab}},p_T,\log(s)) \quad \log(s)$ =  2.5} \\ \hline
$p_T \backslash y_{\textrm{lab}}$ & -1.2 & -1.0 & -0.8 & -0.6 & -0.4 & -0.2 &  0.0 &  0.2 &  0.4 &  0.6 &  0.8 &  1.0 &  1.2 &  1.4 &  1.8 &  2.2 &  2.6 &  3.0\\ \hline
0.05 &  0.02340 &  0.05356 &  0.09602 &  0.14954 &  0.20160 &  0.25475 &  0.32903 &  0.42095 &  0.52039 &  0.63047 &  0.76366 &  0.90364 &  1.05511 &  1.21235 &  1.49540 &  1.70965 &  1.84214 &  1.91000 \\ 0.10 &  0.01013 &  0.03012 &  0.06239 &  0.10743 &  0.15923 &  0.21339 &  0.27721 &  0.35730 &  0.45134 &  0.54595 &  0.67505 &  0.81159 &  0.95224 &  1.10103 &  1.37055 &  1.57251 &  1.70480 &  1.70200 \\ 0.20 &  0.00035 &  0.00326 &  0.01313 &  0.03049 &  0.05681 &  0.09321 &  0.14297 &  0.19888 &  0.27264 &  0.35791 &  0.44786 &  0.55074 &  0.65594 &  0.75658 &  0.93829 &  1.08048 &  1.18114 &  1.18700 \\ 0.30 &  &  &  0.00078 &  0.00521 &  0.01613 &  0.03125 &  0.05381 &  0.08929 &  0.13156 &  0.18262 &  0.24195 &  0.30393 &  0.37501 &  0.43505 &  0.53631 &  0.62194 &  0.68609 &  0.68090 \\ 0.40 &  &  &  &  0.00039 &  0.00262 &  0.00857 &  0.01995 &  0.03603 &  0.06013 &  0.08908 &  0.12196 &  0.16021 &  0.19710 &  0.23308 &  0.29351 &  0.33643 &  0.37352 &  0.37940 \\ 0.50 &  &  &  &  &  0.00022 &  0.00150 &  0.00599 &  0.01368 &  0.02549 &  0.04243 &  0.06202 &  0.08395 &  0.10292 &  0.12236 &  0.15561 &  0.18043 &  0.19839 &  0.20190 \\ 0.60 &  &  &  &  &  &  &  0.00131 &  0.00474 &  0.01078 &  0.02034 &  0.03175 &  0.04294 &  0.05498 &  0.06548 &  0.08341 &  0.09753 &  0.10805 &  0.10980 \\ 0.70 &  &  &  &  &  &  &  0.00019 &  0.00126 &  0.00404 &  0.00888 &  0.01510 &  0.02200 &  0.02850 &  0.03415 &  0.04500 &  0.05308 &  0.05952 &  0.06120 \\ 0.80 &  &  &  &  &  &  &  &  0.00031 &  0.00145 &  0.00363 &  0.00679 &  0.01074 &  0.01448 &  0.01799 &  0.02419 &  0.02958 &  0.03410 &  0.03407 \\ 0.90 &  &  &  &  &  &  &  &  &  0.00044 &  0.00137 &  0.00310 &  0.00531 &  0.00748 &  0.00930 &  0.01294 &  0.01616 &  0.01884 &  0.01941 \\ 1.00 &  &  &  &  &  &  &  &  &  &  0.00055 &  0.00135 &  0.00248 &  0.00379 &  0.00488 &  0.00704 &  0.00900 &  0.01055 &  0.01102 \\ 1.10 &  &  &  &  &  &  &  &  &  &  &  0.00055 &  0.00112 &  0.00180 &  0.00254 &  0.00399 &  0.00540 &  0.00634 &  0.00648 \\ 1.20 &  &  &  &  &  &  &  &  &  &  &  0.00021 &  0.00052 &  0.00090 &  0.00129 &  0.00219 &  0.00301 &  0.00360 &  0.00377 \\ 1.30 &  &  &  &  &  &  &  &  &  &  &  &  0.00022 &  0.00044 &  0.00069 &  0.00123 &  0.00171 &  0.00210 &  0.00216 \\ \hline
\end{tabular}
\begin{tabular}{|ccccccccccccccccccc|}
\hline
\multicolumn{19}{|c|}{$f/\sigma_{\textrm{inel}}(y_{\textrm{lab}},p_T,\log(s)) \quad \log(s)$ =  3.0} \\ \hline
$p_T \backslash y_{\textrm{lab}}$ & -0.8 & -0.6 & -0.4 & -0.2 &  0.0 &  0.2 &  0.4 &  0.6 &  0.8 &  1.0 &  1.2 &  1.4 &  1.6 &  2.0 &  2.4 &  2.8 &  3.2 &  3.6\\ \hline
0.05 &  0.09517 &  0.13520 &  0.18520 &  0.24430 &  0.31840 &  0.42150 &  0.55970 &  0.73880 &  0.92600 &  1.16000 &  1.38300 &  1.58400 &  1.80100 &  2.11200 &  2.34000 &  2.47200 &  2.53000 &  2.55100 \\ 0.10 &  0.05915 &  0.09508 &  0.14220 &  0.19930 &  0.27140 &  0.35940 &  0.46860 &  0.61570 &  0.77780 &  0.97190 &  1.17200 &  1.35800 &  1.55900 &  1.82800 &  1.98000 &  2.05600 &  2.07500 &  2.07500 \\ 0.20 &  0.01306 &  0.02949 &  0.05473 &  0.08868 &  0.13720 &  0.19730 &  0.27140 &  0.36240 &  0.46250 &  0.59160 &  0.72360 &  0.85760 &  0.97050 &  1.17200 &  1.27000 &  1.30000 &  1.30400 &  1.30700 \\ 0.30 &  0.00085 &  0.00519 &  0.01485 &  0.02978 &  0.05262 &  0.08566 &  0.12720 &  0.18480 &  0.24630 &  0.31490 &  0.39030 &  0.46430 &  0.54030 &  0.65520 &  0.71010 &  0.73000 &  0.73000 &  0.73000 \\ 0.40 &  &  0.00037 &  0.00257 &  0.00829 &  0.01875 &  0.03419 &  0.05761 &  0.08574 &  0.12060 &  0.16020 &  0.20120 &  0.24360 &  0.28210 &  0.34150 &  0.37710 &  0.39700 &  0.39700 &  0.39890 \\ 0.50 &  &  &  0.00021 &  0.00156 &  0.00568 &  0.01314 &  0.02436 &  0.03957 &  0.05809 &  0.08031 &  0.10410 &  0.12790 &  0.14910 &  0.18000 &  0.20100 &  0.21050 &  0.21000 &  0.21050 \\ 0.60 &  &  &  &  &  0.00124 &  0.00450 &  0.01013 &  0.01864 &  0.02885 &  0.04066 &  0.05375 &  0.06505 &  0.07750 &  0.09664 &  0.10830 &  0.11230 &  0.11310 &  0.11310 \\ 0.70 &  &  &  &  &  0.00019 &  0.00124 &  0.00396 &  0.00820 &  0.01388 &  0.02043 &  0.02703 &  0.03363 &  0.04003 &  0.05102 &  0.05953 &  0.06340 &  0.06350 &  0.06350 \\ 0.80 &  &  &  &  &  &  0.00030 &  0.00140 &  0.00343 &  0.00636 &  0.00991 &  0.01372 &  0.01762 &  0.02101 &  0.02771 &  0.03256 &  0.03505 &  0.03550 &  0.03550 \\ 0.90 &  &  &  &  &  &  &  0.00042 &  0.00133 &  0.00289 &  0.00487 &  0.00689 &  0.00891 &  0.01088 &  0.01481 &  0.01798 &  0.01990 &  0.02060 &  0.02060 \\ 1.00 &  &  &  &  &  &  &  &  0.00050 &  0.00124 &  0.00231 &  0.00342 &  0.00456 &  0.00575 &  0.00810 &  0.01003 &  0.01138 &  0.01180 &  0.01180 \\ 1.10 &  &  &  &  &  &  &  &  &  0.00052 &  0.00104 &  0.00167 &  0.00231 &  0.00306 &  0.00450 &  0.00577 &  0.00670 &  0.00703 &  0.00710 \\ 1.20 &  &  &  &  &  &  &  &  &  0.00019 &  0.00048 &  0.00082 &  0.00118 &  0.00159 &  0.00244 &  0.00324 &  0.00381 &  0.00410 &  0.00417 \\ 1.30 &  &  &  &  &  &  &  &  &  &  0.00021 &  0.00040 &  0.00061 &  0.00084 &  0.00134 &  0.00187 &  0.00227 &  0.00246 &  0.00247 \\ \hline
\end{tabular}
\begin{tabular}{|ccccccccccccccccccc|}
\hline
\multicolumn{19}{|c|}{$f/\sigma_{\textrm{inel}}(y_{\textrm{lab}},p_T,\log(s)) \quad \log(s)$ =  3.6} \\ \hline
$p_T \backslash y_{\textrm{lab}}$ & -0.8 & -0.6 & -0.4 & -0.2 &  0.0 &  0.2 &  0.4 &  0.6 &  0.8 &  1.0 &  1.4 &  1.8 &  2.2 &  2.6 &  3.0 &  3.4 &  3.8 &  4.2\\ \hline
0.05 &  0.08903 &  0.11700 &  0.15600 &  0.21010 &  0.29410 &  0.42120 &  0.61020 &  0.85530 &  1.12500 &  1.41000 &  2.11000 &  2.70000 &  3.10000 &  3.30000 &  3.33000 &  3.33000 &  3.33000 &  3.33000 \\ 0.10 &  0.05403 &  0.08203 &  0.12010 &  0.17610 &  0.25410 &  0.35510 &  0.50020 &  0.69520 &  0.90230 &  1.17600 &  1.75000 &  2.19000 &  2.48100 &  2.61000 &  2.63000 &  2.63000 &  2.63000 &  2.63000 \\ 0.20 &  0.01241 &  0.02702 &  0.04983 &  0.08104 &  0.12210 &  0.18680 &  0.27010 &  0.37010 &  0.49420 &  0.64420 &  0.96020 &  1.25700 &  1.45000 &  1.52000 &  1.53000 &  1.53000 &  1.53000 &  1.53000 \\ 0.30 &  0.00081 &  0.00486 &  0.01421 &  0.02772 &  0.04703 &  0.07864 &  0.12010 &  0.18010 &  0.24610 &  0.32310 &  0.51110 &  0.67210 &  0.76800 &  0.82200 &  0.83000 &  0.83000 &  0.83000 &  0.83000 \\ 0.40 &  &  0.00034 &  0.00238 &  0.00775 &  0.01681 &  0.03152 &  0.05303 &  0.08224 &  0.11800 &  0.16000 &  0.25510 &  0.33900 &  0.39600 &  0.42400 &  0.43000 &  0.43000 &  0.43000 &  0.43000 \\ 0.50 &  &  &  0.00019 &  0.00144 &  0.00524 &  0.01191 &  0.02201 &  0.03602 &  0.05462 &  0.07703 &  0.13040 &  0.17600 &  0.20400 &  0.21470 &  0.21800 &  0.21700 &  0.21700 &  0.21700 \\ 0.60 &  &  &  &  &  0.00116 &  0.00404 &  0.00921 &  0.01661 &  0.02551 &  0.03802 &  0.06501 &  0.09151 &  0.10700 &  0.11450 &  0.11800 &  0.11800 &  0.11800 &  0.11800 \\ 0.70 &  &  &  &  &  0.00017 &  0.00114 &  0.00361 &  0.00730 &  0.01201 &  0.01821 &  0.03311 &  0.04671 &  0.05700 &  0.06300 &  0.06540 &  0.06550 &  0.06550 &  0.06550 \\ 0.80 &  &  &  &  &  &  0.00028 &  0.00130 &  0.00319 &  0.00570 &  0.00876 &  0.01600 &  0.02420 &  0.03030 &  0.03460 &  0.03710 &  0.03850 &  0.03850 &  0.03850 \\ 0.90 &  &  &  &  &  &  &  0.00038 &  0.00126 &  0.00258 &  0.00416 &  0.00803 &  0.01240 &  0.01650 &  0.01940 &  0.02130 &  0.02170 &  0.02170 &  0.02170 \\ 1.00 &  &  &  &  &  &  &  &  0.00046 &  0.00110 &  0.00196 &  0.00402 &  0.00650 &  0.00880 &  0.01080 &  0.01230 &  0.01270 &  0.01280 &  0.01280 \\ 1.10 &  &  &  &  &  &  &  &  &  0.00045 &  0.00089 &  0.00204 &  0.00336 &  0.00471 &  0.00607 &  0.00707 &  0.00767 &  0.00777 &  0.00777 \\ 1.20 &  &  &  &  &  &  &  &  &  0.00017 &  0.00040 &  0.00102 &  0.00176 &  0.00256 &  0.00340 &  0.00409 &  0.00445 &  0.00454 &  0.00454 \\ 1.30 &  &  &  &  &  &  &  &  &  &  0.00018 &  0.00052 &  0.00092 &  0.00139 &  0.00193 &  0.00235 &  0.00255 &  0.00260 &  0.00260 \\ \hline
\end{tabular}